%% file: TesisArxiv.tex
\documentclass[a4paper, twoside]{thesis} 

\usepackage{a4}
\usepackage{amsmath}
\usepackage{amsfonts}
\usepackage{amssymb}
\usepackage{theorem}
\usepackage[dvips]{color}
\usepackage{bm}
\usepackage{graphicx}

\begin{document}

\frontmatter
\input{TitlePageFO}

\setcounter{secnumdepth}{-1}
\chapter{$Agradecimientos\hspace{4 pt}_{[Acknowledgements]}$}
\input{AgraFO}
\setcounter{secnumdepth}{2}

\tableofcontents

\mainmatter

\pagebreak

\chapter{Introduction}
\input{IntroFO}

\chapter{The quantum harmonic oscillator}
\label{HarmonicOscillator}
\input{HarmonicOscillatorFO}

\chapter[Quantization of the electromagnetic field in an optical cavity]{Quantization of the electromagnetic field \\in an optical cavity}
\label{Quantization}
\input{QuantizationFO}

\chapter{Quantum theory of open cavities}
\label{OpenSystems}
\input{OpenSystemsFO}

\chapter{Detection of the output field}
\label{Detection}
\input{DetectionFO}

\chapter[Quantum description and basic properties of OPOs]{Quantum description and basic properties \\of optical parametric oscillators}
\label{OPOs}
\input{OPOsFO}

\chapter[Basic phenomena in multi--mode OPOs]{Basic phenomena in multi--mode Optical Parametric Oscillators}
\label{MultiOPOs}
\input{MultiOPOsFO}

\chapter[Deep study of spontaneous symmetry breaking through the 2tmDOPO]{Deep study of spontaneous symmetry breaking \\through the two--transverse--mode DOPO}
\label{2tmDOPO}
\input{2tmDOPOFO}

\chapter[The 2tmDOPO with injected signal]{The two--transverse--mode DOPO with injected signal}
\label{2tmDOPOwithIS}
\input{2tmDOPOwithISFO}

\chapter{Type II OPO: Polarization symmetry breaking \\and frequency degeneracy}
\label{TypeIIOPO}
\input{TypeIIOPOsFO}

\chapter{DOPOs tuned to arbitrary transverse families}
\label{FamiliesDOPO}
\input{DOPOfamilyFO}

\chapter{Conclusions and outlook}
\label{Conclusions}
\input{ConclusionsFO}

\appendix

\chapter{Quantum description of physical systems}
\label{QuantumMechanics}
\input{QuantumMechanicsFO}

\chapter[Linear stochastic equations with additive noise]{Linear stochastic equations with additive noise}
\label{LinStoApp}
\input{LinearStochasticFO}

\chapter[Linearization of the 2tmDOPO Langevin equations]{Linearization of the two--transverse--mode DOPO Langevin equations}
\label{Lin2tmDOPO}
\input{Lin2tmDOPOFO}

\chapter{Correlation functions of $\cos\theta(\tau)$ and $\sin\theta(\tau)$}
\label{SinCosCorr}
\input{SinCosCorrFO}

\chapter[Details about the numerical simulation of the 2tmDOPO equations]{Details about the numerical simulation of the two--transverse--mode DOPO equations}
\label{Numerical2tmDOPO}
\input{Numerical2tmDOPOFO}


\bibliographystyle{IEEEtran}
\addcontentsline{toc}{chapter}{Bibliography}
\bibliography{Carl}

\end{document}

%% file: TitlePageFO.tex
\thispagestyle{empty}

\begin{center}


  \begin{figure}[h!]
  \center
  \includegraphics[width=12.21cm]{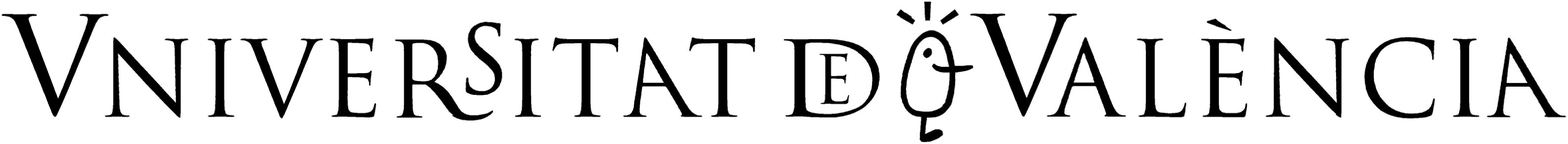}	
  \end{figure}
  
  \vspace{0.5 cm}
  
  {\bf \large Departament d'\`Optica, Facultat de F\'isica}
  
  \vspace{0.6 cm}
  
  \begin{figure}[h!]
  \center
  \includegraphics[width=3.96cm]{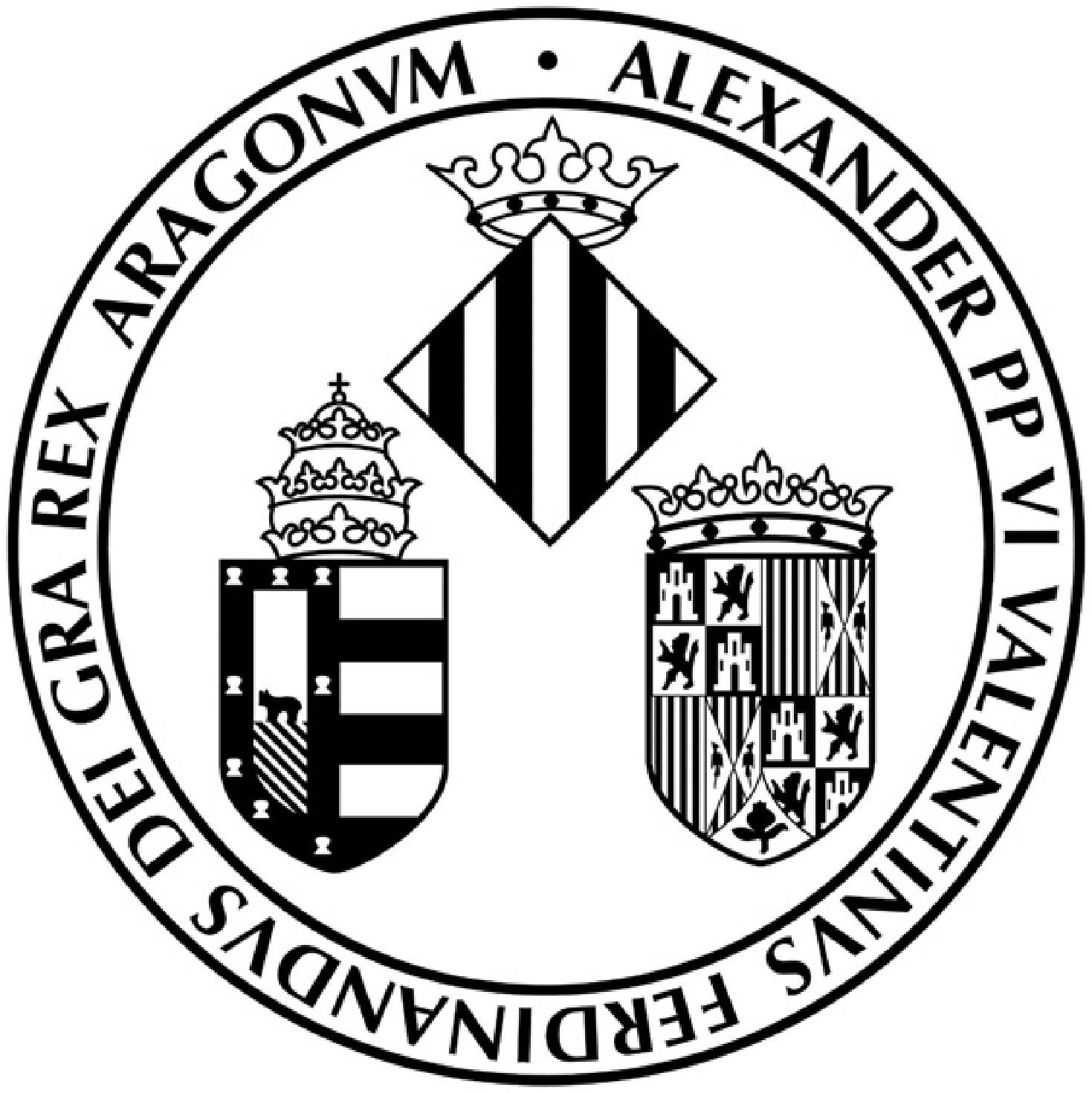}	
  \end{figure}

  \vspace{1.5 cm}

  {\bf \huge Contributions}\\
  \vspace{0.3 cm}
  {\bf \huge to the Quantum Optics of}\\
  \vspace{0.3 cm}
  {\bf \huge Multi--mode Optical Parametric Oscillators}\\
  \vspace{1.3 cm}

\end{center}

\begin{flushleft}

  \hspace{1.5cm}{\bf \large PhD dissertation by}\\
  \vspace{0.4 cm}

  \hspace{1.5cm}{\bf \Large Carlos Navarrete--Benlloch}\\
  \vspace{2 cm}

\end{flushleft}

\begin{flushright}

  {\bf \large Under the supervision of}\\
  \vspace{0.4 cm}
  {\bf \Large Eugenio Rold\'an Serrano}\\
  \vspace{0.2 cm}
  {\bf \large and}\\
  \vspace{0.2 cm}
  {\bf \Large Germ\'an J. de Valc\'arcel Gonzalvo}\\
  \vspace{3.1 cm}

\end{flushright}

\begin{center}

  {\bf \Large Valencia, Spain; December, 2011}

\end{center}


\newpage
\thispagestyle{empty}
\vspace*{18cm}

%% file: AgraFO.tex
Escribir los agradecimientos de esta tesis genera en m\'{\i} una doble
sensaci\'{o}n de\textit{ }felicidad. El primer motivo es obvio: ello significa
que el final de este largo (pero bonito) camino que ha supuesto redactar tan
pensado documento est\'{a} llegando a su fin, lo cual no es poco, ya que meses
atr\'{a}s me parec\'{\i}a que el camino era infinito (tras escribir cada
apartado, siempre ten\'{\i}a la sensaci\'{o}n de que hab\'{\i}a algo m\'{a}s
que decir, algo m\'{a}s sobre lo que incidir, algo m\'{a}s que matizar... la
paradoja de Zen\'{o}n parec\'{\i}a m\'{a}s real que nunca). Sin embargo, la
mayor parte de la alegr\'{\i}a que supone escribir estos agradecimientos viene
de otro sitio:\textit{ }pensar en a qui\'{e}n agradecer y c\'{o}mo, me hace
recordar el camino exacto que me ha llevado a este punto, recordar a las
personas y los momentos que han supuesto (en mayor o menor medida) puntos de
inflexi\'{o}n en mi vida, que me han ayudado a elegir qu\'{e} direcci\'{o}n
tomar en cada bifurcaci\'{o}n; y claro, teniendo en cuenta lo contento que me
ha hecho la f\'{\i}sica y lo que le rodea desde que entr\'{e} en la
universidad, recordar tales cosas no puede suponer m\'{a}s que una tremenda alegr\'{\i}a.

\bigskip

En este sentido, sin duda son mis padres los primeros a los que debo mi
agradecimiento (tanto cronol\'{o}gicamente, fueron los primeros que estuvieron
ah\'{\i} para m\'{\i}, como cuantitativamente). Corren tiempos en los que
siento que la econom\'{\i}a es el motor fundamental de la mayor parte de la
gente, tiempos en los que estamos constantemente bombardeados con la idea de
que el camino a la felicidad es acumular dinero (y poder, a ser posible); pero
tambi\'{e}n siento que he conseguido desligarme de tal \textquotedblleft
movimiento\textquotedblright, que he conseguido encontrar dos cosas en la
vida, la ciencia y la m\'{u}sica, de las cuales puedo plantearme vivir, a la
vez que sentirme feliz y realizado haci\'{e}ndolas. Pues bien, puedo decir con
total certeza que tal logro es fruto de los valores que mis padres me han
inculcado desde peque\~{n}o. Ellos siempre me han proporcionado todos los
medios que estuvieran a su alcance y todo el apoyo posible para hacer aquello
que quisiera, eso s\'{\i}, siempre que consiguiera convencerles de dos cosas:
que lo har\'{\i}a esforz\'{a}ndome al m\'{a}ximo de verdad, y, sobre todo, que
lo har\'{\i}a \'{u}nicamente mientras me hiciera realmente feliz; las palabras
dinero, futuro, o triunfo, no estaban en su discurso.

\bigskip

Quiz\'{a}s quien m\'{a}s ha \textquotedblleft sufrido\textquotedblright\ mi
faceta de cient\'{\i}fico es Susana, mi novia, compa\~{n}era de andanzas en la
vida, y mayor inspiraci\'{o}n durante los \'{u}ltimos 8 a\~{n}os. No puedo
m\'{a}s que agradecerle la paciencia infinita que tiene conmigo; del futuro
s\'{o}lo espero que las cosas buenas que conllevan compartir la vida con un
cient\'{\i}fico--m\'{u}sico superen como hasta ahora la inestabilidad y las
incomodidades (vida en el extranjero, noches y fines de semana de trabajo,
etc...) intr\'{\i}nsecas a dicho \textquotedblleft
t\'{\i}tulo\textquotedblright. Bien sabes que \textquotedblleft te
vull\textquotedblright\ aunque a veces no te lo diga lo suficiente.

\bigskip

En lo que a la parte cient\'{\i}fica se refiere, no cabe ninguna duda de que a
quien m\'{a}s tengo que agradecer es a mis directores Eugenio y Germ\'{a}n;
desde el principio (y hablamos de antes de terminar la carrera de F\'{\i}sica)
me hicieron sentir como un colaborador m\'{a}s con el potencial de aportar
buenas ideas a la investigaci\'{o}n que estuvi\'{e}semos desarrollando, a la
vez que pusieron a mi disposici\'{o}n todos los medios posibles para
transferirme toda la f\'{\i}sica que almacenan (que no es poca). Pero m\'{a}s
all\'{a} de lo puramente cient\'{\i}fico, me siento orgulloso de que me hayan
dejado entrar en sus vidas hasta el punto de haberse convertido en dos de las
personas que m\'{a}s aprecio y en las que m\'{a}s me apoyo a d\'{\i}a de hoy.
Desde luego, Eugenio y Germ\'{a}n, no aspiro m\'{a}s que a convertirme el
d\'{\i}a de ma\~{n}ana en alguien tan digno y merecedor de cari\~{n}o, respeto
y admiraci\'{o}n como vosotros.

\bigskip

Hay gente que piensa que por ser hijo \'{u}nico no tengo hermanos, pero de una
de las cosas que m\'{a}s orgulloso me siento es de haber demostrado que dicha
gente se equivoca: Apa, Pabl, Fons, Alberto y Rika son prueba de ello. Apa y
Pabl han estado ah\'{\i} casi desde que puedo recordar; juntos hemos pasado
por casi todo lo que se puede pasar, y hemos sobrevivido! Mi vida sin Fons,
ciertamente habr\'{\i}a sido completamente distinta; a \'{e}l no le debo
\'{u}nicamente su amistad totalmente desinteresada, sino tambi\'{e}n el
haberme abierto musicalmente, y con ello haber contribuido radicalmente a que
mi vida sea como es ahora. Alberto es el que m\'{a}s tard\'{o} en llegar a mi
vida, pero se ha convertido en quien mejor entiende esta faceta tan importante
de mi vida que es la ciencia; no conozco persona m\'{a}s sincera y falta de
maldad o ego\'{\i}smo (vale, quiz\'{a} Fons!), por lo cual tiene mi m\'{a}s
sincera admiraci\'{o}n. Rika fue mi pilar b\'{a}sico durante los primeros
a\~{n}os de carrera, los cuales dudo que hubieran sido tan productivos y
realizadores sin su presencia; s\'{o}lo puedo darle las gracias por esos
a\~{n}os que vivimos juntos, y decirle cu\'{a}nto echo de menos la
sensaci\'{o}n de tenerle a mi (y estar a su) lado.

\bigskip

Me siento muy afortunado de contar con una familia que siempre me hace sentir
especial, a la vez que me pone los pies en la tierra (lo cual no es sencillo
trabajando en cosas tan abstractas como la F\'{\i}sica Te\'{o}rica, o tan
intensas como la M\'{u}sica). Siento que una parte de cada uno vosotros
est\'{a} en mi interior, tanto de los que me han criado: mi Abuelo Manolo, mi
Yaya Lola, mi T\'{\i}a Loles, y mis T\'{\i}os Pepa y Chani; como de con los
que me he criado: mi Teta Jose y mis Primos Rafa, Patricia, Dani y Carola. No
me gustar\'{\i}a olvidarme tampoco de los nuevos miembros de la familia que
han ido apareciendo con el tiempo: mis sobrinitos Pablo y Alicia, y mis
\textquotedblleft cu\~{n}aos\textquotedblright\ Paco, Sergi, Vicky, y Silvia;
ni tampoco de mis \textquotedblleft t\'{\i}os y primos
postizos\textquotedblright, en especial de mi t\'{\i}a Jose y mi t\'{\i}o
Pedro, que tan buenas palabras guardan siempre para m\'{\i}.

Tambi\'{e}n me gustar\'{\i}a agradecer a mi \textquotedblleft familia por
extensi\'{o}n\textquotedblright, la familia de Su, a la cual tengo muy
presente cada d\'{\i}a y considero una parte muy relevante de mi vida. Muchas
gracias a \textquotedblleft Los Mayores\textquotedblright\ (Paloma, Carlos y
Amparo) y a \textquotedblleft Los J\'{o}venes\textquotedblright\ (Emi, Moni, y
Mir), por abrirme las puertas de sus vidas sin trabas (bueno, aunque a Mir le
cost\'{o} un poco m\'{a}s que los dem\'{a}s... jejeje).

\bigskip

Me es imposible trazar la trayectoria que me ha llevado a este momento sin
recordar a todos los profesores que me han influenciado e inspirado a lo largo
de mi vida. Dentro de \'{e}stos, quiz\'{a}s los m\'{a}s importantes, porque
sin ellos dudo que me hubiese dado cuenta de que la F\'{\i}sica y las
Matem\'{a}ticas son mi vocaci\'{o}n, son mis profesores del instituto (del
\textquotedblleft cole\textquotedblright, como nosotros lo llamamos). Encarna
Giner fue sin duda la que m\'{a}s me abri\'{o} los ojos y a la que con m\'{a}s
cari\~{n}o recuerdo; cuando menos lo esperaba, ella me mostr\'{o} lo bonitas
que pueden llegar a ser las matem\'{a}ticas, y me dej\'{o} maravillado tras
hacerme entender que s\'{o}lo requieren de comprenderlas (y no de
memorizarlas) para utilizarlas e interiorizarlas. Mi mayor suerte fue su
formaci\'{o}n en F\'{\i}sica Te\'{o}rica (a la que yo finalmente me he
dedicado), ya que le permiti\'{o} introducirme en el mundo de la Relatividad
General y la Geometr\'{\i}a Diferencial hacia el final de nuestro tiempo
juntos. Chencho Pascual, mi gran profesor de F\'{\i}sica, fue el primer
profesor que nos plante\'{o} preguntas interesantes, con \'{a}nimo real de que
fu\'{e}ramos nosotros, los estudiantes, los que encontr\'{a}ramos las
respuestas; cuando cre\'{\i}a que las Matem\'{a}ticas eran mi vocaci\'{o}n,
\'{e}l me hizo ver que la F\'{\i}sica es precisamente la aplicaci\'{o}n de
\'{e}stas al mundo real, la forma en que preguntamos a la naturaleza para
obtener respuestas a las preguntas m\'{a}s fundamentales, y con ello me hizo
encontrar mi vocaci\'{o}n real, la F\'{\i}sica Te\'{o}rica. Siendo
cient\'{\i}fico, quiz\'{a}s algo m\'{a}s sorprendente es el hecho de que
cuando pienso en los profesores que m\'{a}s me motivaron en dicha \'{e}poca,
mi profesor de Lengua Castellana, \'{A}ngel Umbert, sea de los primeros que me
viene a la cabeza; gracias a \'{e}l tambi\'{e}n, aunque no sepa verbalizar muy
bien porqu\'{e} exactamente (al margen de que fuera, sin duda, uno de los
mejores profesores que he tenido nunca).

Durante mi \'{u}ltimo a\~{n}o de instituto tuve la suerte de participar en la
Olimpiada de F\'{\i}sica, lo cual me permiti\'{o} recibir \textquotedblleft
clases avanzadas\textquotedblright\ de F\'{\i}sica de mano de varios
profesores de la Universitat de Val\`{e}ncia y la Universidad Polit\'{e}cnica
de Valencia. La experiencia fue inolvidable, y no puedo m\'{a}s que agradecer
a dichos profesores su entrega desinteresada que tanto me motiv\'{o} a coger
el camino de la F\'{\i}sica finalmente con el fin de seguir aprendiendo y
convertirme en un docente como ellos alg\'{u}n d\'{\i}a. Guardo especial
cari\~{n}o a Fernando Tena, que se encarg\'{o} de \textquotedblleft
cuidar\textquotedblright\ a los ol\'{\i}mpicos valencianos durante la fase
espa\~{n}ola, y que fue quien m\'{a}s me inspir\'{o} durante todas las fases,
as\'{\i} como a Amparo Pons, Jos\'{e} Luis Cruz, Miguel Andr\'{e}s y Paco Pomer.

La carrera de F\'{\i}sica cumpli\'{o} sobradamente con mis expectativas en lo
que a disfrutar aprendiendo se refiere; a\~{n}os atr\'{a}s nunca hubiese
pensado que una \textquotedblleft actividad intelectual\textquotedblright%
\ podr\'{\i}a llenarme y motivarme tanto. No obstante, esto no hubiese sido
posible sin el esfuerzo de algunos profesores a los que ahora me gustar\'{\i}a
agradecer expl\'{\i}citamente: Jos\'{e} Mar\'{\i}a Ib\'{a}\~{n}ez (C\'{a}lculo
y Evoluci\'{o}n Estelar), Jos\'{e} Mar\'{\i}a Mart\'{\i} (\'{A}lgebra y
Din\'{a}mica de Flu\'{\i}dos), Ram\'{o}n Cases (Lab. F\'{\i}sica General y
F\'{\i}sica Nuclear y de Part\'{\i}culas), Jos\'{e} Luis Cruz (F\'{\i}sica
General y Ondas Electromagn\'{e}ticas), Domingo Mart\'{\i}nez (F\'{\i}sica
General y Lab. Electromagnetismo), Eugenio Rold\'{a}n (F\'{\i}sica General,
\'{O}ptica Cu\'{a}ntica y \'{O}ptica No lineal), Albert Ferrando (F\'{\i}sica
General), Pas Garc\'{\i}a (\'{O}ptica Geom\'{e}trica y Procesos
Estoc\'{a}sticos en Fot\'{o}nica), Carlos Ferreira (\'{O}ptica
Geom\'{e}trica), Julio Pellicer (Termodin\'{a}mica), Chantal Ferrer
(Mec\'{a}nica y Ondas), Jon Marcaide (Astronom\'{\i}a y Astrof\'{\i}sica),
Juan Z\'{u}\~{n}iga (M\'{e}todos Num\'{e}ricos), Juan Carlos Soriano
(\'{O}ptica y Procesos Estoc\'{a}sticos en Fot\'{o}nica), Jos\'{e} Mar\'{\i}a
Isidro (F\'{\i}sica Cu\'{a}ntica), Vicente Mu\~{n}oz (Electromagnetismo),
Juanfran S\'{a}nchez (Electromagnetismo), Kiko Botella (Lab. F\'{\i}sica
Cu\'{a}ntica y Mec\'{a}nica Cu\'{a}ntica Avanzada), Pepe Navarro Salas
(Geometr\'{\i}a Diferencial, Variedades y Topolog\'{\i}a, y Teor\'{\i}a
Cu\'{a}ntica de Campos en Espacios Curvos), Benito Gimeno
(Electrodin\'{a}mica), Nuria Rius (Mec\'{a}nica Te\'{o}rica y Teor\'{\i}a
Cu\'{a}ntica de Campos a Temperatura Finita), Manuel Vicente Vacas
(Mec\'{a}nica Cu\'{a}ntica), Miguel Andr\'{e}s (Ondas Electromagn\'{e}ticas),
Vladimir Garc\'{\i}a (F\'{\i}sica Estad\'{\i}stica), Germ\'{a}n de
Valc\'{a}rcel (\'{O}ptica Cu\'{a}ntica y \'{O}ptica No Lineal), Fernando Silva
(\'{O}ptica Cu\'{a}ntica), Ram\'{o}n Lapiedra (Relatividad General y
Cosmolog\'{\i}a), Toni Pich (Teor\'{\i}a Cu\'{a}ntica de Campos y
Cromodin\'{a}mica Cu\'{a}ntica), Javier Garc\'{\i}a (Procesos Estoc\'{a}sticos
en Fot\'{o}nica), Pepe Bernabeu (Teor\'{\i}a Electrod\'{e}bil), Vicent
Gim\'{e}nez (Teor\'{\i}a Cu\'{a}ntica de Campos en el Ret\'{\i}culo), as\'{\i}
como a Armando P\'{e}rez y Arcadi Santamaria, a Pedro Fern\'{a}ndez y Javier
Urchegu\'{\i}a (ambos de la Universidad Polit\'{e}cnica de Valencia), y a Juan
Le\'{o}n (del Instituto de F\'{\i}sica Fundamental del CSIC en Madrid), que,
aunque no me dieron clase, contribuyeron a mis ganas de iniciar la carrera
investigadora. Por razones obvias de espacio, me es imposible explicar los
motivos concretos por los cuales cada uno de estos profesores puso su granito
de arena (o saco de arena en muchos casos) para convertirme en el investigador
que soy hoy en d\'{\i}a, pero estoy seguro de que la mayor\'{\i}a de ellos
conocen perfectamente sus contribuciones particulares.

\bigskip

Por otro lado, uno siempre empieza la universidad con un poco de nerviosismo,
sobre todo porque nunca sabe con qu\'{e} tipo de gente se va a encontrar en
clase (especialmente alguien como yo, que estudi\'{o} todos los cursos
anteriores en el mismo centro, con lo cual no hizo el cambio del
\textquotedblleft colegio\textquotedblright\ al \textquotedblleft
instituto\textquotedblright). Por suerte, gran parte de la culpa de que sienta
que la carrera de F\'{\i}sica me llen\'{o} tanto la tienen precisamente los
compa\~{n}eros con los que comenc\'{e} a estudiarla: Rika, Ave, Bego, Diana,
Javi (Alicante), Javi (Aragorn), Susana, Carlos, Ra\'{u}l y V\'{\i}ctor, son a
los que m\'{a}s cari\~{n}o tengo, sobre todo porque son con los que m\'{a}s
contacto sigo teniendo (aunque no me olvido de Juan, Mar\'{\i}a Luisa y Ali!).
Algo m\'{a}s tarde (hacia tercero), llegaron otros compa\~{n}eros a los que
tambi\'{e}n aprecio mucho: Alberto, David y Sergio; y algo m\'{a}s tarde
a\'{u}n, ya en la \'{e}poca del doctorado, conoc\'{\i} a Zahara e Isa, que me
han ayudado de varias formas en la recta final del doctorado. Me alegro de que
supi\'{e}ramos ver que el camino correcto era el de compartir y ayudarnos los
unos a los otros tanto como pudi\'{e}ramos (algo que parece trivial, pero que
no muchas clases hacen); no es casualidad que no sean pocos los profesores que
afirman que hemos sido \textquotedblleft el curso m\'{a}s gratificante que han
dado\textquotedblright, gracias y enhorabuena compa\~{n}eros!

\bigskip

2006 fue un a\~{n}o muy especial: es el a\~{n}o en el que termin\'{e} la
carrera, a la vez que aprend\'{\i}a los fundamentos de la Biolog\'{\i}a
Molecular y la Ingenier\'{\i}a Gen\'{e}tica a trav\'{e}s de mi
participaci\'{o}n en el proyecto iGEM. Aprender esas ramas de la Biolog\'{\i}a
es algo que llevaba a\~{n}os queriendo hacer, pero que nunca pens\'{e} que
podr\'{\i}a realizar de verdad, y me gustar\'{\i}a agradecer en especial a
Arnau, Emilio, Alberto, Cate, Chevi, Gus, Carlos, Cris, Diana, Minerva,
Albert, Pedro, Javi y Jes\'{u}s, por haber generado el ambiente
cient\'{\i}fico y humano perfecto para desarrollar tal labor. Nunca
olvidar\'{e} el a\~{n}o que pasamos juntos y las experiencias que vivimos
(tanto nuestros primeros congresos cient\'{\i}ficos ---en MIT ni m\'{a}s ni
menos, eso s\'{\i} fue \textquotedblleft colarse\textquotedblright\ en el
mundo de la investigaci\'{o}n por la puerta grande---, como las carreras de
sillas en el laboratorio); mi tiempo con vosotros es con lo que comparo
cualquier otro momento de mi carrera investigadora cuando quiero saber si de
verdad estoy siendo feliz en lo que hago, no se puede ir con m\'{a}s
ilusi\'{o}n a un laboratorio de con la que iba en aquella \'{e}poca.

\bigskip

Quiero agradecer tambi\'{e}n a los miembros del grupo de \'{O}ptica
Cu\'{a}ntica y No Lineal en el que trabajo, tanto a los presentes como a los
pasados, por haberme hecho sentir en un ambiente inmejorable durante mi
doctorado; me alegra poder llamar \textquotedblleft amigos\textquotedblright%
\ a la mayor\'{\i}a de ellos. Muchas gracias en especial a Fernando Silva, que
siempre me ayuda con todas sus fuerzas en cada m\'{\i}nima cosa que necesito;
quiero que sepa que considero que, junto a Alberto y Fons (ver arriba), es la
persona con menos maldad que conozco. Muchas gracias tambi\'{e}n por su
inmejorable actitud a los compa\~{n}eros con los que he colaborado en el
grupo: Ferran Garcia (con quien comparto algunas ideas de esta tesis), Robert
H\"{o}ppner (pese a que no consigui\'{e}ramos hacer llegar a buen puerto
nuestras ideas), Alejandro Romanelli (sin el cual no creo que hubiese hecho la
parte num\'{e}rica de la tesis nunca) y Joaqu\'{\i}n Ruiz--Rivas (con quien
acabo de comenzar, pero con quien ya he compartido unas cuantas
alegr\'{\i}as!); me he encontrado muy a gusto trabajando con vosotros. No me
olvido de Amparo Docavo, la persona que hace que las tareas administrativas
que tan mal se me dan parezcan sencillas; gracias por tener siempre una
sonrisa en la cara cuando hablamos. Isabel P\'{e}rez fue la estudiante de
doctorado que allan\'{o} el camino hacia la \'{O}ptica Cu\'{a}ntica (la de
verdad, no esa con el campo cl\'{a}sico, jiji) en el grupo, de forma que yo ya
pude tomar el relevo con \textquotedblleft el chip cambiado\textquotedblright;
le agradezco que me haya tratado desde que estaba en la carrera como un amigo,
y, en lo que a esta tesis se refiere, que se haya tomado la \textquotedblleft
molestia\textquotedblright\ de evaluarla (nadie mejor para hacerlo, por otro
lado). Agradezco tambi\'{e}n a V\'{\i}ctor S\'{a}nchez por tantas
conversaciones interesantes que hemos tenido sobre ciencia y lo que la rodea,
y a Kestutis Staliunas por ayudarme siempre que lo he necesitado. Hago
extensivos estos agradecimientos al resto de gente con la que he tenido el
placer de coincidir de los grupos de Gand\'{\i}a (V\'{\i}ctor Espinosa, Javier
Redondo, Rub\'{e}n Pic\'{o}, Paco Camarena, Joan Mart\'{\i}nez,...) y Terrasa
(Crina Cojocaru, Ram\'{o}n Herrero, Josep Font, Lina Maigyte,...). Por
\'{u}ltimo, me gustar\'{\i}a dar las gracias y transmitir todo mi cari\~{n}o a
Ram\'{o}n Corbal\'{a}n y Ram\'{o}n Vilaseca (Los Ramones!), cuya personalidad
me inspira y tomo como modelo, por haber dejado a mi alrededor a tantos
\textquotedblleft hijos cient\'{\i}ficos de calidad\textquotedblright\ que
tambi\'{e}n son algunas de las mejores personas que conozco.

\bigskip

Everyone who knows me knows that I've never been much of a traveller,
especially when it is for touristic reasons; however, at a very early stage of
my PhD I came to realize that travelling \ for scientific reasons (such as
congresses or visits to research groups) is actually quite rewarding at many
levels. I would have never grown to be the scientist that I am without the
experiences that I've lived and the people that I've met during these
\textquotedblleft scientific travels\textquotedblright. In this sense, I would
like to first thank Ignacio Cirac, Peter Drummond, Jeff Shapiro, and Nicolas
Cerf for allowing me to become a \textquotedblleft temporary
member\textquotedblright\ of their research groups, and for treating me like a
fellow researcher, giving me all the tools needed to become productive under
their watch. As a PhD student it is easier to hear the not--so--nice stories
about the behaviour of top researchers than the nice ones, and meeting them
has been one of the most inspiring experiences on this regard, as now I know
that one can be at the very top of science without compromising his kindness
and integrity.

On the other hand, if I have been any productive at all during my visits to
foreign research centres, it is because I've been lucky enough to collaborate
with some of the nicest people that I have ever met, and who of course I
consider now my friends. I owe so much to In\'{e}s de Vega and Diego Porras,
my very first collaborators (together with Ignacio) outside the shelter of my
research group and my usual research area; thanks for treating me so well and
have confidence in me. Fate decided that I should meet Ra\'{u}l
Garc\'{\i}a--Patr\'{o}n at the best possible time, and, by his hand,
Nicolas\ Cerf. Both Ra\'{u}l and Nicolas came as a blessing during my visit to
MIT, inviting me to collaborate on a very interesting topic, and giving me all
the chances to offer my opinions and ideas concerning it, even though they
knew that I never worked on pure quantum information before. Finally on this
list of collaborators are my dearest friends Giuseppe Patera and Chiara
Molinelli; I am so glad that we are finally collaborating on something
altogether, I am sure that many happy memories will born from this collaboration!

Living in a new city and working in a new environment and topic can be a
little bit overwhelming; however, if I didn't get any anxious in my
three--months visits to Munich, Melbourne, and Boston it is in part because of
the amazing people that I met there. I would like to thank many people for
making me feel welcome and \textquotedblleft part of
something\textquotedblright\ during these visits. At the MPQ, my officemates
(Heike Schwager, Philipp Hauke, and Christine Muschik), \textquotedblleft the
Spanish crowd\textquotedblright\ (apart from Diego and In\'{e}s, Mari Carmen
Ba\~{n}uls, Maria Eckholt, Miguel Aguado, Lucas Lamata, and Oriol Romero),
Veronika Lechner (always a ray of sunshine in the cold Munich!), G\`{e}za
Giedke (pure kindness), Albert Schliesser (oh my god! an experimentalist on
this list!), as well as David N\'{o}voa, Christina Krauss, Mikel Sanz,
S\'{e}bastien Perseguers, Eric Kessler, Fernando Patawski, Matteo Rizzi, and
Maarten van den Nest. In Melbourne I had the best housemate possible, Simon
Gorman, an amazing musician (animateur!), father of the two sweetest/craziest
little blondies in the world (Scarlett and Sapphire) and with the coolest
sister (Rachel!); I also had the chance to meet some of the best musicians
that I know, such as Chris Hale or Gian Slater (thanks to Esther for hooking
me up with Chris!). The working environment at the ACQAO in Swinburne
University was excellent thanks to Tim Vaughan, Margaret Reid, Qionyi He,
Peter Hannaford, Xia--Ji Liu, Hui Hu, and the kind Tatiana Tchernova, who
helped me with the administrative issues. I had the chance to start writing
this dissertation while living halfway between Harvard and MIT, what was
actually an incredibly inspiring experience, highly amplified by spending lots
and lots of time with my dear friend David Adjiashvili in the many restaurants
that the area has to offer, and specially at \textquotedblleft our
office\textquotedblright, the Atomic Bean Cafe, with the best waitress ever,
Sam! I also thank Chelsea Mart\'{\i}nez, my \textquotedblleft orientation
friend\textquotedblright\ Mar\'{\i}a Ram\'{\i}rez, my officemate Niv
Chandrassekaran, and the Quantum Girls (Veronika Stelmakh, Valentina
Schettini, and Maria Tengner), for being such a wonderful people.

\bigskip

Una parte muy importante de mi vida durante los \'{u}ltimos 12 a\~{n}os ha
sido la m\'{u}sica. Al primero que debo mi agradecimiento en este sentido es a
Pabl, mi \textquotedblleft pareja musical\textquotedblright, con quien he
compartido todos y cada uno de los proyectos en los que he participado. Guardo
adem\'{a}s mucho cari\~{n}o hacia la gente con la que Pabl y yo formamos
nuestra primera banda, Desyrius, especialmente a Vero, Fran y Jos\'{e}
\'{A}ngel, as\'{\i} como a la gente de nuestra segunda banda, Fuego Fatuo
(Hada, Empar y Vicent son lo que recuerdo m\'{a}s v\'{\i}vidamente), la cual
nos permiti\'{o} dar el salto hacia el proyecto totalmente instrumental y
abierto a cualquier estilo musical en el que llevamos trabajando ya 7
a\~{n}os, Versus Five. Mis agradecimientos a los miembros originales de Vs5
(Borja, Diego y Dani), por poner tanta ilusi\'{o}n en un proyecto a priori tan
\textquotedblleft friki\textquotedblright, y hacer con ello que empezara a
funcionar. Tambi\'{e}n a la gente que ha ido dejando su huella en este
proyecto tan gratificante a nivel personal (Moli, Marcos, Luis y Ra\'{u}l) y a
los compa\~{n}eros del resto de bandas de Valencia con las que hemos
interaccionado (Vahladian, \textquotedblleft los Groovers\textquotedblright, y
Non Essential en especial). Y, sobre todo, muchas gracias a la gente que ha
participado en la banda en los \'{u}ltimos tiempos: los Versus (Riki, Osvaldo,
Xavi, Javi, V\'{\i}ctor y ---por poco no llegas a esta lista de
agradecimientos--- David!), los \textquotedblleft crew\textquotedblright%
\ (Apa, Alberto, Rika, Ferr\'{a}ndiz y Fons), los ingenieros (Jorge y
Alberto), los dise\~{n}adores (Juanma e Isma) y las colaboraciones (Voro y David).

\bigskip

El final de mi doctorado ha sido (y en parte sigue siendo) un momento de
relativa ansiedad, sobre todo por la incertidumbre acerca de lo que vendr\'{a}
luego, y la perspectiva de un cambio obligatorio respecto a una situaci\'{o}n
en la que me siento muy c\'{o}modo y feliz. No obstante, hay un par de
\textquotedblleft tareas peri\'{o}dicas\textquotedblright\ que me han ayudado
a llevar mejor esta situaci\'{o}n: el front\'{o}n y las \textquotedblleft
cenas de los mi\'{e}rcoles\textquotedblright. Quiero agradecer a la gente con
la que juego o he jugado a front\'{o}n estos \'{u}ltimos dos a\~{n}os
(aproximadamente), especialmente a Alberto, Rafa, Antonio (mi compa\~{n}ero!),
Joaqu\'{\i}n, Juan, Pablo, Manu y los Joses; mucha de la ansiedad que siento
se va en cada raquetazo y en cada conversaci\'{o}n que mantenemos. Igualmente,
agradezco a la gente de las cenas de los mi\'{e}rcoles, en especial a Cyn,
Paty, Carlos, Gonzalo, Bel\'{e}n, Sergio, Amparo y Ana; no sab\'{e}is
cu\'{a}nto me hacen olvidar la ansiedad de la que hablaba las risas que nos pegamos.

\bigskip

Finalmente, muchas gracias a Alfredo Luis (un sol!), Giuseppe Patera,
Isabel\ P\'{e}rez, Alberto Aparici, y mis directores por haberse tomado el tiempo de corregir mi tesis,
as\'{\i} como a Antonio Garrido, que ha puesto tanto esfuerzo en la
edici\'{o}n (impresi\'{o}n y encuadernaci\'{o}n) de esta tesis.

\bigskip

En definitiva, muchas gracias a todos los que hab\'{e}is contribuido en mayor
o menor medida a que llegue hasta el punto en el que me encuentro ahora;
s\'{e} que me arriesgado mucho al confeccionar una lista tan espec\'{\i}fica,
ya que, teniendo en cuenta el poco tiempo que he tenido para hacerla, la
probabilidad de que haya olvidado a alguien es alta. A\'{u}n as\'{\i}, me
gustar\'{\i}a que se entendiera este esfuerzo como una muestra real de mi
gratitud, y si alguien siente que \textquotedblleft le he
olvidado\textquotedblright, tan s\'olo decirle que probablemente tenga raz\'{o}n y que no
se lo tome a mal: lo he intentado!

\bigskip

Un abrazo a todos!

\bigskip

\begin{flushright}
Carlos.
\end{flushright} 

%% file: IntroFO.tex
I have divided this introductory chapter into two blocks. From the research
point of view, this thesis is devoted to the study of multi--mode quantum
phenomena in optical parametric oscillators (OPOs), which, as will become
clear as one goes deeper into the dissertation, means the generation of
squeezed states of light; in the first part of this introduction I review what
these class of states are, and comment on the state of the art of their
generation and applications. On the other hand, from a thematic perspective
this thesis is quite unusual: Two-thirds of the dissertation are devoted to a
self--contained text about the fundamentals of quantum optics as applied to
the field of squeezing and entanglement, and specially to the modeling of
OPOs, while only one-third of it is devoted to the original research that I
have developed during my PhD student years; a detailed discussion about the
organization of the thesis is then tackled in the second part of this introduction.

\section{Squeezed states of light: Generation and applications}

One of the most amazing predictions offered by the quantum theory of light is
what has been called vacuum fluctuations: Even in the absence of photons
(vacuum), the value of the fluctuations of some observables are different from
zero. These fluctuations cannot be removed by improving the experimental
instrumental, and hence, they are a source of non-technical noise (quantum
noise), which seems to establish a limit for the precision of experiments
involving light.

During the late 1970s and mid-1980s, ways for overcoming this fundamental
limit were predicted and experimentally demonstrated \cite{Meystre91book}. In
the case of the quadratures of light (equivalent to the position and momentum
of a harmonic oscillator), the trick was to eliminate (squeeze) quantum noise
from one quadrature at the expense of increasing the noise of its canonically
conjugated one in order to preserve their Heisenberg uncertainty relation.
States with this property are called squeezed states, and they can be
generated by means of nonlinear optical processes. Even though the initial
experiments were performed with materials having third order nonlinearities
\cite{Slusher85}, nowadays the most widely used nonlinear materials are
$\chi^{(2)}$--crystals, whose induced polarization has a quadratic response to
the applied light field \cite{Drummond04book}. Inside such crystals it takes
place the process of parametric--down conversion, in which photons of
frequency $2\omega_{0}$ are transformed into correlated pairs of photons at
lower frequencies $\omega_{1}$ and $\omega_{2}$ such that $2\omega_{0}%
=\omega_{1}+\omega_{2}$; when working at frequency degeneracy $\omega
_{1}=\omega_{2}=\omega_{0}$, the down--converted field can be shown to be in a
squeezed state \cite{Meystre91book}.

In order to increase the nonlinear interaction, it is customary to introduce
the nonlinear material inside an optical cavity; when $\chi^{(2)}$--crystals
are used, such a device is called an optical parametric oscillator (OPO). To
date, the best squeezing ever achieved is a 93\% of noise reduction with
respect to vacuum \cite{Mehmet10} (see also \cite{Vahlbruch08,Takeno07}), and
frequency degenerate optical parametric oscillators (DOPOs) are the systems
holding this record.

Apart from fundamental reasons, improving the quality of squeezed light is an
important task because of its applications. Among these, the most promising
ones appear in the fields of quantum information with continuous variables
\cite{Braunstein05, WeedbrookUN} (as mixing squeezed beams with beam splitters
offers the possibility to generate multipartite entangled beams
\cite{vanLoock00,Aoki03}) and high--precision measurements (such as beam
displacement and pointing measurements \cite{Treps02,Treps03} or gravitational
wave detection \cite{Vahlbruch05,Goda08}).

In this thesis we offer new phenomena with which we hope to help increasing
the capabilities of future optical parametric oscillators as sources of
squeezed light.

\section{Overview of the thesis}

\subsection{The quantum optics behind squeezing, entanglement, and
OPOs.\label{QOManualOverview}}

As I have already commented, most of this dissertation is not dedicated to
actual research, but to a self--contained introduction to the physics of
squeezing, entanglement, and OPOs\footnote{There are several books which talk
about many of the questions that I introduce in this part of the dissertation,
see for example
\cite{Louisell73book,Cohen89book,Carmichael93book,Carmichael99book,Carmichael08book,Mandel95book,Scully97book,Schleich01book,Drummond04book,Gerry05book,Grynberg10book,Walls94book}%
.}. Let me first expose the reasons why I made this decision.

First of all, after four years interiorizing the mathematical language and
physical phenomena of the quantum optics field, I've come to develop a certain
personal point of view of it (note that \textquotedblleft
personal\textquotedblright\ does not necessarily mean \textquotedblleft
novel\textquotedblright, at least not\ for every aspect of the field). I felt
like this dissertation was my chance to proof myself up to what point I have
truly made mine this field I will be supposed to be an expert in (after
completion of my PhD); I believe that evaluation boards could actually
evaluate the PhD candidate's expertise more truthfully with this kind of
dissertation, rather than with one built just by gathering in a coherent,
expanded way the articles published over his/her post--graduate years.

On the other hand, even though there is a lot of research devoted to OPOs and
squeezing in general, it doesn't exist (to my knowledge) any book in which all
the things needed to understand this topic are explained in a fully
self--contained way, specially in the multi--mode regime in which the research
of this thesis focuses. As a PhD student, I have tried to make myself a
self--contained composition of the field, spending quite a long time going
through all the books and articles devoted to it that I've become aware of. I
wanted my thesis to reflect this huge part of my work, which I felt that could
be helpful for researchers that would like to enter the exciting field of
quantum optics.

Let me now make a summary of what the reader will find in this first
two--thirds of the thesis (numbering of this list's items follows the actual
numbering of the dissertation chapters, see the table of contents):

\begin{itemize}
\item[\ref{QuantumMechanics}.] The true starting point of the thesis is
Appendix \ref{QuantumMechanics}. The goal of this chapter is the formulation
of the axioms of quantum mechanics as I feel that are more suited to the
formalism to be used in quantum optics.\newline In order to properly introduce
these, I first review the very basics of classical mechanics making special
emphasis on the Hamiltonian formalism and the mathematical structure that
observable quantities have on it. I then summarize the theory of Hilbert
spaces, putting special care in the properties of the infinite--dimensional
ones, as these are the most relevant ones in quantum optics.\newline After a
brief historical quote about how the quantum theory was built during the first
third of the XX century, I introduce the axioms of quantum mechanics trying to
motivate them as much as possible from three points of view: Mathematical
consistency, capacity to incorporate experimental observations, and
convergence to classical physics in the limits in which we know that it
works.\newline I decided to relegate this part to an appendix, rather to the
first chapter, because I felt that even though some readers might find my
formulation and motivation of the axioms a little bit different than what they
are used to, in essence all what I explain here is supposed to be of common
undergraduate knowledge, and I preferred to start the thesis in some place new
for any person coming from outside the field of quantum optics.

\item[\ref{HarmonicOscillator}.] In the second chapter I introduce the quantum
description of the one--dimensional harmonic oscillator (which I show later to
be of fundamental interest for the quantum theory of light), and study
different properties of it. In particular, I start by showing that its
associated Hilbert space is infinite--dimensional, and that its energy is
quantized. Furthermore, I explain that the energy of the oscillator is not
zero in its quantum mechanical ground state (the vacuum state) owed to
uncertainties on its quadratures, which are just dimensionless versions of its
position and momentum.\newline I then introduce some special quantum
mechanical states of the oscillator. First, coherent states as the states
which allow us to make the correspondence between the quantum and classical
descriptions, that is, the states whose associated experimental statistics
lead to the observations expected for a classical harmonic oscillator; I
discuss in depth how when the oscillator is in this state, its amplitude and
phase are affected by the vacuum fluctuations.\newline I introduce then the
main topic of the thesis: The squeezed states. To motivate their definition, I
first explain how the phase and amplitude fluctuations can destroy the
potential application of oscillators as sensors when the signals that we want
to measure are tiny, and define squeezed states as states which have the
uncertainty of its phase or amplitude below the level that these have in a
coherent or vacuum state. I also show how to generate them via unitary
evolution, that is, by making the oscillator evolve with a particular
Hamiltonian.\newline Entangled states of two oscillators are introduced
immediately after. In order to motivate them, I first show how their
conception lead Einstein, Podolsky, and Rosen to believe that they had proved
the inconsistency of quantum mechanics; their arguments felt so reasonable,
that this apparent paradox was actually a puzzle for several decades. I then
define rigorously the concept of entangled states, explaining that they can be
understood as states in which the oscillators share correlations which go
beyond what is classically allowed, what is accomplished by making use of the
superposition principle, the main difference between the classical and quantum
descriptions. I also show that these states can be generated via unitary
evolution, and build a specially important class of such states: The two--mode
squeezed vacuum states.\newline In this same section I will be able to
introduce the first original result of the thesis
\cite{NavarreteUNc,GarciaPatronUN}: I show how by adding or subtracting
excitations locally on the oscillators, the entanglement of these class of
states can be enhanced; I developed this part of the work at the Massachusetts
Institute of Technology in collaboration with Ra\'{u}l
Garc\'{\i}a--Patr\'{o}n, Nicolas Cerf, Jeff H. Shapiro, and Seth Lloyd during
a three--months visit to that institution in 2010.\newline In the last part of
this chapter I explain how one can make a quantum mechanical formulation of
the harmonic oscillator relying solely on distributions in phase space. This
appears to suggest that quantum mechanics enters the dynamics of classical
systems just as additional noise blurring their trajectories in phase space;
however, I quickly show that, even though this picture can be quite true for
some quantum mechanical states, it is not the case in general, as this quantum
mechanical phase space distributions do not correspond to probability density
functions in the usual sense.\newline I pay particular attention to one of
such distributions, the positive $P$ distribution, as during the thesis I make
extensive use of it. In the last section of the chapter I show how thanks to
it one can reduce the dynamics of the oscillator to a finite set of first
order differential equations with noise (stochastic Langevin equations), which
in general are easier to deal with than the quantum mechanical evolution
equations for the state of the oscillator (von Neumann equation) or its
observables (Heisenberg equations).

\item[\ref{Quantization}.] The third chapter is devoted to the quantum theory
of light. In the first section I review Maxwell's theory of electromagnetism,
showing that in the absence of sources light satisfies a simple wave equation,
and finally defining the concept of spatial modes of light as the independent
solutions of this equation consistent with the physical boundary conditions of
the particular system to be studied. Then I prove that the electromagnetic
field in free space can be described as a mechanical system consisting of a
collection of independent harmonic oscillators ---one for each spatial mode of
the system (which in this case are plane--waves)---, and then proceed to
develop a quantum theory of light by treating quantum mechanically these
oscillators. The concept of photons is then linked to the excitations of these
electromagnetic oscillators.\newline The reminder of the chapter is devoted to
the quantization of optical beams inside an optical cavity. In order to do
this, I first find the spatial modes satisfying the boundary conditions
imposed by the cavity mirrors (the so-called transverse modes), showing that,
in general, modes with different transverse profiles resonate inside the
cavity at different frequencies. In other words, the cavity acts as a filter,
allowing only the presence of optical beams having particular transverse
shapes and frequencies. In the last section I prove that, similarly to free
space, optical beams confined inside the cavity can be described as a
collection of independent harmonic oscillators, and quantize the theory accordingly.

\item[\ref{OpenSystems}.] Real cavities have not perfectly reflecting mirrors,
not because they don't exist, but rather because we need to be able to inject
light inside the resonator, as well as study or use in applications the light
that comes out from it. In this fourth chapter I apply the theory of open
quantum systems to the case of having one partially transmitting
mirror.\newline The first step is to model the open cavity system, what I do
by assuming that the intracavity mode interacts with a continuous set of
external modes having frequencies around the cavity resonance. Then, I study
how the external modes affect the evolution of the intracavity mode both in
the Heisenberg and the Schr\"{o}dinger pictures.\newline In the Heisenberg
picture it is proved that the formal integration of the external modes leads
to a linear damping term in the equations for the intracavity mode, plus an
additional driving term consisting in a combination of external operators, the
so-called input operator; this equation is known as the quantum Langevin
equation (for its similarity with stochastic Langevin equations, as the input
operator can be seen as kind of a quantum noise), and is the optical version
of the fluctuation--dissipation theorem.\newline In the Schr\"{o}dinger
picture, on the other hand, the procedure consists in finding the evolution
equation for the reduced density operator of the intracavity mode. It is shown
that the usual von Neumann equation acquires an additional term which cannot
be written in a Hamiltonian manner, showing that the loss of intracavity
photons through the partially transmitting mirror is not a reversible process.
The resulting equation is known as the master equation of the intracavity
mode. This is actually the approach which we have chosen to use for most of
our research, as in our case it has several advantages over the Heisenberg
approach, as shown all along the research part of this thesis.\newline The
context of open quantum systems will give me the chance to introduce the work
that I develop in collaboration with In\'{e}s de Vega, Diego Porras, and J.
Ignacio Cirac \cite{Navarrete11b}, which started during a three--months visit
to the Max--Planck Institute for Quantum Optics in 2008. I will show that it
is possible to simulate quantum--optical phenomena (such as superradiance)
with cold atoms trapped in optical lattices; what is interesting is that being
highly tunable systems, it is possible to operate them in regimes where some
interesting, but yet to be observed superradiant phenomena have been predicted
to appear.

\item[\ref{Detection}.] The next chapter is devoted to the measurement
techniques that are used to analyze the light coming out from the cavity. To
this aim, I first relate the output field with the intracavity field and the
input field driving the cavity, a relation that can be seen as the boundary
conditions in the mirror. Together with the quantum Langevin equation, this is
known in quantum optics as the input--output theory.\newline Then I use an
idealized version of a photodetector to show how the techniques of
photodetection and balanced homodyne detection are somehow equivalent,
respectively, to a measurement of the photon number and the quadratures of the
detected field.\newline After this intuitive and simplified version of these
detection schemes, I pass to explain how real photodetectors work. The goal of
the section is to analyze which quantity is exactly the one measured via
homodyne detection in real experiments, arriving to the conclusion that it is
the so-called noise spectrum (kind of a correlation spectrum of the
quadratures).\newline I then redefine the concept of squeezing in an
experimentally useful manner, which, although not in spirit, differs a little
from the simplified version introduced in Chapter \ref{HarmonicOscillator} for
the harmonic oscillator. Even though I introduce this new definition of
squeezing by reasoning from what is experimentally accessible, it is obvious
from a theoretical point of view that the simple \textquotedblleft uncertainty
below vacuum or coherent state\textquotedblright\ definition cannot be it for
the output field, as it does not consist on a simple harmonic oscillator mode,
but on a continuous set of modes having different frequencies around the
cavity resonance.

\item[\ref{OPOs}.] Chapter \ref{OPOs} is the last one of this self--contained
introduction to OPOs, and its goal is to develop the quantum model of OPOs,
and to show that they are sources of squeezed and entangled light.\newline
OPOs being an optical cavity with a nonlinear crystal inside, the chapter
starts by giving an overview of the linear and nonlinear properties of
dielectric media within Maxwell's theory of electromagnetism. It is shown in
particular, how second--order nonlinearities of dielectrics give rise to the
phenomenon of parametric down--conversion: When pumped with an optical beam of
frequency $2\omega_{0}$, the polarization of the nonlinear material is able to
generate a pair of beams at frequencies $\omega_{1}$ and $\omega_{2}$ such
that $2\omega_{0}=\omega_{1}+\omega_{2}$. It is then shown that, at the
quantum level, the process can be understood as the annihilation of one pump
photon, and the simultaneous creation of the pair of down--converted photons
(called signal and idler photons).\newline The case of an OPO in which the
signal and idler photons are indistinguishable, that is, they have the same
frequency, polarization, and transverse structure, is then analyzed. Such a
system is known as the (single--mode) degenerate optical parametric oscillator
(DOPO). It is shown that the classical theory predicts that the pump power
must exceed some threshold level in order for the down--converted field to be
generated; quantum theory, on the other hand, predicts that the
down--converted field will have large levels of squeezing when operating the
DOPO close to this threshold.\newline Similarly, it is shown that when signal
and idler are distinguishable either on frequency or polarization (or both),
these form an entangled pair when working close to threshold. On the other
hand, for any pump power above threshold it is shown that the signal and idler
beams have perfectly correlated intensities (photon numbers): They are what is
called twin beams.\newline This chapter is also quite important because most
of the mathematical techniques used to analyze the dynamics of any OPO
configuration in the thesis are introduced here.
\end{itemize}

Even though my main intention has been to stress the physical meaning of the
different topics that I have introduced, I have also tried to at least sketch
all the mathematical derivations needed to go from one result to the next one.
Hence, even though some points are quite concise and dense, I hope the reader
will find this introductory part interesting as well as understandable.

\subsection{Original results: New phenomena in multi--mode OPOs.}

In the last third of the thesis I review most of the research that I have
developed in my host group, the Nonlinear and Quantum Optics group at the
Universitat de Val\`{e}ncia. If I had to summarize the main contribution of
these research in a couple of sentences, I would say something along the
following lines: \textquotedblleft Although one can try to favour only one
down--conversion channel, OPOs are intrinsically multi-mode, and their
properties can be understood in terms of three phenomena: Bifurcation
squeezing, spontaneous symmetry breaking, and pump clamping. Favouring the
multi-mode regime is indeed interesting because one can obtain several modes
showing well marked non-classical features (such as squeezing and
entanglement) at any pump level above threshold\textquotedblright.

Apart from my supervisors Eugenio Rold\'{a}n and Germ\'{a}n J. de
Valc\'{a}rcel, some of the research has been developed in collaboration with
Ferran V. Garcia--Ferrer (another PhD student in the group), and Alejandro
Romanelli (a visiting professor from the Universidad de la Rep\'{u}blica, in
Urugay). An extensive, analytical summary of this part of the thesis can be
found in Chapter \ref{Conclusions}; here, I just want to explain the
organization of this original part of the dissertation without entering too
much into the specifics of each chapter (as in the previous section, numbering
follows that of the actual chapters):

\begin{itemize}
\item[\ref{MultiOPOs}.] In the first completely original chapter I introduce
the concept of multi--mode OPOs as those which have many down--conversion
channels for a given pumped mode. As I argue right at the beginning of the
chapter, I believe that, one way or another, this is actually the way in which
OPOs operate.\newline Then, I explain how both the classical and quantum
properties of such systems (that is, of general OPOs) can be understood in
terms of two fundamental phenomena: Pump clamping \cite{Navarrete09} and
spontaneous symmetry breaking
\cite{PerezArjona06,PerezArjona07,Navarrete08,Navarrete10}. This general
conclusion is what I consider the most important contribution of my
thesis.\newline The most interesting feature of these phenomena is that,
contrary to the usual OPO model with a single down--conversion channel, where
large levels of squeezing or entanglement are found only when working close to
threshold, they allow multi--mode OPOs to generate highly squeezed or
entangled light at any pump level (above threshold); we talk then about a
noncritical phenomenon, as the system parameters don't need to be finely
(critically) tuned in order to find the desired property.

\item[\ref{2tmDOPO},\ref{2tmDOPOwithIS}.] Most of the research I have
developed during my thesis has been devoted to study in depth the phenomenon
of noncritical squeezing induced by spontaneous symmetry breaking
\cite{Navarrete08,Navarrete10,Navarrete09,GarciaFerrer09,GarciaFerrer10,Navarrete11a,NavarreteUNa}%
. In this chapters, and using the most simple DOPO configuration allowing for
the phenomenon (which have called two--transverse--mode DOPO, or 2tmDOPO in
short), I consider several features which are specially important in order to
understand up to what point the phenomenon is experimentally observable, and
offers a real advantage in front of the critical generation of squeezed light
\cite{Navarrete10,Navarrete11a}.

\item[\ref{TypeIIOPO}.] In the previous chapters the phenomenon of spontaneous
symmetry breaking is introduced in the spatial degrees of freedom of the light
field (in particular, the guiding example is the spontaneous breaking of the
system's invariance under rotations around the propagation axis). In the first
part of this chapter the phenomenon is extended to the polarization degrees of
freedom of light \cite{GarciaFerrer10}, by using an OPO in which the
down--converted photons are distinguishable in polarization, but have the same
frequency.\newline It is then explained how obtaining frequency degeneracy has
been only achieved in experiments by introducing birefringent elements inside
the cavity \cite{Laurat05}, which actually break the symmetry of the system,
hence destroying the phenomenon of spontaneous symmetry breaking.
Nevertheless, we argue that some residual noncritical squeezing should
survive, and discuss that it has been indeed observed in a previous experiment
\cite{Laurat05}.\newline The final part of the chapter is devoted to prove
that frequency degeneracy can be also obtained by a different strategy
consisting on the injection of an optical beam at the degenerate frequency
inside the cavity \cite{NavarreteUNa}.

\item[\ref{FamiliesDOPO}.] In the last chapter I consider the system in which
we originally predicted the phenomenon of pump clamping: A DOPO in which
several transverse modes coexist at the down--converted frequency
\cite{Navarrete09}. I show that using clever cavity designs, it is possible to
get large levels of squeezing in many transverse modes at the same time, what
could be interesting for quantum information protocols requiring multipartite
entanglement (quantum correlations shared between more than two parties).
\end{itemize}

I would like to remind the reader that an extended summary of these part can
be found on the concluding chapter. Let me now make a summary of the
publications derived from my PhD research:

\begin{enumerate}
\item C. Navarrete-Benlloch, E. Rold\'{a}n, and G. J. de Valc\'{a}rcel\newline%
\textit{Non-critically squeezed light via spontaneous rotational symmetry
breaking.}\newline Physical Review Letters \textbf{100}, 203601 (2008).

\item C. Navarrete-Benlloch, G. J. de Valc\'{a}rcel, and E.
Rold\'{a}n.\newline\textit{Generating highly squeezed Hybrid Laguerre-Gauss
modes in large Fresnel number degenerate optical parametric oscillators}%
.\newline Physical Review A \textbf{79}, 043820 (2009).

\item F. V. Garcia-Ferrer, C. Navarrete-Benlloch, G. J. de Valc\'{a}rcel, and
E. Rold\'{a}n.\newline\textit{Squeezing via spontaneous rotational symmetry
breaking in a four-wave mixing cavity}.\newline IEEE Journal of Quantum
Electronics \textbf{45}, 1404 (2009).

\item C. Navarrete-Benlloch, A. Romanelli, E. Rold\'{a}n, and G. J. de
Valc\'{a}rcel.\newline\textit{Noncritical quadrature squeezing in
two-transverse-mode optical parametric oscillators}.\newline Physical Review A
\textbf{81}, 043829 (2010).

\item F. V. Garcia-Ferrer, C. Navarrete-Benlloch, G. J. de Valc\'{a}rcel, and
E. Rold\'{a}n.\newline\textit{Noncritical quadrature squeezing through
spontaneous polarization symmetry breaking}.\newline Optics Letters
\textbf{35}, 2194 (2010).

\item C. Navarrete-Benlloch, I. de Vega, D. Porras, and J. I. Cirac.\newline%
\textit{Simulating quantum-optical phenomena with cold atoms in optical
lattices}.\newline New Journal of Physics \textbf{13}, 023024 (2011).

\item C. Navarrete-Benlloch, E. Rold\'{a}n, and G. J. de
Valc\'{a}rcel.\newline\textit{Squeezing properties of a two-transverse-mode
degenerate optical parametric oscillator with an injected signal}.\newline
Physical Review A \textbf{83}, 043812 (2011).
\end{enumerate}

In addition to these published articles, the following ones are in
preparation, close to being submitted:

\begin{enumerate}
\item[8.] C. Navarrete-Benlloch, R. Garc\'{\i}a-Patr\'{o}n, J. H. Shapiro, and
N. Cerf.\newline\textit{Enhancing entanglement by photon addition and
subtraction}.\newline In preparation.

\item[9.] R. Garc\'{\i}a-Patr\'{o}n, C. Navarrete-Benlloch, S. Lloyd, J. H.
Shapiro, and N. Cerf.\newline\textit{A new approach towards proving the
minimum entropy conjecture for bosonic channels}.\newline In preparation.

\item[10.] C. Navarrete-Benlloch, E. Rold\'{a}n, and G. J. de
Valc\'{a}rcel.\newline\textit{Actively-phase-locked type II optical parametric
oscillators: From non-degenerate to degenerate operation. }\newline In preparation.

\item[11.] C. Navarrete-Benlloch and G. J. de
Valc\'{a}rcel.\newline\textit{Effect of anisotropy on the noncritical squeezing properties of two--transverse--mode optical parametric oscillators. }\newline In preparation.
\end{enumerate} 

%% file: HarmonicOscillatorFO.tex
The harmonic oscillator is one of the basic models in physics; it describes
the dynamics of systems close to their equilibrium state, and hence has a wide
range of applications. It is also of special interest for the purposes of this
thesis, as we will see in the next chapter that the electromagnetic field can
be modeled as a collection of one--dimensional harmonic oscillators.

This section is then devoted to the study of this simple system. We first
explain how the one--dimensional harmonic oscillator is described in a
classical context by a trajectory in phase space. The first step in the
quantum description will be finding the Hilbert space by which it is
described. We then show how coherent states reconcile the quantum and
classical descriptions, and allow us to understand the amplitude--phase
properties of the quantum oscillator. Squeezed and entangled states are then
introduced; understanding the properties of these states is of major relevance
for this thesis. In the context of entangled states we will have the chance to
introduce the work developed by the author of the thesis during a
three--months visit to the Massachusetts Institute of Technology in 2010, and
where it is shown how entanglement can be enhanced by adding or subtracting
excitations locally to the oscillators. We finally explain how to build
rigorous phase space representations of quantum states, with special emphasis
in the properties of the positive \textit{P} representation, as we will make
extensive use of it in this thesis.

We would like to stress that a summary of classical mechanics, Hilbert spaces,
and the axioms of quantum mechanics (as well as definitions of the usual
objects like the Hamiltonian, uncertainties, etc...) is exposed in Appendix
\ref{QuantumMechanics}.

\section{Classical analysis of the harmonic oscillator\label{ClassHO}}

Consider the basic mechanical model of a \textit{one--dimensional harmonic
oscillator}: A particle of mass $m$ is at some equilibrium position which we
take as $x=0$; we displace it from this position by some amount $a$, and then
a restoring force $F=-kx$ starts acting trying to bring the particle back to
$x=0$. Newton's equation of motion for the particle is therefore $m\ddot
{x}=-kx$, which together with the initial conditions $x\left(  0\right)  =a$
and $\dot{x}\left(  0\right)  =v$ gives the solution $x\left(  t\right)
=a\cos\omega t+(v/\omega)\sin\omega t$, being $\omega=\sqrt{k/m}$ the
so-called \textit{angular frequency}. Therefore the particle will be bouncing
back and forth between positions $-\sqrt{a^{2}+v^{2}/\omega^{2}}$ and
$\sqrt{a^{2}+v^{2}/\omega^{2}}$ with periodicity $2\pi/\omega$ (hence the name
`harmonic oscillator').

\begin{figure}
[t]
\begin{center}
\includegraphics[
height=2.5892in,
width=2.597in
]%
{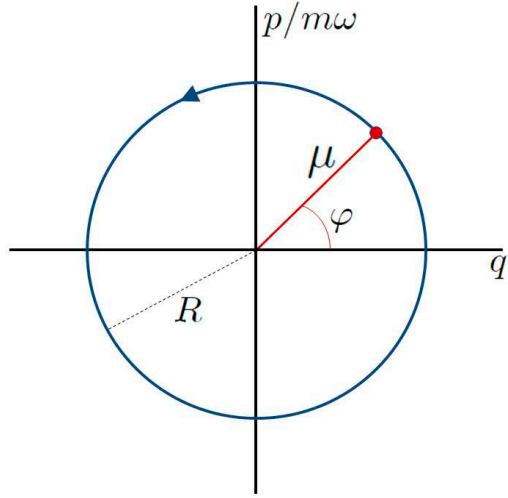}%
\caption{Phase space trajectory of the classical harmonic oscillator.}%
\label{fOsci1}%
\end{center}
\end{figure}

Let us study now the problem from a Hamiltonian point of view. For this
one--dimensional problem with no constraints, we can take the position of the
particle and its momentum as the generalized coordinate and momentum, that is,
$q=x$ and $p=m\dot{x}$. The restoring force derives from a potential $V\left(
x\right)  =kx^{2}/2$, and hence the Hamiltonian takes the form%
\begin{equation}
H=\frac{p^{2}}{2m}+\frac{m\omega^{2}}{2}q^{2}\text{.} \label{ClassHOham}%
\end{equation}
The canonical equations read
\begin{equation}
\dot{q}=\frac{p}{m}\text{ \ \ \ \ and \ \ \ \ }\dot{p}=-m\omega^{2}q,
\end{equation}
which together with the initial conditions $q\left(  0\right)  =a$ and
$p\left(  0\right)  =mv$ give the trajectory%
\begin{equation}
\left(  q,\frac{p}{m\omega}\right)  =\left(  a\cos\omega t+\frac{v}{\omega
}\sin\omega t,\frac{v}{\omega}\cos\omega t-a\sin\omega t\right)  ,
\end{equation}
where we normalize the momentum to $m\omega$ for simplicity. Starting at the
phase space point $\left(  a,v/\omega\right)  $ the system evolves drawing a
circumference of radius $R=\sqrt{a^{2}+v^{2}/\omega^{2}}$ as shown in Figure
\ref{fOsci1}, returning to its initial point at times $t_{k}=2\pi k/\omega$,
with $k\in%
\mathbb{N}
$. This circular trajectory could have been derived without even solving the
equations of motion, as the conservation of the Hamiltonian $H\left(
t\right)  =H\left(  0\right)  $ leads directly to $q^{2}+p^{2}/m^{2}\omega
^{2}=R^{2}$, which is exactly the circumference of Figure \ref{fOsci1}. This
is a simple example of the power of the Hamiltonian formalism.%

There is another useful description of the harmonic oscillator, the so called
\textit{amplitude--phase} or \textit{complex}\ representation. The amplitude
and phase refer to the polar coordinates in phase space, say $\mu=\sqrt
{q^{2}+p^{2}/m^{2}\omega^{2}}$ and $\varphi=\arctan\left(  p/m\omega q\right)
$, as shown in Figure \ref{fOsci1}. In terms of these variables, the
trajectory reads simply $\left(  \mu,\varphi\right)  =\left(  R,\varphi
_{0}-\omega t\right)  $, with $\varphi_{0}=\arctan\left(  v/\omega a\right)
$, so that the evolution is completely described by a linear time variation of
the oscillator's phase. From these variables we can define the complex
variable $\nu=\mu\exp\left(  \mathrm{i}\varphi\right)  =q+\left(
\mathrm{i}/m\omega\right)  p$, in terms of which the trajectory reads
$\nu\left(  t\right)  =R\exp[\mathrm{i}\left(  \varphi_{0}-\omega t\right)
]$, and the Hamiltonian can be written as%
\begin{equation}
H=\frac{m\omega^{2}}{2}\nu^{\ast}\nu. \label{NormalHOham}%
\end{equation}
The pair $\{\nu,\nu^{\ast}\}$ is known as the \textit{normal variables} of the oscillator.

\section{Hilbert space of the harmonic oscillator and number states}

The easiest way to understand the structure of the Hilbert space associated to
the one--dimensional harmonic oscillator is by finding the eigensystem of its
associated Hamiltonian operator. According to the axioms of quantum mechanics
which we review in Appendix \ref{QuantumMechanics}, the operator corresponding
to its classical Hamiltonian (\ref{ClassHOham}) is%
\begin{equation}
\hat{H}=\frac{\hat{p}^{2}}{2m}+\frac{m\omega^{2}}{2}\hat{q}^{2}\text{,}%
\end{equation}
where $\hat{q}$ and $\hat{p}$ are the self--adjoint operators associated to
the position and momentum of the oscillator. At a first sight, this
Hamiltonian may seem difficult to diagonalize because it is a combination of
two non-commuting observables, as according to Axiom III we have $\left[
\hat{q},\hat{p}\right]  =\mathrm{i}\hbar$, see (\ref{CanComPosMom}). However,
we can write it in an easier--looking way as follows. Let us first define
dimensionless versions of the position and momentum operators as%
\begin{equation}
\hat{X}=\sqrt{\frac{2\omega m}{\hbar}}\hat{q}\text{ \ \ \ and \ \ }\hat
{Y}=\sqrt{\frac{2}{\hbar\omega m}}\hat{p}\text{,}%
\end{equation}
whose corresponding observables we will call the \textsf{X}\textit{ }and
\textsf{Y}\ \textit{quadratures} in the following, which satisfy the
commutation relation $[\hat{X},\hat{Y}]=2\mathrm{i}$. The Hamiltonian can be
rewritten then as%
\begin{equation}
\hat{H}=\frac{\hbar\omega}{4}(\hat{X}^{2}+\hat{Y}^{2}).
\end{equation}
Next, we decompose these quadratures as\footnote{Note that this kind of
decompositions are always allowed for any self--adjoint operator; they are
equivalent to write a real number as the addition of a complex number and its
complex conjugate.}%
\begin{equation}
\hat{X}=\hat{a}^{\dagger}+\hat{a}\text{ \ \ \ and \ \ }\hat{Y}=\mathrm{i}%
\left(  \hat{a}^{\dagger}-\hat{a}\right)  , \label{aToX}%
\end{equation}
where the operator $\hat{a}$ and its adjoint $\hat{a}^{\dagger}$ are called
the \textit{annihilation }and\textit{ creation }operators, respectively, and
satisfy the commutation relations $\left[  \hat{a},\hat{a}^{\dagger}\right]
=1$, which we will denote by \textit{canonical commutation relations }in the
following; these operators can be seen as the quantum counterparts of the
normal variables of the classical oscillator. In terms of these operators the
Hamiltonian takes the form\footnote{Note that this Hamiltonian could have been
obtained by following another quantization procedure based on the normal
variables of the oscillator. In particular, we could simetrize the classic
Hamiltonian (\ref{NormalHOham}) respect to the normal variables, writing it
then as
\begin{equation}
H=\frac{m\omega^{2}}{4}\left(  \nu^{\ast}\nu+\nu\nu^{\ast}\right)  ,
\end{equation}
to then make the classical--to--quantum correspondences
\begin{equation}
\nu\rightarrow\sqrt{2\hbar/m\omega}\hat{a}\ \ \text{and}\ \ \nu^{\ast
}\rightarrow\sqrt{2\hbar/m\omega}\hat{a}^{\dagger},
\end{equation}
replacing the Axiom III introduced in Section \ref{Axioms} by $[\hat{a}%
,\hat{a}^{\dagger}]=1$, $[\hat{a},\hat{a}]=[\hat{a}^{\dagger},\hat{a}%
^{\dagger}]=0$. This quantization procedure offers an alternative to the
procedure based on the generalized coordinates and momenta of a mechanical
system.}%
\begin{equation}
\hat{H}=\hbar\omega\left(  \hat{a}^{\dagger}\hat{a}+\frac{1}{2}\right)
\text{,}%
\end{equation}
and hence the problem is simplified to finding the eigensystem of the
self--adjoint operator $\hat{n}=\hat{a}^{\dagger}\hat{a}$, which we will call
the \textit{number operator}.

Let us call $n$ to a generic real number contained in the spectrum of $\hat
{n}$, whose corresponding eigenvector we denote by $|n\rangle$, so that,
$\hat{n}|n\rangle=n|n\rangle$. We normalize the vectors to one by definition,
that is, $\langle n|n\rangle=1$ $\forall n$. The eigensystem of $\hat{n}$ is
readily found from the following two properties:

\begin{itemize}
\item $\hat{n}$ is a positive operator, as for any vector $|\psi\rangle$ it is
satisfied $\langle\psi|\hat{n}|\psi\rangle=\left(  \hat{a}|\psi\rangle,\hat
{a}|\psi\rangle\right)  \geq0$. When applied to its eigenvectors, this
property forbids the existence of negative eigenvalues, that is $n\geq0$.

\item Using the commutation relation\footnote{This is straightforward to find
by using the property $[\hat{A}\hat{B},\hat{C}]=\hat{A}[\hat{B},\hat{C}%
]+[\hat{A},\hat{C}]\hat{B}$, valid for any three operators $\hat{A}$, $\hat
{B}$, and $\hat{C}$.} $[\hat{n},\hat{a}]=-\hat{a}$, it is trivial to show that
the vector $\hat{a}|n\rangle$ is also an eigenvector of $\hat{n}$ with
eigenvalue $n-1$. Similarly, from the commutation relation $[\hat{n},\hat
{a}^{\dagger}]=\hat{a}^{\dagger}$ it is found that the vector $\hat
{a}^{\dagger}|n\rangle$ is an eigenvector of $\hat{n}$ with eigenvalue $n+1$.
\end{itemize}

These two properties imply that the spectrum of $\hat{n}$ is the set of
natural numbers $n\in\left\{  0,1,2,...\right\}  \equiv%
\mathbb{N}
$, and that the eigenvector $|0\rangle$ corresponding to $n=0$ must satisfy
$\hat{a}|0\rangle=0$; otherwise it would be possible to find negative
eigenvalues, hence contradicting the positivity of $\hat{n}$.\ Thus, the set
of eigenvectors $\left\{  |n\rangle\right\}  _{n\in%
\mathbb{N}
}$ is an infinite, countable set. Moreover, using the property $\hat
{a}|0\rangle=0$ and the commutation relations, it is easy to prove that the
eigenvectors corresponding to different eigenvalues are orthogonal, that is,
$\langle n|m\rangle=\delta_{nm}$. Finally, according to the axioms of quantum
mechanics only the vectors normalized to one are physically relevant. Hence,
we conclude that the space generated by the eigenvectors of $\hat{n}$ is
isomorphic to $l^{2}\left(  \infty\right)  $, and hence it is an
infinite--dimensional Hilbert space (see Section \ref{InfiniteHilbert}).

Summarizing, we have been able to prove that the Hilbert space associated to
the one--dimensional harmonic oscillator is infinite--dimensional. In the
process, we have explicitly built an orthonormal basis of this space by using
the eigenvectors $\left\{  |n\rangle\right\}  _{n\in%
\mathbb{N}
}$ of the number operator $\hat{n}$, with the annihilation and creation
operators $\{\hat{a},\hat{a}^{\dagger}\}$ allowing us to move through this set
as
\begin{equation}
\hat{a}|n\rangle=\sqrt{n}|n-1\rangle\text{ \ \ \ \ and \ \ \ \ }\hat
{a}^{\dagger}|n\rangle=\sqrt{n+1}|n+1\rangle, \label{DownUp}%
\end{equation}
the factors in the square roots being easily found from normalization requirements.

Let us now explain some physical consequences of this. The vectors $\left\{
|n\rangle\right\}  _{n\in%
\mathbb{N}
}$ are eigenvectors of the energy (the Hamiltonian) with eigenvalues $\{
E_{n}=\hbar\omega(n+1/2)\}_{n\in%
\mathbb{N}
}$, and hence quantum theory predicts that the energy of the oscillator is
quantized: Only a discrete set of energies separated by $\hbar\omega$ can be
measured in an experiment. The number of \textit{quanta} or
\textit{excitations} is given by $n$, and that's why $\hat{n}$ is called
the\ \textquotedblleft number\textquotedblright\ operator, it `counts' the
number of excitations. Similarly, the creation and annihilation operators
receive their names because they add and subtract excitations. As these
vectors have a well defined number of excitations, $\Delta n=0$, we will call
them \textit{number states}. Consequently, $|0\rangle$ will be called the
\textit{vacuum state} of the oscillator, as it has no quanta.

On the other hand, while in classical mechanics the harmonic oscillator can
have zero energy ---what happens when it is in its equilibrium state---, quantum
mechanics predicts that the minimum energy that the oscillator can have is
$E_{0}=\hbar\omega/2>0$. One way to understand where this \textit{zero--point}
energy comes from is by minimizing the expectation value of the Hamiltonian,
which can be written as
\begin{equation}
\langle\hat{H}\rangle=\frac{\hbar\omega}{4}(\Delta X^{2}+\Delta Y^{2}%
+\left\langle X\right\rangle ^{2}+\left\langle Y\right\rangle ^{2}),
\label{MeanEnergyQuadratures}%
\end{equation}
subject to the constraint $\Delta X\Delta Y\geq1$ imposed by the uncertainty
principle. It is easy to argue that the minimum value of $\langle\hat
{H}\rangle$ is obtained for the state satisfying $\Delta X=\Delta Y=1$ and
$\langle X\rangle=\langle Y\rangle=0$, which corresponds, not surprisingly, to
the vacuum state $|0\rangle$. Hence, the energy present in the ground state of
the oscillator comes from the fact that the uncertainty principle does not
allow its position and momentum to be exactly zero, they have some
fluctuations even in the vacuum state, and this \textit{vacuum fluctuations}
contribute to the energy of the oscillator.

\section{Coherent states}

\subsection{Connection to classical mechanics}

Based on the previous sections, we see that the classical and quantum
formalisms seem completely different in essence: While classically the
oscillator can have any positive value of the energy and has a definite
trajectory in phase space, quantum mechanics allows only for discrete values
of the energy and introduces position and momentum uncertainties which prevent
the existence of well defined trajectories. In this section we show that,
despite their differences, both descriptions are compatible in some limit.

Let us first note that, instead of its position and momentum, from now on we
take the \textsf{X} and \textsf{Y} quadratures of the oscillator as the
observables defining the phase space. Using these variables, the classical
phase space trajectory of the oscillator reads $(X,Y)=\left(  X_{0}\cos\omega
t+Y_{0}\sin\omega t,Y_{0}\cos\omega t-X_{0}\sin\omega t\right)  $, which
defines a circumference of radius $R=\sqrt{X_{0}^{2}+Y_{0}^{2}}$, being
$X_{0}=X\left(  0\right)  $ and $Y_{0}=Y\left(  0\right)  $.

Quantum mechanics is all about predicting the statistics of experiments, see
Section \ref{Axioms}. Hence, a way of connecting it to classical mechanics is
by finding the quantum state which predicts that the statistics obtained in
the experiment will coincide with what is classically expected. The following
two points explain the properties that such a state should have in the case of
the harmonic oscillator:

\begin{itemize}
\item Classically, the energy is a continuous observable. On the hand, the
ratio between the energies of two consecutive number states is $E_{n+1}%
/E_{n}=\left(  n+3/2\right)  /(n+1/2)$; hence, as the number of excitations
increases, the discrete character of the energy becomes barely perceptive,
that is, $E_{n+1}/E_{n}\sim1$ if $n\gg1$. Thus, the state should have a large
number of excitations, that is, $\langle\hat{n}\rangle\gg1$.

\item The expectation value of the quadratures must describe the classical
circular trajectory, while the uncertainties of both quadratures should be
well below the radius of the circumference defined by it, that is,
$(\langle\hat{X}\left(  t\right)  \rangle,\langle\hat{Y}\left(  t\right)
\rangle)=\left(  X_{0}\cos\omega t+Y_{0}\sin\omega t,Y_{0}\cos\omega
t-X_{0}\sin\omega t\right)  $ with $\{\Delta X,\Delta Y\}\ll R$. Hence, at all
effects the experimental outcomes predicted by quantum theory for the
quadratures will coincide with those expected from classical mechanics. Note
that the condition for the uncertainties requires $R\gg1$, as we know that
$\Delta X=\Delta Y=1$ is the minimum simultaneous value that the variances can take.
\end{itemize}

Then, if states satisfying this properties exist, we see that it is indeed
possible to reconcile quantum theory with classical mechanics at least on what
concerns to physical observations.

Our starting point for obtaining these states are the Heisenberg evolution
equations for the quadrature operators $\hat{X}$ and $\hat{Y}$, which read%
\begin{subequations}
\begin{align}
\frac{d}{dt}\hat{X}  & =\frac{1}{\mathrm{i}\hbar}[\hat{X},\hat{H}]=\omega
\hat{Y},\\
\frac{d}{dt}\hat{Y}  & =\frac{1}{\mathrm{i}\hbar}[\hat{Y},\hat{H}]=-\omega
\hat{X}\text{,}%
\end{align}
from which we obtain%
\end{subequations}
\begin{subequations}
\begin{align}
\hat{X}\left(  t\right)    & =\hat{X}\left(  0\right)  \cos\omega t+\hat
{Y}\left(  0\right)  \sin\omega t,\\
\hat{Y}\left(  t\right)    & =\hat{Y}\left(  0\right)  \cos\omega t-\hat
{X}\left(  0\right)  \sin\omega t\text{.}%
\end{align}
\end{subequations}
Hence, the expectation values of the quadratures describe the classical
trajectories by construction, so that in order to reproduce a classical
trajectory with initial quadratures $\left(  X_{0},Y_{0}\right)  $, the state
must satisfy $\langle\hat{X}\left(  0\right)  \rangle=X_{0}$ and $\langle
\hat{Y}\left(  0\right)  \rangle=Y_{0}$. Writing now the quadratures in terms
of the annihilation and creation operators (\ref{aToX}) these last relations
can be recasted as $\left\langle \hat{a}\right\rangle =\left(  X_{0}%
+\mathrm{i}Y_{0}\right)  /2$; being $\left(  X_{0},Y_{0}\right)  $ arbitrary,
this condition shows that the states we are looking for are the eigenvectors
of the annihilation operator, that is, the states $|\alpha\rangle$ satisfying
$\hat{a}|\alpha\rangle=\alpha|\alpha\rangle$ with $\alpha=\left(
X_{0}+\mathrm{i}Y_{0}\right)  /2\in%
\mathbb{C}
$. We will call \textit{coherent states} to the eigenvectors of the
annihilation operator.

It is straightforward to show that, in addition, coherent states have
$\langle\hat{n}\rangle=|\alpha|^{2}$, $R=2|\alpha|$ and $\Delta X=\Delta Y=1$,
so that in the limit $|\alpha|\gg1$ they satisfy all the requisites that we
needed to make the connection with classical mechanics. Note that these
properties hold at any time.

In the following we show that the eigenvectors of the annihilation operator
exist by finding an explicit representation of them in the Hilbert space of
the oscillator, and discuss some of their properties.

\subsection{Properties of the coherent states\label{CohProp}}

It is simple to find an explicit representation of the coherent states in
terms of the basis of number states, that is, to find the coefficients $c_{n}$
of the expansion $|\alpha\rangle=\sum_{n=0}^{\infty}c_{n}|n\rangle$.
Introducing this expansion in the eigenvector equation\footnote{On a second
thought, the existence of states satisfying this equation is kind of amazing:
Consider for example a coherent state with $|\alpha|^{2}\ll1$, so that we are
sure that the mean number of excitations is really below one; this equation
states that we can still annihilate as many excitations as we want without
altering the state!} $\hat{a}|\alpha\rangle=\alpha|\alpha\rangle$, we find the
simple recurrence%
\begin{equation}
c_{n}=\frac{\alpha}{\sqrt{n}}c_{n-1}\Longrightarrow c_{n}=\frac{\alpha^{n}%
}{\sqrt{n!}}c_{0}\text{.}%
\end{equation}
On the other hand, the determination of $c_{0}$ comes from the normalization
condition%
\begin{equation}
\langle\alpha|\alpha\rangle=1=|c_{0}|^{2}\sum_{n=0}^{\infty}|\alpha
|^{2}/n!\Longrightarrow c_{0}=\exp\left(  -\frac{|\alpha|^{2}}{2}\right)  ,
\end{equation}
where we have chosen $c_{0}$ to be a positive real as the global phase of the
state cannot play any physical role\footnote{This is evident since we have
formulated quantum mechanics in terms of the density operator, which for all
pure states $e^{\mathrm{i}\phi}|\psi\rangle$ reads $\hat{\rho}=|\psi
\rangle\langle\psi|$ irrespective of $\phi$ (see Section \ref{Axioms}).}.
Hence, the coherent states are finally written as%
\begin{equation}
|\alpha\rangle=\sum_{n=0}^{\infty}e^{-|\alpha|^{2}/2}\frac{\alpha^{n}}%
{\sqrt{n!}}|n\rangle\text{.} \label{NumToCoh}%
\end{equation}
Note that for $\alpha=0$, coherent states are equivalent to the vacuum state,
that is, $|\alpha=0\rangle=|n=0\rangle$. For any other value of $\alpha$
coherent states do not have a well defined number of excitations; instead, the
number of \textit{quanta} is distributed according to a Poissonian probability
distribution%
\begin{equation}
P_{n}\left(  \alpha\right)  =|\langle n|\alpha\rangle|^{2}=e^{-|\alpha|^{2}%
}\frac{|\alpha|^{2n}}{n!}.
\end{equation}

Let us now define the \textit{displacement operator }%
\begin{equation}
\hat{D}\left(  \alpha\right)  =\exp\left(  \alpha\hat{a}^{\dagger}%
-\alpha^{\ast}\hat{a}\right)  ;
\end{equation}
the general formula $\exp(\hat{A}+\hat{B})=\exp(-[\hat{A},\hat{B}]/2)\exp
(\hat{A})\exp(\hat{B})$, valid for operators $\hat{A}$ and $\hat{B}$ which
commute with their commutator, allows us to write it also as%
\begin{equation}
\hat{D}^{\left(  \mathrm{n}\right)  }\left(  \alpha\right)  =\exp\left(
-|\alpha|^{2}/2\right)  \exp\left(  \alpha\hat{a}^{\dagger}\right)
\exp\left(  -\alpha^{\ast}\hat{a}\right)  \text{,}%
\end{equation}
or%
\begin{equation}
\hat{D}^{\left(  \mathrm{a}\right)  }\left(  \alpha\right)  =\exp\left(
|\alpha|^{2}/2\right)  \exp\left(  -\alpha^{\ast}\hat{a}\right)  \exp\left(
\alpha\hat{a}^{\dagger}\right)  ,
\end{equation}
which we will refer to as its \textit{normal} and \textit{antinormal forms},
respectively\footnote{With full generality, given an operator $\hat{A}\left(
\hat{a},\hat{a}^{\dagger}\right)  $, we denote by $\hat{A}^{\left(
\mathrm{n}\right)  }\left(  \hat{a},\hat{a}^{\dagger}\right)  $ and $\hat
{A}^{\left(  \mathrm{a}\right)  }\left(  \hat{a},\hat{a}^{\dagger}\right)  $
its \textit{normal} and \textit{antinormal} forms. $\hat{A}^{\left(
\mathrm{n}\right)  }$ and $\hat{A}^{\left(  \mathrm{a}\right)  }$ are obtained
by writing $\hat{A}\left(  \hat{a},\hat{a}^{\dagger}\right)  $ with all the
creation operators to the left or to the right, respectively, with the help of
the commutation relations. Hence, for example, the operator $\hat{A}=\hat
{a}\hat{a}^{\dagger}\hat{a}$ has $\hat{A}^{\left(  \mathrm{n}\right)  }%
=\hat{a}^{\dagger}\hat{a}^{2}+\hat{a}$ and $\hat{A}^{\left(  \mathrm{a}%
\right)  }=\hat{a}^{2}\hat{a}^{\dagger}-\hat{a}$ as normal and antinormal
forms.}. Coherent states can then be obtained by \textit{displacing} vacuum%
\begin{equation}
|\alpha\rangle=\hat{D}\left(  \alpha\right)  |0\rangle\text{,}%
\label{DisplacedVacuum}%
\end{equation}
what is trivially proved using the normal form of the displacement operator.
This is interesting because the displacement operator is unitary by
construction, that is, $\hat{D}^{\dagger}\left(  \alpha\right)  \hat{D}\left(
\alpha\right)  =\hat{D}\left(  \alpha\right)  \hat{D}^{\dagger}\left(
\alpha\right)  =\hat{I}$, and hence the coherent state $|\alpha\rangle$ can be
created from a vacuum--state oscillator by making it evolve with a Hamiltonian
$\hat{H}_{D}=\mathrm{i}\hbar(\alpha\hat{a}^{\dagger}-\alpha^{\ast}\hat{a})/T$
during a time $T$. This is in contrast to the number states, which cannot be
generated by unitary evolution of the oscillator in any way.%

Note that when a displacement is applied to the annihilation and creation
operators, we get\footnote{This is trivially proved by using the
Baker--Campbell--Haussdorf lemma%
\begin{equation}
e^{\hat{B}}\hat{A}e^{-\hat{B}}=\sum_{n=0}^{\infty}\frac{1}{n!}\underset
{n}{\underbrace{[\hat{B},[\hat{B},...[\hat{B},}}\hat{A}\underset
{n}{\underbrace{]...]]}}, \label{BCHlemma}%
\end{equation}
valid for two general operators $\hat{A}$ and $\hat{B}$.}%
\begin{equation}
\hat{D}^{\dagger}\left(  \alpha\right)  \hat{a}\hat{D}\left(  \alpha\right)
=\hat{a}+\alpha\text{ \ \ \ and \ \ }\hat{D}^{\dagger}\left(  \alpha\right)
\hat{a}^{\dagger}\hat{D}\left(  \alpha\right)  =\hat{a}^{\dagger}+\alpha
^{\ast}\text{,}%
\end{equation}
which shows where the name `displacement' comes from. Hence, applied to the
quadratures we get%
\begin{equation}
\hat{D}^{\dagger}\left(  \alpha\right)  \hat{X}\hat{D}\left(  \alpha\right)
=\hat{X}+x\text{ \ \ \ and \ \ }\hat{D}^{\dagger}\left(  \alpha\right)
\hat{Y}\hat{D}\left(  \alpha\right)  =\hat{Y}+y\text{,}%
\end{equation}
with $x+\mathrm{i}y=2\alpha$ and $\{x,y\}\in%
\mathbb{R}
$. This transformation changes the mean of the quadratures but not its
variance. Hence, a coherent state has the same uncertainty properties as the
vacuum state.

Note finally that coherent states cannot form a true basis of the Hilbert
space because they do not form a countable set. They cannot form a generalized
continuous basis either (see Section \ref{InfiniteHilbert}) because we have
proved that they can be normalized in the usual sense, and hence, they are
vectors defined inside the Hilbert space. Moreover, the inner product of two
different coherent states $|\alpha\rangle$ and $|\beta\rangle$ reads%
\begin{equation}
\langle\alpha|\beta\rangle=\exp\left(  -\frac{|\alpha|^{2}}{2}+\alpha^{\ast
}\beta-\frac{|\beta|^{2}}{2}\right)  ,
\end{equation}
and hence coherent states are not orthogonal. Despite all these, they do form
a resolution of the identity, as it is easy to prove that%
\begin{equation}
\int_{%
\mathbb{C}
}d^{2}\alpha\frac{|\alpha\rangle\langle\alpha|}{\pi}=\sum_{n=0}^{\infty
}|n\rangle\langle n|=\hat{I},
\end{equation}
where the integral covers the entire complex--$\alpha$ space. Hence, even
though coherent states do not form a basis, they can still be used to
represent any vector or operator: They form an \textit{overcomplete basis}.
Indeed, this overcompleteness can even be an advantage in some circumstances;
for example, it is easy to show that any operator $\hat{L}$ is completely
determined from just its diagonal elements $\langle\alpha|\hat{L}%
|\alpha\rangle$ in the coherent basis, as
\begin{align}
\langle\alpha|\hat{L}|\alpha\rangle & =\sum_{n,m=0}^{\infty}e^{-|\alpha
|^{2}/2}\frac{\alpha^{\ast m}\alpha^{n}}{\sqrt{n!m!}}L_{mn}\\
& \Downarrow\nonumber\\
L_{mn}  & =\frac{1}{\sqrt{n!m!}}\left.  \frac{\partial^{m+n}}{\partial
\alpha^{\ast m}\partial\alpha^{n}}\langle\alpha|\hat{L}|\alpha\rangle
e^{|\alpha|^{2}/2}\right\vert _{\alpha,\alpha^{\ast}=0}\text{.}\nonumber
\end{align}
We will use these \textit{coherent representations} quite a lot during this
thesis. We will come back to them in the last section of this chapter.

\begin{figure}
[t]
\begin{center}
\includegraphics[
width=\textwidth
]%
{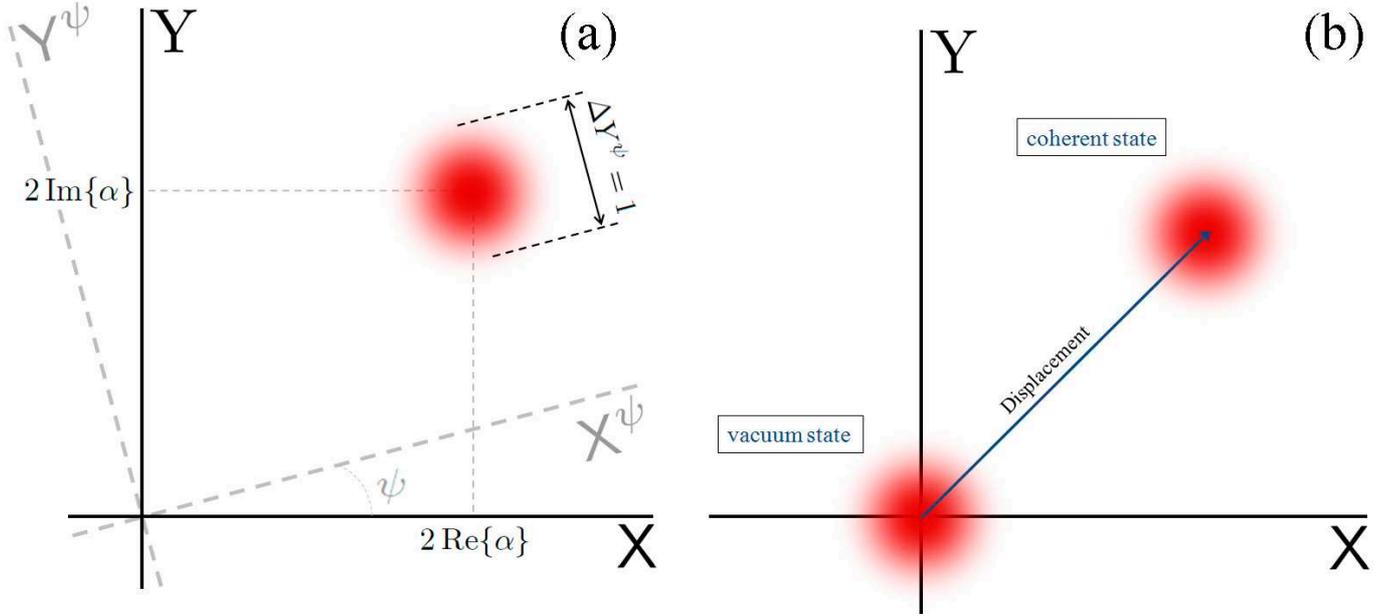}%
\caption{(a) Phase space sketch of a coherent state. (b) A coherent state of
arbitrary amplitude can be generated by applying a displacement to the vacuum
state.}%
\label{fOsci2}%
\end{center}
\end{figure}

\subsection{Phase space sketch of coherent states}

Even though in quantum mechanics systems do not have a defined trajectory on
phase space, it is useful to have some pictorial representation of their
quantum states. The rigorous way of building such representations is discussed
at the end of this chapter; here we just want to motivate this representation
and build it for coherent states from an intuitive point of view. Among other
things, this representation will allow us to understand the importance of
squeezed states, which are described in the next section and are actually the
basic theme of this thesis. It will also help us to understand the amplitude
and phase properties of the quantum harmonic oscillator, two observables which
still lack of a satisfactory description in terms of self--adjoint operators
\cite{Gerry05book,Mandel95book}.

Let us first build a \textit{general quadrature} defined along an arbitrary
direction $\psi$ of phase space as%
\begin{equation}
\hat{X}^{\psi}=\hat{X}\cos\psi+\hat{Y}\sin\psi=e^{-\mathrm{i}\psi}\hat
{a}+e^{\mathrm{i}\psi}\hat{a}^{\dagger}\text{.}%
\end{equation}
We will denote the quadrature defined along its orthogonal direction by
$\hat{Y}^{\psi}=\hat{X}^{\psi+\pi/2}$. These orthogonal quadratures define a
new coordinate system in phase space rotated by an angle $\psi$ respect to the
original \textsf{X--Y} system (see Figure \ref{fOsci2}). They also satisfy the
commutation relation $[\hat{X}^{\psi},\hat{Y}^{\psi}]=2\mathrm{i}$, and hence
must satisfy the uncertainty relation $\Delta X^{\psi}\Delta Y^{\psi}\geq1$.

Coherent states $|\alpha\rangle$ admit a simple, descriptive representation in
phase space (Figure \ref{fOsci2}). The idea for this sketch is to represent
the statistics that would be obtained if a general quadrature is
measured\footnote{We will learn how to perform quadrature measurements in
Chapter 6 for the case of light.}. In the classical limit, a reasonable
representation is then simply a point $(\langle\hat{X}\rangle,\langle\hat
{Y}\rangle)=(2\operatorname{Re}\{\alpha\},2\operatorname{Im}\{\alpha\})$ in
phase space. As quantum mechanics starts showing, uncertainties start playing
a role. It is easy to see that the uncertainty of any quadrature $\hat
{X}^{\psi}$ in a coherent state is $\Delta X^{\psi}=1$, irrespective of $\psi
$, so that the statistics of a measurement of the quadratures will be spread
around the mean equally in any direction of phase space. Hence, the classical
representation can be generalized by drawing a circle of unit radius
representing the quantum uncertainties associated to a measurement of these
quadratures. This is shown in Figure \ref{fOsci2}a.

The special case $\alpha=0$, the vacuum state, is represented in Figure
\ref{fOsci2}b. Note that it is represented by the exact same circle (the
uncertainty $\Delta X^{\psi}$ does not depend on $\alpha$ either), but now
centered at the origin of phase space. The representation of any other
coherent state is then obtained by displacing this uncertainty circle to the
point $(2\operatorname{Re}\{\alpha\},2\operatorname{Im}\{\alpha\})$, and hence
these states can be visualized as classical states carrying with the quantum
vacuum uncertainties.

\begin{figure}
[t]
\begin{center}
\includegraphics[
height=3.1514in,
width=3.851in
]%
{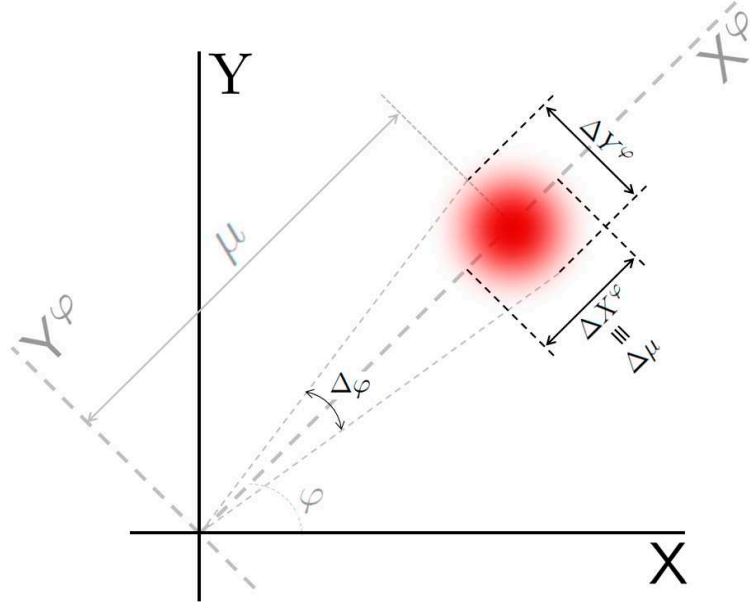}%
\caption{Quantum amplitude--phase properties of a coherent state. $\Delta\mu$
and $\Delta\varphi$ represent the uncertainties in the amplitude and phase of
the oscillator, which are obviously related to the uncertainties of the
amplitude and phase quadratures, $\hat{X}^{\varphi}$ and $\hat{Y}^{\varphi}$,
respectively.}%
\label{fOsci3}%
\end{center}
\end{figure}

This intuitive picture of coherent states allows us to understand its
amplitude and phase properties (Figure \ref{fOsci3}). The mean values of the
quadratures define a phase $\varphi=$ $\arctan\langle\hat{Y}\rangle
/\langle\hat{X}\rangle$ and an amplitude $\mu=\sqrt{\langle\hat{Y}\rangle
^{2}+\langle\hat{X}\rangle^{2}}$; in the classical limit these are exactly the
phase and amplitude that would be measured for the oscillator. When quantum
uncertainties in the quadratures cannot be neglected, it is reasonable to
think that $\varphi$ and $\mu$ will still be the mean values measured for the
phase and amplitude of the oscillator, but now they will be also affected by
some uncertainties. It seems obvious from Figure \ref{fOsci3} that these
amplitude and phase uncertainties are related to the uncertainties of the
$\hat{X}^{\varphi}$ and $\hat{Y}^{\varphi}$ quadratures, which we shall
consequently call the \textit{amplitude} and \textit{phase }quadratures.
Hence, even if at the quantum level the phase and amplitude observables are
not satisfactorily understood, one can somehow relate their properties to
those of the amplitude and phase quadratures, at least for states with a well
defined amplitude, that is, $\mu>\Delta X^{\varphi}$.

\section{Squeezed states}

\subsection{Definition and relevance}

An important application of harmonic oscillators is \textit{sensing}: The
oscillator is put in contact with a system that we want to test, and some
information about this gets encoded as phase or amplitude modulations in the
oscillator. We have seen in the previous section that when the oscillator is
in a coherent state both its amplitude and phase suffer from uncertainties,
and hence the encoded signal cannot be perfectly retrieved from measurements
on the oscillator. When any other source of technical noise is removed, that
is, when the measurement equipment behaves basically as ideal, this
\textit{quantum noise} becomes the main limitation; moreover, when the signal
generated by the system that we want to study is tiny, it can even be
disguised below quantum noise, so that it could not be distinguished at all.
Note that this quantum noise appears even if the oscillator is in its vacuum
state $|\alpha=0\rangle$, as it has its roots in the \textit{vacuum
fluctuations} of the position and momentum of the oscillator.%

\begin{figure}
[t]
\begin{center}
\includegraphics[
width=\textwidth
]%
{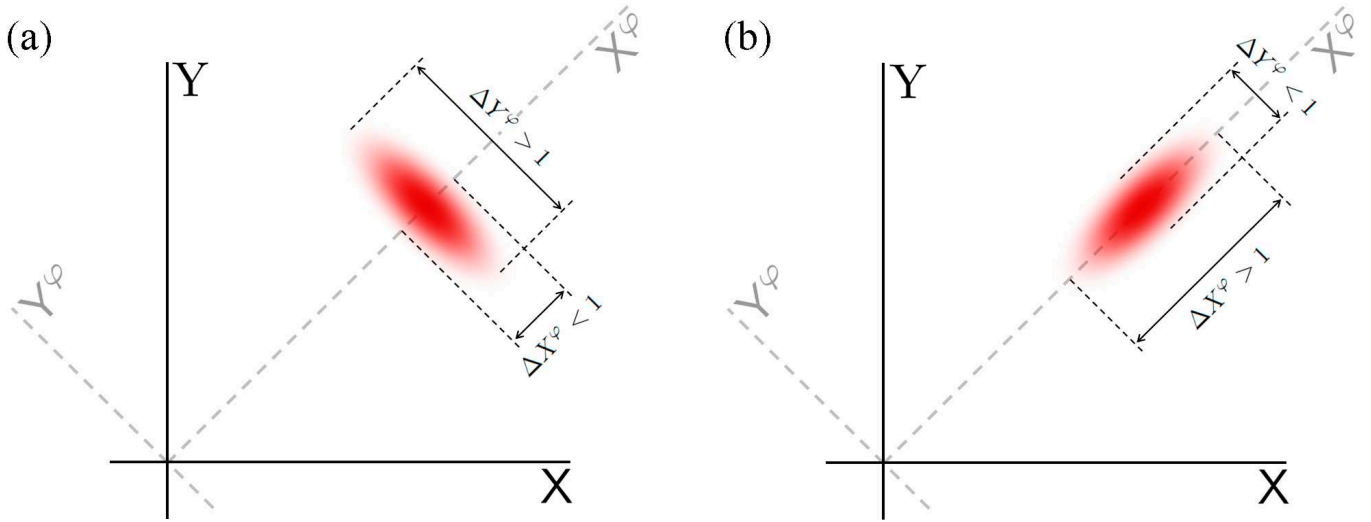}%
\caption{Phase space sketches of the squeezed states. At (a) the uncertainty
of the amplitude quadrature is reduced below the vacuum level at the expense
of increasing the uncertainty of the phase quadrature. At (b), on the other
hand, it is the phase quadrature the one which is squeezed, while the
amplitude quadrature becomes more noisy.}%
\label{fOsci4}%
\end{center}
\end{figure}

Squeezed states are the solution to this problem; the idea is the following.
Suppose that the signal is encoded in the amplitude of the oscillator. In a
coherent state the amplitude and phase quadratures are affected of equal
uncertainties $\Delta X^{\varphi}=\Delta Y^{\varphi}=1$; however, we can
conceive a state of the oscillator in which the uncertainty of the amplitude
quadrature is reduced, while that of the phase quadrature is increased, say
$\Delta X^{\varphi}\ll1$ and $\Delta Y^{\varphi}\gg1$, so that the product of
uncertainties keeps lower bounded by one, $\Delta X^{\varphi}\Delta
Y^{\varphi}\geq1$. The phase space sketch of such state is depicted in Figure
\ref{fOsci4}a. In this case the amplitude quadrature is well defined, and one
can in principle monitor its modulations with arbitrary accuracy. Of course,
this is accomplished at the expense of not being able to retrieve any
information from the phase quadrature, but if we only care about the signal
encoded in the amplitude quadrature that's not a problem.

We will define \textit{squeezed states} as those in which some quadrature, say
$\hat{X}^{\psi}$, has an uncertainty below the vacuum or coherent level, that
is, $\Delta X^{\psi}<1$. We then say that quadrature $\hat{X}^{\psi}$ is
\textit{squeezed.} Together with the \textit{amplitude squeezed state} already
introduced (for which $\psi=\varphi$), we show the phase space sketch of a
\textit{phase squeezed state }(for which $\psi=\varphi+\pi/2$) in Figure
\ref{fOsci4}. Let us now study one specially relevant type of squeezed states.%

\begin{figure}
[t]
\begin{center}
\includegraphics[
width=\textwidth
]%
{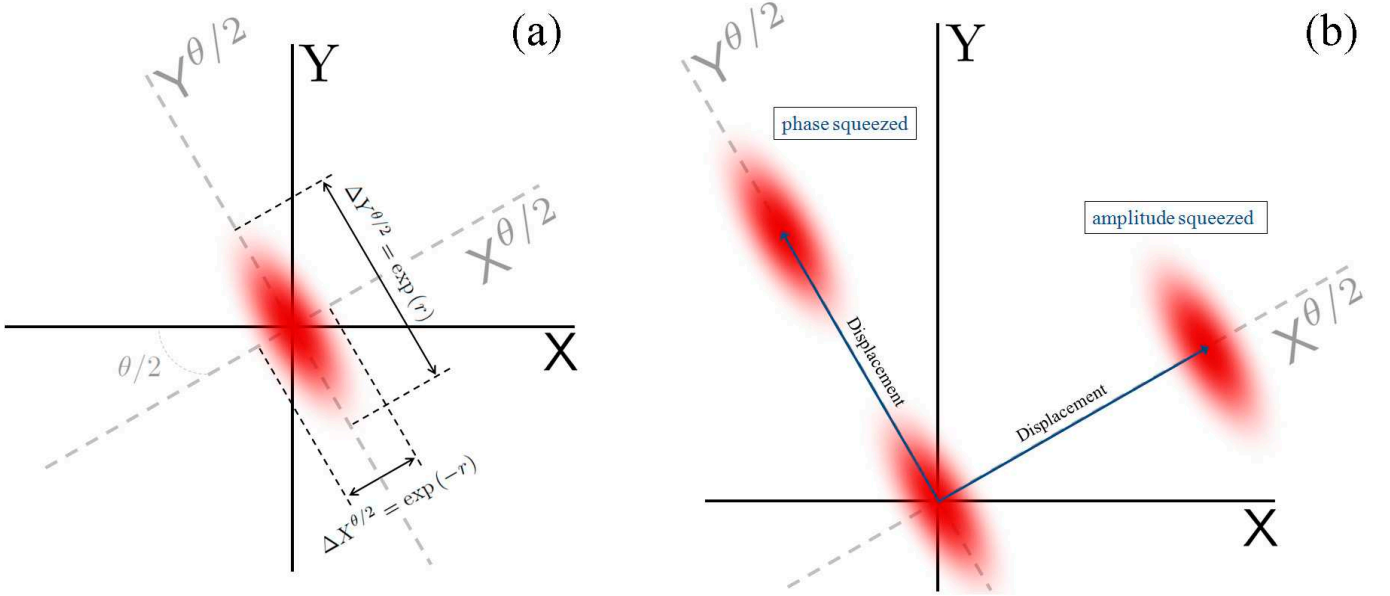}%
\caption{(a) Phase space sketch of a squeezed vacuum state. (b) Applying
displacements in the direction of the squeezed or antisqueezed quadratures,
one obtains amplitude or phase squeezed states, respectively.}%
\label{fOsci5}%
\end{center}
\end{figure}

\subsection{Minimum uncertainty squeezed states}

\textit{Minimum uncertainty states} are states which satisfy the lower bound
of the quadrature uncertainty relation, that is, $\Delta X^{\psi}\Delta
Y^{\psi}=1$ $\forall\psi$. The simplest of these states is the vacuum state
$|0\rangle$; any other number state $|n\neq0\rangle$ is not contained in this
class, as it is easily checked that it satisfies $\Delta X^{\psi}=2n+1$ for
any $\psi$. Coherent states, on the other hand, are minimum uncertainty
states, as they are obtained from vacuum by the displacement transformation,
which does not change the quadrature variances.

It is possible to generate squeezed states of this kind by using the
\textit{squeezing operator}%
\begin{equation}
\hat{S}\left(  z\right)  =\exp\left(  \frac{z^{\ast}}{2}\hat{a}^{2}-\frac
{z}{2}\hat{a}^{\dagger2}\right)  \text{,} \label{Ssq}%
\end{equation}
where $z\in%
\mathbb{C}
$ is called the \textit{squeezing parameter}. Similarly to the displacement
operator, this operator is unitary; hence, it is generated by making the
oscillator evolve with the Hamiltonian $\hat{H}_{S}=\mathrm{i}\hbar(z^{\ast
}\hat{a}^{2}-z\hat{a}^{\dagger2})/2T$ during a time $T$. In this thesis we
will explain how this Hamiltonian can be generated in optical systems.

Applied to the annihilation operator this evolution gives%
\begin{equation}
\hat{S}^{\dagger}\left(  z\right)  \hat{a}\hat{S}\left(  z\right)  =\hat
{a}\cosh r-e^{\mathrm{i}\theta}\hat{a}^{\dagger}\sinh r\text{ ,}%
\end{equation}
where we have written $z$ in the polar form $z=r\exp\left(  \mathrm{i}%
\theta\right)  $, and have used (\ref{BCHlemma}). In terms of quadratures,
these expressions are easily rewritten as%
\begin{equation}
\hat{S}^{\dagger}\left(  z\right)  \hat{X}^{\theta/2}\hat{S}\left(  z\right)
=e^{-r}\hat{X}^{\theta/2}\text{ \ \ \ \ and \ \ \ }\hat{S}^{\dagger}\left(
z\right)  \hat{Y}^{\theta/2}\hat{S}\left(  z\right)  =e^{r}\hat{Y}^{\theta
/2}\text{.} \label{SqueezingTransformation}%
\end{equation}
Suppose now that before the \textit{squeezing transformation} the state of the
system was vacuum, which has the statistical properties $\langle\hat{X}^{\psi
}\rangle=0$ and $\Delta X^{\psi}=1$ for all $\psi$ as already seen. After the
transformation (\ref{SqueezingTransformation}) the mean of any quadrature is
still zero, but the uncertainty of quadrature $\hat{X}^{\theta/2}$ has
decreased to $\Delta X^{\theta/2}=\exp\left(  -r\right)  $, while that of
quadrature $\hat{Y}^{\theta/2}$ has increased to $\Delta Y^{\theta/2}%
=\exp\left(  r\right)  $. Hence, the squeezing operator creates a
\textit{minimum uncertainty squeezed state}, that is, a state in which the
uncertainty of one quadrature is reduced below the vacuum level, while the
quadratures still satisfy the lower bound set by uncertainty relation.

The phase space sketch of this \textit{squeezed vacuum state} is showed in
Figure \ref{fOsci5}a. The uncertainty circle associated to the vacuum state
has turned into an ellipse, showing that the quadrature uncertainty along the
$\theta/2$ direction of phase space is reduced. An amplitude squeezed state
can be then created by applying a subsequent displacement along the $\theta/2$
axis as shown in Figure \ref{fOsci5}b. If the displacement is applied along
the $(\theta+\pi)/2$ direction, then a phase squeezed state is obtained. As
displacements do not change the uncertainty properties of the state, these
amplitude or phase squeezed states are still minimum uncertainty states.

\section{Entangled states}

\subsection{The EPR argument and quantum non-locality}

Even though Einstein is considered one of the founding fathers of quantum
mechanics, he always felt uncomfortable with its probabilistic character.
Fruit of this criticism, in 1935, and together with Podolsky and Rosen, he
came out with an argument which was supposed to tumble down the foundations of
quantum mechanics, showing in particular how the theory was both
\textit{incomplete} and \textit{inconsistent with causality} \cite{Einstein35}%
. Looking from our current perspective, it is quite ironic how the very same
ideas they introduced, far from destroying the theory, are the power source
for some of the most promising present applications of quantum physics.

Before introducing the EPR argument (standing for Einstein, Podolsky, and
Rosen), let us explain the properties of the eigensystem of the quadrature
operators. Consider the self--adjoint operator associated to the \textsf{X}
quadrature, that is $\hat{X}$. It is easy to argue that this operator has a
pure continuous spectrum\footnote{Indeed, from the relation%
\begin{equation}
\exp\left(  \frac{\mathrm{i}}{2}\xi\hat{Y}\right)  \hat{X}\exp\left(
-\frac{\mathrm{i}}{2}\xi\hat{Y}\right)  =\hat{X}+\xi,
\end{equation}
with $\xi\in%
\mathbb{R}
$, which is easily proved with the Baker--Campbell--Haussdorf lemma
(\ref{BCHlemma}), it follows that if $|x\rangle$ is an eigenvector of $\hat
{X}$ with $x$ eigenvalue, then the vector $\exp(-\mathrm{i}\xi\hat
{Y}/2)|x\rangle$ is also an eigenvector of $\hat{X}$ with eigenvalue $x+\xi$.
Now, as this holds for any real $\xi$, we conclude that the spectrum of
$\hat{X}$ is the whole real line.} $\{x\}_{x\in%
\mathbb{R}
}$, with corresponding eigenvectors $\{|x\rangle\}_{x\in%
\mathbb{R}
}$ which are Dirac--normalized, that is, $\langle x|x^{\prime}\rangle
=\delta\left(  x-x^{\prime}\right)  $. Of course, this vectors cannot describe
a physical state of the harmonic oscillator, as they do not belong to its
Hilbert space (they cannot be properly normalized, see Section
\ref{InfiniteHilbert}). In fact, an eigenstate $|x\rangle$ corresponds to the
infinitely squeezed state $|x\rangle=\mathrm{lim}_{r\rightarrow\infty}\hat
{D}\left(  x/2\right)  \hat{S}\left(  r\right)  |0\rangle$, which is
unphysical as it leads to an infinite mean energy of the oscillator, see Eq.
(\ref{MeanEnergyQuadratures}). Similar reasoning applies to the \textsf{Y}
quadrature, whose associated self--adjoint operator $\hat{Y}$ has a continuous
spectrum $\{y\}_{y\in%
\mathbb{R}
}$ with corresponding eigenvectors $\{|y\rangle\}_{y\in%
\mathbb{R}
}$.

The EPR argument starts by considering two harmonic oscillators with Hilbert
spaces $\mathcal{H}_{A}$ and $\mathcal{H}_{B}$, which after interacting for a
while are left in the state%
\begin{equation}
|EPR\rangle=\int dx|x,x\rangle.
\end{equation}
Note that this state can be written also as%
\begin{equation}
|EPR\rangle=\int dy|y,-y\rangle,
\end{equation}
in terms of the eigenstates of the \textsf{Y} quadratures of the oscillators.
Note also that $|EPR\rangle$ cannot be a true state of the oscillators, as it
cannot be normalized, but let us forget about this detail for the sake of the
argument; we will deal later with a realistic situation. The oscillators $A$
and $B$ are then given, respectively, to \textit{Alice} and \textit{Bob}, two
observers placed at distant locations, so that they are not able to interact anymore.

EPR argue then as follows. Imagine that Alice measures $\mathsf{X}$ and
obtains the result\footnote{Again, this is an idealized situation used just
for the sake of argumentation; this is because having a continuous spectrum, a
measurement of $\hat{X}$ cannot give a definite number $x_{0}$.} $x_{0}$;
according to quantum mechanics the state of oscillators collapses to
$|x_{0},x_{0}\rangle$, and hence, any subsequent measurement of $\mathsf{X}$
performed by Bob will reveal that his oscillator has a definite value of this
quadrature, $x_{0}$. However, Alice could have measured $\mathsf{Y}$ instead,
obtaining for example the result $y_{0}$; in this case, quantum mechanics says
that the state would have collapsed to $|y_{0},-y_{0}\rangle$, after which Bob
would have concluded that his oscillator had a definite value of its
\textsf{Y} quadrature, $-y_{0}$. Now, and this is the center of all the
argument,\textit{ assuming that nothing Alice may do can alter the physical
state of Bob's oscillator} (the oscillators are separated, even
\textit{space--like} or \textit{causally} during the life of Alice and Bob if
we like!), one must conclude that the oscillator $B$ must had well defined
values of both its \textsf{X} and \textsf{Y} quadratures from the beginning,
hence \textit{violating the quantum mechanical uncertainty relation} $\Delta
X\Delta Y\geq1$, and showing that quantum mechanics is inconsistent.

Even though it seems a completely reasonable statement (specially at 1935,
just a decade after the true birth of quantum mechanics), the center of their
argument is actually its flaw. The reason is that the state of the system is
not an \textit{element of reality} (in EPR's words), it is just a
mathematically convenient object which describes the statistics that would be
obtained if a physical observable is measured. Consequently, causality does
not apply to it: \textit{The actions of Alice can indeed alter Bob's state},
even if these are causally disconnected. Of course, a completely different
matter is whether Alice and Bob can use this \textit{spooky action at a
distance} (in Einstein's words) to transmit information superluminically. Even
though there is no rigorous proof for the negative answer to this question,
such violation of causality has never been observed or predicted, even in the
most sound and subtle applications of this \textit{quantum non-local effects}
(like \textit{teleportation}), and hence, most physicists believe that despite
non-local effects at the level of states, quantum mechanics cannot violate
causality in any way.

\subsection{Entanglement and the two--mode squeezing operator}

The work of Einstein, Podolsky, and Rosen is the very best example of how one
can make advances in a theory by trying to disprove it. Even if their
motivation was based on wrong ideas, they were the first ones to realize that
in quantum mechanics it is possible to create correlations between systems
which go beyond those admitted in the classical world. Such states were coined
\textit{entangled states} by Schr\"{o}dinger, which was actually supportive of
the EPR ideas, and a strong believer of the incompleteness of quantum
mechanics. After the 60's, physicists stopped looking at these states as the
puzzle EPR suggested they were, and started searching for possible
applications of them to various problems. Bell was the first one who realized
the potential of such states, proving that they could be used to rule out the
incompleteness of quantum mechanics \cite{Bell64} (exactly the opposite of
what EPR created them for!), or, in other words, to prove that the
probabilistic character of quantum mechanics does not come from some missing
information we fail to account for, but from a probabilistic character of
nature itself\footnote{Strictly speaking, he proved that no \textit{local}
\textit{hidden--variables theory} is consistent with the predictions of
quantum mechanics.}. Nowadays, entangled states have been shown to be a
resource for remarkable applications such as the fast\textit{ }performance of
computational tasks that would be impossible to perform classically (like
factorization of large numbers \cite{Shor97}, which is actually at the core of
every present cryptographic system). This section is devoted to understand a
little deeper these kind of states, as well as explaining how they can be
generated in harmonic oscillator systems.

As in the previous section, consider two harmonic oscillators with joint
Hilbert space $\mathcal{H}_{A}\otimes\mathcal{H}_{B}$. If the state of the
joint system is of the type $\hat{\rho}_{AB}^{(\mathrm{t})}=\hat{\rho}%
_{A}\otimes\hat{\rho}_{B}$, that is, a tensor product of two arbitrary density
operators, the actions performed by Alice on the $A$ oscillator won't affect
Bob's oscillator, the statistics of which are given by $\hat{\rho}_{B}$, no
matter the actual state $\hat{\rho}_{A}$ (see Axiom V in Appendix
\ref{QuantumMechanics}). In this case $A$ and $B$ are \textit{uncorrelated}.
For any other type of joint state, $A$ and $B$ will share some kind of correlation.

Correlations are not strange in classical systems; hence, the problem in
quantum mechanics is to distinguish between correlations which can appear at a
classical level, and correlations which are purely quantum, as only in the
latter case one can expect to exploit them in applications requiring
entanglement. Intuitively, a state will have only classical correlations if,
starting from a state of the $\hat{\rho}_{AB}^{(\mathrm{t})}$ type, Alice and
Bob can prepare it by making use only of \textit{local operations} (such as
local unitaries or measurements) and \textit{classical communication }(such as
phone calls). It is not difficult to convince oneself that the most general
state that can be created by such means has the so-called \textit{separable
form}%
\begin{equation}
\hat{\rho}_{AB}^{(\mathrm{s})}=\sum_{k}w_{k}\hat{\rho}_{A}^{(k)}\otimes
\hat{\rho}_{B}^{(k)}, \label{RhoSep}%
\end{equation}
where the $\hat{\rho}^{(k)}$'s are density operators and $\sum_{k}w_{k}=1$.
Any state which cannot be written as a convex mixture of tensor product states
will induce quantum correlations between $A$ and $B$, and will therefore be an
entangled state in the spirit of $|EPR\rangle$.

There is yet another way of justifying that states which cannot be written in
the separable form (\ref{RhoSep}) will make $A$ and $B$ share quantum
correlations. The idea is that the fundamental difference between classical
and quantum mechanics is the superposition principle. Hence, it is intuitive
that correlations will have a quantum nature when the joint state of the
oscillators exploits the concept of \textquotedblleft superposition of joint
states\textquotedblright, that is, when it cannot be written as a tensor
product of two independent states of the oscillators, or as a purely classical
statistical mixture of these, which corresponds exactly to (\ref{RhoSep}).

Given a general state $\hat{\rho}_{AB}\in\mathcal{H}_{A}\otimes\mathcal{H}%
_{B}$ it is hard to find out whether it is separable or not. However, for the
class of so-called \textit{Gaussian states} (to which the states to be
discussed in this thesis belong), a \textit{necessary and} \textit{sufficient}
criterion is known. This criterion was proposed independently by Duan et al.
\cite{Duan00} and Simon \cite{Simon00}, and it can be formulated as
follows\footnote{Even though this criterion is much simpler, it is worth
remarking that for the ten years before its formulation, the only
experimentally testable criterion for entanglement was that of Reid's
\cite{Reid89a}; this criterion was based on how much information one can infer
about the quadratures of one of the oscillators by performing measurements on
the other, and in a sense expanded the EPR ideas to entangled states with
imperfect correlations.}. Consider two orthogonal quadratures $\{\hat{X}%
_{A}^{\varphi},\hat{Y}_{A}^{\varphi}\}$ and $\{\hat{X}_{B}^{\varphi},\hat
{Y}_{B}^{\varphi}\}$ for the oscillators $A$ and $B$, respectively; then their
joint state is separable if and only if%
\begin{equation}
W_{AB}^{\varphi}=V[(\hat{X}_{A}^{\varphi}-\hat{X}_{B}^{\varphi})/\sqrt
{2}]+V[(\hat{Y}_{A}^{\varphi}+\hat{Y}_{B}^{\varphi})/\sqrt{2}]\geq2,
\label{EntanglementCrit}%
\end{equation}
for every $\varphi$. This criterion can be seen as a formalization of the ideas
introduced in the EPR article; just note that a coherent state has $V[(\hat
{X}_{A}^{\varphi}-\hat{X}_{B}^{\varphi})/\sqrt{2}]=V[(\hat{Y}_{A}^{\varphi
}+\hat{Y}_{B}^{\varphi})/\sqrt{2}]=1$ $\forall\varphi$, and hence if
$W_{AB}^{\varphi}<2$ one can be sure that two orthogonal quadratures of the
oscillators are correlated above what is classically allowed. For example, a
maximal violation of this inequality is obtained for the unphysical
$|EPR\rangle$ state, which has $W_{AB}^{0}=0$, showing that there is a perfect
correlation (anticorrelation) between the \textsf{X} (\textsf{Y}) quadratures
of the oscillators. Note that this criterion does not quantify the amount of
entanglement present in the state, it offers just a way to prove whether a
state is separable or not (see the next section for quantitative
entanglement). In practice, however, and incorrectly without further
arguments, it customary to state that the lower $W_{AB}^{\varphi}$ is, the
larger the amount of entanglement is.

Let us now explain how to generate a class of pure entangled states which
coincide with $|EPR\rangle$ in some (unphysical) limit. We will call
\textit{EPR-like states} to such states. To proceed, note that the conditions
$V[(\hat{X}_{A}^{\varphi}-\hat{X}_{B}^{\varphi})/\sqrt{2}]<1$ and $V[(\hat
{Y}_{A}^{\varphi}+\hat{Y}_{B}^{\varphi})/\sqrt{2}]<1$ are actually quite
reminiscent of the quadrature squeezing that we introduced in the previous
section, with the difference that now the squeezing is present in a joint
quadrature. Let us call $\hat{a}$ and $\hat{b}$ to the annihilation operators
for the $A$ and $B$ harmonic oscillators, respectively. Consider the unitary
operator%
\begin{equation}
\hat{S}_{AB}\left(  z\right)  =\exp(z^{\ast}\hat{a}\hat{b}-z\hat{a}^{\dagger
}\hat{b}^{\dagger}),\label{Sab}%
\end{equation}
which we will call the \textit{two--mode squeezing operator}, which is easily
proved to transform the annihilation operators as%
\begin{subequations}
\begin{align}
\hat{S}_{AB}^{\dagger}\left(  z\right)  \hat{a}\hat{S}_{AB}\left(  z\right)
& =\hat{a}\cosh r-e^{\mathrm{i}\theta}\hat{b}^{\dagger}\sinh r\text{,}\\
\hat{S}_{AB}^{\dagger}\left(  z\right)  \hat{b}\hat{S}_{AB}\left(  z\right)
& =\hat{b}\cosh r-e^{\mathrm{i}\theta}\hat{a}^{\dagger}\sinh r\text{,}%
\end{align}
where $z=r\exp(\mathrm{i}\theta)$. Using these expressions and their
conjugates, it is straightforward to prove that if the oscillators are
initially in their vacuum state, then%
\end{subequations}
\begin{subequations}
\begin{align}
V[(\hat{X}_{A}^{\theta/2}+\hat{X}_{B}^{\theta/2})/\sqrt{2}] &  =V[(\hat{Y}%
_{A}^{\theta/2}-\hat{Y}_{B}^{\theta/2})/\sqrt{2}]=\exp(-2r),\\
V[(\hat{X}_{A}^{\theta/2}-\hat{X}_{B}^{\theta/2})/\sqrt{2}] &  =V[(\hat{Y}%
_{A}^{\theta/2}+\hat{Y}_{B}^{\theta/2})/\sqrt{2}]=\exp(2r).
\end{align}
Hence, applied to the vacuum state of the oscillators, the two--mode squeezing
operator generates an entangled state, the so-called \textit{two--mode
squeezed vacuum state}, as $W_{AB}^{(\theta-\pi)/2}=2\exp(-2r)<2$. In
particular, for $\theta=\pi$ and $r\rightarrow\infty$ we have $W_{AB}%
^{0}\rightarrow0$, and hence the $|EPR\rangle$ state is recovered. For a
finite value of $r$ we then get an EPR state with imperfect correlations
between orthogonal quadratures of the oscillators, these correlations being
always above what is classically permitted. In Chapter \ref{OPOs} we will
learn how to generate this type of states.%

\begin{figure}[t]
\begin{center}
\includegraphics[width=\textwidth]{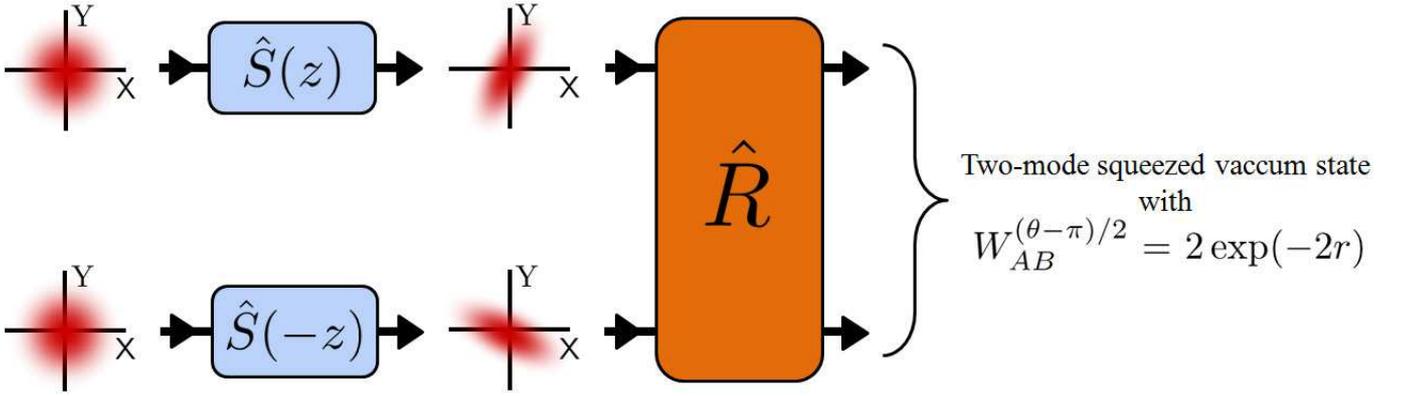}
\caption{Schematic relation between entanglement and squeezing: Two
orthogonally squeezed vacua are connected to a two-mode squeezed vacuum state
by the beam splitter transformation $\hat{R}$.}%
\label{fOsci6}%
\end{center}
\end{figure}

The connection between squeezing and entanglement can be made even more
explicit. To this aim, assume that we induce an evolution of the entangled
oscillators $A$ and $B$ corresponding to the unitary operator%
\end{subequations}
\begin{equation}
\hat{R}=\exp\left[  \frac{\pi}{4}(\hat{a}\hat{b}^{\dagger}-\hat{a}^{\dagger
}\hat{b})\right]  \text{,} \label{Rop}%
\end{equation}
what in the Heisenberg picture means that the boson operators are transformed
as%
\begin{subequations}
\begin{align}
\hat{a}_{1} &  =\hat{R}^{\dagger}\hat{a}\hat{R}=(\hat{b}+\hat{a})/\sqrt{2}\\
\hat{a}_{2} &  =\hat{R}^{\dagger}\hat{b}\hat{R}=(\hat{b}-a)/\sqrt{2}\text{.}%
\end{align}
In terms of the new, independent boson operators $\hat{a}_{1}$ and $\hat
{a}_{2}$, the two-mode squeezing operator is written as%
\end{subequations}
\begin{equation}
\hat{S}_{AB}\left(  z\right)  =\exp\left(  \frac{z^{\ast}}{2}\hat{a}_{1}%
^{2}-\frac{z}{2}\hat{a}_{1}^{\dagger2}-\frac{z^{\ast}}{2}\hat{a}_{2}^{2}%
+\frac{z}{2}\hat{a}_{2}^{\dagger2}\right)  =\hat{S}_{1}\left(  z\right)
\hat{S}_{2}\left(  -z\right)  ,
\end{equation}
that is, as two individual squeezing operators for each of the new modes. Read
backwards, this shows that one can entangle two oscillators by squeezing them
along two orthogonal directions of phase space, and then make them interact
according to the unitary operator (\ref{Rop}), see Figure \ref{fOsci6}. In the
case of light we shall see that the unitary transformation (\ref{Rop})
corresponds to a 50/50 beam splitter. This means that one can generate two
entangled optical beams by mixing in a beam splitter two beams that have been
previously squeezed; this is interesting because, as we shall see in Chapter
\ref{OPOs}, we are now in position to generate light with large levels of
squeezing experimentally.

\subsection{Enhancing entanglement by addition and subtraction of excitations}

In this section we will introduce part of the work initiated during a
three--month visit to the Massachusetts Institute of Technology by the author
of this thesis \cite{NavarreteUNc}. In particular, we have been able to prove
that Alice and Bob can enhance the entanglement of the two--mode squeezed
vacuum state\ by annihilating or creating excitations on their corresponding
oscillators\footnote{These operations correspond to the application of the
annihilation or creation of excitations through the operators $\hat{a}$ and
$\hat{a}^{\dagger}$. Not being unitary, these operators cannot correspond to
any Hamiltonian interaction; nevertheless, by using measurement--based
techniques, they can be simulated probabilistically (that is, conditioned to a
specific outcome in some measurement). Nowadays, it is possible to perform
photon addition and subtraction from light beams
\cite{Wenger04,Zavatta04,Zavatta07,Parigi07,Zavatta09}. In particular, photon
subtraction is accomplished by making the beam pass through a beam splitter
with very high transmittance; whenever a photon is detected in the reflected
beam, it means that photon subtraction has been accomplished in the
transmitted beam \cite{Wenger04}. Photon addition is a little more involved
\cite{Zavatta04}; in these case the beam acts as the seed of the signal mode
in a down-conversion process, and whenever a photon is detected in the
conjugate spatial mode (the idler mode), it means that a photon has been added
to the beam. These operations have been used to make direct tests of things as
fundamental as the bosonic commutation relations \cite{Parigi07,Zavatta09}.}.
Even though this result was known prior to our work, we offer a systematic
analysis of such phenomenon, as well as prove it analytically for the very
first time in a simple (but relevant) scheme that we introduce here.%

\begin{figure}
[ptb]
\begin{center}
\includegraphics[
height=2.0591in,
width=4.0776in
]%
{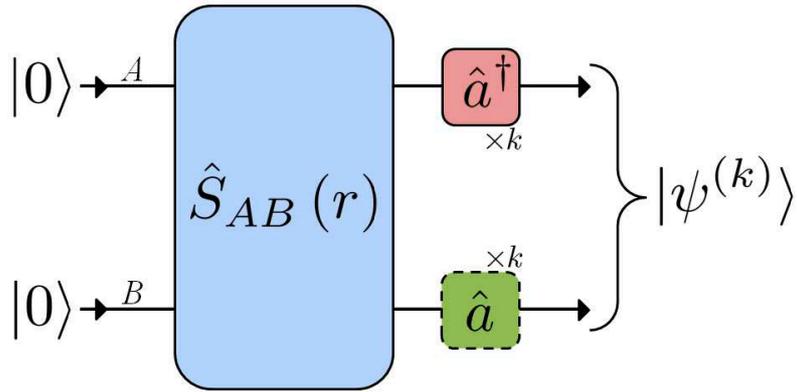}%
\caption{Schematic representation of the example studied in this section.
After the generation of the two--mode squeezed vacuum state, Alice adds $k$
excitations to her oscillator (pink box with solid contour), or Bob subtracts
$k$ excitations out of his (green box with dashed contour); in either case the
final state is $|\psi^{(k)}\rangle$, and what we have been able to prove is
that the entanglement increases with the number of operations, $k$.}%
\label{fOsci7}%
\end{center}
\end{figure}

Let us start by explaining how to quantify the entanglement present in a joint
state of the oscillators. Even though we pretty much understand the conditions
that a good entanglement measure $\mathcal{E}[\hat{\rho}_{AB}]$ must
satisfy\footnote{Technically, it must be an \textit{entanglement monotone},
see \cite{EisertUN} for a clear and concise explanation.}, we haven't been
able to build a satisfactory one for general bipartite states (the ones
introduced so far either do not satisfy all the needed conditions, or can only
be evaluated efficiently for restricted classes of states). Fortunately, when
dealing only with pure states $\hat{\rho}_{AB}=|\psi\rangle_{AB}\langle\psi|$,
it has been proved that there is a \textit{unique} entanglement measure, the
so-called \textit{entanglement entropy}. Its definition is actually quite
intuitive: One just needs to check how mixed the reduced state of one party is
left when we trace the subspace corresponding to the other party.
Operationally this means that, given the reduced states $\hat{\rho}%
_{A}=\mathrm{tr}_{B}\{|\psi\rangle_{AB}\langle\psi|\}$ or $\hat{\rho}%
_{B}=\mathrm{tr}_{A}\{|\psi\rangle_{AB}\langle\psi|\}$, this entanglement
measure is given by%
\begin{equation}
\mathcal{E}[|\psi\rangle_{AB}]\equiv\mathcal{S}[\hat{\rho}_{A}]=\mathcal{S}%
[\hat{\rho}_{B}],
\end{equation}
where $\mathcal{S}[\hat{\rho}]=-\mathrm{tr}\{\hat{\rho}\log\hat{\rho}\}$ is
the so-called \textit{von\ Neumann entropy}, which measures how mixed the
state $\hat{\rho}$ is \cite{Nielsen00book}.

It is particularly simple to evaluate this entanglement measure when the state
is given in its so-called Schmidt form\footnote{It is simple to prove that
every pure bipartite state can be uniquely written in this form
\cite{Nielsen00book}.},%
\begin{equation}
|\psi\rangle_{AB}=\sum_{j=1}^{\infty}\sqrt{p_{j}}|u_{j}\rangle\otimes
|v_{j}\rangle,
\end{equation}
being $\{p_{j}\}_{j=1,...,\infty}$ a probability distribution, and
$\{|u_{j}\rangle\}_{j=1,...,\infty}$ and $\{|v_{j}\rangle\}_{j=1,...,\infty}$
two orthonormal bases. In this case, the reduced density operators are given
by $\hat{\rho}_{A}=\sum_{j=1}^{\infty}p_{j}|u_{j}\rangle\langle u_{j}|$ and
$\hat{\rho}_{B}=\sum_{j=1}^{\infty}p_{j}|v_{j}\rangle\langle v_{j}|$, which
are already in their diagonal form, and hence we have%
\begin{equation}
\mathcal{E}[|\psi\rangle_{AB}]=-\sum_{j=1}^{\infty}p_{j}\log p_{j}.
\label{SchmidtE}%
\end{equation}

Now that we understand how to evaluate the amount of entanglement present in a
pure bipartite state, we can pass to analyze the scheme depicted in Figure
\ref{fOsci7}. The idea is that once the two oscillators (initialized in its
vacuum state) go through the two--mode squeezer (with $\theta=0$ for
simplicity), Alice adds $k$ excitations, that is, she applies $k$ times the
creation operator to her oscillator. It is not difficult to show that in the
number state basis, the two--mode squeezed vacuum state reads
\cite{Gerry05book}%
\begin{equation}
|\mathrm{TMSV}\rangle=\hat{S}_{AB}\left(  r\right)  |0,0\rangle=\sum
_{n=0}^{\infty}\sqrt{1-\lambda^{2}}\lambda^{n}|n,n\rangle,
\end{equation}
with $\lambda=\tanh r$. This state is already in Schmidt form with a
distribution given by $p_{n}=(1-\lambda^{2})\lambda^{2n}$, which plugged in
(\ref{SchmidtE}) leads to the entanglement entropy%
\begin{equation}
\mathcal{E}[|\mathrm{TMSV}\rangle]=\log(1-\lambda^{2})-\frac{2\lambda^{2}%
}{1-\lambda^{2}}\log\lambda.
\end{equation}
It is very simple to check that this function is monotonically increasing with
$r$, just as expected by the preceding section (quantum correlations are
larger the larger is the squeezing, and hence the entanglement should increase
with $r$).

Now we need to find the state after the $k$ additions performed by Alice; it
is completely straightforward to show by using (\ref{DownUp}), that the final
state reads%
\begin{equation}
|\psi^{(k)}\rangle=\sum_{n=0}^{\infty}\sqrt{p_{n}^{(k)}}|n+k,n\rangle,
\end{equation}
with%
\begin{equation}
p_{n}^{(k)}=(1-\lambda^{2})^{k+1}\lambda^{2n}\binom{n+k}{n}.
\end{equation}
It is worth noting that this is exactly the same state that is obtained when
Bob performs $k$ subtractions on his oscillator, that is, Alice adding
excitations is the same as Bob subtracting them (see Figure \ref{fOsci7}).

Just as with the two--mode squeezed vacuum state, this state is already in
Schmidt form (check that $p_{n}^{(k)}$ is a probability distribution in $n$),
and hence the entanglement entropy is easily evaluated as $\mathcal{E}%
^{(k)}=-\sum_{n=0}^{\infty}p_{n}^{(k)}\log p_{n}^{(k)}$. Unfortunately, this
sum is not analytical except for $k=0$; nevertheless, we can prove that it is
an increasing function of $k$ as follows. Using the pascal identity%
\begin{equation}
\binom{n+k+1}{k+1}=\binom{n+k}{k+1}+\binom{n+k}{k},
\end{equation}
we can write%
\begin{equation}
p_{n}^{(k+1)}=\lambda^{2}p_{n-1}^{(k+1)}+(1-\lambda^{2})p_{n}^{(k)},
\end{equation}
where we set $p_{-n}^{(k)}=0$ for $n>0$ by definition. Now, because of the
strict concavity of the function $h\left(  x\right)  =-x\log x$, we have%
\begin{equation}
\sum_{n=0}^{\infty}h\left[  p_{n}^{(k)}\right]  \geq\lambda^{2}\sum
_{n=0}^{\infty}h\left[  p_{n-1}^{(k+1)}\right]  +(1-\lambda^{2})\sum
_{n=0}^{\infty}h\left[  p_{n}^{(k)}\right]  ,
\end{equation}
and since $\sum_{n=0}^{\infty}h\left[  p_{n-1}^{(k+1)}\right]  =\sum
_{n=0}^{\infty}h\left[  p_{n}^{(k+1)}\right]  $, this expression is equivalent
to%
\begin{equation}
\mathcal{E}^{(k+1)}\geq\mathcal{E}^{(k)}\text{.}%
\end{equation}
Hence, we have been able to prove analytically that entanglement increases
with the number of additions/subtractions when acting on one mode of the
entangled pair. The (numerical) discussion concerning the case of Alice and
Bob acting both simultaneously on their respective oscillators can be found in
\cite{NavarreteUNc}, but nothing qualitatively different is found.

In a second work \cite{GarciaPatronUN} we apply similar ideas in an effort to
prove the so-called \textit{minimum entropy conjecture for bosonic channels}
\cite{Giovanetti04a,Giovanetti04b}, which states that input coherent states
are the states which minimize the output entropy of the most relevant class of
optical communication channels (\textit{phase--insensitive Gaussian channels}
\cite{Holevo99,Holevo01}), or, equivalently, that they maximize the capacity
of those channels, that is, the number of bits per second that can be
transmitted reliably through the channel.

\section{The harmonic oscillator in phase space}

\subsection{Phase space distributions}

In the previous sections we have been able to make phase space sketches of
various quantum states of the harmonic oscillator, which in particular allow
us to visualize the statistics that would be obtained when measuring its
quadratures. In this section we show how these sketches can be made in a
rigorous way, by making a correspondence between the quantum state of the
oscillator and a distribution in phase space \cite{Schleich01book}. One can
tend to think that this suggests that quantum mechanics can be formulated just
as noise impregnating the classical phase space trajectories, but the
situation is a lot more subtle as we will show: Either these distributions are
not probability density functions in the usual sense (they can be negative or
strongly divergent), or the way that averages are made with them is not the
usual one.

In the following we consider that the oscillator is in some generic state
$\hat{\rho}$, and review three natural ways of defining a phase space
distribution associated to it.

\textbf{The Wigner distribution}. As we have already argued, the quadratures
$\hat{X}^{\psi}$ posses a continuous spectrum $\{x^{\psi}\}_{x^{\psi}\in%
\mathbb{R}
}$ and hence a continuous set of generalized eigenvectors $\{|x^{\psi}%
\rangle\}_{x^{\psi}\in%
\mathbb{R}
}$. Hence, quantum mechanics assigns a probability density function $\langle
x^{\psi}|\hat{\rho}|x^{\psi}\rangle$ to any point $x^{\psi}=x\cos\psi
+y\sin\psi$ of the phase space $\left(  x,y\right)  $, which completely
describes the statistics of a measurement of the $\hat{X}^{\psi}$ quadrature.

It follows that a natural way of defining a phase space distribution, say
$W\left(  x,y\right)  $, is then as the one whose marginals $\int_{-\infty
}^{+\infty}dyW\left(  x,y\right)  $ and $\int_{-\infty}^{+\infty}dxW\left(
x,y\right)  $ give, respectively, the probability density functions $\langle
x|\hat{\rho}|x\rangle$ and $\langle y|\hat{\rho}|y\rangle$ corresponding to a
measurement of the \textsf{X} and \textsf{Y} quadratures. It is possible to
show that such a distribution, which is known as the \textit{Wigner
distribution} \cite{Wigner32}, can be uniquely defined as%
\begin{equation}
W\left(  x,y\right)  =\frac{1}{4\pi}\int_{%
\mathbb{R}
}d\xi\exp(-\mathrm{i}y\xi/2)\langle x+\xi/2|\hat{\rho}|x-\xi/2\rangle.
\end{equation}
It can also be shown that given an operator $\hat{A}(\hat{X},\hat{Y})$, we can
evaluate its expectation value as%
\begin{equation}
\langle\hat{A}\rangle=\int_{%
\mathbb{R}
^{2}}dxdyW\left(  x,y\right)  A^{\left(  \mathrm{s}\right)  }\left(
x,y\right)  , \label{SymmMean}%
\end{equation}
where $A^{\left(  \mathrm{s}\right)  }$ is obtained by writing $\hat{A}%
(\hat{X},\hat{Y})$ as a symmetric function of $\hat{X}$ and $\hat{Y}$ with the
help of the commutation relations, and then changing these operators by the
real variables $x$ and $y$, respectively.

This expression deserves some comments. Note that when discussing the axioms
of quantum mechanics (Section \ref{Axioms}) we decided that a reasonable
strategy to generate the self--adjoint operators associated to a given
classical observable was to first symmetrize its associated classical phase
space function and then change the position and momenta by the corresponding
self--adjoint operators; hence, it may look like Eq. (\ref{SymmMean}) is
telling us that quantum mechanics appears \textit{just} as noise acting onto
the classical picture of the system, that is, by introducing probability
density function $W\left(  x,y\right)  $ in phase space to which every
observable must be averaged. Even though this intuitive picture might be
useful to understand certain quantum phenomena (like squeezing), this is far
from the end of the story; quantum mechanics is a little more subtle: Even
though $W\left(  x,y\right)  $ is normalized to one and is a real function of
the phase space variables, it can take negative values \cite{Schleich01book},
and hence it is not a true probability density function. In other words, in
general quantum mechanics \textit{cannot} be simulated by adding classical
noise to the system. As an example, for coherent and squeezed states it is
possible to show that the Wigner distribution is positive all over phase space
\cite{Schleich01book}, and hence the interpretation of quantum mechanics as
quantum noise acting onto the classical trajectories is somehow correct;
however, this interpretation breaks down for number states, as in that case
the Wigner function takes negative values \cite{Schleich01book}.

\textbf{The Husimi distribution}. Coherent states offer an alternative way of
building a phase space distribution. We saw that they are specified by a
complex parameter $\alpha$, \textit{coherent amplitude} in the following, and
admit a simple representation in a phase space formed with the variables
$\left(  x,y\right)  =\left(  2\operatorname{Re}\left\{  \alpha\right\}
,2\operatorname{Im}\left\{  \alpha\right\}  \right)  $. These suggests that we
could define another phase space distribution, say $Q\left(  x,y\right)  $, by
asking to the state $\hat{\rho}$ of the oscillator how much does it project
onto the set of coherent states $\left\{  |\alpha\rangle\right\}  _{\alpha\in%
\mathbb{C}
}$, that is,%
\begin{equation}
Q\left(  \alpha\right)  =\frac{\langle\alpha|\hat{\rho}|\alpha\rangle}{\pi
}\text{,}%
\end{equation}
where the factor $\pi^{-1}$ appears to ensure that $Q$ is normalized to one.
This simple distribution is called the \textit{Husimi} $Q$
\textit{distribution} \cite{Husimi40}, and has a very nice property: It is
positive and finite for any state of the oscillator\footnote{Note that, in
particular, it is lower bounded by $0$ and upper bounded by $1/\pi$, as
follows from the positivity and the unit trace of $\hat{\rho}$, respectively},
and hence it is a true phase space probability distribution.

It has however a drawback: Its marginals do not give the proper probability
density functions for quadrature measurements. Nevertheless, we can still
evaluate the expectation value of an operator $\hat{A}\left(  \hat{a},\hat
{a}^{\dagger}\right)  $ with it as%
\begin{equation}
\langle\hat{A}\rangle=\int_{%
\mathbb{C}
}d^{2}\alpha Q\left(  \alpha\right)  A^{\left(  \mathrm{a}\right)  }\left(
\alpha,\alpha^{\ast}\right)  ,
\end{equation}
where $A^{\left(  \mathrm{a}\right)  }$ is obtained by writing $\hat{A}\left(
\hat{a},\hat{a}^{\dagger}\right)  $ in anti--normal order with the help of the
commutation relations, and then changing the annihilation and creation
operators by $\alpha$ and $\alpha^{\ast}$, respectively. Note that this
expression cannot be seen as an average of the classical phase space function
associated to the observable because of the antinormal order.

Note finally that it is possible to characterize the full state $\hat{\rho}$
by only the diagonal elements $\langle\alpha|\hat{\rho}|\alpha\rangle$ because
coherent states form an overcomplete basis, as we already discussed in Section
\ref{CohProp}.

\textbf{The Glauber--Sudarshan distribution}. Instead of asking how much does
our state project onto a coherent state, we can assign a phase space
distribution in the complex--$\alpha$ plane with a different but equally
reasonable question: How do we have to mix the set of coherent states
$\left\{  |\alpha\rangle\right\}  _{\alpha\in%
\mathbb{C}
}$ to generate the state $\hat{\rho}$ of the oscillator? In other words, we
are asking for the distribution $P\left(  \alpha\right)  $ satisfying%
\begin{equation}
\hat{\rho}=\int_{%
\mathbb{C}
}d^{2}\alpha P\left(  \alpha\right)  |\alpha\rangle\langle\alpha|\text{.}
\label{Pgs}%
\end{equation}
If the coherent states were orthogonal, the answer would be the\textit{ }$Q$
distribution; however, as they are not, this $P\left(  \alpha\right)  $, known
as the \textit{Glauber--Sudarshan} $P$ \textit{distribution}
\cite{Glauber63,Sudarshan63}, is a completely different distribution. In fact,
this distribution can be negative and has in general strong divergences such
as derivatives of the delta function. As an example, here we show the $P$
distribution associated to a coherent and a number state%
\begin{subequations}
\begin{align}
P_{|\alpha_{0}\rangle}\left(  \alpha\right)   &  =\delta^{2}(\alpha-\alpha
_{0}),\\
P_{|n\rangle}\left(  \alpha\right)   &  =\frac{\exp\left(  |\alpha
|^{2}\right)  }{n!}\left(  \partial_{\alpha\alpha^{\ast}}^{2}\right)
^{n}\delta^{2}(\alpha),
\end{align}
\end{subequations}
which are not difficult to find by using the general formula\footnote{The
identity%
\begin{equation}
|n\rangle\langle n|=\left.  \frac{1}{n!}\left(  \partial_{\alpha,\alpha^{\ast
}}^{2}\right)  ^{n}\left(  e^{|\alpha|^{2}}|\alpha\rangle\langle
\alpha|\right)  \right\vert _{\alpha=0},
\end{equation}
is also useful in the case of the number state. Note that this identity comes
directly from the representation of coherent states in terms of number states
(\ref{NumToCoh}).}%
\begin{equation}
P\left(  \alpha\right)  =\int_{%
\mathbb{C}
}\frac{d^{2}\beta}{\pi^{2}}\exp\left(  |\beta|^{2}+|\alpha|^{2}+\beta^{\ast
}\alpha-\beta\alpha^{\ast}\right)  \langle-\beta|\hat{\rho}|\beta\rangle,
\end{equation}
which comes from inverting equation (\ref{Pgs}).

Despite these properties which make the $P$ distribution being far from a true
probability distribution, it turns out to be extremely useful to solve quantum
optical problems. This is because we can evaluate the expectation value of an
operator $\hat{A}\left(  \hat{a},\hat{a}^{\dagger}\right)  $ as%
\begin{equation}
\langle\hat{A}\rangle=\int_{%
\mathbb{C}
}d^{2}\alpha P\left(  \alpha\right)  A^{\left(  \mathrm{n}\right)  }\left(
\alpha,\alpha^{\ast}\right)  ,
\end{equation}
where $A^{\left(  \mathrm{n}\right)  }$ is obtained by writing $\hat{A}\left(
\hat{a},\hat{a}^{\dagger}\right)  $ in normal order with the help of the
commutation relations, and then changing $\{\hat{a},\hat{a}^{\dagger}\}$ by
$\{\alpha,\alpha^{\ast}\}$, and, as we shall see in Chapter \ref{Detection},
most of the theoretical predictions for experimental observations involving
light detection are formulated ultimately as the expectation value of some
operator written in normal order.

\bigskip

We will refer to these continuous representations of the density operator as
\textit{coherent representations. }It is possible to show that the three
distributions $W$, $Q$, and $P$ exist for any quantum state $\hat{\rho}$. This
means that all the information concerning the state of the oscillator is
completely contained in any of them, that is, the three of them are equivalent
to the density operator $\hat{\rho}$. Hence, they provide a quantum mechanical
description of the harmonic oscillator in phase space, but they cannot be
regarded as probability distributions in the classical sense, showing that
there are quantum states which cannot be understood within the framework of
classical mechanics.

\subsection{The positive \textit{P} distribution}

The three types of distributions that we have defined have both advantages and
drawbacks: The marginals of the $W$ distribution give the correct statistics
for quadrature measurements, but the distribution itself can be negative; on
the other hand, the $Q$ distribution is a well behaved probability
distribution, but it gives quantum expectation values of operators only in
antinormal order; finally, the $P$ distribution gives expectation values of
operators in normal order (which turns out to be quite useful in quantum
optics), but it usually has singularities beyond those of the Dirac--delta function.

In this section we develop a new probability distribution, the
\textit{positive }$P$\textit{ distribution}, which is always a well behaved
probability distribution and gives expectation values of operators written in
normal order. These are the properties that will be needed for the purposes of
this thesis, but, as we will see, these are accomplished at the expense of
working in a phase space with twice the usual dimensionality. Even so, we will
sketch in the next section (and show explicitly along the next chapters) how
this distribution allows us to make calculations which could be impossible to
perform otherwise.

The positive $P$ distribution was originally defined by Drummond and Gardiner
\cite{Drummond80} via the following non-diagonal coherent state representation
of the density matrix%
\begin{equation}
\hat{\rho}=\int_{%
\mathbb{C}
^{2}}d^{2}\alpha d^{2}\alpha^{+}P\left(  \alpha,\alpha^{+}\right)
\hat{\Lambda}\left(  \alpha,\alpha^{+}\right)  , \label{positiveP}%
\end{equation}
where we have defined the so-called \textit{non-diagonal projector}%
\begin{equation}
\hat{\Lambda}\left(  \alpha,\alpha^{+}\right)  =\frac{|\alpha\rangle
\langle\left(  \alpha^{+}\right)  ^{\ast}|}{\langle\left(  \alpha^{+}\right)
^{\ast}|\alpha\rangle}=\exp\left(  -\alpha^{+}\alpha\right)  \exp\left(
\alpha a^{\dagger}\right)  |0\rangle\langle0|\exp\left(  \alpha^{+}a\right)
\text{.}%
\end{equation}
We use the same notation for this distribution and for the Glauber--Sudarshan
distribution because from now on we will make use of the positive $P$
distribution only, and hence there will be no room for confusion. Note that as
$\alpha$ and $\alpha^{+}$ are complex, independent variables, $P\left(
\alpha,\alpha^{+}\right)  $ lives in what we will call an \textit{extended
phase space }$\left(  \alpha,\alpha^{+}\right)  $, the four--dimensional space
formed by the real and imaginary parts of $\alpha$ and $\alpha^{+}$.

Even though the positive $P$ distribution is not uniquely defined from
(\ref{positiveP}), it can always be chosen as \cite{Drummond80}%
\begin{equation}
P\left(  \alpha,\alpha^{+}\right)  =\frac{1}{4\pi}\exp\left(  -\frac{1}%
{4}|\alpha^{\ast}-\alpha^{+}|^{2}\right)  Q\left[  \frac{1}{2}\left(
\alpha^{\ast}+\alpha^{+}\right)  ^{\ast}\right]  ,
\end{equation}
which shows explicitly that it is a well defined probability distribution in
the extended phase space $\left(  \alpha,\alpha^{+}\right)  $.

Similarly to the Glauber--Sudarshan $P$ distribution, it allows us to find the
expectation value of an operator $\hat{A}\left(  \hat{a},\hat{a}^{\dagger
}\right)  $ as%
\begin{equation}
\langle\hat{A}\rangle=\int_{%
\mathbb{C}
^{2}}d^{2}\alpha d^{2}\alpha^{+}P\left(  \alpha,\alpha^{+}\right)  A^{\left(
\mathrm{n}\right)  }\left(  \alpha,\alpha^{+}\right)  ;
\end{equation}
note that the correspondence $\hat{a}^{\dagger}\rightarrow\alpha^{\ast}$ in
the normal form of the operator $\hat{A}$ is replaced within this
representation by $\hat{a}^{\dagger}\rightarrow\alpha^{+}$.

We would like to finally note that it is completely straightforward to prove
that $\hat{\Lambda}\left(  \alpha,\alpha^{+}\right)  $ is an analytic
function\footnote{Given the complex function $f\left(  z\right)
=f_{\mathrm{R}}\left(  z_{\mathrm{R}},z_{\mathrm{I}}\right)  +\mathrm{i}%
f_{\mathrm{I}}\left(  z_{\mathrm{R}},z_{\mathrm{I}}\right)  $ of a complex
variable $z=z_{\mathrm{R}}+\mathrm{i}z_{\mathrm{I}}$, where $f_{\mathrm{R}}$,
$f_{\mathrm{I}}$, $z_{\mathrm{R}}$, and $z_{\mathrm{I}}$ are all real, we say
that it is analytic if the derivative of the function in the complex--$z$
plane does not depend on the direction along which it is performed. A
necessary and sufficient condition for this is that all the partial
derivatives $\partial f_{k}/\partial z_{l}$ ($k,l=\mathrm{R,I}$) are
continuous and satisfy the \textit{Cauchy--Riemann relations}%
\begin{equation}
\frac{\partial f_{\mathrm{R}}}{\partial z_{\mathrm{R}}}=\frac{\partial
f_{\mathrm{I}}}{\partial z_{\mathrm{I}}}\text{ \ \ and \ \ }\frac{\partial
f_{\mathrm{R}}}{\partial z_{\mathrm{I}}}=-\frac{\partial f_{\mathrm{I}}%
}{\partial z_{\mathrm{R}}}.
\end{equation}
Note that this relations imply that $\partial f/\partial z^{\ast}=0$, and
hence, an analytical complex function can only depend on $z_{\mathrm{R}}$ and
$z_{\mathrm{I}}$ through the combination $z_{\mathrm{R}}+\mathrm{i}%
z_{\mathrm{I}}$, that is, it cannot depend on the complex conjugate of $z$. An
important property of analytical complex functions is that the derivative
respect to the complex variable can be made in two equivalent ways%
\begin{equation}
\frac{df}{dz}=\frac{\partial f}{\partial z_{\mathrm{R}}}=-\mathrm{i}%
\frac{\partial f}{\partial z_{\mathrm{I}}}\text{.}%
\end{equation}
All this discussion is generalized to a complex function of several complex
variables in a natural straightforward manner.} of the complex variables
$\left(  \alpha,\alpha^{+}\right)  $, while the distribution $P\left(
\alpha,\alpha^{+}\right)  $ itself is not, as it has to depend on both
$\left(  \alpha,\alpha^{+}\right)  $ and their complex conjugates in order to
be a real function. We will exploit the analycity of $\hat{\Lambda}\left(
\alpha,\alpha^{+}\right)  $ in the next section and along the upcoming chapters.

\subsection{Fokker--Planck and stochastic Langevin
equations\label{FPandLangevin}}

As a taste of the usefulness of the positive $P$ representation, consider the
following evolution equation for the state of an oscillator:%
\begin{equation} \label{MasterExample}
\frac{d\hat{\rho}}{dt}=\left[  -\mathrm{i}\omega\hat{a}^{\dagger}\hat
{a}+\left(  \kappa\hat{a}^{\dagger2}-\kappa^{\ast}\hat{a}^{2}\right)  +\left(
\mathcal{E}\hat{a}^{\dagger}-\mathcal{E}^{\ast}\hat{a}\right)  ,\hat{\rho
}\right]+\gamma\left(  2\hat{a}\hat{\rho}\hat{a}^{\dagger}-\hat{a}^{\dagger}\hat
{a}\hat{\rho}-\hat{\rho}\hat{a}^{\dagger}\hat{a}\right),
\end{equation}
being $\omega$, $\mathcal{E}$, $\kappa$, and $\gamma$ some parameters
($\omega$ and $\gamma$ are real). The first term includes the regular
Hamiltonian evolution of the oscillator, an squeezing term, and a term
corresponding to an external driving of the oscillator, while the second part
cannot be written in Hamiltonian form, and describes damping on the ideal
harmonic motion of the oscillator. In the forthcoming chapters we will see how
these terms arise naturally in the evolution of light contained in an optical
cavity. In any case, let us assume for the moment that the state of the
harmonic oscillator is subject to this evolution equation. We are going to
show that it can be reduced to a set of 2 complex differential equations with
noise, that is, to a set of \textit{stochastic equations }%
\cite{Gardiner09book}.

One way to solve these operator equation would be to project it onto a
truncated number state basis $\{|0\rangle,|1\rangle,...,|N\rangle\}$,
obtaining then a linear system with $N\times N$ variables (the independent
elements of the representation of the density operator). This method, however,
is usually of little help if we intend to deal with states having a large
number of photons, so that the size of the linear system is too large. It is
in this case when coherent representations can be useful. Let us explain this
for the positive $P$ distribution.

We would like to stress that even though we are using equation
(\ref{MasterExample}) as a guiding example, most of the derivations that we
are going to perform are completely general.

Let us collect the coherent variables into a single vector $\boldsymbol{\alpha
}=\operatorname{col}(\alpha,\alpha^{+})$. It is very simple to prove that the
nondiagonal projector $\hat{\Lambda}\left(  \boldsymbol{\alpha}\right)  $
satisfies the following properties:%
\begin{subequations}
\begin{align}
\hat{a}\hat{\Lambda}\left(  \boldsymbol{\alpha}\right)   &  =\alpha
\hat{\Lambda}\left(  \boldsymbol{\alpha}\right)  \\
\hat{\Lambda}\left(  \boldsymbol{\alpha}\right)  \hat{a} &  =\left(
\alpha+\frac{\partial}{\partial\alpha^{+}}\right)  \hat{\Lambda}\left(
\boldsymbol{\alpha}\right)  \\
\hat{\Lambda}\left(  \boldsymbol{\alpha}\right)  \hat{a}^{\dagger} &
=\alpha^{+}\hat{\Lambda}\left(  \boldsymbol{\alpha}\right)  \\
\hat{a}^{\dagger}\hat{\Lambda}\left(  \boldsymbol{\alpha}\right)   &  =\left(
\alpha^{+}+\frac{\partial}{\partial\alpha}\right)  \hat{\Lambda}\left(
\boldsymbol{\alpha}\right)  ;
\end{align}
therefore, introducing the expansion of the density operator in terms of the
positive $P$ distribution (\ref{positiveP}), these properties allow us to
rewrite the evolution equation (\ref{MasterExample}) as%
\end{subequations}
\begin{equation}  \label{preFP}
\int_{%
\mathbb{C}
^{2}}d^{4}\boldsymbol{\alpha}\hat{\Lambda}\left(  \boldsymbol{\alpha}\right)
\partial_{t}P\left(  \boldsymbol{\alpha};t\right)
=\int_{%
\mathbb{C}
^{2}}d^{4}\boldsymbol{\alpha}d^{2}\alpha^{+}P\left(  \boldsymbol{\alpha
};t\right)  \left[  \sum_{j}A_{j}\left(  \boldsymbol{\alpha}\right)
\partial_{j}+\frac{1}{2}\sum_{jl}D_{jl}\left(  \boldsymbol{\alpha}\right)
\partial_{j}\partial_{l}\right]  \hat{\Lambda}\left(  \boldsymbol{\alpha
}\right)  ,
\end{equation}
where all the indices can take the values $\alpha$ or $\alpha^{+}$, and
\begin{subequations}
\label{ADexample}%
\begin{align}
A_{\alpha}\left(  \boldsymbol{\alpha}\right)   &  =\mathcal{E}-(\gamma
+\mathrm{i}\omega)\alpha+2\kappa\alpha^{+},\\
A_{\alpha^{+}}\left(  \boldsymbol{\alpha}\right)   &  =\mathcal{E}^{\ast
}-(\gamma-\mathrm{i}\omega)\alpha^{+}+2\kappa^{\ast}\alpha,\\
D_{\alpha\alpha}\left(  \boldsymbol{\alpha}\right)   &  =2\kappa=D_{\alpha
^{+}\alpha^{+}}^{\ast}\left(  \boldsymbol{\alpha}\right)  ,\\
D_{\alpha\alpha^{+}}\left(  \boldsymbol{\alpha}\right)   &  =D_{\alpha
^{+}\alpha}\left(  \boldsymbol{\alpha}\right)  =0\text{.}%
\end{align}
The symmetric matrix $D$ can always be decomposed as $D=BB^{T}$ for some
matrix $B$, although this is not unique (note that it has to be a $2\times d$
matrix, though, with $d$ arbitrary). $D$ being diagonal in this particular
example, we could choose for example%
\end{subequations}
\begin{equation}
B=%
\begin{bmatrix}
\sqrt{2\kappa} & 0\\
0 & \sqrt{2\kappa^{\ast}}%
\end{bmatrix}
.\label{Bexample}%
\end{equation}

Define now the real and imaginary parts of the coherent amplitudes
$\alpha=\alpha_{\mathrm{R}}+\mathrm{i}\alpha_{\mathrm{I}}$ and $\alpha
^{+}=\alpha_{\mathrm{R}}^{+}+\mathrm{i}\alpha_{\mathrm{I}}^{+}$, which we
collect in the vectors $\boldsymbol{\alpha}_{\mathrm{R}}=\operatorname{col}%
(\alpha_{\mathrm{R}},\alpha_{\mathrm{R}}^{+})$ and $\boldsymbol{\alpha
}_{\mathrm{I}}=\operatorname{col}(\alpha_{\mathrm{I}},\alpha_{\mathrm{I}}%
^{+})$. Define also the real and imaginary parts of the vector $\mathbf{A}$
and the matrix $B$ by%
\begin{subequations}
\begin{align}
A_{j}\left(  \boldsymbol{\alpha}\right)    & =A_{\mathrm{R}}^{j}%
(\boldsymbol{\alpha}_{\mathrm{R}},\boldsymbol{\alpha}_{\mathrm{I}}%
)+\mathrm{i}A_{\mathrm{I}}^{j}(\boldsymbol{\alpha}_{\mathrm{R}}%
,\boldsymbol{\alpha}_{\mathrm{I}}),\\
B_{jl}\left(  \boldsymbol{\alpha}\right)    & =B_{\mathrm{R}}^{jl}%
(\boldsymbol{\alpha}_{\mathrm{R}},\boldsymbol{\alpha}_{\mathrm{I}}%
)+\mathrm{i}B_{\mathrm{I}}^{jl}(\boldsymbol{\alpha}_{\mathrm{R}}%
,\boldsymbol{\alpha}_{\mathrm{I}})\text{.}%
\end{align}
\end{subequations}
The analyticity of $\hat{\Lambda}\left(  \boldsymbol{\alpha}\right)  $ allows
us to choose the derivatives $\partial_{j}$ as $\partial_{j}^{\mathrm{R}}$ or
$-i\partial_{j}^{\mathrm{I}}$ at will, where we use the notation
$\partial/\partial\alpha_{\mathrm{R}}=\partial_{\alpha}^{\mathrm{R}}$ and
$\partial/\partial\alpha_{\mathrm{I}}=\partial_{\alpha}^{\mathrm{I}}$
(similarly for $\alpha^{+}$), and we choose them so that the differential
operator inside the square brackets reads%
\begin{equation}
\sum_{j}\left(  A_{\mathrm{R}}^{j}\partial_{j}^{\mathrm{R}}+A_{\mathrm{I}}%
^{j}\partial_{j}^{\mathrm{I}}\right)  +\frac{1}{2}\sum_{jlm}\left(
B_{\mathrm{R}}^{jm}B_{\mathrm{R}}^{lm}\partial_{j}^{\mathrm{R}}\partial
_{l}^{\mathrm{R}}+2B_{\mathrm{R}}^{jm}B_{\mathrm{I}}^{lm}\partial
_{j}^{\mathrm{R}}\partial_{l}^{\mathrm{I}}+B_{\mathrm{I}}^{jm}B_{\mathrm{I}%
}^{lm}\partial_{j}^{\mathrm{I}}\partial_{l}^{\mathrm{I}}\right)  \text{,}%
\end{equation}
for future convenience.

The main result of the section is obtained by performing an integration by
parts in (\ref{preFP}) to make the derivatives act onto $P\left(
\boldsymbol{\alpha};t\right)  $ instead of onto the projector $\hat{\Lambda
}\left(  \boldsymbol{\alpha}\right)  $, what leads to%
\begin{equation}
\partial_{t}P\left(  \boldsymbol{\alpha};t\right)    =\left[  -\sum
_{j}\left(  \partial_{j}^{\mathrm{R}}A_{\mathrm{R}}^{j}+\partial
_{j}^{\mathrm{I}}A_{\mathrm{I}}^{j}\right)+\frac{1}{2}\sum_{jlm}\left(  \partial_{j}^{\mathrm{R}}\partial
_{l}^{\mathrm{R}}B_{\mathrm{R}}^{jm}B_{\mathrm{R}}^{lm}+2\partial
_{j}^{\mathrm{R}}\partial_{l}^{\mathrm{I}}B_{\mathrm{R}}^{jm}B_{\mathrm{I}%
}^{lm}+\partial_{j}^{\mathrm{I}}\partial_{l}^{\mathrm{I}}B_{\mathrm{I}}%
^{jm}B_{\mathrm{I}}^{lm}\right)  \right]  P\left(  \boldsymbol{\alpha
};t\right)  ,
\end{equation}
where we have used the fact that physical states must correspond to positive
$P$ distributions which decay to zero at infinity, so that we can get rid of
the total derivative terms appearing upon integrating by parts. Hence, we have
turned the operator equation (\ref{MasterExample}) into a linear partial
differential equation for the $P$ distribution. Indeed, this equation is a
very special one. By defining the vector $\mathcal{\vec{A}}=\operatorname{col}%
(A_{\mathrm{R}}^{\alpha},A_{\mathrm{R}}^{\alpha^{+}},A_{\mathrm{I}}^{\alpha
},A_{\mathrm{I}}^{\alpha^{+}})$ and the matrix%
\begin{equation}
\mathcal{D}=%
\begin{bmatrix}
B_{\mathrm{R}}B_{\mathrm{R}}^{T} & B_{\mathrm{R}}B_{\mathrm{I}}^{T}\\
B_{\mathrm{I}}B_{\mathrm{R}}^{T} & B_{\mathrm{I}}B_{\mathrm{I}}^{T}%
\end{bmatrix}
=\mathcal{BB}^{T}\text{ \ \ with \ \ }\mathcal{B}=%
\begin{bmatrix}
B_{\mathrm{R}} & 0_{2\times2}\\
B_{\mathrm{I}} & 0_{2\times2}%
\end{bmatrix}
,\label{RealDB}%
\end{equation}
we can write it in the more compact way%
\begin{equation}
\partial_{t}P\left(  \mathbf{x};t\right)  =\left[  -\sum_{j=1}^{4}\partial
_{j}\mathcal{A}_{j}\left(  \mathbf{x}\right)  +\frac{1}{2}\sum_{jl=1}%
^{4}\partial_{j}\partial_{l}\mathcal{D}_{jl}\left(  \mathbf{x}\right)
\right]  P\left(  \mathbf{x};t\right)  ,\label{FP}%
\end{equation}
where we have defined the vector $\mathbf{x}=\operatorname{col}(\alpha
_{\mathrm{R}},\alpha_{\mathrm{R}}^{+},\alpha_{\mathrm{I}},\alpha_{\mathrm{I}%
}^{+})$. When the matrix $\mathcal{D}$ is positive semidefinite, as it is in
our case, this is known as the \textit{Fokker--Planck equation }%
\cite{Gardiner09book}; this equation has been subject of study for a very long
time, and we have a lot of different techniques available to solve it or
extract information from it. $\mathcal{\vec{A}}$, $\mathcal{D}$, and
$\mathcal{B}$, are known as the \textit{drift vector}, the \textit{diffusion
matrix}, and the \textit{noise matrix}, respectively.

For this thesis, the most important property of Fokker--Planck equations is
its relation to stochastic equations. In particular, it can be proved that any
Fokker--Planck equation written in the form (\ref{FP}) is equivalent to the
following system of \textit{stochastic} \textit{equations }%
\cite{Gardiner09book}%
\begin{equation}
\mathbf{\dot{x}}=\mathcal{\vec{A}}\left(  \mathbf{x}\right)  +\mathcal{B}%
\left(  \mathbf{x}\right)  \boldsymbol{\eta}\left(  t\right)  \text{,}
\label{RealStochastic}%
\end{equation}
where $\boldsymbol{\eta}\left(  t\right)  =\operatorname{col}[\eta_{1}\left(
t\right)  ,\eta_{2}\left(  t\right)  ,\eta_{3}\left(  t\right)  ,\eta
_{4}\left(  t\right)  ]$ is a vector whose elements are independent, real,
Gaussian noises which satisfy the statistical properties%
\begin{equation}
\left\langle \eta_{j}\left(  t\right)  \right\rangle =0\text{ \ \ \ \ and
\ \ \ \ }\left\langle \eta_{j}\left(  t\right)  \eta_{l}\left(  t^{\prime
}\right)  \right\rangle =\delta_{jl}\delta(t-t^{\prime})\text{.}
\label{RealGaussStat}%
\end{equation}
We will call \textit{stochastic Langevin equations }to these equations. The
equivalence between them and the original Fokker--Planck equation must be
understood in the \textquotedblleft average\textquotedblright\ sense, that is,
given any function $F(\mathbf{x})$, we can find its average either by using
the distribution $P\left(  \mathbf{x};t\right)  $ or the solutions
$\mathbf{x}[\boldsymbol{\eta}(t)]$ of the variables in terms of noise
integrals as%
\begin{equation}
\int_{%
\mathbb{R}
^{4}}d^{4}\mathbf{x}F(\mathbf{x})P\left(  \mathbf{x};t\right)  =\left\langle
F\left\{  \mathbf{x}[\boldsymbol{\eta}(t)]\right\}  \right\rangle .
\end{equation}

Given the simple structure of $\mathcal{B}$, see (\ref{RealDB}), we can
rewrite the Langevin equations in the complex form%
\begin{equation}
\boldsymbol{\dot{\alpha}}=\mathbf{A}\left(  \boldsymbol{\alpha}\right)
+B\left(  \boldsymbol{\alpha}\right)  \boldsymbol{\eta}\left(  t\right)  ,
\label{ComplexLangevin}%
\end{equation}
where $\boldsymbol{\eta}\left(  t\right)  =\operatorname{col}[\eta_{1}\left(
t\right)  ,\eta_{2}\left(  t\right)  ]$ is in this case a two--dimensional
vector. This is an amazingly simple final result: As complex stochastic
variables, $\alpha$ and $\alpha^{+}$ evolve according to a set of Langevin
equations built with the original $\mathbf{A}$ and $D=BB^{T}$. Then, whenever
we are able to rewrite an arbitrary master equation in the form (\ref{preFP})
by using the positive $P$ representation, we can directly write down equations
(\ref{ComplexLangevin}) for the coherent amplitudes of the system.

Applied to our purposes, this means that given the solution
$\boldsymbol{\alpha}=\boldsymbol{\alpha}[\boldsymbol{\eta}\left(  t\right)  ]$
of equation (\ref{ComplexLangevin}), we can evaluate the quantum expectation
value of any operator $\hat{A}\left(  \hat{a},\hat{a}^{\dagger}\right)  $ as%
\begin{equation}
\langle\hat{A}\rangle=\langle A^{\left(  \mathrm{n}\right)  }\left(
\alpha,\alpha^{+}\right)  \rangle_{P},
\end{equation}
where in the following we will denote by $\langle...\rangle_{P}$ to any
stochastic average obtained from the Langevin equations within the positive
$P$ representation.

There is one subtlety that we have not mentioned though. Without entering into
much detail (see \cite{Gardiner09book} for a deep discussion), a stochastic
equation can be interpreted in infinitely many inequivalent ways, which attend
to how the integrals involving noise are defined in the Riemann sense, that
is, as a limit of some series. The connection we made between the
Fokker--Planck equation and the Langevin equations is only valid if the
stochastic equations are interpreted in a very particular form first defined
by Ito (the so-called \textit{Ito interpretation}). The problem with this
interpretation is that integration is defined in such a way that the rules of
calculus as we know them no longer apply, and one needs to use a new form of
calculus called \textit{Ito calculus}. Integrals can be defined in a way which
allows us to use the usual rules of calculus, arriving then to the so-called
\textit{Stratonovich interpretation} of stochastic equations. However, there
is a price to pay: Interpreted in such way, the set of stochastic equations
associated to the Fokker--Planck equation (\ref{FP}) is no longer
(\ref{ComplexLangevin}), but the same type of equation with a different
deterministic part given by%
\begin{equation}
A_{j}^{(\mathrm{St})}\left(  \boldsymbol{\alpha}\right)  =A_{j}\left(
\boldsymbol{\alpha}\right)  -\frac{1}{2}\sum_{lm}B_{lm}\left(
\boldsymbol{\alpha}\right)  \partial_{l}B_{jm}\left(  \boldsymbol{\alpha
}\right)  ,\text{ \ \ \ \ with \ \ \ \ }j,l,m=\alpha,\alpha^{+},
\label{ItoToStrat}%
\end{equation}
that is, the deterministic part of the equations acquires an extra term which
depends on the noise matrix. In this thesis we shall always use this
Stratonovich form. Nevertheless, note that if the noise matrix does not depend
on the coherent amplitudes $\boldsymbol{\alpha}$, both forms are the same,
that is, Ito and Stratonovich interpretations differ only when noise is
multiplicative, not additive. This is indeed the case in our simple example
(\ref{ADexample},\ref{Bexample}), whose associated set of complex Langevin
equations are given by%
\begin{subequations}
\begin{align}
\dot{\alpha}  &  =\mathcal{E}-(\gamma+\mathrm{i}\omega)\alpha+2\kappa
\alpha^{+}+\sqrt{2\kappa}\eta_{1}(t),\\
\dot{\alpha}^{+}  &  =\mathcal{E}^{\ast}-(\gamma-\mathrm{i}\omega)\alpha
^{+}+2\kappa^{\ast}\alpha+\sqrt{2\kappa^{\ast}}\eta_{2}(t),
\end{align}
both within the Ito and Stratonovich interpretations.
\end{subequations} 

%% file: QuantizationFO.tex
The goal of this chapter is the\ quantization of the electromagnetic field
inside an optical cavity with perfectly reflecting mirrors. We will follow a
heuristic, but physically intuitive approach in which the electromagnetic
field is put in correspondence with a collection of harmonic oscillators.
Nevertheless, we would like to note that the quantization of the
electromagnetic field can be carried out in a more rigorous fashion by
generalizing the method that we described for mechanical systems in the
previous chapter and in Appendix \ref{QuantumMechanics}; this procedure relies
on a Lagrangian theory of fields, and is called \textit{canonical quantization
}\cite{Cohen89book}.

In order to show how the heuristic procedure works, we will first apply the
method to a simple, but highly relevant case: The quantization of the
electromagnetic field in free space. This will allow us to introduce
plane--waves as well as the concept of polarization. We then pass to describe
optical beams within the paraxial approximation, and discuss the modal
decomposition of such fields in terms of transverse modes. We finally consider
the electromagnetic field inside a stable optical cavity and proceed to its quantization.

\section{Light as an electromagnetic wave}

Nowadays it feels quite natural to say that \textit{light} is an
\textit{electromagnetic wave}. Arriving to this conclusion, however, was not
trivial at all. The history of such a discovery starts in the first half of
the XIX century with Faraday, who showed that the polarization of light can
change when subject to a magnetic field; this was the first hint suggesting
that there could be a connection between light and electromagnetism, and he
was the first to propose that light could be an electromagnetic disturbance of
some kind, able to propagate without the need of a reference medium. However,
this qualitative idea did not find a rigorous mathematical formulation until
the second half of the century, when Maxwell developed a consistent theory of
electromagnetism, and showed how the theory was able to predict the existence
of electromagnetic waves propagating at a speed which was in agreement with
the speed measured for light at that time \cite{Maxwell865}. A couple of
decades after his proposal, the existence of electromagnetic waves was
experimentally demonstrated by Hertz \cite{Hertz893book}, and the theory of
light as an electromagnetic wave found its way towards being accepted.

Our starting point are Maxwell's equations formulated as partial differential
equations for the \textit{electric} and \textit{magnetic\footnote{As we won't
deal with materials sensitive to the magnetic field, we will use the term
\textquotedblleft magnetic field\textquotedblright\ for the $\mathbf{B}%
$-field, which is usually denoted by \textquotedblleft magnetic induction
field\textquotedblright\ when it needs to be distinguished from the
$\mathbf{H}$-field (which we won't be using in this thesis).}} vector
\textit{fields} $\mathbf{E}\left(  \mathbf{r},t\right)  $ and $\mathbf{B}%
\left(  \mathbf{r},t\right)  $, respectively, where $\mathbf{r}$ ($t$) is the
position (time) where (when) the fields are observed. This formulation is due
to Heaviside \cite{Heaviside894book}, as Maxwell originally proposed his
theory in terms of \textit{quaternions}. The theory consists of four
equations; the first two are called the \textit{homogeneous Maxwell equations}
and read%
\begin{equation}
\boldsymbol{\nabla}\cdot\mathbf{B}=0\text{ \ \ \ and \ \ \ }\boldsymbol{\nabla
}\times\mathbf{E}=-\partial_{t}\mathbf{B}, \label{MaxwellHomoEqs}%
\end{equation}
where $\boldsymbol{\nabla}=\left(  \partial_{x},\partial_{y},\partial
_{z}\right)  $. The other two are called the \textit{inhomogeneous Maxwell
equations} and are written as%
\begin{equation}
\boldsymbol{\nabla}\cdot\mathbf{E}=\rho/\varepsilon_{0}\text{ \ \ \ and
\ \ \ }\boldsymbol{\nabla}\times\mathbf{B}=\mu_{0}\varepsilon_{0}\partial
_{t}\mathbf{E}+\mu_{0}\mathbf{j}, \label{MaxwellInhomoEqs}%
\end{equation}
where any electric or magnetic source is introduced in the theory by a
\textit{charge density} function $\rho\left(  \mathbf{r},t\right)  $ and a
\textit{current distribution} vector $\mathbf{j}\left(  \mathbf{r},t\right)
$, respectively; the parameters $\varepsilon_{0}=8.8\times10^{-12}$
\textrm{F/m} and $\mu_{0}=1.3\times10^{-6}$ \textrm{H/m} are the so-called
\textit{electric permittivity} and the \textit{magnetic permeability} of
vacuum, respectively.

We will show the process of quantization of the electromagnetic field in the
absence of sources ($\rho=0$ and $\mathbf{j}=\mathbf{0}$). Under these
circumstances, the inhomogeneous equations are simplified to%
\begin{equation}
\boldsymbol{\nabla}\cdot\mathbf{E}=0\text{ \ \ \ and \ \ \ }c^{2}%
\boldsymbol{\nabla}\times\mathbf{B}=\partial_{t}\mathbf{E}\text{,}%
\end{equation}
where $c=1/\sqrt{\varepsilon_{0}\mu_{0}}\simeq3\times10^{8}\mathrm{m/s}$.

The homogeneous equations (\ref{MaxwellHomoEqs}) allow us to derive the fields
from a \textit{scalar potential} $\phi\left(  \mathbf{r},t\right)  $ and a
\textit{vector potential} $\mathbf{A}\left(  \mathbf{r},t\right)  $ as%
\begin{equation}
\mathbf{B}=\boldsymbol{\nabla}\times\mathbf{A}\text{ \ \ and \ \ }%
\mathbf{E}=-\boldsymbol{\nabla}\phi-\partial_{t}\mathbf{A}\text{,}
\label{AtoEB}%
\end{equation}
hence reducing to four the degrees of freedom of the electromagnetic field.
These potentials, however, are not unique; we can always use an arbitrary
function $\Lambda\left(  \mathbf{r},t\right)  $ to change them as%
\begin{equation}
\mathbf{A\rightarrow A}+\boldsymbol{\nabla}\Lambda\text{ \ \ and \ \ }%
\phi\rightarrow\phi-\partial_{t}\Lambda\text{,}%
\end{equation}
what is known as the \textit{gauge invariance} of Maxwell's equations.

Introducing (\ref{AtoEB}) into the inhomogeneous equations we get the
equations satisfied by the potentials%
\begin{subequations}
\begin{align}
\left(  c^{2}\boldsymbol{\nabla}^{2}-\partial_{t}^{2}\right)  \mathbf{A}  &
=\partial_{t}\boldsymbol{\nabla}\phi+c^{2}\boldsymbol{\nabla}\left(
\boldsymbol{\nabla}\cdot\mathbf{A}\right) \\
\boldsymbol{\nabla}^{2}\phi+\partial_{t}\boldsymbol{\nabla}\cdot\mathbf{A}  &
=0.
\end{align}
The problem is highly simplified if we exploit the gauge invariance and choose
$\boldsymbol{\nabla}\cdot\mathbf{A}=0$, as the second equation can be shown to
further imply that $\phi=0$ for physical fields vanishing at infinity, and
hence the only equations left are%
\end{subequations}
\begin{equation}
\left(  c^{2}\boldsymbol{\nabla}^{2}-\partial_{t}^{2}\right)  \mathbf{A}%
=\mathbf{0} \label{WaveEq}%
\end{equation}
which are wave equations with speed $c$ for the components of the vector
potential. Note that the condition $\boldsymbol{\nabla}\cdot\mathbf{A}=0$
(known as the \textit{Coulomb condition}) relates the three components of
$\mathbf{A}$, and hence, only two degrees of freedom of the initial six (the
electric and magnetic vector fields) remain.

It is finally important to note that the wave equation (\ref{WaveEq}) has a
unique solution inside a given spatio--temporal region only if both the vector
potential and its derivative along the direction normal to the boundary of the
region are specified at any point of the boundary \cite{Jackson62book}; these
are known as \textit{Dirichlet} and \textit{Neumann} conditions, respectively.
In general, however, physical problems do not impose all those constraints,
and hence there coexist several solutions of the wave equation, which we will
call \textit{spatio--temporal} \textit{modes} of the system.

In all the optical systems that we will treat, we will be able to write these
solutions in the form $\boldsymbol{\varepsilon}A\left(  \mathbf{r},t\right)
$, with $\boldsymbol{\varepsilon}$ a constant unit vector and $A\left(
\mathbf{r},t\right)  $ a function of space and time; this is called the
\textit{scalar} \textit{approximation}, and can be made when the system looks
the same for the two independent components of the vector potential (for
example, both components must be subject to the same boundary conditions). In
addition, it will be convenient to look for solutions separable in space and
time, that is, $A\left(  \mathbf{r},t\right)  =u\left(  \mathbf{r}\right)
\mathcal{A}(t)$. We will see that the connection with the harmonic oscillator
is made through the temporal part of the solutions. As for the spatial part,
introducing the separable ansatz in the wave equation, we find that it must
satisfy the equation%
\begin{equation}
\boldsymbol{\nabla}^{2}u+k^{2}u=0\text{,} \label{Helmholtz}%
\end{equation}
where $k^{2}$ is a suitable \textit{separation constant}. This is a Helmholtz
equation, which yields a unique solution inside a given volume only if the
Dirichlet \textit{or} Neumann conditions are given for $u(\mathbf{r})$
\cite{Jackson62book}. If the boundary conditions are less restrictive than
these, several solutions $\{u_{n}(\mathbf{r})\}$ appear corresponding to the
eigenfunctions of the differential operator $\boldsymbol{\nabla}^{2}$, which
is a self--adjoint differential operator called \textit{Laplacian}, whose
corresponding eigenvalues $\{-k_{n}^{2}\}$ can be shown to be all negative.
These solutions are known as the \textit{spatial modes} of the system, and
play a central role in the heuristic quantization procedure that we will follow.

\section{Quantization in free space: Plane waves and polarization}

\subsection{Quantization in terms of plane waves\label{PlaneQuanti}}

The scalar approximation works fine in free space, as by definition it has no
boundaries. This also means that the Helmholtz equation (\ref{Helmholtz}) has
to be solved in all $%
\mathbb{R}
^{3}$. In this case, the eigenfunctions of the Laplacian form a continuous set
given by $\{\exp\left(  \mathrm{i}\mathbf{k}\cdot\mathbf{r}\right)
\}_{\mathbf{k}\in%
\mathbb{R}
^{3}}$, with corresponding eigenvalues $\{-|\mathbf{k}|^{2}\}_{\mathbf{k}\in%
\mathbb{R}
^{3}}$. Hence, the solutions of the wave equation will be of the type
$\boldsymbol{\varepsilon}\left(  \mathbf{k}\right)  \mathcal{A}(\mathbf{k}%
,t)\exp\left(  \mathrm{i}\mathbf{k}\cdot\mathbf{r}\right)  $. Moreover, the
Coulomb condition imposes that $\mathbf{k}\cdot\boldsymbol{\varepsilon}\left(
\mathbf{k}\right)  =0$, that is, the vector potential is orthogonal to the
mode vectors $\mathbf{k}$; this means that for any $\mathbf{k}$ we can define
two vectors $\boldsymbol{\varepsilon}_{1}\left(  \mathbf{k}\right)  $ and
$\boldsymbol{\varepsilon}_{2}\left(  \mathbf{k}\right)  $, which are called
the \textit{polarization vectors}, satisfying the relations%
\begin{subequations}
\begin{align}
\mathbf{k}\cdot\boldsymbol{\varepsilon}_{\sigma}\left(  \mathbf{k}\right)   &
=0,\text{ \ \ \ }\left(  \sigma=1,2\right) \\
\boldsymbol{\varepsilon}_{\sigma}^{\ast}\left(  \mathbf{k}\right)
\cdot\boldsymbol{\varepsilon}_{\sigma^{\prime}}\left(  \mathbf{k}\right)   &
=\delta_{\sigma\sigma^{\prime}},\text{ \ \ }\left(  \sigma,\sigma^{\prime
}=1,2\right) \\
\boldsymbol{\varepsilon}_{1}^{\ast}\left(  \mathbf{k}\right)  \times
\boldsymbol{\varepsilon}_{2}\left(  \mathbf{k}\right)   &  =\mathbf{k}%
/k\equiv\boldsymbol{\kappa},
\end{align}
with $k=|\mathbf{k}|$, so that we can make the following modal expansion of
the vector potential:%
\end{subequations}
\begin{equation}
\mathbf{A}\left(  \mathbf{r},t\right)  =\sum_{\sigma=1,2}\int_{%
\mathbb{R}
^{3}}d^{3}\mathbf{k}\boldsymbol{\varepsilon}_{\sigma}\left(  \mathbf{k}%
\right)  \mathcal{A}_{\sigma}(\mathbf{k},t)\exp\left(  i\mathbf{k}%
\cdot\mathbf{r}\right)  \text{.} \label{FreeModalExp}%
\end{equation}
Note that the reality of the vector potential imposes that
\begin{equation}
\boldsymbol{\varepsilon}_{\sigma}^{\ast}\left(  \mathbf{k}\right)
\mathcal{A}_{\sigma}^{\ast}(\mathbf{k},t)=\boldsymbol{\varepsilon}_{\sigma
}\left(  -\mathbf{k}\right)  \mathcal{A}_{\sigma}(-\mathbf{k},t).
\end{equation}

From the wave equation, we get that the temporal part of the vector potential
satisfies the equation%
\begin{equation}
\left[  \frac{d^{2}}{dt^{2}}+\omega^{2}\left(  \mathbf{k}\right)  \right]
\mathcal{A}_{\sigma}(\mathbf{k},t)=0,
\end{equation}
with $\omega\left(  \mathbf{k}\right)  =ck$. Being this the equation of motion
of a harmonic oscillator, one might be tempted to directly interpret the modal
amplitudes $\mathcal{A}_{\sigma}(\mathbf{k},t)$ as the positions of harmonic
oscillators associated to the different modes; however, $\mathcal{A}_{\sigma
}(\mathbf{k},t)$ is not real in general, and hence we need another strategy to
make the connection with the harmonic oscillator. The idea is to define%
\begin{equation}
\nu_{\sigma}\left(  \mathbf{k},t\right)  =\frac{1}{\Gamma}\left[
\mathcal{A}_{\sigma}(\mathbf{k},t)+\frac{\mathrm{i}}{\omega_{k}}%
\mathcal{\dot{A}}_{\sigma}(\mathbf{k},t)\right]  , \label{EMnormal}%
\end{equation}
where $\Gamma$ is a real parameter that will be chosen later, which satisfies
the equation%
\begin{equation}
\dot{\nu}_{\sigma}(\mathbf{k},t)=-\mathrm{i}\omega\left(  \mathbf{k}\right)
\nu_{\sigma}(\mathbf{k},t)\Longrightarrow\nu_{\sigma}\left(  \mathbf{k}%
,t\right)  =\nu_{\sigma}\left(  \mathbf{k}\right)  \exp\left[  -\mathrm{i}%
\omega\left(  \mathbf{k}\right)  t\right]  , \label{NormalEvo}%
\end{equation}
where we denote $\nu_{\sigma}\left(  \mathbf{k},0\right)  $ simply by
$\nu_{\sigma}\left(  \mathbf{k}\right)  $; this expression shows that
$\nu_{\sigma}\left(  \mathbf{k},t\right)  $ and $\nu_{\sigma}^{\ast}\left(
\mathbf{k},t\right)  $ are subject to the same evolution as that of the normal
variables of a harmonic oscillator, see Section \ref{ClassHO}.

This approach shows that the modes of the electromagnetic field behave as
harmonic oscillators of frequency $\omega\left(  \mathbf{k}\right)  $ having $\{\nu_{\sigma}\left(  \mathbf{k},t\right)  ,\nu_{\sigma
}^{\ast}\left(  \mathbf{k},t\right)  \}_{\mathbf{k}\in%
\mathbb{R}
^{3}}$ as normal variables. Indeed, we can show that the electromagnetic energy, which is defined
in vacuum as the integral%
\begin{equation}
E_{\mathrm{em}}=\frac{1}{2}\int d^{3}\mathbf{r}\left[  \varepsilon
_{0}\mathbf{E}^{2}\left(  \mathbf{r},t\right)  +\frac{1}{\mu_{0}}%
\mathbf{B}^{2}\left(  \mathbf{r},t\right)  \right]  \text{,} \label{Eem}%
\end{equation}
can be written as the Hamiltonian of this collection of harmonic oscillators.
By using the modal expansion of the vector potential (\ref{FreeModalExp}) and
its relation with the electric and magnetic fields (\ref{AtoEB}), it is
straightforward to show that%
\begin{equation}
E_{\mathrm{em}}=4\pi^{3}\varepsilon_{0}\sum_{\sigma=1,2}\int_{%
\mathbb{R}
^{3}}d^{3}\mathbf{k}\left[  |\mathcal{\dot{A}}_{\sigma}(\mathbf{k}%
,t)|^{2}+\omega_{k}^{2}|\mathcal{A}_{\sigma}(\mathbf{k},t)|^{2}\right]
\text{.}%
\end{equation}
Now, inverting the relations (\ref{EMnormal}) we get%
\begin{subequations}
\begin{align}
\mathcal{A}_{\sigma}(\mathbf{k},t)  &  =\frac{\Gamma}{2}\left[  \nu_{\sigma
}\left(  \mathbf{k},t\right)  +\nu_{\sigma}^{\ast}\left(  -\mathbf{k}%
,t\right)  \right], \\
\mathcal{\dot{A}}_{\sigma}(\mathbf{k},t)  &  =\frac{\omega_{k}\Gamma
}{2\mathrm{i}}\left[  \nu_{\sigma}\left(  \mathbf{k},t\right)  -\nu_{\sigma
}^{\ast}\left(  -\mathbf{k},t\right)  \right]  \text{,}%
\end{align}
from which one trivially obtains%
\end{subequations}
\begin{equation}
E_{\mathrm{em}}=4\pi^{3}\varepsilon_{0}\Gamma^{2}\sum_{\sigma=1,2}\int_{%
\mathbb{R}
^{3}}d^{3}\mathbf{k}\omega^{2}\left(  \mathbf{k}\right)  \nu_{\sigma}^{\ast
}\left(  \mathbf{k},t\right)  \nu_{\sigma}\left(  \mathbf{k},t\right)
\text{.}%
\end{equation}
Hence, by choosing $\Gamma=1/\sqrt{8\pi^{3}\varepsilon_{0}}$, the final
expression for the electromagnetic energy reads%
\begin{equation}
E_{\mathrm{em}}=\sum_{\sigma=1,2}\int_{%
\mathbb{R}
^{3}}d^{3}\mathbf{k}\frac{\omega^{2}\left(  \mathbf{k}\right)  }{2}\nu
_{\sigma}^{\ast}\left(  \mathbf{k}\right)  \nu_{\sigma}\left(  \mathbf{k}%
\right)  \text{,} \label{EnergyNormal}%
\end{equation}
which is exactly a sum of the Hamiltonians for each harmonic oscillator of
frequency $\omega\left(  \mathbf{k}\right)  $ with normal variables
$\{\nu_{\sigma}\left(  \mathbf{k},t\right)  ,\nu_{\sigma}^{\ast}\left(
\mathbf{k},t\right)  \}_{\mathbf{k}\in%
\mathbb{R}
^{3}}$ and unit mass.

These derivations suggest that we can treat the electromagnetic field as a
mechanical system consisting of a (continuous) collection of harmonic
oscillators, and quantize it accordingly, that is, by replacing the normal
variables of each oscillator $\{\nu_{\sigma}(  \mathbf{k},t)  ,\nu_{\sigma}^{\ast
}(  \mathbf{k},t)  \}_{\mathbf{k}\in%
\mathbb{R}
^{3}}$ by the set $\sqrt{2\hbar/\omega\left(
\mathbf{k}\right)  }\{\hat{a}_{\sigma}\left(  \mathbf{k},t\right)  ,\hat
{a}_{\sigma}^{\dagger}\left(  \mathbf{k},t\right)  \}_{\mathbf{k}\in%
\mathbb{R}
^{3}}$ of \textit{boson operators}, where the annihilation and creation operators satisfy the commutation
relations%
\begin{subequations}
\begin{align}
\lbrack\hat{a}_{\sigma}\left(  \mathbf{k},t\right)  ,\hat{a}_{\sigma^{\prime}%
}^{\dagger}\left(  \mathbf{k}^{\prime},t\right)  ]  & =\delta_{\sigma
\sigma^{\prime}}\delta^{3}\left(  \mathbf{k}-\mathbf{k}^{\prime}\right)
\text{,}\label{CanCom}\\
\lbrack\hat{a}_{\sigma}\left(  \mathbf{k},t\right)  ,\hat{a}_{\sigma^{\prime}%
}\left(  \mathbf{k}^{\prime},t\right)  ]  & =[\hat{a}_{\sigma}^{\dagger
}\left(  \mathbf{k},t\right)  ,\hat{a}_{\sigma^{\prime}}^{\dagger}\left(
\mathbf{k}^{\prime},t\right)  ]=0.
\end{align}
\textit{Photons} are then associated to the excitations of each
electromagnetic oscillator $(\sigma,\mathbf{k})$.

According to the quantization scheme that we used in the previous chapter for
the harmonic oscillator, the Hamiltonian for the electromagnetic field in free
space is then found from (\ref{EnergyNormal}) as
\end{subequations}
\begin{equation}
\hat{H}_{\mathrm{em}}=\sum_{\sigma=1,2}\int_{%
\mathbb{R}
^{3}}d^{3}\mathbf{k}\frac{\hbar\omega\left(  \mathbf{k}\right)  }{2}\left[
\hat{a}_{\sigma}^{\dagger}\left(  \mathbf{k}\right)  \hat{a}_{\sigma}\left(
\mathbf{k}\right)  +\hat{a}_{\sigma}\left(  \mathbf{k}\right)  \hat{a}%
_{\sigma}^{\dagger}\left(  \mathbf{k}\right)  \right]  \text{.}%
\end{equation}
Using the commutation relations (\ref{CanCom}) we can rewrite it in the
following normal form%
\begin{equation}
\hat{H}_{\mathrm{em}}=\sum_{\sigma=1,2}\int_{%
\mathbb{R}
^{3}}d^{3}\mathbf{k}\hbar\omega\left(  \mathbf{k}\right)  \hat{a}_{\sigma
}^{\dagger}\left(  \mathbf{k}\right)  \hat{a}_{\sigma}\left(  \mathbf{k}%
\right)  +4\pi\hbar\delta^{3}\left(  \mathbf{0}\right)  \int_{0}^{\infty
}dkk^{3}\equiv:\hat{H}_{\mathrm{em}}:+E_{\infty}\text{.}%
\end{equation}
The first contribution is like $\hat{H}_{\mathrm{em}}$ but where the boson
operators have been reordered in normal order as if they were usual numbers,
that is, without making use of the commutation relations, an operation that we
denote by $:\hat{A}\left(  \hat{a},\hat{a}^{\dagger}\right)  :$ for any
operator in the following. The second contribution, on the other hand, is a
number, but an infinite one. Note however, that it contributes as a global
phase to the time evolution operator, and it is therefore not observable, what
allows us to redefine the origin of the energy such as this contribution is
removed\footnote{There are certain situations (like the \textit{Casimir}
\textit{effect} or the \textit{Lamb shift}) in which this term has observable
consequences, and the theory must be properly \textit{renormalized}. We will
however not find such situations in this thesis, and hence a direct removal of
these infinite quantities leads to the proper results.}.

\begin{figure}
[ptb]
\begin{center}
\includegraphics[
width=0.8\textwidth
]%
{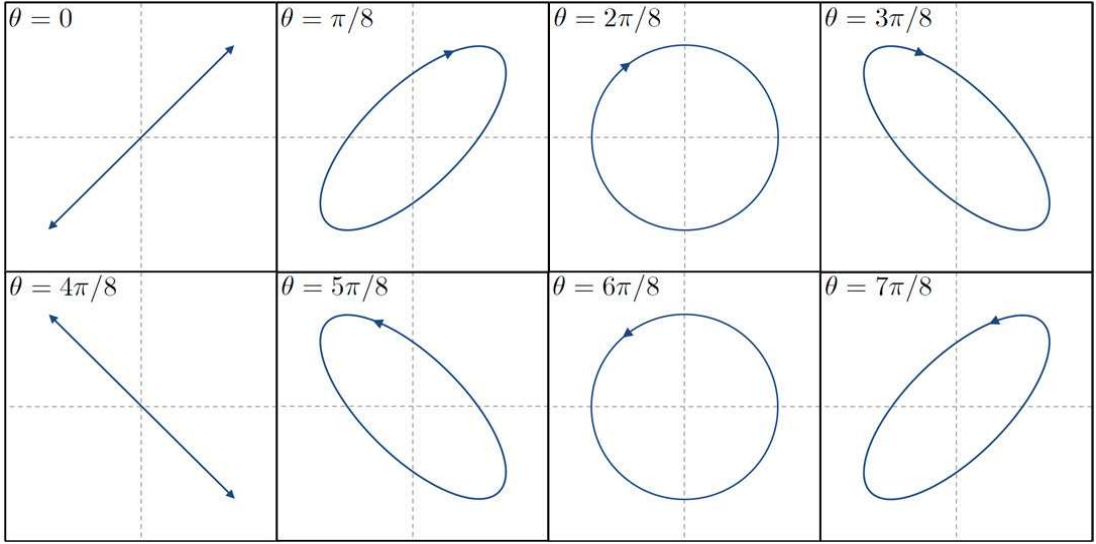}%
\caption{Polarization ellipse for $\varphi=\pi/4$. Note how the polarization
changes from linear to circular and from left to right as a function of
$\theta$, but the ellipse is always oriented along the $\pm\pi/4$ axes.}%
\label{fQuanti2}%
\end{center}
\end{figure}

The vector potential becomes then an operator $\mathbf{\hat{A}}\left(
\mathbf{r},t\right)  =\mathbf{\hat{A}}^{(+)}\left(  \mathbf{r},t\right)
+\mathbf{\hat{A}}^{(-)}\left(  \mathbf{r},t\right)  $, with%
\begin{equation}
\mathbf{\hat{A}}^{(+)}\left(  \mathbf{r},t\right) =\sum_{\sigma=1,2}%
\int_{%
\mathbb{R}
^{3}}d^{3}\mathbf{k}\sqrt{\frac{\hbar}{16\pi^{3}\varepsilon_{0}\omega\left(
\mathbf{k}\right)  }}\boldsymbol{\varepsilon}_{\sigma}\left(  \mathbf{k}%
\right)  \hat{a}_{\sigma}\left(  \mathbf{k}\right)  \exp\left[  -\mathrm{i}%
\omega\left(  \mathbf{k}\right)  t+\mathrm{i}\mathbf{k}\cdot\mathbf{r}\right]=\left[  \mathbf{\hat{A}}^{(-)}\left(  \mathbf{r},t\right)  \right]
^{\dagger}.
\end{equation}
This decomposition of the vector potential in terms of $\mathbf{\hat{A}}%
^{(\pm)}\left(  \mathbf{r},t\right)  $, usually called its
\textit{positive--frequency} and \textit{negative--frequency} parts, will be
used for every field from now on; we will sometimes refer to either part by itself as
\textquotedblleft the field\textquotedblright\. The electric and
magnetic fields are described then by the operators%
\begin{subequations}
\begin{align}
\mathbf{\hat{E}}^{(+)}\left(  \mathbf{r},t\right)   &  =\mathrm{i}\int_{%
\mathbb{R}
^{3}}d^{3}\mathbf{k}\sqrt{\frac{\hbar\omega\left(  \mathbf{k}\right)  }%
{16\pi^{3}\varepsilon_{0}}}\boldsymbol{\varepsilon}_{\sigma}\left(
\mathbf{k}\right)  \hat{a}_{\sigma}\left(  \mathbf{k}\right)  \exp\left[
-\mathrm{i}\omega\left(  \mathbf{k}\right)  t+\mathrm{i}\mathbf{k}%
\cdot\mathbf{r}\right]  ,\\
\mathbf{\hat{B}}^{(+)}\left(  \mathbf{r},t\right)   &  =\mathrm{i}\sum
_{\sigma=1,2}\int_{%
\mathbb{R}
^{3}}d^{3}\mathbf{k}\sqrt{\frac{\hbar}{16\pi^{3}\varepsilon_{0}\omega\left(
\mathbf{k}\right)  }}\left[  \mathbf{k\times}\boldsymbol{\varepsilon}_{\sigma
}\left(  \mathbf{k}\right)  \right]  \hat{a}_{\sigma}\left(  \mathbf{k}%
\right)  \exp\left[  -\mathrm{i}\omega\left(  \mathbf{k}\right)
t+\mathrm{i}\mathbf{k}\cdot\mathbf{r}\right]  ;
\end{align}
this is consistent with quantum mechanics, as being observable quantities, the
fields must correspond to self--adjoint operators. Note that all these
expressions are given within the Heisenberg picture, as the operators evolve
in time.

Note finally that the spatio--temporal modes of the problem have the
functional form $\exp\left[  -\mathrm{i}\omega\left(  \mathbf{k}\right)
t+\mathrm{i}\mathbf{k}\cdot\mathbf{r}\right]  $. These modes are called
\textit{plane--waves}. They are \textquotedblleft plane\textquotedblright%
\ because all the points laying in the plane $\mathbf{k}\cdot\mathbf{r}=const$
take the same value of the mode; they are \textquotedblleft
waves\textquotedblright\ because the mode at the space time point $\left(
\mathbf{r},t\right)  $ \textit{has} the same value that it \textit{had} at
$\left(  \mathbf{r-}c\boldsymbol{\kappa}t,0\right)  $, and hence these modes
can be visualized as a plane propagating at speed $c$ along the direction of
$\boldsymbol{\kappa}$. We will call \textit{wave vector }to $\mathbf{k}$ in
the following. Our eyes are sensitive to electromagnetic waves having
frequencies $\omega\left(  \mathbf{k}\right)  $ between $2.5\times10^{15}%
$\textrm{Hz }(red) to $5\times10^{15}$\textrm{Hz }(violet), or equivalently
\textit{wavelengths} $\lambda=2\pi/k$ between $760\mathrm{nm}$ and
$380\mathrm{nm}$, which are customarily called \textit{light}; in this thesis
we will focus on this region of the \textit{electromagnetic spectrum }(which
sometimes is extended to the near infrared or ultraviolet), which we will
refer to as \textit{optical domain}.%

\subsection{Vector character of the fields and polarization}

Let us explain now a property of the electromagnetic field that we will be
using in several parts of this thesis, the \textit{polarization}. To this aim
we come back to a classical description of the electromagnetic field.

Consider a Cartesian system defined by the triad $\{\mathbf{e}_{x},\mathbf{e}_{y},\mathbf{e}_{z}\}$ of unit length vectors. Consider now an
electromagnetic field described by a plane--wave mode propagating along the
$\mathbf{e}_{z}$ direction, and having definite frequency $\omega$ and
polarization vector $\boldsymbol{\varepsilon}$, whose vector potential is
given by
\end{subequations}
\begin{equation}
\mathbf{A}\left(  \mathbf{r},t\right)  =\frac{A_{0}}{2}\boldsymbol{\varepsilon
}\exp\left(  -\mathrm{i}\omega t+\mathrm{i}kz\right)  +\mathrm{c.c.},
\end{equation}
with $A_{0}$ some amplitude that we take real and positive without loss of
generality. The associated electric and magnetic fields are given by%
\begin{subequations}
\begin{align}
\mathbf{E}\left(  \mathbf{r},t\right)   &  =\mathrm{i}\frac{\omega A_{0}}%
{2}\boldsymbol{\varepsilon}\exp\left(  -\mathrm{i}\omega t+\mathrm{i}%
kz\right)  +\mathrm{c.c.},\\
\mathbf{B}\left(  \mathbf{r},t\right)   &  =\mathrm{i}\frac{kA_{0}}{2}\left[
\mathbf{e}_{z}\times\boldsymbol{\varepsilon}\right]  \exp\left(
-\mathrm{i}\omega t+\mathrm{i}kz\right)  +\mathrm{c.c.}\text{;}%
\end{align}
the first interesting thing that we observe is that the electric and magnetic
fields are not only orthogonal to the propagation direction (we say that they
are \textit{transverse fields}), but also to each other.

\begin{figure}
[t]
\begin{center}
\includegraphics[
width=0.35\textwidth
]%
{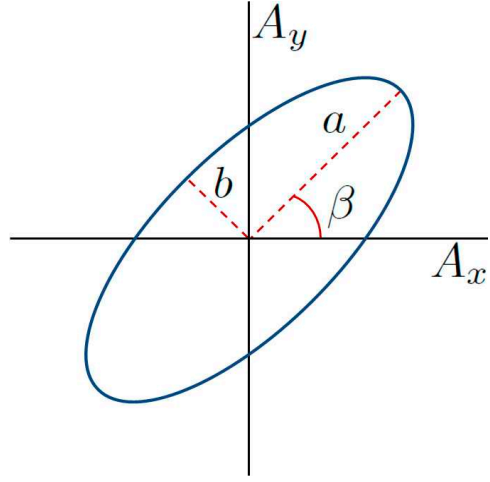}%
\caption{Ellipse described by the components of the vector potential as they
evolve.}%
\label{fQuanti1}%
\end{center}
\end{figure}

Suppose now that we observe the time evolution of the electric field at a
given position which we take as $\mathbf{r}=\mathbf{0}$ to simplify the
upcoming discussion. Let us parametrize with full generality the polarization
vector as%
\end{subequations}
\begin{equation}
\boldsymbol{\varepsilon}(\theta,\varphi)=\mathbf{e}_{x}\exp\left(
-\mathrm{i}\theta\right)  \cos\varphi+\mathbf{e}_{y}\exp\left(  \mathrm{i}%
\theta\right)  \sin\varphi, \label{PolPar}%
\end{equation}
with $\theta\in\lbrack0,\pi]$ and $\varphi\in\lbrack0,\pi/2]$. At the
observation point, the vector potential takes the simple form%
\begin{equation}
\mathbf{A}\left(  t\right)  =\left(
\begin{array}
[c]{c}%
\cos\varphi\cos\left(  \omega t+\theta\right) \\
\sin\varphi\cos\left(  \omega t-\theta\right)
\end{array}
\right)  ,
\end{equation}
where we have taken $A_{0}=1$ for simplicity, and we have represented
$\mathbf{A}$ in our Cartesian system, but obviating the $z$ component which is
always zero given the transversality of the fields. It is not difficult to
realize that, starting at the point $\mathbf{A}(0)=(\cos\varphi\cos\theta
,\sin\varphi\cos\theta)$, this vector describes an ellipse in the $x-y$ plane
as time increases, as shown in Figure \ref{fQuanti1}. Let us call $a$ and $b$
to the \textit{semi--major} and \textit{semi--minor} \textit{axes} of this
ellipse (positive definite), and $\beta\in\lbrack-\pi/2,\pi/2]$ to its
\textit{orientation} (the angle that its major axis forms with the $x$ axis).
The orientation of this ellipse, called the \textit{polarization ellipse}, is
given by%
\begin{equation}
\tan2\beta=\tan2\varphi\cos2\theta\text{,}%
\end{equation}
while defining an auxiliary angle $\chi\in\lbrack-\pi/4,\pi/4]$ from
$\sin2\chi=\sin2\varphi\sin2\theta$, the semi--major and semi--minor axes are
given by%
\[
a^{2}=\frac{1}{1+\tan^{2}\chi}\text{ \ \ and \ \ }b^{2}=\frac{\tan^{2}\chi
}{1+\tan^{2}\chi}\text{.}%
\]
Hence, the eccentricity of the ellipse is%
\begin{equation}
e=\sqrt{1-\tan^{2}\chi};
\end{equation}
for $e=1$ ($\chi=0$) the ellipse degenerates into a straight line, while for
$e=0$ ($\chi=\pm\pi/4$) the ellipse is a circumference. In the first (second)
case we say that light has \textit{linear} (\textit{circular}) polarization.
Of special interest for this thesis is the case $\varphi=\pi/4$. In this case,
the orientation of the ellipse is $\beta=\pm\pi/4$, and the polarization
changes from linear to circular (and from right to left, see below) depending
on the angle $\theta$ as shown in Figure \ref{fQuanti2}.%

The vector potential can then be seen as a point in the polarization ellipse
rotating around it with angular frequency $\omega$ \textit{clockwise} or
\textit{anticlockwise} for $\sin2\theta>0$ and $\sin2\theta<0$, respectively.
In the first (second) case we say that light has \textit{right} (\textit{left}%
) \textit{elicity}.

\section{Paraxial optics and transverse modes}

In the previous section we have quantized the electromagnetic field by using a
modal decomposition based on plane waves propagating in all possible
directions. However, we have seen that plane waves have an infinite transverse
size, and hence cannot describe real fields, which have a finite extent. In
this section we are going to introduce another set of modes which are better
adapted to the optical scenarios that we will treat along this thesis, and can
even be generated in laboratories.

\subsection{Optical beams and the paraxial approximation}

In this thesis we will deal with the electromagnetic field in the form of
\textit{optical beams}, that is, light propagating nearly within a given
direction, which we take as the $z$ axis. The vector potential of any such
beam can be written as%
\begin{equation}
\mathbf{A}\left(  \mathbf{r},t\right)  =\int_{\mathcal{O}}dk_{z}%
\mathbf{\tilde{A}}\left(  k_{z};\mathbf{r}_{\perp},z,t\right)  \exp\left(
\mathrm{i}k_{z}z\right)  ,
\end{equation}
where $\mathcal{O}$\ refers to wave vectors $k_{z}$ such that $\lambda
=2\pi/|k_{z}|$ is in the optical domain, $\mathbf{r}_{\perp}=\left(
x,y\right)  $ are the \textit{transverse coordinates}, and $\mathbf{\tilde{A}%
}\left(  k_{z};\mathbf{r}_{\perp},z,t\right)  $ is an amplitude which,
compared to $|k_{z}|$, varies slowly both along the \textit{longitudinal axis}
and the \textit{transverse plane} generated, respectively, by $\mathbf{e}_{z}$
and $\{\mathbf{e}_{x},\mathbf{e}_{y}\}$, that is,
\begin{subequations}
\label{ParaxialConditions}%
\begin{align}
\mathbf{|}\partial_{z}^{2}\mathbf{\tilde{A}|}  &  \ll\mathbf{|}k_{z}%
\partial_{z}\mathbf{\tilde{A}|}\ll k_{z}^{2}\mathbf{|\tilde{A}|}%
\label{zSlowCond}\\
\mathbf{|}\boldsymbol{\nabla}_{\perp}\cdot\mathbf{\tilde{A}|}  &
\ll\mathbf{|}k_{z}\mathbf{\tilde{A}|}, \label{tSlowCond}%
\end{align}
where $\boldsymbol{\nabla}_{\perp}=\left(  \partial_{x},\partial_{y}\right)
$. These generalizations of plane waves are called \textit{paraxial waves}. It
is easy to prove that the polarization properties of $\mathbf{\tilde{A}%
}\left(  k_{z};\mathbf{r}_{\perp},z,t\right)  $ are just as those of plane
waves (we say that the \textit{polarization degrees of freedom} decouple from
the \textit{spatio--temporal} ones), so that we can still work under the
scalar approximation as in the previous section, and look for solutions of the
wave equation of the type $\boldsymbol{\varepsilon}\left(  k_{z}\right)
\mathcal{A}(k_{z},t)u\left(  k_{z};\mathbf{r}_{\perp},z\right)  \exp\left(
\mathrm{i}kz\right)  $.

$u\left(  k_{z};\mathbf{r}_{\perp},z\right)  $ satisfies the slowly varying
conditions (\ref{ParaxialConditions}), what allows us to rewrite the Helmholtz
equation as%
\end{subequations}
\begin{equation}
\left(  2\mathrm{i}k_{z}\partial_{z}+\boldsymbol{\nabla}_{\perp}^{2}\right)
u\left(  k_{z};\mathbf{r}_{\perp},z\right)  =0. \label{HelmholtzParaxial}%
\end{equation}
The solution of this equation is unique provided that the transverse structure
of $u\left(  k_{z};\mathbf{r}_{\perp},z\right)  $ is given at one plane
$z=z_{0}$; indeed, this solution can be written in the integral form
\cite{Born80book}%
\begin{equation}
u\left(  k_{z};\mathbf{r}_{\perp},z\right)  =-\frac{\mathrm{i}k_{z}}%
{2\pi(z-z_{0})}\int_{%
\mathbb{R}
^{2}}d^{2}\mathbf{r}_{\perp}^{\prime}u\left(  k_{z};\mathbf{r}_{\perp}%
^{\prime},z_{0}\right)  \exp\left[  -\frac{\mathrm{i}k_{z}}{2(z-z_{0})}\left(
\mathbf{r}_{\perp}-\mathbf{r}_{\perp}^{\prime}\right)  ^{2}\right]  \text{.}
\label{IntegralParaxial}%
\end{equation}
Hence, an optical beam can be visualized as a \textit{transverse pattern}
propagating along the longitudinal axis, whose shape is changing owed to the
\textit{diffraction} term $\boldsymbol{\nabla}_{\perp}^{2}$ of equation
(\ref{HelmholtzParaxial}).

\subsection{Transverse modes\label{TransModes}}

The paraxial equation has a very special set of solutions whose transverse
shape is retained upon propagation except for variations of its scale and the
acquisition of a phase, that is, mathematically they satisfy%
\begin{equation}
u\left(  k_{z};\mathbf{r}_{\perp},z\right)  =\sigma(z)\exp\left[
\mathrm{i}\gamma\left(  \mathbf{r}_{\perp},z\right)  \right]  u\left[
k_{z};\sigma\left(  z\right)  \mathbf{r}_{\perp},z_{0}\right]  \text{,}
\label{TransverseModesDefinition}%
\end{equation}
for some real functions $\sigma\left(  z\right)  $ and $\gamma\left(
\mathbf{r}_{\perp},z\right)  $. In a sense, these are eigenfunctions of the
integral operator acting on $u\left(  k_{z};\mathbf{r}_{\perp},z_{0}\right)  $
in (\ref{IntegralParaxial}), the so-called \textit{paraxial propagator}. These
solutions are known as \textit{transverse modes} and form an infinite, but
countable set. Their explicit form depend on the transverse coordinate system
that one uses to express the propagator. In the following we discuss the
structure of such modes in \textit{Cartesian} and \textit{polar} coordinates.%

Let us start by analyzing the propagation of the most simple transverse mode
(which is the same in any coordinate system). The structure of the integral
equation (\ref{IntegralParaxial}) invites us to try a Gaussian ansatz%
\begin{equation}
u\left(  k_{z};\mathbf{r}_{\perp},z_{0}\right)  =u_{0}\exp\left[
\mathrm{i}\frac{k_{z}\mathbf{r}_{\perp}^{2}}{2q\left(  z_{0}\right)  }\right]
, \label{FundIn}%
\end{equation}
where $q\left(  z\right)  $ is a complex parameter which we will call the
\textit{q--parameter} of the beam, and is usually written as%
\begin{equation}
\frac{1}{q\left(  z\right)  }=\frac{1}{R\left(  z\right)  }+\mathrm{i}\frac
{2}{k_{z}w^{2}\left(  z\right)  },
\end{equation}
being $R\left(  z\right)  $ and $w\left(  z\right)  $ real functions called
the \textit{curvature radius} and \textit{spot size} of the mode,
respectively. Introducing this decomposition of the \textit{q}--parameter in
(\ref{FundIn}) we get
\begin{equation}
u\left(  k_{z};\mathbf{r}_{\perp},z_{0}\right)  \exp\left(  \mathrm{i}%
k_{z}z_{0}\right)  =u_{0}\exp\left[  -\frac{\mathbf{r}_{\perp}^{2}}%
{w^{2}\left(  z_{0}\right)  }\right]  \exp\left[  \mathrm{i}k_{z}\left(
z_{0}+\frac{\mathbf{r}_{\perp}^{2}}{2R\left(  z_{0}\right)  }\right)  \right]
,
\end{equation}
which has a very suggestive form: A spherical surface of curvature radius
$\rho$ centered at $\mathbf{r}=\mathbf{0}$ can be written within the paraxial
approximation as%
\begin{equation}
z=\rho\sqrt{1-\frac{\mathbf{r}_{\perp}^{2}}{\rho^{2}}}\simeq\rho
-\frac{\mathbf{r}_{\perp}^{2}}{2\rho},
\end{equation}
and hence the wavefront\footnote{The surface formed by the points of equal
phase.} of our Gaussian ansatz is spherical and has a curvature radius
$R\left(  z_{0}\right)  $. On the other hand, this \textit{spherical wave} is
modulated by a circular Gaussian distribution in the transverse plane of
thickness $w\left(  z_{0}\right)  $. It is straightforward to prove that after
the action of the paraxial propagator, this transverse pattern evolves to%
\begin{equation}
u\left(  k_{z};\mathbf{r}_{\perp},z\right)  =\frac{u_{0}}{w\left(  z\right)
}\exp\left[  \mathrm{i}\frac{k_{z}\mathbf{r}_{\perp}^{2}}{2q\left(  z\right)
}+\mathrm{i}\psi\left(  z\right)  \right]  ,
\end{equation}
where%
\begin{subequations}
\begin{align}
q\left(  z\right)   &  =q\left(  z_{0}\right)  +z-z_{0},\\
\psi\left(  z\right)   &  =\arg\left\{  1+\frac{z-z_{0}}{q\left(
z_{0}\right)  }\right\}  .
\end{align}
It is customary to take the reference plane $z_{0}=0$ as that in which the
wavefront is plane, that is, the plane in which $R\left(  z_{0}=0\right)
=\infty$,$\ $denoting $w\left(  z_{0}=0\right)  $ by $w_{0}$. Particularized
to this election, the previous expressions lead to
\end{subequations}
\begin{subequations}
\label{FreePropLaws}%
\begin{align}
w^{2}\left(  z\right)   &  =w_{0}^{2}\left[  1+\left(  \frac{z}{z_{R}}\right)
^{2}\right]  ,\\
R\left(  z\right)   &  =z\left[  1+\left(  \frac{z_{R}}{z}\right)
^{2}\right]  ,\\
\psi\left(  z\right)   &  =-\arctan\left(  \frac{z}{z_{R}}\right)  \in
\lbrack-\frac{\pi}{2},\frac{\pi}{2}],
\end{align}
where we have defined the \textit{Rayleigh length }$z_{R}=k_{z}w_{0}^{2}/2$
(note that it is not positive definite, it is negative when $k_{z}$ is
negative). $\psi\left(  z\right)  $ is known as the Gouy phase. Note that this
pattern $u\left(  k_{z};\mathbf{r}_{\perp},z\right)  $ is related to $u\left(
k_{z};\mathbf{r}_{\perp},z_{0}\right)  $ in the form
(\ref{TransverseModesDefinition}), and hence, it is indeed a transverse mode;
it is called the \textit{fundamental transverse mode}. The rest of transverse
modes have a structure similar to this Gaussian mode, but modulated by
additional transverse functions.

\begin{figure}
[t]
\begin{center}
\includegraphics[
height=3.8329in,
width=3.9686in
]%
{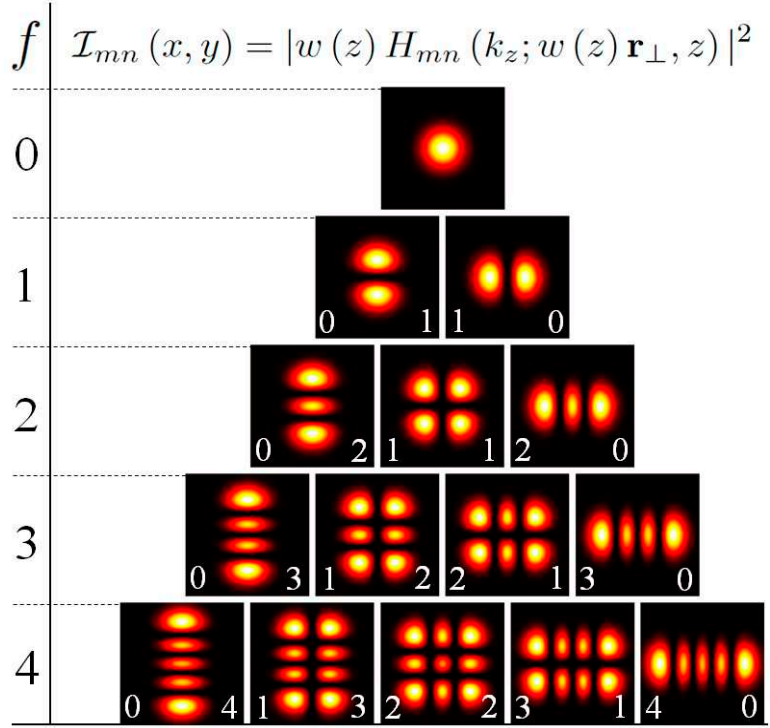}%
\caption{Transverse distribution of the Hermite--Gauss modes.}%
\label{fQuanti3}%
\end{center}
\end{figure}

For Cartesian coordinates $\mathbf{r}_{\perp}=(x,y)$, the transverse modes are
known as the \textit{Hermite--Gauss} (HG) or TEM$_{mn}$\textit{ modes
}$\{H_{mn}\left(  k_{z};\mathbf{r}_{\perp},z\right)  \}_{m,n\in%
\mathbb{N}
}$, and have the form%
\end{subequations}
\begin{equation} \label{Hmn}
H_{mn}\left(  k_{z};\mathbf{r}_{\perp},z\right)=\frac{1}{\sqrt
{2^{m+n-1}\pi m!n!}w\left(  z\right)  }H_{m}\left[  \frac{\sqrt{2}x}{w\left(
z\right)  }\right]  H_{n}\left[  \frac{\sqrt{2}y}{w\left(  z\right)  }\right]\exp\left[  \mathrm{i}k_{z}\frac{x^{2}+y^{2}}{2q\left(  z\right)
}+\mathrm{i}\left(  1+m+n\right)  \psi\left(  z\right)  \right]  ,
\end{equation}
where $H_{m}\left(  x\right)  $ denote the Hermite polynomials\footnote{The
Hermite polynomial $H_{m}\left(  x\right)  $ can be found from the Rodrigues
formula%
\begin{equation}
H_{m}\left(  x\right)  =\left(  -1\right)  ^{m}\exp\left(  x^{2}\right)
\frac{d^{m}}{dt^{m}}\exp\left(  -x^{2}\right)  .
\end{equation}
Hence, for example, $H_{0}\left(  x\right)  =1$ and $H_{1}\left(  x\right)
=2x$.}. The indices $m$ and $n$, called collectively the \textit{transverse
modal indices}, are called individually the \textit{horizontal }and
\textit{vertical} \textit{indices}\ of the mode, respectively. Note that these
modes form an orthonormal set in the transverse plane, that is,%
\begin{equation}
\int_{%
\mathbb{R}
^{2}}d^{2}\mathbf{r}_{\perp}H_{mn}^{\ast}\left(  k_{z};\mathbf{r}_{\perp
},z\right)  H_{m^{\prime}n^{\prime}}\left(  k_{z};\mathbf{r}_{\perp},z\right)
=\delta_{mm^{\prime}}\delta_{nn^{\prime}}\ \forall z.
\end{equation}
Indeed, it is possible to show that they also form a basis for any function
contained in \textrm{L}$^{2}\left(  \mathbf{r}_{\perp}\right)  $, that is, for
any square integrable function in $%
\mathbb{R}
^{2}$; this \textit{completeness} of the modes is expressed mathematically as%
\begin{equation}
\sum_{m,n}H_{mn}^{\ast}\left(  k_{z};\mathbf{r}_{\perp}^{\prime},z\right)
H_{mn}\left(  k_{z};\mathbf{r}_{\perp},z\right)  =\delta^{2}\left(
\mathbf{r}_{\perp}-\mathbf{r}_{\perp}^{\prime}\right)  \text{ }\forall
z\text{.}%
\end{equation}

Let us define the collective index $f=m+n$, which we will call \textit{family
index} for reasons that will become clear when dealing with optical cavities;
we say that modes with the same family parameter $f$ are contained in the
\textit{f}'\textit{th transverse family}. Note that family $f$ contains $f+1$
modes. In Figure \ref{fQuanti3} we plot the transverse distribution
$\mathcal{I}_{mn}\left(  \mathbf{r}_{\perp}\right)  =|w\left(  z\right)
H_{mn}\left(  k_{z};w\left(  z\right)  \mathbf{r}_{\perp},z\right)  |^{2}$ of
these modes. We see that the transverse indices $m$ and $n$ give the number of
zeros present in the $x$ and $y$ directions respectively (the number of
\textit{nodal lines}).%

\begin{figure}
[t]
\begin{center}
\includegraphics[
height=3.8441in,
width=3.9747in
]%
{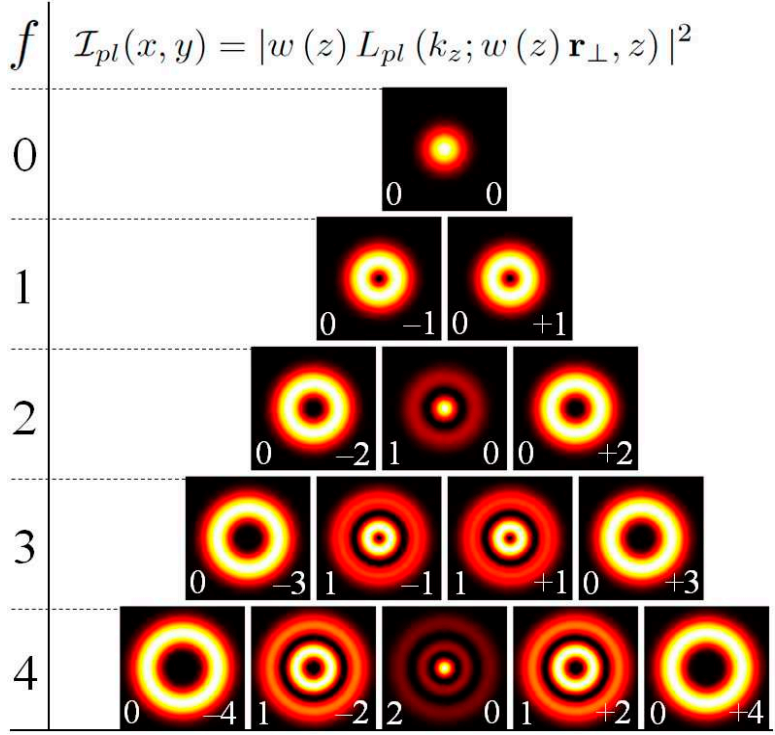}%
\caption{Transverse distribution of the Laguerre--Gauss modes.}%
\label{fQuanti4}%
\end{center}
\end{figure}

On the other hand, for polar coordinates $\mathbf{r}_{\perp}=r(\cos\phi
,\sin\phi)$, the set of transverse modes are known as the
\textit{Laguerre--Gauss }(LG)\textit{ modes }$\{L_{pl}\left(  k_{z}%
;\mathbf{r}_{\perp},z\right)  \}_{p\in%
\mathbb{N}
}^{l\in%
\mathbb{Z}
}$, and have the form%
\begin{equation}\label{Lpl}
L_{pl}\left(  k_{z};\mathbf{r}_{\perp},z\right)=\sqrt{\frac{2p!}%
{\pi\left(  p+|l|\right)  !}}\frac{1}{w\left(  z\right)  }\left[  \frac
{\sqrt{2}r}{w\left(  z\right)  }\right]  ^{|l|}L_{p}^{|l|}\left[  \frac
{2r^{2}}{w^{2}\left(  z\right)  }\right]\exp\left[  \mathrm{i}k_{z}\frac{r^{2}}{2q\left(  z\right)
}+\mathrm{i}\left(  1+2p+|l|\right)  \psi\left(  z\right)  +\mathrm{i}%
l\phi\right]  ,
\end{equation}
where $L_{p}^{|l|}\left(  t\right)  $ denote the modified Laguerre
polynomials\footnote{The modified Laguerre polynomial $L_{p}^{|l|}\left(
t\right)  $ can be found from the Rodrigues formula%
\begin{equation}
L_{p}^{|l|}\left(  t\right)  =\frac{t^{-|l|}\exp\left(  t\right)  }{p!}%
\frac{d^{p}}{dt^{p}}\left[  t^{p+|l|}\exp\left(  -t\right)  \right]  .
\end{equation}
Hence, for example, $L_{0}^{|l|}\left(  t\right)  =1$ and $L_{1}^{|l|}\left(
t\right)  =-t+|l|+1$.}. In this case $p$ and $l$ are known as the
\textit{radial} and \textit{polar indices}, respectively. These modes form
also an orthonormal basis of \textrm{L}$^{2}\left(  \mathbf{r}_{\perp}\right)
$.%

\begin{figure}
[t]
\begin{center}
\includegraphics[
height=3.8744in,
width=4.0058in
]%
{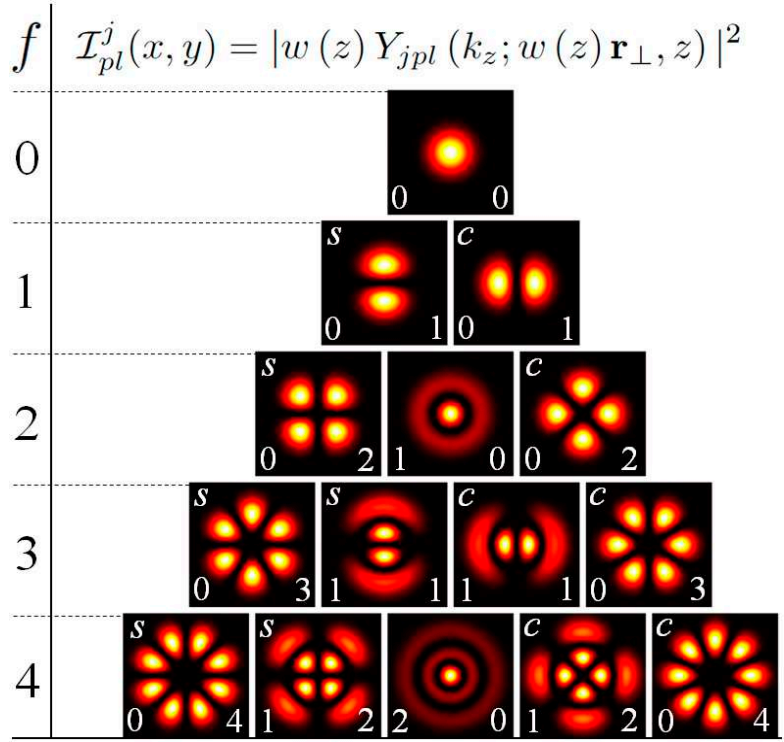}%
\caption{Transverse distribution of the Hybrid Laguerre--Gauss modes. Note
that the modes of its first family are the same as the corresponding
Hermite--Gauss modes.}%
\label{fQuanti5}%
\end{center}
\end{figure}

For the LG modes the family index is defined as $f=2p+|l|$. In Figure
\ref{fQuanti4} we show $\mathcal{I}_{pl}\left(  \mathbf{r}_{\perp}\right)
=|w\left(  z\right)  L_{pl}\left(  k_{z};w\left(  z\right)  \mathbf{r}_{\perp
},z\right)  |^{2}$ just as we did with the HG modes. Note that this intensity
profiles are independent of the polar angle $\phi$, and $p$ gives us the
number of zero--intensity circumferences (the number of \textit{radial nodes}).

Instead of the LG modes which have a complex dependence $\exp\left(
\mathrm{i}l\phi\right)  $ on the polar angle, sometimes it is more convenient
to use the \textit{Hybrid Laguerre--Gauss} (HLG) modes $\{Y_{pl}^{j}\left(
k_{z};\mathbf{r}_{\perp},z\right)  \}_{p,l\in%
\mathbb{N}
}^{j=c,s}$, which are defined as
\begin{subequations}
\label{Ypl}%
\begin{align}
Y_{cpl}\left(  k_{z};\mathbf{r}_{\perp},z\right)   &  =\frac{1}{\sqrt{2}%
}\left[  L_{pl}\left(  k_{z};\mathbf{r}_{\perp},z\right)  +L_{p,-l}\left(
k_{z};\mathbf{r}_{\perp},z\right)  \right]  \propto\cos\left(  l\phi\right),
\label{Ycpl}\\
Y_{spl}\left(  k_{z};\mathbf{r}_{\perp},z\right)   &  =\frac{1}{\sqrt{2}%
i}\left[  L_{pl}\left(  k_{z};\mathbf{r}_{\perp},z\right)  -L_{p,-l}\left(
k_{z};\mathbf{r}_{\perp},z\right)  \right]  \propto\sin\left(  l\phi\right)  ,
\label{Yspl}%
\end{align}
and have a trigonometric dependence on $\phi$. Of course, these relations are
not valid for $l=0$, where the polar dependence disappears and the convention
is to define the Hybrid \textit{cosine} mode as the LG mode and the Hybrid
\textit{sine} mode as zero, that is, $Y_{cp0}\left(  k_{z};\mathbf{r}_{\perp
},z\right)  =L_{p0}\left(  k_{z};\mathbf{r}_{\perp},z\right)  $ and
$Y_{sp0}\left(  k_{z};\mathbf{r}_{\perp},z\right)  =0$. In Figure
\ref{fQuanti5} we show the intensity of these modes. Note that they are like
the LG modes, but modulated by a periodic function of the polar angle; in
particular, given the dependence $\cos\left(  l\phi\right)  $ or $\sin\left(
l\phi\right)  $, $2l$ gives the number of zeros present in any circle around
$r=0$ (the number of \textit{polar nodes}). This set of transverse modes is
usually denoted by TEM$_{pl}^{\ast}$ in the literature.

In the following, in order to make the next derivations general, we define an
arbitrary set of transverse modes $\left\{  T_{\mathbf{n}}\left(
k_{z};\mathbf{r}_{\perp},z\right)  \right\}  _{\mathbf{n}}$ satisfying the
orthonormality and completeness relations%
\end{subequations}
\begin{subequations}
\begin{align}
\int_{%
\mathbb{R}
^{2}}d^{2}\mathbf{r}_{\perp}T_{\mathbf{n}}^{\ast}\left(  k_{z};\mathbf{r}%
_{\perp},z\right)  T_{\mathbf{n}^{\prime}}\left(  k_{z};\mathbf{r}_{\perp
},z\right)    & =\delta_{\mathbf{nn}^{\prime}},\\
\sum_{\mathbf{n}}T_{\mathbf{n}}^{\ast}\left(  k_{z};\mathbf{r}_{\perp}%
^{\prime},z\right)  T_{\mathbf{n}}\left(  k_{z};\mathbf{r}_{\perp},z\right)
& =\delta^{2}\left(  \mathbf{r}_{\perp}-\mathbf{r}_{\perp}^{\prime}\right)
\text{ }\forall z;
\end{align}
for example, when the modes are taken to be the HG, LG, or HLG, the set of
transverse modal indices $\{\mathbf{n\}}$ is $\{n,m\}_{n,m\in%
\mathbb{N}
}$, $\{p,l\}_{p\in%
\mathbb{N}
}^{l\in%
\mathbb{Z}
}$, and $\{j,p,l\}_{p,l\in%
\mathbb{N}
}^{j=c,s}$, respectively.

As for arbitrary $q\left(  z\right)  $ and $\psi\left(  z\right)  $ parameters
the transverse modes form a basis of \textrm{L}$^{2}\left(  \mathbf{r}_{\perp
}\right)  $, which is the space of physical transverse patterns (they cannot
extend to infinity or have singularities), the vector potential associated to
any optical beam propagating along the $z$ axis can be expanded as%
\end{subequations}
\begin{equation}
\mathbf{A}\left(  \mathbf{r},t\right)  =\sum_{\sigma=1,2}\sum_{\mathbf{n}}%
\int_{\mathcal{O}}dk_{z}\boldsymbol{\varepsilon}_{\sigma\mathbf{n}}\left(
k_{z}\right)  \mathcal{A}_{\sigma\mathbf{n}}(k_{z},t)T_{\mathbf{n}}\left(
k_{z};\mathbf{r}_{\perp},z\right)  \exp\left(  \mathrm{i}k_{z}z\right)  .
\label{ParaxialExpansion}%
\end{equation}
The HG, LG, and HLG modes satisfy the relations
\begin{subequations}
\begin{align}
H_{mn}^{\ast}\left(  k_{z};\mathbf{r}_{\perp},z\right)    & =H_{mn}\left(
-k_{z};\mathbf{r}_{\perp},z\right)  ,\\
L_{pl}^{\ast}\left(  k_{z};\mathbf{r}_{\perp},z\right)    & =L_{p,-l}\left(
-k_{z};\mathbf{r}_{\perp},z\right)  ,\\
Y_{jpl}^{\ast}\left(  k_{z};\mathbf{r}_{\perp},z\right)    & =Y_{jpl}\left(
-k_{z};\mathbf{r}_{\perp},z\right)  ;
\end{align}
hence, similarly to what happened with plane waves, the reality of the vector
potential imposes the following restrictions on the polarization and temporal
parts of the modes:%
\end{subequations}
\begin{subequations}
\begin{align}
\boldsymbol{\varepsilon}_{\sigma mn}^{\ast}\left(  k_{z}\right)
\mathcal{A}_{\sigma mn}^{\ast}(k_{z},t)  &  =\boldsymbol{\varepsilon}_{\sigma
mn}\left(  -k_{z}\right)  \mathcal{A}_{\sigma mn}(-k_{z},t),\\
\boldsymbol{\varepsilon}_{\sigma pl}^{\ast}\left(  k_{z}\right)
\mathcal{A}_{\sigma pl}^{\ast}(k_{z},t)  &  =\boldsymbol{\varepsilon}_{\sigma
p,-l}\left(  -k_{z}\right)  \mathcal{A}_{\sigma p,-l}(-k_{z},t),\\
\boldsymbol{\varepsilon}_{\sigma jpl}^{\ast}\left(  k_{z}\right)
\mathcal{A}_{\sigma jpl}^{\ast}(k_{z},t)  &  =\boldsymbol{\varepsilon}_{\sigma
jpl}\left(  -k_{z}\right)  \mathcal{A}_{\sigma jpl}(-k_{z},t)\text{.}%
\end{align}
Note finally that all these transverse bases have two free parameters, e.g.,
the spot size at the waist plane and the position of this plane; in free space
we can choose these parameters as we feel more convenient, and hence this
paraxial basis is actually a continuous family of bases. We will see that
these transverse modes form also a basis for light contained in an optical
cavity, but in this case the parameters of the basis get fixed by the cavity geometry.

\section{Quantization in an optical cavity}

\subsection[Geometrical optics: Propagation through optical elements and stable optical cavities]{Geometrical optics: Propagation through optical elements and stable optical cavities}

Optical cavities are useful for their ability to trap light. In their most
basic configuration, they consist on two spherical mirrors facing each other
(\textit{linear resonators}), so that when an optical beam enters the system,
it stays bouncing back and forth between these. In this section we explain the
conditions that any cavity must satisfy in order to truly confine light.

Consider an optical beam propagating nearly within the $z$ axis. In the
preceding section we studied the propagation of such a beam by taking into
account the diffraction effects of the wave equation. There is yet another
description of the beam which is very useful owed to its simplicity, but works
only for large \textit{Fresnel numbers} $F=d^{2}/\lambda l$, where $d$ is the
transverse size of the beam and $l$ the distance that it has propagated along
the $z$ axis. This approach is known as \textit{geometrical optics}, and it
visualizes the beam as a collection of \textit{collimated} \textit{rays}
propagating as straight lines (hence, it doesn't take into account the changes
in transverse size or shape of the beam). Within this description, any ray is
specified at a given longitudinal position $z$ by a vector%
\end{subequations}
\begin{equation}
\mathbf{v}\left(  z\right)  =%
\begin{bmatrix}
r\left(  z\right) \\
\theta\left(  z\right)
\end{bmatrix}
,
\end{equation}
being $r\left(  z\right)  $ its distance to the $z$ axis and $\theta(z)$ the
angle that it forms with the this axis as it propagates. We will call
\textit{transverse position} and \textit{inclination} to these \textit{ray
coordinates}.

The effect of \textit{optical elements} on the beam is described within this
framework by a so-called ABCD matrix%
\begin{equation}
M=%
\begin{bmatrix}
A & B\\
C & D
\end{bmatrix}
,
\end{equation}
which takes a ray $\mathbf{v}\left(  z_{in}\right)  =\operatorname{col}%
(r,\theta)$ of the beam at its input and transforms it into the ray%
\begin{equation}
\mathbf{v}\left(  z_{out}\right)  =M\mathbf{v}\left(  z_{in}\right)  =%
\begin{bmatrix}
Ar+B\theta\\
Cr+D\theta
\end{bmatrix}
\text{.}%
\end{equation}
The most simple optical operation consists on allowing the beam to propagate
in free space a distance $l$. The resulting ABCD matrix is trivially found to
be%
\begin{equation}
M_{\mathrm{free}}\left(  l\right)  =%
\begin{bmatrix}
1 & l\\
0 & 1
\end{bmatrix}
\text{,} \label{FreeABCD}%
\end{equation}
that is, the inclination of the rays doesn't change, while their transverse
positions are increased by $l\theta$. Another example consists in a spherical
interface with curvature radius\footnote{Some comments on sign conventions are
in order. First, optical elements are described as seen from the ray. The
curvature radius of convex (concave) dielectric surfaces is then taken as
positive (negative), and the opposite for reflecting surfaces.} $R$ separating
two dielectric media with refractive indices\footnote{The refractive index is
rigorously introduced in Chapter 4.} $n_{L}$ and $n_{R}$ (assumed propagation
from \textit{left} to \textit{right}); by using Snell's law in paraxial form
$n_{L}\theta_{L}=n_{R}\theta_{R}$ and simple geometrical arguments, it is
straightforward to prove that its ABCD matrix is%
\begin{equation}
M_{\text{\textrm{interface}}}\left(  n_{L},n_{R},R\right)  =%
\begin{bmatrix}
1 & 0\\
\frac{n_{L}-n_{R}}{n_{R}}\frac{1}{R} & \frac{n_{L}}{n_{R}}%
\end{bmatrix}
, \label{RefractionABCD}%
\end{equation}
that is, in this case the transverse positions of the rays do not change,
while their inclination does.

From these two fundamental ABCD systems (\textit{propagation} and
\textit{refraction}) one can find the ABCD matrices of a whole family of
useful optical elements such as \textit{thin lenses }or \textit{mirrors},
which have%
\begin{equation}
M_{\text{\textrm{lens}}}\left(  f\right)  =%
\begin{bmatrix}
1 & 0\\
-\frac{1}{f} & 1
\end{bmatrix}
\text{ \ \ and \ \ }M_{\text{\textrm{mirror}}}\left(  R\right)  =%
\begin{bmatrix}
1 & 0\\
-\frac{2}{R} & 1
\end{bmatrix}
,
\end{equation}
where $f$ is called the \textit{focal length} of the lens and $R$ is the
curvature radius of the mirror.

Given a general ABCD matrix $M$ connecting two planes with refractive indices
$n_{in}$ and $n_{out}$, one has $\det\{M\}=n_{in}/n_{out}$; to prove this just
note that $M$ can always be decomposed as a combination of propagation
(\ref{FreeABCD}) and refraction (\ref{RefractionABCD}) ABCD matrices, and that
the determinant of a product of matrices equals the product of the individual determinants.

It is very interesting to note that the transverse modes that we just
described in the previous section preserve their transverse shape upon
propagation through an arbitrary ABCD system. In particular, it can be proved
that if we take an arbitrary transverse mode with \textit{q--}parameter
$q\left(  z_{in}\right)  $ and Gouy phase $\psi\left(  z_{in}\right)  $ at the
input face of the ABCD system, it is transformed at the output face of the
optical system into the exact same mode but with new parameters
\begin{subequations}
\label{ABCDlaws}%
\begin{align}
q\left(  z_{out}\right)   &  =\frac{Aq\left(  z_{in}\right)  +B}{Cq\left(
z_{in}\right)  +D},\label{qABDClaw}\\
\psi\left(  z_{out}\right)   &  =\psi\left(  z_{in}\right)  -\arg\left\{
A+\frac{|B|}{q\left(  z_{in}\right)  }\right\}  ; \label{GouyABCDlaw}%
\end{align}
these expressions are known as the \textit{ABCD propagation laws}.

Let us now pass to describe optical cavities from this geometrical viewpoint.
Consider a ray $\mathbf{v}\left(  z_{0}\right)  =\operatorname{col}(r,\theta)$
at some reference plane $z=z_{0}$ of the cavity; after a roundtrip inside the
cavity, it is transformed by an ABCD matrix that we will call%
\end{subequations}
\begin{equation}
M_{\mathrm{roundtrip}}=%
\begin{bmatrix}
A_{\mathrm{rt}} & B_{\mathrm{rt}}\\
C_{\mathrm{rt}} & D_{\mathrm{rt}}%
\end{bmatrix}
,
\end{equation}
which consists in the composition of all the operations performed by the
optical elements inside the cavity until the ray comes back to the reference
plane $z=z_{0}$ with the same propagation direction (hence $\det
\{M_{\mathrm{roundtrip}}\}=1$). After $N$ roundtrips the ray coordinates will
become%
\begin{equation}
\mathbf{v}_{N}\left(  z_{0}\right)=M_{\mathrm{roundtrip}}^{N}%
\mathbf{v}\left(  z_{0}\right)=\frac{1}{\sin\theta}%
\begin{bmatrix}
A_{\mathrm{rt}}\sin N\theta-\sin\left[  \left(  N-1\right)  \theta\right]   &
B_{\mathrm{rt}}\sin N\theta\\
C_{\mathrm{rt}}\sin N\theta & D_{\mathrm{rt}}\sin N\theta-\sin\left(
N-1\right)  \theta
\end{bmatrix}
\mathbf{v}\left(  z_{0}\right)  ,
\end{equation}
with $\cos\theta=(A_{\mathrm{rt}}+D_{\mathrm{rt}})/2$, where the second
equality comes from the \textit{Sylvester theorem} applied to the $N$'th power
of a $2\times2$ matrix with determinant one. The ray will stay confined in the
cavity only if $\theta$ is real, as otherwise the trigonometric functions
become hyperbolic, and the ray coordinates diverge. This means that the
condition for a resonator to confine optical beams is%
\begin{equation}
-1<\frac{A_{\mathrm{rt}}+D_{\mathrm{rt}}}{2}<1\label{ADstability}%
\end{equation}
An optical cavity satisfying this condition is said to be \textit{stable}.
Note that it looks like this condition may depend on the reference plane from
which the roundtrip matrix is evaluated; however, it is possible to show that
the trace of the roundtrip matrix does not depend on the reference plane.

As stated, in this thesis we assume that the optical cavity consists in two
spherical mirrors with curvature radii $R_{1}$ and $R_{2}$ separated a
distance $L$; for reasons that will become clear in Chapter \ref{OPOs}, we
allow for the existence of a plane dielectric slab of length $l_{\mathrm{c}}$
and refractive index $n_{\mathrm{c}}$. Taking the reference plane at mirror
one with the rays propagating away from it, the roundtrip ABCD matrix of this
simple cavity is easily found to be%
\begin{equation}
M_{\mathrm{roundtrip}}=%
\begin{bmatrix}
2g_{2}-1 & 2g_{2}L_{\mathrm{eff}}\\
2\left(  2g_{1}g_{2}-g_{2}-g_{1}\right)  /L_{\mathrm{eff}} & 4g_{1}%
g_{2}-2g_{2}-1
\end{bmatrix}
, \label{rtMatrix}%
\end{equation}
where we have defined the \textit{g--parameters} of the cavity%
\begin{equation}
g_{j}=1-\frac{L_{\mathrm{eff}}}{R_{j}}\text{ \ \ }(j=1,2),
\end{equation}
as well as its \textit{effective length}%
\begin{equation}
L_{\mathrm{eff}}=L-\left(  1-\frac{1}{n_{\mathrm{c}}}\right)  l_{\mathrm{c}}.
\end{equation}
This linear resonator is stable then only if%
\begin{equation}
0<g_{1}g_{2}<1. \label{g1g2Stability}%
\end{equation}
In this thesis we will only work with stable linear cavities.

\subsection{Considerations at a non-reflecting interface}

As stated, throughout the thesis we will work with a linear cavity having a
dielectric medium inside. In this section we discuss some particularities that
must be taken into account for the field due to the presence of this medium.

The details concerning how Maxwell's equations are modified in the presence of
a medium with refractive index $n_{\mathrm{m}}$ will be explained in Chapter
\ref{OPOs}. Here, however, we just want to note that for a homogeneous,
isotropic medium and reasonably weak fields, the only effect of the medium is
to change the wave equation of the vector potential (\ref{WaveEq}) as%
\begin{equation}
\left(  \frac{1}{n_{\mathrm{m}}^{2}}\boldsymbol{\nabla}^{2}-\frac{1}{c^{2}%
}\partial_{t}^{2}\right)  \mathbf{A}=\mathbf{0}\text{.} \label{MediumWaveEq}%
\end{equation}
Therefore, for field configurations of the type we have discussed,
$\boldsymbol{\varepsilon}u\left(  \mathbf{r}\right)  \mathcal{A}(t)$, the
effect of the medium is to change the Helmholtz equation for the spatial part
as%
\begin{equation}
\boldsymbol{\nabla}^{2}u+n_{\mathrm{m}}^{2}k^{2}u=0,
\end{equation}
that is, to multiply by a factor $n_{\mathrm{m}}$ the wave vector of the
spatial modes of the system. Hence, except for this $n_{\mathrm{m}}$ factor,
plane waves (or transverse modes within the paraxial approximation) can still
be considered the spatial modes of a medium which fills the whole $%
\mathbb{R}
^{3}$ space.

A different matter is what happens when the field crosses a plane interface
separating two media with refractive indices $n_{L}$ and $n_{R}$. The first
interesting thing to note is that the shape of the transverse modes is not
modified at all, what is not surprising by continuity arguments: It would make
no sense that the thickness or the radius of the beam changed abruptly upon
crossing the interface. To see this, just note that given the $q$ and $\psi$
parameters of the modes at both sides of the interface, denoted by
$(q_{L},\psi_{L})$ and $(q_{R},\psi_{R})$, the propagation laws
(\ref{ABCDlaws}) together with the refraction ABCD matrix
(\ref{RefractionABCD}) tells us that%
\begin{equation}
\frac{n_{R}}{q_{R}}=\frac{n_{L}}{q_{L}}\text{ \ \ and \ \ }\psi_{R}=\psi_{L};
\end{equation}
hence, the complex Gaussian profile of the modes match each other, that is%
\begin{equation}
\exp\left[  \mathrm{i}n_{R}k_{z}\frac{\mathbf{r}_{\perp}^{2}}{2q_{R}%
}+\mathrm{i}\left(  1+f\right)  \psi_{R}\right]  =\exp\left[  \mathrm{i}%
n_{L}k_{z}\frac{\mathbf{r}_{\perp}^{2}}{2q_{L}}+\mathrm{i}\left(  1+f\right)
\psi_{L}\right]  ,
\end{equation}
and therefore the modes are independent of the refractive index. Note though,
that the spot size and the curvature radius of the modes in a medium with
refractive index $n_{\mathrm{m}}$ are defined by%
\begin{equation}
\frac{1}{q\left(  z\right)  }=\frac{1}{R\left(  z\right)  }+\mathrm{i}\frac
{2}{n_{\mathrm{m}}k_{z}w^{2}\left(  z\right)  }\text{.}%
\end{equation}

Crossing an interface induces further relations between the vector potentials
at both sides of the interface, say $\mathbf{A}_{L}\left(  \mathbf{r}%
,t\right)  $ and $\mathbf{A}_{R}\left(  \mathbf{r},t\right)  $ with%
\begin{equation}
\mathbf{A}_{j}\left(  \mathbf{r},t\right)  =\boldsymbol{\varepsilon}%
A_{j}(t)T\left(  k;\mathbf{r}_{\perp},z_{\mathrm{I}}\right)  \exp\left(
-\mathrm{i}\omega t+\mathrm{i}n_{j}kz_{\mathrm{I}}\right)  +\mathrm{c.c.},
\label{ClassicalVectorPotential}%
\end{equation}
for a single mode with polarization $\boldsymbol{\varepsilon}$, frequency
$\omega=ck$, and transverse profile $T\left(  k;\mathbf{r}_{\perp},z\right)
$. $z_{\mathrm{I}}$ is the longitudinal position of the plane interface and
$A_{j}(t)$ is a slowly varying envelope which remains almost unaltered after
an optical cycle, that is,%
\begin{equation}
\dot{A}_{j}(t)\ll2\pi/\omega.
\end{equation}
The corresponding electric and magnetic fields are given by%
\begin{subequations}
\begin{align}
\mathbf{E}_{j}\left(  \mathbf{r},t\right)   &  =\mathrm{i}\omega
\boldsymbol{\varepsilon}A_{j}(t)T\left(  k;\mathbf{r}_{\perp},z_{\mathrm{I}%
}\right)  \exp\left(  -\mathrm{i}\omega t+\mathrm{i}n_{j}kz_{\mathrm{I}%
}\right)  +\mathrm{c.c.},\\
\mathbf{B}_{j}\left(  \mathbf{r},t\right)   &  =\mathrm{i}\frac{n_{j}\omega
}{c}\left(  \mathbf{e}_{z}\times\boldsymbol{\varepsilon}\right)
A_{j}(t)T\left(  k;\mathbf{r}_{\perp},z_{\mathrm{I}}\right)  \exp\left(
-\mathrm{i}\omega t+\mathrm{i}n_{j}kz_{\mathrm{I}}\right)  +\mathrm{c.c.},
\end{align}
where we have made use of the paraxial approximations
(\ref{ParaxialConditions}). The instantaneous power content in the transverse
plane is given by%
\end{subequations}
\begin{equation}
P_{j}(t)=\frac{1}{\mu_{0}}\frac{\omega}{2\pi}\int_{t-\pi/\omega}^{t+\pi
/\omega}dt^{\prime}\int_{%
\mathbb{R}
^{2}}d^{2}\mathbf{r}_{\perp}\left\vert \mathbf{E}_{j}\left(  \mathbf{r}%
,t^{\prime}\right)  \mathbf{\times B}_{j}\left(  \mathbf{r},t^{\prime}\right)
\right\vert =\frac{2n_{j}\omega^{2}}{c\mu_{0}}\left\vert A_{j}(t)\right\vert
^{2}, \label{Power}%
\end{equation}
that is, by the absolute value of the time averaged Poynting vector,
integrated in the transverse plane. Now, assume that the interface has some
kind of anti--reflecting coating, so that the whole beam is transmitted from
the left medium to the right one; as the power must be conserved, that is,
$P_{L}=P_{R}$, this establishes the relation%
\begin{equation}
\mathbf{A}_{L}\left(  \mathbf{r},t\right)  =\sqrt{\frac{n_{R}}{n_{L}}%
}\mathbf{A}_{R}\left(  \mathbf{r},t\right)  \text{.}%
\end{equation}
Note that even though we have argued with a single mode, this relation holds
for any field as follows from a trivial generalization of the argument.

All these relations will be important to perform a proper quantization of the
field inside the cavity.

\subsection{Modes of an optical resonator\label{CavityModes}}

As in the case of free space, the first step in order to quantize the
electromagnetic field inside an optical resonator is to find the set of
spatial modes satisfying the Helmoltz equation and the boundary conditions
imposed by the cavity mirrors. We are going to work with paraxial optical
beams whose transverse size is small compared with the transverse dimensions
of the cavity; under these conditions, we know that any transverse structure
of the field can be expanded in terms of the paraxial transverse modes
(\ref{ParaxialExpansion}) that we introduced in Section \ref{TransModes}. On
the other hand, the mirrors impose an additional boundary condition on these
transverse modes: After a roundtrip in the cavity, they must exactly match
themselves, that is, not only they need to be shape--preserving with
propagation, but also \textit{self--reproducing }after a roundtrip. This means
that the modes of a general optical cavity can be found by imposing the
condition%
\begin{equation}
T_{\mathbf{n}}\left(  k_{z};\mathbf{r}_{\perp},z_{0}\right)  \exp\left(
\mathrm{i}k_{z}z_{0}\right)  =\left.  T_{\mathbf{n}}\left(  k_{z}%
;\mathbf{r}_{\perp},z\right)  \exp\left(  \mathrm{i}k_{z}z\right)  \right\vert
_{z_{0}\oplus\mathrm{r.t}}, \label{CavityModesDef}%
\end{equation}
where `$z_{0}\oplus\mathrm{r.t}$' stands for `roundtrip propagation from the
reference plane $z=z_{0}$'. Using the ABCD propagation laws (\ref{ABCDlaws}),
this condition is easily proved to be equivalent to%
\begin{subequations}
\begin{align}
q\left(  z_{0}\right)   &  =\frac{A_{\mathrm{rt}}q\left(  z_{0}\right)
+B_{\mathrm{rt}}}{C_{\mathrm{rt}}q\left(  z_{0}\right)  +D_{\mathrm{rt}}%
}\text{,}\\
2q\pi &  =-\left(  1+f\right)  \arg\left\{  A_{\mathrm{rt}}+\frac
{B_{\mathrm{rt}}}{q\left(  z_{0}\right)  }\right\}  +2k_{z}L_{\mathrm{opt}},
\end{align}
with $q\in%
\mathbb{N}
$ the so-called \textit{longitudinal index }(do not confuse it with the
\textit{q}--parameter of the modes, $q\left(  z\right)  $, which is always
written with a label referring to the longitudinal position where it is
evaluated). $L_{\mathrm{opt}}$ is the \textit{optical length} of the cavity,
that is, if rays propagate a longitudinal length $l_{j}$ on the $j$'th optical
element with refractive index $n_{j}$, $L_{\mathrm{opt}}=\sum_{j}n_{j}l_{j}$,
where the sum extends to all the optical elements contained in the cavity
(including free space propagation).

The first condition fixes the \textit{q}--parameter of the modes at the
reference plane to%
\end{subequations}
\begin{equation}
\frac{1}{q\left(  z_{0}\right)  }=\frac{D_{\mathrm{rt}}-A_{\mathrm{rt}}%
}{2B_{\mathrm{rt}}}+\frac{\mathrm{i}}{|B_{\mathrm{rt}}|}\sqrt{1-\left(
\frac{A_{\mathrm{rt}}+D_{\mathrm{rt}}}{2}\right)  ^{2}}\text{.}%
\end{equation}
Surprisingly, this expression leads to the condition $-1<\left(
A_{\mathrm{rt}}+D_{\mathrm{rt}}\right)  /2<1$ for these modes to exist in the
resonator, which is exactly the stability condition (\ref{ADstability}) that
we already found by geometrical means. On the other hand, the second condition
discretizes the frequencies at which transverse modes can \textit{resonate}
inside the cavity; in particular, the modes exists at the cavity only at
frequencies%
\begin{equation}
\omega_{qf}=\Omega_{\mathrm{FSR}}\left[  q+\frac{1+f}{2\pi}\zeta\right]  ,
\end{equation}
where%
\begin{equation}
\zeta=\arg\left\{  \left(  \frac{A_{\mathrm{rt}}+D_{\mathrm{rt}}}{2}\right)
+\mathrm{i}\sqrt{1-\left(  \frac{A_{\mathrm{rt}}+D_{\mathrm{rt}}}{2}\right)
^{2}}\right\}
\end{equation}
is the Gouy phase accumulated after the roundtrip, $\Omega_{\mathrm{FSR}}=\pi
c/L_{\mathrm{opt}}$ is the so-called \textit{free spectral range} of the
resonator, and $f$ is the family index that we defined in Section
\ref{TransModes}. Note that these frequencies are independent of the reference
plane $z=z_{0}$ because they depend only on the trace of the roundtrip matrix.
Note also that modes having the same family index $f$ `live' inside the cavity
at the same frequencies, what gives sense to the name of this index.%

\begin{figure}
[ptb]
\begin{center}
\includegraphics[
height=2.3653in,
width=4.8084in
]%
{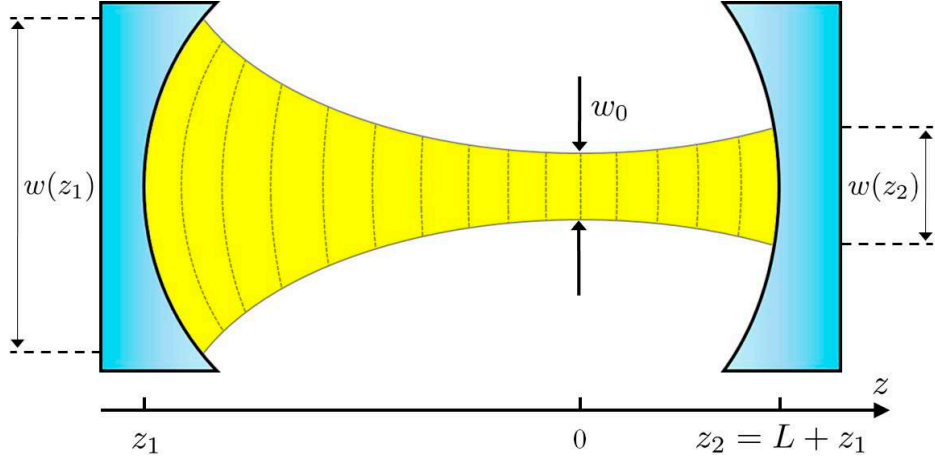}%
\caption{Schematic representation of the wavefront and the thickness of the
modes in a linear cavity with two concave mirrors. Note that in this case the
waist plane is between the mirrors.}%
\label{fQuanti6}%
\end{center}
\end{figure}

For the particular case of the linear resonator that we introduced above,
which has an optical length%
\begin{equation}
L_{\mathrm{opt}}=L+\left(  n_{\mathrm{c}}-1\right)  l_{\mathrm{c}},
\end{equation}
we have%
\begin{equation}
\zeta=2\arccos\pm\sqrt{g_{1}g_{2}}\in\lbrack0,\pi],
\end{equation}
where the `+' and `--' signs hold for $g_{2}>0$ and $g_{2}<0$, respectively.
Hence the resonance distribution of the different families inside the cavity
is fixed by the product $g_{1}g_{2}$.

On other hand, calling $z_{1}$ to the longitudinal position of mirror 1, whose
plane we took as the reference when evaluating the roundtrip matrix
(\ref{rtMatrix}), we get%
\begin{subequations}
\begin{align}
R\left(  z_{1}\right)   &  =-R_{1},\\
w_{qf}\left(  z_{1}\right)   &  =\left(  \frac{L_{\mathrm{eff}}\lambda_{qf}%
}{\pi}\right)  ^{1/2}\left[  \frac{g_{2}}{g_{1}\left(  1-g_{1}g_{2}\right)
}\right]  ^{1/4},
\end{align}
where $\lambda_{qf}=2\pi c/\omega_{qf}$. Note that even though the spot size
depends on the particular longitudinal and family indices $(q,f)$ of the mode,
the \textit{q}--parameter is the same for all of them, as it depends on
$k_{z}w^{2}$ which is independent of $(q,f)$. Note also that the spherical
wavefront of the modes is adapted to mirror 1 (the sign just means that the
wave is propagating away from the mirror). We could have taken the position
$z=z_{2}$ of mirror 2 as the longitudinal reference and would have found that
the wavefront is also adapted to mirror 2 at $z=z_{2}$. Hence, upon
propagation from mirror 1 to mirror 2, the mode is reshaped as shown in
Figures \ref{fQuanti6} and \ref{fQuanti7}. For a cavity with two concave
mirrors (Figure \ref{fQuanti6}) the wavefront of the modes has to change the
sign of its radius with the propagation, and hence it becomes plane at a
longitudinal position inside the cavity. This is not the case if one of the
mirrors is convex (Figure \ref{fQuanti7}), a situation leading to a waist
plane outside the cavity.%

\begin{figure}
[ptb]
\begin{center}
\includegraphics[
height=2.7605in,
width=4.7772in
]%
{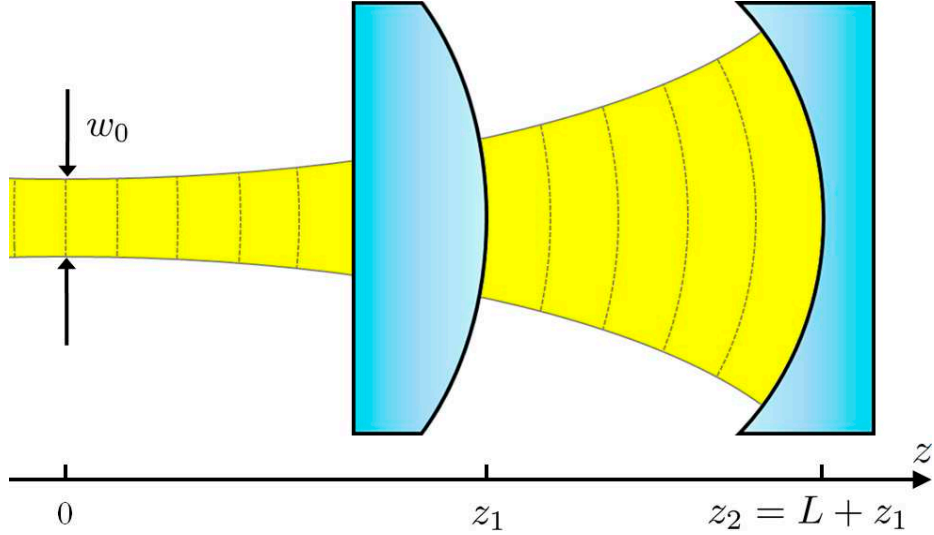}%
\caption{Schematic representation of the wavefront and the thickness of the
modes in a linear cavity having one convex mirror. In this case the waist
plane is outside resonator.}%
\label{fQuanti7}%
\end{center}
\end{figure}

Having characterized the modes in the plane $z=z_{1}$, we can now find them at
any other plane by using the ABCD propagation laws. In particular, we would
like to refer the modes to the plane in which the wavefronts are plane, that
is, $R\left(  z=0\right)  =\infty$, which we take as the longitudinal origin
$z=0$; we will denote this plane by \textit{beam waist} or \textit{waist
plane}, as it is the plane where the spot size of the modes is minimum. Let us
denote by $w_{0,qf}$ the spot size of a mode with indices $(q,f)$ at this
plane, and by $z_{1,\mathrm{eff}}$ the longitudinal position of mirror 1 (as
we allow an intracavity plane dielectric slab, we have to use effective
lengths, and then transform to geometric length depending on where the slab is
placed). The ABCD propagation law (\ref{qABDClaw}) allows us to write
$q\left(  z_{0}\right)  =q\left(  z_{1}\right)  +z_{1,\mathrm{eff}}$, which
after imposing $q\left(  z_{0}\right)  =k_{z}w_{0}^{2}/2\mathrm{i}$, leads us
to%
\end{subequations}
\begin{subequations}
\begin{align}
z_{1,\mathrm{eff}}  &  =L_{\mathrm{eff}}\frac{(1-g_{1})|g_{2}|}{|g_{1}%
+g_{2}-2g_{1}g_{2}|},\label{z1}\\
w_{0,qf}  &  =\left(  \frac{\lambda_{qf}L_{\mathrm{eff}}}{\pi}\right)
^{1/2}\left[  \frac{g_{1}g_{2}(1-g_{1}g_{2})}{(g_{1}+g_{2}-2g_{1}g_{2})^{2}%
}\right]  ^{1/4}\text{.} \label{w0}%
\end{align}
The first thing to note is that the waist plane is the same for all the modes,
even though they have different spot sizes in general. Taking this waist plane
as our reference, the modes at any other plane (or even outside the cavity if
it is allowed to leave, what we will do in the next section) are given by
(\ref{FreePropLaws}), but with $z$ being an effective longitudinal coordinate
when dielectric plane slabs are involved, that is,
\end{subequations}
\begin{subequations}
\label{PropModePar}%
\begin{align}
w_{qf}^{2}\left(  z\right)   &  =w_{0,qf}^{2}\left[  1+\left(  \frac
{z_{\mathrm{eff}}}{z_{R}}\right)  ^{2}\right]  ,\\
R\left(  z\right)   &  =z_{\mathrm{eff}}\left[  1+\left(  \frac{z_{R}%
}{z_{\mathrm{eff}}}\right)  ^{2}\right]  ,\\
\psi\left(  z\right)   &  =-\arctan\left(  \frac{z_{\mathrm{eff}}}{z_{R}%
}\right)  \in\lbrack-\frac{\pi}{2},\frac{\pi}{2}].
\end{align}

To fix ideas, let us consider the example an empty symmetric resonator, that
is, a resonator satisfying $g_{1}=g_{2}\equiv g=1-L/R$, in which the cavity
resonances and Rayleigh length read%
\end{subequations}
\begin{subequations}
\begin{align}
\omega_{qf}  &  =\Omega_{\mathrm{FSR}}\left[  q+\frac{1+f}{\pi}\arccos
g\right]  ,\\
|z_{R}|  &  =\frac{L}{2}\sqrt{\frac{1+g}{1-g}}.
\end{align}
Consider now the following extreme situations:
\end{subequations}
\begin{itemize}
\item $R\rightarrow\infty$ (near--planar resonator). In this case the
longitudinal frequencies of the different transverse families are $\omega
_{qf}=\Omega_{\mathrm{FSR}}q$. Hence, at every cavity resonance we can find
all the transverse modes. In this case the Rayleigh length is by far larger
than the cavity length, what means that the modes are basically plane waves.

\item $L=2R$ (concentric resonator). In this case $\omega_{qf}=\Omega
_{\mathrm{FSR}}(q+f+1)$, what means that $\omega_{qf}=\omega_{q-1,f+1}%
=\omega_{q-2,f+2}=...$, and hence, we find again all the transverse modes at a
given resonance. In this case the Rayleigh length is zero, and hence the modes
are point--like at the waist plane.

\item $L=R$ (confocal resonator). In this case $\omega_{qf}=\Omega
_{\mathrm{FSR}}(q+f/2+1/2)$, which implies $\omega_{qf}=\omega_{q-1,f+2}%
=\omega_{q-2,f+4}=...$, and hence even and odd families are separated inside
the cavity by half a free spectral range. As for the Rayleigh length, it is
half the cavity's effective length.
\end{itemize}

To get an idea of the numbers, let us consider the example of an empty,
confocal, symmetric resonator with lengths varying from 5 $\mathrm{mm}$ to 500
\ \textrm{mm}. The free spectral range varies then between 200 \textrm{GHz}
and 2 \textrm{GHz}. On the other hand, the spot size at the waist varies from
20 $\mathrm{\mu m}$ to 205 $\mathrm{\mu m}$ at a wavelength of 532 \textrm{nm
}(frequency of $3.5\times10^{15}$ $\mathrm{Hz}$).

\subsection{Quantization of the electromagnetic field inside a
cavity\label{CavityQuanti}}

\begin{figure}
[ptb]
\begin{center}
\includegraphics[
height=1.5636in,
width=3.0035in
]%
{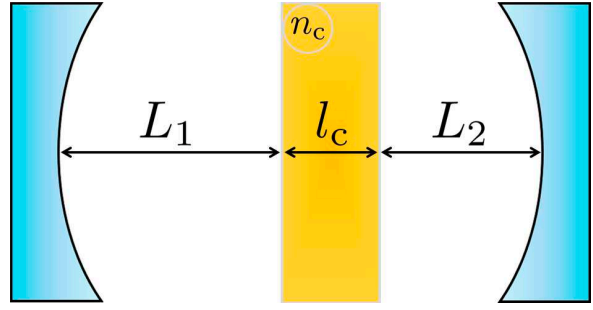}%
\caption{We quantize the electromagnetic field inside a cavity containing a
dielectric slab.}%
\label{fQuanti8}%
\end{center}
\end{figure}

In the previous discussion we have shown that the modes of the resonator are
the transverse modes that we studied in the previous section, but with a
\textit{q}--parameter fixed by the cavity geometry. In addition, instead of a
continuous set of optical wave vectors $k_{z}$ only a discrete set $\left\{
k_{z,qf}=\pm\omega_{qf}/c\right\}  _{q\in%
\mathbb{Z}
}$ can exist for the modes contained in a given family $f$. Hence, the vector
potential inside the linear cavity depicted in Figure \ref{fQuanti8} can be
written as%
\begin{align}
\mathbf{A}\left(  \mathbf{r},t\right)  =\sum_{\sigma=1,2}\sum_{\mathbf{n}}%
\sum_{k_{z}^{qf}\in\mathcal{O}}\frac{1}{\sqrt{n(z)}}\boldsymbol{\varepsilon
}_{\sigma\mathbf{n}}\left(  k_{z,qf}\right)    & \mathcal{A}_{\sigma
\mathbf{n}}(k_{z,qf},t)T_{\mathbf{n}}\left(  k_{z,qf};\mathbf{r}_{\perp
},z\right)  \\
\times & \exp\left[  \mathrm{i}n(z)k_{z,qf}z\right]  ,\nonumber
\end{align}
where we have introduced the $z$--dependent refractive index%
\begin{equation}
n(z)=\left\{
\begin{array}
[c]{cc}%
1 & z_{1}<z<z_{1}+L_{1}\text{ \ \ or \ \ }z_{1}+L_{1}+l_{\mathrm{c}}%
<z<z_{1}+L\\
n_{\mathrm{c}} & z_{1}+L_{1}<z<z_{1}+L_{1}+l_{\mathrm{c}}%
\end{array}
\right.  .
\end{equation}
Note that the refractive index does not appear in $\boldsymbol{\varepsilon
}_{\sigma\mathbf{n}}\left(  k_{z,qf}\right)  $ or $\mathcal{A}_{\sigma
\mathbf{n}}(k_{z,qf},t)$ because $k_{z,qf}$ acts in that case just as a mode label.%

Quantization is made following exactly the same procedure as in free space
(see Section \ref{PlaneQuanti}) but with some particularities that we discuss
now. We define first the electromagnetic normal variables for each mode%
\begin{equation}
\nu_{\sigma\mathbf{n}}(k_{z,qf},t)=\frac{1}{\Gamma}\left[  \mathcal{A}%
_{\sigma\mathbf{n}}(k_{z,qf},t)+\frac{\mathrm{i}}{\omega_{qf}}\mathcal{\dot
{A}}_{\sigma\mathbf{n}}(k_{z,qf},t)\right]  ,
\end{equation}
which again evolve as the normal variables of harmonic oscillators
(\ref{NormalEvo}) with the corresponding frequency. In this case the
dielectric medium contributes to the electromagnetic energy contained in
system, which cannot be written as (\ref{Eem}) in its presence, but in the
modified form \cite{Jackson62book,Griffiths99book}%
\begin{equation}
E_{\mathrm{em}}=\frac{1}{2}\int d^{3}\mathbf{r}\left[  \varepsilon_{0}%
n^{2}(z)\mathbf{E}^{2}\left(  \mathbf{r},t\right)  +\frac{1}{\mu_{0}%
}\mathbf{B}^{2}\left(  \mathbf{r},t\right)  \right]  .
\end{equation}
In order to simplify this expression we use some approximations. First, the
paraxial condition (\ref{zSlowCond}) allow us to perform the approximation%
\begin{equation}
\int_{z_{0}}^{z_{0}+l}dz\int_{%
\mathbb{R}
^{2}}d^{2}\mathbf{r}_{\perp}T_{\mathbf{n}}\left(  k_{z,qf};\mathbf{r}_{\perp
},z\right)  T_{\mathbf{n}^{\prime}}\left(  k_{z,qf}^{\prime};\mathbf{r}%
_{\perp},z\right)\exp\left[  \mathrm{i}n(z)\left(  k_{z,qf}%
+k_{z,qf}^{\prime}\right)  z\right]\simeq  l\delta_{\mathbf{nn}^{\prime}}\delta_{k_{z,qf},-k_{z,qf}^{\prime}},
\end{equation}
valid for any reference plane $z=z_{0}$ as long as $l\gg\{|k_{z,qf}%
|^{-1},|k_{z,qf}^{\prime}|^{-1}\}$, that is, as long as light is able to
undergo many cycles between $z=z_{0}$ and $z=z_{0}+l$. Then, the second
paraxial condition (\ref{tSlowCond}) leads to the approximation%
\begin{equation}
\boldsymbol{\nabla}\left\{  T_{\mathbf{n}}\left(  k_{z,qf};\mathbf{r}_{\perp
},z\right)  \exp\left[  \mathrm{i}n(z)k_{z,qf}z\right]  \right\}
\simeq\mathrm{i}n(z)k_{z,qf}\mathbf{e}_{z}T_{\mathbf{n}}\left(  k_{z,qf}%
;\mathbf{r}_{\perp},z\right)\exp\left[  \mathrm{i}n(z)k_{z,qf}z\right]  \text{.}
\end{equation}
With this approximations at hand, it is straightforward to write the
electromagnetic energy in terms of the normal variables as%
\begin{equation}
E_{\mathrm{em}}=\sum_{\sigma=1,2}\sum_{\mathbf{n}}\sum_{k_{z}^{qf}%
\in\mathcal{O}}\frac{\varepsilon_{0}L_{\mathrm{opt}}\Gamma^{2}\omega_{qf}^{2}%
}{2}\nu_{\sigma\mathbf{n}}^{\ast}(k_{z,qf})\nu_{\sigma\mathbf{n}}(k_{z,qf}).
\end{equation}
The last particularity of optical cavities is that there is no sense in making
a distinction between modes propagating to the right or to the left, because
they cannot be excited separately. Hence, we just define collective normal
variables $\nu_{\sigma q\mathbf{n}}=\nu_{\sigma\mathbf{n}}(\pm|k_{z,qf}|)$, so
that the modes get completely specified inside the cavity by their
polarization $\sigma$, and their longitudinal and transverse indices
$(q,\mathbf{n})$. Therefore, the electromagnetic energy can be finally written
as%
\begin{equation}
E_{\mathrm{em}}=\sum_{\sigma q\mathbf{n}}\frac{\omega_{qf}^{2}}{2}\nu_{\sigma
q\mathbf{n}}^{\ast}\nu_{\sigma q\mathbf{n}},
\end{equation}
where we have chosen $\Gamma=1/\sqrt{2\varepsilon_{0}L_{\mathrm{opt}}}$ so
that this expression coincides with that of a collection of harmonic
oscillators as before. Quantization is therefore introduced by making a
correspondence between the normal variables $\{\nu_{\sigma q\mathbf{n}}%
(t),\nu_{\sigma q\mathbf{n}}^{\ast}(t)\}$ and boson operators $\sqrt
{2\hbar/\omega_{qf}}\{\hat{a}_{\sigma q\mathbf{n}}(t),\hat{a}_{\sigma
q\mathbf{n}}^{\dagger}(t)\}$ satisfying the commutation relations%
\begin{subequations}
\begin{align}
\lbrack\hat{a}_{\sigma q\mathbf{n}}\left(  t\right)  ,\hat{a}_{\sigma^{\prime
}q^{\prime}\mathbf{n}^{\prime}}^{\dagger}\left(  t\right)  ]  & =\delta
_{\sigma\sigma^{\prime}}\delta_{qq^{\prime}}\delta_{\mathbf{nn}^{\prime}%
}\text{,}\\
\lbrack\hat{a}_{\sigma q\mathbf{n}}\left(  t\right)  ,\hat{a}_{\sigma^{\prime
}q^{\prime}\mathbf{n}^{\prime}}\left(  t\right)  ]  & =[\hat{a}_{\sigma
q\mathbf{n}}^{\dagger}\left(  t\right)  ,\hat{a}_{\sigma^{\prime}q^{\prime
}\mathbf{n}^{\prime}}^{\dagger}\left(  t\right)  ]=0.
\end{align}
The Hamiltonian of the electromagnetic field inside the cavity is then written
as%
\end{subequations}
\begin{equation}
\hat{H}_{\mathrm{em}}=\sum_{\sigma q\mathbf{n}}\hbar\omega_{qf}\hat{a}_{\sigma
q\mathbf{n}}^{\dagger}\hat{a}_{\sigma q\mathbf{n}},
\end{equation}
where we have already removed the infinite contribution appearing when using
the commutation relations to write it in normal order.

Finally, the vector potential inside the resonator is given by the operator%
\begin{equation}\label{Acav}
\mathbf{\hat{A}}^{(+)}\left(  \mathbf{r},t\right)=\sum_{\sigma
q\mathbf{n}}\frac{1}{2}\sqrt{\frac{\hbar}{n(z)\varepsilon_{0}L_{\mathrm{opt}%
}\omega_{qf}}}\boldsymbol{\varepsilon}_{\sigma q\mathbf{n}}\hat{a}_{\sigma
q\mathbf{n}}(t)\left\{  T_{\mathbf{n}}\left(  k_{qf};\mathbf{r}_{\perp},z\right)
\exp\left[  \mathrm{i}n(z)k_{qf}z\right]  +\mathrm{c.c.}\right\} ,
\end{equation}
where we have defined the positive definite modal wave vector $k_{qf}%
=\omega_{qf}/c$, while the electric and magnetic fields correspond then to%
\begin{subequations}
\begin{align}
\mathbf{\hat{E}}^{(+)}\left(  \mathbf{r},t\right)&=\mathrm{i}\sum_{\sigma
q\mathbf{n}}\frac{1}{2}\sqrt{\frac{\hbar\omega_{qf}}{n(z)\varepsilon
_{0}L_{\mathrm{opt}}}}\boldsymbol{\varepsilon}_{\sigma q\mathbf{n}}\hat
{a}_{\sigma q\mathbf{n}}(t)\left\{  T_{\mathbf{n}}\left(  k_{qf};\mathbf{r}_{\perp},z\right)
\exp\left[  \mathrm{i}n(z)k_{z,qf}z\right]  +\mathrm{c.c.}\right\},\label{Ecav}
\\
\mathbf{\hat{B}}^{(+)}\left(  \mathbf{r},t\right)=&\mathrm{i}\sum_{\sigma
q\mathbf{n}}\frac{1}{2}\sqrt{\frac{n(z)\hbar\omega_{qf}}{c^{2}\varepsilon
_{0}L_{\mathrm{opt}}}}\boldsymbol{\varepsilon}_{\sigma q\mathbf{n}}\hat
{a}_{\sigma q\mathbf{n}}(t)\left\{  T_{\mathbf{n}}\left(  k_{qf};\mathbf{r}_{\perp},z\right)
\exp\left[  \mathrm{i}n(z)k_{z,qf}z\right]  -\mathrm{c.c.}\right\},
\end{align}
where the expressions emphasize the parts of the fields propagating to the
right and to the left, which will be of later convenience even if their
associated boson operators are the same.
\end{subequations} 

%% file: OpenSystemsFO.tex
In the previous chapter we have been able to quantize the electromagnetic
inside an optical cavity formed by perfectly reflecting mirrors. However, in
reality optical cavities must have at least one partially transmitting mirror
allowing us to both inject light inside it and observe the intracavity
processes by studying the light that comes out of it. In this section we
explain how to deal with such an open cavity within the quantum formalism. To
this aim we first build a model for an open cavity with only one partially
transmitting mirror, and then proceed to derive the evolution equations
satisfied by the intracavity field both in the Heisenberg and Schr\"{o}dinger pictures.

In the last section, and as an example of another type of phenomenology that
can be studied with the formalism of open quantum systems, we show how
quantum--optical phenomena can be simulated with ultra--cold atoms trapped in
optical lattices \cite{deVega08,Navarrete11b} (a work developed by the author
of the thesis at the Max--Planck Institute for Quantum Optics).

\section{The open cavity model}

For simplicity, we consider only one cavity mode with frequency $\omega
_{\mathrm{c}}$, polarization $\boldsymbol{\varepsilon}$, and in the transverse
mode $T(\omega_{\mathrm{c}}/c;\mathbf{r}_{\perp},z)$, whose boson operators we
denote by $\{\hat{a},\hat{a}^{\dagger}\}$. The following discussion is
straightforwardly generalized to an arbitrary number of cavity modes.

We propose a model in which the cavity mode is coupled through the partially
transmitting mirror to the external modes matching it in polarization and
transverse shape. These external modes can be modeled as the modes of a second
cavity which shares the partially transmitting mirror with the real cavity,
but has the second mirror placed at infinity. According to the previous
chapter, the field corresponding to such cavity can be written as%
\begin{equation}
\mathbf{\hat{A}}\left(  \mathbf{r},t\right)=\lim_{L\rightarrow\infty
}\sum_{q}\frac{\boldsymbol{\varepsilon}}{2}\sqrt{\frac{\hbar}{\varepsilon
_{0}L\omega_{qf}}}\hat{b}_{q}(t)\left\{  T\left(  \omega_{q}/c;\mathbf{r}%
_{\perp},z\right)  \exp\left[  \mathrm{i}\omega_{q}z/c\right]+T\left(  -\omega_{q}/c;\mathbf{r}_{\perp},z\right)  \exp\left[
-\mathrm{i}\omega_{q}z/c\right]  \right\}  +\mathrm{H.c.},
\end{equation}
where the boson operators satisfy the commutation relations $[\hat{b}%
_{q}(t),\hat{b}_{q^{\prime}}(t)]=\delta_{qq^{\prime}}$, and $\omega_{q}=(\pi
c/L)q+\Delta\omega^{\perp}$, being $\Delta\omega^{\perp}$ a contribution
coming from the transverse structure of the modes, which is the same for all
the longitudinal modes. Now, as the length of this auxiliary cavity goes to
infinity, the set of longitudinal modes becomes infinitely dense in frequency
space, so that the sum over $q$ can be replaced by the following integral:%
\begin{equation}
\lim_{L\rightarrow\infty}\sum_{q}=\frac{L}{\pi c}\int_{\mathcal{O}}d\omega.
\end{equation}
Accordingly, the Kronecker delta converges to a Dirac delta as%
\begin{equation}
\lim_{L\rightarrow\infty}\delta_{qq^{\prime}}=\frac{\pi c}{L}\delta
(\omega-\omega^{\prime})\text{,}%
\end{equation}
so that defining new continuous boson operators by%
\begin{equation}
\hat{b}(\omega)=\sqrt{\frac{L}{\pi c}}\lim_{L\rightarrow\infty}\hat{b}_{q},
\end{equation}
which satisfy the commutation relations $[\hat{b}(\omega),\hat{b}^{\dagger
}(\omega^{\prime})]=\delta(\omega-\omega^{\prime})$, we can finally write the
vector potential of the field outside the cavity as%
\begin{equation}\label{Aext}
\mathbf{\hat{A}}\left(  \mathbf{r},t\right)=\int_{\mathcal{O}}%
d\omega\sqrt{\frac{\hbar}{4\pi c\varepsilon_{0}\omega}}\boldsymbol{\varepsilon
}\hat{b}(\omega;t)\left\{  T\left(  \omega/c;\mathbf{r}_{\perp},z\right)
\exp\left[  \mathrm{i}\omega z/c\right]+T\left(  -\omega/c;\mathbf{r}_{\perp},z\right)  \exp\left[
-\mathrm{i}\omega z/c\right]  \right\}  +\mathrm{H.c.}.
\end{equation}

From the previous sections, we know that the free evolution of the cavity mode
and the external modes is ruled by the Hamiltonian $\hat{H}_{0}=\hat
{H}_{\mathrm{cav}}+\hat{H}_{\mathrm{ext}}$ with%
\begin{equation}
\hat{H}_{\mathrm{cav}}=\hbar\omega_{\mathrm{c}}\hat{a}^{\dagger}\hat{a}%
+\hat{H}_{\mathrm{c}}\text{ \ \ \ \ \ and \ \ \ \ \ }\hat{H}_{\mathrm{ext}%
}=\int_{\mathcal{O}}d\omega\hbar\omega\hat{b}^{\dagger}(\omega)\hat{b}%
(\omega)\text{,}%
\end{equation}
where $\hat{H}_{\mathrm{c}}$ stands for any other intracavity process that we
decide to introduce.

The cavity and external modes are coupled through the mirror via a beam-splitter Hamiltonian
\begin{equation}
\hat{H}_{\mathrm{int}}=\mathrm{i}\hbar\int_{\mathcal{O}}d\omega g(\omega)[\hat{b}^{\dagger}(\omega)\hat{a}-\hat{a}^\dagger\hat{b}(\omega)],
\end{equation}
where the parameter $g(\omega)$ depends basically on the transmitivity of the
mirror at the corresponding frequency, and satisfies $g^{2}(\omega)\ll
\omega_{\mathrm{c}}$ for optical frequencies. On the other hand, it is to be expected that only frequencies around the cavity frequency will contribute to the
interaction (\textit{resonant interaction}), what allows us to rewrite
$g(\omega)=\sqrt{\gamma/\pi}$ in terms of a frequency--independent constant
$\gamma$, as the transmitivity of mirrors is a slowly varying function of
frequency at least within the interval $[\omega_{\mathrm{c}}-\gamma
,\omega_{\mathrm{c}}+\gamma]$, as well as extend the integration limits to
$[-\infty,+\infty]$, so that finally%
\begin{equation}
\hat{H}_{\mathrm{int}}=\mathrm{i}\hbar\sqrt{\frac{\gamma}{\pi}}\int_{-\infty
}^{+\infty}d\omega\lbrack\hat{b}^{\dagger}(\omega)\hat{a}-\hat{a}^{\dagger
}\hat{b}(\omega)].
\end{equation}

This is the basic model that we will use to `open the cavity'. In the
following we show how to deduce reduced evolution equations for the
intracavity mode only both at the level of operators (Heisenberg picture) and
states (Schr\"{o}dinger picture).%

\section{Heisenberg picture approach: The quantum Langevin
equation\label{HeisenbergOpen}}

The Heisenberg equations of motion of the annihilation operators are%
\begin{subequations}
\begin{align}
\frac{d\hat{a}}{dt}  &  =-\mathrm{i}\omega_{\mathrm{c}}\hat{a}+\left[  \hat
{a},\frac{\hat{H}_{\mathrm{c}}}{\mathrm{i}\hbar}\right]  -\sqrt{\frac{\gamma
}{\pi}}\int_{-\infty}^{+\infty}d\omega\hat{b}(\omega),\\
\frac{d\hat{b}(\omega)}{dt}  &  =-\mathrm{i}\omega\hat{b}(\omega)+\sqrt
{\frac{\gamma}{\pi}}\hat{a}\text{.}%
\end{align}
We can formally integrate the second equation as%
\end{subequations}
\begin{equation}
b(\omega;t)=b_{0}(\omega)e^{-\mathrm{i}\omega t}+\sqrt{\frac{\gamma}{\pi}}%
\int_{0}^{t}dt^{\prime}e^{\mathrm{i}\omega(t^{\prime}-t)}\hat{a}(t^{\prime}),
\label{bTOa}%
\end{equation}
where $b_{0}(\omega)$ are the external annihilation operators at the initial
time. Introducing this solution into the evolution equation for $\hat{a}$,
using $\int_{\tau_{0}}^{\tau_{1}}d\tau f(\tau)\delta(\tau-\tau_{1})=f(\tau
_{1})/2$, and defining the \textit{input operator}%
\begin{equation}
\hat{b}_{\mathrm{in}}(t)=-\frac{1}{\sqrt{2\pi}}\int_{-\infty}^{+\infty}d\omega
e^{-\mathrm{i}\omega t}\hat{b}_{0}(\omega), \label{bin}%
\end{equation}
which is easily shown to satisfy the commutation relations
\begin{subequations}
\begin{align}
\lbrack\hat{b}_{\mathrm{in}}(t),\hat{b}_{\mathrm{in}}^{\dagger}(t^{\prime})]
&  =\delta(t-t^{\prime}),\\
\lbrack\hat{b}_{\mathrm{in}}(t),\hat{b}_{\mathrm{in}}(t^{\prime})]  &
=[\hat{b}_{\mathrm{in}}^{\dagger}(t),\hat{b}_{\mathrm{in}}^{\dagger}%
(t^{\prime})]=0,
\end{align}
we get%
\end{subequations}
\begin{equation}
\frac{d\hat{a}}{dt}=-(\gamma+\mathrm{i}\omega_{\mathrm{c}})\hat{a}+\left[
\hat{a},\frac{\hat{H}_{\mathrm{c}}}{\mathrm{i}\hbar}\right]  +\sqrt{2\gamma
}\hat{b}_{\mathrm{in}}(t),
\end{equation}
as the evolution equation for the intracavity mode.

\begin{figure}
[t]
\begin{center}
\includegraphics[
height=1.4676in,
width=3.3572in
]%
{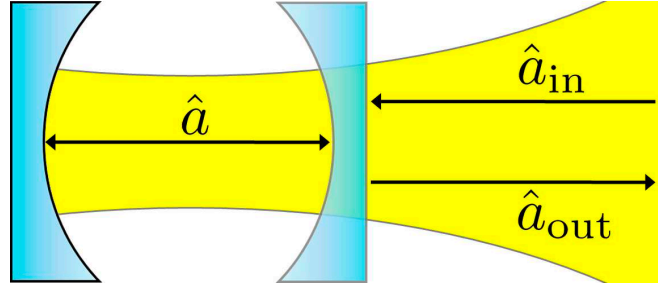}%
\caption{Scheme of the open cavity model. One of the mirrors is not perfectly
reflecting, what allows the external modes to drive the cavity via a
collective operator $\hat{a}_{\mathrm{in}}$. Similarly, the field coming out
from the cavity can be described by a single annihilation operator $\hat
{a}_{\mathrm{out}}$, as we show in the next chapter.}%
\label{fOpen1}%
\end{center}
\end{figure}

This equation is known as the \textit{quantum Langevin equation} for its
similarity with the stochastic Langevin equation, the operator $b_{\mathrm{in}%
}(t)$ playing the role of the stochastic noise. $b_{\mathrm{in}}(t)$ is
interpreted as the operator accounting for the input field driving the cavity
at each instant (see Figure \ref{fOpen1}). Indeed, note that when the external
modes are in the vacuum state initially, we have%
\begin{equation}
\langle\hat{b}_{\mathrm{in}}(t)\rangle=\langle\hat{b}_{\mathrm{in}}^{\dagger
}(t)\hat{b}_{\mathrm{in}}(t^{\prime})\rangle=0\text{ \ \ \ \ and
\ \ \ \ }\langle\hat{b}_{\mathrm{in}}(t)\hat{b}_{\mathrm{in}}^{\dagger
}(t^{\prime})\rangle=\delta(t-t^{\prime}), \label{VacCorr}%
\end{equation}
so much like complex, Gaussian noises in stochastic Langevin equations.

In this thesis we will also consider injection of a (ideal) monochromatic
laser at frequency $\omega_{\mathrm{L}}$, so that%
\begin{equation}
\langle\hat{b}_{0}(\omega)\rangle=\alpha_{\mathrm{L}}\delta(\omega
-\omega_{\mathrm{L}})\text{ \ \ \ \ and \ \ \ \ }\langle\hat{b}_{0}^{\dagger
}(\omega)\hat{b}_{0}(\omega^{\prime})\rangle=|\alpha_{\mathrm{L}}|^{2}%
\delta(\omega-\omega_{\mathrm{L}})\delta(\omega^{\prime}-\omega_{\mathrm{L}}),
\end{equation}
being $\alpha_{\mathrm{L}}$ the amplitude of the coherent injection. In this
case, it is recommendable to define a new input operator%
\begin{equation}
\hat{a}_{\mathrm{in}}(t)=\hat{b}_{\mathrm{in}}(t)-\langle\hat{b}_{\mathrm{in}%
}(t)\rangle,
\end{equation}
which can be shown to satisfy the vacuum correlations (\ref{VacCorr}), in
terms of which the quantum Langevin equation for the intracavity mode reads%
\begin{equation}
\frac{d\hat{a}}{dt}=\mathcal{E}e^{-i\omega_{\mathrm{L}}t}-(\gamma
+\mathrm{i}\omega_{\mathrm{c}})\hat{a}+\left[  \hat{a},\frac{\hat
{H}_{\mathrm{c}}}{\mathrm{i}\hbar}\right]  +\sqrt{2\gamma}\hat{a}%
_{\mathrm{in}}(t), \label{ReducedHeisenberg}%
\end{equation}
with $\mathcal{E}=\alpha_{\mathrm{L}}\sqrt{\gamma/\pi}$. Hence, the injection
of a coherent, monochromatic field at frequency $\omega_{\mathrm{L}}$ is
equivalent to the addition of the following new time--dependent term in the
Hamiltonian%
\begin{equation}
\hat{H}_{\mathrm{inj}}(t)=\mathrm{i}\hbar\left(  \mathcal{E}a^{\dagger
}e^{-\mathrm{i}\omega_{\mathrm{L}}t}-\mathcal{E}^{\ast}ae^{\mathrm{i}%
\omega_{\mathrm{L}}t}\right)  , \label{Hinj}%
\end{equation}
while considering the external modes in vacuum.

Note finally that the solution of this quantum Langevin equation for the case
of an empty cavity ($\hat{H}_{\mathrm{c}}=0$), reads%
\begin{equation}\label{aempty}
\hat{a}(t)=\hat{a}(t_{0})e^{-(\gamma+\mathrm{i}\omega_{\mathrm{c}}%
)t}+\frac{\mathcal{E}}{\gamma+\mathrm{i}(\omega_{\mathrm{c}}-\omega
_{\mathrm{L}})}\left[  e^{-\mathrm{i}\omega_{\mathrm{L}}t}-e^{-(\gamma
+\mathrm{i}\omega_{\mathrm{c}})t}\right]+\sqrt{2\gamma}\int_{0}^{t}dt^{\prime}\hat{a}_{\mathrm{in}}(t^{\prime
})e^{(\gamma+\mathrm{i}\omega_{\mathrm{c}})(t^{\prime}-t)}\text{,}
\end{equation}
which for $t\gg\gamma^{-1}$ is simplified to%
\begin{equation}
\hat{a}(t)=\frac{\mathcal{E}}{\gamma+\mathrm{i}(\omega_{\mathrm{c}}%
-\omega_{\mathrm{L}})}e^{-\mathrm{i}\omega_{\mathrm{L}}t}+\sqrt{2\gamma}%
\int_{0}^{t}dt^{\prime}\hat{a}_{\mathrm{in}}(t^{\prime})e^{(\gamma
+\mathrm{i}\omega_{\mathrm{c}})(t^{\prime}-t)}\text{.}\label{aSLempty}%
\end{equation}
This time limit is known as the \textit{stationary limit}, as it is possible
to show that (\ref{aSLempty}) predicts the expectation value of any operator
(or function of operators) to be invariant under translations of the time
origin, what defines a quantum state as \textit{stationary}
\cite{Mandel95book}. It is possible to show that this condition is satisfied
if and only if the density operator describing the state of the system
commutes with the Hamiltonian in the Schr\"{o}dinger picture.

\section{Schr\"{o}dinger picture approach: The master
equation\label{SchrodingerOpen}}

Consider now the density operator $\hat{\rho}$ corresponding to the state of
the whole system \textquotedblleft cavity mode + external
modes\textquotedblright, which evolves according to the von Neumann equation%
\begin{equation}
\mathrm{i}\hbar \frac{d\hat{\rho}}{dt}=[\hat{H}_{0}+\hat{H}_{\mathrm{int}}+\hat
{H}_{\mathrm{inj}}(t),\hat{\rho}].
\end{equation}
Note that following the discussion in the previous section, we have included
the coherent injection in the Hamiltonian, so that the initial state of the
external modes can be taken as vacuum, that is, $\hat{\rho}(t=0)=\hat{\rho
}_{\mathrm{c}}(t=0)\otimes\hat{\rho}_{\mathrm{vac}}$, with%
\begin{equation}
\langle\hat{b}(\omega)\rangle_{\mathrm{vac}}=\langle\hat{b}^{\dagger}%
(\omega)\hat{b}(\omega^{\prime})\rangle_{\mathrm{vac}}=\langle\hat{b}%
(\omega)\hat{b}(\omega^{\prime})\rangle_{\mathrm{vac}}=0, \label{VacExpExt}%
\end{equation}
where it is worth reminding that the expectation value of any external
operator $\hat{A}_{\mathrm{ext}}$ is evaluated as $\langle\hat{A}%
_{\mathrm{ext}}\rangle_{\mathrm{vac}}=\mathrm{tr}_{\mathrm{ext}}\{\hat{\rho
}_{\mathrm{vac}}\hat{A}_{\mathrm{ext}}\}$.

We are going to work in the interaction picture (see Section \ref{Axioms})
defined by the transformation operator $\hat{U}_{\mathrm{I}}=\hat{U}_{1}%
\hat{U}_{2}$, which is a sequence of two transformations defined as $\hat
{U}_{j}=\exp[\hat{H}_{j}t/\mathrm{i}\hbar]$ with%
\begin{equation}
\hat{H}_{1}=\hbar\omega_{\mathrm{L}}\left[  \hat{a}^{\dagger}\hat{a}%
+\int_{-\infty}^{+\infty}d\omega\hat{b}^{\dagger}(\omega)\hat{b}%
(\omega)\right]  ,
\end{equation}
and%
\begin{equation}
\hat{H}_{2}=\hat{U}_{1}^{\dagger}\left[  \hat{H}_{0}+\hat{H}%
_{\mathrm{inj}}(t)\right]  \hat{U}_{1}-\hat{H}_{1}=\hbar\Delta_{\mathrm{c}}\hat{a}^{\dagger}\hat{a}+\hat{H}_{\mathrm{c}}%
+\int_{-\infty}^{+\infty}d\omega\hbar\Delta(\omega)\hat{b}^{\dagger}%
(\omega)\hat{b}(\omega)+i\hbar\left(  \mathcal{E}a^{\dagger}-\mathcal{E}%
^{\ast}a\right)  \text{.}
\end{equation}
In this expression $\Delta_{\mathrm{c}}=\omega_{\mathrm{c}}-\omega
_{\mathrm{L}}$, $\Delta(\omega)=\omega-\omega_{\mathrm{L}}$, and we have
assumed that $[\hat{H}_{\mathrm{c}},\hat{U}_{1}]=0$ (physically, this means
that we consider only intracavity processes that conserve the number of
photons). The first transformation rotates the modes to the laser frequency,
and is performed in order to eliminate the time dependence of the Hamiltonian
(of $\hat{H}_{\mathrm{inj}}$ in particular). The second one is made in order
to remove all the pieces but $\hat{H}_{\mathrm{int}}$ from the Hamiltonian, as
a crucial incoming approximation requires the Hamiltonian to be `small'.

In this new picture, the state $\rho_{\mathrm{I}}=\hat{U}_{2}^{\dagger}\hat
{U}_{1}^{\dagger}\hat{\rho}\hat{U}_{1}\hat{U}_{2}$ evolves according to
$\mathrm{i}\hbar d\hat{\rho}_{\mathrm{I}}/dt=[\hat{H}_{\mathrm{I}},\hat{\rho
}_{\mathrm{I}}]$, with $\hat{H}_{\mathrm{I}}=\hat{U}_{2}^{\dagger}\hat{U}%
_{1}^{\dagger}\hat{H}_{\mathrm{int}}\hat{U}_{1}\hat{U}_{2}$. Let us define the
interaction picture intracavity annihilation operator%
\begin{equation}
\hat{a}_{\mathrm{I}}(t)=\hat{U}_{2}^{\dagger}\hat{U}_{1}^{\dagger}\hat{a}%
\hat{U}_{1}\hat{U}_{2},
\end{equation}
so that taking into account that%
\begin{equation}
\hat{U}_{2}^{\dagger}\hat{U}_{1}^{\dagger}\hat{b}(\omega)\hat{U}_{1}\hat
{U}_{2}=e^{-\mathrm{i}\omega t}\hat{b}(\omega),
\end{equation}
the interaction picture Hamiltonian can be rewritten as%
\begin{equation}
\hat{H}_{\mathrm{I}}(t)=\mathrm{i}\hbar\sqrt{\frac{\gamma}{\pi}}\int_{-\infty
}^{+\infty}d\omega\left[  \hat{a}_{\mathrm{I}}(t)\hat{b}^{\dagger}%
(\omega)e^{\mathrm{i}\omega t}-\hat{a}_{\mathrm{I}}^{\dagger}(t)\hat{b}%
(\omega)e^{-\mathrm{i}\omega t}\right]  .
\end{equation}

We seek now for the evolution equation of the reduced state $\hat{\rho
}_{\mathrm{CI}}=\mathrm{tr}_{\mathrm{ext}}\{\hat{\rho}_{\mathrm{I}}\}$
corresponding to the intracavity mode alone. To do so, we first integrate
formally the von Neumann equation for the whole system as%
\begin{equation}
\hat{\rho}_{\mathrm{I}}(t)=\hat{\rho}_{\mathrm{I}}(t_{0})+\frac{1}%
{\mathrm{i}\hbar}\int_{0}^{t}dt^{\prime}[\hat{H}_{\mathrm{I}}(t^{\prime}%
),\hat{\rho}_{\mathrm{I}}(t^{\prime})];
\end{equation}
reintroducing this expression into the von Neumann equation, making the
partial trace over the external modes taking into account that $\mathrm{tr}%
_{\mathrm{ext}}\{[\hat{H}_{\mathrm{I}}(t),\hat{\rho}_{\mathrm{I}}%
(0)]\}\sim\langle\hat{b}\rangle_{\mathrm{vac}}=0$, and making the variable
change $t^{\prime}=t-\tau$ in the time integral, we get the following
integro--differential equation for the reduced density operator $\hat{\rho
}_{\mathrm{CI}}$:%
\begin{equation}
\frac{d\hat{\rho}_{\mathrm{CI}}(t)}{dt}=-\frac{1}{\hbar^{2}}\int_{0}^{t}%
d\tau\mathrm{tr}_{\mathrm{ext}}\{[\hat{H}_{\mathrm{I}}(t),[\hat{H}%
_{\mathrm{I}}(t-\tau),\hat{\rho}_{\mathrm{I}}(t-\tau)]]\}.
\label{ExactIntegroDiff}%
\end{equation}
This equation is exact, but now we are going to introduce two important
approximations which lead to huge simplifications. The first is the
\textit{Born approximation}, which states that the external modes are
basically unaffected by the intracavity dynamics, that is, $\hat{\rho
}_{\mathrm{I}}(t)=\hat{\rho}_{\mathrm{CI}}(t)\otimes\hat{\rho}_{\mathrm{vac}}%
$. Then we perform the \textit{Markov approximation}, which states that memory
effects can be neglected, that is, we can take $\hat{\rho}_{\mathrm{I}}%
(t-\tau)=\hat{\rho}_{\mathrm{I}}(t)$ in the integral kernel. It can be shown
that this approximations are quite good in quantum optics thanks to the fact
that $\gamma/\omega_{\mathrm{c}}\ll1$ \cite{Breuer02book}.

Introducing these approximations in (\ref{ExactIntegroDiff}), and performing
the partial traces using (\ref{VacExpExt}), it is completely straightforward
to arrive to the following linear differential equation for the reduced
intracavity state%
\begin{equation}
\frac{d\hat{\rho}_{\mathrm{CI}}}{dt}=2\gamma\hat{a}_{\mathrm{I}}\hat{\rho
}_{\mathrm{CI}}\hat{a}_{\mathrm{I}}^{\dagger}-\gamma\hat{a}_{\mathrm{I}%
}^{\dagger}\hat{a}_{\mathrm{I}}\hat{\rho}_{\mathrm{CI}}-\gamma\hat{\rho
}_{\mathrm{CI}}\hat{a}_{\mathrm{I}}^{\dagger}\hat{a}_{\mathrm{I}},
\label{MasterInt}%
\end{equation}
where all the operators are evaluated at the same time. Finally, coming back
to the Schr\"{o}dinger picture, we finally arrive to%
\begin{equation}
\frac{d\hat{\rho}_{\mathrm{C}}}{dt}=\left[  \frac{\hat{H}_{\mathrm{cav}}%
+\hat{H}_{\mathrm{inj}}(t)}{\mathrm{i}\hbar},\hat{\rho}_{\mathrm{C}}\right]
+\gamma\left(  2\hat{a}\hat{\rho}_{\mathrm{C}}\hat{a}^{\dagger}-\hat
{a}^{\dagger}\hat{a}\hat{\rho}_{\mathrm{C}}-\hat{\rho}_{\mathrm{C}}\hat
{a}^{\dagger}\hat{a}\right)  ,
\end{equation}
where $\hat{\rho}_{\mathrm{C}}=\mathrm{tr}_{\mathrm{ext}}\{\hat{\rho}\}$ is
the reduced state of the intracavity mode in the Schr\"{o}dinger picture. This
equation is known as the \textit{master equation} of the intracavity mode.

In order to evaluate intracavity moments of observables,\ in this thesis we
will choose this Schr\"{o}dinger approach in general. The reason is that
having a master equation for the intracavity state only, we can apply the
positive \textit{P} techniques that we introduced at the end of Chapter
\ref{HarmonicOscillator} for the harmonic oscillator, what will allow us to
deal with stochastic equations instead of operator equations of the type
(\ref{ReducedHeisenberg}).

\section{Relation of the model parameters to physical parameters}

In the previous sections we have introduced the basic model that we will use
for the evolution of a light mode inside an open cavity. The model turned out
to have two basic parameters, $\mathcal{E}$ and $\gamma$, describing the rates
at which coherent light is injected into the cavity and intracavity light is
lost through the partially transmitting mirror, respectively. In this section
we connect these model parameters to relevant physical parameters like the
transmitivity of the mirror and the power of the injected laser beam.

This connection is easily done by following classical arguments. Consider the
classical electromagnetic field associated to the intracavity mode, whose
corresponding vector potential can be written as%
\begin{equation}\label{ClassicalIntracavityField}
\mathbf{A}\left(  \mathbf{r},t\right)=\frac{1}{2}\sqrt{\frac
{1}{2n(z)\varepsilon_{0}L_{\mathrm{opt}}}}\boldsymbol{\varepsilon}%
\nu(t)\{T(k_{\mathrm{c}};\mathbf{r}_{\perp},z)\exp\left[  \mathrm{i}%
n(z)k_{\mathrm{c}}z\right]+T^{\ast}(k_{\mathrm{c}};\mathbf{r}_{\perp},z)\exp\left[  -\mathrm{i}%
n(z)k_{\mathrm{c}}z\right]  \}+\mathrm{c.c.},
\end{equation}
in terms of the normal variable $\nu$ of the equivalent harmonic oscillator as
we showed in Section \ref{CavityQuanti}. In this expression we have defined
$k_{\mathrm{c}}=\omega_{\mathrm{c}}/c$. Now, remember that in Chapter
\ref{HarmonicOscillator} we learned that the mean values of quantum
observables provide the behavior of their classical counterparts as long as
quantum fluctuations can be neglected (see also the discussion at the
beginning of Section \ref{ClassiDOPO}). Applied to the cavity mode, this means
that its classical normal variable evolves as $\nu(t)=\sqrt{2\hbar
/\omega_{\mathrm{c}}}\langle\hat{a}(t)\rangle$. Then, for an empty cavity
($\hat{H}_{\mathrm{c}}=0$) and injecting \textit{on resonance} for simplicity,
that is, $\omega_{\mathrm{L}}=\omega_{\mathrm{c}}$, we have%
\begin{equation}
\nu(t)=\nu_{0}e^{-(\gamma+\mathrm{i}\omega_{\mathrm{c}})t}+\sqrt{\frac{2\hbar
}{\omega_{\mathrm{c}}}}\frac{\mathcal{E}}{\gamma}\left[  e^{-\mathrm{i}%
\omega_{\mathrm{c}}t}-e^{-(\gamma+\mathrm{i}\omega_{\mathrm{c}})t}\right]
,\label{NVempty}%
\end{equation}
with $\nu_{0}=\nu(t=0)$, as follows from (\ref{aempty}).

\textbf{Loss rate}. We can connect the damping rate $\gamma$ with the mirror
transmitivity $\mathcal{T}$ (reflectivity $\mathcal{R}=1-\mathcal{T}$) by
noting that in the absence of injection, $\mathcal{E}=0$, a fraction
$\mathcal{R}^{1/2}$ of the field amplitude is lost through the mirror at every
roundtrip; hence, after $m\in%
\mathbb{N}
$ roundtrips, we will have the relation%
\begin{equation}
\mathbf{A}\left(  \mathbf{r}_{\perp},z\oplus_{m}\mathrm{r.t.},t_{m}\right)
=\mathcal{R}^{m/2}\mathbf{A}\left(  \mathbf{r}_{\perp},z,t_{m}\right)  ,
\end{equation}
where `$\oplus_{m}\mathrm{r.t.}$' stands for `plus $m$--roundtrips' and
$t_{m}=2mL_{\mathrm{opt}}/c$. Introducing the solution (\ref{NVempty}) for
$\nu(t)$ in this expression, and taking into account the definition of the
cavity modes (\ref{CavityModesDef}), it is straightforward to get%
\begin{equation}
\gamma=-\frac{c}{4L_{\mathrm{opt}}}\ln\mathcal{R}\simeq\frac{c\mathcal{T}%
}{4L_{\mathrm{opt}}}\text{,} \label{Gamma}%
\end{equation}
where in the last equality we have assumed that $\mathcal{R}\approx1$. Hence,
we see that for small mirror transmitivities, there exists a linear relation
between the model parameter $\gamma$ and the actual mirror transmitivity
$\mathcal{T}$.

In order to understand the magnitude of this parameter let us consider a
cavity with lengths varying between 5 $\mathrm{mm}$ to 500 \textrm{mm} as in
the previous chapter; the damping rate $\gamma$ varies then between 150
\textrm{MHz} and 1.5 \textrm{MHz} for $\mathcal{R}=0.99$, and 1.5 \textrm{GHz}
and 0.015 \textrm{GHz} for $\mathcal{R}=0.9$, which are well below optical
frequencies. It is customary to call \textit{quality factor} to the ratio
between the free spectral range and the damping rate, $Q=\Omega_{\mathrm{FSR}%
}/\gamma$. This parameter is very important as it measures the ratio between
the separation of the longitudinal resonances of a given transverse family and
the width of these. In other words, when the quality factor is large one can
distinguish the different longitudinal resonances, while when it is small the
Lorentzians associated to different resonances overlap, and the resonances are
no longer distinguishable. In this thesis we will always work with cavities
having large enough quality factors as can be deduced from the numerical
examples that we have introduced.

\textbf{Injection parameter}. Consider now the part of the intracavity vector
potential propagating along the $-\mathbf{e}_{z}$ direction. According to
(\ref{ClassicalIntracavityField}) and (\ref{NVempty}), for a finite injection
parameter $\mathcal{E}$ and in the stationary limit $t\gg\gamma^{-1}$, we can
write it as%

\begin{equation}
\mathbf{A}_{\leftarrow}\left(  \mathbf{r},t\right)  =\frac{1}{2}\sqrt
{\frac{\hbar}{n(z)\varepsilon_{0}L_{\mathrm{opt}}\omega_{\mathrm{c}}}}%
\frac{|\mathcal{E}|}{\gamma}\boldsymbol{\varepsilon}T^{\ast}(k_{\mathrm{c}%
};\mathbf{r}_{\perp},z)e^{-\mathrm{i}n(z)k_{\mathrm{c}}%
z-\mathrm{i}\omega_{\mathrm{c}}t+\mathrm{i}\varphi}+\mathrm{c.c.},
\label{Astationary}%
\end{equation}
being $\varphi$ the phase of the injected laser beam. On the other hand, based
on (\ref{ClassicalVectorPotential}) and (\ref{Power}), the vector potential
associated to the laser inciding the cavity is written in terms of its power
$P_{\mathrm{inj}}$ as%
\begin{equation}
\mathbf{A}_{\mathrm{inj}}\left(  \mathbf{r},t\right)  =\sqrt{\frac
{P_{\mathrm{inj}}}{2\varepsilon_{0}c\omega_{\mathrm{c}}^{2}}}%
\boldsymbol{\varepsilon}T^{\ast}\left(  k_{\mathrm{c}};\mathbf{r}_{\perp},z\right)  \exp\left(  -\mathrm{i}k_{\mathrm{c}}z-\mathrm{i}\omega_{\mathrm{c}%
}t+\mathrm{i}\varphi\right)  +\mathrm{c.c.},
\end{equation}
We can obtain an expression of the field inside the cavity by looking at it as
an interferometer: The part of the injection which is transmitted inside the
cavity at a given time interferes with the fields that have been bouncing back
and forth for several roundtrips. In this sense, the complete field will be a
superposition of the fields%
\begin{subequations}
\begin{align}
\mathbf{A}^{(0)}\left(  \mathbf{r},t\right)   &  =\mathcal{T}^{1/2}%
\mathbf{A}_{\mathrm{inj}}\left(  \mathbf{r},t\right)  \text{,}\\
\mathbf{A}^{(1)}\left(  \mathbf{r},t\right)   &  =\mathcal{T}^{1/2}%
\mathcal{R}^{1/2}\mathbf{A}_{\mathrm{inj}}\left(  \mathbf{r}_{\perp}%
,z\oplus\mathrm{r.t.},t\right)  \text{,}\\
\mathbf{A}^{(2)}\left(  \mathbf{r},t\right)   &  =\mathcal{T}^{1/2}%
\mathcal{R}\mathbf{A}_{\mathrm{inj}}\left(  \mathbf{r}_{\perp},z\oplus
_{2}\mathrm{r.t.},t\right)  \text{,}\\
&  \vdots\nonumber\\
\mathbf{A}^{(m)}\left(  \mathbf{r},t\right)   &  =\mathcal{T}^{1/2}%
\mathcal{R}^{m/2}\mathbf{A}_{\mathrm{inj}}\left(  \mathbf{r}_{\perp}%
,z\oplus_{m}\mathrm{r.t.},t\right)  \text{,}%
\end{align}
and hence, assuming that the time is large enough as to take the
$m\rightarrow\infty$ limit we get%
\end{subequations}
\begin{equation}
\mathbf{A}_{\leftarrow}\left(  \mathbf{r},t\right)  =\sqrt{\frac
{P_{\mathrm{inj}}}{2n(z)\varepsilon_{0}c}}\mathcal{T}^{1/2}\left(  \sum
_{m=0}^{\infty}\mathcal{R}^{m/2}\right)  \boldsymbol{\varepsilon}T^{\ast
}\left(  k_{\mathrm{c}};\mathbf{r}_{\perp},z\right)  e^{-\mathrm{i}k_{\mathrm{c}}z-\mathrm{i}\omega_{\mathrm{c}}t+\mathrm{i}%
\varphi} +\mathrm{c.c.},
\end{equation}
where we have used again the definition of the cavity modes
(\ref{CavityModesDef}). Comparing (\ref{Astationary}) with this expression,
and performing the geometric sum we finally get%
\begin{equation}
|\mathcal{E}|=\gamma\sqrt{\frac{2LP_{\mathrm{inj}}}{\hbar\omega_{\mathrm{c}}%
c}}\frac{\mathcal{T}^{1/2}}{1-\mathcal{R}^{1/2}}\simeq\sqrt{\frac{2\gamma
}{\hbar\omega_{\mathrm{c}}}P_{\mathrm{inj}}}\text{,} \label{PtoE}%
\end{equation}
where in the last equality we have assumed that $\mathcal{R}\approx1$ and have
made use of (\ref{Gamma}).

\section{Quantum--optical phenomena with optical lattices}

In connection with the theory of open quantum systems that we have
particularized to optical cavities in this chapter, we now introduce the work
that the author of this thesis started at the Max--Planck--Institute for
Quantum Optics during a three--months visit in 2008, and that ended up
published at the beginning of 2011 \cite{Navarrete11b}.

Ultracold atoms trapped in optical lattices are well known as quite versatile
simulators of a large class of many--body condensed-matter
Hamiltonians\footnote{One of the first proposals in this direction was the
article by Jacksch et al. \cite{Jaksch98}, where it was proved that the
dynamics of cold atoms trapped in optical lattices is described by the
Bose--Hubbard Hamiltonian provided that certain conditions are satisfied;
shortly after, the superfluid--to--Mott insulator phase transition
characteristic of this Hamiltonian was observed in the laboratory
\cite{Greiner02}. Since then, the broad tunability of the lattice parameters,
and the increasing ability to trap different kind of particles (like bosonic
and fermionic atoms with arbitrary spin or polar molecules), has allowed
theoreticians to propose optical lattices as promising simulators for
different types of generalized Bose--Hubbard and spin models which are in
close relation to important condensed--matter phenomena
\cite{Jaksch04rev,Lewenstein07rev,Bloch08rev}. Recent experiments have shown
that optical lattices can be used to address open problems in physics like,
e.g., high--$T_{\mathrm{c}}$ superconductivity \cite{Schneider08}, to study
phenomena in low dimensions such as the Berezinskii--Kosterlitz--Thouless
transition \cite{Hadzibabic06}, or to implement quantum computation schemes
\cite{Anderlini07}.} (see \cite{Jaksch04rev,Lewenstein07rev,Bloch08rev} for
extensive reviews on this subject). In this work, and exploiting currently
available technology only, we have been able to show that they can also be
used to simulate phenomena traditionally linked to quantum--optical systems,
that is, to light--matter interactions \cite{deVega08,Navarrete11b}. In this
brief exposition of the work, we explain in particular how to observe
phenomena arising from the collective spontaneous emission of atomic and
harmonic oscillator ensembles such as sub/superradiance, including some
phenomena which still lack experimental observation.

The concept of superradiance was introduced by Dicke in 1954 when studying the
spontaneous emission of a collection of two--level atoms \cite{Dicke54} (see
\cite{Gross82rev} for a review). He showed that certain collective states
where the excitations are distributed symmetrically over the whole sample have
enhanced emission rates. Probably the most stunning example is the
single--excitation symmetric state (now known as the symmetric Dicke state),
which instead of decaying with the single--atom decay rate $\Gamma_{0}$, was
shown to decay with $N\Gamma_{0}$, $N$ being the number of atoms. He also
suggested that the emission rate of the state having all the atoms excited
should be enhanced at the initial steps of the decay process, which was a most
interesting prediction from the experimental point of view, as this state is
in general easier to prepare. However, Dicke used a very simplified model in
which all the atoms interact with a common radiation field within the dipolar
approximation; almost 15 years later, and motivated by the new atom--inversion
techniques, several authors showed that dipolar interactions impose a
threshold value for the atom density, that is, for the number of interacting
atoms, in order for superradiance to appear \cite{Ernst68,Agarwal70,Rehler71}.

It is worth noting that together with atomic ensembles, spontaneous emission
of collections of harmonic oscillators were also studied at that time
\cite{Agarwal70}. Two interesting features of this system were reported: (i)
an initial state with all the harmonic oscillators excited does not show rate
enhancement, on the contrary, most excitations remain within the sample in the
steady state, while (ii) the state having all the oscillators in the same
coherent state has a superradiant rate. This predictions have not been
observed in the laboratory yet to our knowledge.

\subsection{The basic idea}

\begin{figure}
[t]
\begin{center}
\includegraphics[
width=0.4\textwidth
]%
{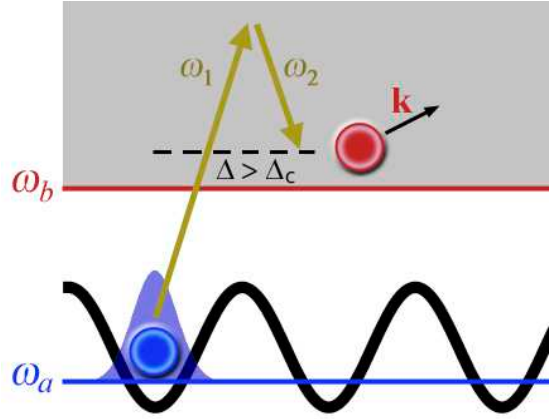}%
\caption{Scheme of our proposed setup. Atoms in state $a$ are trapped in an
optical lattice, while atoms in state $b$ are free, and can thus have any
momentum. An external pair of Raman lasers connect the two levels with some
detuning $\Delta$. We will show that above some critical value $\Delta
=\Delta_{\mathrm{c}}$ the atoms in state $b$ are able to leave the trap
(left), and behave as the photons emitted in superradiant quantum--optical
systems. The case $\Delta<\Delta_{\mathrm{c}}$ and more quantum--optical
phenomenology can be found in \cite{Navarrete11b}.}%
\label{fOpen2}%
\end{center}
\end{figure}

The basic setup that we introduce, and that was already presented in
\cite{deVega08}, is depicted in figure \ref{fOpen2}. Consider a collection of
bosonic atoms with two relevant internal states labeled by $a$ and $b$ (which
may correspond to hyperfine ground--state levels, see for example
\cite{Jaksch04rev}). Atoms in state $a$ are trapped by a deep optical lattice
in which the localized wavefunction of traps at different lattice sites do not
overlap (preventing hopping of atoms between sites), while atoms in state $b$
are not affected by the lattice, and hence behave as free particles. A pair of
lasers forming a Raman scheme drive the atoms from the trapped state to the
free one \cite{Jaksch04rev}, providing an effective interaction between the
two types of particles. We consider the situation of having non-interacting
bosons in the lattice \cite{Chin10rev}, as well as hard--core bosons in the
collisional blockade regime, where only one or zero atoms can be in a given
lattice site \cite{Paredes04}. In the first regime, the lattice consists of a
collection of harmonic oscillators placed at the nodes of the lattice; in the
second regime, two--level systems replace the harmonic oscillators, the two
levels corresponding to the absence or presence of an atom in the lattice
site. Therefore, it is apparent that this system is equivalent to a collection
of independent emitters (harmonic oscillators or atoms) connected only through
a common radiation field, the role of this radiation field being played by the
free atoms. This system is therefore the cold--atom analog of the systems
which are usually considered in the quantum description of light--matter
interaction, with the difference that the radiated particles are massive, and
hence have a different dispersion relation than that of photons in vacuum.%

Let us introduce the model for such system. Our starting point is the
Hamiltonian of the system in second quantization \cite{Greiner96book}. We will
denote by $\left\vert a\right\rangle $ and $\left\vert b\right\rangle $ the
trapped and free atomic states, respectively (having internal energies
$\hbar\omega_{a}^{0}$ and $\hbar\omega_{b}^{0}$). Two--body interactions for
the trapped atoms are included with the usual contact-like pseudopotential
\cite{Bloch08rev, Chin10rev}, but we neglect the collisions for the free
atoms. The Hamiltonian is then written as $\hat{H}=\hat{H}_{0}+\hat{H}_{a-b}$,
with%
\begin{subequations}
\begin{align}
\hat{H}_{0}  &  =\sum_{j=a,b}\int d^{3}\mathbf{r}\,\hat{\Psi}_{j}^{\dagger
}\left(  \mathbf{r}\right)  \left(  H_{j}+\hbar\omega_{j}^{0}\right)
\hat{\Psi}_{j}\left(  \mathbf{r}\right)  +\frac{g}{2}\int d^{3}\mathbf{r}%
\,\hat{\Psi}_{a}^{\dagger2}\left(  \mathbf{r}\right)  \hat{\Psi}_{a}%
^{2}\left(  \mathbf{r}\right)  ,\label{H0}\\
\hat{H}_{a-b}  &  =\hbar\Omega\int d^{3}\mathbf{r}\,e^{\mathrm{i}%
\left(  \mathbf{k}_{L}\cdot\mathbf{r}-\omega_{L}t\right)  }\hat{\Psi}%
_{a}\left(  \mathbf{r}\right)  \hat{\Psi}_{b}^{\dagger}\left(  \mathbf{r}%
\right)  +\mathrm{H.c.}; \label{Hab}%
\end{align}
$\hat{H}_{0}$ contains the individual dynamics of the atoms, $H_{j}$ being the
first--quantized motion Hamiltonian of the atom in the corresponding state,
and $g=4\pi\hbar^{2}a_{s}/m$, where $a_{s}$ is the s--wave scattering length
of the trapped atoms (which have mass $m$). $\hat{H}_{a-b}$ contains the Raman
coupling between the atomic states, $\mathbf{k}_{L}=\mathbf{k}_{1}%
-\mathbf{k}_{2}$ and $\omega_{L}=\omega_{1}-\omega_{2}$ (laser wave vector and
frequency in the following) being the relative wave vector and frequency of
the two lasers involved in the Raman scheme (see figure \ref{fOpen2}), with
$\Omega$ the corresponding two--photon Rabi frequency.

For atoms in state $\left\vert a\right\rangle $, $H_{a}=-\left(  \hbar
^{2}/2m\right)  \mathbf{\nabla}^{2}+V_{\mathrm{opt}}\left(  \mathbf{r}\right)
$, where $V_{\mathrm{opt}}\left(  \mathbf{r}\right)  $ corresponds to a
3--dimensional optical lattice with cubic geometry and lattice period $d_{0}$.
We work with ultracold atoms under conditions such that their wavefunctions
can be described by the set of first--band Wannier functions localized around
the nodes of the lattice \cite{Jaksch04rev}. The traps of the optical lattice
are approximated by isotropic harmonic potentials \cite{Jaksch04rev}, what
allows us to write the Wannier functions as
\end{subequations}
\begin{equation}
w_{0}\left(  \mathbf{r}-\mathbf{r}_{\mathbf{j}}\right)  =\frac{1}{\pi
^{3/4}X_{0}^{3/2}}\exp\left[  -\left(  \mathbf{r}-\mathbf{r}_{\mathbf{j}%
}\right)  ^{2}/2X_{0}^{2}\right]  ,
\end{equation}
where $\mathbf{r}_{\mathbf{j}}=d_{0}\mathbf{j}$ is the position of the
$\mathbf{j}\in\mathbb{Z}^{3}$ lattice site (we consider $M$ sites in each
orthogonal direction defining the cubic lattice), and $X_{0}^{2}=\hbar
/m\omega_{0}$, being $\omega_{0}$ the frequency of the harmonic trap. The
energy associated to these wave functions is $E_{0}=3\hbar\omega_{0}/2$. We
will assume that Wannier functions localized at different lattice sites do not
overlap, hence preventing tunneling between sites.

On the other hand, atoms in state $\left\vert b\right\rangle $ can move freely
in every direction of space according to $H_{b}=-\left(  \hbar^{2}/2m\right)
\mathbf{\nabla}^{2}$, and hence the plane waves $\psi_{\mathbf{k}}\left(
\mathbf{r}\right)  =e^{\mathrm{i}\mathbf{k}\cdot\mathbf{r}}/\sqrt{V}%
$, with energy $E_{\mathbf{k}}=\hbar^{2}k^{2}/2m$, are their motion
eigenfunctions ($V$ is the total available volume for the free atoms, which we
might take as infinite for calculations).

We consider two opposite regimes for the interaction between trapped atoms.
The first limit consists in neglecting the interactions, which might be
accomplished by, e.g., tuning the scattering length with an additional
magnetic field through a Feshbach resonance \cite{Chin10rev}. Let us expand
the quantum fields as%
\begin{equation}
\hat{\Psi}_{a}\left(  \mathbf{r}\right)  =\sum_{\mathbf{j}}w_{0}\left(
\mathbf{r}-\mathbf{r}_{\mathbf{j}}\right)  \hat{a}_{\mathbf{j}}\text{
\ \ \ \ and \ \ \ \ }\hat{\Psi}_{b}\left(  \mathbf{r}\right)  =\sum
_{\mathbf{k}}\psi_{\mathbf{k}}\left(  \mathbf{r}\right)  \hat{b}_{\mathbf{k}},
\end{equation}
where the operators $\{\hat{a}_{\mathbf{j}},\hat{a}_{\mathbf{j}}^{\dagger}\}$
and $\{\hat{b}_{\mathbf{k}},\hat{b}_{\mathbf{k}}^{\dagger}\}$ satisfy
canonical bosonic commutation relations, and create or annihilate an atom at
lattice site $\mathbf{j}$ and a free atom with momentum $\mathbf{k}$,
respectively. Working in the interaction picture defined by the transformation
operator $\hat{U}_{\mathrm{I}}=\exp[\hat{H}_{0}t/\mathrm{i}\hbar]$, we get the
Hamiltonian
\begin{equation}
\hat{H}_{\mathrm{I}}=\sum_{\mathbf{j},\mathbf{k}}g_{k}\exp[\mathrm{i}%
\Delta_{k}t-\mathrm{i}\left(  \mathbf{k}-\mathbf{k}_{L}\right)  \cdot
\mathbf{r}_{\mathbf{j}}]\hat{a}_{\mathbf{j}}\hat{b}_{\mathbf{k}}^{\dagger
}+\mathrm{H.c.}, \label{Hoscis}%
\end{equation}
with
\begin{equation}
g_{k}=\hbar\Omega\sqrt{\frac{8\pi^{3/2}X_{0}^{3}}{V}}\exp\left[  -\frac{1}%
{2}X_{0}^{2}\left(  \mathbf{k}-\mathbf{k}_{L}\right)  ^{2}\right]
,\qquad\Delta_{k}=\frac{\hbar k^{2}}{2m}-\Delta, \label{Hpar}%
\end{equation}
$\Delta=\omega_{L}-\left(  \omega_{b}-\omega_{a}\right)  $ being the detuning
of the laser frequency respect to the $\left\vert a\right\rangle
\rightleftharpoons\left\vert b\right\rangle $ transition ($\omega_{a}%
=\omega_{a}^{0}+3\omega_{0}/2$ and $\omega_{b}=\omega_{b}^{0}$).

As for the second limit, we assume that the on--site repulsive atom--atom
interaction is the dominant energy scale, and hence the trapped atoms behave
as hard--core bosons in the collisional blockade regime, what prevents the
presence of two atoms in the same lattice site \cite{Paredes04}; this means
that the spectrum of $\hat{a}_{\mathbf{j}}^{\dagger}\hat{a}_{\mathbf{j}}$ can
be restricted to the first two states $\left\{  \left\vert 0\right\rangle
_{\mathbf{j}},\left\vert 1\right\rangle _{\mathbf{j}}\right\}  $, having 0 or
1 atoms at site $\mathbf{j}$, and then the boson operators $\left\{  \hat
{a}_{\mathbf{j}}^{\dagger},\hat{a}_{\mathbf{j}}\right\}  $ can be changed by
spin-like ladder operators $\left\{  \hat{\sigma}_{\mathbf{j}}^{\dagger}%
,\hat{\sigma}_{\mathbf{j}}\right\}  =\left\{  |1\rangle_{\mathbf{j}}%
\langle0|,|0\rangle_{\mathbf{j}}\langle1|\right\}  $. In this second limit the
Hamiltonian reads
\begin{equation}
\hat{H}_{\mathrm{I}}=\sum_{\mathbf{j},\mathbf{k}}g_{k}e%
^{\mathrm{i}\Delta_{k}t-\mathrm{i}\left(  \mathbf{k}-\mathbf{k}_{L}\right)
\cdot\mathbf{r}_{\mathbf{j}}}\hat{\sigma}_{\mathbf{j}}\hat{b}_{\mathbf{k}%
}^{\dagger}+\mathrm{H.c.}. \label{Hspins}%
\end{equation}

Hamiltonians (\ref{Hoscis}) and (\ref{Hspins}) show explicitly how this system
mimics the dynamics of collections of harmonic oscillators or atoms,
respectively, interacting with a common radiation field. Note finally that in
order to satisfy that the trapped atoms are within the first Bloch band, it is
required that $\omega_{0}\gg\Delta,\Omega$.%

\begin{figure}
[t]
\begin{center}
\includegraphics[
width=\textwidth
]%
{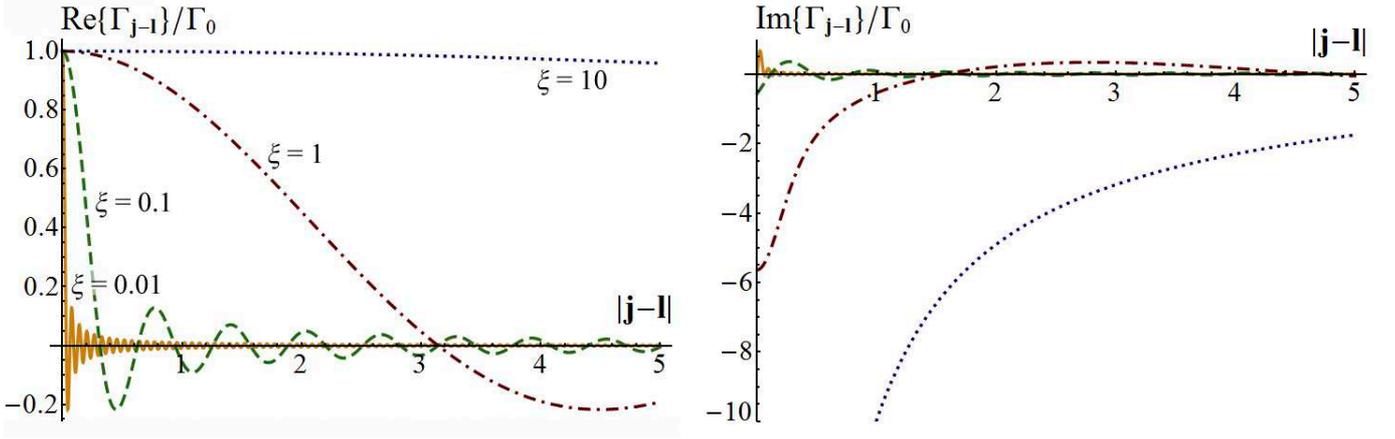}%
\caption{We plot the Markov couplings (real and imaginary parts at left and
right, respectively) as a function of the distance between sites for
$\tilde{\Delta}>0$. We have chosen $d_{0}/X_{0}=10$ and $\mathbf{k}%
_{L}=\mathbf{0}$, and plotted 4 different values of $\xi$. It can be
appreciated that this parameter controls the spatial range of the
interactions.}%
\label{fOpen3}%
\end{center}
\end{figure}

We are going to analyze the system by considering the free atoms as an
environment for the trapped atoms, similarly to the optical cavity, in which
the external modes act as an environment for the intracavity modes. Following
the same approach as that of Section \ref{SchrodingerOpen}, that is, assuming
that the environment is in a non-evolving vacuum (Born--Markov approximation)
and tracing out its associated Hilbert space, it is simple \cite{Navarrete11b}
to get the following master equation for the reduced density operator of the
trapped atoms
\begin{equation}
\frac{d\hat{\rho}}{dt}=\sum_{\mathbf{j},\mathbf{l}}\Gamma_{\mathbf{j}%
-\mathbf{l}}\hat{a}_{\mathbf{l}}\hat{\rho}\hat{a}_{\mathbf{j}}^{\dagger
}-\Gamma_{\mathbf{j}-\mathbf{l}}\hat{a}_{\mathbf{j}}^{\dagger}\hat
{a}_{\mathbf{l}}\hat{\rho}+\mathrm{H.c.}; \label{Master}%
\end{equation}
a similar equation is obtained for the hard--core bosons but replacing the
boson operators by the corresponding spin operators. Note the similarity
between this equation, and the equation of the damped mode inside the cavity
(\ref{MasterInt}), their main difference being that now the different modes of
the system\ interact via the Markov couplings%
\begin{equation}
\Gamma_{\mathbf{j}-\mathbf{l}}=\mathrm{i}\exp\left(  -\mathrm{i}\mathbf{k}%
_{L}\cdot\mathbf{r}_{\mathbf{j}-\mathbf{l}}\right)  \frac{\Gamma_{0}\xi
}{\left\vert \mathbf{j}-\mathbf{l}\right\vert }\left[  1-\mathrm{erf}\left(
\frac{d_{0}}{2X_{0}}\left\vert \mathbf{j}-\mathbf{l}\right\vert \right)
-\exp\left(  -\nu\frac{\left\vert \mathbf{j}-\mathbf{l}\right\vert }{\xi
}\right)  \right]  ,
\end{equation}
where $\xi=1/d_{0}k_{0}$ with $k_{0}=X_{0}^{-1}\sqrt{2|\tilde{\Delta}%
|/\omega_{0}}$ measures the ratio between the lattice spacing and the
characteristic wavelength of the radiated atoms, $\tilde{\Delta}%
=\Delta-4\Omega^{2}/\omega_{0}$, and $\nu=1$ $(-\mathrm{i})$ for
$\tilde{\Delta}<0$ $(\tilde{\Delta}>0)$. The error function is defined as
$\mathrm{erf}\left(  x\right)  =\left(  2/\sqrt{\pi}\right)  \int_{0}%
^{x}\mathrm{d}u\exp\left(  -u^{2}\right)  $. Note that the term
`$1-\mathrm{erf}\left(  d_{0}\left\vert \mathbf{j}-\mathbf{l}\right\vert
/2X_{0}\right)  $' is basically zero for $\mathbf{j}\neq\mathbf{l}$, and
therefore the $\xi$ parameter dictates the spatial range of the interactions
as can be appreciated in figure \ref{fOpen3}. Finally, $\Gamma_{0}=4\Omega
^{2}\sqrt{2\pi|\tilde{\Delta}|/\omega_{0}^{3}}$ is the single--emitter decay
rate, that is, the rate at which the atoms are emitted by the lattice when the
sites emit independently (see \cite{Navarrete11b} for details).

\subsection{Superradiant phenomenology \label{Superradiance}}

In order to prove that quantum--optical phenomena can be found in this system
we now show that the superradiant phenomenology described at the beginning of
the section appears on it. For $\mathbf{k}_{L}=0$ and $\tilde{\Delta}>0$, the
Markov couplings are complex in general, and therefore the master equation of
the system takes the form
\begin{equation}
\frac{d\hat{\rho}}{dt}=\frac{1}{\mathrm{i}\hbar}\left[  \hat{H}_{\mathrm{d}%
},\hat{\rho}\right]  +\mathcal{D}\left[  \hat{\rho}\right]  ,
\label{Master Eq}%
\end{equation}
with a dissipation term given by
\begin{equation}
\mathcal{D}\left[  \hat{\rho}\right]  =\sum_{\mathbf{j},\mathbf{l}}%
\gamma_{\mathbf{j}-\mathbf{l}}\left(  2\hat{a}_{\mathbf{l}}\hat{\rho}\hat
{a}_{\mathbf{j}}^{\dagger}-\hat{a}_{\mathbf{j}}^{\dagger}\hat{a}_{\mathbf{l}%
}\hat{\rho}-\hat{\rho}\hat{a}_{\mathbf{l}}^{\dagger}\hat{a}_{\mathbf{j}%
}\right)  , \label{Diss}%
\end{equation}
having collective decay rates
\begin{equation}
\gamma_{\mathbf{j}-\mathbf{l}}=\mathrm{Re}\left\{  \Gamma_{\mathbf{j}%
-\mathbf{l}}\right\}  =\Gamma_{0}\mathrm{sinc}\left(  \frac{\left\vert
\mathbf{j}-\mathbf{l}\right\vert }{\xi}\right)  ,
\end{equation}
and a reversible term corresponding to inhomogeneous dephasing with
Hamiltonian
\begin{equation}
\hat{H}_{\mathrm{d}}=\sum_{\mathbf{j},\mathbf{l}}\hbar\Lambda_{\mathbf{j}%
-\mathbf{l}}\hat{a}_{\mathbf{j}}^{\dagger}\hat{a}_{\mathbf{l}}, \label{Hd}%
\end{equation}
being
\begin{equation}
\Lambda_{\mathbf{j}-\mathbf{l}}=\mathrm{Im}\left\{  \Gamma_{\mathbf{j}%
-\mathbf{l}}\right\}  =\frac{\Gamma_{0}\xi}{\left\vert \mathbf{j}%
-\mathbf{l}\right\vert }\left[  1-\mathrm{erf}\left(  \frac{d_{0}}{2X_{0}%
}\left\vert \mathbf{j}-\mathbf{l}\right\vert \right)  -\cos\left(
\frac{\left\vert \mathbf{j}-\mathbf{l}\right\vert }{\xi}\right)  \right]  .
\label{Lambda}%
\end{equation}
The same holds for hard--core bosons but replacing the boson operators by the
corresponding spin operators.

For $\xi\ll1$ the Markov couplings do not connect different lattice sites,
that is $\Gamma_{\mathbf{j}-\mathbf{l}}\simeq\Gamma_{\mathbf{0}}%
\delta_{\mathbf{j},\mathbf{l}}$, and the sites emit independently. On the
other hand, when $\xi\gg M$ the collective decay rates become homogeneous,
$\gamma_{\mathbf{j}-\mathbf{l}}\simeq\Gamma_{0}$, and we enter the Dicke
regime. Hence, we expect to observe the superradiant phase--transition in our
system by varying the parameter $\xi$.

Note that in the Dicke regime the dephasing term cannot be neglected and
connects the sites inhomogeneously with $\Lambda_{\mathbf{j}\neq\mathbf{l}%
}\simeq\Gamma_{0}\xi/\left\vert \mathbf{j}-\mathbf{l}\right\vert $. This term
appears in the optical case too, although it was inappropriately neglected in
the original work by Dicke \cite{Dicke54} when assuming the dipolar
approximation in his initial Hamiltonian, and slightly changes his original
predictions as pointed out in \cite{Friedberg72,Friedberg74} (see also
\cite{Gross82rev}).

In the following we analyze the superradiant behavior of our system by
studying the evolution of the total number of particles in the lattice
$n_{\mathrm{T}}=\sum_{\mathbf{j}}\left\langle \hat{a}_{\mathbf{j}}^{\dagger
}\hat{a}_{\mathbf{j}}\right\rangle $ and the rate of emitted atoms
\begin{equation}
\mathcal{R}\left(  t\right)  =\sum_{\mathbf{k}}\frac{d}{dt}\left\langle
\hat{b}_{\mathbf{k}}^{\dagger}\hat{b}_{\mathbf{k}}\right\rangle =-\dot
{n}_{\mathrm{T}};
\end{equation}
in the last equality we have used that the total number operator
$\sum_{\mathbf{k}}\hat{b}_{\mathbf{k}}^{\dagger}\hat{b}_{\mathbf{k}}%
+\sum_{\mathbf{j}}\hat{a}_{\mathbf{j}}^{\dagger}\hat{a}_{\mathbf{j}}$ is a
constant of motion.

\subsubsection{Hard--core bosons: Atomic superradiance. \label{AtomicSuper}}

Let us start by analyzing the case of a lattice in an initial Mott phase
having one atom per site in the collisional blockade regime, which is the
analog of an ensemble of excited atoms \cite{deVega08}. As explained in the
Introduction, Dicke predicted that superradiance should appear in this system
as an enhancement of the emission rate at early times \cite{Dicke54}, although
this was later proved to happen only if the effective number of interacting
emitters exceeds some threshold value \cite{Ernst68,Agarwal70,Rehler71}: This
is the superradiant phase transition.

In our system, the number of interacting spins is governed by the parameter
$\xi$ (see figure \ref{fOpen3}), and the simplest way to show that the
superradiant phase transition appears by varying it, is by evaluating the
initial slope of the rate which can be written as\footnote{Note that the
evolution equation of the expectation value of any operator $\hat{O}$ can be
written as
\begin{equation}
\frac{d}{dt}\left\langle \hat{O}\left(  t\right)  \right\rangle =\mathrm{tr}%
\left\{  \frac{d\hat{\rho}}{dt}\hat{O}\right\}  =-\sum_{\mathbf{m},\mathbf{l}%
}\left\{  \Gamma_{\mathbf{m}-\mathbf{l}}\left\langle \left[  \hat{O},\hat
{a}_{\mathbf{m}}^{\dagger}\right]  \hat{a}_{\mathbf{l}}\right\rangle
+\Gamma_{\mathbf{m}-\mathbf{l}}^{\ast}\left\langle \hat{a}_{\mathbf{l}%
}^{\dagger}\left[  \hat{a}_{\mathbf{m}},\hat{O}\right]  \right\rangle
\right\}  , \label{ExpEvo}%
\end{equation}
and similarly the hard--core bosons in terms of the spin ladder operators.}
\begin{equation}
\left.  \frac{d}{dt}\mathcal{R}\right\vert _{t=0}=-4M^{3}\Gamma_{0}^{2}\left[
1-\sum_{\mathbf{m}\neq\mathbf{j}}\frac{\mathrm{sinc}^{2}\left(  \left\vert
\mathbf{j}-\mathbf{m}\right\vert /\xi\right)  }{M^{3}}\right]  .
\label{HardCoreRateDerIni}%
\end{equation}

\begin{figure}
[t]
\begin{center}
\includegraphics[
width=\textwidth
]%
{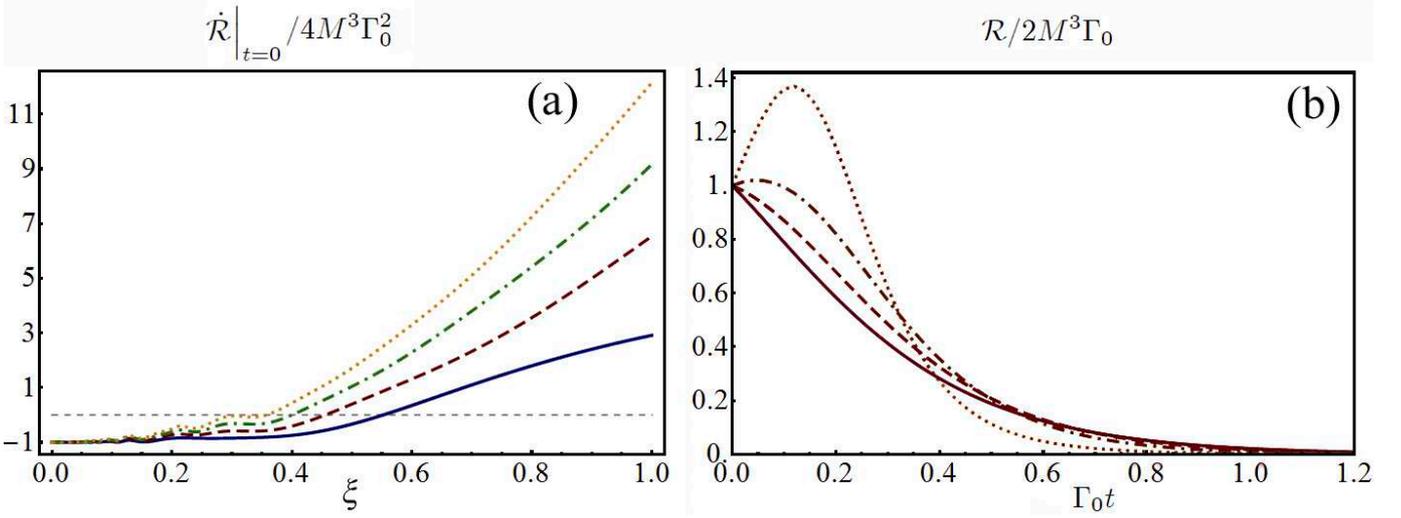}%
\caption{Collective emission properties for an initial Mott state of
hard--core bosons. (a) Time derivative of the rate at $t=0$ as a function of
the range of the interactions $\xi$, see (\ref{HardCoreRateDerIni}). The
values $M^{3}=8$ (solid, blue), $27$ (dashed, red), $64$ (dashed--dotted,
green), and $125$ (dotted, yellow) are considered. It can be appreciated that
there exists a critical value of $\xi$ above which the rate is enhanced at the
initial times. (b) Rate as a function of time for $M^{3}=27$. The values
$\xi=0.01$ (solid), $0.5$ (dashed), $1$ (dashed--dotted), and $10$ (dotted)
are considered. As expected from (a), the maximum of the rate is delayed above
some critical $\xi$ value. Note that both the rate and its derivative have
been normalized to the values expected for independent emitters, which are
$4M^{3}\Gamma_{0}^{2}$ and $2M^{3}\Gamma_{0}$, respectively.}%
\label{fOpen4}%
\end{center}
\end{figure}

This expression has a very suggestive form: The term corresponding to the rate
associated to independent emitters is balanced by a collective contribution
arising from the interactions between them. In figure \ref{fOpen4}a we show
the dependence of this derivative with $\xi$ for various values of the number
of sites $M^{3}$. It can be appreciated that there exists a critical value of
$\xi$ above which the sign of the derivative is reversed; hence, the rate
increases at the initial time and we expect its maximum to be no longer at
$t=0$, which is a signature of superradiance.%

The time evolution of the rate for a cubic lattice with $M^{3}=27$ sites is
shown in figure \ref{fOpen4}b for different values of $\xi$. We can appreciate
how above some critical $\xi$ value the maximum rate of emission is delayed as
expected. In order to find $\mathcal{R}(t)$ we have simulated the evolution
equations for the coherences $c_{\mathbf{jl}}=\langle\hat{\sigma}_{\mathbf{j}%
}^{\dagger}\hat{\sigma}_{\mathbf{l}}\rangle$ and the populations
$s_{\mathbf{j}}=\left\langle \hat{\sigma}_{\mathbf{j}}^{3}\right\rangle $,
which we close by using the semiclassical approximation $\left\langle
\hat{\sigma}_{\mathbf{m}}^{\dagger}\hat{\sigma}_{\mathbf{j}}^{3}\hat{\sigma
}_{\mathbf{l}}\right\rangle =\left\langle \hat{\sigma}_{\mathbf{m}}^{\dagger
}\hat{\sigma}_{\mathbf{l}}\right\rangle \left\langle \hat{\sigma}_{\mathbf{j}%
}^{3}\right\rangle -2\delta_{\mathbf{jl}}\left\langle \hat{\sigma}%
_{\mathbf{m}}^{\dagger}\hat{\sigma}_{\mathbf{l}}\right\rangle $; they read
(\ref{ExpEvo})%
\begin{subequations}
\begin{align}
\dot{c}_{\mathbf{jl}}  &  =-4\Gamma_{0}c_{\mathbf{jl}}+\sum_{\mathbf{m}}%
\Gamma_{\mathbf{l}-\mathbf{m}}c_{\mathbf{jm}}s_{\mathbf{l}}+\Gamma
_{\mathbf{j}-\mathbf{m}}^{\ast}c_{\mathbf{ml}}s_{\mathbf{j}},\\
\dot{s}_{\mathbf{j}}  &  =-2\sum_{\mathbf{l}}\Gamma_{\mathbf{j}-\mathbf{l}%
}c_{\mathbf{jl}}+\Gamma_{\mathbf{j}-\mathbf{l}}^{\ast}c_{\mathbf{lj}}.
\label{HardCoreSemiclassEqs}%
\end{align}
\end{subequations}
We have checked the validity of these semiclassical equations by comparing
them with a direct simulation of the master equation for small number of sites
in 1D and 2D geometries; except for small quantitative deviations, they offer
the same results.

\subsubsection{Non-interacting bosons: Harmonic oscillators superradiance.
\label{HarmonicSuper}}

Let us analyze now the case of having non-interacting bosons in the lattice,
which is equivalent to a collection of harmonic oscillators as discussed
before. In previous works on superradiance this system was studied in parallel
to its atomic counterpart \cite{Agarwal70}, and here we show how our system
offers a physical realization of it. We will show that superradiant effects
can be observed in the evolution of the total number of atoms in the lattice,
both for initial Mott and superfluid phases\footnote{Let us note that a
superfluid state with $N$ excitations distributed over the entire lattice is
more easily defined in the discrete Fourier--transform basis
\begin{equation}
\hat{f}_{\mathbf{q}}=\frac{1}{M^{3/2}}\sum_{\mathbf{j}}\exp\left(  \frac
{2\pi\mathrm{i}}{M}\mathbf{q}\cdot\mathbf{j}\right)  \hat{a}_{\mathbf{j}},
\label{DFT}%
\end{equation}
with $\mathbf{q}=\left(  q_{x},q_{y},q_{z}\right)  $ and $q_{x,y,z}%
=0,1,2,...M-1$, as the state having $N$ excitations in the zero--momentum
mode, that is, $\left\vert \mathrm{SF}\right\rangle _{N}=\left(  N!\right)
^{-1/2}\hat{f}_{\mathbf{0}}^{\dagger N}\left\vert 0\right\rangle $.}.

The evolution of the total number of atoms in the lattice is given by --see
(\ref{ExpEvo})--
\begin{equation}
\dot{n}_{\mathrm{T}}=-2\sum_{\mathbf{j},\mathbf{l}}\gamma_{\mathbf{j}%
-\mathbf{l}}\mathrm{Re}\left\{  \left\langle \hat{a}_{\mathbf{j}}^{\dagger
}\hat{a}_{\mathbf{l}}\right\rangle \right\}  ,
\end{equation}
and hence depends only on the real part of the Markov couplings. Therefore, we
restrict our analysis to the dissipative term $\mathcal{D}\left[  \hat{\rho
}\right]  $ of the master equation (\ref{Master Eq}).

By diagonalizing the real, symmetric collective decay rates with an orthogonal
matrix $S$ such that $\sum_{\mathbf{jl}}S_{\mathbf{pj}}\gamma_{\mathbf{j}%
-\mathbf{l}}S_{\mathbf{ql}}=\bar{\gamma}_{\mathbf{p}}\delta_{\mathbf{pq}}$,
one can find a set of modes $\left\{  \hat{c}_{\mathbf{p}}=\sum_{\mathbf{j}%
}S_{\mathbf{pj}}\hat{a}_{\mathbf{j}}\right\}  $ with definite decay
properties. Then, it is completely straightforward to show that the total
number of atoms can be written as a function of time as
\begin{equation}
n_{\mathrm{T}}\left(  t\right)  =\sum_{\mathbf{p}}\left\langle \hat
{c}_{\mathbf{p}}^{\dagger}\hat{c}_{\mathbf{p}}\left(  0\right)  \right\rangle
\exp\left(  -2\bar{\gamma}_{\mathbf{p}}t\right)  . \label{ntt}%
\end{equation}

In general, $\gamma_{\mathbf{j}-\mathbf{l}}$ requires numerical
diagonalization. However, in the limiting cases $\xi\ll1$ and $\xi\gg M$, its
spectrum becomes quite simple. Following the discussion after (\ref{Lambda}),
in the $\xi\ll1$ limit $\gamma_{\mathbf{j}-\mathbf{l}}=\Gamma_{0}%
\delta_{\mathbf{jl}}$ is already diagonal and proportional to the identity.
Hence, any orthogonal matrix $S$ defines an equally suited set of modes all
decaying with rate $\Gamma_{0}$. Therefore, if the initial number of atoms in
the lattice is $N$, this will evolve as
\begin{equation}
n_{\mathrm{T}}\left(  t\right)  =N\exp\left(  -2\Gamma_{0}t\right)  ,
\label{TotalNumberIndependent}%
\end{equation}
irrespective of the particular initial state of the lattice (e.g., Mott or
superfluid). The emission rate $\mathcal{R}=2\Gamma_{0}N\exp\left(
-2\Gamma_{0}t\right)  $, corresponds to the independent decay of the $N$ atoms
as expected in this regime having no interaction between the emitters.%

\begin{figure}
[ptb]
\begin{center}
\includegraphics[
width=\textwidth
]%
{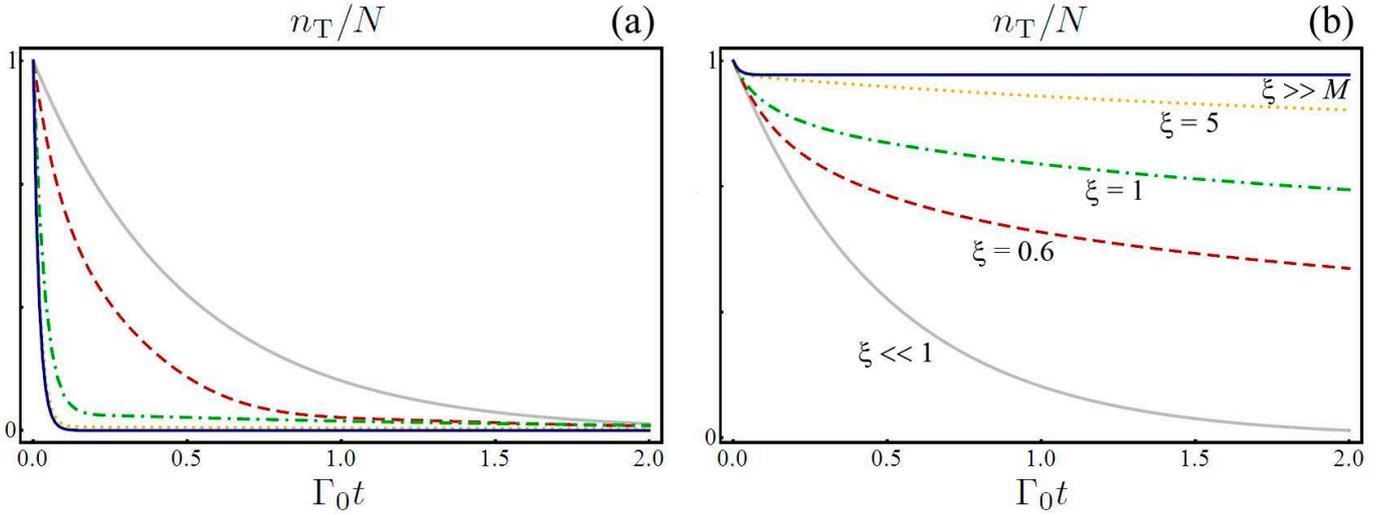}%
\caption{Evolution of the total number of atoms in a lattice having $M^{3}=27$
sites for initial superfluid (a) and Mott (b) phases with $N$ initial
non-interacting atoms. The solid curves correspond to the limits $\xi\ll1$
(grey) and $\xi\gg M$ (dark--blue). Note how the `evaporation time' is reduced
for the initial superfluid state as $\xi$ increases (a). Equivalently, note
how for an initial Mott state the atoms tend to stay in the lattice as $\xi$
increases (b).}%
\label{fOpen5}%
\end{center}
\end{figure}

Let us consider now the opposite limit $\xi\gg M$; in this case $\gamma
_{\mathbf{j}-\mathbf{l}}=\Gamma_{0}$ $\forall$ $\left(  \mathbf{j}%
,\mathbf{l}\right)  $, and the dissipative term can be written in terms of the
symmetrical discrete Fourier--transform mode only as $\mathcal{D}\left[
\hat{\rho}\right]  =M^{3}\Gamma_{0}\left(  2\hat{f}_{\mathbf{0}}\hat{\rho}%
\hat{f}_{\mathbf{0}}^{\dagger}-\hat{f}_{\mathbf{0}}^{\dagger}\hat
{f}_{\mathbf{0}}\hat{\rho}-\hat{f}_{\mathbf{0}}^{\dagger}\hat{f}_{\mathbf{0}%
}\hat{\rho}\right)  $, see (\ref{DFT}). Hence, the discrete Fourier--transform
basis diagonalizes the problem, and shows that all the modes have zero decay
rate except the symmetrical one, which has an enhanced rate proportional to
the number of emitters. Therefore, starting with a superfluid state, the $N$
initial atoms will decay exponentially with initial rate $NM^{3}\Gamma_{0}$,
that is
\begin{equation}
n_{\mathrm{T}}\left(  t\right)  =N\exp\left(  -2M^{3}\Gamma_{0}t\right)  .
\label{TotalNumberSuperfluid}%
\end{equation}
On the other hand, if the initial state corresponds to a Mott phase, most of
the atoms will remain in the lattice, as only the component which projects
onto the symmetric mode will be emitted; concretely, from (\ref{ntt}) and
(\ref{DFT}) the number of atoms in the lattice will evolve for this particular
initial state as
\begin{equation}
n_{\mathrm{T}}\left(  t\right)  =\left(  N-\frac{N}{M^{3}}\right)  +\frac
{N}{M^{3}}\exp\left(  -2M^{3}\Gamma_{0}t\right)  . \label{TotalNumberMott}%
\end{equation}

Hence, according to this picture, superradiant collective effects can be
observed in our system by two different means. Calling $t_{0}$ the time needed
to radiate the atoms in the absence of collective effects, one could start
with a superfluid phase and measure this `evaporation time' as a function of
$\xi$; this should go from $t_{0}$ for $\xi\ll1$, to a much shorter time
$t_{0}/M^{3}$ for $\xi\gg M$ (see figure \ref{fOpen5}a). Alternatively, one
could start with a Mott phase, and measure the number of atoms left in the
lattice in the steady state as a function of $\xi$; in this case, it should go
from $n_{\mathrm{T,steady}}=0$ after a time $t_{0}$ for $\xi\ll1$, to
$n_{\mathrm{T,steady}}=N-N/M^{3}$ after a time $t_{0}/M^{3}$ for $\xi\gg M$
(see figure \ref{fOpen5}b).

In order to find the evolution of $n_{\mathrm{T}}$ we have simulated the
equations satisfied by the coherences $c_{\mathbf{jl}}=\langle\hat
{a}_{\mathbf{j}}^{\dagger}\hat{a}_{\mathbf{l}}\rangle$, which read --see
(\ref{ExpEvo})--
\begin{equation}
\dot{c}_{\mathbf{jl}}=-\sum_{\mathbf{m}}\left[  \Gamma_{\mathbf{l}-\mathbf{m}%
}c_{\mathbf{jm}}+\Gamma_{\mathbf{j}-\mathbf{m}}^{\ast}c_{\mathbf{ml}}\right]
. \label{cSystem}%
\end{equation}
Note that in this case the equations are closed without the need of a
semiclassical approximation, and hence they are exact. Note that they
reproduce the analytic evolution of $n_{\mathrm{T}}$ as given by
(\ref{TotalNumberIndependent}), (\ref{TotalNumberMott}), and
(\ref{TotalNumberSuperfluid}) in the corresponding limits (see figure
\ref{fOpen5}).

Our results connect directly to those found by Agarwal some decades ago
\cite{Agarwal70}. Working with a Dicke-like model, he showed that if all the
oscillators start in the same coherent state $\left\vert \alpha\right\rangle
$, the initial number of excitations, which in that case is given by
$N=M^{3}\left\vert \alpha\right\vert ^{2}$, decays following
(\ref{TotalNumberSuperfluid}). This is not a coincidence, but rather a
consequence that, if $N$ is large enough, a multi--coherent state of that kind
is a good approximation of a superfluid state with that number of excitations.
He also predicted that if the oscillators start in a number state, most of the
excitations would remain in the steady state as follows from
(\ref{TotalNumberMott}). 

%% file: DetectionFO.tex
In the previous chapter we developed a model to study the dynamics of the
intracavity modes of an open cavity from a quantum viewpoint. However, it is
the field coming out of the cavity the one which is customarily studied or
used in applications, not the intracavity modes. The first part of this
chapter is devoted to relate this output field with the intracavity field. We
then explain the basic detection schemes which are used to characterize this
output field, namely direct photodetection and balanced homodyne detection,
which ideally give us access to the intensity (photon number) and the
quadratures of light, respectively. Understanding how the field quadratures
are analyzed in real experiments will allow us to reintroduce squeezing in an
experimentally useful manner at the end of the chapter.

\section{The output field}

Using the expression of the vector potential outside the resonator
(\ref{Aext}), we see that the field coming out from the cavity can be written
as%
\begin{equation}
\mathbf{\hat{A}}_{\mathrm{out}}^{(+)}\left(  \mathbf{r},t\right)
=\int_{-\infty}^{+\infty}d\omega\sqrt{\frac{\hbar}{4\pi c\varepsilon_{0}%
\omega}}\boldsymbol{\varepsilon}T\left(  \omega/c;\mathbf{r}_{\perp},z\right)
\hat{b}(\omega;t)\exp\left(  \mathrm{i}\omega z/c\right)  ,
\end{equation}
where the operator $\hat{b}(\omega;t)$ is given by (\ref{bTOa}) in terms of
the initial operators $\hat{b}_{0}(\omega)$ and the intracavity mode $\hat{a}%
$. Now, given that only the frequencies around the cavity resonance contribute
to the dynamics as we argued in the previous chapter, we can replace the
slowly varying function of the frequency $T\left(  \omega/c;\mathbf{r}_{\perp
},z\right)  /\sqrt{\omega}$ by its value at $\omega_{\mathrm{c}}$, arriving to%
\begin{equation}
\mathbf{\hat{A}}_{\mathrm{out}}^{(+)}\left(  \mathbf{r},t\right)  =\sqrt
{\frac{\hbar}{4\pi c\varepsilon_{0}\omega_{\mathrm{c}}}}%
\boldsymbol{\varepsilon}T\left(  k_{\mathrm{c}};\mathbf{r}_{\perp},z\right)
\int_{-\infty}^{+\infty}d\omega\hat{b}(\omega;t)\exp\left(  \mathrm{i}\omega
z/c\right)  \text{.}%
\end{equation}
Introducing the solution (\ref{bTOa}) for $\hat{b}(\omega;t)$ in this
equation, we can write the output field as the sum of two terms, namely,
$\mathbf{\hat{A}}_{\mathrm{out}}^{(+)}\left(  \mathbf{r},t\right)
=\mathbf{\hat{A}}_{\mathrm{in}}^{(+)}\left(  \mathbf{r},t\right)
+\mathbf{\hat{A}}_{\mathrm{source}}^{(+)}\left(  \mathbf{r},t\right)  $ with%
\begin{subequations}
\begin{align}
\mathbf{\hat{A}}_{\mathrm{in}}^{(+)}\left(  \mathbf{r},t\right)   &
=\sqrt{\frac{\hbar}{4\pi c\varepsilon_{0}\omega_{\mathrm{c}}}}%
\boldsymbol{\varepsilon}T\left(  k_{\mathrm{c}};\mathbf{r}_{\perp},z\right)
\int_{-\infty}^{+\infty}d\omega\hat{b}_{0}(\omega)\exp\left[  -\mathrm{i}%
\omega(t_{R}-t_{0})\right]  ,\\
\mathbf{\hat{A}}_{\mathrm{source}}^{(+)}\left(  \mathbf{r},t\right)   &
=\sqrt{\frac{\hbar\gamma}{4\pi^{2}c\varepsilon_{0}\omega_{\mathrm{c}}}%
}\boldsymbol{\varepsilon}T\left(  k_{\mathrm{c}};\mathbf{r}_{\perp},z\right)
\int_{-\infty}^{+\infty}d\omega\int_{0}^{t}dt^{\prime}\hat{a}(t^{\prime}%
)\exp\left[  \mathrm{i}\omega(t^{\prime}-t_{R})\right]  ,
\end{align}
where we have defined the \textit{retarded time}\textbf{ }$t_{R}=t-z/c$.
Recalling the definition of the input operator (\ref{bin}), the first term can
be rewritten as%
\end{subequations}
\begin{equation}
\mathbf{\hat{A}}_{\mathrm{in}}^{(+)}\left(  \mathbf{r},t\right)  =-\sqrt
{\frac{\hbar}{2c\varepsilon_{0}\omega_{\mathrm{c}}}}\boldsymbol{\varepsilon
}T\left(  k_{\mathrm{c}};\mathbf{r}_{\perp},z\right)  \hat{a}_{\mathrm{in}%
}(t_{R}),
\end{equation}
and hence it is identified with a contribution coming from the input field
driving the cavity. On the other hand, performing the frequency integral in
the source term, we can rewrite it as%
\begin{equation}
\mathbf{\hat{A}}_{\mathrm{source}}^{(+)}\left(  \mathbf{r},t\right)  =\left\{
\begin{array}
[c]{cc}%
\sqrt{\frac{\hbar\gamma}{c\varepsilon_{0}\omega_{\mathrm{c}}}}%
\boldsymbol{\varepsilon}T\left(  k_{\mathrm{c}};\mathbf{r}_{\perp},z\right)
\hat{a}(t_{R}) & z<ct\\
\sqrt{\frac{\hbar\gamma}{4c\varepsilon_{0}\omega_{\mathrm{c}}}}%
\boldsymbol{\varepsilon}T\left(  k_{\mathrm{c}};\mathbf{r}_{\perp},z\right)
\hat{a}(t_{R}) & z=ct\\
0 & z>ct
\end{array}
\right.  ,
\end{equation}
which is therefore identified with a contribution of the intracavity field,
hence the name \textquotedblleft source\textquotedblright. We will always work
in the stationary limit $t\gg\gamma^{-1}$, and near the cavity, so that $z$
will always be smaller than $ct$, what allows us to even neglect the $z/c$
delay in the retarded time, that is, $t_{R}\simeq t$ (in any case, this
doesn't have any fundamental relevance, it will just simplify future derivations).

Hence, defining the \textit{output operator}%
\begin{equation}
\hat{a}_{\mathrm{out}}(t)=\sqrt{2\gamma}\hat{a}(t)-\hat{a}_{\mathrm{in}%
}(t)\text{,} \label{Out-CavIn}%
\end{equation}
the output field is finally written as%
\begin{equation}
\mathbf{\hat{A}}_{\mathrm{out}}^{(+)}\left(  \mathbf{r},t\right)  =\sqrt
{\frac{\hbar}{2c\varepsilon_{0}\omega_{\mathrm{c}}}}\boldsymbol{\varepsilon
}T\left(  k_{\mathrm{c}};\mathbf{r}_{\perp},z\right)  \hat{a}_{\mathrm{out}%
}(t)\text{.}%
\end{equation}
Hence the field coming out from the cavity is a superposition of the
intracavity field leaking through the partially transmitting mirror and the
part of the input field which is reflected, just as expected by the classical
boundary conditions at the mirror (see Figure \ref{fOpen1}).

It is interesting for future purposes to generalize this expression to the
case of several modes resonating at the same frequency inside the cavity. The
corresponding output field can be written in this case as%
\begin{equation}
\mathbf{\hat{A}}_{\mathrm{out}}^{(+)}\left(  \mathbf{r},t\right)  =\sqrt
{\frac{\hbar}{2c\varepsilon_{0}\omega_{\mathrm{c}}}}\sum_{\sigma\mathbf{n}%
}\boldsymbol{\varepsilon}_{\sigma\mathbf{n}}T_{\mathbf{n}}\left(
k_{\mathrm{c}};\mathbf{r}_{\perp},z\right)  \hat{a}_{\sigma\mathbf{n}%
,\mathrm{out}}(t)\text{,} \label{AoutMulti}%
\end{equation}
where each cavity mode with annihilation operator $\hat{a}_{\sigma\mathbf{n}}$
is driven by an independent input operator $\hat{a}_{\sigma\mathbf{n}%
,\mathrm{in}}$ so that
\begin{equation}
\hat{a}_{\sigma\mathbf{n},\mathrm{out}}(t)=\sqrt{2\gamma_{\sigma\mathbf{n}}%
}\hat{a}_{\sigma\mathbf{n}}(t)-\hat{a}_{\sigma\mathbf{n},\mathrm{in}}(t),
\end{equation}
being $\gamma_{\sigma\mathbf{n}}$ the cavity loss rate for each mode.

In the case of an empty cavity, it is straightforward to prove that this
operators satisfy the commutation relations%
\begin{subequations}
\begin{align}
\lbrack\hat{a}_{\sigma\mathbf{n},\mathrm{out}}(t),\hat{a}_{\sigma^{\prime
}\mathbf{n}^{\prime},\mathrm{out}}^{\dagger}(t^{\prime})]  &  =\delta
_{\sigma\sigma^{\prime}}\delta_{\mathbf{nn}^{\prime}}\delta(t-t^{\prime}),\\
\lbrack\hat{a}_{\sigma\mathbf{n},\mathrm{out}}(t),\hat{a}_{\sigma^{\prime
}\mathbf{n}^{\prime},\mathrm{out}}(t^{\prime})]  &  =[\hat{a}_{\sigma
\mathbf{n},\mathrm{out}}^{\dagger}(t),\hat{a}_{\sigma^{\prime}\mathbf{n}%
^{\prime},\mathrm{out}}^{\dagger}(t^{\prime})]=0.
\end{align}
When other intracavity processes are included one has in general to check how
this commutation relations are modified.

\begin{figure}
[ptb]
\begin{center}
\includegraphics[
height=1.2548in,
width=2.9888in
]%
{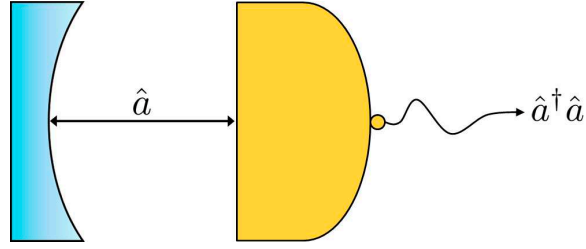}%
\caption{Schematic representation of Mollow's ideal single--mode detection.}%
\label{fDetec1}%
\end{center}
\end{figure}

\section[Ideal detection: An intuitive picture of photo- and homodyne detection]{Ideal detection: An intuitive picture of photo- and homodyne detection}

Photodetection is the most fundamental measurement technique for light. As we
shall see with a particular example (homodyne detection), any other scheme
used for measuring different properties of light makes use of photodetection
as a part of it.

This technique is based on the photoelectric effect or variations of it. The
idea is that when the light beam that we want to detect impinges a metallic
surface, it is able to release some of the bound electrons of the metal, which
are then collected by an anode. The same happens if light incides on a
semiconductor surface, though in this case instead of becoming free, valence
electrons are promoted to the conduction band. The most widely used metallic
photodetectors are known as \textit{photo--multiplier tubes}, while those
based on semiconducting films are the so-called \textit{avalanche photo
diodes}. In both cases, each photon is able to create one single electron,
whose associated current would be equally difficult to measure by electronic
means; for this reason, each photoelectron is accelerated towards a series of
metallic plates at increasing positive voltages, releasing then more electrons
which contribute to generate a measurable electric pulse, the
\textit{photopulse}.

It is customarily said that counting photopulses is equivalent to counting
photons, and hence, photodetection is equivalent to a measurement of the
number of photons of the light field. This is a highly idealized situation,
valid only in some limits which are actually not the ones in which we usually
work in the field of squeezing. Nevertheless, it is interesting to analyze
this ideal limit first to gain some intuition about the photodetection
process, and this is what we will do in this section. Using this idealized
picture of photodetectors, we will study a detection scheme called homodyne
detection, which will be shown to be equivalent to a measurement of the
quadratures of light.

Consider the following model for a perfectly efficient detection scheme. A
single--mode field with boson operators $\{\hat{a},\hat{a}^{\dagger}\}$
initially in some state $\hat{\rho}$ is kept in continuous interaction with a
photodetector during a time interval $T$. The intuitive picture of such a
scenario is shown in Figure \ref{fDetec1}: A cavity formed by the
photodetector itself and an extra perfectly reflecting mirror contains a
single mode. By developing a microscopic model of the detector and its
interaction with the light mode, Mollow was able to show that the probability
of generating $n$ photoelectrons (equivalently, the probability of observing
$n$ photopulses) during the time interval $T$ is given by \cite{Mollow68}%
\end{subequations}
\begin{equation}
p_{n}=\left\langle :\frac{\left(  1-e^{-\kappa T}\right)  ^{n}\hat{a}^{\dagger
n}\hat{a}^{n}}{n!}\exp\left[  -(1-e^{-\kappa T})\hat{a}^{\dagger}\hat
{a}\right]  :\right\rangle ,
\end{equation}
where the expectation value has to be evaluated in the initial state
$\hat{\rho}$ of the light mode, and $\kappa$ is some parameter accounting for
the light--detector interaction. Using the operator identity :$\exp\left[
-(1-e^{-\lambda})\hat{a}^{\dagger}\hat{a}\right]  $: $=\exp(-\lambda\hat
{a}^{\dagger}\hat{a})$ \cite{Louisell73book}, and the help of the number state
basis $\{|n\rangle\}_{n\in%
\mathbb{N}
}$, it is straightforward to get%
\begin{equation}
p_{n}=\sum_{m=n}^{\infty}\langle n|\hat{\rho}|n\rangle\frac{m!}{n!(m-n)!}%
\left(  1-e^{-\kappa T}\right)  ^{n}\left(  e^{-\kappa T}\right)
^{m-n}\underset{T\gg\kappa^{-1}}{\longrightarrow}\langle n|\hat{\rho}%
|n\rangle,
\end{equation}
and hence, for large enough detection times the number of observed pulses
follows the statistics of the number of photons. In other words, this ideal
photodetection scheme is equivalent to measuring the number operator $\hat
{a}^{\dagger}\hat{a}$ as already commented.%

\begin{figure}
[t]
\begin{center}
\includegraphics[
height=3.4817in,
width=3.9781in
]%
{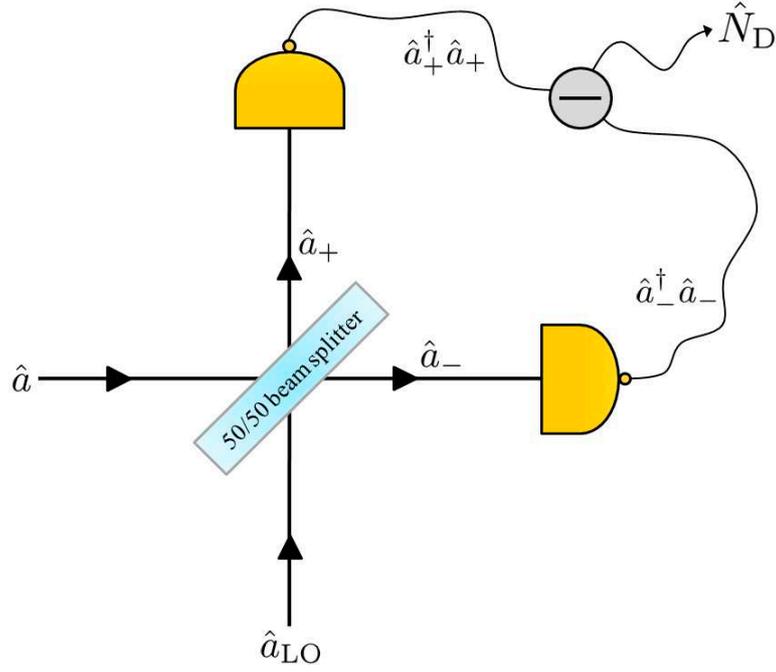}%
\caption{Homodyne detection scheme with ideal photodetectors. When the local
oscillator is in a strong coherent state, this setup gives access to the
quadratures of light.}%
\label{fDetec2}%
\end{center}
\end{figure}

Even though the output of the photodetectors can take only integer values
(number of recorded photopulses), they can be arranged to approximately
measure the quadratures of light, which we remind are continuous observables.
This arrangement is called homodyne detection. The basic scheme is shown in
Figure \ref{fDetec2}. The mode we want to measure is mixed in a beam splitter
with another mode, called the \textit{local oscillator}, which is in a
coherent state $|\alpha_{\mathrm{LO}}\rangle$. When the beam splitter is 50/50
the homodyne scheme is said to be \textit{balanced}, and the annihilation
operators of the modes leaving its output ports are given by%
\begin{equation}
\hat{a}_{\pm}=\frac{1}{\sqrt{2}}\left(  \hat{a}\pm\hat{a}_{\mathrm{LO}%
}\right)  \text{,}%
\end{equation}
being $\hat{a}_{\mathrm{LO}}$ the annihilation operator of the local
oscillator mode. These modes are measured with independent photodetectors, and
then the corresponding signals are subtracted. Based on the idealized
photodetection picture, this scheme is analogous to a measurement of the
photon number difference%
\begin{equation}
\hat{N}_{\mathrm{D}}=\hat{a}_{+}^{\dagger}\hat{a}_{+}-\hat{a}_{-}^{\dagger
}\hat{a}_{-}=\hat{a}_{\mathrm{LO}}^{\dagger}\hat{a}+\hat{a}_{\mathrm{LO}}%
\hat{a}^{\dagger}\text{.}%
\end{equation}
Taking into account that the local oscillator is in a coherent state with
amplitude $\alpha_{\mathrm{LO}}=|\alpha_{\mathrm{LO}}|\exp(i\varphi)$, and is
not correlated with our measured mode, it is not difficult to show that the
first moments of this operator can be written as%
\begin{subequations}
\begin{align}
\left\langle \hat{N}_{\mathrm{D}}\right\rangle  &  =|\alpha_{\mathrm{LO}%
}|\left\langle \hat{X}^{\varphi}\right\rangle \\
\left\langle \hat{N}_{\mathrm{D}}^{2}\right\rangle  &  =|\alpha_{\mathrm{LO}%
}|^{2}\left[  \left\langle \hat{X}^{\varphi2}\right\rangle +\frac{\left\langle
\hat{a}^{\dagger}\hat{a}\right\rangle }{|\alpha_{\mathrm{LO}}|^{2}}\right]  .
\end{align}
Hence, in the \textit{strong local oscillator limit} $|\alpha_{\mathrm{LO}%
}|^{2}\gg$ $\left\langle \hat{a}^{\dagger}\hat{a}\right\rangle $, the output
signal of the homodyne scheme has the mean of a quadrature $\hat{X}^{\varphi}$
of the analyzed mode (the one selected by the phase of the local oscillator),
as well as its same quantum statistics.

Even though this picture offers all the basic ingredients that one has to
understand about light detection, it is in a sense far from how light is
observed in real experiments. In the following sections we discuss
photodetection and homodyne detection scenarios which are closer to the
experimental ones.

\section[Real photodetection: The photocurrent and its power spectrum]{Real photodetection: The photocurrent and its power spectrum\label{RealPhotodetection}}

In the detection schemes that we will consider throughout this thesis, the
photodetectors will be measuring light beams with a high flux of photons (for
example, in homodyne detection we have seen that the local oscillator is a
strong coherent field). Under these circumstances several photopulses can be
created during a time interval comparable to the width of a single pulse, so
that it is not possible to count pulses anymore. Instead, one sees the signal
coming out from the detector basically as a continuous current; it is the goal
of this section to model the statistics of this \textit{photocurrent} and
explain the way in which it is analyzed experimentally.%

\begin{figure}
[ptb]
\begin{center}
\includegraphics[
height=2.7415in,
width=4.1831in
]%
{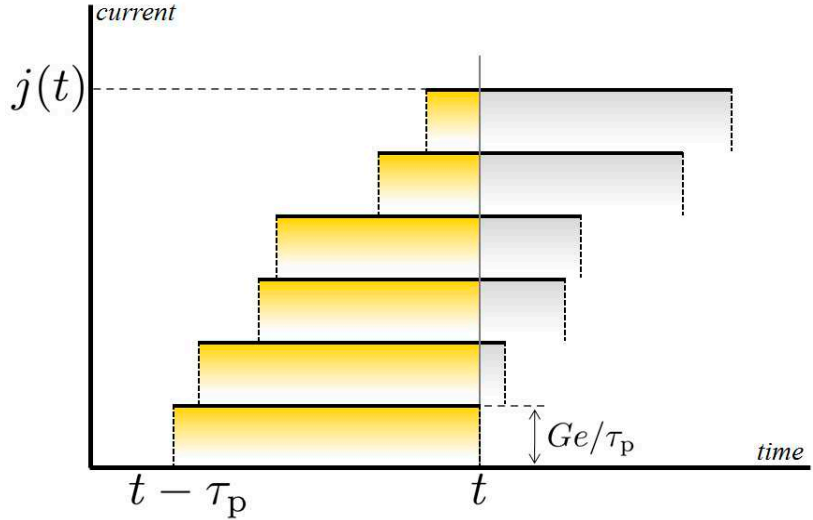}%
\caption{Carmichael's picture of how the continuous photocurrent is built up
from the individual photopulses.}%
\label{fDetec3}%
\end{center}
\end{figure}

For simplicity\footnote{In the next sections we follow closely the treatment
of photodetection and homodyne detection proposed by Carmichael
\cite{Carmichael87,Carmichael93book,Carmichael08book}}, let us assume that the
pulses generated after the amplification stage of each photoelectron are
square like and have a temporal width $\tau_{\mathrm{p}}$. This width is
equivalent to the time response of the detector, which we will consider to be
instantaneous eventually, that is, $\tau_{\mathrm{p}}\rightarrow0$. As shown
in Figure \ref{fDetec3}, the photocurrent $j(t)$ detected at time $t$ is a
superposition of all the pulses generated in the time interval $[t-\tau
_{\mathrm{p}},t]$, that is,%
\end{subequations}
\begin{equation}
j(t)=\frac{Ge}{\tau_{\mathrm{p}}}n(t-\tau_{\mathrm{p}},t),
\end{equation}
where $e$ is the electron charge, $G$ is the number of electrons generated in
the amplification stage from the initial photoelectron, and $n(t-\tau
_{\mathrm{p}},t)$ is the number of photoelectrons generated in the interval
$[t-\tau_{\mathrm{p}},t]$. As shown by Kelley and Kleiner \cite{Kelley64}, who
generalized the semiclassical formula previously obtained by Mandel
\cite{Mandel58, Mandel59}, for a field such as the one coming out from a
cavity (\ref{AoutMulti}), the statistics of the number of pulses
$n(t-\tau_{\mathrm{p}},t)$ are dictated by the distribution
\cite{Mandel95book}%
\begin{equation}
p_{n}(t-\tau_{\mathrm{p}},t)=\left\langle :\frac{\hat{J}^{n}(t,\tau
_{\mathrm{p}})\exp\left[  -\hat{J}(t,\tau_{\mathrm{p}})\right]  }%
{n!}:\right\rangle ,
\end{equation}
where \textquotedblleft: :\textquotedblright\ must be understood from now on
as normal order and \textit{time order} (time increases to the right for
products of creation operators and to the right for products of annihilation
operators), and we have defined the \textit{photocurrent} \textit{operator}%
\begin{equation}
\hat{J}(t,\tau_{\mathrm{p}})=\eta\frac{2c\varepsilon_{0}\omega
_{\mathrm{c}}}{\hbar}\int_{t-\tau_{\mathrm{p}}}^{t}dt^{\prime}\int_{%
\mathbb{R}
^{2}}d^{2}\mathbf{r}_{\perp}\mathbf{\hat{A}}_{\mathrm{out}}^{(-)}\left(
\mathbf{r}_{\perp},z_{\mathrm{d}},t^{\prime}\right)  \mathbf{\hat{A}%
}_{\mathrm{out}}^{(+)}\left(  \mathbf{r}_{\perp},z_{\mathrm{d}},t^{\prime
}\right)\simeq\eta\tau_{\mathrm{p}}\sum_{\sigma\mathbf{n}}\hat{n}_{\sigma
\mathbf{n},\mathrm{out}}(t),
\end{equation}
being $\hat{n}_{\sigma\mathbf{n},\mathrm{out}}=$ $\hat{a}_{\sigma
\mathbf{n},\mathrm{out}}^{\dagger}\hat{a}_{\sigma\mathbf{n},\mathrm{out}}$,
$\eta\in\lbrack0,1]$ the photon--photoelectron conversion factor, and
$z_{\mathrm{d}}$ the longitudinal position of the photodetector. In loose
terms, the operator $\hat{n}_{\sigma\mathbf{n},\mathrm{out}}$ `counts' the
total number of output photons detected during the time that a photopulse
lasts. Note that we have assumed that the transverse size of the detector is
larger than the transverse thickness of the output beam, and the last equality
follows from assuming that $\hat{n}_{\sigma\mathbf{n},\mathrm{out}}(t)$
doesn't change too much in the brief interval $[t-\tau_{\mathrm{p}},t]$.

Let us evaluate the mean and the second factorial moment of the number of
pulses $n(t-\tau_{\mathrm{p}},t)$:%
\begin{subequations}
\begin{align}
\overline{n(t-\tau_{\mathrm{p}},t)}&=\sum_{n=0}^{\infty}np_{n}%
(t-\tau_{\mathrm{p}},t)=\left\langle :\hat{J}(t,\tau_{\mathrm{p}})\sum_{n=1}^{\infty}\frac{\hat
{J}^{n-1}(t,\tau_{\mathrm{p}})}{(n-1)!}e^{-\hat{J}(t,\tau
_{\mathrm{p}})} :\right\rangle=\left\langle :\hat{J}(t,\tau_{\mathrm{p}}):\right\rangle
\\
\overline{n(t-\tau_{\mathrm{p}},t)\left[  n(t-\tau_{\mathrm{p}},t)-1\right]  }&=\sum_{n=0}^{\infty}n(n-1)p_{n}(t-\tau_{\mathrm{p}},t)=\left\langle :\hat{J}^{2}(t,\tau_{\mathrm{p}})\sum_{n=2}^{\infty}\frac{\hat{J}^{n-2}(t,\tau_{\mathrm{p}})}{(n-2)!}e^{-\hat{J}(t,\tau_{\mathrm{p}})}:\right\rangle=\left\langle :\hat{J}^{2}(t,\tau_{\mathrm{p}}):\right\rangle,
\end{align}
where we use the overbar to stress the difference between quantum averages and
averages concerning the random electric signal coming out from the experiment.
These expressions allow us to write the first moments of the photocurrent as
\end{subequations}
\begin{subequations}
\label{PhotoMoments}%
\begin{align}
\overline{j(t)} &  =\frac{Ge}{\tau_{\mathrm{p}}}\left\langle :\hat{J}%
(t,\tau_{\mathrm{p}}):\right\rangle =\eta Ge\sum_{\sigma\mathbf{n}%
}\left\langle \hat{n}_{\sigma\mathbf{n},\mathrm{out}}(t)\right\rangle \\
\overline{j^{2}(t)} &  =\frac{G^{2}e^{2}}{\tau_{\mathrm{p}}^{2}}\left[
\left\langle :\hat{J}^{2}(t,\tau_{\mathrm{p}}):\right\rangle +\left\langle
:\hat{J}(t,\tau_{\mathrm{p}}):\right\rangle \right]  \\
&  =\eta^{2}G^{2}e^{2}\left[  \sum_{\sigma\sigma^{\prime}\mathbf{nn}^{\prime}%
}\left\langle \hat{a}_{\sigma\mathbf{n},\mathrm{out}}^{\dagger}\hat{a}%
_{\sigma^{\prime}\mathbf{n}^{\prime},\mathrm{out}}^{\dagger}\hat{a}%
_{\sigma\mathbf{n},\mathrm{out}}\hat{a}_{\sigma^{\prime}\mathbf{n}^{\prime
},\mathrm{out}}(t)\right\rangle +\frac{1}{\eta\tau_{\mathrm{p}}}\sum
_{\sigma\mathbf{n}}\left\langle \hat{n}_{\sigma\mathbf{n},\mathrm{out}%
}(t)\right\rangle \right]  ,\nonumber
\end{align}
Note that any mode in a vacuum state does not contribute to the statistics of
the photocurrent. This is due to the normal ordering appearing in
$p_{n}(t-\tau_{\mathrm{p}},t)$.%

\begin{figure}
[ptb]
\begin{center}
\includegraphics[
height=2.757in,
width=4.7841in
]%
{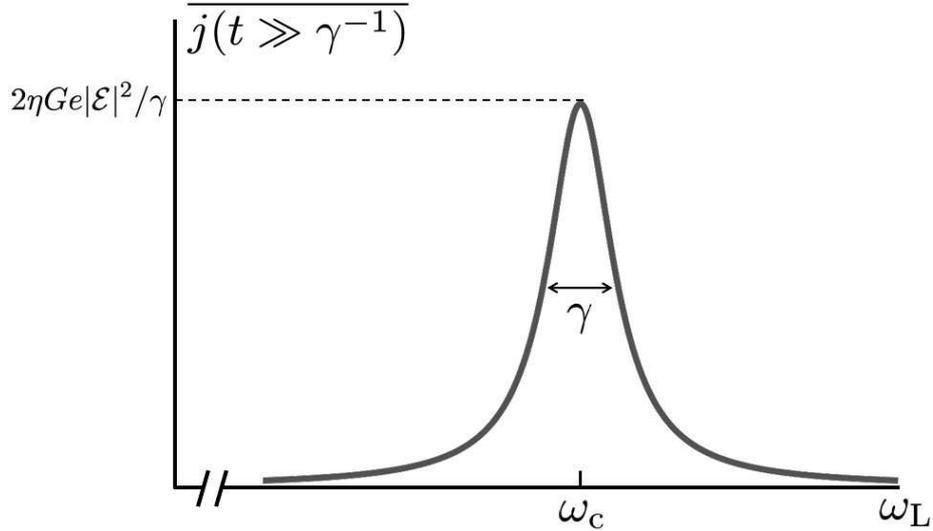}%
\caption{Mean of the photocurrent measured for a single--mode field coming
out from an empty cavity as a function of the frequency of the laser which
drives it. Maximum transmission is obtained when the laser is tuned to the
cavity resonance.}%
\label{fDetec4}%
\end{center}
\end{figure}

To gain some insight, consider the output of an empty single--mode cavity in
the stationary limit, which can be obtained from (\ref{Out-CavIn}), with
$\hat{a}(t)$ given by (\ref{aempty}) and the input operator $\hat
{a}_{\mathrm{in}}(t)$ satisfying%
\end{subequations}
\begin{equation}
\left\langle \hat{a}_{\mathrm{in}}^{\dagger}(t_{1})\hat{a}_{\mathrm{in}%
}^{\dagger}(t_{2})...\hat{a}_{\mathrm{in}}^{\dagger}(t_{j})\hat{a}%
_{\mathrm{in}}(t_{1}^{\prime})\hat{a}_{\mathrm{in}}(t_{2}^{\prime})...\hat
{a}_{\mathrm{in}}(t_{l}^{\prime})\right\rangle =0\text{ }\forall(j,l),
\end{equation}
because the external modes are assumed to be in vacuum initially. This
property allows us to relate moments of the output operators with intracavity
moments in a very simple way:%
\begin{align}\label{Out-Intra}
\left\langle \hat{a}_{\mathrm{out}}^{\dagger}(t_{1})\hat{a}_{\mathrm{out}%
}^{\dagger}(t_{2})...\hat{a}_{\mathrm{out}}^{\dagger}(t_{j})\hat
{a}_{\mathrm{out}}(t_{1}^{\prime})\hat{a}_{\mathrm{out}}(t_{2}^{\prime
})...\hat{a}_{\mathrm{out}}(t_{l}^{\prime})\right\rangle=(2\gamma)^{(j+l)/2}\left\langle \hat{a}^{\dagger}(t_{1})\hat{a}^{\dagger
}(t_{2})...\hat{a}^{\dagger}(t_{j})\hat{a}(t_{1}^{\prime})\hat{a}%
(t_{2}^{\prime})...\hat{a}(t_{l}^{\prime})\right\rangle \text{ }%
\forall(j,l)
\end{align}
Then, in terms of the intracavity operators, the photocurrent moments can be
written as%
\begin{subequations}
\begin{align}
\overline{j(t)} &  =2\gamma\eta Ge\left\langle \hat{n}(t)\right\rangle
=2\pi\eta Ge|\mathcal{E}|^{2}\frac{\gamma/\pi}{\gamma^{2}+(\omega_{\mathrm{c}%
}-\omega_{\mathrm{L}})^{2}}\equiv\overline{j_{0}}\\
\overline{j^{2}(t)} &  =4\gamma^{2}\eta^{2}G^{2}e^{2}\left\langle \hat{n}%
^{2}(t)\right\rangle +\frac{2\gamma\eta G^{2}e^{2}}{\tau_{\mathrm{p}}%
}(1-2\gamma\eta\tau_{\mathrm{p}})\left\langle \hat{n}(t)\right\rangle ,
\end{align}
where $\hat{n}=\hat{a}^{\dagger}\hat{a}$ is the number operator for the
photons inside the cavity. The mean of the photocurrent is proportional to the
mean of the number of photons inside the cavity, similarly to the idealized
photodetection description discussed in the previous section. However, the
second moment of the photocurrent is not directly proportional to the second
moment of $\hat{n}$, it has a deviation proportional to the mean, that is, an
additional Poissonian contribution which shows that the photocurrent signal is
affected by extra noise, and hence does not offer 100\% truthful information
about the intracavity number fluctuations. The situation is even more subtle
as we show now.

\begin{figure}
[t]
\begin{center}
\includegraphics[
width=\textwidth
]%
{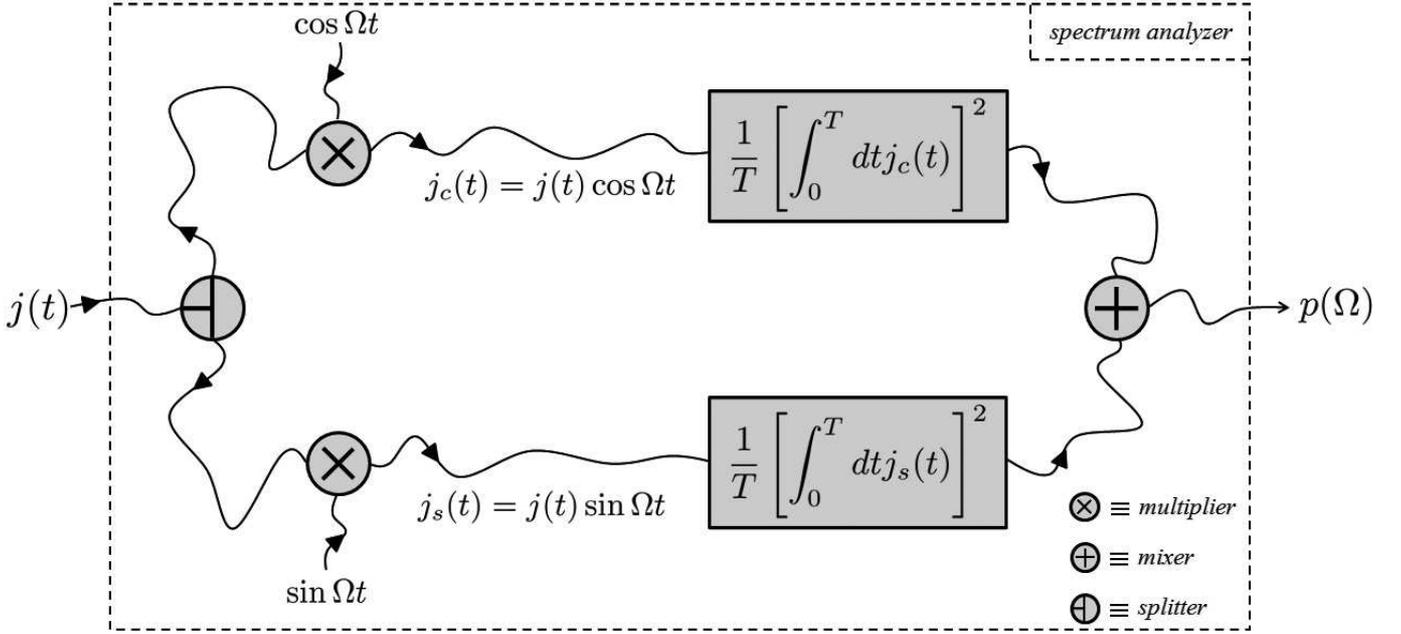}%
\caption{Schematic representation of the action of a spectrum analyzer onto an
incoming electric current. After dividing it and multiply the separated
currents by a sine and a cosine oscillating at frequency $\Omega$, the signals
are accumulated during a time $T$ (\textit{detection time}); the squares of
these quantities (normalized to $T$) are added, finally building the
spectral signal $p(\Omega)$.}%
\label{fDetec5}%
\end{center}
\end{figure}

Note that as a function of the frequency $\omega_{\mathrm{L}}$ of the laser
which `scans' the cavity, the mean photocurrent (or the mean intracavity
photon number) is a Lorentzian of width $\gamma$ centered at the cavity
resonance $\omega_{\mathrm{c}}$ (see Figure \ref{fDetec4}), which read in
reverse shows that only external modes within the interval $[\omega
_{\mathrm{c}}-\gamma,\omega_{\mathrm{c}}+\gamma]$ can be excited as we
explained in the previous chapter. However, the variance associated to this
photocurrent can be written as%
\end{subequations}
\begin{equation}
V\left[  j(t)\right]  =\overline{j^{2}(t)}-\overline{j(t)}^{2}=\frac{Ge}%
{\tau_{\mathrm{p}}}\overline{j_{0}}\text{,}%
\end{equation}
showing that the noise in the photocurrent dominates the output signal when
the photodetector has a fast response, that is, $\tau_{\mathrm{p}}%
\rightarrow0$. Hence, instantaneous photocurrent signals cannot be usually
studied in experiments. Instead, one studies the \textit{power spectrum} of
the photocurrent, which can be obtained at real time by introducing the
electric signal coming out from the photodetector into a \textit{spectrum
analyzer}. The scheme of a spectrum analyzer is shown in Figure \ref{fDetec5};
the mean of the new output signal is then%
\begin{equation}
P(\Omega)=\frac{1}{T}\overline{\left[  \int_{0}^{T}dtj(t)\cos\Omega t\right]
^{2}+\left[  \int_{0}^{T}dtj(t)\sin\Omega t\right]  ^{2}},
\end{equation}
which can be easily rewritten as%
\begin{equation}
P(\Omega)=\frac{1}{T}\int_{0}^{T}dt\int_{0}^{T}dt^{\prime}\cos\left[
\Omega(t-t^{\prime})\right]  \overline{j(t)j(t^{\prime})},
\end{equation}
in terms of the two--time correlation function of the photocurrent. When this
correlation depends solely on the time difference $|t-t^{\prime}|$, that is, when
$\overline{j(t)j(t^{\prime})}=\mathcal{J}(|t-t^{\prime}|)$, this expression
can be simplified even further as%
\begin{equation}
P(\Omega)=\int_{-\infty}^{+\infty}d\tau\mathcal{J}(|\tau|)e^{-\mathrm{i}%
\Omega\tau},
\end{equation}
in the $T\rightarrow\infty$ limit. In such case we say that $j(t)$ corresponds
to a \textit{stationary process} \cite{Mandel95book}.%

Using similar techniques as in the previous derivations, it is lengthy but
simple to show that the correlation function of the photocurrent can be
written in terms of the photocurrent operator as%
\begin{align}
\overline{j(t)j(t^{\prime})}=\frac{G^{2}e^{2}}{\tau_{\mathrm{p}}^{2}%
}\left[  \left\langle :\hat{J}(t,\tau_{\mathrm{p}})\hat{J}(t^{\prime}%
,\tau_{\mathrm{p}}):\right\rangle \right.  \label{PhotoCorrelation}\left.  +H(\tau_{\mathrm{p}}-|t^{\prime}-t|)\left\langle :\hat{J}%
(\min\{t,t^{\prime}\},\tau_{\mathrm{p}}-|t^{\prime}-t|):\right\rangle \right]
\end{align}
where%
\begin{equation}
H(x)=\left\{
\begin{array}
[c]{cc}%
1 & x>0\\
0 & x\leq0
\end{array}
\right.  ,
\end{equation}
is a step function. Particularizing this expression for the empty single--mode
cavity in the stationary limit, it is straightforward to show that its
associated power spectrum reads%
\begin{equation}
P(\Omega)=2\pi\overline{j_{0}}^{2}\left[  \delta(\Omega)+\frac{Ge}{2\pi
}\overline{j_{0}}^{2}\right]  ,
\end{equation}
where we have assumed a large detection time and a fast time response of the
photodetector, that is, $T\rightarrow\infty$ and $\tau_{\mathrm{p}}%
\rightarrow0$. This shows that the massive noise masking the signal $j(t)$ at
all times is concentrated in frequency space at $\Omega=0$. For any other
frequency (technically for $\Omega\gg T^{-1}$) the power spectrum is finite.

\section[Real homodyne detection: Squeezing and the noise spectrum]{Real homodyne detection: Squeezing and the noise spectrum\label{RealHomodyneDetection}}

Now that we understand how to treat the signal coming out of a real
photodetector, we can analyze as well the signal coming out from a homodyne
detection scheme as is usually done in experiments. This will allow us to
introduce squeezing in a manner which is useful for experimental matters
through the so--called squeezing spectrum.%

\begin{figure}
[t]
\begin{center}
\includegraphics[
height=3.1349in,
width=4.7469in
]%
{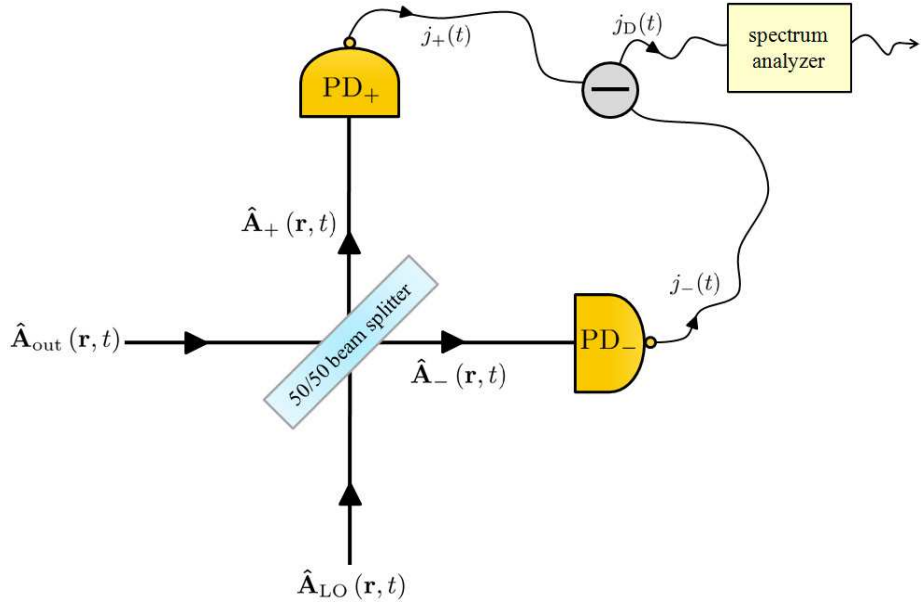}%
\caption{Balanced homodyne detection scheme. The idea is exactly as in Figure
\ref{fDetec2}, but now the setup is analyzed for a field coming out from a
cavity, and considering the realistic picture of photodetection introduced in
the previous section.}%
\label{fDetec6}%
\end{center}
\end{figure}

The situation is depicted in Figure \ref{fDetec6}. Our output multi--mode
field (\ref{AoutMulti}) is mixed in a 50/50 beam splitter with a \textit{local
oscillator field} coming out from a laser whose central frequency
$\omega_{\mathrm{LO}}$ is close to $\omega_{\mathrm{c}}$, and is assumed to
emit in a well defined mode with polarization $\boldsymbol{\varepsilon
}_{\Lambda}$ and transverse profile $T_{\Lambda}(k_{\mathrm{c}};\mathbf{r}%
_{\perp},z)$, that is,%
\begin{equation}
\mathbf{\hat{A}}_{\mathrm{LO}}^{(+)}\left(  \mathbf{r},t\right)  =\sqrt
{\frac{\hbar}{2c\varepsilon_{0}\omega_{\mathrm{c}}}}\boldsymbol{\varepsilon
}_{\Lambda}T_{\Lambda}\left(  k_{\mathrm{c}};\mathbf{r}_{\perp},z\right)
\hat{a}_{\mathrm{LO}}(t);
\end{equation}
the state of this laser field is modelled as a stationary coherent state, and
hence its associated output operator $\hat{a}_{\mathrm{LO}}(t)$ satisfies%
\begin{align}
\left\langle \hat{a}_{\mathrm{LO}}^{\dagger}(t_{1})\hat{a}_{\mathrm{LO}%
}^{\dagger}(t_{2})...\hat{a}_{\mathrm{LO}}^{\dagger}(t_{j})\hat{a}%
_{\mathrm{LO}}(t_{1}^{\prime})\hat{a}_{\mathrm{LO}}(t_{2}^{\prime})...\hat
{a}_{\mathrm{LO}}(t_{l}^{\prime})\right\rangle=e^{\mathrm{i}\omega_{\mathrm{LO}}(t_{1}+t_{2}+...+t_{j}%
-t_{1}^{\prime}-t_{2}^{\prime}-...-t_{l}^{\prime})} \alpha
_{\mathrm{LO}}^{\ast j}\alpha_{\mathrm{LO}}^{l}\text{ }\forall(j,l),
\end{align}
where we will write $\alpha_{\mathrm{LO}}=|\alpha_{\mathrm{LO}}|\exp
(i\varphi)$.

On the other hand, the fields coming out from the beam splitter $\mathbf{\hat
{A}}_{\pm}^{(+)}=\left[  \mathbf{\hat{A}}_{\mathrm{out}}^{(+)}\pm
\mathbf{\hat{A}}_{\mathrm{LO}}^{(+)}\right]  /\sqrt{2}$ can be written as%
\begin{align}
\mathbf{\hat{A}}_{\pm}^{(+)}\left(  \mathbf{r},t\right)=\sqrt
{\frac{\hbar}{2c\varepsilon_{0}\omega_{\mathrm{c}}}}\left\{  \frac{1}{\sqrt
{2}}\boldsymbol{\varepsilon}_{\Lambda}T_{\Lambda}\left(  k_{\mathrm{c}%
};\mathbf{r}_{\perp},z\right)  \left[  \hat{a}_{\Lambda,\mathrm{out}}%
(t)+\hat{a}_{\mathrm{LO}}(t)\right]+\sum_{\sigma\mathbf{n}\neq\Lambda}\boldsymbol{\varepsilon}%
_{\sigma\mathbf{n}}T_{\mathbf{n}}\left(  k_{\mathrm{c}};\mathbf{r}_{\perp
},z\right)  \hat{a}_{\sigma\mathbf{n},\mathrm{out}}(t)\right\}  .
\end{align}
The photodetectors $\mathrm{PD}_{\pm}$ are statistically independent, and
hence, based on the previous section, the probability of observing $n_{+}$ and
$n_{-}$ photopulses in the time intervals $[t_{+}-\tau_{\mathrm{p}},t_{+}]$
and $[t_{-}-\tau_{\mathrm{p}},t_{-}]$, respectively, is given by%
\begin{equation}
p(n_{+},t_{+}-\tau_{\mathrm{p}},t_{+};n_{-},t_{-}-\tau_{\mathrm{p}}%
,t_{-})=p_{n_{+}}(t_{+}-\tau_{\mathrm{p}},t_{+})p_{n_{-}}(t_{-}-\tau
_{\mathrm{p}},t_{-}),
\end{equation}
where the photocurrent operators associated to the photodetectors can be
written as%
\begin{equation}
\hat{J}_{\pm}(t,\tau_{\mathrm{p}})=\frac{\eta\tau_{\mathrm{p}}}{2}\left[
\hat{n}_{\mathrm{LO}}(t)+\hat{n}_{\Lambda,\mathrm{out}}(t)\pm\hat{\chi
}(t)+2\sum_{\sigma\mathbf{n}\neq\Lambda}\hat{n}_{\sigma\mathbf{n}%
,\mathrm{out}}(t)\right]  ,\label{Jpm}%
\end{equation}
and we have defined the operator%
\begin{equation}
\hat{\chi}(t)=\hat{a}_{\mathrm{LO}}^{\dagger}(t)\hat{a}_{\Lambda,\mathrm{out}%
}(t)+\hat{a}_{\mathrm{LO}}(t)\hat{a}_{\Lambda,\mathrm{out}}^{\dagger}(t).
\end{equation}
Note that the mean and two--time correlation functions of this operator are
proportional to the ones of the \textit{output quadrature operator} selected
by the local oscillator%
\begin{equation}
\hat{X}_{\Lambda,\mathrm{out}}^{\varphi-\omega_{\mathrm{LO}}t}%
(t)=e^{-\mathrm{i}\varphi+\mathrm{i}\omega_{\mathrm{LO}}t}\hat{a}%
_{\Lambda,\mathrm{out}}(t)+e^{\mathrm{i}\varphi-\mathrm{i}\omega_{\mathrm{LO}%
}t}\hat{a}_{\Lambda,\mathrm{out}}^{\dagger}(t),
\end{equation}
that is,%
\begin{subequations}
\begin{align}
\left\langle \hat{\chi}(t)\right\rangle  & =|\alpha_{\mathrm{LO}}|\langle
\hat{X}_{\Lambda,\mathrm{out}}^{\varphi-\omega_{\mathrm{LO}}t}(t)\rangle\\
\left\langle \hat{\chi}(t)\hat{\chi}(t^{\prime})\right\rangle  &
=|\alpha_{\mathrm{LO}}|^{2}\langle\hat{X}_{\Lambda,\mathrm{out}}%
^{\varphi-\omega_{\mathrm{LO}}t}(t)\hat{X}_{\Lambda,\mathrm{out}}%
^{\varphi-\omega_{\mathrm{LO}}t^{\prime}}(t^{\prime})\rangle\text{.}%
\end{align}
The fast optical oscillations of the $\hat{a}_{\Lambda,\mathrm{out}}(t)$
operator are canceled by the rapidly oscillating function $\exp(i\omega
_{\mathrm{LO}}t)$ added by the local oscillator. In order to simplify the
derivations, it is then customary to move to an interaction picture defined by
the transformation operator $\hat{U}_{0}=\exp[\hat{H}_{0}t/i\hbar]$ with%
\end{subequations}
\begin{equation}
\hat{H}_{0}=\sum_{\sigma\mathbf{n}}\hbar\omega_{\mathrm{LO}}\left[  \hat
{a}_{\sigma\mathbf{n}}^{\dagger}\hat{a}_{\sigma\mathbf{n}}+\int_{-\infty
}^{+\infty}d\omega\hat{b}_{\sigma\mathbf{n}}^{\dagger}(\omega)\hat{b}%
_{\sigma\mathbf{n}}(\omega)\right]  ,
\end{equation}
so that%
\begin{equation}
\hat{U}_{0}^{\dagger}\hat{X}_{\Lambda,\mathrm{out}}^{\varphi-\omega
_{\mathrm{LO}}t}\hat{U}_{0}=\hat{X}_{\Lambda,\mathrm{out}}^{\varphi},
\end{equation}
that is, the fast linear variation of the quadrature's phase is removed. From
now on, we assume to work in this picture, that is, quantum expectation values
are to be taken with the interaction picture state $\hat{\rho}_{\mathrm{I}%
}(t)=\hat{U}_{0}^{\dagger}\hat{\rho}(t)\hat{U}_{0}$. We shall see how useful
this picture is in the next chapter, when analyzing the squeezing properties
of optical parametric oscillators.

It is not difficult to prove that in the strong local oscillator regime, the
difference photocurrent $j_{\mathrm{D}}(t)=j_{+}(t)-j_{-}(t)$ carries
information about the $\hat{\chi}(t)$ operator only; any other possible
contribution to this photocurrent is balanced out thanks to the 50/50 beam
splitter. To this aim we evaluate the first moments of the difference
photocurrent, which, given the statistical independency of the photodetectors,
can be written in terms of the moments of $j_{\pm}(t)$ as%
\begin{equation}
\overline{j_{\mathrm{D}}(t)}=\overline{j_{+}(t)}-\overline{j_{-}(t)}\text{
\ \ \ \ and \ \ \ \ }\overline{j_{\mathrm{D}}^{2}(t)}=\overline{j_{+}^{2}%
(t)}+\overline{j_{-}^{2}(t)}-2\overline{j_{+}(t)}\times\overline{j_{-}(t)}.
\end{equation}
Making use of the relation (\ref{PhotoMoments}) between the moments of
$j_{\pm}(t)$ and the quantum expectation values of their corresponding
photocurrent operators (\ref{Jpm}), it is straightforward to show after some
algebra that the mean and variance of the difference photocurrent are given by%
\begin{subequations}
\begin{align}
\overline{j_{\mathrm{D}}(t)} &  =\eta Ge|\alpha_{\mathrm{LO}}|\left\langle
\hat{X}_{\Lambda,\mathrm{out}}^{\varphi}\right\rangle, \label{JdifMean}
\\
V\left[  j_{\mathrm{D}}(t)\right]   &  =\frac{\eta^{2}G^{2}e^{2}%
|\alpha_{\mathrm{LO}}|^{2}}{\tau_{\mathrm{p}}}\left(  1+\sum_{\sigma
\mathbf{n}}\frac{\left\langle \hat{n}_{\sigma\mathbf{n},\mathrm{out}%
}\right\rangle }{|\alpha_{\mathrm{LO}}|^{2}}\right)+\eta^{3}G^{2}e^{2}|\alpha_{\mathrm{LO}}|^{2}\left[  \left\langle :\left(
\delta\hat{X}_{\Lambda,\mathrm{out}}^{\varphi}\right)  ^{2}:\right\rangle
+\sum_{\sigma\mathbf{n}}\frac{\left\langle :\delta\hat{n}_{\sigma
\mathbf{n},\mathrm{out}}^{2}:\right\rangle }{|\alpha_{\mathrm{LO}}|^{2}%
}\right]  ,
\end{align}
\end{subequations}
where we remind that we make use of
the notation $\delta\hat{B}=\hat{B}-\langle\hat{B}\rangle$. Taking the strong
local oscillator limit%
\begin{equation}
|\alpha_{\mathrm{LO}}|^{2}\gg\left\{  \left\langle \hat{n}_{\sigma
\mathbf{n},\mathrm{out}}\right\rangle ,\left\langle :\delta\hat{n}%
_{\sigma\mathbf{n},\mathrm{out}}^{2}:\right\rangle \right\}  \text{ }%
\forall(\sigma\mathbf{n})\text{,}%
\end{equation}
the variance of the difference photocurrent is written as%
\begin{equation}
V\left[  j_{\mathrm{D}}(t)\right]  =\eta^{2}G^{2}e^{2}|\alpha_{\mathrm{LO}%
}|^{2}\left\{  \frac{1}{\tau_{\mathrm{p}}}+\eta\langle:[\delta\hat{X}%
_{\Lambda,\mathrm{out}}^{\varphi}(t)]^{2}:\rangle\right\}  .
\end{equation}
Expression (\ref{JdifMean}) shows that the mean of $j_{\mathrm{D}}(t)$ carries
information about the quadrature of the mode selected by the local oscillator,
but similarly to the direct detection case, in the fast response limit
$\tau_{\mathrm{p}}\rightarrow0$ this signal is masked by massive noise at all times.

Hence, also in homodyne detection one is forced to consider the power spectrum
of the difference photocurrent. As there are no contributions from any of the
modes with $(\sigma,\mathbf{n)}\neq\Lambda$, we will remove the local
oscillator mode index $\Lambda$ from now on. Using again the statistical
independency of the signals coming out from the photodetectors $\mathrm{PD}%
_{\pm}$, one can write the two--time correlation function of $j_{\mathrm{D}%
}(t)$ as%
\begin{equation}
\overline{j_{\mathrm{D}}(t)j_{\mathrm{D}}(t^{\prime})}=\overline{j_{+}%
(t)j_{+}(t^{\prime})}+\overline{j_{-}(t)j_{-}(t^{\prime})}-\overline{j_{+}%
(t)}\times\overline{j_{-}(t^{\prime})}-\overline{j_{+}(t^{\prime})}%
\times\overline{j_{-}(t)}.
\end{equation}
Using now the general expression (\ref{PhotoCorrelation}) for the two--time
correlation function of a photocurrent, it is lengthy but trivial to arrive to
the following expression for the power spectrum of $j_{\mathrm{D}}(t)$%
\begin{equation}
P_{\mathrm{D}}(\Omega)=\frac{1}{T}\int_{0}^{T}dt\int_{0}^{T}dt^{\prime}%
\cos\left[  \Omega(t-t^{\prime})\right]  \overline{j_{\mathrm{D}%
}(t)j_{\mathrm{D}}(t^{\prime})}=P_{\mathrm{shot}}+P(\Omega),
\label{PdSpectrum}%
\end{equation}
being%
\begin{equation}
P_{\mathrm{shot}}=\eta G^{2}e^{2}|\alpha_{\mathrm{LO}}|^{2},
\end{equation}
a contribution to the spectrum which does not depend either on the state of
the system or in the frequency (the so-called \textit{shot noise}
contribution), and
\begin{equation}
P(\Omega)=\frac{\eta^{2}G^{2}e^{2}}{T}\int_{0}^{T}dt\int_{0}^{T}dt^{\prime
}\cos\left[  \Omega(t-t^{\prime})\right]  \left\langle :\delta\hat
{X}_{\mathrm{out}}^{\varphi}(t)\delta\hat{X}_{\mathrm{out}}^{\varphi
}(t^{\prime}):\right\rangle
\end{equation}
a contribution which does depend on these.

It is customary to define a normalized version of this quantity called the
\textit{noise spectrum}%
\begin{equation}
V^{\mathrm{out}}(\hat{X}^{\varphi};\Omega)=\frac{P_{\mathrm{D}}(\Omega
)}{P_{\mathrm{shot}}}=1+S^{\mathrm{out}}(\hat{X}^{\varphi};\Omega)
\end{equation}
where%
\begin{equation}
S^{\mathrm{out}}\left(  \hat{X}^{\varphi};\Omega\right)  =\frac{\eta}{T}%
\int_{0}^{T}dt\int_{0}^{T}dt^{\prime}\cos\left[  \Omega(t-t^{\prime})\right]
\langle:\delta\hat{X}_{\mathrm{out}}^{\varphi}(t)\delta\hat{X}_{\mathrm{out}%
}^{\varphi}(t^{\prime}):\rangle,
\end{equation}
is the so-called \textit{squeezing spectrum}, which, when $\langle:\delta
\hat{X}_{\mathrm{out}}^{\varphi}(t)\delta\hat{X}_{\mathrm{out}}^{\varphi
}(t^{\prime}):\rangle$ depends only on $|t-t^{\prime}|$, that is, for
stationary states, can be written as%
\begin{equation}
S^{\mathrm{out}}\left(  \hat{X}^{\varphi};\Omega\right)  =\eta\int_{-\infty
}^{+\infty}dt^{\prime}\langle:\delta\hat{X}_{\mathrm{out}}^{\varphi}%
(t)\delta\hat{X}_{\mathrm{out}}^{\varphi}(t+t^{\prime}):\rangle e^{-\mathrm{i}%
\Omega t^{\prime}}.
\end{equation}
Accordingly, we will call \textit{noise frequency }to $\Omega$, which needs
not to be confused with the optical frequencies.

Note that the multi--time moments of the output operators still satisfy
(\ref{Out-Intra}), and hence, in terms of the quadratures of the intracavity
mode the squeezing spectrum reads%
\begin{equation}
S^{\mathrm{out}}(\hat{X}^{\varphi};\Omega)=\frac{2\gamma\eta}{T}\int_{0}%
^{T}dt\int_{0}^{T}dt^{\prime}\cos\left[  \Omega(t-t^{\prime})\right]
\langle:\delta\hat{X}^{\varphi}(t)\delta\hat{X}^{\varphi}(t^{\prime}):\rangle,
\label{GenSqSpectrum}%
\end{equation}
and similarly for the stationary expression. Hence, knowing the state of the
intracavity mode, we can evaluate the noise spectrum measured for the field
coming out from the cavity. In the upcoming chapters we will assume unit
conversion efficiency\footnote{Real experiments are not that far from this
situation, arriving to efficiencies above 90\%.}, that is, $\eta=1$.

In Chapter \ref{HarmonicOscillator} we introduced squeezed states of a single
harmonic oscillator as those in which one of its quadratures had an
uncertainty below that of vacuum or a coherent state. However, we have seen
that this simple intuitive object (the uncertainty of a single--mode
quadrature) is not the relevant quantity in real experiments designed to
measure the quadratures of light, and hence we need to come back a little bit
and redefine the concept of squeezing in a way which is interesting from the
experimental point of view.

To this aim, let us proceed similarly to how we did in Chapter
\ref{HarmonicOscillator}. Consider a quasiperiodic modulation of frequency
$\Omega_{\mathrm{signal}}$ encoded in some quadrature of a cavity mode. It is
to be expected that such modulation will appear in the noise spectrum of the
corresponding output mode as a peak centered at the $\Omega=\Omega
_{\mathrm{signal}}$. The problem is that if this modulation is very small, the
corresponding peak in the spectrum can be disguised by the basal shot noise
contribution, and it will be impossible to observe it. This is indeed the case
when the state of the light mode is coherent (like vacuum), as then the
squeezing spectrum is zero for any quadrature and at any noise frequency, so
that the noise spectrum has only the flat $V^{\mathrm{out}}(\hat{X}^{\varphi
};\Omega)=1$ shot noise contribution. We can then define squeezed states of
light in an experimentally useful manner as those states in which
$V^{\mathrm{out}}(\hat{X}^{\varphi};\Omega)<1$ for some quadrature and some frequency.

It is not difficult to show that the noise spectra associated to two
orthogonal quadratures satisfy the Heisenberg uncertainty relation%
\begin{equation}
V^{\mathrm{out}}(\hat{X}^{\varphi};\Omega)V^{\mathrm{out}}(\hat{X}%
^{\varphi+\pi/2};\Omega)\geq1.
\end{equation}
Hence, if the shot noise contribution is balanced out for some quadrature, it
will be increased accordingly for its orthogonal quadrature, similarly to what
happened with the uncertainty of single--mode quadratures in the discussion of
Chapter \ref{HarmonicOscillator}. 

%% file: OPOsFO.tex
In the previous chapters we have developed the basic tools which allow us to
study within the quantum formalism the light generated in optical cavities. It
is now time to apply these to a specific system, optical parametric
oscillators, which will be shown to be ideal candidates for the generation of
squeezed and entangled light.

Such systems belong to the class of so-called nonlinear optical cavities, that
is, resonators containing a nonlinear optical medium. For this reason, the
first section of the chapter contains a brief introduction to the optics of
dielectric media, with special emphasis on nonlinear uniaxial crystals and the
three--wave mixing processes that occur inside them. We then introduce the
model for optical parametric oscillators that will be used all along this
thesis, and study the quantum properties of the light generated by them in
different regimes.

\section{Dielectric media and nonlinear optics}

Roughly speaking \cite{Jackson62book,Griffiths99book}, from the point of view
of their electromagnetic properties most materials fit either into the class
of \textit{conductors} or the class of \textit{dielectrics}. A conducting
material has charges which are free to move through it, and hence the
application of an electromagnetic field can induce an electric current on it.
On the other hand, in dielectric media every charge is strongly bounded to one
specific atom or molecule fixed in some position of the material, and hence
these can be seen as \textit{insulators}. However, when an electromagnetic
field is applied to a dielectric, the electron cloud surrounding each nucleus
can be slightly displaced, what means that the center of mass of the positive
and negative charge distributions of the atoms or molecules get slightly
separated, and then the medium acquires a \textit{polarization}, that is, a
dipole moment per unit volume. In this section we briefly introduce the
description of the electromagnetic field in such media from a classical
viewpoint. We will work only with dielectrics which are sensitive to the
electric field, but not to the magnetic field.

In principle, in order to describe the effects appearing in the field when
propagating in the dielectric material one needs to develop a microscopic
model for the medium, introduce it in the inhomogeneous Maxwell's equations
via a charge distribution $\rho(\mathbf{r},t)$ and a current distribution
$\mathbf{j}(\mathbf{r},t)$, and then solve the resulting field--medium coupled
equations. However, for the type of media we will work with and for weak
enough electromagnetic fields, one can assume that the polarization
$\mathbf{P}\left(  \mathbf{r},t\right)  $ acquired by the medium is a low
order polynomial of the applied electric field, say\footnote{To lighten the
notation of the sections to come, we use the convention that summation over
repeated indices is understood.}%
\begin{equation}
P_{j}\left(  \mathbf{r},t\right)  =\varepsilon_{0}\chi_{jk}^{(1)}E_{k}\left(
\mathbf{r},t\right)  +\varepsilon_{0}\chi_{jkl}^{(2)}E_{k}\left(
\mathbf{r},t\right)  E_{l}\left(  \mathbf{r},t\right)  +..., \label{GenPol}%
\end{equation}
so that the medium can be treated as a non-dynamical system whose information
is all contained the $\overleftrightarrow{\chi}^{(n)}$ coefficients
(\textit{tensors} of order $n+1$ technically\footnote{In the following, we
denote a tensor of order $k$ with components $T_{j_{1}j_{2}...j_{k}}$ by
$\overleftrightarrow{T}$.}), called the \textit{n--order}
\textit{susceptibilities}. In such circumstances one can define a
\textit{displacement field}%
\begin{equation}
\mathbf{D}\left(  \mathbf{r},t\right)  =\varepsilon_{0}\mathbf{E}\left(
\mathbf{r},t\right)  +\mathbf{P}\left(  \mathbf{r},t\right)  \text{,}%
\end{equation}
and replace the inhomogeneous Maxwell's equations in vacuum
(\ref{MaxwellInhomoEqs}) by the so-called \textit{macroscopic Maxwell's
equations} \cite{Jackson62book,Griffiths99book}%
\begin{equation}
\boldsymbol{\nabla}\cdot\mathbf{D}=0\text{ \ \ \ and \ \ \ }\boldsymbol{\nabla
}\times\mathbf{B}=\mu_{0}\partial_{t}\mathbf{D}\text{.}%
\end{equation}
Note that we assume to work with homogeneous media, as the susceptibilities
are independent of the position vector. Moreover, it is simple to show that
materials with an inversion symmetry have null second--order susceptibility.
These include any isotropic medium, as well as 11 of the 32 classes of
crystals \cite{Boyd03book}, whose lowest-order nonlinear susceptibility is
then $\overleftrightarrow{\chi}^{(3)}$.

In the following we discuss the effects of the linear and quadratic
polarizations separately.

\subsection{Linear dielectrics and the refractive index}

Let us consider only the linear term in the polarization (\ref{GenPol}), and
define the \textit{permittivity tensor}%
\begin{equation}
\overleftrightarrow{\varepsilon}=\varepsilon_{0}(1+\overleftrightarrow{\chi
}^{(1)})\text{.}%
\end{equation}
As in vacuum, Maxwell's equations can be highly simplified with the help of
the scalar and vector potentials (\ref{AtoEB}). Choosing now the gauge
condition $\boldsymbol{\nabla}\overleftrightarrow{\varepsilon}\mathbf{A}=0$,
it is simple to prove that we can still select $\phi=0$, and that the vector
potential satisfies the modified wave equation%
\begin{equation}
\left[  \boldsymbol{\nabla}^{2}-(\mu_{0}\overleftrightarrow{\varepsilon
})\partial_{t}^{2}\right]  \mathbf{A}=\boldsymbol{0}, \label{GenWaveEq}%
\end{equation}
where we have neglected a $\boldsymbol{\nabla}\left(  \boldsymbol{\nabla}%
\cdot\mathbf{A}\right)  $ term, what can be safely done within the paraxial
approximation as is easily proved.

When the medium is isotropic we have $\overleftrightarrow{\chi}^{(1)}%
=\chi^{(1)}\overleftrightarrow{I}$, being $\overleftrightarrow{I}$ the
identity matrix and $\chi^{(1)}\geq0$, and hence we recover the wave equation
that we introduced in (\ref{MediumWaveEq}),%
\begin{equation}
\left[  \frac{c^{2}}{n^{2}}\boldsymbol{\nabla}^{2}-\partial_{t}^{2}\right]
\mathbf{A}=\mathbf{0},
\end{equation}
being $n=\sqrt{1+\chi^{(1)}}$ the so called \textit{refractive index} of the
medium. Note that the speed of the waves is no longer $c$ but $c/n$, and
hence, light slows down in a dielectric medium.

As this thesis will focus on second--order nonlinearities, we won't be working
with isotropic media though; we will consider instead dielectric crystals
having an intrinsic coordinate system defined by the orthonormal triad of real
vectors $\{\mathbf{e}_{\mathrm{e}},\mathbf{e}_{\mathrm{o}},\mathbf{e}%
_{\mathrm{o}}^{\prime}\}$ in which the linear susceptibility tensor takes the
diagonal form%
\begin{equation}
\overleftrightarrow{\chi}^{(1)}=\left[
\begin{array}
[c]{ccc}%
\chi_{\mathrm{e}} & 0 & 0\\
0 & \chi_{\mathrm{o}} & 0\\
0 & 0 & \chi_{\mathrm{o}}%
\end{array}
\right]  ;
\end{equation}
such birefringent crystals are called \textit{uniaxial} because the refractive
index is $n_{\mathrm{e}}=\sqrt{1+\chi_{\mathrm{e}}}$ along the
\textit{extraordinary} direction $\mathbf{e}_{\mathrm{e}}$, while it is
$n_{\mathrm{o}}=\sqrt{1+\chi_{\mathrm{o}}}$ along the \textit{ordinary}
directions $\{\mathbf{e}_{\mathrm{o}},\mathbf{e}_{\mathrm{o}}^{\prime}\}$
\cite{Boyd03book}. In the following, we will assume that the orientation of
the crystal is always such that one of the ordinary axes coincides with the
$z$--axis, say $\mathbf{e}_{\mathrm{o}}^{\prime}\equiv\mathbf{e}_{z}$, so that
the extraordinary and ordinary linear polarization components of the field
propagating along the $z$--axis feel different refractive indices inside the
crystal. To see this, just note that writing the vector potential as%
\begin{equation}
\mathbf{A}\left(  \mathbf{r},t\right)  =A_{\mathrm{e}}\left(  \mathbf{r}%
,t\right)  \mathbf{e}_{\mathrm{e}}+A_{\mathrm{o}}\left(  \mathbf{r},t\right)
\mathbf{e}_{\mathrm{o}},
\end{equation}
the components $A_{j}\left(  \mathbf{r},t\right)  $ satisfy the wave equations%
\begin{equation}
\left[  \frac{c^{2}}{n_{\mathrm{e}}^{2}}\boldsymbol{\nabla}^{2}-\partial
_{t}^{2}\right]  A_{\mathrm{e}}=0\text{, \ \ \ \ and \ \ \ \ }\left[
\frac{c^{2}}{n_{\mathrm{o}}^{2}}\boldsymbol{\nabla}^{2}-\partial_{t}%
^{2}\right]  A_{\mathrm{o}}=0\text{,}%
\end{equation}
according to (\ref{GenWaveEq})\footnote{Note that when quantizing the
electromagnetic field in Chapter 2 we have assumed the dielectric slab inside
the cavity to be isotropic. The procedure there showed is naturally
generalized to the case of working with an uniaxial medium: One just needs to
take the two allowed polarization vectors $\{\boldsymbol{\varepsilon}_{\sigma
q\mathbf{n}}\}_{\sigma=1,2}$ along the ordinary and extraordinary directions
for all the modes, and replace the refractive indices accompanying those
components of the fields by the corresponding ones.}.

An important property of dielectric materials is called \textit{dispersion}.
To see what this is, note that when writing (\ref{GenPol}) we have assumed
that the response of the medium to the electric field is instantaneous.
Specially in the case of the linear term, this assumption is wrong for most
materials, and one has to write more generally
\begin{equation}
P_{j}\left(  \mathbf{r},t\right)  =\varepsilon_{0}\int_{0}^{\infty}d\tau
\chi_{jk}^{(1)}(\tau)E_{k}\left(  \mathbf{r},t-\tau\right)  ,
\end{equation}
where the temporal shape of the first--order susceptibility determines exactly
how the polarization at a given time depends on the history of the electric
field. Defining a spectral decomposition for a general vector field
$\mathbf{F}\left(  \mathbf{r},t\right)  $ as\footnote{We choose to introduce
the results of this section and the next one in terms of continuous spectral
decompositions just because the notation is more compact. Exactly the same
expressions apply when only a discrete set of frequencies exists for the
fields, as happens inside an optical cavity.}%
\begin{equation}
\mathbf{\tilde{F}}\left(  \mathbf{r},t\right)  =\int_{-\infty}^{+\infty
}d\omega\mathbf{\tilde{F}}\left(  \mathbf{r},\omega\right)  \exp
(-\mathrm{i}\omega t),
\end{equation}
with $\mathbf{F}\left(  \mathbf{r},-\omega\right)  =\mathbf{F}^{\ast}\left(
\mathbf{r},\omega\right)  $, one can write the previous expression as%
\begin{equation}
\tilde{P}_{j}\left(  \mathbf{r},\omega\right)  =\varepsilon_{0}\chi_{jk}%
^{(1)}(\omega;\omega)\tilde{E}_{k}\left(  \mathbf{r},\omega\right)  ,
\end{equation}
where%
\begin{equation}
\chi_{jk}^{(1)}(\omega;\omega)=\int_{0}^{\infty}d\tau\chi_{jk}^{(1)}(\tau
)\exp(-\mathrm{i}\omega\tau).
\end{equation}
Hence, a non-instantaneous response of the medium is translated into a
frequency dependence of the refractive index, which is the phenomenon known as
dispersion. Note that while $\overleftrightarrow{\chi}^{(1)}(\tau)$ is real,
$\overleftrightarrow{\chi}^{(1)}(\omega;\omega)$ can be complex, although it
has to satisfy $\overleftrightarrow{\chi}^{(1)}(-\omega;-\omega)=\left[
\overleftrightarrow{\chi}^{(1)}(\omega;\omega)\right]  ^{\ast}$ because of the
reality of the fields. In this thesis we will ignore any light absorption by
the dielectric medium; under such circumstances one can prove that the
$\chi_{jk}^{(1)}(\omega;\omega)$ coefficients are real, and furthermore, that
the refractive index is a monotonically increasing function of the frequency
\cite{Boyd03book} (what is known as \textit{normal dispersion}).%

\begin{figure}
[ptb]
\begin{center}
\includegraphics[
height=5.4016in,
width=3.8251in
]%
{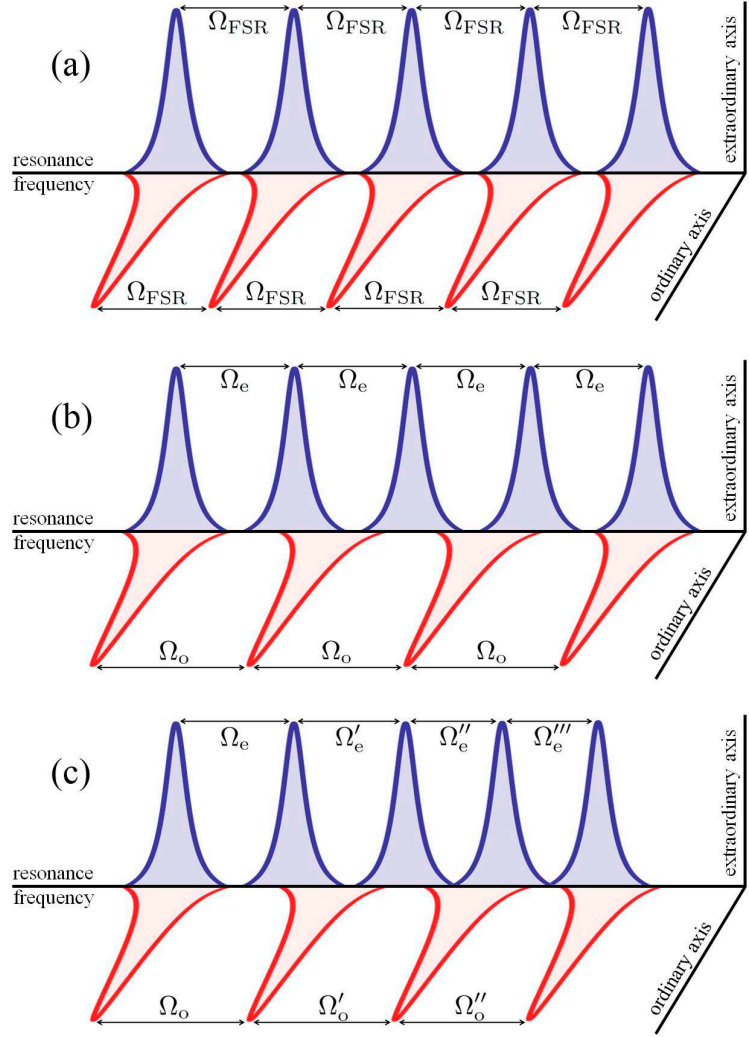}%
\caption{Longitudinal resonances of a given transverse mode inside an optical
cavity. (a) Without taking into account anisotropy and dispersion, the
separation between resonances is given by $\Omega_{\mathrm{FSR}}$ for any
polarization of the modes. (b) The birefringence of the crystal introduces
different refractive indices for the ordinary and extraordinary polarization
components, and therefore these have no longer the same free spectral range.
(c) The normal dispersion of the crystal introduces further corrections; in
particular it increases the optical length of the cavity as frequency
increases, making the free spectral range decrease accordingly, that is,
$\Omega_{j}>\Omega_{j}^{\prime}>\Omega_{j}^{\prime\prime}>\Omega_{j}%
^{\prime\prime\prime}>...$ ($j=\mathrm{e,o}$). }%
\label{fOPO1}%
\end{center}
\end{figure}

It is interesting to analyze how the anisotropy and dispersion of the crystal
affects the structure of the resonances inside an optical cavity. In Figure
\ref{fOPO1} we show the longitudinal resonances associated to a given
transverse mode polarized along $\mathbf{e}_{\mathrm{e}}$ or $\mathbf{e}%
_{\mathrm{o}}$. As we showed in Section \ref{CavityModes}, without considering
anisotropy or dispersion effects, the longitudinal modes are separated by the
free spectral range $\Omega_{\mathrm{FSR}}$, being this the same for every
polarization (Figure \ref{fOPO1}a). An uniaxic crystal introduces different
refractive indices for the ordinary and extraordinary polarization components
of the field; accordingly, the free spectral range associated to modes
polarized along the ordinary and extraordinary directions will no longer be
same, as it is inversely proportional to the optical length and hence to the refractive index (Figure \ref{fOPO1}b). On the other hand,
the normal dispersion of the crystal makes the refractive index increase with
frequency, and hence, the free spectral range will decrease as the frequency
increases (Figure \ref{fOPO1}c). Nevertheless, this last effect can be
compensated for when needed by adding further dispersive elements inside the
cavity such as prisms.

\subsection{Second order nonlinear dielectrics and frequency
conversion\label{SecOrderNonLinearity}}

In a sense, the linear term of the polarization contributes to the dynamics of
the field in a trivial way: It just modifies the speed of the electromagnetic
waves propagating inside it. We are going to show now that, on the other hand,
the nonlinear terms of the polarization induce nontrivial phenomena such as
the excitation of frequencies not present in the field prior to its
interaction with the dielectric \cite{Boyd03book}. In this thesis we are
concerned only with second--order effects, and hence, we will forget about
nonlinear terms other than%
\begin{equation}
\varepsilon_{0}\chi_{jkl}^{(2)}E_{k}\left(  \mathbf{r},t\right)  E_{l}\left(
\mathbf{r},t\right)  \equiv P_{j}^{(2)}\left(  \mathbf{r},t\right)  ,
\label{P(2)}%
\end{equation}
in (\ref{GenPol}). In the following we will simply denote $\mathbf{P}%
^{(2)}\left(  \mathbf{r},t\right)  $ and $\overleftrightarrow{\chi}^{(2)}$ by
\textit{nonlinear polarization} and \textit{nonlinear susceptibility tensor},
respectively, as no other nonlinear contribution will ever appear. Again, this
expression assumes an instantaneous response of the medium, and it must be
generalized as%
\begin{equation}
P_{j}^{(2)}\left(  \mathbf{r},t\right)  =\varepsilon_{0}\int_{0}^{\infty}%
d\tau_{1}\int_{0}^{\infty}d\tau_{1}\chi_{jkl}^{(2)}(\tau_{1},\tau_{2}%
)E_{k}\left(  \mathbf{r},t-\tau_{1}\right)  E_{l}\left(  \mathbf{r},t-\tau
_{2}\right)  ,
\end{equation}
or in frequency space as%
\begin{equation}
\tilde{P}_{j}^{(2)}\left(  \mathbf{r},\omega\right)  =\varepsilon_{0}%
\int_{-\infty}^{+\infty}d\omega_{1}\int_{-\infty}^{+\infty}d\omega_{2}%
\chi_{jkl}^{(2)}(\omega;\omega_{1},\omega_{2})\tilde{E}_{k}\left(
\mathbf{r},\omega_{1}\right)  \tilde{E}_{l}\left(  \mathbf{r},\omega
_{2}\right)  ,
\end{equation}
where we have defined%
\begin{equation}
\chi_{jkl}^{(2)}(\omega;\omega_{1},\omega_{2})=\delta(\omega-\omega_{1}%
-\omega_{2})\int_{0}^{\infty}d\tau\int_{0}^{\infty}d\tau^{\prime}\chi
_{jkl}^{(2)}(\tau,\tau^{\prime})\exp\left[  \mathrm{i}\left(  \omega_{1}%
\tau_{1}+\omega_{2}\tau_{2}\right)  \right]  \text{.} \label{Chi(2)}%
\end{equation}
Because of the delta--function, this expression states that the $\omega$
frequency component of the nonlinear polarization receives contributions from
any pair $(\omega_{1},\omega_{2})$ of frequency components of the electric
field satisfying $\omega_{1}+\omega_{2}=\omega$; this delta--function in
$\overleftrightarrow{\chi}^{(2)}(\omega;\omega_{1},\omega_{2})$ appears even
in the instantaneous case (\ref{P(2)}), and hence it is not a consequence of
the dependence of $\mathbf{P}^{(2)}\left(  \mathbf{r},t\right)  $ on the
history of the electric field. Actually, the effect that this has on
$\overleftrightarrow{\chi}^{(2)}$ is that, for example, given two pairs
$(\omega_{1},\omega_{2})$ and $(\Omega_{1},\Omega_{2})$ satisfying
$\omega=\omega_{1}+\omega_{2}=\Omega_{1}+\Omega_{2}$, one has in general%
\begin{equation}
\chi_{jkl}^{(2)}(\omega;\omega_{1},\omega_{2})\neq\chi_{jkl}^{(2)}%
(\omega;\Omega_{1},\Omega_{2}),
\end{equation}
which is the generalization of the linear dispersion that we introduced in the
previous section. We note, however, that this \textit{nonlinear dispersion} is
not as relevant as the normal dispersion appearing in the linear case, and one
can safely ignore it within a fairly broad spectral region \cite{Seres01}.

Similarly to the linear susceptibility, $\overleftrightarrow{\chi}%
^{(2)}(\omega;\omega_{1},\omega_{2})$ might be complex, although it also has
to satisfy\newline%
\begin{equation}
\overleftrightarrow{\chi}^{(2)}(-\omega;-\omega_{1},-\omega_{2}%
)=[\overleftrightarrow{\chi}^{(2)}(\omega;\omega_{1},\omega_{2})]^{\ast}.
\end{equation}
In any case, when absorption is neglected it can be proved to be real
\cite{Boyd03book}. A number of general properties of the $\chi_{jkl}%
^{(2)}(\omega;\omega_{1},\omega_{2})$ coefficients upon permutation of its
Cartesian or frequency arguments can also be proved \cite{Boyd03book}. For the
case of a lossless medium that we will consider, the most general of these is
the one called \textit{full permutation symmetry}, which states that
$\chi_{jkl}^{(2)}(\omega;\omega_{1},\omega_{2})$ is left unchanged upon
permutation of two frequency arguments, as long as the Cartesian indices are
permuted accordingly, that is,%
\begin{equation}
\chi_{jkl}^{(2)}(\omega;\omega_{1},\omega_{2})=\chi_{jlk}^{(2)}(\omega
;\omega_{2},\omega_{1})=\chi_{lkj}^{(2)}(\omega_{2};\omega_{1},-\omega)=...;
\end{equation}
these relations will be useful in the next section.

It is time now to see how the phenomenon of frequency conversion appears. To
this aim, let us build now the wave equation satisfied by the electric field,
which is easily found to be%
\begin{equation}
\left[  \boldsymbol{\nabla}^{2}-\mu_{0}\overleftrightarrow{\varepsilon
}\partial_{t}^{2}\right]  \mathbf{E}\left(  \mathbf{r},t\right)  =-\mu
_{0}\partial_{t}^{2}\mathbf{P}^{(2)}\left(  \mathbf{r},t\right)  ,
\label{EwaveEq}%
\end{equation}
from the \textquotedblleft rotational Maxwell's equations\textquotedblright%
\ (the equations involving $\boldsymbol{\nabla}\times\mathbf{E}$ and
$\boldsymbol{\nabla}\times\mathbf{B}$). Imagine that we introduce a wave of
frequency $\omega_{0}$ in the medium. As we have shown, this wave induces a
nonlinear polarization $P_{j}^{(2)}\left(  \mathbf{r},2\omega_{0}\right)
=\varepsilon_{0}\chi_{jkl}^{(2)}(2\omega_{0};\omega_{0},\omega_{0}%
)E_{k}\left(  \mathbf{r},\omega_{0}\right)  E_{l}\left(  \mathbf{r},\omega
_{0}\right)  $ in the medium oscillating at frequency $2\omega_{0}$, and hence
the wave equation for the field component oscillating at that frequency will
be%
\begin{equation}
\left[  \boldsymbol{\nabla}^{2}+4\omega_{0}^{2}\mu_{0}\overleftrightarrow
{\varepsilon}\right]  E_{j}\left(  \mathbf{r},2\omega_{0}\right)  =4\omega
_{0}^{2}\mu_{0}\tilde{P}_{j}^{(2)}\left(  \mathbf{r},2\omega_{0}\right)  ;
\end{equation}
this clearly shows that $\mathbf{\tilde{P}}^{(2)}\left(  \mathbf{r}%
,2\omega_{0}\right)  $ acts as a source for a new wave which will oscillate at
frequency $2\omega_{0}$. This phenomenon is known as \textit{second harmonic
generation}. It is a particular case of the phenomenon known as
\textit{frequency sum generation}, in which the wave oscillating at frequency
$2\omega_{0}$ is generated from two waves of frequencies $\omega_{1}$ and
$\omega_{2}$ satisfying $\omega_{1}+\omega_{2}=2\omega_{0}$. All these effects
are known as \textit{up--conversion}, as the generated wave has a frequency
larger than the initial ones.

One can think about the opposite kind of process, \textit{down--conversion},
in which a wave of frequency $\omega_{0}$ is generated from an initial wave
oscillating at frequency $2\omega_{0}$. Note however, that this process is
mediated by the nonlinear polarization term $\tilde{P}_{j}^{(2)}\left(
\mathbf{r},\omega_{0}\right)  =\varepsilon_{0}\chi_{jkl}^{(2)}(\omega
_{0};2\omega_{0},-\omega_{0})\tilde{E}_{k}\left(  \mathbf{r},2\omega
_{0}\right)  \tilde{E}_{l}^{\ast}\left(  \mathbf{r},\omega_{0}\right)  $,
which is zero if the $\omega_{0}$ wave $\mathbf{\tilde{E}}\left(
\mathbf{r},\omega_{0}\right)  $ is zero initially. Hence, the phenomenon of
down--conversion cannot happen without some kind of initial seed. When this
initial seed is not an externally injected wave, but rather some initial
fluctuations in the system (e.g., some small amount of thermal photons), the
process is known as \textit{spontaneous parametric down--conversion} (the term
\textquotedblleft parametric\textquotedblright\ has a rather obscure origin
that we won't worry about).

In general, the down--conversion process might not be frequency degenerate,
and hence two waves oscillating at frequencies $\omega_{1}$ and $\omega_{2}$
are spontaneously generated from the initial wave of frequency $2\omega
_{0}=\omega_{1}+\omega_{2}$. In a down--conversion context, it is customary to
call \textit{signal} and \textit{idler} to these two waves\footnote{Which one
is the signal or idler is not relevant. This nomenclature comes from the case
in which one of the waves, say that of frequency $\omega_{1}$, is injected as
a seed, and hence the other is generated through the
\textit{frequency--difference generation }process $2\omega_{0}-\omega
_{1}\rightarrow\omega_{2}$. In this case, the injected wave is the
\textit{signal}, while the nonlinearly generated one is the \textit{idler}.},
and \textit{pump} to the $2\omega_{0}$ wave. We will stick with this
nomenclature, as most of the thesis (and certainly this chapter) deals with
down--conversion processes.

As we will see when we develop their quantum theory, all this processes
(collectively known as \textit{three--wave mixing} \textit{processes}) can be
intuitively understood as a conversion of a pair of photons of frequencies
$\{\omega_{\mathrm{s}},\omega_{\mathrm{i}}\}$ into one photon of frequency
$\omega_{\mathrm{p}}=\omega_{\mathrm{s}}+\omega_{\mathrm{i}}$, or vice versa.
The dielectric medium acts just as a host, it doesn't take any active
involvement in the conversion process; hence, the energy and momentum of the
photons must be conserved. Conservation of the energy is granted by the
delta--function appearing in (\ref{Chi(2)}). On the other hand, conservation
of the linear momentum requires one further condition, namely%
\begin{equation}
n_{\mathrm{p}}k_{\mathrm{p}}=n_{\mathrm{s}}k_{\mathrm{s}}+n_{\mathrm{i}%
}k_{\mathrm{i}},
\end{equation}
being $n_{j}$ the refractive index of the medium for the corresponding wave
and $k_{j}=\omega_{j}/c$ its wave vector (in the following we assume that the
three waves propagate in the same direction, what is known as \textit{collinear
three--wave mixing}, but everything we say applies for any other scheme). This
condition is known as the \textit{phase--matching condition}, and we will show
how it appears rigorously in the quantum description of the process (next section).

Consider a situation in which the three waves (pump, signal, and idler) have
the same polarization, or in which the nonlinear medium does not show
birefringence\footnote{Actually, there is only one example of a crystalline
class which does not have an inversion center (and hence, has second--order
susceptibility), but does not show birefringence: the class \={4}2m
\cite{Boyd03book}, to which for example Gallium arsenide belongs.}, so that
all the different polarization components of the field feel the same
refractive index; in such cases the phase--matching condition can be recasted
as%
\begin{equation}
\frac{\omega_{\mathrm{i}}}{\omega_{\mathrm{s}}}=-\frac{n(\omega_{\mathrm{p}%
})-n(\omega_{\mathrm{s}})}{n(\omega_{\mathrm{p}})-n(\omega_{\mathrm{i}})}.
\label{IsoCond}%
\end{equation}
The right hand side of this expression is always negative because, as follows from the normal dispersion commented above, $n(\omega_{\mathrm{p}})>\{n(\omega_{\mathrm{s}}),n(\omega_{\mathrm{i}})\}$; hence in order to
conserve linear momentum the three waves involved in the down--conversion
process cannot have the same polarization, and the medium must show birefringence.

As commented above, we will work with uniaxial media, although the
phase--matching condition can also be satisfied in biaxial media too. For
definiteness, and without loss of generalization, let us assume that
$\omega_{\mathrm{s}}>\omega_{\mathrm{i}}$ (we choose the higher frequency wave
of the signal--idler pair as the signal), and that the extraordinary
refractive index is larger than the ordinary one \footnote{In general, an
uniaxial crystal is called \textit{positive} if $n_{\mathrm{e}}>n_{\mathrm{o}%
}$, and \textit{negative} in the opposite case. Although we assume to work
with positive uniaxial crystals, everything that we say is readily generalized
to the negative case.}, that is, $n_{\mathrm{e}}(\omega)>n_{\mathrm{o}}%
(\omega)$; in this case the pump (signal) wave has to be polarized along the
ordinary (extraordinary) axis so that the numerator on the right hand side of
(\ref{IsoCond}) can become positive, that is, the phase--matching condition becomes%
\begin{equation}
\frac{n_{\mathrm{e}}(\omega_{\mathrm{s}})-n_{\mathrm{o}}(\omega_{\mathrm{p}}%
)}{n_{\mathrm{o}}(\omega_{\mathrm{p}})-n_{\mathrm{i}}}=\frac{\omega
_{\mathrm{i}}}{\omega_{\mathrm{s}}}\text{.} \label{NDcond}%
\end{equation}
We can distinguish then two types of processes. In a \textit{type I} process
both signal and idler have the same polarization (extraordinary with our
conventions), while in a \textit{type II} process they have orthogonal
polarizations (extraordinary the signal and ordinary the idler with our conventions).

Note that when signal and idler are frequency degenerate, the phase matching
conditions are reduced to
\begin{subequations}
\label{Dcond}%
\begin{align}
\text{\lbrack type I] \ \ \ \ }n_{\mathrm{e}}(\omega_{0})  &  =n_{\mathrm{o}%
}(2\omega_{0}),\label{DIcond}\\
\text{\lbrack type II] \ \ \ \ }n_{\mathrm{e}}(\omega_{0})  &  =2n_{\mathrm{o}%
}(2\omega_{0})-n_{\mathrm{o}}(\omega_{0}).
\end{align}
Hence, we see that the conditions for frequency degenerate processes are quite
critical, they require a fine tuning of the crystal parameters (as well as a
proper stabilization of the cavity resonances in the case of working inside an
optical resonator, as we will).

Note finally that it may seem that the frequencies of the signal and idler
waves are completely fixed by the frequency we choose for the pump wave and
the refractive indices of the crystal, so that in order to change to different
down--conversion frequencies we may need to change either the pump frequency,
or the crystal we are working with. Fortunately, the conditions (\ref{NDcond})
and (\ref{Dcond}) are much more flexible than they look, because the
refractive indices of the crystal depend on the temperature in a known (or at
least tabulated) way. Hence, even for a fixed pump frequency, one can still
achieve the phase--matching condition for a desired signal--idler frequency
pair by tuning the refractive indices of the crystal\footnote{Another well
known method to tune the phase--matching condition is by making a rotation
$\theta$ of the crystal around the ordinary direction $\mathbf{e}_{\mathrm{o}%
}$. Under these circumstances, the polarization component of the field
parallel to this direction will still feel a refractive index $n_{\mathrm{o}}%
$. However, the polarization component orthogonal to this direction won't be
parallel to the extraordinary direction anymore, and hence it will feel an
effective refractive index $n_{\mathrm{e}}^{\prime}$ given by
\cite{Boyd03book}%
\[
\frac{1}{n_{\mathrm{e}}^{\prime}}=\frac{\cos^{2}\theta}{n_{\mathrm{e}}}%
+\frac{\sin^{2}\theta}{n_{\mathrm{o}}}%
\]
For reasons that will become obvious along the next chapters, we will assume
during most of this thesis that no \textit{angular tuning} is needed.} via
\textit{temperature tuning}.%

\section[Quantum model for a general OPO]{Quantum model for a general optical parametric oscillator\label{GeneralOPOModel}}

In order to optimize the frequency conversion process, it is customary to
embed the nonlinear crystal in an optical cavity, so that, e.g. for
down--conversion, if one pump photon is not transformed into a signal--idler
pair when it first crosses the crystal, it is reflected back in order to have
one more chance to be down--converted. This system consisting of a $\chi
^{(2)}$--crystal inside a resonator is known as \textit{optical parametric
oscillator} (OPO).

From the previous chapters we know how to treat quantum mechanically an
optical cavity. The only question left in order to analyze OPOs within that
framework is which intracavity Hamiltonian $\hat{H}_{\mathrm{c}}$ (in the
notation of Chapter \ref{OpenSystems}) accounts for the three--wave mixing
processes occurring inside the nonlinear crystal. An intuitive way to proceed
is as follows.

The energy associated to an electromagnetic field interacting with an electric
(point--like) dipole $\mathbf{d}$ is given by $-\mathbf{d\cdot E}%
(\mathbf{r}_{\mathrm{d}})$, being $\mathbf{r}_{\mathrm{d}}$ the position of
the dipole. Now, assume for a moment that the cavity field and the nonlinear
polarization appearing in the crystal are independent; then, one would assign
the Hamiltonian%
\end{subequations}
\begin{equation}
\hat{H}_{\mathrm{c}}=-\int_{\mathrm{crystal}}d^{3}\mathbf{r\hat{P}}%
^{(2)}\left(  \mathbf{r}\right)  \cdot\mathbf{\hat{E}}\left(  \mathbf{r}%
\right)  , \label{PE}%
\end{equation}
to account for their interaction. Of course, the field and the nonlinear
polarization are not independent at all (the former is indeed induced by the
later!), but keeping in mind that the nonlinear polarization is rather small,
the three--wave mixing Hamiltonian will be small compared to the free
Hamiltonian, and hence this rude approximation can be justified as the first
term of a perturbative expansion of the true Hamiltonian\footnote{Note that
this is not true for the linear part of the induced polarization, that is, the
refractive index, which is not small at all, and hence one needs to take it
into account when developing the quantum theory of the electromagnetic field
of the corresponding system. Note also that this is exactly what we did when
quantizing the field inside a cavity containing a dielectric medium.}. We
would like to stress that, even though rigorous canonical quantization
procedures are known for the Maxwell equations in nonlinear dielectric media
\cite{Hillery09,Drummond04book,Drummond01}, such intuitive perturbative
expansion leading to (\ref{PE}) has not been performed to our knowledge, and
hence, rigorously, (\ref{PE}) cannot be regarded other than as a
phenomenological, albeit very successful model.

\begin{figure}
[t]
\begin{center}
\includegraphics[
height=3.7438in,
width=3.9885in
]%
{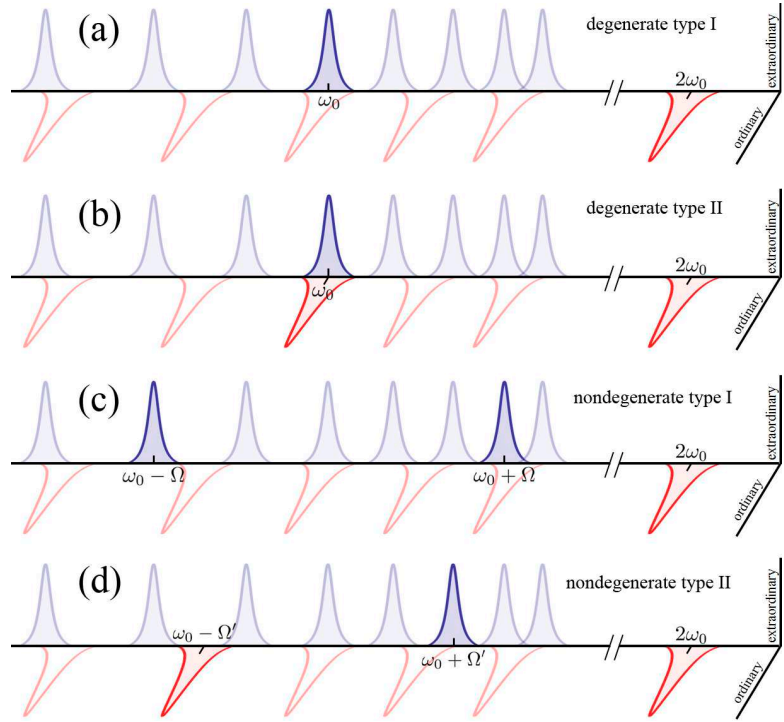}%
\caption{Three--wave mixing processes inside an optical cavity. In this
example the structure of the resonances inside the cavity is such that the 4
types of processes are can appear, as they are all energy conserving. Hence,
when the pump field has frequency $2\omega_{0}$ signal and idler can be
indistinguishable (a), or distinguishable in polarization (b), in frequency
(c), or in them both (d). }%
\label{fOPO2}%
\end{center}
\end{figure}

Let us evaluate now this Hamiltonian for the different three--wave mixing
processes that can occur inside the resonator. Consider the cavity resonances
depicted in Figure \ref{fOPO2}. In the previous section we showed that one can
distinguish between four types of three--wave mixing processes attending to
the degeneracies that signal and idler may have in frequency and polarization.
If we choose some cavity resonance $2\omega_{0}$ as the pump mode (with
ordinary polarization according to the conventions of the previous section),
then one can operate the OPO in four different regimes: \textit{degenerate
type I OPO}, in which the signal and idler are indistinguishable and we talk
then about a single mode, the signal mode, having frequency $\omega_{0}$ and
extraordinary polarization (Figure \ref{fOPO2}a); \textit{degenerate type II
OPO}, in which the signal and idler have orthogonal polarizations but the same
frequency (Figure \ref{fOPO2}b); \textit{non-degenerate type I}, in which the
signal and idler have the same polarization but correspond to two opposite
frequency sidebands around $\omega_{0}$ (Figure \ref{fOPO2}c); and
\textit{non-degenerate type II}, in which the signal and idler are
distinguishable both in frequency and polarization (Figure \ref{fOPO2}d).

We can evaluate the Hamiltonian (\ref{PE}) for the four processes altogether.
To do so, let us write the electric field as%
\begin{equation}
\mathbf{\hat{E}}\left(  \mathbf{r}\right)  =\mathbf{\hat{E}}_{\mathrm{p}%
}^{(+)}\left(  \mathbf{r}\right)  +(1-\delta_{\mathrm{si}}/2)\mathbf{\hat{E}%
}_{\mathrm{s}}^{(+)}\left(  \mathbf{r}\right)  +(1-\delta_{\mathrm{si}%
}/2)\mathbf{\hat{E}}_{\mathrm{i}}^{(+)}\left(  \mathbf{r}\right)
+\mathrm{H.c.},
\end{equation}
where $\mathbf{\hat{E}}_{\mathrm{p,s,i}}^{(+)}\left(  \mathbf{r}\right)  $ are
the (Schr\"{o}dinger picture) electric fields associated to the pump, signal,
and idler cavity modes involved in the three--wave mixing process, and the
factors $(1-\delta_{\mathrm{si}}/2)$ are introduced to include also the case
in which signal and idler are indistinguishable (in which case $\delta
_{\mathrm{si}}=1$), that is, the degenerate type I process.

The pump field has always $\omega_{\mathrm{p}}=2\omega_{0}$ and ordinary
polarization (with the conventions of the previous section); on the other
hand, the signal field has always extraordinary polarization, but its
frequency is different for the degenerate (Figures \ref{fOPO2}a and
\ref{fOPO2}b) and non-degenerate (Figures \ref{fOPO2}c and \ref{fOPO2}d)
schemes; finally, the polarization and frequency of the idler field depend on
the particular three--wave mixing scheme. Hence, based on (\ref{Ecav}) we can
write the different frequency components of the field as
\begin{subequations}
\label{psiFields}%
\begin{align}
\mathbf{\hat{E}}_{\mathrm{p}}^{(+)}\left(  \mathbf{r}\right)   &
=\mathrm{i}\sqrt{\frac{\hbar\omega_{\mathrm{p}}}{4\varepsilon_{0}%
n_{\mathrm{p}}L_{\mathrm{p}}}}\mathbf{e}_{\mathrm{o}}\hat{a}_{\mathrm{p}%
}\left[  G(k_{\mathrm{p}};\mathbf{r}_{\perp},z)e^{\mathrm{i}n_{\mathrm{p}%
}k_{\mathrm{p}}z}+\mathrm{c.c.}\right],\\
\mathbf{\hat{E}}_{\mathrm{s}}^{(+)}\left(  \mathbf{r}\right)   &
=\mathrm{i}\sqrt{\frac{\hbar\omega_{\mathrm{s}}}{4\varepsilon_{0}%
n_{\mathrm{s}}L_{\mathrm{s}}}}\mathbf{e}_{\mathrm{e}}\sum_{\mathbf{n}}\hat
{a}_{\mathrm{s},\mathbf{n}}\left[  T_{\mathbf{n}}(k_{\mathrm{s}}%
;\mathbf{r}_{\perp},z)e^{\mathrm{i}n_{\mathrm{s}}k_{\mathrm{s}}z}%
+\mathrm{c.c.}\right],\\
\mathbf{\hat{E}}_{\mathrm{i}}^{(+)}\left(  \mathbf{r}\right)   &
=\mathrm{i}\sqrt{\frac{\hbar\omega_{\mathrm{i}}}{4\varepsilon_{0}%
n_{\mathrm{i}}L_{\mathrm{i}}}}\boldsymbol{\varepsilon}_{\mathrm{i}}%
\sum_{\mathbf{n}}\hat{a}_{\mathrm{i},\mathbf{n}}\left[  T_{\mathbf{n}%
}(k_{\mathrm{i}};\mathbf{r}_{\perp},z)e^{\mathrm{i}n_{\mathrm{i}}%
k_{\mathrm{i}}z}+\mathrm{c.c.}\right],
\end{align}
where $L_{j}$ is the optical length of the cavity evaluated for the refractive
index seen by the corresponding mode, and the polarization of the idler mode
$\boldsymbol{\varepsilon}_{\mathrm{i}}$ is left unspecified. Note that we have
assumed that at the cavity resonance $2\omega_{0}$ there exists only a
TEM$_{00}$ mode, which is denoted by $G(k;\mathbf{r}_{\perp},z)$, while
several transverse modes $\{T_{\mathbf{n}}(k;\mathbf{r}_{\perp}%
,z)\}_{\mathbf{n}}$ can resonate at the signal and idler frequencies (the same
for both, as this is the situation that we will find along this thesis). We
denote by $\hat{a}_{\mathrm{p}}$, $\hat{a}_{\mathrm{s},\mathbf{n}}$, and
$\hat{a}_{\mathrm{i},\mathbf{n}}$, the intracavity annihilation operators for
pump, signal, and idler photons, respectively.

As for the nonlinear polarization, we can also write it as%
\end{subequations}
\begin{equation}
\mathbf{\hat{P}}^{(2)}\left(  \mathbf{r},t\right)  =\mathbf{\hat{P}%
}_{\mathrm{p}}^{(+)}\left(  \mathbf{r}\right)  +(1-\delta_{\mathrm{si}%
}/2)\mathbf{\hat{P}}_{\mathrm{s}}^{(+)}\left(  \mathbf{r}\right)
+(1-\delta_{\mathrm{si}}/2)\mathbf{\hat{P}}_{\mathrm{i}}^{(+)}\left(
\mathbf{r}\right)  +\mathrm{H.c.},
\end{equation}
where, based on the previous section, we have\footnote{The factors 2 come
because, for example $\mathbf{\hat{P}}_{\mathrm{s}}^{(+)}\left(
\mathbf{r}\right)  $, receives contributions both from $\chi_{jlm}%
^{(2)}(\omega_{\mathrm{s}};\omega_{\mathrm{p}},-\omega_{\mathrm{i}})$ and
$\chi_{jlm}^{(2)}(\omega_{\mathrm{s}};-\omega_{\mathrm{i}},\omega_{\mathrm{p}%
})$; this contributions are easily shown to be the same by making use of full
permutation symmetry.}%
\begin{subequations}
\begin{align}
\mathbf{\hat{P}}_{\mathrm{p}}^{(+)}\left(  \mathbf{r}\right)   &
=(2-\delta_{\omega_{\mathrm{s}},\omega_{\mathrm{i}}})\varepsilon_{0}\chi
_{jlm}^{(2)}(\omega_{\mathrm{p}};\omega_{\mathrm{s}},\omega_{\mathrm{i}%
})\mathbf{e}_{j}[\mathbf{\hat{E}}_{\mathrm{s}}^{(+)}\left(  \mathbf{r}\right)
]_{l}[\mathbf{\hat{E}}_{\mathrm{i}}^{(+)}\left(  \mathbf{r}\right)  ]_{m},\\
\mathbf{\hat{P}}_{\mathrm{s}}^{(+)}\left(  \mathbf{r}\right)   &
=2\varepsilon_{0}\chi_{jlm}^{(2)}(\omega_{\mathrm{s}};\omega_{\mathrm{p}%
},-\omega_{\mathrm{i}})\mathbf{e}_{j}[\mathbf{\hat{E}}_{\mathrm{p}}%
^{(+)}\left(  \mathbf{r}\right)  ]_{l}[\mathbf{\hat{E}}_{\mathrm{i}}%
^{(-)}\left(  \mathbf{r}\right)  ]_{m},\\
\mathbf{\hat{P}}_{\mathrm{i}}^{(+)}\left(  \mathbf{r}\right)   &
=2\varepsilon_{0}\chi_{jlm}^{(2)}(\omega_{\mathrm{i}};\omega_{\mathrm{p}%
},-\omega_{\mathrm{s}})\mathbf{e}_{j}[\mathbf{\hat{E}}_{\mathrm{p}}%
^{(+)}\left(  \mathbf{r}\right)  ]_{l}[\mathbf{\hat{E}}_{\mathrm{s}}%
^{(-)}\left(  \mathbf{r}\right)  ]_{m},
\end{align}
or in terms of the boson operators%
\end{subequations}
\begin{subequations}
\begin{align}
\mathbf{\hat{P}}_{\mathrm{p}}^{(+)}\left(  \mathbf{r}\right)   &
=-(2-\delta_{\omega_{\mathrm{s}},\omega_{\mathrm{i}}})\sqrt{\frac{\hbar
^{2}\omega_{\mathrm{s}}\omega_{\mathrm{i}}}{16n_{\mathrm{s}}n_{\mathrm{i}%
}L_{\mathrm{s}}L_{\mathrm{i}}}}\varepsilon_{0}\chi_{j\mathrm{e}m}^{(2)}%
(\omega_{\mathrm{p}};\omega_{\mathrm{s}},\omega_{\mathrm{i}})\mathbf{e}%
_{j}\left[  \boldsymbol{\varepsilon}_{\mathrm{i}}\right]  _{m}\sum
_{\mathbf{nm}}\hat{a}_{\mathrm{s},\mathbf{n}}\hat{a}_{\mathrm{i},\mathbf{m}%
}
\\
&  \times\left[  T_{\mathbf{n}}(k_{\mathrm{s}};\mathbf{r}_{\perp
},z)T_{\mathbf{m}}(k_{\mathrm{i}};\mathbf{r}_{\perp},z)e^{\mathrm{i}\left(
n_{\mathrm{s}}k_{\mathrm{s}}+n_{\mathrm{i}}k_{\mathrm{i}}\right)  z}+T_{\mathbf{n}}(k_{\mathrm{s}};\mathbf{r}_{\perp},z)T_{\mathbf{m}%
}^{\ast}(k_{\mathrm{i}};\mathbf{r}_{\perp},z)e^{\mathrm{i}\left(
n_{\mathrm{s}}k_{\mathrm{s}}-n_{\mathrm{i}}k_{\mathrm{i}}\right)
z}+\mathrm{c.c.}\right]  ,\nonumber
\\
\mathbf{\hat{P}}_{\mathrm{s}}^{(+)}\left(  \mathbf{r}\right)   &
=2\sqrt{\frac{\hbar^{2}\omega_{\mathrm{p}}\omega_{\mathrm{i}}}{16n_{\mathrm{p}%
}n_{\mathrm{i}}L_{\mathrm{p}}L_{\mathrm{i}}}}\varepsilon_{0}\chi
_{j\mathrm{o}m}^{(2)}(\omega_{\mathrm{s}};\omega_{\mathrm{p}},-\omega
_{\mathrm{i}})\mathbf{e}_{j}\left[  \boldsymbol{\varepsilon}_{\mathrm{i}%
}\right]  _{m}\sum_{\mathbf{n}}\hat{a}_{\mathrm{p}}\hat{a}_{\mathrm{i}%
,\mathbf{n}}^{\dagger}
\\
&  \times\left[  G(k_{\mathrm{p}};\mathbf{r}_{\perp},z)T_{\mathbf{n}}^{\ast
}(k_{\mathrm{i}};\mathbf{r}_{\perp},z)e^{\mathrm{i}\left(  n_{\mathrm{p}%
}k_{\mathrm{p}}-n_{\mathrm{i}}k_{\mathrm{i}}\right)  z}+G(k_{\mathrm{p}};\mathbf{r}_{\perp},z)T_{\mathbf{n}}%
(k_{\mathrm{i}};\mathbf{r}_{\perp},z)e^{\mathrm{i}\left(  n_{\mathrm{p}%
}k_{\mathrm{p}}+n_{\mathrm{i}}k_{\mathrm{i}}\right)  z}+\mathrm{c.c.}\right]
,\nonumber
\\
\mathbf{\hat{P}}_{\mathrm{i}}^{(+)}\left(  \mathbf{r}\right)   &
=2\sqrt{\frac{\hbar^{2}\omega_{\mathrm{p}}\omega_{\mathrm{s}}}{16n_{\mathrm{p}%
}n_{\mathrm{s}}L_{\mathrm{p}}L_{\mathrm{s}}}}\varepsilon_{0}\chi
_{j\mathrm{oe}}^{(2)}(\omega_{\mathrm{i}};\omega_{\mathrm{p}},-\omega
_{\mathrm{s}})\mathbf{e}_{j}\sum_{\mathbf{n}}\hat{a}_{\mathrm{p}}\hat
{a}_{\mathrm{s},\mathbf{n}}^{\dagger}
\\
&  \times\left[  G(k_{\mathrm{p}};\mathbf{r}_{\perp},z)T_{\mathbf{n}}^{\ast
}(k_{\mathrm{s}};\mathbf{r}_{\perp},z)e^{\mathrm{i}\left(  n_{\mathrm{p}%
}k_{\mathrm{p}}-n_{\mathrm{s}}k_{\mathrm{s}}\right)  z}+G(k_{\mathrm{p}};\mathbf{r}_{\perp},z)T_{\mathbf{n}}%
(k_{\mathrm{s}};\mathbf{r}_{\perp},z)e^{\mathrm{i}\left(  n_{\mathrm{p}%
}k_{\mathrm{p}}+n_{\mathrm{s}}k_{\mathrm{s}}\right)  z}+\mathrm{c.c.}\right]
.\nonumber
\end{align}
Neglecting counter--rotating terms such as $\mathbf{\hat{P}}_{j}^{(+)}\left(
\mathbf{r}\right)  \cdot\mathbf{\hat{E}}_{l}^{(+)}\left(  \mathbf{r}\right)  $
or $\mathbf{\hat{P}}_{\mathrm{p}}^{(-)}\left(  \mathbf{r}\right)
\cdot\mathbf{\hat{E}}_{\mathrm{s}}^{(+)}\left(  \mathbf{r}\right)  $, and
taking into account that by full permutation symmetry%
\end{subequations}
\begin{equation}
\chi_{\mathrm{oe}j}^{(2)}(\omega_{\mathrm{p}};\omega_{\mathrm{s}}%
,\omega_{\mathrm{i}})=\chi_{\mathrm{eo}j}^{(2)}(-\omega_{\mathrm{s}}%
;-\omega_{\mathrm{p}},\omega_{\mathrm{i}})=\chi_{j\mathrm{oe}}^{(2)}%
(-\omega_{\mathrm{i}};-\omega_{\mathrm{p}},\omega_{\mathrm{s}}),
\end{equation}
it is not difficult to write the Hamiltonian as%
\begin{equation}
\hat{H}_{\mathrm{c}}=i\hbar\sum_{\mathbf{nm}}\chi_{\mathbf{nm}}\hat
{a}_{\mathrm{p}}\hat{a}_{\mathrm{s},\mathbf{n}}^{\dagger}\hat{a}%
_{\mathrm{i},\mathbf{m}}^{\dagger}+\mathrm{H.c.},
\end{equation}
where the coupling parameters are given by%
\begin{equation}
\chi_{\mathbf{nm}}=\chi_{\mathrm{oe}j}^{(2)}(\omega_{\mathrm{p}}%
;\omega_{\mathrm{s}},\omega_{\mathrm{i}})\left[  \boldsymbol{\varepsilon
}_{\mathrm{i}}\right]  _{j}\sqrt{\frac{\hbar\omega_{\mathrm{p}}\omega
_{\mathrm{s}}\omega_{\mathrm{i}}}{16\varepsilon_{0}n_{\mathrm{p}}%
n_{\mathrm{s}}n_{\mathrm{i}}L_{\mathrm{p}}L_{\mathrm{s}}L_{\mathrm{i}}}%
}(3-\delta_{\mathrm{si}}-\delta_{\omega_{\mathrm{s}},\omega_{\mathrm{i}}%
}/2)I_{\mathbf{nm}},
\end{equation}
and we have defined the \textit{three--mode overlapping integral}%
\begin{align}
I_{\mathbf{nm}}=\int_{\mathrm{crystal}}d^{3}\mathbf{r}\left[  G(k_{\mathrm{p}%
};\mathbf{r}_{\perp},z)T_{\mathbf{n}}^{\ast}(k_{\mathrm{s}};\mathbf{r}_{\perp
},z)T_{\mathbf{m}}^{\ast}(k_{\mathrm{i}};\mathbf{r}_{\perp},z)e^{\mathrm{i}%
\Delta kz}+G(k_{\mathrm{p}};\mathbf{r}_{\perp},z)T_{\mathbf{n}}^{\ast}(k_{\mathrm{s}%
};\mathbf{r}_{\perp},z)T_{\mathbf{m}}(k_{\mathrm{i}};\mathbf{r}_{\perp
},z)e^{\mathrm{i}\Delta k_{1}z}\right.
\\
\left.+G^{\ast}(k_{\mathrm{p}};\mathbf{r}_{\perp},z)T_{\mathbf{n}}^{\ast
}(k_{\mathrm{s}};\mathbf{r}_{\perp},z)T_{\mathbf{m}}(k_{\mathrm{i}}%
;\mathbf{r}_{\perp},z)e^{\mathrm{i}\Delta k_{2}z}+G^{\ast}(k_{\mathrm{p}};\mathbf{r}_{\perp},z)T_{\mathbf{n}}^{\ast
}(k_{\mathrm{s}};\mathbf{r}_{\perp},z)T_{\mathbf{m}}^{\ast}(k_{\mathrm{i}%
};\mathbf{r}_{\perp},z)e^{\mathrm{i}\Delta k_{3}z}\right]   &  +\mathrm{c.c.}%
,\nonumber
\end{align}
with
\begin{align}
\Delta k &  =n_{\mathrm{p}}k_{\mathrm{p}}-n_{\mathrm{s}}k_{\mathrm{s}%
}-n_{\mathrm{i}}k_{\mathrm{i}},\text{ \ \ \ \ \ \ }\Delta k_{1}=n_{\mathrm{p}%
}k_{\mathrm{p}}-n_{\mathrm{s}}k_{\mathrm{s}}+n_{\mathrm{i}}k_{\mathrm{i}},\\
\Delta k_{2} &  =-n_{\mathrm{p}}k_{\mathrm{p}}-n_{\mathrm{s}}k_{\mathrm{s}%
}+n_{\mathrm{i}}k_{\mathrm{i}},\text{ \ \ \ \ }\Delta k_{3}=-n_{\mathrm{p}%
}k_{\mathrm{p}}-n_{\mathrm{s}}k_{\mathrm{s}}-n_{\mathrm{i}}k_{\mathrm{i}%
}.\nonumber
\end{align}
Just as we stated in the previous section, this Hamiltonian shows how the
three--wave mixing process is described from a quantum viewpoint as the
annihilation of a pump photon and the simultaneous creation of a signal--idler
pair, or vice versa. In order for the process to be efficient, we need the
three--mode--overlapping integral to be as large as possible; let us now
simplify the model a little bit in order to understand under which conditions
$I_{\mathbf{nm}}$ is maximized.

Assume that the center of the crystal coincides with the cavity's waist plane,
and that its length $l_{\mathrm{c}}$ is much smaller than the Rayleigh length
of the resonator, that is, $l_{\mathrm{c}}\ll z_{R}$, see after
(\ref{FreePropLaws}). Under these circumstances, the longitudinal variation of
the transverse modes can be neglected inside the crystal, see
(\ref{PropModePar}), and therefore all the longitudinal integrals in
$I_{\mathbf{nm}}$ are of the form%
\begin{equation}
\int_{-l_{\mathrm{c}}/2}^{l_{\mathrm{c}}/2}dze^{\mathrm{i}\Delta k_{j}%
z}=l_{\mathrm{c}}\mathrm{\operatorname{sinc}}\left(  \frac{\Delta
k_{j}l_{\mathrm{c}}}{2}\right)  \text{;}%
\end{equation}
hence, we see that in order to maximize the conversion process, some of the
$\Delta k_{j}$'s must satisfy $\left\vert \Delta k_{j}\right\vert \ll
l_{\mathrm{c}}^{-1}$. On the other hand, $\Delta k=0$ is precisely the phase
matching condition that we introduced in the previous section using linear
momentum conservation arguments, and hence, we assume now that $\Delta k\ll
l_{\mathrm{c}}^{-1}$; under these conditions, it is completely trivial to show
that the rest of $\Delta k_{j}$'s are all well above $l_{\mathrm{c}}^{-1}$, so
that their contribution to $I_{\mathbf{nm}}$ can be neglected.

The Hamiltonian can be even more simplified if we choose the Laguerre--Gauss
basis for the transverse modes (\ref{Lpl}), which are written in the cavity
waist as%
\begin{equation}
L_{pl}\left(  k;\mathbf{r}_{\perp},z=0\right)  =\mathcal{R}_{p}^{|l|}%
(k;r)\exp\left[  \mathrm{i}l\phi\right]  ,
\end{equation}
with%
\begin{equation}
\mathcal{R}_{p}^{|l|}(k;r)=\sqrt{\frac{2p!}{\pi\left(  p+|l|\right)  !}}%
\frac{1}{w}\left(  \frac{\sqrt{2}r}{w}\right)  ^{|l|}L_{p}^{|l|}\left(
\frac{2r^{2}}{w^{2}}\right)  \exp\left(  -\frac{r^{2}}{w^2}\right)  ,
\end{equation}
being $w$ the spot size of the modes at the cavity waist. The interesting
thing about these modes is that they are eigenfunctions of the orbital angular
momentum operator, that is, $-i\partial_{\phi}L_{pl}\left(  k;\mathbf{r}%
_{\perp},z\right)  =lL_{pl}\left(  k;\mathbf{r}_{\perp},z\right)  $, and hence
conservation of this quantity in the conversion process should appear
explicitly in this basis. In particular, taking into account the phase
matching condition $\Delta k\ll l_{\mathrm{c}}^{-1}$ that we have already
selected, in the Laguerre--Gauss basis the three--mode overlapping integral
reads%
\begin{equation}
I_{pl,p^{\prime}l^{\prime}}=2l_{\mathrm{c}}\delta_{-ll^{\prime}}\int
_{0}^{+\infty}rdrG(k_{\mathrm{p}};r)\mathcal{R}_{p}^{|l|}(k_{\mathrm{s}%
};r)\mathcal{R}_{p^{\prime}}^{|l^{\prime}|}(k_{\mathrm{i}};r),
\end{equation}
and hence the three--wave mixing Hamiltonian takes the final form%
\begin{equation}
\hat{H}_{\mathrm{c}}=\mathrm{i}\hbar\sum_{pp^{\prime},l\geq0}\chi_{pp^{\prime
}}^{l}\hat{a}_{\mathrm{p}}\hat{a}_{\mathrm{s},p,l}^{\dagger}\hat
{a}_{\mathrm{i},p^{\prime},-l}^{\dagger}+\mathrm{H.c.}%
,\label{GenTWMHamiltonian}%
\end{equation}\label{ChiGen}
where the sum extend over all the opposite orbital angular momentum pairs
present at the signal--idler cavity resonances, and%
\begin{equation}
\chi_{pp^{\prime}}^{l} =\frac{3-\delta_{\mathrm{si}}-\delta_{\omega
_{\mathrm{s}},\omega_{\mathrm{i}}}/2}{1+\delta_{l,0}}l_{\mathrm{c}}%
\chi_{\mathrm{oe}j}^{(2)}(\omega_{\mathrm{p}};\omega_{\mathrm{s}}%
,\omega_{\mathrm{i}})\left[  \boldsymbol{\varepsilon}_{\mathrm{i}}\right]
_{j}\sqrt{\frac{\hbar\omega_{\mathrm{p}}\omega_{\mathrm{s}}\omega
_{\mathrm{i}}}{\varepsilon_{0}n_{\mathrm{p}}n_{\mathrm{s}}n_{\mathrm{i}%
}L_{\mathrm{p}}L_{\mathrm{s}}L_{\mathrm{i}}}}\int_{0}^{+\infty}%
rdrG(k_{\mathrm{p}};r)\mathcal{R}_{p}^{l}(k_{\mathrm{s}};r)\mathcal{R}%
_{p^{\prime}}^{l}(k_{\mathrm{i}};r).
\end{equation}

Let us remark at this point that, ideally, we can decide which of the possible
intracavity three--wave mixing processes is selected by operating the OPO in
the corresponding phase matching condition. In most of the thesis we will
assume that this is the case. In practice, however, one can select between
type I and II at will, but additional care must be taken in order to ensure
working with a desired pair of signal--idler frequencies around $\omega_{0}$,
or the degenerate frequency $\omega_{0}$ itself. We shall come back to this
issue in the next chapter.

We have now at our disposal a quantum model for a general OPO. Note that in
the case of indistinguishable signal and idler, $\hat{H}_{\mathrm{c}}$ is
similar to the Hamiltonian that we used to generate squeezed states of the
harmonic oscillator, see (\ref{Ssq}). Similarly, when signal and idler are
distinguishable, the Hamiltonian resembles the one which produced entangled
states (\ref{Sab}). Hence, the OPO seems a perfect candidate for the
generation of squeezed and entangled states of light. The rest of the chapter
is devoted to analyze these properties of the light generated by the OPO.

\section{Squeezing properties of the DOPO\label{DOPO}}

In this section we study the squeezing properties of degenerate type I OPOs
(\textit{DOPO}s). To this purpose, let us work with the simplest transverse
mode configuration, in which we consider only a TEM$_{00}$ mode resonating at
the signal frequency. Squeezing in OPOs was originally predicted by using this
\textit{single--mode DOPO} model \cite{Collet84}\footnote{The story of such
prediction was actually quite intricate \cite{Meystre91book}. The problem was
that the first analyses of squeezing in OPOs were performed not in terms of
the noise spectrum of the field coming out from the cavity, but in terms of
the uncertainties of the intracavity modes, which showed only a 50\% noise
reduction even in the most ideal situation \cite{Milburn81,McNeil83}. Hence,
these initial predictions were a little disappointing, since such small levels
of squeezing couldn't help much for the applications of squeezing proposed at
that time \cite{Yuen78,Shapiro79,Yuen80,Caves81}. The puzzle was finally
solved by Collet and Gardiner \cite{Collet84}, who applied the input--output
formalism previously introduced by Yurke and Denker in electronic circuits
\cite{Yurke84b}, to analyze the field coming out from the parametric
oscillator in terms of its squeezing spectrum.}, and throughout this thesis we
will learn under which conditions it is valid.

The analysis performed in this section is of major importance for this thesis,
as any other OPO configuration will be analyzed with simple adaptations of it.

\subsection{The DOPO within the positive $P$ representation\label{ModelDOPO}}

The single--mode DOPO deals then with two cavity modes only: The pump mode,
which resonates at frequency $\omega_{\mathrm{p}}=2\omega_{0}$ and has
ordinary polarization, and the signal mode, which resonates at frequency
$\omega_{\mathrm{s}}=\omega_{0}$ and has extraordinary polarization (Figure
\ref{fOPO2}a). Both are in a TEM$_{00}$ transverse mode. The pump mode is
driven by an external laser beam of frequency $2\omega_{\mathrm{L}}$ close to
$2\omega_{0}$, while no injection is used for the signal mode; hence, signal
photons are expected to appear from the pump field via spontaneous parametric down--conversion.%

\begin{figure}
[t]
\begin{center}
\includegraphics[
height=3.2007in,
width=4.804in
]%
{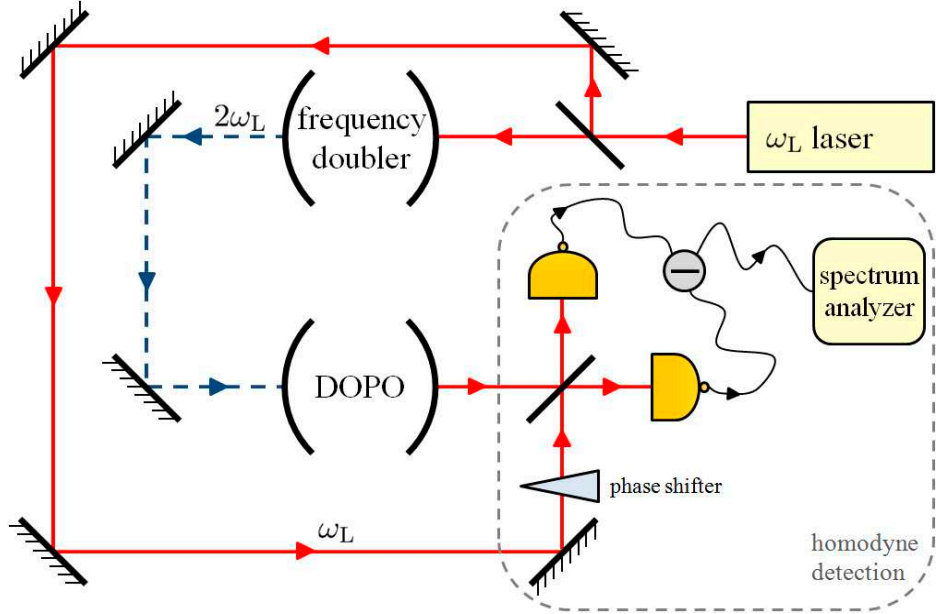}%
\caption{Schematic representation of a typical DOPO experiment. Note that the
same laser is used both as pump injection (after doubling its frequency) and
as local oscillator of the homodyne detection with which the signal field is
studied. This is a very convenient scheme, as the phases of all the involved
fields are more or less locked to the initial laser phase.}%
\label{fOPO3}%
\end{center}
\end{figure}

In Figure \ref{fOPO3} it is shown the basic experimental scheme used to pump
and detect the DOPO. Note that the same laser is used both as the injection
for the cavity (after doubling its frequency) and as the local oscillator
field, which ensures that all the phases of the system are more or less
locked. Based on the analysis of homodyne detection introduced in the previous
chapter, we move then to the interaction picture defined by the transformation
operator%
\begin{equation}
\hat{U}_{\mathrm{0}}=\exp[\hat{H}_{\mathrm{0}}t/\mathrm{i}\hbar]\text{
\ \ \ \ with \ \ \ \ }\hat{H}_{\mathrm{0}}=\hbar\omega_{\mathrm{L}}\left(
2\hat{a}_{\mathrm{p}}^{\dagger}\hat{a}_{\mathrm{p}}+\hat{a}_{\mathrm{s}%
}^{\dagger}\hat{a}_{\mathrm{s}}\right)  ,
\end{equation}
that is, we \textquotedblleft rotate\textquotedblright\ the system to the
local oscillator (or injection) frequency. The state $\hat{\rho}$ of the
intracavity modes is written as $\hat{\rho}_{\mathrm{I}}=\hat{U}_{\mathrm{0}%
}^{\dagger}\hat{\rho}\hat{U}_{\mathrm{0}}$ in this picture, and satisfies the
master equation%
\begin{equation}\label{DOPOmaster}
\frac{d\hat{\rho}_{\mathrm{I}}}{dt}=\left[  -2\mathrm{i}\Delta\hat
{a}_{\mathrm{p}}^{\dagger}\hat{a}_{\mathrm{p}}-\mathrm{i}\Delta\hat
{a}_{\mathrm{s}}^{\dagger}\hat{a}_{\mathrm{s}}+\frac{\chi}{2}\hat
{a}_{\mathrm{p}}\hat{a}_{\mathrm{s}}^{\dagger2}+\mathcal{E}_{\mathrm{p}%
}a_{\mathrm{p}}^{\dagger}+\mathrm{H.c.},\hat{\rho}_{\mathrm{I}}\right]
+\sum_{j=\mathrm{p,s}}\gamma_{j}(2\hat{a}_{j}\hat{\rho}_{\mathrm{I}}\hat
{a}_{j}^{\dagger}-\hat{a}_{j}^{\dagger}\hat{a}_{j}\hat{\rho}_{\mathrm{I}}%
-\hat{\rho}_{\mathrm{I}}\hat{a}_{j}^{\dagger}\hat{a}_{j}),
\end{equation}
where $\Delta=\omega_{0}-\omega_{\mathrm{L}}$ is the detuning of the laser
field with respect to the cavity resonance. We take $\mathcal{E}_{\mathrm{p}}$
as a positive real, which is as taking the phase of the injected laser beam as
the reference for any other phase of the system. The nonlinear coupling
parameter $\chi$ is found from (\ref{ChiGen}) particularized to the
configuration that we are treating; taking into account that%
\begin{equation}
\int_{0}^{+\infty}rdrG(k_{\mathrm{p}};r)G^{2}(k_{\mathrm{s}};r)=\left(
\frac{2}{\pi}\right)  ^{3/2}\int_{0}^{+\infty}dr\frac{r}{w_{\mathrm{p}%
}w_{\mathrm{s}}^{2}}\exp\left[  -\left(  \frac{1}{w_{\mathrm{p}}^{2}}+\frac
{2}{w_{\mathrm{s}}^{2}}\right)  r^{2}\right]=\frac{1}{2\pi^{3/2}w_{\mathrm{s}}},
\end{equation}
where we have used that $w_{\mathrm{p}}^{2}/w_{\mathrm{s}}^{2}=\lambda
_{\mathrm{p}}/\lambda_{\mathrm{s}}=1/2$, see (\ref{w0}), we get%
\begin{equation}
\chi=3\frac{l_{\mathrm{c}}}{w_{\mathrm{s}}}\chi_{\mathrm{oee}}^{(2)}%
(2\omega_{0};\omega_{0},\omega_{0})\sqrt{\frac{\hbar\omega_{0}^{3}}{8\pi
^{3}\varepsilon_{0}n_{\mathrm{c}}^{3}L_{\mathrm{opt}}^{3}}},
\end{equation}
where pump and signal feel the same refractive index $n_{\mathrm{c}%
}=n_{\mathrm{e}}(\omega_{0})=n_{\mathrm{o}}(2\omega_{0})$, as follows from the
phase--matching condition (\ref{DIcond}).

Our goal is to evaluate the noise spectrum of the field coming out from the
cavity, which was shown in the previous chapter to be given by%
\begin{equation}
V^{\mathrm{out}}(\hat{X}_{\mathrm{s}}^{\varphi};\Omega)=1+S(\hat
{X}_{\mathrm{s}}^{\varphi};\Omega),
\end{equation}
with%
\begin{equation}
S(\hat{X}_{\mathrm{s}}^{\varphi};\Omega)=2\gamma_{\mathrm{s}}\int_{-\infty
}^{+\infty}dt^{\prime}\exp(-\mathrm{i}\Omega t^{\prime})\langle:\delta\hat
{X}_{\mathrm{s}}^{\varphi}\left(  t\right)  \delta\hat{X}_{\mathrm{s}%
}^{\varphi}\left(  t+t^{\prime}\right)  :\rangle, \label{SqSpectrumSignal}%
\end{equation}
when the state of the system is stationary, what will be shown to happen in
the DOPO for $t\gg\gamma_{\mathrm{s}}^{-1}$. Recall that the expectation value
in this expression is to be evaluated within the interaction picture, that is,
via the density operator $\hat{\rho}_{\mathrm{I}}$ satisfying the master
equation (\ref{DOPOmaster}).

We use the positive $P$ representation in order to evaluate the squeezing
spectrum (\ref{SqSpectrumSignal}). In particular, following the same steps as
those described in Section \ref{FPandLangevin}, it is completely
straightforward to show that the stochastic Langevin equations associated to
the master equation (\ref{DOPOmaster}) are
\begin{subequations}
\label{DOPOlangevin}%
\begin{align}
\dot{\alpha}_{\mathrm{p}}  &  =\mathcal{E}_{\mathrm{p}}-(\gamma_{\mathrm{p}%
}+\mathrm{i}\Delta)\alpha_{\mathrm{p}}-\frac{\chi}{2}\alpha_{\mathrm{s}}%
^{2},\\
\dot{\alpha}_{\mathrm{p}}^{+}  &  =\mathcal{E}_{\mathrm{p}}-(\gamma
_{\mathrm{p}}-\mathrm{i}\Delta)\alpha_{\mathrm{p}}^{+}-\frac{\chi}{2}%
\alpha_{\mathrm{s}}^{+2},\\
\dot{\alpha}_{\mathrm{s}}  &  =-(\gamma_{\mathrm{s}}+\mathrm{i}\Delta
)\alpha_{\mathrm{s}}+\chi\alpha_{\mathrm{p}}\alpha_{\mathrm{s}}^{+}+\sqrt
{\chi\alpha_{\mathrm{p}}}\eta(t),\\
\dot{\alpha}_{\mathrm{s}}^{+}  &  =-(\gamma_{\mathrm{s}}-\mathrm{i}%
\Delta)\alpha_{\mathrm{s}}^{+}+\chi\alpha_{\mathrm{p}}^{+}\alpha_{\mathrm{s}%
}+\sqrt{\chi\alpha_{\mathrm{p}}^{+}}\eta^{+}(t),
\end{align}
where $\eta(t)$ and $\eta^{+}(t)$ are independent real Gaussian noises which
satisfy the usual statistical properties (\ref{RealGaussStat}). Even though
noise is multiplicative in these equations, it is easy to show that they have
the same form both within the Ito and Stratonovich interpretations, and hence
we can directly use the usual rules of calculus to analyze them.

In order to better appreciate the parameters needed to model the system, we
will make the following variable change%
\end{subequations}
\begin{equation}
\tau=\gamma_{\mathrm{s}}t,\text{ \ \ \ \ }\beta_{m}(\tau)=\frac{\chi}%
{\gamma_{\mathrm{s}}\sqrt{\gamma_{\mathrm{p}}/\gamma_{m}}}\alpha_{m}(t),\text{
\ \ \ \ }\zeta_{m}(\tau)=\frac{1}{\sqrt{\gamma_{\mathrm{s}}}}\eta_{m}(t),
\end{equation}
where with these definitions the new noises $\zeta(\tau)$ and $\zeta^{+}%
(\tau)$ satisfy the same statistical properties as $\eta(t)$ and $\eta^{+}%
(t)$, but now with respect to the dimensionless time $\tau$. In terms of these
\textit{scaled variables}, the Langevin equations (\ref{DOPOlangevin}) read
(derivatives with respect to $\tau$ are understood)
\begin{subequations}
\label{scaledDOPOlangevin}%
\begin{align}
\dot{\beta}_{\mathrm{p}}  &  =\kappa\lbrack\sigma-(1+\mathrm{i}\tilde{\Delta
})\beta_{\mathrm{p}}-\beta_{\mathrm{s}}^{2}/2]\\
\dot{\beta}_{\mathrm{p}}^{+}  &  =\kappa\lbrack\sigma-(1-\mathrm{i}%
\tilde{\Delta})\beta_{\mathrm{p}}^{+}-\beta_{\mathrm{s}}^{+2}/2]\\
\dot{\beta}_{\mathrm{s}}  &  =-(1+\mathrm{i}\tilde{\Delta})\beta_{\mathrm{s}%
}+\beta_{\mathrm{p}}\beta_{\mathrm{s}}^{+}+g\sqrt{\beta_{\mathrm{p}}}%
\zeta(\tau)\\
\dot{\beta}_{\mathrm{s}}^{+}  &  =-(1-\mathrm{i}\tilde{\Delta})\beta
_{\mathrm{s}}^{+}+\beta_{\mathrm{p}}^{+}\beta_{\mathrm{s}}+g\sqrt
{\beta_{\mathrm{p}}^{+}}\zeta^{+}(\tau),
\end{align}
which have only four dimensionless parameters%
\end{subequations}
\begin{equation}
\kappa=\gamma_{\mathrm{p}}/\gamma_{\mathrm{s}}\text{, \ \ \ \ }\tilde{\Delta
}=\Delta/\gamma_{\mathrm{s}}\text{, \ \ \ \ }\sigma=\chi\mathcal{E}%
_{\mathrm{p}}/\gamma_{\mathrm{p}}\gamma_{\mathrm{s}}\text{, \ \ \ \ }%
g=\chi/\sqrt{\gamma_{\mathrm{p}}\gamma_{\mathrm{s}}}, \label{ScaledParameters}%
\end{equation}
which are, respectively, normalized versions of the pump's damping rate, the
detuning, the pump injection parameter, and the nonlinear coupling. In the
following we will set $\Delta=0$ for simplicity; we will analyze the effects
of detuning in Section \ref{DetuningAndPump}.

These equations allow us to evaluate the squeezing spectrum as%
\begin{equation}
S(\hat{X}_{\mathrm{s}}^{\varphi};\Omega)=\frac{2}{g^{2}}\int_{-\infty
}^{+\infty}d\tau^{\prime}\exp(-\mathrm{i}\tilde{\Omega}\tau^{\prime}%
)\langle\delta x_{\mathrm{s}}^{\varphi}\left(  \tau\right)  \delta
x_{\mathrm{s}}^{\varphi}\left(  \tau+\tau^{\prime}\right)  \rangle_{P},
\label{ScaledSqSpectrum}%
\end{equation}
where we have defined normalized versions of the noise frequency
$\tilde{\Omega}=\Omega/\gamma_{\mathrm{s}}$ and the quadratures within the
positive $P$ phase representation%
\begin{equation}
x_{\mathrm{s}}^{\varphi}\left(  \tau\right)  =e^{-\mathrm{i}\varphi}%
\beta_{\mathrm{s}}\left(  \tau\right)  +e^{\mathrm{i}\varphi}\beta
_{\mathrm{s}}^{+}\left(  \tau\right)  . \label{ScaledQuadratures}%
\end{equation}

Hence, in order to evaluate the squeezing spectrum we need the solutions
$\beta_{m}(\tau)$ of (\ref{scaledDOPOlangevin}) in terms of the noises, so
that from them we can evaluate the corresponding stochastic averages. However,
these equations are a set of coupled nonlinear stochastic equations, and hence
it is impossible to perform this task analytically without introducing further
approximations. Let us first show the behavior of the DOPO in the classical
limit, and then we will come back to the problem of solving
(\ref{scaledDOPOlangevin}).

\subsection{Classical analysis of the DOPO\label{ClassiDOPO}}

Before studying the quantum properties of any OPO, it is convenient to analyze
what classical optics has to say about its dynamics. In the next section we
will see that this is not only convenient, but also mandatory in order to
extract analytic information from its quantum model. Moreover, along the
thesis, and starting in the next section, we will learn how to predict the
quantum properties of OPOs just from a qualitative understanding of their
classical behavior.

In principle, the classical behavior of the field inside an OPO must be
studied by using the nonlinear wave equation of the electric field
(\ref{EwaveEq}) with appropriate boundary conditions accounting for the (open)
cavity. There is a simpler route though. From the chapter where we studied the
harmonic oscillator, we learned that quantum predictions become equivalent to
the classical description of the system in a certain limit. We introduced the
coherent states of the oscillator as those which allowed for these
quantum--to--classical transition; the interesting thing to note about
coherent states is that there are no quantum correlations on them, that is,
given two oscillators with annihilation operators $\hat{a}_{1}$ and $\hat
{a}_{2}$ both in some coherent state, it is satisfied%
\begin{equation}
\left\langle \hat{a}_{1}^{2}\right\rangle =\left\langle \hat{a}_{1}%
\right\rangle \left\langle \hat{a}_{1}\right\rangle \text{ \ \ \ \ and
\ \ \ \ }\left\langle \hat{a}_{1}\hat{a}_{2}\right\rangle =\left\langle
\hat{a}_{1}\right\rangle \left\langle \hat{a}_{2}\right\rangle \text{,}
\label{MeanFieldDecorr}%
\end{equation}
for example. Hence, one way to retrieve the classical equations from the
evolution equations for the operators of the system is to take expectation
values, and decorrelate any product of operators following
(\ref{MeanFieldDecorr}), which is as using a coherent ansatz for the state of
the system. The quantities $\nu_{j}=\sqrt{2\hbar/\omega_{j}}\left\langle
\hat{a}_{j}\right\rangle $ are then identified with the normal variables of
the various harmonic oscillators that form the system, being $\omega_{j}$
their natural oscillation frequency. This procedure is called the \textit{mean
field limit}.

Equipped with an stochastic quantum model for the system, the classical limit
is even simpler to devise. The idea now is that the classical behavior of the
system (its classical phase--space trajectory) should be recovered when
quantum noise is negligible. Hence, the classical equations of an OPO should
be recovered by neglecting the noise terms on its corresponding Langevin
equations, and identifying $\nu_{j}=\sqrt{2\hbar/\omega_{j}}\alpha_{j}$ with
the normal variables of the cavity modes. When using the positive $P$
representation there is one more prescription though: In order to come back to
a classical phase space, the variables $\alpha_{j}^{+}$ of the extended phase
space have to be identified with $\alpha_{j}^{\ast}$. This procedure is in
agreement with (and completely equivalent to) the mean field scenario
discussed above.

Applying these ideas to our DOPO model (\ref{scaledDOPOlangevin}), we get the
following classical equations
\begin{subequations}
\label{ClassicalDOPOeqs}%
\begin{align}
\dot{\beta}_{\mathrm{p}}  &  =\kappa\left(  \sigma-\beta_{\mathrm{p}}%
-\beta_{\mathrm{s}}^{2}/2\right)  ,\\
\dot{\beta}_{\mathrm{s}}  &  =-\beta_{\mathrm{s}}+\beta_{\mathrm{p}}%
\beta_{\mathrm{s}}^{\ast}.
\end{align}
This is a set of first--order nonlinear differential equations, which, again,
cannot be solved analytically. Fortunately, there is actually no need for
doing so for our purposes. One is usually interested only on the long time
term solutions that these equations predict for the system. In particular, the
first type of solutions that one needs to look for are the so-called
\textit{stationary} or \textit{steady state} solutions, in which the variables
do not depend on time. These solutions are found by setting $\dot{\beta
}_{\mathrm{p}}=\dot{\beta}_{\mathrm{s}}=0$, then arriving to a simple
algebraic system
\end{subequations}
\begin{subequations}
\label{StationaryDOPOeqs}%
\begin{align}
\sigma &  =\beta_{\mathrm{p}}+\beta_{\mathrm{s}}^{2}/2,\\
\beta_{\mathrm{s}}  &  =\beta_{\mathrm{p}}\beta_{\mathrm{s}}^{\ast}.
\end{align}
This system admits two types of such solutions. In the first type the signal
mode is \textit{switched off}, and hence we have%
\end{subequations}
\begin{equation}
\bar{\beta}_{\mathrm{p}}=\sigma\text{ \ \ \ \ and \ \ \ \ }\bar{\beta
}_{\mathrm{s}}=0, \label{BelowThresholdSol}%
\end{equation}
where the overbar is used to denote \textquotedblleft classical long time term
solution\textquotedblright. In the second type, the signal mode is
\textit{switched on}, and in order to find the explicit form of the solution
we use the following amplitude--phase representation of the variables:
$\beta_{j}=\rho_{j}\exp(\mathrm{i}\varphi_{j})$. From the second equation we
trivially get $\rho_{\mathrm{p}}=1$ and $\exp(\mathrm{i}\varphi_{\mathrm{p}%
})=\exp(2\mathrm{i}\varphi_{\mathrm{s}})$, which introduced in the first
equation leads to $\rho_{\mathrm{s}}=\sqrt{\sigma-1}$ and $\varphi
_{\mathrm{s}}=\{0,\pi\}$, that is,
\begin{equation}
\bar{\beta}_{\mathrm{p}}=1\text{ \ \ \ \ and \ \ \ \ }\bar{\beta}_{\mathrm{s}%
}=\pm\sqrt{2\left(  \sigma-1\right)  }. \label{AboveThresholdSol}%
\end{equation}
Note that this second solution only exists for $\sigma\geq1$; we say that the
\textit{domain of existence} of the solutions are all the parameter space
$\{\sigma,\kappa\}$ and the region $\{\sigma\geq1,\kappa\}$ for the first and
second solutions, respectively. Note also that at $\sigma=1$ both solutions coincide.

Knowing the possible solutions of (\ref{StationaryDOPOeqs}) in the long time
term is not enough. It can happen that a solution exists but is
\textit{unstable}, that is, an arbitrarily small perturbation can make the
system move out from that solution and fall into another solution. It is not
difficult to analyze the stability of a solution. With full generality,
consider the equation%
\begin{equation}
\boldsymbol{\dot{\beta}}=\mathbf{f}\left(  \boldsymbol{\beta}\right)  ,
\end{equation}
where all the complex variables of the system are collected as
\begin{equation}
\boldsymbol{\beta}=(\beta_{1},\beta_{1}^{\ast},\beta_{2},\beta_{2}^{\ast
},...,\beta_{N},\beta_{N}^{\ast}),
\end{equation}
and one stationary solution
$\boldsymbol{\bar{\beta}}$ satisfying $\mathbf{f}\left(  \boldsymbol{\bar
{\beta}}\right)  =\mathbf{0}$. In order to see whether this solution is stable
or not, we just introduce a small perturbation $\mathbf{b}\left(  \tau\right)
$ in the system and analyze how it grows. Writing $\boldsymbol{\beta}%
(\tau)=\boldsymbol{\bar{\beta}}+\mathbf{b}\left(  \tau\right)  $, and making a
series expansion of $\mathbf{f}\left(  \boldsymbol{\beta}\right)  $ up to
linear order in $\mathbf{b}$, it is straightforward to show that the evolution
equation of the perturbations is%
\begin{equation}
\mathbf{\dot{b}}=\mathcal{L}\mathbf{b}\text{ \ \ \ \ with \ \ \ \ }%
\mathcal{L}_{jl}=\left.  \frac{\partial f_{j}}{\partial\beta_{l}}\right\vert
_{\boldsymbol{\bar{\beta}}}, \label{LinearStability}%
\end{equation}
and hence the linear evolution of the perturbations is governed by matrix
$\mathcal{L}$, which we will call the \textit{stability matrix}. Being a
simple linear system, the growth of this perturbation is easily analyzed.
Suppose that we diagonalize $\mathcal{L}$, so that we know the set
$\{\lambda_{j},\mathbf{w}_{j}\}_{j=1,...,2N}$ satisfying $\mathbf{w}_{j}%
^{\ast}\mathcal{L}=\lambda_{j}\mathbf{w}_{j}^{\ast}$; projecting the linear
system (\ref{LinearStability}) onto the $\mathbf{w}_{j}$ vectors, and defining
the projections $c_{j}(t)=\mathbf{w}_{j}^{\ast}\cdot\mathbf{b}(t)$, we get the
simple solutions%
\begin{equation}
c_{j}(t)=c_{j}(0)\exp(\lambda_{j}\tau)\text{.}%
\end{equation}
Hence, the growth of the perturbations is characterized by the real part of
the eigenvalues $\lambda_{j}$. In particular, if \textit{all} of them have
negative real part, the perturbations will tend to decay, and after a time
$\tau\gg\max[|\operatorname{Re}\{\lambda_{j}\}|^{-1}]_{j=1,...,2N}$ the
stationary solution $\boldsymbol{\bar{\beta}}$ will be restored; on the other
hand, if \textit{any} of them has a positive real part, it will mean that
perturbations tend to grow in some direction of phase space, and then the
system won't come back to $\boldsymbol{\bar{\beta}}$. Accordingly, we say in
the first case that $\boldsymbol{\bar{\beta}}$ is a stable, stationary
solution, while it is unstable in the second case.

We can apply these \textit{stability analysis} to our DOPO equations
(\ref{ClassicalDOPOeqs}) and the corresponding stationary solutions
(\ref{BelowThresholdSol},\ref{AboveThresholdSol}). Ordering the different amplitudes as
$\boldsymbol{\beta}=(\beta_{\mathrm{p}},\beta_{\mathrm{p}}^{\ast}%
,\beta_{\mathrm{s}},\beta_{\mathrm{s}}^{\ast})$, the general stability matrix
associated to equations (\ref{ClassicalDOPOeqs}) is%
\begin{equation}
\mathcal{L}=%
\begin{bmatrix}
-\kappa & 0 & -\kappa\bar{\beta}_{\mathrm{s}} & 0\\
0 & -\kappa & 0 & -\kappa\bar{\beta}_{\mathrm{s}}^{\ast}\\
\bar{\beta}_{\mathrm{s}}^{\ast} & 0 & -1 & \bar{\beta}_{\mathrm{p}}\\
0 & \bar{\beta}_{\mathrm{s}} & \bar{\beta}_{\mathrm{p}}^{\ast} & -1
\end{bmatrix}
,
\end{equation}
which has eigenvalues%
\begin{subequations}
\begin{align}
\lambda_{\pm}^{(1)}  &  =\frac{1}{2}\left[  \bar{\rho}_{\mathrm{p}}%
-\kappa-1\pm\sqrt{\left(  \bar{\rho}_{\mathrm{p}}+\kappa-1\right)
^{2}-4\kappa\bar{\rho}_{\mathrm{s}}^{2}}\right]  ,\\
\lambda_{\pm}^{(2)}  &  =\frac{1}{2}\left[  -\bar{\rho}_{\mathrm{p}}%
-\kappa-1\pm\sqrt{\left(  \bar{\rho}_{\mathrm{p}}-\kappa+1\right)
^{2}-4\kappa\bar{\rho}_{\mathrm{s}}^{2}}\right]  .
\end{align}
Particularizing these eigenvalues to the first solution
(\ref{BelowThresholdSol}) it is simple to see that $\lambda_{+}^{(1)}%
=\sigma-1$, and hence this solution becomes unstable when $\sigma>1$. On the
other hand, it is simple to check that all the eigenvalues have negative real
part when particularized to the second solution (\ref{AboveThresholdSol}) in
all its domain of existence.%

\begin{figure}
[t]
\begin{center}
\includegraphics[
height=2.373in,
width=3.6729in
]%
{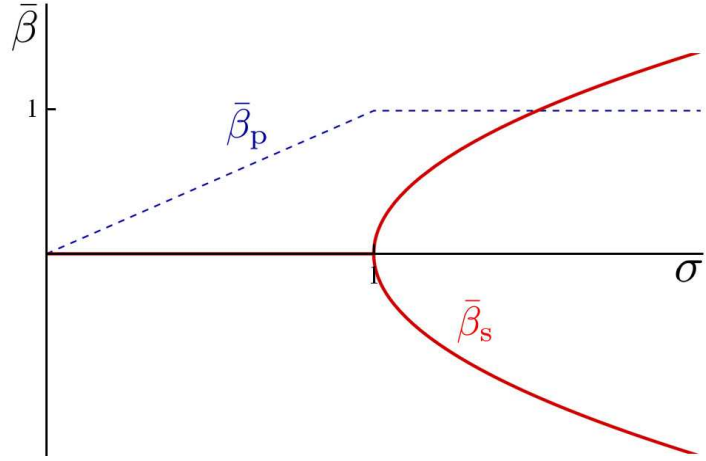}%
\caption{Classical \textit{bifurcation diagram} of the DOPO, that is, the
amplitude of the different fields as a function of the system parameters.
Below threshold ($\sigma<1$) the pump amplitude (blue, dashed line) increases
linearly with the pump parameter $\sigma$, and the signal field is off (red,
solid line). Above this threshold the pump gets `frozen' to a given value, and
the signal field is switched on and can take any of two opposite phases ($0$
or $\pi$) owed to the symmetry $\alpha_{\mathrm{s}}\rightarrow-\alpha
_{\mathrm{s}}$ of equations (\ref{ClassicalDOPOeqs}). Hence the system shows
bistability above threshold.}%
\label{fOPO4}%
\end{center}
\end{figure}

This allows us to understand the classical behavior of the system in the
following way (see Figure \ref{fOPO4}): while the pump injection increases
from $\sigma=0$ to $\sigma=1$, the signal mode remains switched off, and all
the injection is used to excite the pump mode as $\bar{\beta}_{\mathrm{p}%
}=\sigma$; on the other hand, when the point $\sigma=1$ is crossed the pump
mode gets \textquotedblleft frozen\textquotedblright\ to a fix value
$\bar{\beta}_{\mathrm{p}}=1$, while the signal mode starts oscillating inside
the DOPO as $\bar{\beta}_{\mathrm{s}}=\sqrt{\sigma-1}$. The point $\sigma=1$
is called a \textit{bifurcation} of the system, because the system changes
from one long time term stable solution to another. Physically, this means
that the DOPO is a \textit{threshold system}: The injected laser must be above
some threshold power in order for the signal field to start oscillating inside
the cavity. According to this picture, the solutions with the signal mode
switched off (\ref{BelowThresholdSol}) and on (\ref{AboveThresholdSol}) are
denoted by \textit{below} and \textit{above threshold} \textit{solutions}, respectively.

\subsection{Quantum analysis of the DOPO: linearization and noise spectrum\label{QuantumDOPO}}

How does knowing the classical behavior of the DOPO help us to solve the
nonlinear stochastic equations (\ref{scaledDOPOlangevin})? The idea is that
the noise term is multiplied by $g$, which is a very small
quantity\footnote{For example, considering a symmetric, confocal cavity, and
using the expressions we have given for the different model parameters in
terms of physical parameters, it is easy to write%
\begin{equation}
g=\frac{12\chi^{(2)}l_{\mathrm{c}}}{\lambda_{0}^{2}}\sqrt{\frac{2\pi\hbar
c}{L_{\mathrm{eff}}L_{\mathrm{opt}}\mathcal{T}_{\mathrm{p}}\mathcal{T}%
_{\mathrm{s}}\varepsilon_{0}n_{\mathrm{c}}}},%
\end{equation}
where $\lambda_{0}$ is the wavelength of the signal mode, $\chi^{(2)}$ is the
relevant nonlinear susceptibility, and $\mathcal{T}_{\mathrm{p}}%
=\mathcal{T}(\lambda_{0}/2)$ and $\mathcal{T}_{\mathrm{s}}=\mathcal{T}%
(\lambda_{0})$ are the transmittivities of the output mirror at the pump and
signal wavelengths, respectively. If we consider the numerical values (typical
of OPO experiments)%
\begin{equation}%
\begin{array}
[c]{ccc}%
n_{\mathrm{c}}=2, & \lambda_{0}=1064\mathrm{nm,} & \mathcal{T}_{\mathrm{p}%
}=0.1,\\
\chi^{(2)}=2.5\mathrm{pm/V,} & l_{\mathrm{c}}=L/10, & \mathcal{T}_{\mathrm{s}%
}=0.01,
\end{array}%
\end{equation}
we get $g\approx4\times10^{-6}$.} in usual OPOs, that is, $g\ll1$, and hence,
for most values of the system parameters quantum fluctuations act just as a
small perturbation around the classical solution. One can then solve the
Langevin equations perturbatively
\cite{Drummond04book,Chaturvedi02,Drummond02}, that is, by writing the
stochastic amplitudes as%
\end{subequations}
\begin{subequations}
\begin{align}
\beta_{j}  &  =\bar{\beta}_{j}+g\beta_{j}^{(1)}+g^{2}\beta_{j}^{(2)}+...,\\
\beta_{j}^{+}  &  =\bar{\beta}_{j}^{\ast}+g\beta_{j}^{+(1)}+g^{2}\beta
_{j}^{+(2)}+...,
\end{align}
where the $\beta_{j}^{(k)}$ amplitudes are quantities of order $g^{0}$, and
then solve iteratively order by order the Langevin equation. For the purposes
of this thesis it will be enough to consider the first quantum correction,
that is, the order $g$, although we will explain the effect of further quantum
corrections. Let us then write%
\end{subequations}
\begin{equation}
\boldsymbol{\beta}(\tau)=\boldsymbol{\bar{\beta}}+\mathbf{b}\left(
\tau\right)  \label{QflucToAmps}%
\end{equation}
being $\boldsymbol{\bar{\beta}}=\operatorname{col}(\bar{\beta}_{\mathrm{p}%
},\bar{\beta}_{\mathrm{p}}^{\ast},\bar{\beta}_{\mathrm{s}},\bar{\beta
}_{\mathrm{s}}^{\ast})$ a vector containing the classical steady state
solution of the DOPO and $\mathbf{b}\left(  \tau\right)  =\operatorname{col}%
\left[  b_{\mathrm{p}}(\tau),b_{\mathrm{p}}^{+}(\tau),b_{\mathrm{s}}%
(\tau),b_{\mathrm{s}}^{+}(\tau)\right]  $ a vector with the corresponding
quantum fluctuations, which we assume to be order $g$ as explained. Retaining
up to order $g$ terms in the Langevin equation, we get
\begin{equation}
\mathbf{\dot{b}}=\mathcal{L}\mathbf{b}+g\boldsymbol{\zeta}(\tau)\mathbf{,}%
\end{equation}
where $\mathcal{L}$ is the stability matrix already defined in the previous
section, and $\boldsymbol{\zeta}(\tau)=\operatorname{col}[0,0,\sqrt{\bar
{\beta}_{\mathrm{p}}}\zeta(\tau),\sqrt{\bar{\beta}_{\mathrm{p}}^{\ast}}%
\zeta^{+}(\tau)]$. We will call the \textit{linearized Langevin equations} of
the DOPO to this linear system. These can be solved easily by diagonalizing
the stability matrix, after what the solution can be used to find the
squeezing spectrum (\ref{ScaledSqSpectrum}). Let us show this process below
and above the DOPO's threshold separately.

\begin{figure}
[t]
\begin{center}
\includegraphics[
height=1.7677in,
width=4.8032in
]%
{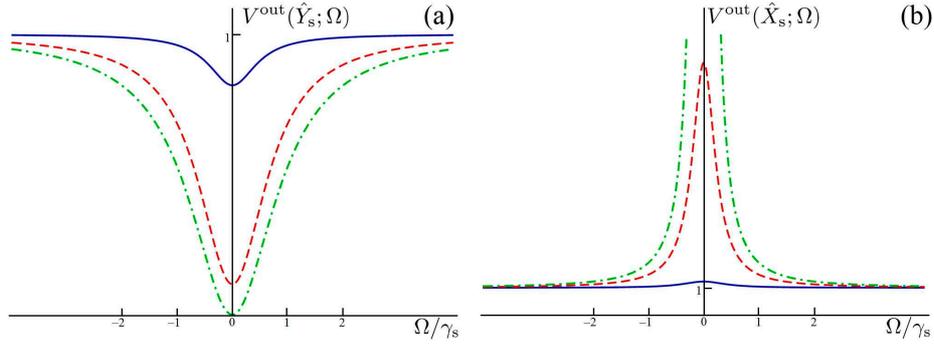}%
\caption{Noise spectrum of the \textsf{Y} (a) and \textsf{X}\ (b) quadratures
of the signal mode as a function of the noise frequency for $\sigma=0.05$
(blue, solid), $0.5$ (red, dashed), and $0.8$ (green, dotted--dashed). }%
\label{fOPO5}%
\end{center}
\end{figure}

\textbf{Quantum properties of the DOPO below threshold. }Particularized to the
below threshold solution (\ref{BelowThresholdSol}), the stability matrix reads%
\begin{equation}
\mathcal{L}=%
\begin{bmatrix}
-\kappa & 0 & 0 & 0\\
0 & -\kappa & 0 & 0\\
0 & 0 & -1 & \sigma\\
0 & 0 & \sigma & -1
\end{bmatrix}
,
\end{equation}
whose block--diagonal form shows that the (linear) properties of pump and
signal are independent. We will learn throughout the thesis that this
decoupling of the classically unexcited modes is a rather general property of
OPOs. As the linearized Langevin equations for the pump modes have no noise
terms, the pump stays in a coherent state. We can focus then on the quantum
properties of the signal mode, which are given by the simple linear system%
\begin{equation}
\mathbf{\dot{b}}_{\mathrm{s}}=\mathcal{L}_{\mathrm{s}}\mathbf{b}_{\mathrm{s}%
}+g\sqrt{\sigma}\boldsymbol{\zeta}_{\mathrm{s}}(\tau),
\end{equation}
with $\mathbf{b}_{\mathrm{s}}\left(  \tau\right)  =\operatorname{col}\left[
b_{\mathrm{s}}(\tau),b_{\mathrm{s}}^{+}(\tau)\right]  $, $\boldsymbol{\zeta
}_{\mathrm{s}}(\tau)=\operatorname{col}\left[  \zeta(\tau),\zeta^{+}%
(\tau)\right]  $, being%
\begin{equation}
\mathcal{L}_{\mathrm{s}}=%
\begin{bmatrix}
-1 & \sigma\\
\sigma & -1
\end{bmatrix}
,
\end{equation}
a symmetric, real matrix with orthonormal eigensystem%
\begin{equation}
\left\{  \lambda_{\pm}=-\left(  1\mp\sigma\right)  ,\mathbf{v}_{\pm}=\frac
{1}{\sqrt{2}}\operatorname{col}(1,\pm1)\right\}  . \label{eigenDOPO}%
\end{equation}
Projecting the linear system onto this eigenvectors and defining the
projections $c_{\pm}=\mathbf{v}_{\pm}\cdot\mathbf{b}_{\mathrm{s}}$, we get%
\begin{equation}
\dot{c}_{j}=\lambda_{j}c_{j}+g\sqrt{\sigma}\zeta_{j}(\tau),
\label{ProjLinLanBelowDOPO}%
\end{equation}
where $\zeta_{\pm}(\tau)=\left[  \zeta(\tau)\pm\zeta^{+}(\tau)\right]
/\sqrt{2}$ are new independent real noises satisfying the usual statistical
properties. This type of one--variable, linear stochastic equations with
additive noise will be appearing all along the rest of the thesis, and we have
included an appendix (Appendix \ref{LinStoApp}) in order to explain how to
find their solutions $c_{j}(\tau)$. We also explain there how to evaluate
their associated two--time correlation function $\langle c_{j}(\tau)c_{j}%
(\tau+\tau^{\prime})\rangle$, which are shown to depend only on $|\tau
^{\prime}|$ in the stationary limit $\tau\gg|\operatorname{Re}\{\lambda
_{j}\}|^{-1}$, and their corresponding spectrum $\tilde{C}_{j}(\tilde{\Omega
})=\int_{-\infty}^{+\infty}d\tau^{\prime}\langle c_{j}(\tau)c_{j}(\tau
+\tau^{\prime})\rangle_{P}\exp(-\mathrm{i}\tilde{\Omega}\tau^{\prime})$.

Note that by writing the projections in terms of the fluctuations we get
\begin{subequations}
\label{ProjToQuadDOPO}%
\begin{align}
c_{+}(\tau) &  =\left[  b_{\mathrm{s}}(\tau)+b_{\mathrm{s}}^{+}(\tau)\right]
/\sqrt{2}=\delta x_{\mathrm{s}}(\tau)/\sqrt{2},\\
c_{-}(\tau) &  =\left[  b_{\mathrm{s}}(\tau)-b_{\mathrm{s}}^{+}(\tau)\right]
/\sqrt{2}=\mathrm{i}\delta y_{\mathrm{s}}(\tau)/\sqrt{2},
\end{align}
and hence the properties of the \textsf{X} and \textsf{Y}\ quadratures of the
signal field are linked to $c_{\pm}$, respectively. Hence, we can evaluate the
noise spectrum of these quadratures as
\end{subequations}
\begin{subequations}
\label{SpectraDOPO}%
\begin{align}
V^{\mathrm{out}}(\hat{X}_{\mathrm{s}};\Omega) &  =1+\frac{4}{g^{2}}\tilde
{C}_{+}(\tilde{\Omega})=\frac{(1+\sigma)^{2}+\tilde{\Omega}^{2}}%
{(1-\sigma)^{2}+\tilde{\Omega}^{2}},\\
V^{\mathrm{out}}(\hat{Y}_{\mathrm{s}};\Omega) &  =1-\frac{4}{g^{2}}\tilde
{C}_{-}(\tilde{\Omega})=\frac{(1-\sigma)^{2}+\tilde{\Omega}^{2}}%
{(1+\sigma)^{2}+\tilde{\Omega}^{2}},
\end{align}
where we have made use of the results in Appendix \ref{LinStoApp}. These
expressions are plotted in Figure \ref{fOPO5} as a function of the
dimensionless noise frequency $\tilde{\Omega}$ for different values of the
pump parameter $\sigma$. Note first that they predict that the state of the
signal modes is a minimum uncertainty state, because
\end{subequations}
\begin{equation}
V^{\mathrm{out}}(\hat{X}_{\mathrm{s}};\Omega)V^{\mathrm{out}}(\hat
{Y}_{\mathrm{s}};\Omega)=1.
\end{equation}
As expected, for $\sigma=0$ (no injection) $V^{\mathrm{out}}(\Omega)=1$ for
both quadratures, that is, the external modes are in a vacuum state. On the
other hand, as $\sigma$ increases, quantum noise gets reduced in the
\textsf{Y} quadrature (and correspondingly increased in the \textsf{X}
quadrature), and hence the signal mode is in a squeezed state. Note that
maximum squeezing is obtained at zero noise frequency, while at $|\Omega
|=(1+\sigma)\gamma_{\mathrm{s}}$ the squeezing level is reduced by a factor
two, that is, $S^{\mathrm{out}}[\hat{Y}_{\mathrm{s}};\pm(1+\sigma
)\gamma_{\mathrm{s}}]=S^{\mathrm{out}}(\hat{Y}_{\mathrm{s}};0)/2$; hence, the
Lorentzian shape sets a bandwidth to the noise frequencies\ for which
squeezing is obtained (see the end of Section \ref{RealHomodyneDetection}).

\begin{figure}
[t]
\begin{center}
\includegraphics[
height=2.2771in,
width=3.1462in
]%
{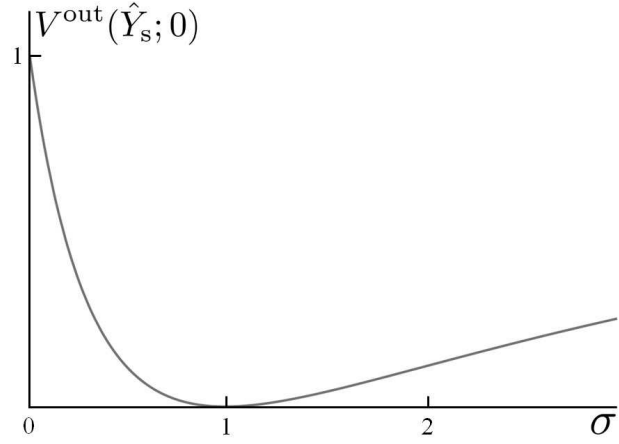}%
\caption{Zero frequency noise spectrum for the \textsf{Y} quadrature of the
signal mode as a function of the pump parameter. As explained in the text, the
linearized theory predicts perfect squeezing only at threshold.}%
\label{fOPO6}%
\end{center}
\end{figure}

Complete noise reduction is predicted at zero noise frequency exactly at
threshold, that is, $V_{\sigma\rightarrow1}^{\mathrm{out}}(\hat{Y}%
_{\mathrm{s}};0)=0$ (see Figure \ref{fOPO6}). This is an unphysical result (as
stressed in Chapter \ref{HarmonicOscillator}, perfect squeezing requires
infinite energy) which shows that the linearization procedure that we have
carried breaks down very close to threshold. Indeed, by going further in the
perturbative expansion of the Langevin equations, one can show that as the
system approaches $\sigma=1$, the different orders in $g$ kick in, and the
noise spectrum becomes finite. In any case, all these contributions are rather
small (e.g., $V^{\mathrm{out}}(\hat{Y}_{\mathrm{s}};0)=3g^{4/3}/4$ for
$\sigma=1-g^{2/3}$ and $\kappa\gg g^{2/3}$), and in practice one can argue as
if squeezing would be perfect at threshold
\cite{Drummond04book,Chaturvedi02,Drummond02}.%

\textbf{Quantum properties of the DOPO above threshold. }We have proved that
large levels of squeezing can be obtained in the signal field coming out from
the DOPO as we approach its threshold $\sigma=1$ from below. Now we are going
to show that above threshold the situation is similar, that is, squeezing in
the signal field decreases as one moves away from threshold.

The situation in this case is, however, much more complicated, because above
threshold the stability matrix involves both the pump and the signal modes,
and even more, not being symmetric, it does not posses a simple orthonormal
eigensystem, though it can be seen to be diagonalizable. In order to avoid all
these complications, let us first work in the $\kappa\gg1$ limit, where the
pump can be adiabatically eliminated, as its time scale $\gamma_{\mathrm{p}%
}^{-1}$ is much shorter than that of the signal, $\gamma_{\mathrm{s}}^{-1}$.

Was (\ref{scaledDOPOlangevin}) a simple differential equation, the procedure
to adiabatically eliminate the pump variables would be clear: One just sets
the corresponding time derivatives to zero, that is,%
\begin{subequations}
\begin{align}
\dot{\beta}_{\mathrm{p}}  &  =0\Longrightarrow\beta_{\mathrm{p}}=\sigma
-\beta_{\mathrm{s}}^{2}/2,\\
\dot{\beta}_{\mathrm{p}}^{+}  &  =0\Longrightarrow\beta_{\mathrm{p}}%
^{+}=\sigma-\beta_{\mathrm{s}}^{+2}/2,
\end{align}
and then substitutes the resulting expressions for the pump variables as a
function of the signal variables in the remaining equations, arriving to
\end{subequations}
\begin{subequations}
\label{adiabaticDOPOlangevin}%
\begin{align}
\dot{\beta}_{\mathrm{s}}  &  =-\beta_{\mathrm{s}}+\left(  \sigma
-\beta_{\mathrm{s}}^{2}/2\right)  \beta_{\mathrm{s}}^{+}+g\sqrt{\sigma
-\beta_{\mathrm{s}}^{2}/2}\zeta(\tau)\\
\dot{\beta}_{\mathrm{s}}^{+}  &  =-\beta_{\mathrm{s}}^{+}+\left(  \sigma
-\beta_{\mathrm{s}}^{+2}/2\right)  \beta_{\mathrm{s}}+g\sqrt{\sigma
-\beta_{\mathrm{s}}^{+2}/2}\zeta^{+}(\tau).
\end{align}
Unfortunately, when dealing with stochastic equations there appears a subtle
complication: While the original set of stochastic equations
(\ref{scaledDOPOlangevin}) is the same either in the Ito or the Stratonovich
forms, the resulting equations (\ref{adiabaticDOPOlangevin}) after applying
this na\"{\i}ve procedure are not. To see this, just note that the noise
matrix associated to these equations is%
\end{subequations}
\begin{equation}
B(\boldsymbol{\alpha})=g%
\begin{bmatrix}
\sqrt{\sigma-\beta_{\mathrm{s}}^{2}/2} & 0\\
0 & \sqrt{\sigma-\beta_{\mathrm{s}}^{+2}/2}%
\end{bmatrix}
,
\end{equation}
so that the terms which connect the Ito and Stratonovich forms
(\ref{ItoToStrat}) read%
\begin{subequations}
\begin{align}
\frac{1}{2}\sum_{lm}B_{lm}\left(  \boldsymbol{\beta}_{\mathrm{s}}\right)
\partial_{l}B_{\beta_{\mathrm{s}}m}\left(  \boldsymbol{\beta}_{\mathrm{s}%
}\right)   &  =\frac{1}{2}B_{\beta_{\mathrm{s}}\beta_{\mathrm{s}}}\left(
\boldsymbol{\beta}_{\mathrm{s}}\right)  \partial_{\beta_{\mathrm{s}}}%
B_{\beta_{\mathrm{s}}\beta_{\mathrm{s}}}\left(  \boldsymbol{\beta}%
_{\mathrm{s}}\right)  =-\frac{g^{2}}{4}\beta_{\mathrm{s}},\\
\frac{1}{2}\sum_{lm}B_{lm}\left(  \boldsymbol{\beta}_{\mathrm{s}}\right)
\partial_{l}B_{\beta_{\mathrm{s}}^{+}m}\left(  \boldsymbol{\beta}_{\mathrm{s}%
}\right)   &  =\frac{1}{2}B_{\beta_{\mathrm{s}}^{+}\beta_{\mathrm{s}}^{+}%
}\left(  \boldsymbol{\beta}_{\mathrm{s}}\right)  \partial_{\beta_{\mathrm{s}%
}^{+}}B_{\beta_{\mathrm{s}}^{+}\beta_{\mathrm{s}}^{+}}\left(
\boldsymbol{\beta}_{\mathrm{s}}\right)  =-\frac{g^{2}}{4}\beta_{\mathrm{s}%
}^{+},
\end{align}
which are different from zero. Hence it is fair to ask within which
interpretation is this \textquotedblleft deterministic\textquotedblright%
\ procedure correct, if it is correct at all. Using the techniques explained
in \cite{Gardiner09book} to perform the adiabatic elimination in the
Fokker--Planck equation (where there are no problems of interpretation), it is
possible to show that this procedure is indeed correct within Ito's
interpretation. Hence, the corresponding Stratonovich equations have
additional terms in the deterministic part, and in particular they read%
\end{subequations}
\begin{subequations}
\begin{align}
\dot{\beta}_{\mathrm{s}}  &  =-(1-g^{2}/4)\beta_{\mathrm{s}}+\left(
\sigma-\beta_{\mathrm{s}}^{2}/2\right)  \beta_{\mathrm{s}}^{+}+g\sqrt
{\sigma-\beta_{\mathrm{s}}^{2}/2}\zeta(\tau),\\
\dot{\beta}_{\mathrm{s}}^{+}  &  =-\left(  1-g^{2}/4\right)  \beta
_{\mathrm{s}}^{+}+\left(  \sigma-\beta_{\mathrm{s}}^{+2}/2\right)
\beta_{\mathrm{s}}+g\sqrt{\sigma-\beta_{\mathrm{s}}^{+2}/2}\zeta^{+}(\tau).
\end{align}
Note that the difference between the Ito and Stratonovich forms of the
equations is of order $g^{2}$, and then disappears when making the
linearization procedure; in particular, the linearized equations read in this
case%
\end{subequations}
\begin{equation}
\mathbf{\dot{b}}_{\mathrm{s}}=\mathcal{L}_{\mathrm{s}}\mathbf{b}_{\mathrm{s}%
}+g\boldsymbol{\zeta}_{\mathrm{s}}(\tau), \label{LinLanAbove}%
\end{equation}
with $\mathbf{b}_{\mathrm{s}}\left(  \tau\right)  =\operatorname{col}\left[
b_{\mathrm{s}}(\tau),b_{\mathrm{s}}^{+}(\tau)\right]  $, $\boldsymbol{\zeta
}_{\mathrm{s}}(\tau)=\operatorname{col}\left[  \zeta(\tau),\zeta^{+}%
(\tau)\right]  $, being%
\begin{equation}
\mathcal{L}_{\mathrm{s}}=%
\begin{bmatrix}
1-2\sigma & 1\\
1 & 1-2\sigma
\end{bmatrix}
,
\end{equation}
a symmetric, real matrix with orthonormal eigensystem%
\begin{subequations}
\begin{align}
\lambda_{+}=-2\left(  \sigma-1\right)  ,  &  \text{ \ \ \ \ }\mathbf{v}%
_{+}=\frac{1}{\sqrt{2}}\operatorname{col}(1,1),\\
\lambda_{-}=-2\sigma,  &  \text{ \ \ \ \ }\mathbf{v}_{-}=\frac{1}{\sqrt{2}%
}\operatorname{col}(1,-1).
\end{align}
Note also that the eigenvectors are the same as the ones below threshold, and
hence they have the same relation to the quadrature fluctuations of the signal
modes (\ref{ProjToQuadDOPO}). Then, by following the same procedure as before,
that is, by projecting the linear system (\ref{LinLanAbove}) onto the
eigensystem of $\mathcal{L}_{\mathrm{s}}$ and evaluating the correlation
spectrum of the projections $c_{\pm}=\mathbf{v}_{\pm}\cdot\mathbf{b}%
_{\mathrm{s}}$, it is straightforward to find the following expressions for
the noise spectrum of the \textsf{X} and \textsf{Y}\ quadratures of the signal
field%
\end{subequations}
\begin{subequations}
\begin{align}
V^{\mathrm{out}}(\hat{X}_{\mathrm{s}};\Omega)  &  =1+\frac{1}{(\sigma
-1)^{2}+\Omega^{2}/4},\\
V^{\mathrm{out}}(\hat{Y}_{\mathrm{s}};\Omega)  &  =1-\frac{1}{\sigma
^{2}+\Omega^{2}/4}.
\end{align}
These have again a Lorentzian shape, and, as already stated at the beginning
of the present section, they predict perfect squeezing in the \textsf{Y}
quadrature at zero noise frequency only at threshold (as before, the complete
noise reduction comes from the linearization, and nonlinear corrections in the
$g$ parameter give rise to finite squeezing levels \cite{Drummond02}). As
$\sigma$ increases this squeezing is lost (see Figure \ref{fOPO6}), with the
particularity that now the state of the signal is no longer a minimum
uncertainty state, because%
\end{subequations}
\begin{equation}
V^{\mathrm{out}}(\hat{X}_{\mathrm{s}};\Omega)V^{\mathrm{out}}(\hat
{Y}_{\mathrm{s}};\Omega)>1,
\end{equation}
for $\sigma>1$.

\subsection{Effect of signal detuning\label{DetuningAndPump}}

In the previous sections we have analyzed the classical and quantum behavior
of the DOPO when everything is on resonance, that is, the frequency
$\omega_{\mathrm{L}}$ of the external laser and its second harmonic coincide
with the phase--matched cavity resonances $\omega_{0}$ and $2\omega_{0}$, the
signal and pump frequencies. Here we just want to show the effect that
detuning has on the different properties of the DOPO by considering the simple
scheme shown in Figure \ref{fOPO7}: The signal field does not resonate at
frequency $\omega_{0}$ but at a slightly shifted frequency $\omega_{0}+\Delta
$. In this conditions one can still drive the pump mode with an external laser
at resonance, but then the signal mode will feel a detuning $\Delta$, leading
to the following Langevin equations for the system
\begin{subequations}
\label{scaledDOPOlangevinDetuning}%
\begin{align}
\dot{\beta}_{\mathrm{p}}  &  =\kappa\lbrack\sigma-\beta_{\mathrm{p}}%
-\beta_{\mathrm{s}}^{2}/2]\\
\dot{\beta}_{\mathrm{p}}^{+}  &  =\kappa\lbrack\sigma-\beta_{\mathrm{p}}%
^{+}-\beta_{\mathrm{s}}^{+2}/2]\\
\dot{\beta}_{\mathrm{s}}  &  =-(1+\mathrm{i}\tilde{\Delta})\beta_{\mathrm{s}%
}+\beta_{\mathrm{p}}\beta_{\mathrm{s}}^{+}+g\sqrt{\beta_{\mathrm{p}}}%
\zeta(\tau)\\
\dot{\beta}_{\mathrm{s}}^{+}  &  =-(1-\mathrm{i}\tilde{\Delta})\beta
_{\mathrm{s}}^{+}+\beta_{\mathrm{p}}^{+}\beta_{\mathrm{s}}+g\sqrt
{\beta_{\mathrm{p}}^{+}}\zeta^{+}(\tau),
\end{align}
where $\tilde{\Delta}=\Delta/\gamma_{\mathrm{s}}$.%

\begin{figure}
[t]
\begin{center}
\includegraphics[
height=1.7781in,
width=4.7867in
]%
{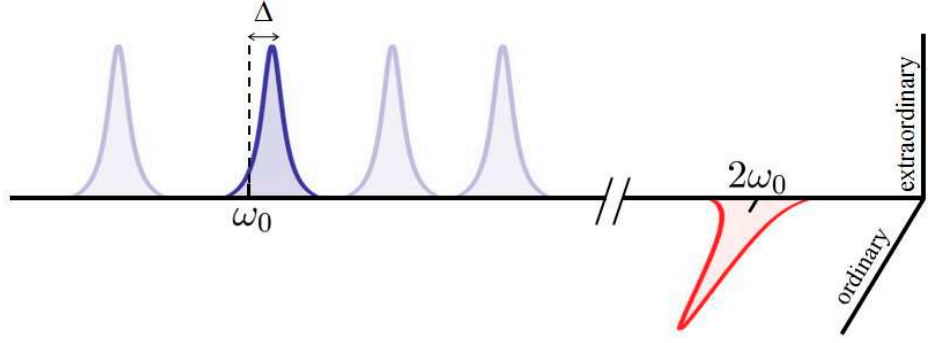}%
\caption{This scheme shows a configuration of the cavity resonances leading to
a DOPO with some detuning in the signal field.}%
\label{fOPO7}%
\end{center}
\end{figure}

The fundamental difference between this case and the resonant one appears at
the classical level. Just as before, the classical equations admit two types
of stationary solutions, one with the signal field switched off, and one with
the signal field switched on. As for the first type, it has exactly the same
form as before (\ref{BelowThresholdSol}), and has an associated stability
matrix%
\end{subequations}
\begin{equation}
\mathcal{L}=%
\begin{bmatrix}
-\kappa & 0 & 0 & 0\\
0 & -\kappa & 0 & 0\\
0 & 0 & -(1+\mathrm{i}\tilde{\Delta}) & \sigma\\
0 & 0 & \sigma & -(1-\mathrm{i}\tilde{\Delta})
\end{bmatrix}
.
\end{equation}
Amongst the eigensystem of this matrix, one can find three negative definite
eigenvalues, plus the following one: $\lambda=-1+\sqrt{\sigma^{2}%
-\tilde{\Delta}^{2}}$. Hence, this solution becomes unstable when
$\sigma>\sqrt{1+\tilde{\Delta}^{2}}$. The solution with the signal field
switched on is slightly different than before; with the notation $\beta
_{j}=\rho_{j}\exp(\mathrm{i}\varphi_{j})$ it is easy to show after some
algebra that it is given by
\begin{subequations}
\label{AboveThresholdDetuned}%
\begin{align}
\rho_{\mathrm{p}}  &  =\sqrt{1+\tilde{\Delta}^{2}}\text{, \ \ \ \ }\sin
\varphi_{\mathrm{p}}=\frac{\tilde{\Delta}}{\sigma}\left[  1-(\sigma^{2}%
-\tilde{\Delta}^{2})^{-1/2}\right]  \text{,} \label{AboveThresholdDetunedPump}%
\\
\rho_{\mathrm{s}}^{2}  &  =2\left(  \sqrt{\sigma^{2}-\tilde{\Delta}^{2}%
}-1\right)  \text{, \ \ \ \ }\varphi_{\mathrm{s}}=(\varphi_{\mathrm{p}}%
-\psi_{\Delta})/2, \label{AboveThresholdDetunedSignal}%
\end{align}
with $\psi_{\Delta}=\arctan\tilde{\Delta}$ defined in the interval
$[-\pi/2,\pi/2]$. The stability of this solution is much more difficult to
analyze than in the previous case; here, however, we are just interested in
one simple fact: The solution only exists for $\sigma>\sqrt{1+\tilde{\Delta
}^{2}}$, which is exactly the point where the previous solution with the
signal mode switched off becomes unstable. It is possible to show that indeed
this solution is stable in all its domain of existence\footnote{When also the
pump mode has some detuning, the situation is much more complicated. One can
find regions of the parameter space where two stable solutions coexist at the
same time (bistable behavior), or new instabilities where the solution becomes
unstable via an eigenvalue with non-zero imaginary part (typical of the
so-called Hopf bifurcations) what signals that the new stable solution to be
born at that point will be no longer stationary, but time--dependent. The
details can be found in \cite{Fabre90}.} \cite{Fabre90}. Hence, the only real
difference between this case and the resonant one is the fact the threshold is
increased to $\sigma=\sqrt{1+\tilde{\Delta}^{2}}$, and the pump mode, whose
phase was locked to that of the external laser when no detuning was present,
gains now a phase that depends on the system parameters --as does the signal
mode through the relation (\ref{AboveThresholdDetunedSignal})--.

As for the squeezing properties of the signal field, nothing is fundamentally
changed: It is simple to show by following the procedure developed in the
previous sections that there exists a squeezed quadrature which shows perfect
noise reduction exactly at threshold, the squeezing level degrading as one
moves away from this point \cite{Fabre90}. The only subtlety which appears is
that phase of the squeezed quadrature is no longer $\pi/2$, but depends on the
$\sigma$ and $\tilde{\Delta}$ parameters \cite{Fabre90}.

\section{Entanglement properties of the OPO\label{OPO}}

We study now the quantum properties of an OPO having distinguishable signal
and idler modes (in polarization or/and frequency); from now on we will call
OPO (in contrast to DOPO) to any such configuration. To this aim, let us work
again with the simplest situation: only a fundamental TEM$_{00}$ transverse
mode is present at the signal and idler resonances, so we are dealing with a
so-called \textit{two--mode OPO}.

We will see that, similarly to the DOPO, the OPO is also a threshold system.
Quantum properties can then be divided in two classes:
\end{subequations}
\begin{enumerate}
\item[(i)] Below threshold the signal and idler modes are in an EPR-like
entangled state. Although below threshold OPOs were studied before (see
\cite{McNeil83,Savage87} for example), Reid and Drummond were the first ones
to notice that the EPR \textit{gedanken} experiment could actually be
implemented with such devices \cite{Reid88,Reid89a,Reid90}. The first
experiment along this line was performed shortly after their proposal
\cite{Ou92}, and of course confirmed the quantum mechanical predictions.

\item[(ii)] Above threshold the signal and idler modes leak out of the cavity
as two bright beams with (ideally) perfectly correlated intensities, usually
referred to as \textit{twin beams}. Such intensity correlations between
macroscopically occupied beams were first predicted \cite{Reynaud87} and
experimentally observed \cite{Heidmann87} by Fabre et al. in Paris. Together
with the intensity correlations, and in order to preserve the corresponding
Heisenberg uncertainty principle, the phase difference between the twin beams
is diffusing. In many articles concerning above threshold OPOs such phase
diffusion is neglected; its effects on various properties of OPOs were first
studied by Lane, Reid, and Walls \cite{Lane88} and by Reid and Drummond
\cite{Reid88,Reid89b}. In order to introduce twin beams in simple terms,\ we
will not consider phase diffusion in this chapter; such effect will be treated
\ rigorously in Chapter \ref{2tmDOPO}\footnote{Indeed, this phenomenon offers
the perfect example of the usefulness of positive $P$ techniques based on the
master equation of the system (see Sections \ref{FPandLangevin} and
\ref{SchrodingerOpen}), as it cannot be studied by dealing with operator
techniques such as that introduced in Section \ref{HeisenbergOpen} because an
operator corresponding to the phase of the cavity modes is not available
\cite{Gerry05book,Mandel95book}.}.
\end{enumerate}

\subsection{The OPO within the positive $P$ representation\label{OPOmodel}}

Let us choose degenerate type II operation, as this is the ideal situation
from the experimental viewpoint, because then the signal and idler fields can
be analyzed using the same local oscillator field. In any case, the results
that we are going to show are independent of the particular type of OPO. The
two--mode OPO we are considering deals then with three cavity modes: The pump
mode which resonates at frequency $\omega_{\mathrm{p}}=2\omega_{0}$ and has
ordinary polarization, and the signal and idler modes which resonate at
frequency $\omega_{\mathrm{s}}=\omega_{\mathrm{i}}=\omega_{0}$ and have
extraordinary and ordinary polarization, respectively (Figure \ref{fOPO2}b).
Every mode is assumed to be in a TEM$_{00}$ transverse mode. The pump mode is
driven by an external, resonant laser beam, while no injection is used for the
signal and idler modes, whose associated photons are then generated via
spontaneous parametric down--conversion.

Similarly to the degenerate case, we move to the interaction picture defined
by the transformation operator%
\begin{equation}
\hat{U}_{\mathrm{0}}=\exp[\hat{H}_{\mathrm{0}}t/\mathrm{i}\hbar]\text{
\ \ \ \ with \ \ \ \ }\hat{H}_{\mathrm{0}}=\hbar\omega_{\mathrm{L}}(2\hat
{a}_{\mathrm{p}}^{\dagger}\hat{a}_{\mathrm{p}}+\hat{a}_{\mathrm{s}}^{\dagger
}\hat{a}_{\mathrm{s}}+\hat{a}_{\mathrm{i}}^{\dagger}\hat{a}_{\mathrm{i}}).
\label{IntPicOPO}%
\end{equation}
In this picture, the state $\hat{\rho}$ of the intracavity modes is written as
$\hat{\rho}_{\mathrm{I}}=\hat{U}_{\mathrm{0}}^{\dagger}\hat{\rho}\hat
{U}_{\mathrm{0}}$, and satisfies the master equation%
\begin{equation}
\frac{d\hat{\rho}_{\mathrm{I}}}{dt}=\left[  \chi\hat{a}_{\mathrm{p}}\hat
{a}_{\mathrm{s}}^{\dagger}\hat{a}_{\mathrm{i}}^{\dagger}+\mathcal{E}%
_{\mathrm{p}}a_{\mathrm{p}}^{\dagger}+\mathrm{H.c.},\hat{\rho}_{\mathrm{I}%
}\right]  +\sum_{j=\mathrm{p,s,i}}\gamma_{j}(2\hat{a}_{j}\hat{\rho
}_{\mathrm{I}}\hat{a}_{j}^{\dagger}-\hat{a}_{j}^{\dagger}\hat{a}_{j}\hat{\rho
}_{\mathrm{I}}-\hat{\rho}_{\mathrm{I}}\hat{a}_{j}^{\dagger}\hat{a}_{j}),
\label{OPOmaster}%
\end{equation}
where we take $\gamma_{\mathrm{s}}=\gamma_{\mathrm{i}}$, what is a reasonable
approximation from the experimental viewpoint. We take $\mathcal{E}%
_{\mathrm{p}}$ as a positive real as before. Once again, the nonlinear
coupling parameter $\chi$ is found from (\ref{ChiGen}) particularized to the
configuration that we are treating, which do not write explicitly because no
real simplifications are obtained in this case.

We use the positive $P$ representation for the density operator, what in this
case leads to the stochastic Langevin equations
\begin{subequations}
\label{OPOlangevin}%
\begin{align}
\dot{\alpha}_{\mathrm{p}}  &  =\mathcal{E}_{\mathrm{p}}-\gamma_{\mathrm{p}%
}\alpha_{\mathrm{p}}-\chi\alpha_{\mathrm{s}}\alpha_{\mathrm{i}},\\
\dot{\alpha}_{\mathrm{p}}^{+}  &  =\mathcal{E}_{\mathrm{p}}-\gamma
_{\mathrm{p}}\alpha_{\mathrm{p}}^{+}-\chi\alpha_{\mathrm{s}}^{+}%
\alpha_{\mathrm{i}}^{+},\\
\dot{\alpha}_{\mathrm{s}}  &  =-\gamma_{\mathrm{s}}\alpha_{\mathrm{s}}%
+\chi\alpha_{\mathrm{p}}\alpha_{\mathrm{i}}^{+}+\sqrt{\chi\alpha_{\mathrm{p}}%
}\xi(t),\\
\dot{\alpha}_{\mathrm{s}}^{+}  &  =-\gamma_{\mathrm{s}}\alpha_{\mathrm{s}}%
^{+}+\chi\alpha_{\mathrm{p}}^{+}\alpha_{\mathrm{i}}+\sqrt{\chi\alpha
_{\mathrm{p}}^{+}}\xi^{+}(t),\\
\dot{\alpha}_{\mathrm{i}}  &  =-\gamma_{\mathrm{s}}\alpha_{\mathrm{i}}%
+\chi\alpha_{\mathrm{p}}\alpha_{\mathrm{s}}^{+}+\sqrt{\chi\alpha_{\mathrm{p}}%
}\xi^{\ast}(t),\\
\dot{\alpha}_{\mathrm{i}}^{+}  &  =-\gamma_{\mathrm{s}}\alpha_{\mathrm{i}}%
^{+}+\chi\alpha_{\mathrm{p}}^{+}\alpha_{\mathrm{s}}+\sqrt{\chi\alpha
_{\mathrm{p}}^{+}}[\xi^{+}(t)]^{\ast},
\end{align}
where now $\xi(t)$ and $\xi^{+}(t)$ are independent complex Gaussian noises
with zero mean and non-zero correlations%
\end{subequations}
\begin{equation}
\langle\xi(t),\xi^{\ast}(t^{\prime})\rangle_{P}=\langle\xi^{+}(t),[\xi
^{+}(t^{\prime})]^{\ast}\rangle_{P}=\delta(t-t^{\prime}).
\label{ComplexNoiseStat}%
\end{equation}
Once again, it is easy to show that these equations have the same form both
within the Ito and Stratonovich interpretations, and hence we can directly use
the usual rules of calculus to analyze them.

In order to simplify the equations, we will make the same variable change as
in the DOPO,%
\begin{equation}
\tau=\gamma_{\mathrm{s}}t,\text{ \ \ \ \ }\beta_{m}(\tau)=\frac{\chi}%
{\gamma_{\mathrm{s}}\sqrt{\gamma_{\mathrm{p}}/\gamma_{m}}}\alpha_{m}(t),\text{
\ \ \ \ }\zeta_{m}(\tau)=\frac{1}{\sqrt{\gamma_{\mathrm{s}}}}\xi_{m}(t),
\label{ScaledVarOPO}%
\end{equation}
where with these definitions the new noises $\zeta(\tau)$ and $\zeta^{+}%
(\tau)$ satisfy the same statistical properties as $\xi(t)$ and $\xi^{+}(t)$,
but now with respect to the dimensionless time $\tau$. In terms of these
normalized variables, the Langevin equations (\ref{OPOlangevin}) read
(derivatives with respect to $\tau$ are understood)
\begin{subequations}
\label{scaledOPOlangevin}%
\begin{align}
\dot{\beta}_{\mathrm{p}}  &  =\kappa\lbrack\sigma-\beta_{\mathrm{p}}%
-\beta_{\mathrm{s}}\beta_{\mathrm{i}}],\\
\dot{\beta}_{\mathrm{p}}^{+}  &  =\kappa\lbrack\sigma-\beta_{\mathrm{p}}%
^{+}-\beta_{\mathrm{s}}^{+}\beta_{\mathrm{i}}^{+}],\\
\dot{\beta}_{\mathrm{s}}  &  =-\beta_{\mathrm{s}}+\beta_{\mathrm{p}}%
\beta_{\mathrm{i}}^{+}+g\sqrt{\beta_{\mathrm{p}}}\zeta(\tau),\\
\dot{\beta}_{\mathrm{s}}^{+}  &  =-\beta_{\mathrm{s}}^{+}+\beta_{\mathrm{p}%
}^{+}\beta_{\mathrm{i}}+g\sqrt{\beta_{\mathrm{p}}^{+}}\zeta^{+}(\tau),\\
\dot{\beta}_{\mathrm{i}}  &  =-\beta_{\mathrm{i}}+\beta_{\mathrm{p}}%
\beta_{\mathrm{s}}^{+}+g\sqrt{\beta_{\mathrm{p}}}\zeta^{\ast}(\tau),\\
\dot{\beta}_{\mathrm{i}}^{+}  &  =-\beta_{\mathrm{i}}^{+}+\beta_{\mathrm{p}%
}^{+}\beta_{\mathrm{s}}+g\sqrt{\beta_{\mathrm{p}}^{+}}[\zeta^{+}(\tau)]^{\ast
},
\end{align}
where the $\sigma$, $\kappa$, and $g$ parameters have the same definition as
before (\ref{ScaledParameters}).

In order to extract information from these equations we will proceed as
before, that is, by linearizing them around the classical stable solution.

\subsection{Classical analysis of the OPO\label{ClassiOPO}}

Using the same procedure as in Section (\ref{ClassiDOPO}), we retrieve the
classical evolution equations of the OPO, which read
\end{subequations}
\begin{subequations}
\begin{align}
\dot{\beta}_{\mathrm{p}}  &  =\kappa\lbrack\sigma-\beta_{\mathrm{p}}%
-\beta_{\mathrm{s}}\beta_{\mathrm{i}}],\\
\dot{\beta}_{\mathrm{s}}  &  =-\beta_{\mathrm{s}}+\beta_{\mathrm{p}}%
\beta_{\mathrm{i}}^{\ast},\\
\dot{\beta}_{\mathrm{i}}  &  =-\beta_{\mathrm{i}}+\beta_{\mathrm{p}}%
\beta_{\mathrm{s}}^{\ast}.
\end{align}
Similarly to (\ref{ClassicalDOPOeqs}), these equations admit two types of
solutions, one with the signal--idler modes switched off, and another with all
the modes switched on. Again, the first type exists for all the parameters,
but becomes unstable for $\sigma>1$, while the second type exists only for
$\sigma\geq1$ and is stable in all its domain of existence. Note however, that
in this case these equations have the continuous symmetry $\{\beta
_{\mathrm{s}}\rightarrow\exp(-\mathrm{i}\theta)\beta_{\mathrm{s}}%
,\beta_{\mathrm{i}}\rightarrow\exp(\mathrm{i}\theta)\beta_{\mathrm{i}}\}$, and
hence the phase difference between the signal and idler modes is not fixed by
the classical equations (a result of paramount importance for the results we
derive later).

Explicitly, the stationary solution below threshold is
\end{subequations}
\begin{equation}
\bar{\beta}_{\mathrm{p}}=\sigma,\text{ \ \ \ \ }\bar{\beta}_{\mathrm{s,i}}=0,
\label{BelowThresholdSolOPO}%
\end{equation}
while now the above threshold stationary solution reads%
\begin{equation}
\bar{\beta}_{\mathrm{p}}=1,\text{ \ \ \ \ }\bar{\beta}_{\mathrm{s,i}}%
=\sqrt{\sigma-1}\exp(\mp\mathrm{i}\theta), \label{AboveThresholdSolOPO}%
\end{equation}
being $\theta$ arbitrary. The stability of these solutions is easily analyzed
following the same steps as in (\ref{ClassiDOPO}), that is, analyzing the sign
of the real part of the eigenvalues of the stability matrix, and leads to the
scenario already commented above.

Let us finally discuss one point regarding what we could expect in real
experiments. Assume that we operate the OPO above threshold. Bright signal and
idler beams are then expected to start oscillating inside the cavity, their
relative phase $\theta$ chosen randomly according to the particular
fluctuations from which the beams are built up. We say then that $\theta$ is
chosen by \textit{spontaneous symmetry breaking}. In this chapter we assume
that after this spontaneous symmetry breaking process, the value of $\theta$
remains fixed; in reality, any source of noise (thermal, electronic, etc...)
can make it drift. We can assume all the classical sources of noise to be
properly stabilized (at least within the time--scale of the experiment), but
even then, quantum noise will remain. Nevertheless, for the purposes of this
introduction to the properties of OPOs we will neglect this \textit{quantum
phase diffusion} phenomenon, which will be properly addressed along the next chapters.

\subsection{The OPO below threshold: Signal--idler entanglement}

Linearizing now the Langevin equations (\ref{scaledOPOlangevin}) around the
below threshold classical solution (\ref{BelowThresholdSolOPO}) we can prove
that the quadratures of the signal and idler modes have correlations of the
EPR type. In particular, we will prove that for $\sigma>0$ these modes satisfy%
\begin{equation}
V^{\mathrm{out}}[(\hat{X}_{\mathrm{s}}-\hat{X}_{\mathrm{i}})/\sqrt{2}%
;\Omega]<1\text{ \ \ \ \ and \ \ \ \ }V^{\mathrm{out}}[(\hat{Y}_{\mathrm{s}%
}+\hat{Y}_{\mathrm{i}})/\sqrt{2};\Omega]<1, \label{PromisedJointSpectra}%
\end{equation}
which is to be regarded as the generalization of the entanglement criterion
(\ref{EntanglementCrit}) to the free--space fields leaking out of the cavity.
Moreover, we will show that these EPR like correlations become perfect exactly
at threshold.

To this aim, let us expand the stochastic amplitudes of the modes as
(\ref{QflucToAmps}); it is straightforward to see that the linear evolution of
the signal/idler fluctuations is decoupled from the pump fluctuations, and
obey the linear system%
\begin{equation}
\mathbf{\dot{b}}_{\mathrm{si}}=\mathcal{L}_{\mathrm{si}}\mathbf{b}%
_{\mathrm{si}}+g\sqrt{\sigma}\boldsymbol{\zeta}_{\mathrm{si}}(\tau),
\end{equation}
with\begin{subequations}
\begin{align}
\mathbf{b}_{\mathrm{si}}\left(  \tau\right)    & =\operatorname{col}\left[
b_{\mathrm{s}}(\tau),b_{\mathrm{s}}^{+}(\tau),b_{\mathrm{i}}(\tau
),b_{\mathrm{i}}^{+}(\tau)\right]  ,\\
\boldsymbol{\zeta}_{\mathrm{si}}(\tau)  & =\operatorname{col}\{\zeta
(\tau),\zeta^{+}(\tau),\zeta^{\ast}(\tau),[\zeta^{+}(\tau)]^{\ast}\},
\end{align}
being%
\end{subequations}
\begin{equation}
\mathcal{L}_{\mathrm{s}}=%
\begin{bmatrix}
-1 & 0 & 0 & \sigma\\
0 & -1 & \sigma & 0\\
0 & \sigma & -1 & 0\\
\sigma & 0 & 0 & -1
\end{bmatrix}
,
\end{equation}
a symmetric, real matrix with orthonormal eigensystem%
\begin{subequations}
\begin{align}
\lambda_{1}  &  =-1-\sigma,\text{ \ \ \ \ }\mathbf{v}_{1}=\frac{1}%
{2}\operatorname{col}(1,1,-1,-1),\\
\lambda_{2}  &  =-1-\sigma,\text{ \ \ \ \ }\mathbf{v}_{2}=\frac{1}%
{2}\operatorname{col}(1,-1,1,-1),\\
\lambda_{3}  &  =1-\sigma,\text{ \ \ \ \ \ \ }\mathbf{v}_{3}=\frac{1}%
{2}\operatorname{col}(1,1,1,1),\\
\lambda_{4}  &  =1-\sigma,\text{ \ \ \ \ \ \ }\mathbf{v}_{4}=\frac{1}%
{2}\operatorname{col}(1,-1,-1,1).
\end{align}
Once again, this linear system is solved by defining the projections
$c_{j}(\tau)=\mathbf{v}_{j}\cdot\mathbf{b}_{\mathrm{si}}\left(  \tau\right)
$, which obey the following simple linear stochastic equation with additive
noise
\end{subequations}
\begin{subequations}
\label{ProjLinLanBelowOPO}%
\begin{align}
\dot{c}_{1,4}  &  =\lambda_{1,4}c_{1,4}+\mathrm{i}g\sqrt{\sigma}\zeta
_{1,4}(\tau)\\
\dot{c}_{2,3}  &  =\lambda_{2,3}c_{2,3}+g\sqrt{\sigma}\zeta_{2,3}(\tau),
\end{align}
where the new real noises $\{\zeta_{j}(\tau)\}_{j=1,2,3,4}$ satisfy the
standard statistical properties (\ref{RealGaussStat}); the correlation
spectrum associated to each projection can be found then by particularizing
the corresponding expression in Appendix \ref{LinStoApp}.

In this case, the relation between the quadrature fluctuations of the signal
and idler modes, and the projections are easily found to be%
\end{subequations}
\begin{subequations}
\begin{align}
c_{1}(\tau)  &  =(\delta x_{\mathrm{s}}-\delta x_{\mathrm{i}})/2,\\
c_{2}(\tau)  &  =\mathrm{i}(\delta y_{\mathrm{s}}+\delta y_{\mathrm{i}})/2,\\
c_{3}(\tau)  &  =(\delta x_{\mathrm{s}}+\delta x_{\mathrm{i}})/2,\\
c_{4}(\tau)  &  =\mathrm{i}(\delta y_{\mathrm{s}}-\delta y_{\mathrm{i}})/2,
\end{align}
\end{subequations}
where the normalized quadratures are defined as in (\ref{ScaledQuadratures}).
Hence, we can evaluate the noise spectra of the different combinations of
signal--idler quadratures as%
\begin{subequations}
\begin{align}
V^{\mathrm{out}}[(\hat{X}_{\mathrm{s}}-\hat{X}_{\mathrm{i}})/\sqrt{2};\Omega]
&  =1+\frac{4}{g^{2}}\tilde{C}_{1}(\tilde{\Omega})=\frac{(1-\sigma)^{2}%
+\tilde{\Omega}^{2}}{(1+\sigma)^{2}+\tilde{\Omega}^{2}},\\
V^{\mathrm{out}}[(\hat{Y}_{\mathrm{s}}+\hat{Y}_{\mathrm{i}})/\sqrt{2};\Omega]
&  =1-\frac{4}{g^{2}}\tilde{C}_{2}(\tilde{\Omega})=\frac{(1-\sigma)^{2}%
+\tilde{\Omega}^{2}}{(1+\sigma)^{2}+\tilde{\Omega}^{2}},\\
V^{\mathrm{out}}[(\hat{X}_{\mathrm{s}}+\hat{X}_{\mathrm{i}})/\sqrt{2};\Omega]
&  =1+\frac{4}{g^{2}}\tilde{C}_{3}(\tilde{\Omega})=\frac{(1+\sigma)^{2}%
+\tilde{\Omega}^{2}}{(1-\sigma)^{2}+\tilde{\Omega}^{2}},\\
V^{\mathrm{out}}[(\hat{Y}_{\mathrm{s}}-\hat{Y}_{\mathrm{i}})/\sqrt{2};\Omega]
&  =1-\frac{4}{g^{2}}\tilde{C}_{4}(\tilde{\Omega})=\frac{(1+\sigma)^{2}%
+\tilde{\Omega}^{2}}{(1-\sigma)^{2}+\tilde{\Omega}^{2}}.
\end{align}
\end{subequations}
Note that the two first (last) expressions coincide with the noise spectrum of
the squeezed (anti--squeezed) quadrature of the DOPO below threshold, see
(\ref{SpectraDOPO}). Hence, we have proved exactly what we promised
(\ref{PromisedJointSpectra}): Below threshold, the signal and idler fields
have correlated \textsf{X} quadratures, and anticorrelated \textsf{Y}
quadratures. This correlations become perfect at zero noise frequency when
working exactly at threshold. Once again, it can be proved that this is a flaw
of the linearization procedure, and nonlinear corrections in the $g$ parameter
make them become finite \cite{Dechoum04}.

\subsection{The OPO above threshold: Twin beams\label{TwinBeams}}

\begin{figure}
[t]
\begin{center}
\includegraphics[
height=2.3583in,
width=4.8231in
]%
{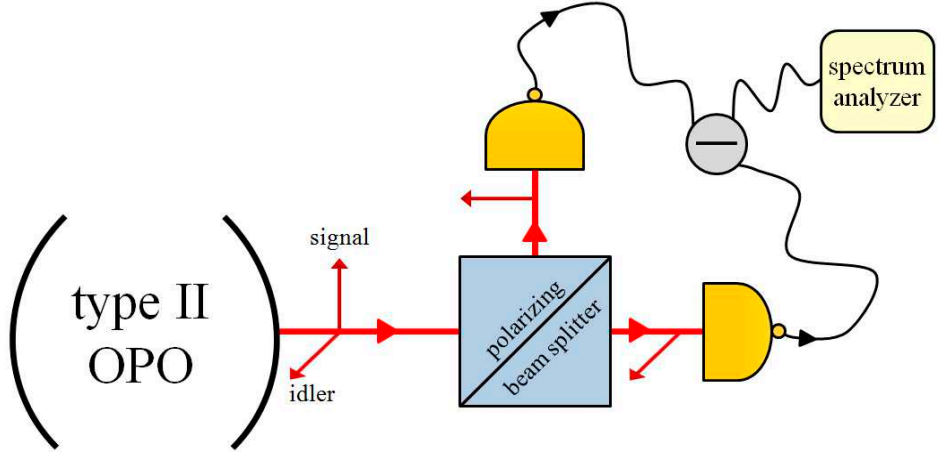}%
\caption{Detection scheme designed for the observation of the intensity
correlations between the signal and idler beams coming out from a type II OPO
above threshold.}%
\label{fOPO8}%
\end{center}
\end{figure}

Consider now the OPO operating above threshold. We are going to prove that the
signal and idler beams, which have now a non-zero mean field, have perfectly
correlated intensities; they are what we will call twin beams. In Figure
\ref{fOPO8} we show the scheme of a typical experiment designed to look for
such intensity correlations. The signal and idler beams are first separated by
a polarizing beam splitter, and then measured with independent photodetectors;
the corresponding electric signals are then subtracted, and the power spectrum
of this \textit{difference photocurrent} is finally analyzed.%

Using the approach that we developed in Section \ref{RealPhotodetection}, it
is not difficult to show that, similarly to the homodyne detection spectrum
(\ref{PdSpectrum}), in this case the power spectrum of the difference
photocurrent is also formed by a shot noise contribution, plus a part which
depends on the state of the system (and goes to zero for vacuum or coherent
states). In particular, assuming the state of the system to be stationary, one
easily gets%
\begin{equation}
\frac{P_{\mathrm{D}}(\Omega)}{P_{\mathrm{shot}}}=1+\frac{1}{\langle\hat
{n}_{\mathrm{s,out}}\rangle+\langle\hat{n}_{\mathrm{i,out}}\rangle}%
\int_{-\infty}^{+\infty}dt^{\prime}\langle:\delta\hat{n}_{\mathrm{D,out}%
}(t)\delta\hat{n}_{\mathrm{D,out}}(t+t^{\prime}):\rangle e^{-\mathrm{i}\Omega
t^{\prime}}, \label{PidSpectrum}%
\end{equation}
where we have defined the number--difference operator%
\begin{equation}
\hat{n}_{\mathrm{D,out}}=\hat{n}_{\mathrm{s,out}}-\hat{n}_{\mathrm{i,out}%
}\text{.}%
\end{equation}
We can rewrite this expression in terms of intracavity operators by using
relations (\ref{Out-Intra}). After doing this, and given that all the
expectation values are in normal order, we can directly use the positive $P$
representation to evaluate them, just as we did in the previous sections.
Taking also into account the definition of the normalized variables
(\ref{ScaledVarOPO}), this leads to%
\begin{equation}
\frac{P_{\mathrm{D}}(\Omega)}{P_{\mathrm{shot}}}=1+\frac{2}{g^{2}(\langle
n_{\mathrm{s}}\rangle+\langle n_{\mathrm{i}}\rangle)}\int_{-\infty}^{+\infty
}dt^{\prime}\langle\delta n_{\mathrm{D}}(t)\delta n_{\mathrm{D}}(t+t^{\prime
})\rangle_{P}e^{-\mathrm{i}\Omega t^{\prime}}, \label{NormPidSpectrum}%
\end{equation}
where we have defined the normalized versions of the different operators
within the positive $P$ representation:%
\begin{equation}
n_{\mathrm{s}}=\beta_{\mathrm{s}}^{+}\beta_{\mathrm{s}}\text{, \ \ \ \ }%
n_{\mathrm{i}}=\beta_{\mathrm{i}}^{+}\beta_{\mathrm{i}}\text{, \ \ \ \ and
\ \ \ \ }n_{\mathrm{D}}=\beta_{\mathrm{s}}^{+}\beta_{\mathrm{s}}%
-\beta_{\mathrm{i}}^{+}\beta_{\mathrm{i}}\text{.}%
\end{equation}

The last manipulation that we need is to consider the linearized version of
this expression. To this aim, we can expand the coherent amplitudes as usual
(\ref{QflucToAmps}), that is, as some quantum fluctuations around the
classical stationary solution (\ref{AboveThresholdSolOPO}), and then expand
the spectrum (\ref{NormPidSpectrum}) up to the lowest order in the
fluctuations. The final result is actually very intuitive:
\begin{equation}
\frac{P_{\mathrm{D}}(\Omega)}{P_{\mathrm{shot}}}=V^{\mathrm{out}}[(\hat
{X}_{\mathrm{s}}^{-\theta}-\hat{X}_{\mathrm{i}}^{\theta})/\sqrt{2};\Omega],
\end{equation}
that is, the linearized version of the spectrum (\ref{PidSpectrum})
corresponds to the noise spectrum of the difference between the amplitude
quadratures of the signal and idler modes. This result comes from the fact
that the linearized version of the number--difference fluctuations itself is
proportional to such quadrature--difference fluctuations, that is,%
\begin{equation}
\delta n_{\mathrm{D}}\simeq\bar{\beta}_{\mathrm{s}}^{\ast}b_{\mathrm{s}}%
+\bar{\beta}_{\mathrm{s}}b_{\mathrm{s}}^{+}-\bar{\beta}_{\mathrm{i}}^{\ast
}b_{\mathrm{i}}-\bar{\beta}_{\mathrm{i}}b_{\mathrm{i}}^{+}=\sqrt{\sigma
-1}(\delta x_{\mathrm{s}}^{-\theta}-\delta x_{\mathrm{i}}^{\theta});
\end{equation}
this actually makes sense, as we already saw in Chapter
\ref{HarmonicOscillator} that the number and phase fluctuations of the
harmonic oscillator are related to the fluctuations of its amplitude and phase
quadratures, provided that the mean value of the oscillator's amplitude is
large enough (large enough $\sigma$ in this case).

Hence, proving that the joint quadrature $(\hat{X}_{\mathrm{s}}^{-\theta}%
-\hat{X}_{\mathrm{i}}^{\theta})/\sqrt{2}$ is squeezed is equivalent to proving
intensity correlations. To do so we follow the procedure of the previous
sections. First, we write the linear system obeyed by the quantum
fluctuations, which for the above threshold OPO reads%
\begin{equation}
\mathbf{\dot{b}}=\mathcal{L}\mathbf{b}+g\boldsymbol{\zeta}(\tau),
\label{LinLanAboveOPO}%
\end{equation}
where $\mathbf{b}=\operatorname{col}(b_{\mathrm{p}},b_{\mathrm{p}}%
^{+},b_{\mathrm{s}},b_{\mathrm{s}}^{+},b_{\mathrm{i}},b_{\mathrm{i}}^{+})$,
$\boldsymbol{\zeta}(\tau)=\operatorname{col}\{0,0,\zeta(\tau),\zeta^{+}%
(\tau),\zeta^{\ast}(\tau),[\zeta^{+}(\tau)]^{\ast}\}$, and the linear
stability matrix reads in this case%
\begin{equation}
\mathcal{L}=%
\begin{bmatrix}
-\kappa & 0 & -\kappa\rho e^{\mathrm{i}\theta} & 0 & -\kappa\rho
e^{-\mathrm{i}\theta} & 0\\
0 & -\kappa & 0 & -\kappa\rho e^{-\mathrm{i}\theta} & 0 & -\kappa\rho
e^{\mathrm{i}\theta}\\
\rho e^{-\mathrm{i}\theta} & 0 & -1 & 0 & 0 & 1\\
0 & \rho e^{\mathrm{i}\theta} & 0 & -1 & 1 & 0\\
\rho e^{\mathrm{i}\theta} & 0 & 0 & 1 & -1 & 0\\
0 & \rho e^{-\mathrm{i}\theta} & 1 & 0 & 0 & -1
\end{bmatrix}
,
\end{equation}
where $\rho=\sqrt{\sigma-1}$. Even though this matrix is not Hermitian, it is
simple to check that the vector%
\begin{equation}
\mathbf{v}=\frac{1}{2}\operatorname{col}(0,0,e^{-\mathrm{i}\theta
},e^{\mathrm{i}\theta},-e^{\mathrm{i}\theta},-e^{-\mathrm{i}\theta}),
\label{v}%
\end{equation}
is an eigenvector of it with eigenvalue $\lambda=-2$ irrespective of the
system parameters, that is, $\mathbf{v}^{\ast}\mathcal{L}=-2\mathbf{v}^{\ast}%
$. On the other hand, the projection of the fluctuations $\mathbf{b}$ onto
this eigenvector is%
\begin{equation}
c(\tau)=\mathbf{v}^{\ast}\cdot\mathbf{b}(\tau)=[\delta x_{\mathrm{s}}%
^{-\theta}(\tau)-\delta x_{\mathrm{i}}^{\theta}(\tau)]/2;
\end{equation}
then, by projecting the linear system (\ref{LinLanAboveOPO}) onto $\mathbf{v}$
we get the evolution equation for this projection, which reads%
\begin{equation}
\dot{c}=-2c+\mathrm{i}g\eta(t),
\end{equation}
where $\eta(t)$ is a real noise which satisfies the usual statistics
(\ref{RealGaussStat}). As usual this is a linear stochastic equation with
additive noise, see Appendix \ref{LinStoApp}. Hence, the noise spectrum we are
looking for can be finally written as%
\begin{equation}
V^{\mathrm{out}}[(\hat{X}_{\mathrm{s}}^{-\theta}-\hat{X}_{\mathrm{i}}^{\theta
})/\sqrt{2};\Omega]=1+\frac{4}{g^{2}}\tilde{C}(\tilde{\Omega})=\frac
{\tilde{\Omega}^{2}}{4+\tilde{\Omega}^{2}},
\end{equation}
where once again we have used the results of Appendix \ref{LinStoApp} for the
correlation spectrum of $c(\tau)$.

As stated, the signal and idler modes have perfect amplitude quadrature
correlations above threshold (for any value of the system parameters) and at
zero noise frequency, what is equivalent to perfect intensity correlations as
explained above. The shape of the spectrum is the usual Lorentzian that we
have found for all the previous perfectly squeezed quadratures; however, as we
show in Section ..., in this case this shape is not an unrealistic artifact of
the linearization: the intensities (or amplitude quadratures) of the generated
twin beams can be indeed perfectly correlated.

Let us finally remark that the vector (\ref{v}) is not the only one with a
definite eigenvalue in all parameter space; it is simple to check that the
vector%
\begin{equation}
\mathbf{v}_{0}=\frac{\mathrm{i}}{2}\operatorname{col}(0,0,e^{-\mathrm{i}%
\theta},-e^{\mathrm{i}\theta},-e^{\mathrm{i}\theta},e^{-\mathrm{i}\theta}),
\end{equation}
satisfies $\mathbf{v}_{0}^{\ast}\mathcal{L}=0$, and hence has zero eigenvalue.
This zero eigenvalue is an indicator that there is a direction of phase space
along which quantum noise can act freely. It is actually easy to realize which
is the corresponding variable of the system affected by this undamped quantum
noise; one just needs to evaluate the projection of the quantum fluctuations
$\mathbf{b}$ onto $\mathbf{v}_{0}$, what leads to
\begin{equation}
c_{0}(\tau)=\mathbf{v}_{0}^{\ast}\cdot\mathbf{b}(\tau)=[\delta y_{\mathrm{s}%
}^{-\theta}(\tau)-\delta y_{\mathrm{i}}^{\theta}(\tau)]/2,
\end{equation}
that is, the difference between the phase quadratures of the signal and idler
modes is the variable we are looking for. In other words, as within the
linearized description the fluctuations of the phase quadrature are related to
the fluctuations of the actual phase of the mode, the phase difference
$\theta$ that we have taken as a fixed quantity of the above threshold
solution, is actually dynamical, it should diffuse with time owed to quantum
noise (as already advanced above). Fortunately, it can be proved that adding
this phase diffusion doesn't change the results we have shown concerning
intensity correlations \cite{Lane88}; it however affects other aspects of
above threshold OPOs that we haven't tackled in this introduction, and we will
need to treat the phase diffusion in depth in the next chapters. 

%% file: MultiOPOsFO.tex
In the previous chapter we developed the quantum model of an optical
parametric oscillator, showing that it is a threshold system capable of
generating squeezed and entangled states of light. In order to prove so, we
simplified the system as much as possible, particularly assuming that only a
single three--wave mixing process is selected, that is, that pump photons can
only be \textquotedblleft broken\textquotedblright\ into one particular
signal--idler pair (an extraordinary, frequency degenerate, TEM$_{00}$ pair of
photons in the case of the DOPO, for example). This is however not general
enough, as pump photons may have many different \textquotedblleft decay
channels\textquotedblright. We will call \textit{multi--mode OPOs} to such
OPOs in which pump photons have not a single decay channel.

The first type of multi--mode OPOs that were ever studied were DOPOs in which
several transverse modes could resonate at the signal frequency, just as we
allowed in the derivation of the three--wave mixing Hamiltonian in Section
\ref{GeneralOPOModel}. In this case, the different decay channels available
for the pump photons correspond to opposite orbital angular momentum
signal--idler pairs, as can be appreciated in expression
(\ref{GenTWMHamiltonian}). The study of these \textit{spatial multi--mode
OPOs} was first motivated either by the desire to understand the spatial
correlations of the signal beam \cite{Kolobov89a,Kolobov89b,Kolobov91}, to
model real experiments \cite{LaPorta91}, or to describe quantum mechanically
the phenomenon of pattern formation \cite{Lugiato92}. Shortly after these
preliminary studies, Lugiato and collaborators
\cite{Lugiato93,Grynberg93,Gatti95,Lugiato95} came to realize that the
phenomenology appearing in this system was far more rich than that of their
single--mode counterpart, and their seminal ideas gave rise to the field of
\textit{quantum imaging}.

A different type of multi--mode OPOs are the so-called \textit{temporal
multi--mode OPOs}. In order to understand which are the different decay
channels in this case, let us consider the following example. Assume that the
OPO is tuned so that the frequency degenerate type I process (Figure
\ref{fOPO2}a) is the one with optimal phase matching, that is $c\Delta
k_{\mathrm{D-I}}=2\omega_{0}\left[  n_{\mathrm{o}}(2\omega_{0})-n_{\mathrm{e}%
}(\omega_{0})\right]  =0$. We can evaluate the \textit{phase mismatch}
associated to the nondegenerate process (Figure \ref{fOPO2}c) as%
\begin{equation}
c\Delta k_{\mathrm{ND-I}}=2\omega_{0}n_{\mathrm{o}}(2\omega_{0})-(\omega
_{0}+\Omega)n_{\mathrm{e}}(\omega_{0}+\Omega)-(\omega_{0}-\Omega
)n_{\mathrm{e}}(\omega_{0}-\Omega)\simeq-2\Omega^{2}\left.  \frac{dn_{\mathrm{e}}}{d\omega}\right\vert
_{\omega=\omega_{0}},
\end{equation}
where we have assumed $\Omega\ll\omega_{0}$ and made a series expansion of the
refractive indices around $\omega=\omega_{0}$. It is not difficult to show
that for typical values of the dispersion $\left.  dn_{\mathrm{e}}%
/d\omega\right\vert _{\omega=\omega_{0}}$ and taking into account that
$\Omega$ can be of the order of the free spectral range of the cavity, $\Delta
k_{\mathrm{ND-I}}^{-1}$ is usually much larger than the typical lengths of
nonlinear crystals. In other words, even though not optimally, the phase
matching condition $\Delta k_{\mathrm{ND-I}}\ll l_{\mathrm{c}}^{-1}$ can still
be satisfied by nondegenerate processes, and hence pump photons have many
decay channels open: The degenerate and all the nondegenerate ones allowed by
energy and momentum (phase--matching) conservation. Even though temporal
multi--mode parametric down--conversion was studied in free space a long time
ago \cite{Slusher87,Shelby92,Serkland95}, it wasn't until a few years ago that
these kind of processes were considered in cavity systems
\cite{Valcarcel06,Patera10,Nielsen07}.

This thesis is devoted to the study of multi--mode OPOs. Throughout the rest
of the thesis we will show that both their quantum and classical properties
can be understood in terms of two basic phenomena: \textit{pump clamping} and
\textit{spontaneous symmetry breaking}. In this Chapter we shall introduce
these two phenomena in a general, intuitive way by making use of simple examples.

\section{Pump clamping as a resource for noncritically squeezed
light\label{OPOfreezing}}

In this section we would like to introduce the phenomenon of pump clamping
which was first predicted in \cite{Navarrete09}.

\subsection{Introducing the phenomenon through the simplest
model\label{OPOfreezing2Modes}}

In order to introduce the phenomenon of pump clamping we will consider the
most simple multi--mode scenario: Pump photons can decay through two
degenerate type I channels. As we will argue at the end, the results are
trivially generalized to any other combination of decay channels. Hence,
consider three intracavity modes: The pump mode, with annihilation operator
denoted by $\hat{a}_{0}$, and two signal modes whose annihilation operators we
denote by $\hat{a}_{1}$ and $\hat{a}_{2}$; these are connected via the
Hamiltonian%
\begin{equation}
\hat{H}_{\mathrm{c}}=\mathrm{i}\hbar(\chi_{1}\hat{a}_{0}\hat{a}_{1}^{\dagger
2}+\chi_{2}\hat{a}_{0}\hat{a}_{2}^{\dagger2})+\mathrm{H.c.}%
\end{equation}
The dependence of the coupling parameters $\chi_{j}$ on the system parameters
depends on the particular processes underlaying this basic scheme\footnote{For
example, if the cavity is confocal, so that even and odd families are
spectrally separated by half the free spectral range inside the cavity, we
could assume that the even families resonate at the signal frequency, and
modes 1 and 2 would correspond then to the $L_{00}(k_{\mathrm{s}}%
;\mathbf{r}_{\perp},z)$ and the $L_{10}(k_{\mathrm{s}};\mathbf{r}_{\perp},z)$
transverse modes, as long as we could forget about the rest of the transverse
modes.}; we will assume here that $\chi_{1}>\chi_{2}$ for definiteness, that
is, that the decay channel corresponding to mode 1 is favoured. Moreover, we
will assume that the decay rates for the signal modes are the same, that is,
$\gamma_{1}=\gamma_{2}=\gamma_{\mathrm{s}}$, and that the system is pumped by
the external laser exactly at resonance. Working in the interaction picture
defined in the previous chapter, and scaling the variables as we did there
(see Section \ref{ModelDOPO}), it is simple to show that the positive $P$
Langevin equations associated to this system read
\begin{subequations}
\label{2channelDOPOeqs}%
\begin{align}
\dot{\beta}_{0}  &  =\kappa\lbrack\sigma-\beta_{0}-\beta_{1}^{2}/2-r\beta
_{2}^{2}/2],\\
\dot{\beta}_{0}^{+}  &  =\kappa\lbrack\sigma-\beta_{0}^{+}-\beta_{1}%
^{+2}/2-r\beta_{2}^{+2}/2],\\
\dot{\beta}_{1}  &  =-\beta_{1}+\beta_{0}\beta_{1}^{+}+g\sqrt{\beta_{0}}%
\zeta_{1}(\tau),\\
\dot{\beta}_{1}^{+}  &  =-\beta_{1}^{+}+\beta_{0}^{+}\beta_{1}+g\sqrt
{\beta_{0}^{+}}\zeta_{1}^{+}(\tau),\\
\dot{\beta}_{2}  &  =-\beta_{2}+r\beta_{0}\beta_{2}^{+}+g\sqrt{r\beta_{0}%
}\zeta_{2}(\tau),\\
\dot{\beta}_{2}^{+}  &  =-\beta_{2}^{+}+r\beta_{0}^{+}\beta_{2}+g\sqrt
{r\beta_{0}^{+}}\zeta_{2}^{+}(\tau),
\end{align}
where we have defined the parameter $r=\chi_{2}/\chi_{1}<1$, and all the noises are real, independent, and satisfy the usual statistical properties
(\ref{RealGaussStat}).

\subsubsection{Classical emission}

Let's first analyze the classical emission properties of the system. In
particular, let us study its classical stationary solutions. As we argued in
Section \ref{ClassiDOPO}, the classical equations corresponding to the
Langevin equations (\ref{2channelDOPOeqs}) are
\end{subequations}
\begin{subequations}
\label{2channelDOPOclasseqs}%
\begin{align}
\dot{\beta}_{0}  &  =\kappa\lbrack\sigma-\beta_{0}-\beta_{1}^{2}/2-r\beta
_{2}^{2}/2],\\
\dot{\beta}_{1}  &  =-\beta_{1}+\beta_{0}\beta_{1}^{\ast}%
,\label{2channelDOPOclasseqsB}\\
\dot{\beta}_{2}  &  =-\beta_{2}+r\beta_{0}\beta_{2}^{\ast}.
\label{2channelDOPOclasseqsC}%
\end{align}
Just as the single--mode DOPO, these equations posses a stationary solution
with the signal modes switched off which is given by%
\end{subequations}
\begin{equation}
\bar{\beta}_{0}=\sigma\text{, \ \ \ \ }\beta_{1}=\beta_{2}=0.
\label{2ChannelDOPObelow}%
\end{equation}
On the other hand, the first particularity that we can see in this case is
that a stationary solution with the modes 1 and 2 switched on at the same time
does not exists, as it would imply $|\bar{\beta}_{0}|=1$ and $|\bar{\beta}%
_{0}|=r$ simultaneously, what is not possible (the pump amplitude cannot take
two different values at the same time!). Hence, we are left with two different
stationary solutions, one given by%
\begin{equation}
\bar{\beta}_{0}=1\text{, \ \ \ \ }\bar{\beta}_{1}=\sqrt{2\left(
\sigma-1\right)  }\text{, \ \ \ \ }\bar{\beta}_{2}=0,
\label{2ChannelDOPOabove1}%
\end{equation}
which has the second mode switched off, and another given by%
\begin{equation}
\bar{\beta}_{0}=r^{-1}\text{, \ \ \ \ }\bar{\beta}_{2}=\sqrt{2\left(
\sigma-r^{-1}\right)  }\text{, \ \ \ \ }\bar{\beta}_{1}=0,
\label{2ChannelDOPOabove2}%
\end{equation}
which has the first mode switched off. Note that these solutions have
thresholds located at $\sigma=1$ and $\sigma=r^{-1}>1$, respectively, and
hence the lower threshold corresponds to the solution with mode 1 switched on.

On the other hand, the stability matrix associated to equations
(\ref{2channelDOPOclasseqs}) is given by%
\begin{equation}
\mathcal{L}=%
\begin{bmatrix}
-\kappa & 0 & -\kappa\bar{\beta}_{1} & 0 & -\kappa r\bar{\beta}_{2} & 0\\
0 & -\kappa & 0 & -\kappa\bar{\beta}_{1}^{\ast} & 0 & -\kappa r\bar{\beta}%
_{2}^{\ast}\\
\bar{\beta}_{1}^{\ast} & 0 & -1 & \bar{\beta}_{0} & 0 & 0\\
0 & \bar{\beta}_{1} & \bar{\beta}_{0}^{\ast} & -1 & 0 & 0\\
r\bar{\beta}_{2}^{\ast} & 0 & 0 & 0 & -1 & r\bar{\beta}_{0}\\
0 & r\bar{\beta}_{2} & 0 & 0 & r\bar{\beta}_{0}^{\ast} & -1
\end{bmatrix}
. \label{2ChannelDOPOstability}%
\end{equation}
It is then trivial to show that the eigenvalues of this matrix particularized
to the trivial solution (\ref{2ChannelDOPObelow}) are all real and negative
except for the following two: $\lambda_{1}=-1+\sigma$ and $\lambda
_{2}=-1+r\sigma$. Hence, for $\sigma>1$ this solution becomes unstable, just
as expected from the previous chapter: (\ref{2ChannelDOPObelow}) is the
so-called below threshold solution. Similarly, it is very simple to check that
all the eigenvalues of $\mathcal{L}$ particularized to the solution
(\ref{2ChannelDOPOabove1}) have negative real part for $\sigma>1$, and hence,
this solution is stable in all its domain of existence. Finally, it is
straightforward to find that the solution (\ref{2ChannelDOPOabove2}) is always
unstable, as $\mathcal{L}$ has in this case the eigenvalue $\lambda=-1+r^{-1}$
which is always positive.

Hence, we arrive to the conclusion that only one of the two modes can be
switched on by increasing the external pump injection, the one with the
largest coupling to the pump mode. In other words, the nonlinear competition
is won by the mode with the lowest threshold. Indeed, there is a simple
physical explanation for this. The fluctuations of the signal modes are not
fed directly through the injection parameter $\sigma$, but through the pump
amplitude $\bar{\beta}_{0}$, as can be appreciated in
(\ref{2channelDOPOclasseqsB},\ref{2channelDOPOclasseqsC}). Now, once we get to
$\sigma=1$ ---the threshold of mode 1--- the pump mode gets clamped to
$\bar{\beta}_{0}=1$ no matter how much we increase $\sigma$ (all the remaining
pump injection is used to make the signal mode oscillate). Hence, as mode 2
feels changes in the system only through the pump mode, it will feel as if the
system is frozen to $\sigma=1$, no matter how much we increase the actual
$\sigma$. Hence, the instability point leading to its activation, which
requires $\bar{\beta}_{0}=r^{-1}$, cannot be reached, and hence it will stay
switched off forever.

\subsubsection{Quantum properties}

Classically, the pump mode and mode 1 behave, respectively, as the pump and
signal modes of the single--mode DOPO (see Section \ref{ClassiDOPO}).
Following the analysis of Section \ref{QuantumDOPO}, it is straightforward to
show that the quantum properties of these modes are also just like those that
we found for the single--mode DOPO, that is, the pump mode is always in a
coherent state, while mode 1 is squeezed only when working near to the
threshold $\sigma=1$. Hence we focus on the quantum properties of the
non-amplified mode, which is indeed the one showing the new phenomenon
appearing in the multi--mode case: No matter how much we increase $\sigma$,
from the point of view of mode 2 the system is frozen to its state at
$\sigma=1$. As we are about to see, this is a very interesting effect from the
quantum point of view.

In order to study the quantum properties of mode 2, we use the linearization
procedure that we introduced in the previous chapter, that is, we expand all
the stochastic amplitudes as $\beta_{j}(\tau)=\bar{\beta}_{j}+b_{j}(\tau)$ and
$\beta_{j}^{+}(\tau)=\bar{\beta}_{j}^{\ast}+b_{j}^{+}(\tau)$, and assume that
the $b$'s are order $g$ quantities. Then, given that $\bar{\beta}_{2}=0$
irrespective of the system parameters, the stability matrix
(\ref{2ChannelDOPOstability}) shows that\ the fluctuations of mode 2 get
decoupled from the rest of the modes; in particular, and taking into account
that $\bar{\beta}_{0}$ is real and positive for all the system parameters,
they satisfy the linear system%
\begin{equation}
\mathbf{\dot{b}}_{2}=\mathcal{L}_{2}\mathbf{b}_{2}+g\sqrt{r\bar{\beta}_{0}%
}\boldsymbol{\zeta}_{2}(\tau), \label{2ChannelDOPOlinLan}%
\end{equation}
with $\mathbf{b}_{2}\left(  \tau\right)  =\operatorname{col}\left[  b_{2}%
(\tau),b_{2}^{+}(\tau)\right]  $, $\boldsymbol{\zeta}_{2}(\tau
)=\operatorname{col}\left[  \zeta_{2}(\tau),\zeta_{2}^{+}(\tau)\right]  $,
being%
\begin{equation}
\mathcal{L}_{2}=%
\begin{bmatrix}
-1 & r\bar{\beta}_{0}\\
r\bar{\beta}_{0} & -1
\end{bmatrix}
,
\end{equation}
a real, symmetric matrix with orthonormal eigensystem%
\begin{equation}
\left\{  \lambda_{\pm}=-\left(  1\mp r\bar{\beta}_{0}\right)  ,\mathbf{v}%
_{\pm}=\frac{1}{\sqrt{2}}\operatorname{col}(1,\pm1)\right\}  .
\end{equation}
Note that, as commented, the evolution of the fluctuations of mode 2 does not
depend directly on $\sigma$, but on the stationary value of the pump mode
$\bar{\beta}_{0}$. Also, note that the eigenvectors $\mathbf{v}_{\pm}$ are
exactly the same as those which appeared in the single--mode DOPO below
threshold (\ref{eigenDOPO}), and hence the projections $c_{\pm}=\mathbf{v}%
_{\pm}\cdot\mathbf{b}_{2}$ are related to the quadratures of mode 2 by
\begin{subequations}
\label{ProjToQuadDOPO}%
\begin{align}
c_{+}(\tau)  &  =\left[  b_{2}(\tau)+b_{2}^{+}(\tau)\right]  /\sqrt{2}=\delta
x_{2}(\tau)/\sqrt{2},\\
c_{-}(\tau)  &  =\left[  b_{2}(\tau)-b_{2}^{+}(\tau)\right]  /\sqrt
{2}=\mathrm{i}\delta y_{2}(\tau)/\sqrt{2},
\end{align}
so that the noise spectrum of the \textsf{X} and \textsf{Y}\ quadratures of
mode 2 can be evaluated as%
\end{subequations}
\begin{subequations}
\begin{align}
V^{\mathrm{out}}(\hat{X}_{2};\Omega)  &  =1+\frac{4}{g^{2}}\tilde{C}%
_{+}(\tilde{\Omega}),\\
V^{\mathrm{out}}(\hat{Y}_{2};\Omega)  &  =1-\frac{4}{g^{2}}\tilde{C}%
_{-}(\tilde{\Omega}).
\end{align}
On the other hand, projecting the linear system (\ref{2ChannelDOPOlinLan})
onto the eigenvectors $\mathbf{v}_{\pm}$ we get the following evolution
equations for the projections:%
\end{subequations}
\begin{equation}
\dot{c}_{j}=\lambda_{j}c_{j}+g\sqrt{r\bar{\beta}_{0}}\zeta_{j}(\tau),
\label{2ChannelDOPOprojLinLanBelow}%
\end{equation}
where $\zeta_{\pm}(\tau)=\left[  \zeta_{2}(\tau)\pm\zeta_{2}^{+}(\tau)\right]
/\sqrt{2}$ are new independent real noises satisfying the usual statistical
properties. The general treatment of this type of equations can be found in
Appendix \ref{LinStoApp}. Taking in particular the expression of their
associated correlation spectrum (\ref{CorrSpectrum}), we can finally write
\begin{subequations}
\label{2ChannelDOPOnoiseVar}%
\begin{align}
V^{\mathrm{out}}(\hat{X}_{2};\Omega)  &  =\frac{(1+r\bar{\beta}_{0}%
)^{2}+\tilde{\Omega}^{2}}{(1-r\bar{\beta}_{0})^{2}+\tilde{\Omega}^{2}},\\
V^{\mathrm{out}}(\hat{Y}_{2};\Omega)  &  =\frac{(1-r\bar{\beta}_{0}%
)^{2}+\tilde{\Omega}^{2}}{(1+r\bar{\beta}_{0})^{2}+\tilde{\Omega}^{2}}.
\label{2ChannelDOPOnoiseVarY}%
\end{align}
\end{subequations}
Note first that mode 2 is in a minimum uncertainty state as
\begin{equation}
V^{\mathrm{out}%
}(\hat{X}_{2};\Omega)V^{\mathrm{out}}(\hat{Y}_{2};\Omega)=1
\end{equation}.
On the other hand, for $\sigma<1$ we have $\bar{\beta}_{0}=\sigma$, while for $\sigma>1$
the pump mode is clamped to $\bar{\beta}_{0}=1$. Hence, as $\sigma$ increases,
quantum noise gets reduced in the \textsf{Y} quadrature (and correspondingly
increased in the \textsf{X} quadrature), so that mode 2 is in a squeezed
state. Maximum squeezing is obtained at zero noise frequency and for
$\sigma>1$, where we have%
\begin{equation}
V^{\mathrm{out}}(\hat{Y}_{2};0)=(1-r)^{2}/(1+r)^{2}<1.
\label{2ChannelDOPOZeroNoiseY}%
\end{equation}
The interesting feature about this squeezing is that, owed to the fact that
the system is frozen from the point of view of mode 2, it is independent of
the system parameters (above threshold), a reason why we call it
\textit{noncritical squeezing}; this is in contrast to the squeezing
appearing in the single--mode DOPO, which needs a fine tuning of the system
parameters, in particular to the instability point $\sigma=1$. It is
interesting to note that the squeezing levels predicted by
(\ref{2ChannelDOPOZeroNoiseY}) are above 90\% as long as $r<0.5$; this is
actually very good news, as it means that even if mode 2 has its threshold two
times above the threshold of mode 1, it will still be highly squeezed.

\subsection{Generalization to many down--conversion channels: The OPO output as a multi--mode non-classical field}

The results we have just found for the simple situation of having two
degenerate type I decay channels are trivially generalized to an arbitrary
number of decay channels of any kind.

The idea is that when pumped at frequency $2\omega_{0}$, the OPO can have many
available down--conversion channels (corresponding to degenerate or
non-degenerate, type I or type II processes, and even channels having some
detuning), but one of them will have the lowest threshold, say $\sigma
=\sigma_{0}$. Classically, this channel, and only this, is the one that will
be amplified above threshold, while the rest of the channels will remain
switched off irrespective of the injection $\sigma$.

As for the quantum properties, from the point of view of the non-amplified
channels the OPO will stay frozen at an injection parameter $\sigma=\sigma
_{0}$, and they will show the levels of squeezing or entanglement predicted
for that parameter. In particular, all the channels with indistinguishable
(distinguishable) signal and idler modes will show squeezing (entanglement),
the level of which will depend on the distance between $\sigma_{0}$ and their
corresponding thresholds.

Hence, we predict that a general OPO with a single pump will emit an
intrinsically non-classical field formed by a superposition of various
squeezed or entangled vacua plus the bright amplified mode above threshold.
This prediction was first introduced in \cite{Navarrete09} and it has been
checked experimentally very recently \cite{Chalopin10}.

\section{Noncritically squeezed light via spontaneous symmetry breaking}

We introduce now the phenomenon of noncritically squeezed light generation
through spontaneous symmetry breaking, which will occupy indeed most of the
original research to be found in this thesis.

\subsection{The basic idea\label{BasicSSB}}

The idea behind this phenomenon can be put in an abstract way as follows.
Suppose that we work with an OPO which is invariant under changes of some
continuous degree of freedom of the down--converted field, say $\epsilon$,
which we might call \textit{free parameter} (FP) in the following (we already
found an example of such system, the frequency degenerate type II OPO, in
which the phase difference between signal and idler was a symmetry of the
system). Above threshold the classical or mean field value of the
down--converted field, say $\mathbf{\bar{E}}_{\epsilon}(\mathbf{r},t)$ ---we
explicitly introduce the FP in the field---, is not zero, and when the OPO
starts emitting it, a particular value of the FP is chosen through spontaneous
symmetry breaking, as commented in Section \ref{ClassiOPO}.

Let's think now about what quantum theory might introduce to this classical
scenario. First, as variations of the FP do not affect the system, quantum
fluctuations can be expected to make it fluctuate without opposition,
eventually making it become completely undetermined. Then, invoking now the
uncertainty principle, the complete indetermination of a system's variable
allows for the perfect determination of its corresponding momentum, meaning
this that we could expect perfect squeezing in the quadrature selected by the
local oscillator $-$\textrm{i}$\partial_{\epsilon}\mathbf{\bar{E}}_{\epsilon
}(\mathbf{r},t)$ in an homodyne detection scheme.

\begin{figure}
[t]
\begin{center}
\includegraphics[
height=0.9072in,
width=4.8862in
]%
{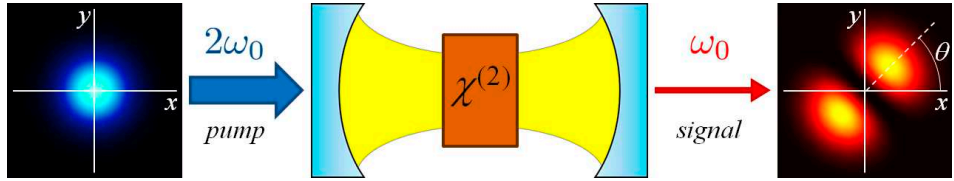}%
\caption{Schematic representation of a DOPO tuned to the first family of
transverse modes at the signal frequency. When pumped above threshold it emits
a TEM$_{10}$ mode whose orientation $\theta$ respect to the $x$ axis is
arbitrary. This is one of the simplest examples of spontaneous symmetry
breaking, and it is actually the one that we will be using to study this
phenomenon in depth.}%
\label{fMulti1}%
\end{center}
\end{figure}

The first proof of such an intuitive reasoning was offered through the study
of a DOPO with plane mirrors \cite{PerezArjona06,PerezArjona07}, where
transverse patterns as well as cavity solitons have been predicted to appear
above threshold
\cite{Oppo94a,Oppo94b,Staliunas95,Valcarcel96,Staliunas97,Longhi97}, hence
breaking the translational symmetry of the system. The main difficulty of this
model was that it is not too close to current experimental setups and such
transverse structures have not even been observed yet in DOPOs. This was one
of the main motivations to study the phenomenon with a simpler, more realistic
system, consisting on a DOPO with spherical mirrors tuned to the first family
of transverse modes at the subharmonic \cite{Navarrete08,Navarrete10} (see
Figure \ref{fMulti1}). We will introduce in this section the phenomenon
through this example, which was actually the first original work originated
from this thesis \cite{Navarrete08}; we will show that it is the rotational
symmetry in the transverse plane the one which is broken in this case, a
symmetry that was also exploited later in another type of nonlinear cavities
which make use of $\chi^{(3)}$ nonlinearities \cite{GarciaFerrer09}.

The phenomenon can be extended to systems in which the FP is not a spatial
degree of freedom in the transverse plane. The first of such generalizations
was made in \cite{GarciaFerrer10}, where the FP was a polarization degree of
freedom of a degenerate type II OPO. This case will be properly addressed in
Chapter \ref{TypeIIOPO}. Continuing with a generalization to different
symmetries, in the outlook section of the concluding chapter we will propose a
new scheme in which the FP is associated to a temporal degree of freedom,
hence closing the types of symmetries that the electromagnetic field has to offer.%

\subsection{The two--transverse--mode DOPO and spontaneous rotational symmetry
breaking\label{SRSB}}

\subsubsection{The two--transverse--mode DOPO model}

In order to introduce the phenomenon of squeezing induced by spontaneous
symmetry breaking we consider a DOPO tuned to the first family of transverse
modes at the signal frequency\footnote{See
\cite{Lassen09,Janousek09,Chalopin10} for related experimental
configurations.}; the DOPO is pumped by a TEM$_{00}$ beam resonant with the
pump mode, that is, $\omega_{\mathrm{L}}=2\omega_{0}$ in the nomenclature of
the previous chapters. We would like to remind that the first family of
transverse modes contains the Laguerre--Gauss modes $L_{0,\pm1}(k_{\mathrm{s}%
};\mathbf{r}_{\perp},z)$ which is the reason why we will call
\textit{two--transverse--mode DOPO} (2tmDOPO) to this system. Most of this
thesis has been devoted to study several aspects of this system
\cite{Navarrete08,Navarrete10,Navarrete11a}, as we have used it as a
playground where analyzing spontaneous symmetry breaking at the quantum level.

Let us denote by $\hat{a}_{0}$ the annihilation operator for the pump mode,
and by $\hat{a}_{\pm1}$ the ones associated to the $L_{0,\pm1}(k_{\mathrm{s}%
};\mathbf{r}_{\perp},z)$ signal modes. Instead of using the Laguerre--Gauss
basis, we can work in the most usual TEM$_{mn}$ basis. Denoting by
$H_{10}^{\psi}(k_{\mathrm{s}};\mathbf{r}_{\perp},z)$ a TEM$_{10}$ mode rotated
an angle $\psi$ with respect to the $x$ axis and by $H_{01}^{\psi
}(k_{\mathrm{s}};\mathbf{r}_{\perp},z)$ to its orthogonal, these are given by%
\begin{subequations}
\begin{align}
H_{10}^{\psi}(k_{\mathrm{s}};\mathbf{r}_{\perp},z) &  =\frac{1}{\sqrt{2}%
}\left[  e^{-\mathrm{i}\psi}L_{0,+1}(k_{\mathrm{s}};\mathbf{r}_{\perp
},z)+e^{\mathrm{i}\psi}L_{0,-1}(k_{\mathrm{s}};\mathbf{r}_{\perp},z)\right]=\sqrt{2}|L_{0,\pm1}(k_{\mathrm{s}};\mathbf{r}_{\perp},z)|\cos(\phi
-\psi),
\\
H_{01}^{\psi}(k_{\mathrm{s}};\mathbf{r}_{\perp},z) &  =\frac{1}{\sqrt
{2}\mathrm{i}}\left[  e^{-\mathrm{i}\psi}L_{0,+1}(k_{\mathrm{s}}%
;\mathbf{r}_{\perp},z)-e^{\mathrm{i}\psi}L_{0,-1}(k_{\mathrm{s}}%
;\mathbf{r}_{\perp},z)\right]=\sqrt{2}|L_{0,\pm1}(k_{\mathrm{s}};\mathbf{r}_{\perp},z)|\sin(\phi
-\psi).
\end{align}
These relations can be inverted as%
\end{subequations}
\begin{equation}
L_{0,\pm1}(k_{\mathrm{s}};\mathbf{r}_{\perp},z)=\frac{1}{\sqrt{2}}%
e^{\pm\mathrm{i}\psi}\left[  H_{10}^{\psi}(k_{\mathrm{s}};\mathbf{r}_{\perp
},z)\pm\mathrm{i}H_{01}^{\psi}(k_{\mathrm{s}};\mathbf{r}_{\perp},z)\right]
.\label{HtoL}%
\end{equation}
The part of the signal field propagating along $+\mathbf{e}_{z}$ can be
written as (Schr\"{o}dinger picture is understood)%
\begin{equation}
\mathbf{\hat{E}}_{\mathrm{s},\rightarrow}^{(+)}\left(  \mathbf{r}\right)
=\mathrm{i}\sqrt{\frac{\hbar\omega_{\mathrm{s}}}{4\varepsilon_{0}%
n_{\mathrm{s}}L_{\mathrm{s}}}}\mathbf{e}_{\mathrm{e}}\left[  \hat{a}%
_{+1}L_{0,+1}(k_{\mathrm{s}};\mathbf{r}_{\perp},z)+\hat{a}_{-1}L_{0,-1}%
(k_{\mathrm{s}};\mathbf{r}_{\perp},z)\right]  e^{\mathrm{i}n_{\mathrm{s}%
}k_{\mathrm{s}}z},\label{SignalField}%
\end{equation}
Introducing the relation (\ref{HtoL}) between the Laguerre--Gauss and the
Hermite--Gauss modes in this expression, we find the relation between their
corresponding annihilation operators:
\begin{subequations}
\label{LGtoHGboson}%
\begin{align}
\hat{a}_{10,\psi} &  =\frac{1}{\sqrt{2}}\left(  e^{\mathrm{i}\psi}\hat{a}%
_{+1}+e^{-\mathrm{i}\psi}\hat{a}_{-1}\right)  ,\\
\hat{a}_{01,\psi} &  =\frac{\mathrm{i}}{\sqrt{2}}\left(  e^{\mathrm{i}\psi
}\hat{a}_{+1}-e^{-\mathrm{i}\psi}\hat{a}_{-1}\right)  .
\end{align}

The quantum model for this OPO configuration was already introduced in Section
\ref{GeneralOPOModel}; we just need to particularize the three--wave mixing
Hamiltonian (\ref{GenTWMHamiltonian}) to the case of operating in frequency
degenerate type I conditions with only the $L_{0,\pm1}(k_{\mathrm{s}%
};\mathbf{r}_{\perp},z)$ present at the signal frequency\footnote{From the
previous section we know now that the only important feature is that this
process is the one with the highest gain, it doesn't matter if other processes
are present (such as frequency non-degenerate processes or processes involving
other transverse families), as long as these have a higher threshold.}.
Working as usual in a picture rotated to the frequency of the injected laser,
which in this case is described by the operator%
\end{subequations}
\begin{equation}
\hat{U}_{0}=\exp[\hat{H}_{0}t/\mathrm{i}\hbar]\text{ \ \ \ \ with
\ \ \ \ }\hat{H}_{0}=\hbar\omega_{0}(2\hat{a}_{0}^{\dagger}\hat{a}_{0}+\hat
{a}_{+1}^{\dagger}\hat{a}_{+1}+\hat{a}_{-1}^{\dagger}\hat{a}_{-1}),
\end{equation}
the master equation of the system can be written in this case as%
\begin{equation}
\frac{d\hat{\rho}_{\mathrm{I}}}{dt}=\left[  \chi\hat{a}_{0}\hat{a}%
_{+1}^{\dagger}\hat{a}_{-1}^{\dagger}+\mathcal{E}_{\mathrm{p}}a_{0}^{\dagger
}+\mathrm{H.c.},\hat{\rho}_{\mathrm{I}}\right]  +\sum_{j=0,\pm1}\gamma
_{j}(2\hat{a}_{j}\hat{\rho}_{\mathrm{I}}\hat{a}_{j}^{\dagger}-\hat{a}%
_{j}^{\dagger}\hat{a}_{j}\hat{\rho}_{\mathrm{I}}-\hat{\rho}_{\mathrm{I}}%
\hat{a}_{j}^{\dagger}\hat{a}_{j}),
\end{equation}
where we will denote $\gamma_{0}$ by $\gamma_{\mathrm{p}}$ to match our
previous notation, and will assume that $\gamma_{+1}=\gamma_{-1}%
=\gamma_{\mathrm{s}}$ which is as to assume that the DOPO is invariant under
rotations around the longitudinal axis, that is, rotationally symmetric in the
transverse plane. As usual, we take $\mathcal{E}_{\mathrm{p}}$ as a positive
real. On the other hand, the nonlinear coupling parameter $\chi$ can be found
by particularizing (\ref{ChiGen}) to the current situation; taking into
account that%
\begin{align}
\int_{0}^{+\infty}rdrG(k_{\mathrm{p}};r)\left[  \mathcal{R}_{0}^{1}%
(k_{\mathrm{s}};r)\right]  ^{2}  & =\left(  \frac{2}{\pi}\right)  ^{3/2}%
\int_{0}^{+\infty}dr\frac{2r^{3}}{w_{\mathrm{p}}w_{\mathrm{s}}^{4}}\exp\left[
-\left(  \frac{1}{w_{\mathrm{p}}^{2}}+\frac{2}{w_{\mathrm{s}}^{2}}\right)
r^{2}\right]=\frac{1}{4\pi^{3/2}w_{\mathrm{s}}}%
\end{align}
where we have used that $w_{\mathrm{p}}^{2}/w_{\mathrm{s}}^{2}=\lambda
_{\mathrm{p}}/\lambda_{\mathrm{s}}=1/2$, see (\ref{w0}), we get%
\begin{equation}
\chi=\frac{3}{2}\frac{l_{\mathrm{c}}}{w_{\mathrm{s}}}\chi_{\mathrm{oee}}%
^{(2)}(2\omega_{0};\omega_{0},\omega_{0})\sqrt{\frac{\hbar\omega_{0}^{3}}%
{8\pi^{3}\varepsilon_{0}n_{\mathrm{c}}^{3}L_{\mathrm{opt}}^{3}}}.
\end{equation}
The above master equation has exactly the same form as that modeling the OPO,
see (\ref{OPOmaster}); this makes perfect sense, as even though the
down--converted photons are indistinguishable in frequency and polarization,
they are not in orbital angular momentum, and hence the 2tmDOPO is not a truly
degenerate system. From an operational point of view this is very fortunate,
as we can then directly take the Langevin equations that we already derived
for the OPO, and rewrite them to match the notation of the current system.
Using also the same scaling for the coherent amplitudes, the time, the noises,
and the parameters as in Section \ref{OPOmodel}, see (\ref{ScaledVarOPO}) and
(\ref{ScaledParameters}), this allows us to write
\begin{subequations}
\label{ScaledLangevin2tmDOPO}%
\begin{align}
\dot{\beta}_{0} &  =\kappa\lbrack\sigma-\beta_{0}-\beta_{+1}\beta_{-1}],\\
\dot{\beta}_{0}^{+} &  =\kappa\lbrack\sigma-\beta_{0}^{+}-\beta_{+1}^{+}%
\beta_{-1}^{+}],\\
\dot{\beta}_{+1} &  =-\beta_{+1}+\beta_{0}\beta_{-1}^{+}+g\sqrt{\beta_{0}%
}\zeta(\tau),\\
\dot{\beta}_{+1}^{+} &  =-\beta_{+1}^{+}+\beta_{0}^{+}\beta_{-1}+g\sqrt
{\beta_{0}^{+}}\zeta^{+}(\tau),\\
\dot{\beta}_{-1} &  =-\beta_{-1}+\beta_{0}\beta_{+1}^{+}+g\sqrt{\beta_{0}%
}\zeta^{\ast}(\tau),\\
\dot{\beta}_{-1}^{+} &  =-\beta_{-1}^{+}+\beta_{0}^{+}\beta_{+1}+g\sqrt
{\beta_{0}^{+}}[\zeta^{+}(\tau)]^{\ast}.
\end{align}
Let us finally remind that the squeezing spectrum of any signal mode can be
evaluated from this normalized variables as%
\end{subequations}
\begin{equation}
S(\hat{X}_{m}^{\varphi};\Omega)=\frac{2}{g^{2}}\int_{-\infty}^{+\infty}%
d\tau^{\prime}\exp(-\mathrm{i}\tilde{\Omega}\tau^{\prime})\langle\delta
x_{m}^{\varphi}\left(  \tau\right)  \delta x_{m}^{\varphi}\left(  \tau
+\tau^{\prime}\right)  \rangle_{P},\label{Susual}%
\end{equation}
with%
\begin{equation}
x_{m}^{\varphi}\left(  \tau\right)  =e^{-\mathrm{i}\varphi}\beta_{m}\left(
\tau\right)  +e^{\mathrm{i}\varphi}\beta_{m}^{+}\left(  \tau\right)
,\label{mScaledQuadrature}%
\end{equation}
$m$ referring to any of transverse modes resonating at the signal frequency.

\subsubsection{Classical emission: Spontaneous rotational symmetry breaking}

We now show that above threshold the two--transverse--mode DOPO shows the
phenomenon of spontaneous symmetry breaking of its rotational invariance in
the transverse plane.

Following the approach introduced in the previous chapter, the classical
equations of the system are retrieved from the Langevin equations by
neglecting the noise terms and making the correspondences $\beta_{j}%
^{+}\rightarrow\beta_{j}^{\ast}$ to come back to the classical phase space:
\begin{subequations}
\begin{align}
\dot{\beta}_{0} &  =\kappa\lbrack\sigma-\beta_{0}-\beta_{+1}\beta_{-1}],\\
\dot{\beta}_{+1} &  =-\beta_{+1}+\beta_{0}\beta_{-1}^{\ast},\\
\dot{\beta}_{-1} &  =-\beta_{-1}+\beta_{0}\beta_{+1}^{\ast}.
\end{align}
Now, from our previous treatment of the OPO (see Section \ref{ClassiOPO}) we
can directly understand the classical emission properties of the current
2tmDOPO. For $\sigma\leq1$ the signal modes are switched off, that is, its
only long time term stable solution is%
\end{subequations}
\begin{equation}
\bar{\beta}_{0}=\sigma\text{, \ \ \ \ }\bar{\beta}_{+1}=\bar{\beta}%
_{-1}=0\text{.}%
\end{equation}
On the other hand, for $\sigma>1$ the only stable solution is%
\begin{equation}
\bar{\beta}_{0}=1\text{, \ \ \ \ }\bar{\beta}_{\pm1}=\rho\exp(\mp
\mathrm{i}\theta)\text{,}\label{2tmDOPOabove}%
\end{equation}
with $\rho=\sqrt{\sigma-1}$ and where $\theta$ is arbitrary, as the classical
equations are invariant under the change $\beta_{\pm1}\longrightarrow\exp
(\mp\mathrm{i}\theta)\beta_{\pm1}$, that is, $\theta$ is the FP of the
2tmDOPO. Taking the expectation value of expression (\ref{SignalField}), and
using\footnote{Note that this result comes from the following chain of
equalities:%
\begin{equation}
\langle\hat{a}_{\pm1}(t)\rangle=\mathrm{tr}\{\hat{\rho}_{\mathrm{I}}\hat
{U}_{0}^{\dagger}\hat{a}_{\pm1}\hat{U}_{0}\}=\exp(-\mathrm{i}\omega
_{0}t)\mathrm{tr}\{\hat{\rho}_{\mathrm{I}}\hat{a}_{\pm1}\}=\exp(-\mathrm{i}%
\omega_{0}t)\langle\alpha_{\pm1}\rangle_{P}=\exp(-\mathrm{i}\omega
_{0}t)\langle\beta_{\pm1}\rangle_{P}/g
\end{equation}
}\ $\langle\hat{a}_{\pm1}(t)\rangle=\exp(-\mathrm{i}\omega_{0}t)\bar{\beta
}_{\pm1}/g$ within the classical limit, we get the form of the signal field as
predicted by classical optics%
\begin{align}
\mathbf{\bar{E}}_{\mathrm{s},\rightarrow}^{(+)}\left(  \mathbf{r},t\right)
=\mathrm{i}\sqrt{\frac{\hbar\omega_{\mathrm{s}}(\sigma-1)}{4\varepsilon
_{0}n_{\mathrm{s}}L_{\mathrm{s}}g^{2}}}\mathbf{e}_{\mathrm{e}}  & \left[
e^{-\mathrm{i}\theta}L_{0,+1}(k_{\mathrm{s}};\mathbf{r}_{\perp}%
,z)+e^{\mathrm{i}\theta}L_{0,-1}(k_{\mathrm{s}};\mathbf{r}_{\perp},z)\right]e^{-\mathrm{i}\omega_{0}t+\mathrm{i}n_{\mathrm{s}}k_{\mathrm{s}}%
z}\propto H_{10}^{\theta}(k_{\mathrm{s}};\mathbf{r}_{\perp},z),
\end{align}
that is, classical emission takes place in the form of a TEM$_{10}$ mode
forming an angle $\theta$ with respect to the $x$ axis (see Figure
\ref{fMulti1}). Hence, the FP is identified in this case with the orientation
of the transverse pattern which is formed above threshold, what finally shows
that the rotational transverse symmetry of the 2tmDOPO is spontaneously broken.

In the following, we will call \textit{bright mode} to $H_{10}^{\theta
}(k_{\mathrm{s}};\mathbf{r}_{\perp},z)$ ---as it is classically excited--- and
\textit{dark mode} to its orthogonal $H_{01}^{\theta}(k_{\mathrm{s}%
};\mathbf{r}_{\perp},z)$ ---as it is classically empty of photons---, and will
define the collective indices $\mathrm{b}=\left(  10,\theta\right)  $ and
$\mathrm{d}=\left(  01,\theta\right)  $ to simplify the notation.

\subsubsection{Quantum properties: Pattern diffusion and noncritical
squeezing\label{Quantum2tmDOPO}}

\textbf{The linearized Langevin equations}. Just as we did with the DOPO, we
are going to discuss the quantum properties of the down--converted field by
inspection of the quantum Langevin equations (\ref{ScaledLangevin2tmDOPO}) in
the limit $\gamma_{\mathrm{p}}\gg\gamma_{\mathrm{s}}$, i.e., $\kappa\gg1$,
where the pump variables can be adiabatically eliminated. We will show in the
next chapter that all the important properties found in this limit are valid
in general.

As we already commented in Section \ref{QuantumDOPO}, the deterministic
procedure for adiabatically eliminating the pump modes (which consists in
assuming that $\dot{\beta}_{0}=\dot{\beta}_{0}^{+}=0$) is still valid in the
stochastic case, but only if the equations are interpreted in the Ito sense.
Hence, after applying this procedure, the Langevin equations read in
Stratonovich form:
\begin{subequations}
\label{AdElEqs}%
\begin{align}
\dot{\beta}_{+1}  &  =-\left(  1-g^{2}/4\right)  \beta_{+1}+(\sigma-\beta
_{+1}\beta_{-1})\beta_{-1}^{+}+g\sqrt{\sigma-\beta_{+1}\beta_{-1}}\zeta\left(
\tau\right)  ,\\
\dot{\beta}_{+1}^{+}  &  =-\left(  1-g^{2}/4\right)  \beta_{+1}^{+}%
+(\sigma-\beta_{+1}^{+}\beta_{-1}^{+})\beta_{-1}+g\sqrt{\sigma-\beta_{+1}%
^{+}\beta_{-1}^{+}}\zeta^{+}\left(  \tau\right)  ,\\
\dot{\beta}_{-1}  &  =-\left(  1-g^{2}/4\right)  \beta_{-1}+(\sigma-\beta
_{+1}\beta_{-1})\beta_{+1}^{+}+g\sqrt{\sigma-\beta_{+1}\beta_{-1}}\zeta^{\ast
}\left(  \tau\right)  ,\\
\dot{\beta}_{-1}^{+}  &  =-\left(  1-g^{2}/4\right)  \beta_{-1}^{+}%
+(\sigma-\beta_{+1}^{+}\beta_{-1}^{+})\beta_{+1}+g\sqrt{\sigma-\beta_{+1}%
^{+}\beta_{-1}^{+}}\left[  \zeta^{+}\left(  \tau\right)  \right]  ^{\ast}.
\end{align}

In order to find analytic predictions from these equations we are going to
linearize them. As explained in the previous chapter, the usual procedure
begins by writing the amplitudes as $\beta_{m}=\bar{\beta}_{m}+\delta\beta
_{m}$ and $\beta_{m}^{+}=\bar{\beta}_{m}^{\ast}+\delta\beta_{m}^{+}$, and
treat the fluctuations as order $g$ perturbations. This is in particular how
we proceeded with the OPO. However, as already stated in the previous chapter
and emphasized in the present one, we expect quantum noise to rotate the
generated TEM$_{10}$ mode (that is, to make the phase difference between the
opposite angular momentum modes diffuse), and hence fluctuations of the fields
in an arbitrary direction of phase space could not be small (i.e., order $g$).
Nevertheless, equations (\ref{AdElEqs}) can be linearized if the amplitudes
are expanded as
\end{subequations}
\begin{subequations}
\label{LinExp}%
\begin{align}
\beta_{\pm1}\left(  \tau\right)   &  =\left[  \rho+b_{\pm1}\left(
\tau\right)  \right]  e^{\mp\mathrm{i}\theta\left(  \tau\right)  }\\
\beta_{\pm1}^{+}\left(  \tau\right)   &  =\left[  \rho+b_{\pm1}^{+}\left(
\tau\right)  \right]  e^{\pm\mathrm{i}\theta\left(  \tau\right)  },
\end{align}
because as we will prove $\theta\left(  \tau\right)  $ carries the larger part
of the fluctuations, while the $b$'s and $\dot{\theta}$ remain as order $g$
quantities. In addition, expanding the fields in this way allows us to track
the evolution of the classical pattern's orientation, as we take $\theta$ as
an explicit quantum variable. Then, writing equations (\ref{AdElEqs}) up to
order $g$, we arrive to the following linear system (arriving to this
expression is not as straightforward as it might seem, there are some
subtleties that we clarify in Appendix \ref{Lin2tmDOPO})%
\end{subequations}
\begin{equation}
-2i\mathrm{i}\rho\mathbf{w}_{0}\dot{\theta}+\mathbf{\dot{b}=}\mathcal{L}%
\mathbf{b}+g\boldsymbol{\zeta}\left(  \tau\right)  , \label{LinLan}%
\end{equation}
with%
\begin{equation}
\mathbf{b}=%
\begin{pmatrix}
b_{+1}\\
b_{+1}^{+}\\
b_{-1}\\
b_{-1}^{+}%
\end{pmatrix}
\text{\ , }\boldsymbol{\zeta}\left(  \tau\right)  =%
\begin{pmatrix}
\zeta\left(  \tau\right) \\
\zeta^{+}\left(  \tau\right) \\
\zeta^{\ast}\left(  \tau\right) \\
\left[  \zeta^{+}\left(  \tau\right)  \right]  ^{\ast}%
\end{pmatrix}
,
\end{equation}
and where $\mathcal{L}$ is a real, symmetric matrix given by%
\begin{equation}
\mathcal{L}=-%
\begin{pmatrix}
\sigma & 0 & \sigma-1 & -1\\
0 & \sigma & -1 & \sigma-1\\
\sigma-1 & -1 & \sigma & 0\\
-1 & \sigma-1 & 0 & \sigma
\end{pmatrix}
,
\end{equation}
with the following eigensystem%
\begin{equation}%
\begin{array}
[c]{ll}%
\lambda_{0}=0, & \mathbf{w}_{0}=\frac{1}{2}\operatorname{col}\left(
1,-1,-1,1\right) \\
\lambda_{1}=-2, & \mathbf{w}_{1}=\frac{1}{2}\operatorname{col}\left(
1,1,-1,-1\right) \\
\lambda_{2}=-2\left(  \sigma-1\right)  , & \mathbf{w}_{2}=\frac{1}%
{2}\operatorname{col}\left(  1,1,1,1\right) \\
\lambda_{3}=-2\sigma, & \mathbf{w}_{3}=\frac{1}{2}\operatorname{col}\left(
1,-1,1,-1\right)  .
\end{array}
\label{Eigensystem}%
\end{equation}

Defining the projections $c_{m}\left(  \tau\right)  =\mathbf{w}_{m}%
\cdot\mathbf{b}\left(  \tau\right)  $, and projecting the linear system
(\ref{LinLan}) onto these eigenmodes, we find the following set of decoupled
linear equations ($c_{0}$ is set to zero, as otherwise it would just entail a
redefinition of $\theta$)
\begin{subequations}
\label{ProjLinLan}%
\begin{align}
\dot{\theta}  &  =\frac{g}{2\rho}\eta_{0}\left(  \tau\right)  \label{ThetaEvo}%
\\
\dot{c}_{1}  &  =-2c_{1}+\mathrm{i}g\eta_{1}\left(  \tau\right)
\label{c1evo}\\
\dot{c}_{2}  &  =-2\left(  \sigma-1\right)  c_{2}+g\eta_{2}\left(  \tau\right)
\label{c2evo}\\
\dot{c}_{3}  &  =-2\sigma c_{3}+g\eta_{3}\left(  \tau\right)  , \label{c3evo}%
\end{align}
where the following real noises have been defined%
\end{subequations}
\begin{subequations}
\begin{align}
\eta_{0}\left(  \tau\right)   &  =\mathrm{i}\mathbf{w}_{0}\cdot
\boldsymbol{\zeta}\left(  \tau\right)  =\operatorname{Im}\left\{  \zeta
^{+}\left(  \tau\right)  -\zeta\left(  \tau\right)  \right\} \\
\eta_{1}\left(  \tau\right)   &  =-\mathrm{i}\mathbf{w}_{1}\cdot
\boldsymbol{\zeta}\left(  \tau\right)  =\operatorname{Im}\left\{  \zeta
^{+}\left(  \tau\right)  +\zeta\left(  \tau\right)  \right\} \\
\eta_{2}\left(  \tau\right)   &  =\mathbf{w}_{2}\cdot\boldsymbol{\zeta}\left(
\tau\right)  =\operatorname{Re}\left\{  \zeta\left(  \tau\right)  +\zeta
^{+}\left(  \tau\right)  \right\} \\
\eta_{3}\left(  \tau\right)   &  =\mathbf{w}_{3}\cdot\boldsymbol{\zeta}\left(
\tau\right)  =\operatorname{Re}\left\{  \zeta\left(  \tau\right)  -\zeta
^{+}\left(  \tau\right)  \right\}  ,
\end{align}
which satisfy the usual statistical properties (\ref{RealGaussStat}).

Note finally that in the long time term the solutions for the projections
$c_{j}$ will be of order $g$ (see Appendix \ref{LinStoApp}), and hence so will
be the $b$'s (note that this is not the case for $\theta$, whose initial value
is completely arbitrary, although its variation $\dot{\theta}$ is indeed of
order $g$). This is consistent with the initial assumptions about the orders
in $g$ of the involved quantities.

\textbf{Quantum diffusion of the classical pattern. }Just as we commented in
the previous sections, equation (\ref{ThetaEvo}) shows that the orientation of
the bright mode (the FP of the system) diffuses with time ruled by quantum
noise. How fast this diffusion is can be measured by evaluating the variance
of $\theta$. Using the statistical properties of noise (\ref{RealGaussStat}),
it is straightforward to obtain the following result%
\end{subequations}
\begin{equation}
V_{\theta}\left(  \tau\right)  =\left\langle \delta\theta^{2}\left(
\tau\right)  \right\rangle _{P}=D\tau\text{,} \label{Theta Var}%
\end{equation}
where $D=d/\left(  \sigma-1\right)  $ with
\begin{equation}
d=g^{2}/4=\chi^{2}/4\gamma_{\mathrm{p}}\gamma_{\mathrm{s}}. \label{d}%
\end{equation}
Note that we have assumed that $V_{\theta}\left(  0\right)  =0$, which is as
saying that every time the 2tmDOPO is switched on, that is, on any stochastic
realization, the bright mode appears with the same orientation; in Chapter
\ref{2tmDOPOwithIS} we will show that the pattern can be locked to a desired
orientation by injecting a low--power TEM$_{10}$ seed at the signal frequency
inside the cavity, and hence our assumption can be satisfied experimentally by
switching on the 2tmDOPO with this seed present, and turning it off at one
point so that quantum noise can start making the bright mode rotate arbitrarily.

Passing to quantitative matters, Eq. (\ref{Theta Var}) shows that the
delocalization of the pattern's orientation increases as time passes by,
though for typical system parameters (see Section \ref{QuantumDOPO}, where we
argued that $g\approx4\times10^{-6}$) one finds $d\simeq4\times10^{-12}$, and
hence the rotation of the pattern will be fast only when working extremely
close to threshold.

\textbf{Independent quadratures and noncritical squeezing. }The first step
towards analyzing the squeezing properties of the field within the framework
presented above, is identifying a set of independent quadratures. As showed in
the previous analysis, these appear naturally within our approach, as the
eigenmodes $\{\mathbf{w}_{m}\}_{m=0,1,2,3}$ of the stability matrix give us a
set of quadratures with well defined squeezing properties. In particular, from
(\ref{LinExp}), (\ref{LGtoHGboson}), and (\ref{mScaledQuadrature}), it is easy
to find the following relations%
\begin{subequations}
\begin{align}
x_{\mathrm{b}}\left(  \tau\right)   &  =2\sqrt{2}\rho+\sqrt{2}c_{2}\left(
\tau\right) \\
y_{\mathrm{b}}\left(  \tau\right)   &  =-\mathrm{i}\sqrt{2}c_{3}\left(
\tau\right) \\
x_{\mathrm{d}}\left(  \tau\right)   &  =\mathrm{i}\sqrt{2}c_{0}\left(
\tau\right) \\
y_{\mathrm{d}}\left(  \tau\right)   &  =\sqrt{2}c_{1}\left(  \tau\right)  ,
\end{align}
where $\{x_{\mathrm{b}},y_{\mathrm{b}}\}$ and $\{x_{\mathrm{d}},y_{\mathrm{d}%
}\}$ are the (normalized) \textsf{X} and \textsf{Y} quadratures of the bright
and dark modes, $H_{10}^{\theta}(k_{\mathrm{s}};\mathbf{r}_{\perp},z)$ and
$H_{01}^{\theta}(k_{\mathrm{s}};\mathbf{r}_{\perp},z)$, respectively.

The evolution of these quadratures can therefore be found from the equations
satisfied by the projections (\ref{ProjLinLan}), which are solved in Appendix
\ref{LinStoApp} (remember that $c_{0}=0$). In particular, using the expression
for the correlation spectrum of the projections (\ref{CorrSpectrum}) as usual,
it is straightforward to find the following results
\end{subequations}
\begin{subequations}
\label{NoiseSpectra}%
\begin{align}
V^{\mathrm{out}}\left(  \hat{X}_{\mathrm{b}};\Omega\right)   &  =1+\frac
{1}{\left(  \sigma-1\right)  ^{2}+\tilde{\Omega}^{2}/4}\label{XcSpectrum}\\
V^{\mathrm{out}}\left(  \hat{Y}_{\mathrm{b}};\Omega\right)   &  =1-\frac
{1}{\sigma^{2}+\tilde{\Omega}^{2}/4}\label{YcSpectrum}\\
V^{\mathrm{out}}\left(  \hat{X}_{\mathrm{d}};\Omega\right)   &
=1\label{XsSpectrum}\\
V^{\mathrm{out}}\left(  \hat{Y}_{\mathrm{d}};\Omega\right)   &  =1-\frac
{1}{1+\tilde{\Omega}^{2}/4}. \label{YsSpectrum}%
\end{align}

We see that the quadratures of the bright mode have the same behavior as those
of the single mode DOPO: The quadrature \textsf{Y}$_{\mathrm{b}}$ is perfectly
squeezed $\left(  V^{\mathrm{out}}=0\right)  $ at zero noise frequency
$\left(  \tilde{\Omega}=0\right)  $ only at the bifurcation $\left(
\sigma=1\right)  $.

On the other hand, the dark mode has perfect squeezing in its \textsf{Y}
quadrature at zero noise frequency. What is interesting is that this result is
independent of the distance from threshold, and thus, it is a noncritical phenomenon.

This result was first shown in \cite{Navarrete08}, and we can understand it by
following the reasoning given in above: As the orientation $\theta$ of the
classically excited pattern is undetermined in the long--time limit, its
orbital angular momentum must be fully determined at low noise frequencies. On
the other hand, the orbital angular momentum of the bright mode $H_{10}%
^{\theta}(k_{\mathrm{s}};\mathbf{r}_{\perp},z)$ is nothing but its $\pi/2$
phase shifted orthogonal Hermite--Gauss mode, i.e., $-\mathrm{i}\partial
_{\phi}H_{10}^{\theta}(k_{\mathrm{s}};\mathbf{r}_{\perp},z=\mathrm{i}%
H_{01}^{\theta}(k_{\mathrm{s}};\mathbf{r}_{\perp},z)$, which used as a local
oscillator in a homodyne detection experiment would lead to the observation of
the \textsf{Y}$_{\mathrm{d}}$ quadrature fluctuations.

\end{subequations} 

%% file: 2tmDOPOFO.tex
In the previous chapter we have learned that spontaneous symmetry breaking
allows for the obtention of noncritically squeezed light in OPOs. We have
introduced this phenomenon by studying a particular example, the
two--transverse--mode DOPO, in which the rotational invariance in the
transverse plane is the symmetry broken by the TEM$_{10}$ signal field
generated above threshold; the fundamental results at the quantum level have
been that (i) this bright pattern rotates randomly in the transverse plane
owed to quantum noise, and (ii) the dark mode consisting the the TEM$_{10}$
mode orthogonal to the generated one shows perfect squeezing on its \textsf{Y}
quadrature irrespective of the distance to threshold.

In this chapter (and the next one) we answer some questions that appear
naturally after the results of the previous chapter, and which are important
both to get a clear understanding of the phenomenon, and to evaluate up to
what point it is experimentally observable.

\section{On canonical pairs and noise transfer\label{OnCan}}

Although the noncritical squeezing of the dark mode coincide with what was
intuitively expected after the arguments given in Section \ref{BasicSSB}, a
strange, unexpected result has appeared in (\ref{NoiseSpectra}): The
quadratures of the dark mode seem to violate the uncertainty principle as
$V^{\mathrm{out}}(\hat{X}_{\mathrm{d}};\Omega=0)V^{\mathrm{out}}(\hat
{Y}_{\mathrm{d}};\Omega=0)=0$.

Moreover, the noise spectrum of an arbitrary quadrature of the dark mode,
which can be written in terms of its \textsf{X} and \textsf{Y} quadratures as
$\hat{X}_{\mathrm{d}}^{\varphi}=\hat{X}_{\mathrm{d}}\cos\varphi+\hat
{Y}_{\mathrm{d}}\sin\varphi$, is%
\begin{equation}
V^{\mathrm{out}}\left(  \hat{X}_{\mathrm{d}}^{\varphi};\Omega\right)
=1-\frac{\sin^{2}\varphi}{1+\tilde{\Omega}^{2}/4}\text{,}%
\end{equation}
what shows that all the quadratures of the dark mode (except its \textsf{X}
quadrature) are squeezed. Thus two orthogonal quadratures cannot form a
canonical pair as they satisfy the relation
\begin{equation}
V^{\mathrm{out}}(\hat{X}_{\mathrm{d}}^{\varphi};\Omega)V^{\mathrm{out}}%
(\hat{Y}_{\mathrm{d}}^{\varphi};\Omega)<1,
\end{equation}
in clear violation of the uncertainty principle. This might look surprising,
but one needs to keep in mind that the bright and dark modes are not true
modes, as their orientation is not a fixed quantity, but a dynamical variable
of the problem\footnote{This gives rise to an even more important question: If
the modes depend on the system variables, can we actually select them with a
local oscillator? One can try to picture a detection scheme using some kind of
feed--forward information in which the orientation of the local oscillator is
locked somehow to the orientation of the bright or dark modes, but it seems
highly unlikely. Nevertheless, we show along the next sections that this does
not destroy the possibility of observing the phenomenon, or even of using it
in applications.}.

The natural question now is: Where does the excess of noise go if it is not
transferred from one quadrature to its orthogonal one? The intuitive answer is
that it goes to the pattern orientation, which is actually fully undetermined
in the long term as we showed above (\ref{Theta Var}). This section is devoted
to prove this statement.

In particular we will prove that two orthogonal quadratures of the dark mode
do not form a canonical pair, while the orientation $\theta$ is the canonical
pair of all the squeezed quadratures. One way to prove this would be to
evaluate the commutator between the corresponding quantum operators. However,
$\theta$ is half the phase difference between the opposite orbital angular
momentum modes $\hat{a}_{\pm1}$, whose associated operator has a very
difficult expression \cite{Luis93,Yu97}, making the calculation of the needed
commutators quite hard. Nevertheless, in \cite{Navarrete08b} we gave evidences
of this via a simple procedure based on the close relation between the quantum
commutators and the classical Poisson brackets introduced in Section
\ref{ClassicalMechanics}, and this is the one we present here.

Let us then come back to a classical description of the system in which a pair
of normal variables $\{\nu_{m},\nu_{m}^{\ast}\}$ is associated to each mode of
the field. The Poisson bracket between two phase space functions $f\left(
\nu_{m},\nu_{m}^{\ast}\right)  $ and $h\left(  \nu_{m},\nu_{m}^{\ast}\right)
$ can be rewritten in terms of the normal variables (instead of the position
and momenta) as%
\begin{equation}
\left\{  f,h\right\}  =%
{\displaystyle\sum\limits_{m}}
\frac{2}{\mathrm{i}\omega_{m}}\left(  \frac{\partial f}{\partial\nu_{m}}%
\frac{\partial h}{\partial\nu_{m}^{\ast}}-\frac{\partial f}{\partial\nu
_{m}^{\ast}}\frac{\partial h}{\partial\nu_{m}}\right)  , \label{PBwithNormal}%
\end{equation}
where the sum covers all the cavity modes, being $\omega_{m}$ their
corresponding frequencies.

As an example, the Poisson bracket between two orthogonal quadratures of a
given mode $X_{m}=\sqrt{\omega_{m}/2\hbar}(\nu_{m}+\nu_{m}^{\ast})$ and
$Y_{m}=-\sqrt{\omega_{m}/2\hbar}\mathrm{i}\left(  \nu_{m}-\nu_{m}^{\ast
}\right)  $ is found to be%
\begin{equation}
\left\{  X_{m},Y_{m}\right\}  =2/\hbar,
\end{equation}
which is consistent with the quantum commutator $[\hat{X}_{m},\hat{Y}%
_{m}]=2\mathrm{i}$ given the correspondence $[\hat{X}_{m},\hat{Y}%
_{m}]=\mathrm{i}\hbar\left\{  X_{m},Y_{m}\right\}  $.

Now, we say that two observables $\hat{A}$ and $\hat{B}$ are canonically
related when their commutator is not an operator, but an imaginary number,
that is, when $[\hat{A},\hat{B}]=\mathrm{i}K$ with $K\in%
\mathbb{R}
$, as therefore they satisfy an uncertainty relation of the type $\Delta
A\Delta B\geq K^{2}/4$. In an analogous way, given the relation between the
commutators and the Poisson brackets, we will say that two classical
observables $f\left(  \nu_{m},\nu_{m}^{\ast}\right)  $ and $h\left(  \nu
_{m},\nu_{m}^{\ast}\right)  $ form a canonical pair if their Poisson bracket
is a real number, that is, $\left\{  f,h\right\}  \in%
\mathbb{R}
$.

In our case, the functions we are interested in are the classical counterparts
of the dark mode quadratures, which using (\ref{LGtoHGboson}) with
$\psi=\theta$ are written in terms of the normal variables of the
Laguerre--Gauss modes as
\begin{equation}
X_{\mathrm{d}}^{\varphi}=\mathrm{i}\sqrt{\frac{\omega_{\mathrm{s}}}{4\hbar}%
}\left[  e^{-\mathrm{i}\varphi}\left(  e^{\mathrm{i}\theta}\nu_{+1}%
-e^{-\mathrm{i}\theta}\nu_{-1}\right)  \right]  +\mathrm{c.c.},
\end{equation}
with the orientation given by%
\begin{equation}
\theta=\frac{1}{2\mathrm{i}}\ln\left(  \frac{\nu_{+1}^{\ast}\nu_{-1}}%
{|\nu_{+1}||\nu_{-1}|}\right)  ,
\end{equation}
which is just half the phase difference between $\nu_{+1}$ and $\nu_{-1}$.

Now, using the definition of the Poisson brackets (\ref{PBwithNormal}) with
$m=\pm1$ and after some algebra, it is simple to show that%
\begin{equation}
\left\{  X_{\mathrm{d}}^{\varphi},Y_{\mathrm{d}}^{\varphi}\right\}
=-\frac{\left(  \left\vert \nu_{+1}\right\vert -\left\vert \nu_{-1}\right\vert
\right)  ^{2}}{2\hbar\left\vert \nu_{+1}\right\vert \left\vert \nu
_{-1}\right\vert },
\end{equation}
and%
\begin{equation}
\left\{  X_{\mathrm{d}}^{\varphi},\theta\right\}  =\frac{\exp(-\mathrm{i}%
\varphi)}{4i\sqrt{\hbar\omega_{\mathrm{s}}}}\frac{\left(  \left\vert \nu
_{+1}\right\vert +\left\vert \nu_{-1}\right\vert \right)  \sqrt{\left\vert
\nu_{+1}\right\vert \left\vert \nu_{-1}\right\vert }}{\left\vert \nu
_{+1}\right\vert ^{3/2}\left\vert \nu_{-1}\right\vert ^{3/2}}+\mathrm{c.c.}%
\end{equation}

On the other hand, in the 2tmDOPO the number of photons with opposite orbital
angular momentum is sensibly equal, that is, $\left\vert \nu_{+1}\right\vert
\approx\left\vert \nu_{-1}\right\vert $; in particular, the steady state
solution of the system above threshold (\ref{2tmDOPOabove}) states that
$\left\vert \nu_{\pm1}\right\vert =\sqrt{2\hbar/\omega_{\mathrm{s}}}%
|\langle\hat{a}_{\pm1}\rangle|=\sqrt{2\hbar/\omega_{\mathrm{s}}}|\bar{\beta
}_{\pm1}/g|=\sqrt{2\hbar/\omega_{\mathrm{s}}}\rho/g$, and thus the dominant
term of the previous Poisson brackets will be%
\begin{equation}
\left\{  X_{\mathrm{d}}^{\varphi},X_{\mathrm{d}}^{\varphi+\pi/2}\right\}
\approx0,
\end{equation}
and%
\begin{equation}
\left\{  X_{\mathrm{d}}^{\varphi},\theta\right\}  \approx-\frac{g\sin\varphi
}{\sqrt{2}\hbar\rho}\text{.}%
\end{equation}

Hence, at least at the classical level two orthogonal dark quadratures
$X_{\mathrm{d}}^{\varphi}$ and $X_{\mathrm{d}}^{\varphi+\pi/2}$ do not form a
canonical pair (moreover, they commute), while $X_{\mathrm{d}}^{\varphi}$ and
$\theta$ do. This can be seen as an evidence of the same conclusion for the
corresponding quantum operators.

\section{Homodyne detection with a fixed local oscillator\label{FixOsci}}

So far we have considered the situation in which one is able to detect
independently the bright and dark modes. However, as shown by Eq.
(\ref{ThetaEvo}), these modes are rotating randomly, what means that a local
oscillator field following that random rotation should be used in order to
detect them separately. This might be a really complicated, if not impossible,
task, so we analyze now the more realistic situation in which the local
oscillator is matched to the orthogonal orientation of the emerging pattern
only at the initial time, remaining then with the same orientation during the
observation time $T$. We will show that even in this case, and as the rotation
of the modes is quite slow (\ref{Theta Var}), large levels of noise reduction
can be obtained.

Without loss of generality, we suppose that the bright mode emerges from the
resonator at some initial time $\tau=0$ oriented within the $x$ axis, i.e.,
$\theta\left(  0\right)  =0$. Hence, by using a TEM$_{01}$ local oscillator
with a phase $\varphi$, the quadrature \textsf{X}$_{01}^{\varphi}$ of a fixed
$H_{01}(k_{\mathrm{s}};\mathbf{r}_{\perp},z)$ mode (the initially dark mode)
will be measured. In terms of the Gauss-Laguerre modes, the amplitude
$\beta_{01}$ of the this mode is given by (\ref{LGtoHGboson}) with $\psi=0$.
Then, by using the expansion (\ref{LinExp}) of the amplitudes $\beta_{m}$ as
functions of the fluctuations $b_{m}$ and the orientation $\theta$, the
normalized quadrature $x_{01}^{\varphi}$ of this mode can be rewritten as%
\begin{align}
x_{01}^{\varphi}=2\sqrt{2}\rho\cos\varphi\sin\theta+\sqrt{2}c_{2}%
\cos\varphi\sin\theta+\sqrt{2}c_{1}\sin\varphi\cos\theta-\mathrm{i}\sqrt{2}c_{3}\sin\varphi
\sin\theta\text{.}
\end{align}
Hence, the two--time correlation function of $x_{01}^{\varphi}$ yields (note
that (\ref{ProjLinLan}) show that $c_{1}$, $c_{2}$, $c_{3}$, $\sin\theta$, and
$\cos\theta$ are uncorrelated, as their corresponding noises are independent)%
\begin{align}\label{StillCorrelation}
\left\langle x_{01}^{\varphi}\left(  \tau_{1}\right)  x_{01}^{\varphi}\left(
\tau_{2}\right)  \right\rangle =  & 2\cos^{2}\varphi\left[  4\rho^{2}%
+C_{2}\left(  \tau_{1},\tau_{2}\right)  \right]  S\left(  \tau_{1},\tau
_{2}\right)+2\sin^{2}\varphi\left[  C_{1}\left(  \tau_{1},\tau_{2}\right)  C\left(
\tau_{1},\tau_{2}\right)  -C_{3}\left(  \tau_{1},\tau_{2}\right)  S\left(
\tau_{1},\tau_{2}\right)  \right]  ,
\end{align}
with
\begin{subequations}
\label{StillCorr}%
\begin{align}
S\left(  \tau_{1},\tau_{2}\right)   &  =\left\langle \sin\theta\left(
\tau_{1}\right)  \sin\theta\left(  \tau_{2}\right)  \right\rangle \\
C\left(  \tau_{1},\tau_{2}\right)   &  =\left\langle \cos\theta\left(
\tau_{1}\right)  \cos\theta\left(  \tau_{2}\right)  \right\rangle \\
C_{m}\left(  \tau_{1},\tau_{2}\right)   &  =\left\langle c_{m}\left(  \tau
_{1}\right)  c_{m}\left(  \tau_{2}\right)  \right\rangle .
\end{align}

As shown in Appendices \ref{LinStoApp} and \ref{SinCosCorr}, the latter
correlation functions can be evaluated by using the linear evolution equations
of the projections $c_{m}$ and $\theta$ -see equations (\ref{Ccorr}),
(\ref{SinCorr}), and (\ref{CosCorr})-; then from (\ref{StillCorrelation}) the
squeezing spectrum of a general quadrature of the $H_{01}(k_{\mathrm{s}%
};\mathbf{r}_{\perp},z)$ mode can be found by using the general expression
(\ref{GenSqSpectrum}), which in terms of the normalized quadratures is written
as%
\end{subequations}
\begin{equation}
S(\hat{X}_{01}^{\varphi};\Omega)=\frac{2}{\tilde{T}g^{2}}\int_{0}^{\tilde{T}%
}d\tau\int_{0}^{\tilde{T}}d\tau^{\prime}\cos\left[  \tilde{\Omega}(\tau
-\tau^{\prime})\right]  \langle\delta x_{01}^{\varphi}\left(  \tau\right)
\delta x_{01}^{\varphi}\left(  \tau+\tau^{\prime}\right)  \rangle_{P},
\end{equation}
with $\tilde{T}=\gamma_{\mathrm{s}}T$, as in this case the stationary
expression (\ref{Susual}) cannot be used because $S\left(  \tau_{1},\tau
_{2}\right)  $ and $C\left(  \tau_{1},\tau_{2}\right)  $ don't reach a
stationary state, see (\ref{SinCorr}) and (\ref{CosCorr}).

\begin{figure}
[t]
\begin{center}
\includegraphics[
height=3.096in,
width=4.2367in
]%
{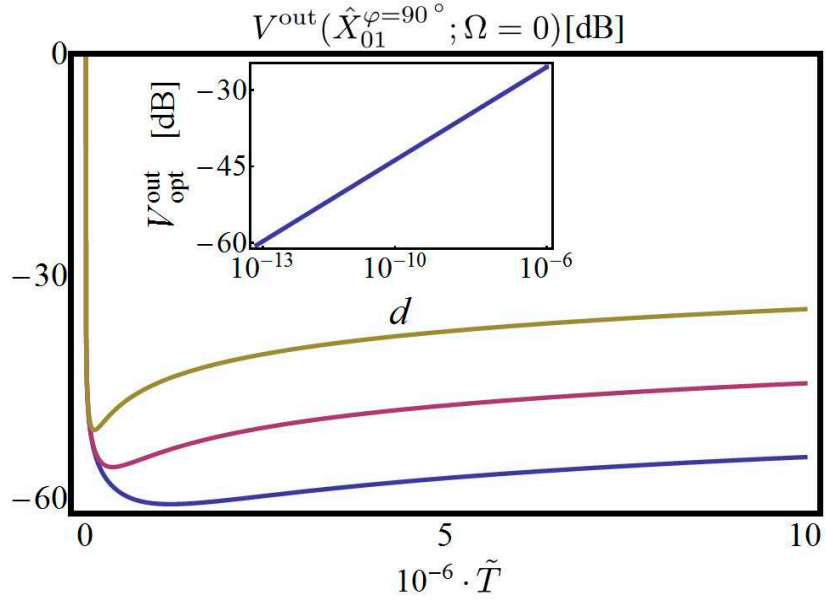}%
\caption{Zero-frequency noise spectrum of the \textsf{Y} quadrature
corresponding to a fixed TEM$_{01}$ mode as a function of the detection time
$\tilde{T}$. Three different values of $d$ are considered ($10^{-11}$,
$10^{-12}$ and $10^{-13}$ from top to bottom). The inset shows also this
spectrum at zero noise frequency, but evaluated at the optimum detection time
$T_{\mathrm{opt}}$ and as a function of $d$ (note that the $d$ axis is in
logarithmic scale). As mentioned in the text $\sigma=\sqrt{2}$.}%
\label{f2tmDOPO1}%
\end{center}
\end{figure}

This integral is easily performed; however the resulting expression for
$S(\hat{X}_{01}^{\varphi};\Omega)$ is too lengthy, and a more compact
approximated expression can be straightforwardly found in the limit of small
$d$, leading to the following expression for the noise spectrum%
\begin{equation}
V^{\mathrm{out}}(\hat{X}_{01}^{\varphi};\Omega)=1+S_{01}^{0}\left(
\Omega\right)  \cos^{2}\varphi+S_{01}^{\pi/2}\left(  \Omega\right)  \sin
^{2}\varphi, \label{FixedNoiseSpectrum}%
\end{equation}
with%
\begin{equation}
S_{01}^{0}=\frac{8}{\tilde{\Omega}^{2}}\left(  1-\operatorname{sinc}%
\tilde{\Omega}T\right)  -\frac{4d\tilde{T}}{\tilde{\Omega}^{2}\left(
\sigma-1\right)  }\cdot\frac{6\left(  \sigma-1\right)  ^{2}+\tilde{\Omega}%
^{2}}{4\left(  \sigma-1\right)  ^{2}+\tilde{\Omega}^{2}}%
\end{equation}
and%
\begin{equation}
S_{01}^{\pi/2}=\frac{8-2\tilde{\Omega}^{2}}{\tilde{T}\left(  4+\tilde{\Omega
}^{2}\right)  ^{2}}-\frac{4}{4+\tilde{\Omega}^{2}}+\frac{8d\tilde{T}\left[
2\left(  \sigma^{2}+1\right)  +\tilde{\Omega}^{2}\right]  }{\left(
\sigma-1\right)  \left(  4+\tilde{\Omega}^{2}\right)  \left(  4\sigma
^{2}+\tilde{\Omega}^{2}\right)  },
\end{equation}
where $\operatorname{sinc}x=\sin\left(  x\right)  /x$. In the following we
will fix $\sigma$ to $\sqrt{2}$ (pump power twice above threshold), as the
results are almost independent of its value as far as it is far enough from
threshold. Hence, the free parameters will be the detection parameters
$\tilde{T}$, $\tilde{\Omega}$ and $\varphi$, and the diffusion $d$ which
depends on the system parameters.%

In this section, the results for $V^{\mathrm{out}}(\hat{X}_{01}^{\varphi
};\Omega)$ are presented in \textrm{dB} units, defined through the relation
$V^{\mathrm{out}}\left[  \mathrm{dB}\right]  =10\log V^{\mathrm{out}}$ (hence,
e.g., $-10$ \textrm{dB} and $-\infty$ \textrm{dB }correspond to 90\% of noise
reduction ---$V^{\mathrm{out}}=0.1$--- and complete noise reduction
---$V^{\mathrm{out}}=0$--- respectively.)

\begin{figure}
[t]
\begin{center}
\includegraphics[
width=\textwidth
]%
{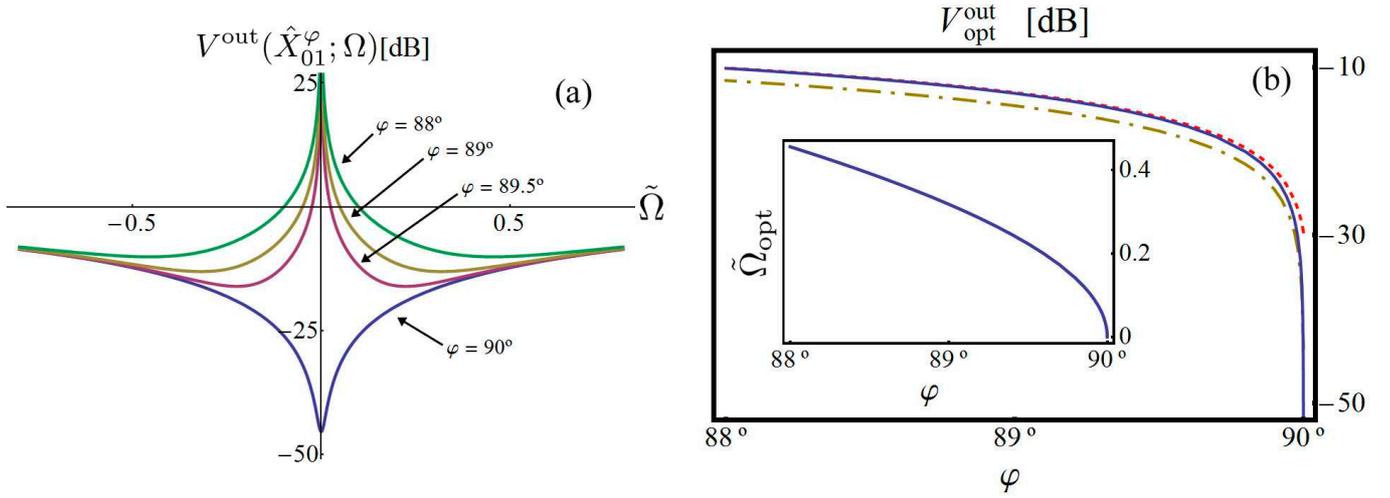}%
\caption{(a) Noise spectrum of the fixed TEM$_{01}$ mode as a function of the
noise frequency $\tilde{\Omega}$, and for four different values of the local
oscillator phase ($\varphi=88$, $89$, $89.5$, and $90$ from top to bottom
curves). It can be appreciated how the infinite fluctuations of $S_{01}^{0}$
at zero noise frequency enter in the spectrum for any $\varphi\neq90$. The
rest of parameters are $\sigma=\sqrt{2}$, $d=10^{-10}$, and $T=T_{\mathrm{opt}%
}$, but the same behavior appears for any other election of the parameters.
(b) Noise spectrum of the fixed TEM$_{01}$ mode evaluated for the optimum
parameters $\tilde{\Omega}_{\mathrm{opt}}$ and $\tilde{T}_{\mathrm{opt}}$ as a
function of $\varphi$ ($d=10^{-13}$ for the blue-solid curve and $10^{-6}$ for
the red-dashed one, having both $\sigma=\sqrt{2}$). In addition, it is plotted
the analogous curve for the single-mode DOPO (gold, dashed-dotted curve). The
inset shows the dependence of the optimum frequency $\tilde{\Omega
}_{\mathrm{opt}}$ with the phase of the local oscillator $\varphi$.}%
\label{f2tmDOPO2}%
\end{center}
\end{figure}

From expression (\ref{FixedNoiseSpectrum}) we see that the maximum level of
squeezing is obtained at $\tilde{\Omega}=0$ (see also Figure \ref{f2tmDOPO2}a)
and when the phase of the local oscillator is tuned exactly to $\pi/2$. In
Figure \ref{f2tmDOPO1} we show the noise spectrum (\ref{FixedNoiseSpectrum})
for these parameters as a function of the detection time $\tilde{T}$ for 3
different values of $d$. We see that in all cases there exist an optimum
detection time for which squeezing is maximum. Minimizing Eq.
(\ref{FixedNoiseSpectrum}) with $\tilde{\Omega}=0$ and $\varphi=\pi/2$ with
respect to $\tilde{T}$, it is straightforward to find that this optimum
detection time is given by%
\begin{equation}
\tilde{T}_{\mathrm{opt}}=\sqrt{\frac{\sigma^{2}\left(  \sigma-1\right)
}{d\left(  \sigma^{2}+1\right)  }}, \label{Topt}%
\end{equation}
with an associated noise spectrum $V_{\mathrm{opt}}^{\mathrm{out}}=1/\tilde
{T}_{\mathrm{opt}}$ (shown in the inset of Figure \ref{f2tmDOPO1} as function
of $d$).

These results show that large levels of noise reduction are obtained for the
\textsf{Y} quadrature of the fixed TEM$_{01}$ mode, even for values of the
diffusion parameter $d$ as large as $10^{-6}$ (remember that $10^{-12}$ is a
more realistic value). However, in real experiments it is not possible to
ensure that $\varphi=90%
\operatorname{{{}^\circ}}%
$ with an uncertainty below approximately $1.5%
\operatorname{{{}^\circ}}%
$ \cite{Vahlbruch08,Mehmet10,Takeno07}, and hence we proceed now to
investigate the level of noise reduction predicted by
(\ref{FixedNoiseSpectrum}) when the local oscillator phase is different from
$\varphi=90%
\operatorname{{{}^\circ}}%
$.%

Of course, when $\varphi\neq90%
\operatorname{{{}^\circ}}%
$ the noise frequency with maximum squeezing is no longer $\tilde{\Omega}=0$,
as in this case the infinite fluctuations of $S_{01}^{0}$ at zero noise
frequency, due to the rotation noise, enter the noise spectrum (see Figure
\ref{f2tmDOPO2}a). By numerical minimization of $V^{\mathrm{out}}(\hat{X}%
_{01}^{\varphi};\Omega)$ with respect to $\tilde{\Omega}$ and $\tilde{T}$ for
different values of $\varphi$ and $d$, it is possible to show that the optimum
value of the detection time is almost independent of $\varphi$ for small
deviations of this from $90%
\operatorname{{{}^\circ}}%
$, and hence it is still given to a good approximation by (\ref{Topt}),
although in this case this minimum is less pronounced than in the $\varphi=90%
\operatorname{{{}^\circ}}%
$ case shown in Figure \ref{f2tmDOPO1} (i.e., the curve is almost horizontal
around $\tilde{T}_{\mathrm{opt}}$). On the other hand, the optimum noise
frequency $\tilde{\Omega}_{\mathrm{opt}}$ is independent of $d$ and depends on
$\varphi$ as shown in the inset of Figure \ref{f2tmDOPO2}b.

As for the squeezing level, in Figure \ref{f2tmDOPO2}b we show the noise
spectrum evaluated at $\tilde{T}_{\mathrm{opt}}$ and $\tilde{\Omega
}_{\mathrm{opt}}$ as a function of $\varphi$ for 2 different values of the
diffusion $d$. Together with these curves, we have plotted the noise spectrum
of the single--mode DOPO\footnote{In the single--mode DOPO the best levels of
squeezing are found below threshold, where its noise spectrum can be written
as $V^{\mathrm{out}}\left(  \hat{X}_{\mathrm{s}}^{\varphi};\Omega\right)
=1+S_{+}\left(  \Omega\right)  \cos^{2}\varphi-S_{-}\left(  \Omega\right)
\sin^{2}\varphi$, with $S_{\pm}\left(  \Omega\right)  =4\sigma/\left[  \left(
\sigma\pm1\right)  ^{2}+\tilde{\Omega}^{2}\right]  $, see (\ref{SpectraDOPO}%
).} evaluated for its optimum parameters (in this case it is optimized respect
to $\sigma$ and $\tilde{\Omega}$) as a function of $\varphi$. We see that the
noise reduction is independent of $d$ as $\varphi$ is taken apart from $90%
\operatorname{{{}^\circ}}%
$. On the other hand, the squeezing level is similar to that of the
single--mode DOPO, as the maximum difference between them are 1.5 \textrm{dB}
(a factor 1.4 in the noise spectrum) in favor of the single--mode DOPO, with
the advantage that in the 2tmDOPO this level is independent of the distance
from threshold.

Therefore we see that the phenomenon of noncritical squeezing through
spontaneous rotational symmetry breaking could be observed in the 2tmDOPO
without the need of following the random rotation of the generated pattern,
which makes its experimental realization feasible with currently available technology.

\section{Beyond the considered approximations\label{BeyondApp}}

\subsection{Beyond the adiabatic elimination of the pump}

The first assumption made in the search for the quantum properties of the
system was that $\gamma_{\mathrm{p}}\gg\gamma_{\mathrm{s}}$, a limit that
allowed the adiabatic elimination of the pump field. Now we are going to show
analytically that the phenomenon of squeezing induced by spontaneous
rotational symmetry breaking is still present even without this assumption,
but still working within the linearized theory (in the next section we will
check it via a numerical simulation of the complete nonlinear stochastic equations).

The way to show this is quite simple; starting from the complete equations
(\ref{ScaledLangevin2tmDOPO}), we expand the amplitudes $\beta_{m}$ around the
classical stationary solution (\ref{2tmDOPOabove}) as we made in
(\ref{LinExp}), but adding now a similar expression for the pump amplitudes:
$\beta_{0}=1+b_{0}$ and $\beta_{0}^{+}=1+b_{0}^{+}$ (note that for the pump
modes the $b$'s are directly small as the phase of this mode is locked to that
of the injection $\mathcal{E}_{\mathrm{p}}$). Then, linearizing these
equations for the fluctuations and noises, we arrive to a linear system
formally equal to (\ref{LinLan}), but with%
\begin{equation}
\mathbf{b}=%
\begin{pmatrix}
b_{0}\\
b_{0}^{+}\\
b_{+1}\\
b_{+1}^{+}\\
b_{-1}\\
b_{-1}^{+}%
\end{pmatrix}
\text{, }\boldsymbol{\zeta}=%
\begin{pmatrix}
0\\
0\\
\zeta\left(  \tau\right) \\
\zeta^{+}\left(  \tau\right) \\
\zeta^{\ast}\left(  \tau\right) \\
\left[  \zeta^{+}\left(  \tau\right)  \right]  ^{\ast}%
\end{pmatrix}
,
\end{equation}
and a linear matrix%
\begin{equation}
\mathcal{L}=%
\begin{pmatrix}
-\kappa & 0 & -\rho & 0 & -\rho & 0\\
0 & -\kappa & 0 & -\rho & 0 & -\rho\\
\rho & 0 & -1 & 0 & 0 & 1\\
0 & \rho & 0 & -1 & 1 & 0\\
\rho & 0 & 0 & 1 & -1 & 0\\
0 & \rho & 1 & 0 & 0 & -1
\end{pmatrix}
.
\end{equation}
Although in this case $\mathcal{L}$ is not Hermitian, it can be checked that
it possess a biorthonormal eigenbasis\footnote{This means that there exist a
set of eigenvectors $\mathbf{v}_{m}$ satisfying $\mathcal{L}\mathbf{v}%
_{m}=\lambda_{m}\mathbf{v}_{m}$, and another set $\mathbf{w}_{m}$ satisfying
$\mathcal{L}^{\dagger}\mathbf{w}_{m}=\lambda_{m}^{\ast}\mathbf{w}_{m}$, such
that $\mathbf{w}_{m}^{\ast}\cdot\mathbf{v}_{m}=\delta_{mn}$.}, and in
particular, the following two vectors are present in its eigensystem:
$\mathbf{w}_{0}^{\prime}=\frac{1}{2}\operatorname{col}\left(
0,0,1,-1,-1,1\right)  $ and $\mathbf{w}_{1}^{\prime}=\frac{1}{2}%
\operatorname{col}\left(  0,0,1,1,-1,-1\right)  $, with corresponding
eigenvalues $\lambda_{0}^{\prime}=0$ and $\lambda_{1}^{\prime}=-2$. These
eigenvectors have null projection onto the pump subspace, and coincide with
$\mathbf{w}_{0}$ and $\mathbf{w}_{1}$ in what concerns to the signal subspace,
see (\ref{Eigensystem}). Hence, all the properties derived from these vectors
are still present without any change. In particular, as they account for the
diffusion of $\theta$ and the squeezing properties of the dark mode, we can
conclude that these properties are still present when working out of the limit
$\gamma_{\mathrm{p}}\gg\gamma_{\mathrm{s}}$.

\subsection{Numerical simulation of the nonlinear equations}

In this section we will show that the diffusion of the orientation and the
associated noncritical squeezing of the dark mode, which have been found by
linearizing the Langevin equations, are also present when we consider the full
nonlinear problem. To do so, we will solve numerically the complete stochastic
equations (\ref{ScaledLangevin2tmDOPO}) using the semi--implicit algorithm
developed by Drummond and Mortimer in \cite{Drummond91}.

\begin{figure}
[t]
\begin{center}
\includegraphics[
height=2.0842in,
width=4.8966in
]%
{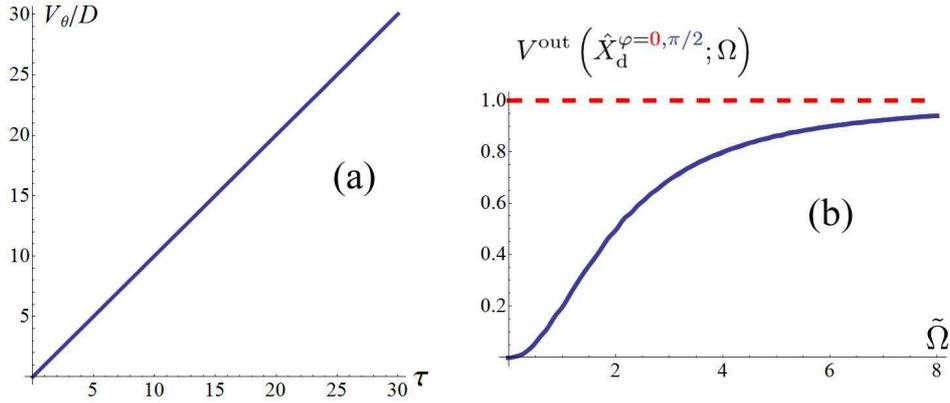}%
\caption{(a) Evolution of the variance of the orientation $\theta$ given by
the numerical simulation. It has been divided by the slope $D$ predicted by
the linearized theory (\ref{Theta Var}), so the straight line obtained is in
perfect agreement with this linear result. Quantitatively, a linear regression
analysis shows that the slope obtained by the data is $0.99985$, with standard
error $7\times10^{-5}$. (b) Noise spectrum of the \textsf{X} (red-dashed
curve) and \textsf{Y} (blue-solid curve) quadratures of the dark mode as
obtained by the numerical simulation. The results are in perfect agreement
with the ones predicted by the linearized theory (\ref{NoiseSpectra}). In
particular, a nonlinear regression analysis of the numerical data respect to
the function $\left(  \omega/2\right)  ^{2}/\left[  b+a\left(  \omega
/2\right)  ^{2}\right]  $, shows that $\left(  a,b\right)  =\left(
0.9996,1.0071\right)  $ are the best fit parameters having standard errors
$\left(  1\times10^{-4},4\times10^{-4}\right)  $, which is in good agreement
with the values $\left(  a,b\right)  =\left(  1,1\right)  $ predicted by
(\ref{NoiseSpectra}). Note finally that the small oscillations of the
blue-solid line would disappear by increasing the number of stochastic
trajectories $\Sigma$.}%
\label{f2tmDOPO3}%
\end{center}
\end{figure}

The details of the numerical simulation are explained in Appendix
\ref{Numerical2tmDOPO}. Here we just want to point out that the important
parameters of the simulation are the step size $\Delta\tau$ used to arrive
from $\tau=0$ to the final integration time $\tau_{\mathrm{end}}$, and the
number of stochastic trajectories, say $\Sigma$, which are used to evaluate
stochastic averages. The initial conditions $\beta_{m}\left(  0\right)  $ are
not relevant as the results in the stationary limit are independent of them.
The system parameters which have been chosen for the simulation are
$\sigma=\sqrt{2}$, $\kappa=1$ (to show also numerically that the adiabatic
elimination has nothing to do with the phenomenon), and $g=10^{-3}$. We
haven't chosen a smaller value for $g$ (like $10^{-6}$ as followed from the
physical parameters considered along the thesis) because such a small number
can make the simulation fail; nevertheless all the results we are going to
show should be independent of $g$ and $\sigma$, and we have also checked that
the same results are obtained for other values of these.%

It is also important to note that we have defined a general quadrature of the
rotating dark mode as (directly from (\ref{LGtoHGboson}) with $\beta=\theta$)%
\begin{align}
x_{\mathrm{d}}^{\varphi}  & =\frac{\mathrm{i}}{\sqrt{2}}\left[  e^{-\mathrm{i}%
\varphi}\left(  e^{\mathrm{i}\theta}\beta_{+1}-e^{-\mathrm{i}\theta}\beta
_{-1}\right)  \right]-\frac{\mathrm{i}}{\sqrt{2}}\left[  e^{\mathrm{i}\varphi}\left(
e^{-\mathrm{i}\theta}\beta_{+1}^{+}-e^{\mathrm{i}\theta}\beta_{-1}^{+}\right)
\right]  ,
\end{align}
with $\theta$ defined within the positive \textit{P} representation through%
\begin{equation}
e^{2\mathrm{i}\theta}=\frac{\beta_{-1}\beta_{+1}^{+}}{\left\vert \beta
_{-1}\right\vert \left\vert \beta_{+1}^{+}\right\vert }.
\end{equation}

Now let us show the results evaluated for the following simulation parameters:
$\Delta\tau=3\cdot10^{-3}$, $\tau_{\mathrm{end}}=30$ and $\Sigma
=7.5\cdot10^{6}$. This simulation has been compared with other ones having
different values of these parameters to ensure convergence.

In Figure \ref{f2tmDOPO3}a we show the variance of $\theta$ as a function of
time. The result has been normalized to $D$, so that the linear result
(\ref{Theta Var}) predicts a straight line forming 45$%
\operatorname{{{}^\circ}}%
$ with respect to the time axis. It can be appreciated that this is indeed
what the simulation shows (see also the caption of the figure).

In Figure \ref{f2tmDOPO3}b, we show the numerical results for the noise
spectrum associated to the quadratures of the dark mode. Only times above
$\tau=10$ have been considered in the correlation function to ensure being
working in the stationary limit. Again, the results shown in Figure
\ref{f2tmDOPO3}b are in reasonable agreement with the linear predictions
(\ref{XsSpectrum}) and (\ref{YsSpectrum}), as explained quantitatively in the
caption of Figure \ref{f2tmDOPO3}.

These results show that the phenomenon of noncritically squeezed light via
spontaneous rotational symmetry breaking is not a product of the linearization. 

%% file: 2tmDOPOwithISFO.tex
In this chapter we keep studying the properties of the 2tmDOPO, analyzing in
particular how the injection of a TEM$_{10}$ mode at the signal frequency
modifies its properties \cite{Navarrete11a}. The study of this \textit{signal
seed} is motivated by two experimental issues. First, the arbitrariness of the
initial orientation of the above--threshold pattern is highly inconvenient,
since one cannot know in advance within which orientation it will rise at a
particular realization of the experiment, what makes the matching of the local
oscillator to the initial dark mode quite difficult; hence some means of
fixing this orientation is called for, and we will show that this is exactly
what the signal seed accomplishes. On the other hand, it is customary to
inject a low intensity signal field in experiments to keep the cavity stable
and locked to the desired frequency.

Note that this signal seed would break the rotational invariance of the system
and hence a degradation of the squeezing level is expected. The goal of this
chapter is then to determine the impact of the signal injection on the quantum
properties of the 2tmDOPO.

In what follows, we will refer to this configuration as \textit{driven
}2tmDOPO, denoting then by \textit{free--running }2tmDOPO the system without
the injection at the signal frequency.

We will show that, fortunately, the 2tmDOPO still exhibits large levels of
noncritical squeezing, even larger than those exhibited by the usual
single-mode DOPO with injected signal, a result that is connected with the
squeezing due to the spontaneous breaking of the rotational symmetry that
exists in the absence of injection, but also to the existence of a new
bifurcation in the system.

The current system is closely related to that analyzed by Protsenko \textit{et
al.} \cite{Protsenko94} who considered a single--mode DOPO with injected
signal (see also \cite{dosSantos05,dosSantos07}). Indeed our model contains,
as a limit, the model studied in \cite{Protsenko94}, but we shall see that the
existence of a second signal mode substantially modifies the properties of the system.

\section{Model of the 2tmDOPO with injected signal}

The system we are dealing with is exactly the same as the one we have already
studied, with the exception that now the signal modes have also an injection
term. In the previous chapters we wrote the system's equations in terms of the
Laguerre--Gauss modes; now, however, it is quite recommendable to work in
Hermite--Gauss basis, as we are injecting one of these modes. To this aim, we
first use the relations (\ref{LGtoHGboson}) between the Laguerre--Gauss and
the Hermite--Gauss boson operators to write the following relation between the
corresponding normalized coherent amplitudes:
\begin{equation}
\beta_{\pm1}=\left(  \beta_{x}\mp\mathrm{i}\beta_{y}\right)  /\sqrt{2},
\label{boson}%
\end{equation}
where $\beta_{x/y}$ are the coherent amplitudes associated to the
$H_{10/01}(k_{\mathrm{s}};\mathbf{r}_{\perp},z)$ modes; then, we use these
relations to rewrite the Langevin equations (\ref{ScaledLangevin2tmDOPO})
which model the 2tmDOPO in terms of the $\beta_{x/y}$ amplitudes, arriving to
\begin{subequations}
\label{LangevinINJ}%
\begin{align}
\dot{\beta}_{0}  &  =\kappa\left[  \sigma-\beta_{0}-\left(  \beta_{x}%
^{2}+\beta_{y}^{2}\right)  /2\right]  ,\\
\dot{\beta}_{0}^{+}  &  =\kappa\left[  \sigma-\beta_{0}^{+}-\left(  \beta
_{x}^{+2}+\beta_{y}^{+2}\right)  /2\right]  ,\\
\dot{\beta}_{x}  &  =\varepsilon_{\mathrm{i}}-\beta_{x}+\beta_{0}\beta_{x}%
^{+}+g\sqrt{\beta_{0}}\zeta_{x}\left(  \tau\right)  ,\\
\dot{\beta}_{x}^{+}  &  =\varepsilon_{\mathrm{i}}^{\ast}-\beta_{x}^{+}%
+\beta_{0}^{+}\beta_{x}+g\sqrt{\beta_{0}^{+}}\zeta_{x}^{+}\left(  \tau\right)
,\\
\dot{\beta}_{y}  &  =-\beta_{y}+\beta_{0}\beta_{y}^{+}+g\sqrt{\beta_{0}}%
\zeta_{y}\left(  \tau\right)  ,\\
\dot{\beta}_{y}^{+}  &  =-\beta_{y}^{+}+\beta_{0}^{+}\beta_{y}+g\sqrt
{\beta_{0}^{+}}\zeta_{y}^{+}\left(  \tau\right)  ,
\end{align}
where in addition to the usual dimensionless parameters, we have introduced
the normalized injection parameter for the signal TEM$_{10}$ mode%
\end{subequations}
\begin{equation}
\varepsilon_{\mathrm{i}}=\frac{\chi\mathcal{E}_{\mathrm{s}}}{\gamma
_{\mathrm{s}}\sqrt{\gamma_{\mathrm{p}}\gamma_{\mathrm{s}}}}, \label{param}%
\end{equation}
being $\mathcal{E}_{\mathrm{s}}$ the injection rate parameter appearing in the
Hamiltonian piece corresponding to this signal injection, see (\ref{Hinj}).

\section{Classical steady states and their stability\label{ClassiIS}}

As discussed in depth in previous sections, the classical limit is recovered
from the Langevin equations (\ref{LangevinINJ}) by neglecting the noise terms
and by making the identification $\beta_{m}^{+}\rightarrow\beta_{m}^{\ast}$
($m=0,x,y$), arriving to
\begin{subequations}
\label{MODEL}%
\begin{align}
\dot{\beta}_{0}  &  =\kappa\left[  \sigma-\beta_{0}-\left(  \beta_{x}%
^{2}+\beta_{y}^{2}\right)  /2\right]  ,\\
\dot{\beta}_{x}  &  =\varepsilon_{\mathrm{i}}-\beta_{x}+\beta_{0}\beta
_{x}^{\ast},\\
\dot{\beta}_{y}  &  =-\beta_{y}+\beta_{0}\beta_{y}^{\ast}.
\end{align}

In this section we study the steady states of this classical model, as well as
their stability properties. Notice that by setting $\beta_{y}=0$, equations
(\ref{MODEL}) become those for a single--mode DOPO with injected signal. Such
model (generalized by the presence of detunings) was studied in
\cite{Drummond79} for the case of real $\varepsilon$ and then extended to
arbitrary injection phases in \cite{Protsenko94}.

A straightforward inspection of (\ref{MODEL}) leads to the conclusion that for
$\varepsilon\neq0$ there are only two types of classical solutions, either
with $\beta_{y}=0$ or with $\beta_{y}\neq0$, both having $\beta_{x}\neq0$. In
the first case the intracavity signal field is single--mode and has the same
shape as the injection; on the other hand, when $\beta_{y}\neq0$ the
intracavity field is in a coherent superposition of both transverse modes. We
will refer to these two types of states as the \textit{one--mode} and
\textit{two--mode} solutions, respectively.

In the following we decompose the fields into modulus and phase as
\end{subequations}
\begin{equation}
\varepsilon_{\mathrm{i}}=\sqrt{\mathcal{I}_{\mathrm{i}}}e^{\mathrm{i}%
\varphi_{\mathrm{i}}},\ \ \beta_{m}=\sqrt{I_{m}}e^{\mathrm{i}\varphi_{m}},
\label{mod-phase}%
\end{equation}
when needed.%

\begin{figure}
[t]
\begin{center}
\includegraphics[
height=3.4912in,
width=3.5838in
]%
{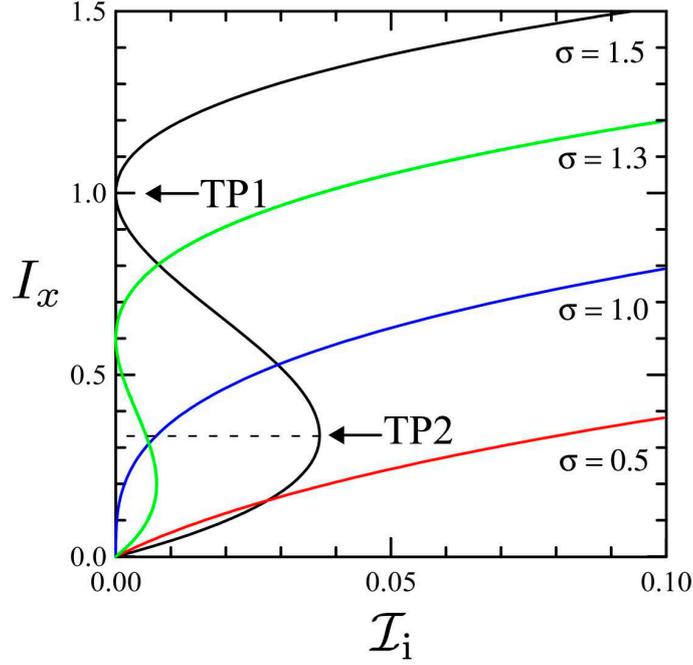}%
\caption{Intensity of the TEM$_{10}$ mode for the `one--mode' solution as a
function of the injected field intensity $\mathcal{I}_{\mathrm{i}}$ for
$\varphi_{\mathrm{i}}=0$ and the four indicated values of $\sigma$. Note that
for $\sigma>1$ the solution is three--valued. We have indicated the turning
points TP1 and TP2 for the case $\sigma=1.5$.}%
\label{f2tmDOPOwithIS1}%
\end{center}
\end{figure}

\subsection{The one--mode solution\label{1Msol}}

We look first for a solution in which the non-injected mode $H_{01}%
(k_{\mathrm{s}};\mathbf{r}_{\perp},z)$ is off ($\beta_{y}=0$). This solution
was the one analyzed in \cite{Protsenko94}, and we refer to that work for the
details. Using our notation, the intensity $I_{x}$ of mode $H_{10}%
(k_{\mathrm{s}};\mathbf{r}_{\perp},z)$ verifies the implicit relation
\begin{equation}
\mathcal{I}_{\mathrm{i}}=\frac{2\left(  \sigma^{2}-X^{2}\right)  ^{2}\left(
X-1\right)  }{\sigma^{2}+X^{2}+2\sigma X\cos2\varphi_{\mathrm{i}}},
\label{Ix_1M}%
\end{equation}
where $X=1+\tfrac{1}{2}I_{x}$, which is a quintic polynomial in $I_{x}$, while
its phase is given by%
\begin{equation}
e^{\mathrm{i}\varphi_{x}}=\pm\frac{\sigma e^{-\mathrm{i}\varphi_{\mathrm{i}}%
}+Xe^{\mathrm{i}\varphi_{\mathrm{i}}}}{\sqrt{\sigma^{2}+X^{2}+2\sigma
X\cos2\varphi_{\mathrm{i}}}}, \label{phix_1M}%
\end{equation}
where the plus and minus signs\ correspond, respectively, to cases $X>\sigma$
and $X<\sigma$. Finally, the intracavity pump field is given by
\begin{subequations}
\label{campo0}%
\begin{align}
I_{0}  &  =\tfrac{1}{4}I_{x}^{2}+\sigma^{2}-\sigma I_{x}\cos2\varphi
_{x},\label{I0sol}\\
\varphi_{0}  &  =\arg\left(  2\sigma-I_{x}e^{2\mathrm{i}\varphi_{x}}\right)  .
\label{fi0off}%
\end{align}

In Figure \ref{f2tmDOPOwithIS1} we show the dependence of the intensity
$I_{x}$ on the injected intensity $\mathcal{I}_{\mathrm{i}}$ in the case
$\varphi_{\mathrm{i}}=0$ (amplification regime) for various values of the
pumping level $\sigma$ as indicated. For $\sigma<1$ (below the free--running
2tmDOPO threshold) the curve is single--valued, while for $\sigma>1$ the curve
is multivalued for small injection intensities. The location of the turning
points (marked as TP1 and TP2) existing for $\sigma>1$ can be obtained from
Eq. (\ref{Ix_1M}) by solving $\partial\mathcal{I}_{\mathrm{i}}/\partial
I_{x}=0$ (see \cite{Protsenko94} for more details). One of them is analytical,%
\end{subequations}
\begin{equation}
I_{x}^{\mathrm{TP1}}=2\left(  \sigma-1\right)  , \label{TP1}%
\end{equation}
which exists only if $\sigma>1$, as expected. For $I_{x}=I_{x}^{\mathrm{TP1}}%
$, $\mathcal{I}_{\mathrm{i}}=0$ as follows from (\ref{Ix_1M}), and then
$I_{x}^{\mathrm{TP1}}$ corresponds to the upper turning points in Figure
\ref{f2tmDOPOwithIS1}. The intensity $I_{x}^{\mathrm{TP2}}$ at the lower
turning points in Figure \ref{f2tmDOPOwithIS1} is given by a fourth order
polynomial, which we do not present here as it gives no analytical information
and, moreover, because it plays no effective role in the system behavior as is
shown below in the stability analysis.

The final conclusion is simple: For $\sigma<1$ Eq. (\ref{Ix_1M}) has a single
real and positive solution, while for $\sigma>1$ the solution is three--valued
for small injection intensities.

\begin{figure}
[t]
\begin{center}
\includegraphics[
height=2.5581in,
width=4.804in
]%
{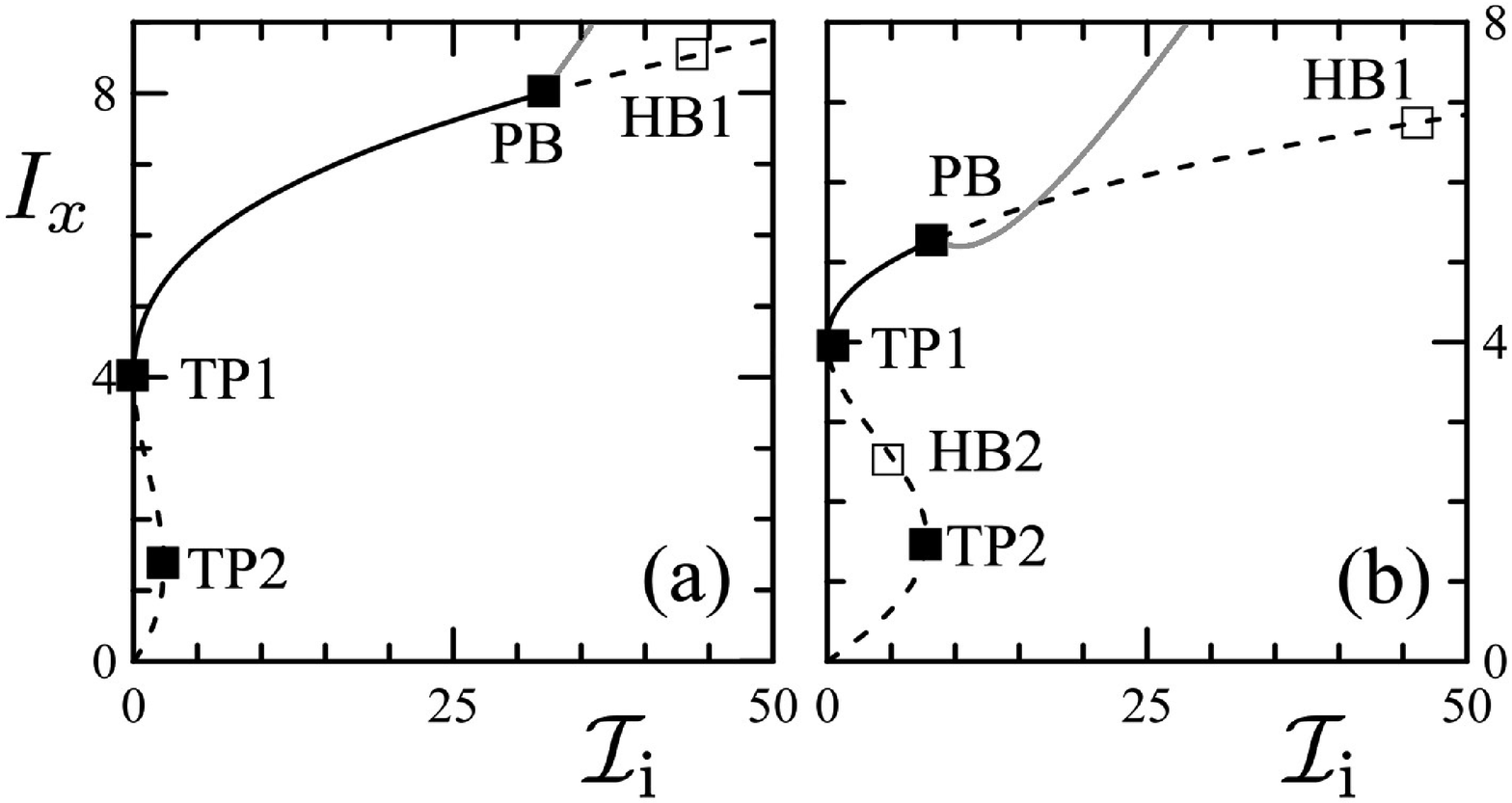}%
\caption{Steady states of the system on the plane $\left\langle \mathcal{I}%
_{\mathrm{i}},I_{10}\right\rangle $. Continuous and dashed lines correspond to
stable and unstable solutions, respectively. The black and grey lines
correspond to the one--mode and two--mode solutions, respectively. The
parameter values are $\sigma=3$ and $\varphi_{\mathrm{i}}=0$, $\kappa=0.25$ in
(a) and $\varphi_{\mathrm{i}}=\pi/3$, $\kappa=1$ in (b). Notice that for
$\mathcal{I}_{\mathrm{i}}\neq0$ the pitchfork bifurcation where the TEM$_{01}$
mode is switched on (marked as PB) is the only relevant bifurcation. TP1 and
TP2 denote the turning points (see Figure \ref{f2tmDOPOwithIS1}) and HB1 and
HB2 denote two different Hopf bifurcations (which are always preceded by the
bifurcations occurring at points TP1 and PB, see text).}%
\label{f2tmDOPOwithIS2}%
\end{center}
\end{figure}

As for the stability of this one--mode solution (see Figure
\ref{f2tmDOPOwithIS2} for a graphical summary of the discussion to follow),
the stability matrix $\mathcal{L}$\ turns out to be block-diagonal (as usual,
the switched off mode is decoupled from the rest). One submatrix affects only
the subspace $\left(  \beta_{y},\beta_{y}^{\ast}\right)  $ and then governs
the possible switching-on of mode $H_{01}(k_{\mathrm{s}};\mathbf{r}_{\perp
},z)$, while the other affects the rest of variables and is the same as the
one considered in \cite{Protsenko94}. Accordingly, the characteristic
polynomial is factorized into two.

The first polynomial, quadratic in $\lambda$, governs the evolution of the
perturbations of the variables $\{\beta_{y},\beta_{y}^{\ast}\}$ as commented,
and gives the following two eigenvalues%

\begin{equation}
\lambda_{1,2}=-1\pm\sqrt{I_{0}}, \label{autov1}%
\end{equation}
with $I_{0}$ given by Eq. (\ref{I0sol}). Solving $\lambda_{1}=0$ gives two
solutions, namely $I_{x}=I_{x}^{\mathrm{TP1}}$, Eq. (\ref{TP1}), which occurs
at $\mathcal{I}_{\mathrm{i}}=0$, and
\begin{equation}
I_{x}=2\sqrt{1+\sigma^{2}+2\sigma\cos2\varphi_{\mathrm{i}}}\equiv
I_{x}^{\mathrm{PB}}, \label{PitchBif}%
\end{equation}
which occurs at, using (\ref{Ix_1M}),%
\begin{equation}
\mathcal{I}_{\mathrm{i}}=4\left(  1+\sigma\cos2\varphi_{\mathrm{i}}%
+\sqrt{1+2\sigma\cos2\varphi_{\mathrm{i}}+\sigma^{2}}\right)  \equiv
\mathcal{I}_{\mathrm{i}}^{\mathrm{PB}}. \label{Ii_PB}%
\end{equation}
In a moment it will become clear why we denote by PB this new bifurcation.

The analyzed solution is stable (i.e., $\lambda_{1}<0$) for $I_{x}%
^{\mathrm{TP1}}<I_{x}<I_{x}^{\mathrm{PB}}$. For $I_{x}>I_{x}^{\mathrm{PB}}$ or
$I_{x}<I_{x}^{\mathrm{TP1}}$, $\lambda_{1}>0$ and the one--mode solution is
unstable. For $\sigma<1$ the turning points TP1 and TP2 don't exist and the
one--mode solution is stable all along its unique branch until $I_{x}%
=I_{x}^{\mathrm{PB}}$. On the other hand, for $\sigma>1$ the one--mode
solution is stable only in the portion of its upper branch going from the
turning point TP1 to the bifurcation PB; hence the branches lying below the
turning point TP1 are unstable, including the turning point TP2 as anticipated.

Let us now understand the meaning of these two instabilities, having in mind
that they affect the subspace corresponding to the (non--injected) TEM$_{01}$ mode
$H_{01}(k_{\mathrm{s}};\mathbf{r}_{\perp},z)$, which is off. The existence of
an instability at the turning point TP1 is due to the fact that its
corresponding injection intensity is null, i.e., $\mathcal{I}_{\mathrm{i}%
}^{\mathrm{TP1}}=0$, Figure \ref{f2tmDOPOwithIS1}. In the absence of injection
the system is rotationally symmetric in the transverse plane, and the 2tmDOPO
can generate equally both modes $H_{10}(k_{\mathrm{s}};\mathbf{r}_{\perp},z)$
and $H_{01}(k_{\mathrm{s}};\mathbf{r}_{\perp},z)$; hence, at $\mathcal{I}%
_{\mathrm{i}}=0$ the solution with $\beta_{y}=0$ is marginally unstable. On
the other hand, the instability at PB is of a different nature as for
injection intensities $\mathcal{I}_{\mathrm{i}}>\mathcal{I}_{\mathrm{i}%
}^{\mathrm{PB}}$ the one--mode solution is no more stable. As this instability
governs the growth of mode $H_{01}(k_{\mathrm{s}};\mathbf{r}_{\perp},z)$ we
conclude that at PB a pitchfork bifurcation takes place giving rise to a new
steady state branch (the two--mode steady state, see Section \ref{TMsol}).
This instability is analogous to the one predicted in intracavity type II
second harmonic generation, where the field orthogonal\ (in the polarization
sense) to the injected one is generated at a pitchfork bifurcation
\cite{Ou94,Eschmann94,Jack96,Andersen03a,Andersen03b,Zhai04,Zhai05,Luo05}.

We show in the reminder of the section that TP1 and PB are the only relevant
instabilities in the 2tmDOPO with injected signal.%

As for the second polynomial it is, as commented, the one analyzed in
\cite{Protsenko94}. We summarize next the results of that analysis and refer
to \cite{Protsenko94} for the details. When $\sigma>1$ an eigenvalue is null
in correspondence with the turning point TP2 in Figure \ref{f2tmDOPOwithIS1},
as usual, but this point (as well as the whole lower and intermediate
branches) is already unstable as stated above. On the other hand, a pair of
complex--conjugate eigenvalues become purely imaginary (then with null real
part) at an intensity $I_{x}=I_{x}^{\mathrm{HB1}}$ existing at any value of
$\sigma$, thus signalling a so-called Hopf bifurcation. In the case $\sigma>1$
the Hopf bifurcation is located on the upper branches in Figure
\ref{f2tmDOPOwithIS1}, and it is then relevant to ask if this bifurcation can
play a role in the 2tmDOPO with injected signal. The expression for
$I_{x}^{\mathrm{HB1}}$ is not analytical in general; however in the
amplification regime ($\varphi_{\mathrm{i}}=0$), $I_{x}^{\mathrm{HB1}%
}=2\left(  \kappa+\sigma+1\right)  >2\left(  \sigma+1\right)  =I_{x}%
^{\mathrm{PB}}$, see (\ref{PitchBif}). As $I_{x}^{\mathrm{HB1}}>I_{x}%
^{\mathrm{PB1}}$, the Hopf bifurcation is always preceded by the pitchfork
bifurcation and then the former never comes into play (note that
$I_{x}^{\mathrm{HB1}}\rightarrow I_{x}^{\mathrm{PB1}}$ for $\kappa
\rightarrow0$ though). On the other hand a second Hopf bifurcation (HB2)
exists in the case $\sigma>1$, but it is always located below the upper
turning point TP1 in Figure \ref{f2tmDOPOwithIS1} \cite{Protsenko94} and then
affects an already unstable solution. In other words, the Hopf bifurcations of
the one--mode solution, which coincide with those of the single--mode DOPO
with injected signal, are always preceded by other bifurcations genuine of the
driven 2tmDOPO, and then play no role in our case. While we have not extended
this analytical proof to arbitrary values of $\varphi_{\mathrm{i}}$, we have
convinced ourselves through a numerical study, that its validity is
general:\ The only bifurcations affecting the one--mode solution are the
turning point bifurcation TP1, and the pitchfork bifurcation PB that gives
rise to the switch on of the mode $H_{01}(k_{\mathrm{s}};\mathbf{r}_{\perp
},z)$. In Figure \ref{f2tmDOPOwithIS2} we exemplify these results for two sets
of parameters.

\subsection{The two--mode solution \label{TMsol}}

Next we consider the case $\beta_{y}\neq0$, which is genuine of the present
driven 2tmDOPO model. After simple but tricky algebra on finds $\beta
_{0}=-\exp\left(  2\mathrm{i}\varphi_{\mathrm{i}}\right)  $ (the intracavity
pump intensity $\left\vert \beta_{0}\right\vert ^{2}=1$ is clamped as usual in
OPOs after a bifurcation is crossed),%
\begin{equation}
\varphi_{y}=\varphi_{\mathrm{i}}\pm\frac{\pi}{2},\ \ \ \ \tan\left(
\varphi_{x}-\varphi_{\mathrm{i}}\right)  =-\frac{4\sigma}{\mathcal{I}%
_{\mathrm{i}}}\sin\left(  2\varphi_{\mathrm{i}}\right)  , \label{phi2M}%
\end{equation}
with the constrain $\left(  \varphi_{x}-\varphi_{\mathrm{i}}\right)
\in\left[  -\frac{\pi}{2},\frac{\pi}{2}\right]  $, and
\begin{subequations}
\label{2M}%
\begin{align}
I_{x}  &  =\frac{\mathcal{I}_{\mathrm{i}}}{4}+\frac{4\sigma^{2}}%
{\mathcal{I}_{\mathrm{i}}}\sin^{2}2\varphi_{\mathrm{i}},\label{Ix_2M}\\
I_{y}  &  =\frac{\mathcal{I}_{\mathrm{i}}}{4}-\frac{4\sigma^{2}}%
{\mathcal{I}_{\mathrm{i}}}\sin^{2}2\varphi_{\mathrm{i}}-2\left[  1+\sigma
\cos2\varphi_{\mathrm{i}}\right]  . \label{Iy_2M}%
\end{align}
Note that the phase of the mode $H_{01}(k_{\mathrm{s}};\mathbf{r}_{\perp},z)$,
$\varphi_{y}$, can take any of two opposite values (this residual discrete
symmetry survives even with the injection of the TEM$_{10}$ mode), see
(\ref{phi2M}); in other words, the sign of $\beta_{y}$ can be either, as usual
in DOPOs above threshold.

For this solution to exist $I_{y}$ must be a positive real ---note that
$I_{x}>0$ always, as follows from (\ref{Ix_2M})---. From (\ref{Iy_2M}) the
condition $I_{y}>0$ is seen to be equivalent to $I_{x}>I_{x}^{\mathrm{PB}}$
or, alternatively, $\mathcal{I}_{\mathrm{i}}>\mathcal{I}_{\mathrm{i}%
}^{\mathrm{PB}}$, as expected: At the pitchfork bifurcation the two--mode
solution is born.%

\begin{figure}
[t]
\begin{center}
\includegraphics[
height=3.2508in,
width=3.4791in
]%
{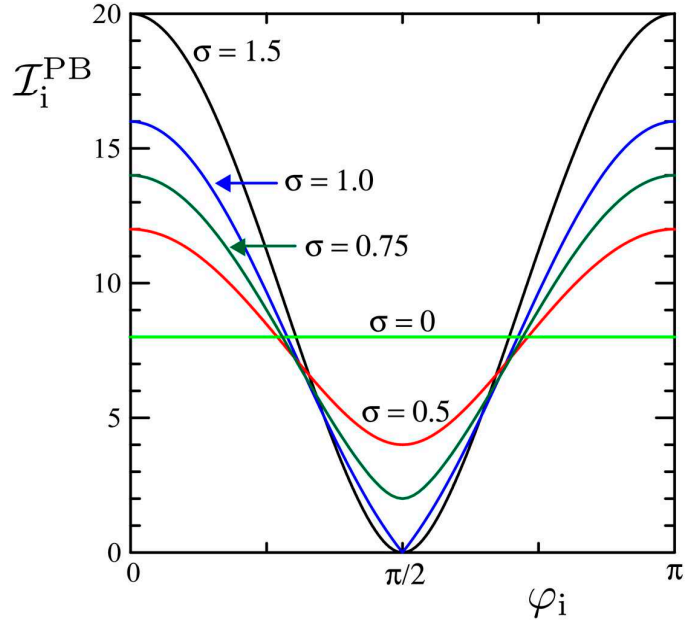}%
\caption{$\mathcal{I}_{\mathrm{i}}^{\mathrm{PB}}$ as a function of the
injected field phase $\varphi$ for the indicated values of $\sigma$, see
(\ref{PitchBif}). Above this value of the injected intensity the non-injected
TEM$_{01}$ mode is switched on. Note that for $\sigma>1$ and $\varphi=\pi/2$
there is no threshold for the generation of the TEM$_{01}$ mode.}%
\label{f2tmDOPOwithIS3}%
\end{center}
\end{figure}

In Figure \ref{f2tmDOPOwithIS3} we show the dependence on $\varphi
_{\mathrm{i}}$ of $\mathcal{I}_{\mathrm{i}}^{\mathrm{PB}}$ for several values
of $\sigma$. In the absence of pump ($\sigma=0$) $\mathcal{I}_{\mathrm{i}%
}^{\mathrm{PB}}$ is independent of the injection's phase $\varphi_{\mathrm{i}%
}$ as no reference phase exists in this case. For $\sigma\neq0$ the threshold
has a maximum in the case of amplification ($\varphi_{\mathrm{i}}=0$) and a
minimum in the case of attenuation ($\varphi_{\mathrm{i}}=\pi/2$).
Interestingly, this minimum is zero when the external pump is above the
free--running 2tmDOPO threshold ($\sigma>1$). These results are actually quite
intuitive: In the attenuation case, $\varphi_{\mathrm{i}}=\pi/2$, the injected
mode $H_{x}\left(  \mathbf{r}\right)  $ is depleted as it has the `wrong'
phase, a fact that makes easier the amplification of the orthogonal mode
$H_{01}(k_{\mathrm{s}};\mathbf{r}_{\perp},z)$, while the amplification case
$\varphi_{\mathrm{i}}=0$ is obviously detrimental for mode $H_{01}%
(k_{\mathrm{s}};\mathbf{r}_{\perp},z)$ as mode $H_{10}(k_{\mathrm{s}%
};\mathbf{r}_{\perp},z)$ is being amplified.

This two--mode solution does not correspond to a HG mode as it consists of the
superposition of two orthogonal HG modes with different amplitudes and phases.
Curiously, in the special case $\varphi_{\mathrm{i}}=\pi/2$ and $\sigma=1$,
equations (\ref{2M}) imply that $\beta_{x}=\pm\mathrm{i}\beta_{y}$, with
$\beta_{x}=\frac{1}{2}\sqrt{\mathcal{I}_{\mathrm{i}}}e^{\mathrm{i}%
\varphi_{\mathrm{i}}}$. This is also true for any $\varphi_{\mathrm{i}}$ value
for sufficiently strong injection $\mathcal{I}_{\mathrm{i}}$. This means that
in these cases the emitted mode is a pure LG mode with $+1$ or $-1$ OAM, see
Eq. (\ref{HtoL}), which is a somewhat unexpected result.

As for the linear stability analysis of this two--mode solution one gets a
sixth order characteristic polynomial for the eigenvalue $\lambda$ from which
no conclusions can be drawn in general. However in the special cases
$\varphi_{\mathrm{i}}=0$ and $\varphi_{\mathrm{i}}=\pi/2$ the polynomial gets
factorized giving rise to two cubic equations of the form $0=P_{1,2}\left(
\lambda\right)  $, with%
\end{subequations}
\begin{equation}
P_{1}\left(  \lambda\right)  =\lambda^{3}+\left(  \kappa+2\right)  \lambda
^{2}+\frac{\kappa}{2}\left(  \mathcal{I}_{\mathrm{i}}\mp4\sigma\right)
\lambda+\frac{\kappa}{2}\mathcal{I}_{\mathrm{i}},
\end{equation}
and $P_{2}\left(  \lambda\right)  =P_{1}\left(  \lambda\right)  \mp
4\kappa\left(  \sigma+1\right)  $, where the upper (lower)\ signs of the
polynomials correspond to the case $\varphi_{\mathrm{i}}=0$ ($\varphi
_{\mathrm{i}}=\pi/2$). It is now easy to demonstrate that all eigenvalues have
negative real part when $\mathcal{I}_{\mathrm{i}}>\mathcal{I}_{\mathrm{i}%
}^{\mathrm{PB}}$: By taking $\lambda=\mathrm{i}\Omega$ in the above
polynomials one obtains the frequency at the bifurcations (if any) and the
values of the injected signal strength $\mathcal{I}_{\mathrm{i}}$ leading to
instabilities. It is easy to see that these bifurcations occur for
$\mathcal{I}_{\mathrm{i}}<\mathcal{I}_{\mathrm{i}}^{\mathrm{PB}}$ (where the
solution does not exist) and hence the two--mode solution is stable within its
whole domain of existence. Through a numerical study of the characteristic
polynomials for arbitrary injection phase $\varphi_{\mathrm{i}}$ we have
convinced ourselves that this conclusion holds always.

\subsection{Summary}

In order to have a global picture of the steady states and their stability,
what is necessary in order to perform the quantum analysis, we summarize the
general results obtained up to now. After all, the picture is very simple, see
Figure \ref{f2tmDOPOwithIS2}.

On one hand the one--mode solution (with mode $H_{01}(k_{\mathrm{s}%
};\mathbf{r}_{\perp},z)$ off) is stable from the null injection point
$\mathcal{I}_{\mathrm{i}}=0$ till the pitchfork bifurcation point at
$\mathcal{I}_{\mathrm{i}}=\mathcal{I}_{\mathrm{i}}^{\mathrm{PB}}$. For
$\sigma<1$ this comprises the single branch existing between these points
(Figure \ref{f2tmDOPOwithIS1}), while for $\sigma>1$ the stable domain extends
from the turning point TP1 along the upper branch till the pitchfork
bifurcation point PB (Figure \ref{f2tmDOPOwithIS1}). No other instabilities
affect this solution. On the other hand the two--mode solution is born at the
pitchfork bifurcation point $\mathcal{I}_{\mathrm{i}}=\mathcal{I}_{\mathrm{i}%
}^{\mathrm{PB}}$ and is stable for any injection intensity $\mathcal{I}%
_{\mathrm{i}}>\mathcal{I}_{\mathrm{i}}^{\mathrm{PB}}$.

\section{Quantum properties of the non-injected mode\label{QuantumIS}}

In this section we study the squeezing properties of the system by linearizing
the Langevin equations (\ref{LangevinINJ}). We are specially interested in the
squeezing properties of the TEM$_{01}$ mode (as it corresponds to the dark
mode in the free--running configuration). Moreover, once we move above the
threshold for its classical generation, we expect it to approach a coherent
state and hence we focus on the region where its mean field is still zero,
that is, we focus on the one--mode solution (Section \ref{1Msol}).%

\begin{figure}
[t]
\begin{center}
\includegraphics[
height=3.7446in,
width=3.998in
]%
{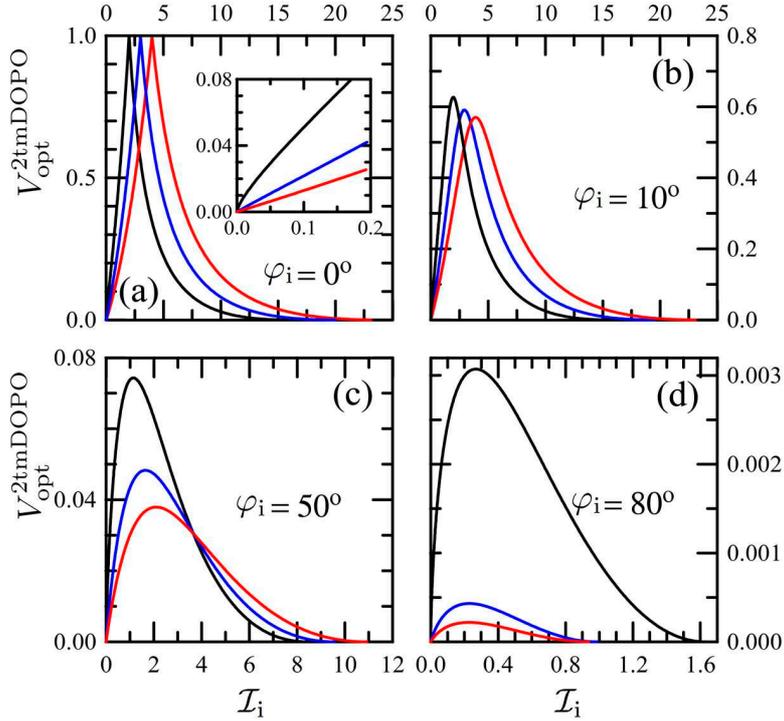}%
\caption{Dependence of the optimal noise spectrum $V_{\mathrm{opt}%
}^{\mathrm{2DOPO}}$, Eq. (\ref{Vopt}), on the injected intensity
$\mathcal{I}_{\mathrm{i}}$ for the four values of the injection phase
$\varphi_{\mathrm{i}}$ indicated at each figure. For each of $\varphi
_{\mathrm{i}}$, three values of $\sigma$ have been plotted, namely $\sigma=1$
(black), $1.5$ (blue), and $2$ (red), which correspond to the upper, middle
and lower curves at the left of the maxima, respectively. Notice the different
scales in each figure.}%
\label{f2tmDOPOwithIS4}%
\end{center}
\end{figure}

As usual, we write $\beta_{m}=\bar{\beta}_{m}+b_{m}$, where the overbar
indicates steady state and $b_{m}$ is an order $g$ fluctuation. Then,
considering the one--mode solution ($\bar{\beta}_{y}=0$), the equations
(\ref{LangevinINJ}) can linearized as%
\begin{subequations}
\begin{align}
\mathbf{\dot{b}}_{0x}  &  =\mathcal{L}_{0x}\mathbf{b}_{0x}+g\sqrt{I_{0}%
}\mathbf{\zeta}_{0x}\left(  t\right)  ,\\
\mathbf{\dot{b}}_{y}  &  =\mathcal{L}_{y}\mathbf{b}_{y}+g\sqrt{I_{0}%
}\mathbf{\zeta}_{y}\left(  t\right)  ,
\end{align}
where $\mathbf{b}_{0x}=\operatorname{col}\left(  b_{0},b_{0}^{+},b_{x}%
,b_{x}^{+}\right)  $, $\mathbf{b}_{y}=\operatorname{col}\left(  b_{y}%
,b_{y}^{+}\right)  $,%
\end{subequations}
\begin{subequations}
\begin{align}
\mathcal{L}_{x}  &  =\left(
\begin{array}
[c]{cccc}%
-\kappa & 0 & -\kappa\bar{\beta}_{x} & 0\\
0 & -\kappa & 0 & -\kappa\bar{\beta}_{x}^{\ast}\\
\bar{\beta}_{x}^{\ast} & 0 & -1 & \bar{\beta}_{0}\\
0 & \bar{\beta}_{x} & \bar{\beta}_{0}^{\ast} & -1
\end{array}
\right)  ,\\
\mathcal{L}_{y}  &  =\left(
\begin{array}
[c]{cc}%
-1 & \bar{\beta}_{0}\\
\bar{\beta}_{0}^{\ast} & -1
\end{array}
\right)  ,
\end{align}
and the noise vectors read%
\end{subequations}
\begin{subequations}
\begin{align}
\mathbf{\zeta}_{0x}  &  =\operatorname{col}\left(  0,0,e^{\mathrm{i}%
\varphi_{0}/2}\zeta_{x},e^{-\mathrm{i}\varphi_{0}/2}\zeta_{x}^{+}\right)  ,\\
\mathbf{\zeta}_{y}  &  =\operatorname{col}\left(  e^{\mathrm{i}\varphi_{0}%
/2}\zeta_{y},e^{-\mathrm{i}\varphi_{0}/2}\zeta_{y}^{+}\right)  .
\end{align}

The procedure we follow below is the same we have used elsewhere in the
previous chapters: We project quantum fluctuations onto the eigenvectors of
the Hermitian matrix $\mathcal{L}_{y}$, as these projections are, up to a
constant factor, the relevant quadratures of the output TEM$_{01}$ mode (those
that are maximally squeezed or antisqueezed). The eigensystem of matrix
$\mathcal{L}_{y}$ ($\mathcal{L}_{y}\cdot\mathbf{w}_{i}=\lambda_{i}%
\mathbf{w}_{i}$) comprises the eigenvalues $\lambda_{1,2}$ given in
(\ref{autov1}) and the associated eigenvectors
\end{subequations}
\begin{equation}
\mathbf{w}_{1,2}=\left(  e^{\mathrm{i}\varphi_{0}/2},\mp e^{-\mathrm{i}%
\varphi_{0}/2}\right)  /\sqrt{2},
\end{equation}
with $\varphi_{0}$ given by Eq. (\ref{fi0off}). Defining the projections
$c_{y}^{\left(  j\right)  }=\mathbf{w}_{j}^{\ast}\cdot\mathbf{b}_{y}$, we
obtain the following decoupled equations%
\begin{equation}
\dot{c}_{y}^{\left(  j\right)  }=\lambda_{j}c_{y}^{\left(  j\right)  }%
+g\sqrt{I_{0}}\zeta_{y}^{\left(  j\right)  }\left(  t\right)  , \label{dcdt}%
\end{equation}
where we defined the standard real white Gaussian noises $\zeta_{y}^{\left(
1,2\right)  }=\left(  \zeta_{y}\mp\zeta_{y}^{+}\right)  /\sqrt{2}$. It is then
trivial to obtain%
\begin{equation}
x_{y}^{\varphi_{0}/2}=\sqrt{2}c_{y}^{\left(  2\right)  },\ \ \ \ \ x_{y}%
^{\varphi_{0}/2+\pi/2}=-\mathrm{i}\sqrt{2}c_{y}^{\left(  1\right)  }.
\end{equation}
After integration of (\ref{dcdt}) ---see Appendix \ref{LinStoApp}--- and
substitution into the stationary squeezing spectrum (\ref{ScaledSqSpectrum}),
we obtain%
\begin{subequations}
\begin{align}
V^{\mathrm{out}}\left(  \hat{X}_{y}^{\varphi_{0}/2};\Omega\right)   &
=1+\frac{4\sqrt{I_{0}}}{\left(  1-\sqrt{I_{0}}\right)  ^{2}+\tilde{\Omega}%
^{2}},\\
V^{\mathrm{out}}\left(  \hat{Y}_{y}^{\varphi_{0}/2};\Omega\right)   &
=1-\frac{4\sqrt{I_{0}}}{\left(  1+\sqrt{I_{0}}\right)  ^{2}+\tilde{\Omega}%
^{2}}. \label{Vopt}%
\end{align}

The only quadrature that can be perfectly squeezed ($V^{\mathrm{out}}=0$) is
$Y_{y}^{\varphi_{0}/2}$ at $\Omega=0$ when $I_{0}=1$ (quadrature
$X_{y}^{\varphi_{0}/2}$ exhibits antisqueezing). According to (\ref{campo0})
and (\ref{Ix_1M}) this happens only in two cases: Either at the turning point
TP1 ($\mathcal{I}_{\mathrm{i}}=0$, no injection, where there exists perfect
squeezing because of the rotational symmetry breaking occurring in the
free--running 2tmDOPO at any $\sigma>1$), or when $\mathcal{I}_{\mathrm{i}%
}=\mathcal{I}_{\mathrm{i}}^{\mathrm{PB}}$, which corresponds to the pitchfork
bifurcation where mode $H_{01}(k_{\mathrm{s}};\mathbf{r}_{\perp},z)$ is
switched on. Figure \ref{f2tmDOPOwithIS4} shows the dependence on the injected
intensity $\mathcal{I}_{\mathrm{i}}$ of the optimal squeezing level
$V_{\mathrm{opt}}^{\mathrm{2tmDOPO}}\equiv V^{\mathrm{out}}(\hat{Y}%
_{y}^{\varphi_{0}/2};\Omega=0)$, see Eq. (\ref{Vopt}), for several values of
the pump level $\sigma$ and the injection phase $\varphi_{\mathrm{i}}$.%

\begin{figure}
[t]
\begin{center}
\includegraphics[
width=10.7 cm
]%
{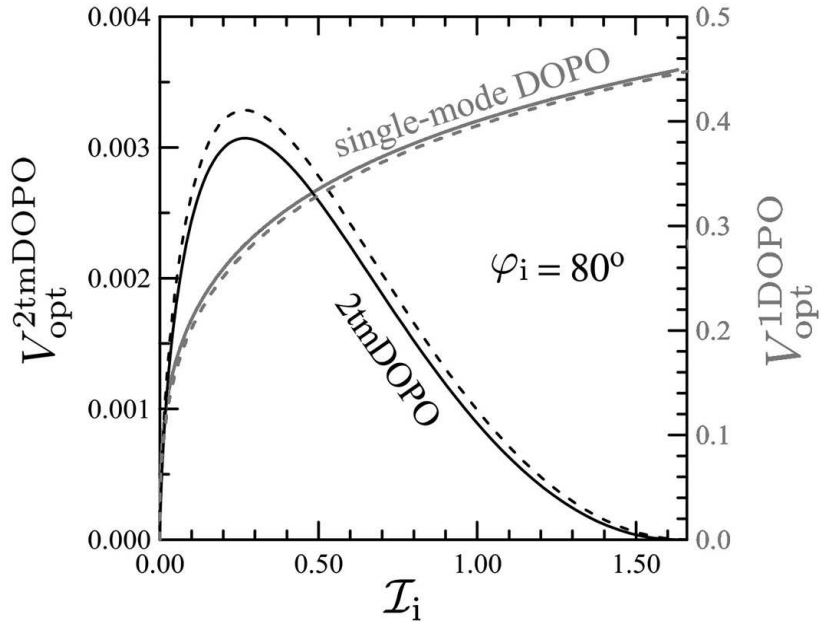}%
\caption{Comparison between the optimum squeezing levels of the 2tmDOPO with
injected signal, $V_{\mathrm{opt}}^{\mathrm{2tmDOPO}}$ (black lines, left
vertical axis), and that of the single--mode DOPO, $V_{\mathrm{opt}%
}^{\mathrm{1DOPO}}$ (grey lines, right vertical axis), as a function of the
injected intensity $\mathcal{I}_{\mathrm{i}}$. We have taken $\varphi
_{\mathrm{i}}=80^{\mathrm{o}}$ and two values of $\sigma$: $\sigma=1$ (full
lines) and $\sigma=0.99$ (dashed lines).}%
\label{f2tmDOPOwithIS5}%
\end{center}
\end{figure}

In terms of the injection's external power, see (\ref{mod-phase}),
(\ref{param}) and (\ref{PtoE}), the injection intensity parameter
$\mathcal{I}_{\mathrm{i}}$ reads%
\end{subequations}
\begin{equation}
\mathcal{I}_{\mathrm{i}}=\frac{\chi^{2}\left\vert \mathcal{E}_{\mathrm{s}%
}\right\vert ^{2}}{\gamma_{\mathrm{s}}^{3}\gamma_{\mathrm{p}}}=2\frac
{P_{\mathrm{s}}}{P_{\mathrm{p,thr}}},
\end{equation}
where $P_{\mathrm{p,thr}}$ is the usual DOPO\ threshold pump power, that is,
the pump power needed for making the signal field oscillate when no injection
is used ($\mathcal{I}_{\mathrm{i}}=0$), which can be found from (\ref{PtoE})
with $\left\vert \mathcal{E}_{\mathrm{p,thr}}\right\vert ^{2}=\left(
\gamma_{\mathrm{s}}\gamma_{\mathrm{p}}/\chi\right)  ^{2}$ as follows from the
threshold condition $\sigma=1$; $P_{\mathrm{s}}$ is the actual injection
power. In usual experiments, when the injected seed is used for active locking
purposes, $P_{\mathrm{s}}$ is a small fraction of $P_{\mathrm{p,thr}}$ (say
$0.001-0.1$); this is the reason why we show a zoom of the region
$\mathcal{I}_{\mathrm{i}}\in\left[  0,0.2\right]  $ in Figure
\ref{f2tmDOPOwithIS4}a (the enlargement is representative of what happens in
the rest of cases represented in Figures \ref{f2tmDOPOwithIS4}). It is
remarkable the very small squeezing degradation that the injected field
induces in the 2tmDOPO model, specially close to attenuation ($\varphi
_{\mathrm{i}}=\pi/2$).

A relevant issue is how this squeezing level compares with the usual
single--mode DOPO with injected signal, whose optimal squeezing level is
calculated in \cite{Protsenko94}, and is in particular given by Eq. (187) in
that paper under the following substitutions, in order to adapt the expression
to our notation: $\left\{  A_{0},A_{1},\Delta_{0},\Delta_{1},E,e,\gamma
,\right\}  \rightarrow\left\{  \beta_{0},\beta_{x}/\sqrt{2},0,0,\sigma
,\varepsilon_{\mathrm{i}}/\sqrt{2},\kappa^{-1}\right\}  $. This comparison is
made in Figure \ref{f2tmDOPOwithIS5}, where we represent in the same plot the
optimal squeezing level both for the 2tmDOPO and the single--mode DOPO with
injected signal. That of the single--mode DOPO, $V_{\mathrm{opt}%
}^{\mathrm{1DOPO}}$, is shown in the limit $\kappa\rightarrow\infty$, which is
a common limit from the experimental viewpoint and yields its best noise
reduction (the 2tmDOPO squeezing level, on the other hand, does not depend on
$\kappa$). As compared with the usual single--mode DOPO we observe that the
squeezing level of the 2tmDOPO is quite insensitive to the injection (note the
different scales for both quantities). As well as that, by increasing
$\mathcal{I}_{\mathrm{i}}$ from zero, the squeezing first degrades but, in the
2tmDOPO, it improves again unlike the single--mode DOPO case. We want to
stress that the squeezing level in a single--mode DOPO is highly sensitive to
the pumping level $\sigma$ (only very close to $\sigma=1$ the squeezing is
high), while in the case of the 2tmDOPO large squeezing levels can be found
for any $\sigma$. This phenomenon is directly related to the rotational
symmetry breaking which is the origin of the perfect, noncritical squeezing
occurring for any $\sigma>1$ for null injection. We can conclude that adding a
small injection to a 2tmDOPO could be useful to improve the levels of
squeezing available with single--mode DOPOs. 

%% file: TypeIIOPOsFO.tex
In the previous chapters we have studied in depth the phenomenon of
spontaneous symmetry breaking via a particular example: The rotational
symmetry breaking which happens in the transverse plane of a 2tmDOPO (as well
as any DOPO tuned to an odd family at the signal frequency, as proved in the
next chapter). We have analyzed many features which are important both from
the theoretical and experimental points of view, showing the phenomenon to be
quite robust. It is then natural to try to generalize the phenomenon to other
types of symmetries, in particular to symmetries whose associated free
parameters are in the polarization or temporal degrees of freedom of the light
coming out of the OPO.

In this chapter we introduce the phenomenon of spontaneous polarization
symmetry breaking (the temporal one will be outlined in the next and final
chapter, where we will discuss the outlook of the research developed in this
thesis). The phenomenon will be analyzed in a system that we already dealt
with: The frequency degenerate type II OPO introduced in Section \ref{OPO}. We
will show it to be completely analogous to the 2tmDOPO, just changing the
$\pm1$ orbital angular momentum modes by the two linearly polarized modes
within the ordinary and extraordinary axes; the free parameter in this case
will correspond to a continuous parameter in the polarization ellipse of the
outgoing signal field. So much as in the 2tmDOPO, we will then be able to
define bright and dark polarization modes, and rephrase all the properties
that we found for them in the 2tmDOPO to the case of the frequency degenerate
type II OPO.

The study of this system will also allow us to tackle a major problem in OPOs:
It is quite difficult to ensure working at exact frequency degeneracy above
threshold (both in type I and II), as we already discussed briefly in Section
\ref{SecOrderNonLinearity}. In \cite{Longchambon04a} it was proposed a way of
locking the frequencies of the signal and idler modes in type II OPOs by
introducing a wave--plate with its fast axis rotated respect to the
ordinary--extraordinary axes (see also \cite{Longchambon04b} for the quantum
analysis of the system and \cite{Laurat05} for the experiments); we will show
that the same locking can be accomplished by injecting an external laser field
with the right polarization.

\section{Spontaneous polarization symmetry breaking\label{SPSB}}

In Section \ref{OPO} we already studied the properties of OPOs in which signal
and idler are distinguishable, as is the case of frequency degenerate type II
OPOs, where signal and idler have orthogonal linear polarizations. In that
section we showed that signal and idler share EPR-like correlations below
threshold, which become perfect exactly at threshold. Above threshold, we only
studied the signal--idler intensity correlations, what we did by neglecting
the quantum diffusion of their phase difference, arguing that it does not
affect the results on intensity correlations at all. On the other hand, in
Chapter \ref{MultiOPOs}, and Section \ref{SRSB} in particular, we developed a
method for studying this phase diffusion when analyzing the 2tmDOPO, which
actually has the same evolution equations as type II OPOs ---just compare
(\ref{ScaledLangevin2tmDOPO}) with (\ref{scaledOPOlangevin})---. In this
section we are going to translate all the results found in the 2tmDOPO to the
frequency degenerate type II OPO, showing how their properties can be
interpreted in terms of spontaneous polarization symmetry breaking.

Let us first write the part of the down--converted (signal+idler) field
propagating to the right as (Schr\"{o}dinger picture is understood)%
\begin{equation}
\mathbf{\hat{E}}_{\mathrm{DC},\rightarrow}^{(+)}\left(  \mathbf{r}\right)
=\mathrm{i}\sqrt{\frac{\hbar\omega_{\mathrm{s}}}{4\varepsilon_{0}%
n_{\mathrm{s}}L_{\mathrm{s}}}}(\mathbf{e}_{\mathrm{e}}\hat{a}_{\mathrm{s}%
}+\mathbf{e}_{\mathrm{o}}\hat{a}_{\mathrm{i}})G(k_{\mathrm{s}};\mathbf{r}%
_{\perp},z)e^{\mathrm{i}n_{\mathrm{s}}k_{\mathrm{s}}z}, \label{EDC}%
\end{equation}
where we have made the (unrealistic) approximation $n_{\mathrm{s}%
}=n_{\mathrm{i}}$ for simplicity, see (\ref{psiFields}); in a moment it will
be clear that this approximation is not relevant at all, except for some
quantitative considerations that we will explain later on.

Taking into account that the classical solution of the system above threshold
is given by (\ref{AboveThresholdSolOPO})%
\begin{equation}
\bar{\beta}_{\mathrm{p}}=1,\text{ \ \ \ \ }\bar{\beta}_{\mathrm{s,i}}%
=\sqrt{\sigma-1}\exp(\mp\mathrm{i}\theta),
\end{equation}
where $\theta$ is arbitrary, that is, it is not fixed by the classical
equations of motion, the form of the field in the classical limit can be
obtained by taking the expectation value of (\ref{EDC}) and making the
correspondence $\langle\hat{a}_{\mathrm{s},\mathrm{i}}(t)\rangle
=\exp(-\mathrm{i}\omega_{0}t)\bar{\beta}_{\mathrm{s},\mathrm{i}}/g$, arriving
to%
\begin{align}
\mathbf{\bar{E}}_{\mathrm{DC},\rightarrow}^{(+)}\left(  \mathbf{r},t\right)
& =\mathrm{i}\sqrt{\frac{\hbar\omega_{\mathrm{s}}(\sigma-1)}{4\varepsilon
_{0}n_{\mathrm{s}}L_{\mathrm{s}}g}}(e^{-\mathrm{i}\theta}\mathbf{e}%
_{\mathrm{e}}+e^{+\mathrm{i}\theta}\mathbf{e}_{\mathrm{o}})G(k_{\mathrm{s}%
};\mathbf{r}_{\perp},z)e^{-\mathrm{i}\omega_{0}t+\mathrm{i}n_{\mathrm{s}%
}k_{\mathrm{s}}z}\propto\boldsymbol{\varepsilon}(\theta,\varphi=\pi/4),
\end{align}
where the general parametrization of the polarization vector
$\boldsymbol{\varepsilon}(\theta,\varphi)$ is given in (\ref{PolPar}). This
shows that the down--converted field can arise with any of the polarizations
depicted in Figure \ref{fQuanti2}, that is, polarized along the $+45%
{{}^o}%
$ or $-45%
{{}^o}%
$ axis, and with arbitrary eccentricity and elicity. Hence, the system posses
an invariance related to a polarization degree of freedom\footnote{When
$n_{\mathrm{s}}>n_{\mathrm{i}}$ is taken into account, the situation is
similar qualitatively, but not quantitatively. In particular, the polarization
of the mean field is in this case%
\begin{equation}
\boldsymbol{\varepsilon}(\theta_{\mathrm{si}},\varphi_{\mathrm{si}%
})=\mathbf{e}_{\mathrm{e}}\exp[-\mathrm{i}\theta_{\mathrm{si}}(z)]\cos
\varphi_{\mathrm{si}}+\mathbf{e}_{\mathrm{o}}\exp[\mathrm{i}\theta
_{\mathrm{si}}(z)]\sin\varphi_{\mathrm{si}},
\end{equation}
with $\theta_{\mathrm{si}}(z)=\theta-(n_{\mathrm{s}}-n_{\mathrm{i}})k_{0}z/2$
and $\tan\varphi_{\mathrm{si}}=\sqrt{n_{\mathrm{s}}L_{\mathrm{s}%
}/n_{\mathrm{i}}L_{\mathrm{i}}}$ ($\varphi_{\mathrm{si}}\in\lbrack0,\pi/2]$).
This means that the polarization has still a free parameter, $\theta$, but is
more complicated than the one explained in the text. Indeed, owed to
$z-$dependence of $\theta_{\mathrm{si}}$, the polarization changes depending
on the exact longitudinal point that we observe inside the crystal (this is
actually a quite expected result, as the crystal is birefringent and hence the
ordinary and extraordinary components of the field acquire different phases
upon propagation through the crystal).}, which is spontaneously broken once
the OPO starts emitting the down--converted field, just as commented above.

Just as in the 2tmDOPO, we can thus define a bright polarization mode
$\boldsymbol{\varepsilon}(\theta,\varphi=\pi/4)=(e^{-\mathrm{i}\theta
}\mathbf{e}_{\mathrm{e}}+e^{+\mathrm{i}\theta}\mathbf{e}_{\mathrm{o}}%
)/\sqrt{2}$ in which mean field emission takes place, and a dark polarization
mode $\boldsymbol{\varepsilon}(\theta+\pi/2,\varphi=\pi/4)=-\mathrm{i}%
(e^{-\mathrm{i}\theta}\mathbf{e}_{\mathrm{e}}-e^{+\mathrm{i}\theta}%
\mathbf{e}_{\mathrm{o}})/\sqrt{2}$ which is empty of photons at the classical level.

As for the quantum properties of the down--converted field, they can also be
directly enunciated without the need of further calculations by exploiting the
analogy with the 2tmDOPO: (\textit{i}) Quantum noise makes $\theta$, and hence
the polarization of the bright mode, diffuse (see Section \ref{Quantum2tmDOPO}%
); (\textit{ii}) the dark mode has all the properties described in Section
\ref{Quantum2tmDOPO} and Chapter \ref{2tmDOPO}, for example, its \textsf{Y}
quadrature ---which is the one detected by a local oscillator with
polarization \textrm{i}$\partial_{\theta}\boldsymbol{\varepsilon}%
(\theta,\varphi=\pi/4)$--- is perfectly squeezed irrespective of the system
parameters. All these results are then in concordance with the general picture
of spontaneous symmetry breaking introduced in Section \ref{BasicSSB}.

This spontaneous polarization symmetry breaking phenomenon was introduced in
\cite{GarciaFerrer10}, where we proposed its observation also in a Kerr
resonator, where the generation of frequency degenerate cross--polarized twin
beams has been accomplished without the need of breaking any invariance of the
system \cite{Vallet90}, as opposed to the state of the art in OPOs, see below.

\section{From nondegenerate to degenerate operation\label{degNONdeg}}

In the previous section we have assumed that the type II OPO works at exact
frequency degeneracy. As explained in \ref{SecOrderNonLinearity} the
phase--matching conditions ensuring that it is the degenerate process the one
with larger gain (lowest threshold) are quite critical; for example, in the
case of \cite{Feng03}, where the authors are able to make the frequency
difference between signal and idler as small a 150 \textrm{kHz} for a cavity
with an 8 \textrm{GHz} free spectral range and a 6 \textrm{MHz} linewidth
(loss rate), variations of the cavity length on the order of the nanometer can
change the oscillation frequencies of signal and idler (mode--hopping).

In practice, this means that it is not possible to work at exact frequency
degeneracy without additional locking techniques that break the phase
invariance of the OPO. The pioneering example of such locking techniques was
performed by Fabre and collaborators
\cite{Longchambon04a,Longchambon04b,Laurat05}. Their idea was to introduce in
the resonator a $\lambda/4$ plate with its fast axis misaligned respect to the
extraordinary axis of the nonlinear crystal. For small misalignments, the
effect of this plate is to introduce a coupling between the signal and idler
modes described by the Hamiltonian%
\begin{equation}
\hat{H}_{\mathrm{P}}=i\hbar(\mu_{\mathrm{P}}\hat{a}_{\mathrm{s}}\hat
{a}_{\mathrm{i}}^{\dagger}-\mu_{\mathrm{P}}^{\ast}\hat{a}_{\mathrm{s}%
}^{\dagger}\hat{a}_{\mathrm{i}}), \label{HP}%
\end{equation}
where $\mu_{\mathrm{P}}$ is some complex parameter. It was then shown in
\cite{Longchambon04a} that in a given region of the parameter space (in
particular of the detunings of signal and idler) the frequencies of signal and
idler get locked to half the pump frequency; this OPO is known as the
\textit{self--phase--locked OPO}, which was already tested experimentally in
\cite{Laurat05}.

Note that, as mentioned above, this self--locking effect is accomplished by
breaking the symmetry $\{\hat{a}_{\mathrm{s}},\hat{a}_{\mathrm{i}%
}\}\longrightarrow\{\exp(-\mathrm{i}\theta)\hat{a}_{\mathrm{s}},\exp
(\mathrm{i}\theta)\hat{a}_{\mathrm{i}}\}$ of the OPO ---check that the
Hamiltonian (\ref{HP}) does not posses this symmetry---, and hence a
degradation of the signal--idler intensity correlations, as well as of the
noncritical squeezing induced by spontaneous polarization symmetry breaking
described above are to be expected. For example, in \cite{Laurat05} the
intensity--difference fluctuations were reduced by an 89\% respect to the
vacuum level prior to the introduction of the plate; then, after obtaining
frequency degeneracy through the self--phase--locking mechanism this noise
reduction fell down to a more humble 65\%.%

\begin{figure}
[t]
\begin{center}
\includegraphics[
width=\textwidth
]%
{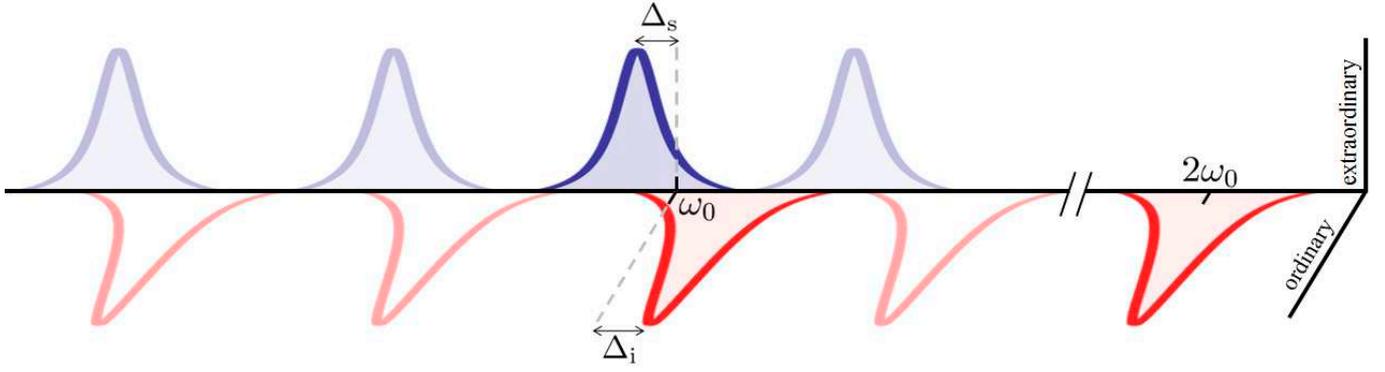}%
\caption{Resonance scheme of the type II OPO in which we propose to obtain
frequency degeneracy via an external injection at the degenerate frequency
$\omega_{0}$.}%
\label{fTypeII1}%
\end{center}
\end{figure}

In this section we introduce an alternative frequency--locking mechanism,
based on the injection of an external beam at the degenerate frequency
\cite{NavarreteUNa}. We will call \textit{actively--phase--locked OPO} to such
OPO configuration. The model for this system follows from the resonance scheme
sketched in Figure \ref{fTypeII1}. The signal and idler modes are already
close to frequency degeneracy \cite{Feng03} and we inject an external laser
field at the degenerate frequency $\omega_{0}$ with arbitrary polarization
$\boldsymbol{\varepsilon}_{\mathrm{L}}=e^{-\mathrm{i}\theta_{\mathrm{L}}}%
\cos\varphi_{\mathrm{L}}\mathbf{e}_{\mathrm{e}}+e^{+\mathrm{i}\theta
_{\mathrm{L}}}\sin\varphi_{\mathrm{L}}\mathbf{e}_{\mathrm{o}}$, see
(\ref{PolPar}). In terms of the signal and idler boson operators, the
corresponding injection Hamiltonian is given by%
\begin{equation}
\hat{H}_{\mathrm{inj-si}}=\mathrm{i}\hbar(\mathcal{E}_{\mathrm{s}%
}a_{\mathrm{s}}^{\dagger}+\mathcal{E}_{\mathrm{i}}a_{\mathrm{i}}^{\dagger
})\exp(-\mathrm{i}\omega_{0}t)+\mathrm{H.c.,}%
\end{equation}
with $\mathcal{E}_{\mathrm{s}}=\mathcal{E}_{\mathrm{L}}e^{-\mathrm{i}%
\theta_{\mathrm{L}}}\cos\varphi_{\mathrm{L}}$ and $\mathcal{E}_{\mathrm{i}%
}=\mathcal{E}_{\mathrm{L}}e^{+\mathrm{i}\theta_{\mathrm{L}}}\sin
\varphi_{\mathrm{L}}$, where the injection parameter for the mode with polarization
$\boldsymbol{\varepsilon}_{\mathrm{L}}$ (which is fed by the whole power
$P_{\mathrm{inj}}$ of the injected field) is denoted by $\mathcal{E}_{\mathrm{L}}=\sqrt{2\gamma
_{\mathrm{s}}P_{\mathrm{inj}}/\hbar\omega_{0}}\exp(\mathrm{i}\phi_{\mathrm{L}%
})$. $\phi_{\mathrm{L}}$ is the phase of
this signal injection relative to the pump injection.%

\begin{figure}
[h]
\begin{center}
\includegraphics[
width=0.8\textwidth
]%
{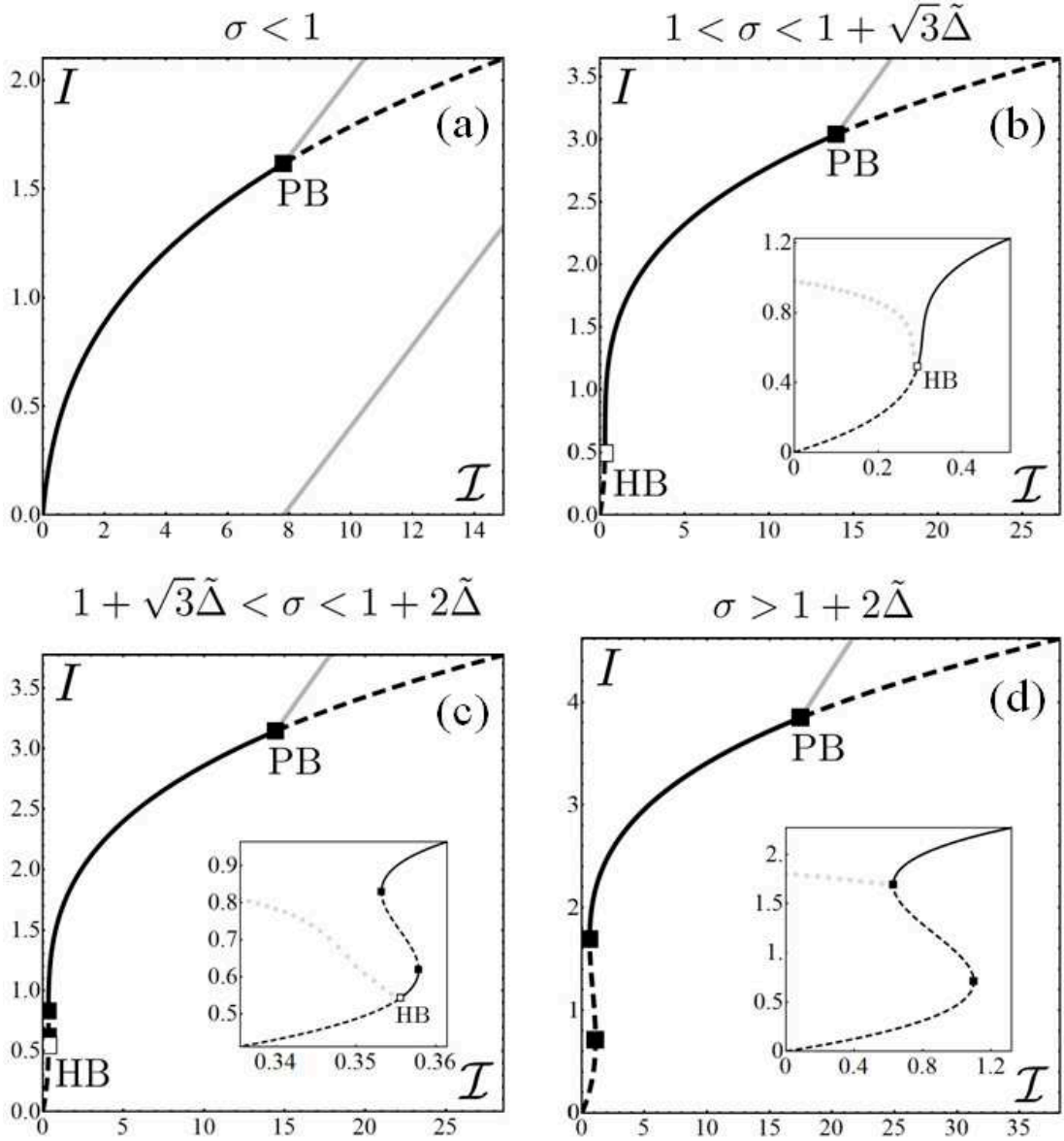}%
\caption{Bifurcation diagrams
of the type II OPO with an injected signal at the degenerate frequency, for
the symmetric configuration $\tilde{\Delta}_{\mathrm{s}}=-\tilde{\Delta
}_{\mathrm{i}}\equiv\tilde{\Delta}$ and $\varepsilon_{\mathrm{s}}%
=\varepsilon_{\mathrm{i}}\equiv\sqrt{\mathcal{I}}$; the value $\tilde{\Delta
}=0.6$ is chosen for all the figures (the same behavior is found for any other
choice), while we set $\sigma$ to 0.5 in (a), 1.98 in (b), 2.09 in (c), and 2.8 in (d). The black lines
correspond to the intensity $I$ of the stationary symmetric solution
(\ref{SymmSol}), the solid or dashed character of the lines meaning that this
solution is stable or unstable, respectively. The upper and lower grey solid
lines correspond to the values of $|\bar{\beta}_{\mathrm{b}}|^{2}/2$ and
$|\bar{\beta}_{\mathrm{d}}|^{2}/2$, respectively, that is to half the
intensity of the bright and dark (only showed in the $\sigma<1$ case) modes;
these lines have been found numerically, and show how above the pitchfork
bifurcation (marked as PB in the figures) the symmetric solution
(\ref{SymmSol}) becomes unstable, and a new asymmetric solution is born. As
explained in the text, for $\sigma>1$ it is possible to find periodic
solutions connecting the $\mathcal{I}=0$ axis with the Hopf bifurcation
(marked as HB in the figures); we have checked numerically that this periodic
orbits exist, and moreover they are \textquotedblleft
symmetric\textquotedblright, that is, $\beta_{\mathrm{s}}(t)=\beta
_{\mathrm{i}}^{\ast}(t)$. The grey circles correspond to the mean value of
$|\beta_{\mathrm{s}}|^{2}$ (half the sum between its maximum and its minimum
of oscillation). Note that stable stationary solutions and periodic orbits
coexist for $1+\sqrt{3}\tilde{\Delta}<\sigma<1+2\tilde{\Delta}$. Note also
that after the Hopf bifurcation is extinguished ($\sigma>1+2\tilde{\Delta}$)
the periodic orbits are connected directly to the upper turning point of the
S--shaped curve (it is to be expected that periodic orbits connecting the
upper and lower turning points exist but become unstable when $\sigma
>1+2\tilde{\Delta}$, although we haven't checked it numerically yet).}%
\label{fTypeII2}%
\end{center}
\end{figure}

Moving to the interaction picture defined by the transformation operator%
\begin{equation}
\hat{U}_{\mathrm{0}}=\exp[\hat{H}_{\mathrm{0}}t/\mathrm{i}\hbar]\text{
\ \ \ \ with \ \ \ \ }\hat{H}_{\mathrm{0}}=\hbar\omega_{\mathrm{0}}(2\hat
{a}_{\mathrm{p}}^{\dagger}\hat{a}_{\mathrm{p}}+\hat{a}_{\mathrm{s}}^{\dagger
}\hat{a}_{\mathrm{s}}+\hat{a}_{\mathrm{i}}^{\dagger}\hat{a}_{\mathrm{i}}),
\end{equation}
the master equation of the system can be written as%
\begin{align}
\frac{d\hat{\rho}_{\mathrm{I}}}{dt}  &  =\left[  -\mathrm{i}\Delta
_{\mathrm{s}}\hat{a}_{\mathrm{s}}^{\dagger}\hat{a}_{\mathrm{s}}+\mathrm{i}%
\Delta_{\mathrm{i}}\hat{a}_{\mathrm{i}}^{\dagger}\hat{a}_{\mathrm{i}}+\chi
\hat{a}_{\mathrm{p}}\hat{a}_{\mathrm{s}}^{\dagger}\hat{a}_{\mathrm{i}%
}^{\dagger}+\mathcal{E}_{\mathrm{p}}a_{\mathrm{p}}^{\dagger}+\mathcal{E}%
_{\mathrm{s}}a_{\mathrm{s}}^{\dagger}+\mathcal{E}_{\mathrm{i}}a_{\mathrm{i}%
}^{\dagger}+\mathrm{H.c.},\hat{\rho}_{\mathrm{I}}\right]+\sum_{j=\mathrm{p,s,i}}\gamma_{j}(2\hat{a}_{j}\hat{\rho}_{\mathrm{I}}%
\hat{a}_{j}^{\dagger}-\hat{a}_{j}^{\dagger}\hat{a}_{j}\hat{\rho}_{\mathrm{I}%
}-\hat{\rho}_{\mathrm{I}}\hat{a}_{j}^{\dagger}\hat{a}_{j}),
\end{align}
where we take $\gamma_{\mathrm{s}}=\gamma_{\mathrm{i}}$ as in (\ref{OPOmaster}%
), and the detunings $\Delta_{\mathrm{s}}=\omega_{0}-\omega_{\mathrm{s}}$ and
$\Delta_{\mathrm{i}}=\omega_{\mathrm{i}}-\omega_{0}$ are taken as positive by
convention. As usual, we take $\mathcal{E}_{\mathrm{p}}$ as a positive real.
The stochastic Langevin equations associated to this master equation are
\begin{subequations}
\begin{align}
\dot{\alpha}_{\mathrm{p}}  &  =\mathcal{E}_{\mathrm{p}}-\gamma_{\mathrm{p}%
}\alpha_{\mathrm{p}}-\chi\alpha_{\mathrm{s}}\alpha_{\mathrm{i}},\\
\dot{\alpha}_{\mathrm{p}}^{+}  &  =\mathcal{E}_{\mathrm{p}}-\gamma
_{\mathrm{p}}\alpha_{\mathrm{p}}^{+}-\chi\alpha_{\mathrm{s}}^{+}%
\alpha_{\mathrm{i}}^{+},\\
\dot{\alpha}_{\mathrm{s}}  &  =\mathcal{E}_{\mathrm{s}}-\left(  \gamma
_{\mathrm{s}}+\mathrm{i}\Delta_{\mathrm{s}}\right)  \alpha_{\mathrm{s}}%
+\chi\alpha_{\mathrm{p}}\alpha_{\mathrm{i}}^{+}+\sqrt{\chi\alpha_{\mathrm{p}}%
}\xi(t),\\
\dot{\alpha}_{\mathrm{s}}^{+}  &  =\mathcal{E}_{\mathrm{s}}-\left(
\gamma_{\mathrm{s}}-\mathrm{i}\Delta_{\mathrm{s}}\right)  \alpha_{\mathrm{s}%
}^{+}+\chi\alpha_{\mathrm{p}}^{+}\alpha_{\mathrm{i}}+\sqrt{\chi\alpha
_{\mathrm{p}}^{+}}\xi^{+}(t),\\
\dot{\alpha}_{\mathrm{i}}  &  =\mathcal{E}_{\mathrm{i}}-\left(  \gamma
_{\mathrm{s}}-\mathrm{i}\Delta_{\mathrm{i}}\right)  \alpha_{\mathrm{i}}%
+\chi\alpha_{\mathrm{p}}\alpha_{\mathrm{s}}^{+}+\sqrt{\chi\alpha_{\mathrm{p}}%
}\xi^{\ast}(t),\\
\dot{\alpha}_{\mathrm{i}}^{+}  &  =\mathcal{E}_{\mathrm{i}}-\left(
\gamma_{\mathrm{s}}+\mathrm{i}\Delta_{\mathrm{i}}\right)  \alpha_{\mathrm{i}%
}^{+}+\chi\alpha_{\mathrm{p}}^{+}\alpha_{\mathrm{s}}+\sqrt{\chi\alpha
_{\mathrm{p}}^{+}}[\xi^{+}(t)]^{\ast},
\end{align}
which are just as (\ref{OPOlangevin}), but including the parameters associated
to the external injection.

We will be interested only in the classical behavior of the system, as our
main intention is to show that signal--idler frequency--locking can be
accomplished with this scheme. Moreover, we will be working in the
$\gamma_{\mathrm{p}}\gg\gamma_{\mathrm{s}}$ limit, where the pump variables
can be adiabatically eliminated. Let us then directly write the classical
equations governing the evolution of the system in this limit, which are
obtained from the Langevin equations by setting the noises and the pump time
derivatives to zero, and making the identifications $\alpha_{j}^{+}%
\rightarrow\alpha_{j}^{\ast}$. As usual, we also redefine time as $\tau
=\gamma_{\mathrm{s}}t$ and the coherent amplitudes as%
\end{subequations}
\begin{equation}
\beta_{\mathrm{s}}=g\alpha_{\mathrm{s}}\exp(\mathrm{i}\theta_{\mathrm{L}%
})\text{ \ \ \ \ and \ \ \ \ }\beta_{\mathrm{i}}=g\alpha_{\mathrm{i}}%
\exp(-\mathrm{i}\theta_{\mathrm{L}}),
\end{equation}
so that the final equations read
\begin{subequations}
\label{InjOPOeqs}%
\begin{align}
\dot{\beta}_{\mathrm{s}}  &  =\varepsilon_{\mathrm{s}}e^{\mathrm{i}%
\phi_{\mathrm{L}}}-\left(  1+\mathrm{i}\tilde{\Delta}_{\mathrm{s}}\right)
\beta_{\mathrm{s}}+(\sigma-\beta_{\mathrm{s}}\beta_{\mathrm{i}})\beta
_{\mathrm{i}}^{\ast},\\
\dot{\beta}_{\mathrm{i}}  &  =\varepsilon_{\mathrm{i}}e^{\mathrm{i}%
\phi_{\mathrm{L}}}-\left(  1-\mathrm{i}\tilde{\Delta}_{\mathrm{i}}\right)
\beta_{\mathrm{i}}+(\sigma-\beta_{\mathrm{s}}\beta_{\mathrm{i}})\beta
_{\mathrm{s}}^{\ast},
\end{align}
where we have defined the parameters%
\end{subequations}
\begin{equation}
\varepsilon_{\mathrm{s}}=\frac{g}{\gamma_{\mathrm{s}}}|\mathcal{E}%
_{\mathrm{L}}|\cos\varphi_{\mathrm{L}}\text{,\ \ \ \ \ }\varepsilon
_{\mathrm{i}}=\frac{g}{\gamma_{\mathrm{s}}}|\mathcal{E}_{\mathrm{L}}%
|\sin\varphi_{\mathrm{L}}\text{, \ \ \ \ and \ \ \ }\tilde{\Delta}_{j}%
=\Delta_{j}/\gamma_{\mathrm{s}}\text{,}%
\end{equation}
while the parameters $g$ and $\sigma$ are defined as usual, see
(\ref{ScaledParameters}). Note that this equations are invariant under changes
of $\theta_{\mathrm{L}}$, and hence, the dynamics of the system are only
sensitive to the parameter $\varphi_{\mathrm{L}}$ of the injection's polarization.

In order to get some analytic insight, we are going to simplify the problem to
what we will call \textit{symmetric configuration} of the
actively--phase--locked OPO: We assume the detunings to be equal,
$\tilde{\Delta}_{\mathrm{s}}=\tilde{\Delta}_{\mathrm{i}}=\tilde{\Delta}$, and
inject with $\varphi_{\mathrm{L}}=\pi/4$ (arbitrary polarization ellipse along
the $\pm45%
{{}^o}%
$ axis), so that signal and idler get equally pumped, $\varepsilon
_{\mathrm{s}}=\varepsilon_{\mathrm{i}}\equiv\sqrt{\mathcal{I}}$. We also
choose to inject in phase with the pump injection, that is, $\phi_{\mathrm{L}%
}=0$. With these simplifications equations (\ref{InjOPOeqs}) are reduced to
\begin{subequations}
\label{InjOPOeqsNorm}%
\begin{align}
\dot{\beta}_{\mathrm{s}}  &  =\sqrt{\mathcal{I}}-\left(  1+\mathrm{i}%
\tilde{\Delta}\right)  \beta_{\mathrm{s}}+(\sigma-\beta_{\mathrm{s}}%
\beta_{\mathrm{i}})\beta_{\mathrm{i}}^{\ast},\\
\dot{\beta}_{\mathrm{i}}  &  =\sqrt{\mathcal{I}}-\left(  1-\mathrm{i}%
\tilde{\Delta}\right)  \beta_{\mathrm{i}}+(\sigma-\beta_{\mathrm{s}}%
\beta_{\mathrm{i}})\beta_{\mathrm{s}}^{\ast}.
\end{align}
These equations have the symmetry $\{\beta_{\mathrm{s}}\rightarrow
\beta_{\mathrm{i}}^{\ast},\beta_{\mathrm{i}}\rightarrow\beta_{\mathrm{s}%
}^{\ast}\}$, what allows us look for symmetric stationary solutions of the
type%
\end{subequations}
\begin{equation}
\bar{\beta}_{\mathrm{s}}=\bar{\beta}_{\mathrm{i}}^{\ast}=\sqrt{I}%
\exp(\mathrm{i}\varphi). \label{SymmSol}%
\end{equation}
Note that whenever this solution exists (and is stable), the classical
down--converted field emitted by the OPO will be%
\begin{equation}
\mathbf{\bar{E}}_{\mathrm{DC},\rightarrow}^{(+)}\left(  \mathbf{r}\right)
=\mathrm{i}\sqrt{\frac{\hbar\omega_{\mathrm{s}}I}{4\varepsilon_{0}%
n_{\mathrm{s}}L_{\mathrm{s}}g^{2}}}(e^{-\mathrm{i}\left(  \theta_{\mathrm{L}%
}-\varphi\right)  }\mathbf{e}_{\mathrm{e}}+e^{\mathrm{i}\left(  \theta
_{\mathrm{L}}-\varphi\right)  }\mathbf{e}_{\mathrm{o}})G(k_{\mathrm{s}%
};\mathbf{r}_{\perp},z)e^{-\mathrm{i}\omega_{0}t+\mathrm{i}n_{\mathrm{s}%
}k_{\mathrm{s}}z},
\end{equation}
where we have used (\ref{EDC}) and made the correspondence $\langle\hat
{a}_{\mathrm{s},\mathrm{i}}(t)\rangle=\exp(-\mathrm{i}\omega_{0}t)\bar{\beta
}_{\mathrm{s},\mathrm{i}}/g$. As expected, this corresponds to a field
oscillating at the degenerate frequency $\omega_{0}$. Moreover, the
polarization of this field (which is always within the $\pm45%
{{}^o}%
$ axis) can always be chosen as linear by selecting a proper $\theta
_{\mathrm{L}}$ tuned to $\varphi$. In the remaining of this section we study
the conditions under which this solution exists and is stable.

It is completely trivial to show from (\ref{InjOPOeqsNorm}) that the intensity
$I$ of the symmetric solution satisfies the third order polynomial%
\begin{equation}
\mathcal{I}=[(I+1-\sigma)^{2}+\tilde{\Delta}^{2}]I, \label{Ipol}%
\end{equation}
while its phase $\varphi$ is uniquely determined from $I$ as%
\begin{equation}
\varphi=\arg\{I+1-\sigma-\mathrm{i}\tilde{\Delta}\}.
\end{equation}
Now, so much as happened in the previous chapter (see Figure
\ref{f2tmDOPOwithIS1}), the polynomial (\ref{Ipol}) sometimes has a single
positive definite solution, while sometimes its three roots are positive
definite. By solving the equation $\partial\mathcal{I}/\partial I=0$, it is
simple to show that the turning points $I_{\pm}$ have the expression%
\begin{equation}
I_{\pm}=\frac{2}{3}(\sigma-1)\pm\frac{1}{3}\sqrt{(\sigma-1)^{2}-3\tilde
{\Delta}^{2}}%
\end{equation}
and hence, they exist only for $\sigma>1+\sqrt{3}\tilde{\Delta}$. For
$\sigma\leq1+\sqrt{3}\tilde{\Delta}$~the solution is therefore single--valued.

In order to analyze the stability of this symmetric solution, we will change
to a new polarization basis%
\begin{equation}
\boldsymbol{\varepsilon}_{\mathrm{b}}=\frac{1}{\sqrt{2}}(e^{-\mathrm{i}\left(
\theta_{\mathrm{L}}-\varphi\right)  }\mathbf{e}_{\mathrm{e}}+e^{\mathrm{i}%
\left(  \theta_{\mathrm{L}}-\varphi\right)  }\mathbf{e}_{\mathrm{o}})\text{
\ \ \ \ and \ \ \ \ }\boldsymbol{\varepsilon}_{\mathrm{d}}=\frac{1}{\sqrt
{2}\mathrm{i}}(e^{-\mathrm{i}\left(  \theta_{\mathrm{L}}-\varphi\right)
}\mathbf{e}_{\mathrm{e}}-e^{\mathrm{i}\left(  \theta_{\mathrm{L}}%
-\varphi\right)  }\mathbf{e}_{\mathrm{o}}),\label{OEtoBD}%
\end{equation}
where $\boldsymbol{\varepsilon}_{\mathrm{b}}$ corresponds to the polarization
mode excited by the symmetric solution (\ref{SymmSol}) and
$\boldsymbol{\varepsilon}_{\mathrm{d}}$ to its orthogonal, that is, to the
bright and dark modes of the system. The corresponding coherent amplitudes are
written as%
\begin{equation}
\beta_{\mathrm{b}}=\frac{1}{\sqrt{2}}(e^{-\mathrm{i}\varphi}\beta_{\mathrm{s}%
}+e^{\mathrm{i}\varphi}\beta_{\mathrm{i}})\text{ \ \ \ \ and \ \ \ \ }%
\beta_{\mathrm{d}}=\frac{\mathrm{i}}{\sqrt{2}}(e^{-\mathrm{i}\varphi}%
\beta_{\mathrm{s}}-e^{\mathrm{i}\varphi}\beta_{\mathrm{i}}),
\end{equation}
and satisfy the evolution equations%
\begin{subequations}
\begin{align}
\dot{\beta}_{\mathrm{b}} &  =\sqrt{2\mathcal{I}}\cos\varphi-\beta_{\mathrm{b}%
}+\tilde{\Delta}\beta_{\mathrm{d}}+(\sigma-\beta_{\mathrm{b}}^{2}%
/2-\beta_{\mathrm{d}}^{2}/2)\beta_{\mathrm{b}}^{\ast},\\
\dot{\beta}_{\mathrm{d}} &  =-\sqrt{2\mathcal{I}}\sin\varphi-\beta
_{\mathrm{d}}-\tilde{\Delta}\beta_{\mathrm{b}}+(\sigma-\beta_{\mathrm{b}}%
^{2}/2-\beta_{\mathrm{d}}^{2}/2)\beta_{\mathrm{d}}^{\ast}.
\end{align}
In this new basis the symmetric solution reads%
\end{subequations}
\begin{equation}
\bar{\beta}_{\mathrm{b}}=\sqrt{2I},\text{ \ \ \ }\bar{\beta}_{\mathrm{d}}=0,
\end{equation}
and its associated stability matrix is%
\begin{equation}
\mathcal{L}=%
\begin{bmatrix}
-1-2I & \sigma-I & \tilde{\Delta} & 0\\
\sigma-I & -1-2I & 0 & \tilde{\Delta}\\
-\tilde{\Delta} & 0 & -1 & \sigma-I\\
0 & -\tilde{\Delta} & \sigma-I & -1
\end{bmatrix}
.
\end{equation}
The characteristic polynomial of this stability matrix can be factorized into
two second order polynomials, namely $P_{\mathrm{I}}(\lambda)=(\lambda
+1+\sigma)^{2}+\tilde{\Delta}^{2}-I^{2}$ and $P_{\mathrm{II}}(\lambda
)=(\lambda+1-\sigma+2I)^{2}+\tilde{\Delta}^{2}-I^{2}$. The bifurcation
diagrams for the different parameter regions are shown in Figure
\ref{fTypeII2}; now we discuss them in depth.

Let us start by studying the instabilities predicted by the first polynomial,
whose roots are given by%
\begin{equation}
\lambda_{\pm}^{\mathrm{I}}=-(1+\sigma)\pm\sqrt{I^{2}-\tilde{\Delta}^{2}}.
\end{equation}

Therefore, the condition $\operatorname{Re}\{\lambda_{\pm}^{\mathrm{I}}\}=0$
can only be satisfied for%
\begin{equation}
I=\sqrt{(1+\sigma)^{2}+\tilde{\Delta}^{2}}\equiv I_{\mathrm{PB}}\text{.}%
\end{equation}
The fact that the instability appears without imaginary part in the
$\lambda_{\pm}^{\mathrm{I}}$, and it is located in the upper branch of the
S--shaped curve ($I_{\mathrm{PB}}>I_{+}$ for any value of the parameters),
signals that it corresponds to a Pitchfork bifurcation where a non-symmetric
stationary solution $\{\bar{\beta}_{\mathrm{s}}=\sqrt{I_{\mathrm{s}}}%
\exp(\mathrm{i}\varphi_{\mathrm{s}}),\bar{\beta}_{\mathrm{i}}=\sqrt
{I_{\mathrm{i}}}\exp(\mathrm{i}\varphi_{\mathrm{i}})\}$ with $I_{\mathrm{s}%
}\neq I_{\mathrm{i}}$ borns (as we have checked numerically, see the grey
lines in Figure \ref{fTypeII2}. This bifurcation is
equivalent to the one introduced in the previous chapter when studying the
effects of a signal injection in the 2tmDOPO, and can be understood as a
switching on of the dark mode.

As for the second polynomial, its roots are given by%
\begin{equation}
\lambda_{\pm}^{\mathrm{II}}=\sigma-1-2I\pm\sqrt{I^{2}-\tilde{\Delta}^{2}}.
\end{equation}
Note that $\lambda_{\pm}^{\mathrm{II}}=0$ for $I=I_{\pm}$, that is, the
turning points of the S--shaped curve signal an instability. It is then simple
to check (for example numerically) that the whole middle branch connecting
this instability points is unstable, as intuition says (see Figures
\ref{fTypeII2}c,d).

But $\lambda_{\pm}^{\mathrm{II}}$ has yet one more instability when%
\begin{equation}
I=\frac{\sigma-1}{2}\equiv I_{\mathrm{HB}}.
\end{equation}
At this instability the eigenvalues become purely imaginary, in particular,
$\lambda_{\pm}^{\mathrm{II}}=\pm\mathrm{i}\omega_{\mathrm{HB}}$ with
$\omega_{\mathrm{HB}}=\sqrt{\tilde{\Delta}^{2}-(\sigma-1)^{2}/4}$, and hence
it corresponds to a Hopf bifurcation. Note that $I_{\mathrm{HB}}$ is negative
for $\sigma<1$, while $\omega_{\mathrm{HB}}$ becomes imaginary for
$\sigma>1+2\tilde{\Delta}$, and hence the Hopf bifurcation only exists in the
region $1<\sigma<1+2\tilde{\Delta}$. It is simple to check that
$I_{\mathrm{HB}}$ is always below $I_{\mathrm{PB}}$ and $I_{-}$; in
particular, it borns at $I=0$ for $\sigma=1$, and climbs the $\mathcal{I}-I$
curve as $\sigma$ increases until it dies at $I=I_{-}$ for $\sigma
=1+2\tilde{\Delta}$ (see Figures \ref{fTypeII2}b,c,d).
The portion of the curve with $I<I_{\mathrm{HB}}$ is unstable, and no
stationary solutions can be found there, as the stable states correspond in
this case to periodic orbits (as we have checked numerically, see Figures
\ref{fTypeII2}b,c,d). This is also quite intuitive
because when no injection is present, that is, for $\mathcal{I}=0$, we know
that the stable states of the OPO above threshold are the ones with the signal
and idler beams oscillating at the non-degenerate frequencies $\omega
_{\mathrm{s}}=\omega_{0}+\Delta$ and $\omega_{\mathrm{i}}=\omega_{0}-\Delta$,
which in the picture we are working on means $\{\beta_{\mathrm{s}}%
(\tau)\propto\exp(-\mathrm{i}\tilde{\Delta}\tau),\beta_{\mathrm{i}}%
(\tau)\propto\exp(\mathrm{i}\tilde{\Delta}\tau)\}$.

This analysis proves that there exist regions in the parameter space where the
frequencies of the signal and idler beams are locked to the degenerate one,
and hence active--locking can be a good alternative to the self--locking
technique already proposed for type II OPOs
\cite{Longchambon04a,Longchambon04b,Laurat05}. 

%% file: DOPOfamilyFO.tex
So far we have studied DOPOs tuned to a fundamental TEM$_{00}$ mode (Section
\ref{DOPO}) and to the first family of transverse modes at the signal
frequency (Section \ref{SRSB} and Chapters \ref{2tmDOPO} and
\ref{2tmDOPOwithIS}). The former was introduced as an example of a system
showing bifurcation squeezing, while we proposed the later as a playground
where studying the phenomenon of noncritical squeezing induced by spontaneous
symmetry breaking. It is then natural to generalize the study\ to a DOPO tuned
to an arbitrary family of transverse modes at the signal frequency, and this
is what this final chapter is intended to do. We remind that a given family,
say family $f$, posses $f+1$ transverse modes having orbital angular momenta
$\{\pm f,\pm(f-2),...,\pm l_{0}\}$, with $l_{0}$ equal to $0$ for even
families and $1$ for the odd ones, so that the part signal field propagating
along $+\mathbf{e}_{z}$ can be written in the Laguerre--Gauss basis as (from
now on all the sums over $l$ are assumed to run over the set $l\in
\{f,f-2,...,l_{0}\}$)%
\begin{equation}
\mathbf{\hat{E}}_{\mathrm{s},\rightarrow}^{(+)}\left(  \mathbf{r}\right)
=\mathrm{i}\sqrt{\frac{\hbar\omega_{\mathrm{s}}}{4\varepsilon_{0}%
n_{\mathrm{s}}L_{\mathrm{s}}}}\mathbf{e}_{\mathrm{e}}\sum_{\pm l}\hat{a}%
_{l}L_{(f-|l|)/2,l}\left(  k_{\mathrm{s}};\mathbf{r}_{\perp},z\right)
e^{\mathrm{i}n_{\mathrm{s}}k_{\mathrm{s}}z};
\end{equation}
hence, for $f\geq2$ this DOPO has more than one parametric down--conversion
channel available, that is, a pump photon can decay into any pair of opposite
angular momentum signal photons. Armed with the things that we have learned so
far (bifurcation squeezing, pump clamping, and spontaneous symmetry breaking),
the properties of such system will be quite intuitive; in fact, this is the
system in which we originally predicted the phenomenon of OPO clamping
\cite{Navarrete09} that was introduced in a general way in Section
\ref{OPOfreezing}.

What is interesting about this particular implementation of a multi--mode DOPO
that we may call \textit{f--transverse--family DOPO} is that, as we will show
in Section \ref{TuningChi}, by tuning the thickness of the pump mode (what can
be done by using monolithic designs in which the cavities for the pump and
signal modes are independent, or singly--resonant OPOs) one can find many
transverse modes with large levels of squeezing, hence creating a highly
nonclassical and multi--mode beam.

\section{The model}

As in the previous cases, the general model for intracavity down--conversion
that we developed in Section \ref{GeneralOPOModel} covers also the
configuration that we are dealing with in this chapter. In particular, the
down--conversion Hamiltonian can be obtained by particularizing
(\ref{GenTWMHamiltonian}) to the current scenario in which the set of transverse modes resonating at
the signal frequency is the corresponding to the Laguerre--Gauss
modes $\{L_{(f-l)/2,\pm l}\left(  k;\mathbf{r}_{\perp},z\right)
\}_{l\in\{f,f-2,...,l_{0}\}}$. Denoting by $\hat{a}_{\mathrm{p}}$ and $\hat{a}_{l}$ the
annihilation operators for pump photons and signal photons with orbital
angular momentum $l$, respectively, we get
\begin{equation}
\hat{H}_{\mathrm{c}}=\mathrm{i}\hbar\sum_{l}\frac{\chi_{l}}{1+\delta_{0,l}%
}\hat{a}_{\mathrm{p}}\hat{a}_{l}^{\dagger}\hat{a}_{-l}^{\dagger}%
+\mathrm{H.c.}, \label{FamTWMHamiltonian}%
\end{equation}
where%
\begin{equation}
\chi_{l}=3l_{\mathrm{c}}\chi_{\mathrm{oee}}^{(2)}(2\omega_{0};\omega
_{0},\omega_{0})\sqrt{\frac{\hbar\omega_{0}^{3}}{2\varepsilon_{0}%
n_{\mathrm{c}}^{3}L_{\mathrm{opt}}^{3}}}\int_{0}^{+\infty}rdrG(k_{\mathrm{p}%
};r)\left[  \mathcal{R}_{(f-l)/2}^{l}(k_{\mathrm{s}};r)\right]  ^{2}.
\label{Chil}%
\end{equation}

Note that the larger is the OAM of the down--converted pair, the smaller is
the overlapping between the square of $\mathcal{R}_{(f-l)/2}^{l}%
(k_{\mathrm{s}};r)$ and the Gaussian pump profile $G(k_{\mathrm{p}};r)$, what
comes from the fact that the modes are thicker the larger their angular OAM
is, and hence the following ordering of the couplings holds%
\begin{equation}
\chi_{f}<\chi_{f-2}<\cdots<\chi_{l_{0}},
\end{equation}
that is, the lowest OAM modes are the ones coupled more strongly to the pump.

The master equation associated to the $f$--transverse--family DOPO in a
picture rotated to the laser frequency and in which everything is assumed to
be resonant reads then%
\begin{align}\label{FamDOPOmaster}
\frac{d\hat{\rho}_{\mathrm{I}}}{dt}  & =\left[  \sum_{l}\frac{\chi_{l}%
}{1+\delta_{0,l}}\hat{a}_{\mathrm{p}}\hat{a}_{l}^{\dagger}\hat{a}%
_{-l}^{\dagger}+\mathcal{E}_{\mathrm{p}}a_{\mathrm{p}}^{\dagger}%
+\mathrm{H.c.},\hat{\rho}_{\mathrm{I}}\right]+\sum_{j=\mathrm{p},\pm l}\gamma_{j}(2\hat{a}_{j}\hat{\rho}_{\mathrm{I}}%
\hat{a}_{j}^{\dagger}-\hat{a}_{j}^{\dagger}\hat{a}_{j}\hat{\rho}_{\mathrm{I}%
}-\hat{\rho}_{\mathrm{I}}\hat{a}_{j}^{\dagger}\hat{a}_{j});
\end{align}
now, assuming that all the signal modes have the same loss rate $\gamma
_{\mathrm{s}}$, we can map this master equation into the following Langevin
equations within the positive $P$ representation:
\begin{subequations}
\label{famDOPOlangevin}%
\begin{align}
\dot{\alpha}_{\mathrm{p}}  &  =\mathcal{E}_{\mathrm{p}}-\gamma_{\mathrm{p}%
}\alpha_{\mathrm{p}}-\sum_{l}\frac{\chi_{l}}{1+\delta_{0,l}}\alpha_{l}%
\alpha_{-l},\\
\dot{\alpha}_{\mathrm{p}}^{+}  &  =\mathcal{E}_{\mathrm{p}}-\gamma
_{\mathrm{p}}\alpha_{\mathrm{p}}^{+}-\sum_{l}\frac{\chi_{l}}{1+\delta_{0,l}%
}\alpha_{l}^{+}\alpha_{-l}^{+},\\
\dot{\alpha}_{l}  &  =-\gamma_{\mathrm{s}}\alpha_{l}+\chi_{l}\alpha
_{\mathrm{p}}\alpha_{-l}^{+}+\sqrt{\chi_{l}\alpha_{\mathrm{p}}}\xi_{l}(t),\\
\dot{\alpha}_{l}^{+}  &  =-\gamma_{\mathrm{s}}\alpha_{l}^{+}+\chi_{l}%
\alpha_{\mathrm{p}}^{+}\alpha_{-l}+\sqrt{\chi_{l}\alpha_{\mathrm{p}}^{+}}%
\xi_{l}^{+}(t),\\
\dot{\alpha}_{-l}  &  =-\gamma_{\mathrm{s}}\alpha_{-l}+\chi_{l}\alpha
_{\mathrm{p}}\alpha_{l}^{+}+\sqrt{\chi_{l}\alpha_{\mathrm{p}}}\xi_{l}^{\ast
}(t),\\
\dot{\alpha}_{-l}^{+}  &  =-\gamma_{\mathrm{s}}\alpha_{-l}^{+}+\chi_{l}%
\alpha_{\mathrm{p}}^{+}\alpha_{l}+\sqrt{\chi_{l}\alpha_{\mathrm{p}}^{+}%
}\left[  \xi_{l}^{+}(t)\right]  ^{\ast},
\end{align}
where all the various complex noises satisfy the usual statistical properties
(\ref{ComplexNoiseStat}) and are independent.

As usual, we now rewrite the equations in terms of the following normalized
variables%
\end{subequations}
\begin{equation}
\tau=\gamma_{\mathrm{s}}t,\text{ \ \ \ \ }\beta_{j}(\tau)=\frac{\chi_{l_{0}}%
}{\gamma_{\mathrm{s}}\sqrt{\gamma_{\mathrm{p}}/\gamma_{j}}}\alpha
_{j}(t),\text{ \ \ \ \ }\zeta_{j}(\tau)=\frac{1}{\sqrt{\gamma_{\mathrm{s}}}%
}\xi_{j}(t),
\end{equation}
arriving to
\begin{subequations}
\label{famDOPOescaledLan}%
\begin{align}
\dot{\beta}_{\mathrm{p}}  &  =\kappa\left[  \sigma-\beta_{\mathrm{p}}-\sum
_{l}\frac{r_{l}}{1+\delta_{0,l}}\beta_{l}\beta_{-l}\right]  ,\\
\dot{\beta}_{\mathrm{p}}^{+}  &  =\kappa\left[  \sigma-\beta_{\mathrm{p}}%
^{+}-\sum_{l}\frac{r_{l}}{1+\delta_{0,l}}\beta_{l}^{+}\beta_{-l}^{+}\right]
,\\
\dot{\beta}_{l}  &  =-\beta_{l}+r_{l}\beta_{\mathrm{p}}\beta_{-l}^{+}%
+g\sqrt{r_{l}\beta_{\mathrm{p}}}\zeta_{l}(\tau),\\
\dot{\beta}_{l}^{+}  &  =-\beta_{l}^{+}+r_{l}\beta_{\mathrm{p}}^{+}\beta
_{-l}+g\sqrt{r_{l}\beta_{\mathrm{p}}^{+}}\zeta_{l}^{+}(\tau),\\
\dot{\beta}_{-l}  &  =-\beta_{-l}+r_{l}\beta_{\mathrm{p}}\beta_{l}^{+}%
+\sqrt{r_{l}\alpha_{\mathrm{p}}}\zeta_{l}^{\ast}(\tau),\\
\dot{\beta}_{-l}^{+}  &  =-\beta_{-l}^{+}+r_{l}\beta_{\mathrm{p}}^{+}\beta
_{l}+\sqrt{r_{l}\beta_{\mathrm{p}}^{+}}\left[  \zeta_{l}^{+}(\tau)\right]
^{\ast},
\end{align}
where in this case the parameters are defined as%
\end{subequations}
\begin{equation}
\kappa=\gamma_{\mathrm{p}}/\gamma_{\mathrm{s}}\text{, \ \ \ \ }\sigma
=\chi_{l_{0}}\mathcal{E}_{\mathrm{p}}/\gamma_{\mathrm{p}}\gamma_{\mathrm{s}%
}\text{, \ \ \ \ }g=\chi_{l_{0}}/\sqrt{\gamma_{\mathrm{p}}\gamma_{\mathrm{s}}%
}\text{, \ \ \ \ }r_{l}=\chi_{l}/\chi_{l_{0}}. \label{FamilyParameters}%
\end{equation}
Note that $0<r_{l}\leq1$, where the equality holds only for $l=l_{0}$.

It will be convenient for future purposes to remind that instead of the
Laguerre--Gauss basis, one could describe the system through the Hybrid
Laguerre--Gauss modes $\{Y_{j,(f-l)/2,l}\left(  k_{\mathrm{s}};\mathbf{r}%
_{\perp},z\right)  \}_{l\in\{f,f-2,...,l_{0}\}}^{j=c,s}$, see (\ref{Ypl}). For
$l=0$ the HLG and LG modes coincide and are rotationally symmetric, while for
$l\neq0$ the HLG modes have a well defined orientation in the transverse plane
(see Figure \ref{fQuanti5}). The annihilation operators for HLG photons are
related to the LG ones by
\begin{equation}
\hat{a}_{c,l}=(\hat{a}_{+l}+\hat{a}_{-l})/\sqrt{2},\text{ \ \ \ \ }\hat
{a}_{s,l}=\mathrm{i}(\hat{a}_{+l}-\hat{a}_{-l})/\sqrt{2}.
\end{equation}

\section{Classical emission\label{ClassiFamilies}}

Based on the phenomenon of pump clamping that we introduced in Section
\ref{OPOfreezing}, it is completely trivial to understand the classical
behavior of the $f$--transverse--family DOPO; let's see this.

The solution with the $\pm l$ couple of OAM modes switched on requires
$\sigma>r_{l}^{-1}$ to exist; hence, the process involving the lowest OAM
modes $\pm l_{0}$ is the one with the lowest oscillation threshold, which is
obtained for $\sigma=1$. As $\sigma$ increases all the injected power is
transferred then to the $\pm l_{0}$ couple, while the pump mode gets clamped
to its value at threshold; the rest of the signal modes remain then switched
off and feel that the DOPO is frozen at $\sigma=1$.%

\begin{figure}
[t]
\begin{center}
\includegraphics[
height=3.5648in,
width=3.5648in
]%
{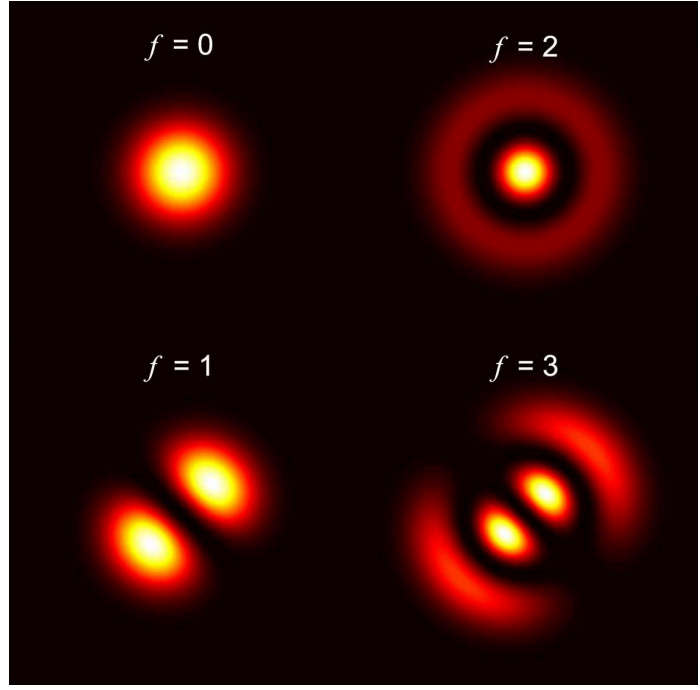}%
\caption{Transverse profile of the signal field above threshold when the DOPO
is tuned to the different families at the signal frequency.}%
\label{fFamilies1}%
\end{center}
\end{figure}

Being more specific, the system has only two types of classical solutions: The
below threshold solution%
\begin{equation}
\bar{\beta}_{\mathrm{p}}=\sigma\text{, \ \ \ \ }\bar{\beta}_{l}=0\text{
}\forall l,
\end{equation}
which is stable for $\sigma<1$, and the above threshold solution, which reads%
\begin{equation}
\bar{\beta}_{\mathrm{p}}=1\text{, \ \ \ \ }\bar{\beta}_{0}=\pm\sqrt
{2(\sigma-1)}\text{, \ \ \ }\bar{\beta}_{\pm l}=0\text{ }\forall l\neq0,
\end{equation}
for even families, and%
\begin{equation}
\bar{\beta}_{\mathrm{p}}=1\text{, \ \ \ \ }\bar{\beta}_{\pm1}=\sqrt{\sigma
-1}\exp(\mp\mathrm{i}\theta)\text{, \ \ \ }\bar{\beta}_{\pm l}=0\text{
}\forall l\neq1,
\end{equation}
with $\theta$ arbitrary, for odd families.

Above threshold, the corresponding classical down--converted fields will be%
\begin{subequations}
\begin{align}
\mathbf{\bar{E}}_{\mathrm{s},\rightarrow}^{(+)}\left(  \mathbf{r}\right)   &
=\mathrm{i}\sqrt{\frac{\hbar\omega_{\mathrm{s}}(\sigma-1)}{2\varepsilon
_{0}n_{\mathrm{s}}L_{\mathrm{s}}g^{2}}}\mathbf{e}_{\mathrm{e}}L_{f/2,0}\left(
k_{\mathrm{s}};\mathbf{r}_{\perp},z\right)  e^{-\mathrm{i}\omega
_{0}t+\mathrm{i}n_{\mathrm{s}}k_{\mathrm{s}}z},\tag{even}\\
\mathbf{\bar{E}}_{\mathrm{s},\rightarrow}^{(+)}\left(  \mathbf{r}\right)   &
=\mathrm{i}\sqrt{\frac{\hbar\omega_{\mathrm{s}}(\sigma-1)}{2\varepsilon
_{0}n_{\mathrm{s}}L_{\mathrm{s}}g^{2}}}\mathbf{e}_{\mathrm{e}}Y_{c,(f-1)/2,1}%
\left(  k_{\mathrm{s}};r,\phi-\theta,z\right)  e^{-\mathrm{i}\omega
_{0}t+\mathrm{i}n_{\mathrm{s}}k_{\mathrm{s}}z},\tag{odd}%
\end{align}
\end{subequations}
for even or odd families, respectively. The transverse profiles of these
classical fields are plotted in Figure \ref{fFamilies1} for the first four
transverse families. It can be appreciated that for even families the field
preserves the rotational symmetry of the system, as $L_{f/2,0}\left(
k_{\mathrm{s}};\mathbf{r}_{\perp},z\right)  $ is rotationally symmetric in the
transverse plane. On the other hand, for odd families the field takes the form
of a HLG mode with $l=1$ \ and arbitrary orientation in the transverse plane,
and hence the phenomenon of spontaneous rotational symmetry breaking appears
in this case.%

\begin{figure}
[t]
\begin{center}
\includegraphics[
width=\textwidth
]%
{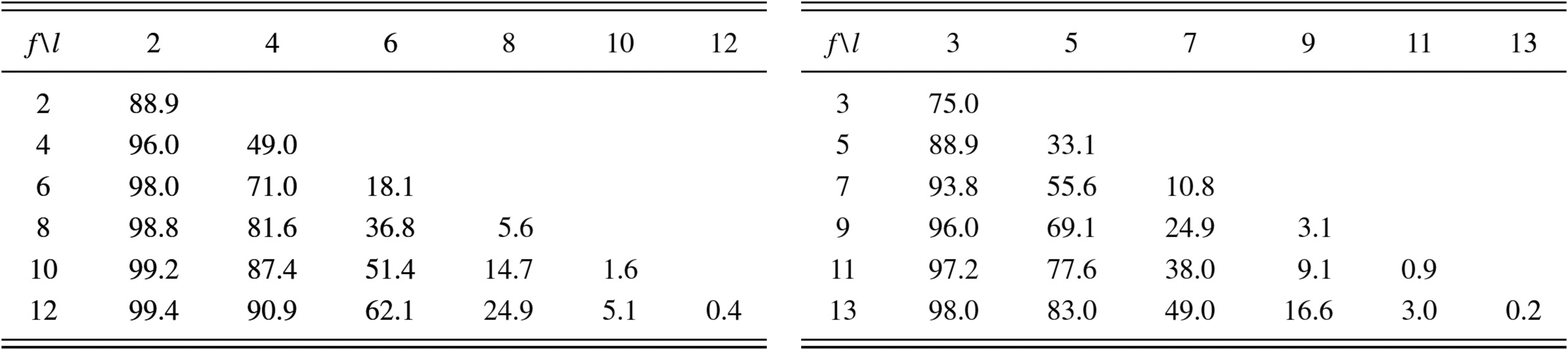}%
\caption{Percentage of noise reduction for the nonamplified hybrid modes
$l\neq l_{0}$ lying in even (left table) and odd (right table) families. Note that large levels of squeezing are
obtained for the lower angular momentum modes.}%
\label{fFamiliesTable}%
\end{center}
\end{figure}

\section{Quantum properties\label{QuantumFamilies}}

Armed with all the background that we have learned in the previous chapters,
the quantum properties of the $f$--transverse--family DOPO are as intuitive as
the classical properties that we just found. First, remember that the dynamics
of non-amplified down--conversion channels are decoupled (within the
linearized theory) from the properties of the classically excited modes, and
then we can discuss their properties separately.

For even families, the $l=0$ mode will behave as the signal field of the
single--mode DOPO, that is, it shows bifurcation squeezing. In particular, its
\textsf{Y} quadrature is perfectly squeezed at zero noise frequency only at
threshold, the squeezing level degrading as one moves far above or below this
instability point (see Section \ref{DOPO}).

For odd families, the $l=\pm1$ modes behave as the signal modes of the
2tmDOPO, that is, their properties can be understood in terms of noncritical
squeezing induced by spontaneous rotational symmetry breaking. Concretely, one
can show that the orientation of the classically excited HLG mode diffuses
with time driven by quantum noise, while the \textsf{Y} quadrature of the mode
orthogonal to this one, $Y_{s,(f-1)/2,1}\left(  k_{\mathrm{s}};r,\phi
-\theta,z\right)  $, is perfectly squeezed irrespective of the system
parameters above threshold (see Section \ref{SRSB}). All the properties that
we have studied in Chapters \ref{2tmDOPO} and \ref{2tmDOPOwithIS} concerning
the phenomenon of spontaneous symmetry breaking also apply to this case.%

As for the modes with $l\neq l_{0}$, let us change to the HLG basis in order
to formulate their quantum properties. The three--wave--mixing Hamiltonian
(\ref{FamTWMHamiltonian}) can be rewritten in this basis as%
\begin{equation}
\hat{H}_{\mathrm{c}}=\mathrm{i}\hbar\sum_{l\neq l_{0}}\chi_{l}\hat
{a}_{\mathrm{p}}\left(  \hat{a}_{c,l}^{\dagger2}+\hat{a}_{s,l}^{\dagger
2}\right)  /2+\mathrm{H.c.},
\end{equation}
where we have not considered the lowest angular momentum modes, as their
properties have been already discussed. Once the threshold $\sigma=1$ is
reached, these modes feel the DOPO as frozen at that point no matter how much
we keep increasing $\sigma$. Hence, and given that the Hamiltonian in the HLG
basis is a combination of down--conversion channels of indistinguishable pairs
of photons, just as the case studied in Section \ref{OPOfreezing2Modes}, we
can directly asses that the noise spectra of the different modes will be%
\begin{subequations}
\begin{align}
V^{\mathrm{out}}(\hat{X}_{c,l};\Omega)  &  =V^{\mathrm{out}}(\hat{X}%
_{s,l};\Omega)=\frac{(1+r_{l}\bar{\beta}_{\mathrm{p}})^{2}+\tilde{\Omega}^{2}%
}{(1-r_{l}\bar{\beta}_{\mathrm{p}})^{2}+\tilde{\Omega}^{2}},\\
V^{\mathrm{out}}(\hat{Y}_{c,l};\Omega)  &  =V^{\mathrm{out}}(\hat{Y}%
_{s,l};\Omega)=\frac{(1-r_{l}\bar{\beta}_{\mathrm{p}})^{2}+\tilde{\Omega}^{2}%
}{(1+r_{l}\bar{\beta}_{\mathrm{p}})^{2}+\tilde{\Omega}^{2}},
\end{align}
see (\ref{2ChannelDOPOnoiseVar}), and hence, all the $\{Y_{j,(f-l)/2,l}\left(
k_{\mathrm{s}};\mathbf{r}_{\perp},z\right)  \}_{l\in\{f,f-2,...,l_{0}%
+2\}}^{j=c,s}$ modes have squeezing on their \textsf{Y} quadrature; in
particular, the maximum levels of squeezing are obtained above threshold
($\bar{\beta}_{\mathrm{p}}=1$) and at zero noise frequency ($\tilde{\Omega}%
=0$), and read%
\end{subequations}
\begin{equation}
V^{\mathrm{out}}(\hat{Y}_{c,l};\Omega=0)=V^{\mathrm{out}}(\hat{Y}_{s,l}%
;\Omega=0)=\frac{(1-r_{l})^{2}}{(1+r_{l})^{2}}\text{.} \label{VYl}%
\end{equation}
Recall that as long as $r_{l}>0.5$, this expression predicts more than 90\% of
noise reduction. Moreover, by defining the integral%
\begin{equation}
I_{l}=\frac{[(f-l)/2]!}{[(f+l)/2]!}\int_{0}^{+\infty}due^{-2u^{2}}%
u^{2l+1}\left[  L_{(f-l)/2}^{l}(u^{2})\right]  ^{2},
\end{equation}
in terms of which the nonlinear couplings are written as (remember that
$w_{\mathrm{p}}^{2}/w_{\mathrm{s}}^{2}=\lambda_{\mathrm{p}}/\lambda
_{\mathrm{s}}=1/2$)%
\begin{equation}
\chi_{l}=\frac{3}{2}\frac{l_{\mathrm{c}}}{w_{\mathrm{s}}}\chi_{\mathrm{oee}%
}^{(2)}(2\omega_{0};\omega_{0},\omega_{0})\sqrt{\frac{8\hbar\omega_{0}^{3}%
}{\pi^{3}\varepsilon_{0}n_{\mathrm{c}}^{3}L_{\mathrm{opt}}^{3}}}I_{l},
\label{ChilIl}%
\end{equation}
the actual ratios $r_{l}$ can be evaluated for the different modes as%
\begin{equation}
r_{l}=\text{\ \ }I_{l}/I_{l_{0}}\text{;}%
\end{equation}
in the tables of Figure \ref{fFamiliesTable} we show the noise reduction obtained for the different OAM pairs of the first
families. Notice that the largest squeezing levels occur for large values of
$f$ and small values of $l$.

It is to be remarked that both the Hybrid mode $Y_{c,(f-l)/2,l}\left(  k_{\mathrm{s}%
};\mathbf{r}_{\perp},z\right)  $ and its orthogonal $Y_{s,(f-l)/2,l}\left(
k_{\mathrm{s}};\mathbf{r}_{\perp},z\right)  $ have the same squeezing
properties. This means that the orientation of the mode is irrelevant, as an
hybrid mode rotated an arbitrary angle $\psi$ respect to the $x$ axis, which
is given by%
\begin{align}
Y_{c,(f-l)/2,l}\left(  k_{\mathrm{s}};r,\phi-\psi,z\right)    &
=Y_{c,(f-l)/2,l}\left(  k_{\mathrm{s}};\mathbf{r}_{\perp},z\right)  \cos
l\psi+Y_{s,(f-l)/2,l}\left(  k_{\mathrm{s}};\mathbf{r}_{\perp},z\right)  \sin
l\psi,
\end{align}
also has the same squeezing properties. This is in clear contrast to the
perfectly squeezed mode in the $l=1$ case which has not an arbitrary
orientation but is orthogonal to the classically excited mode at every instant.

\section{Tuning squeezing through the pump shape\label{TuningChi}}

In this last section of the chapter we would like to discuss how we can take
$r_{l}$ closer to one (and hence increase the squeezing levels of the
non-amplified modes) by modifying the shape of the pump mode.%

\begin{figure}
[t]
\begin{center}
\includegraphics[
height=2.1439in,
width=2.4146in
]%
{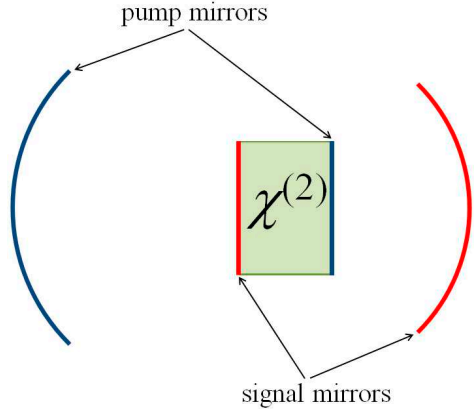}%
\caption{Monolithic design in which the cavities for the pump and signal
fields are independent. The pump (signal) mirrors are made completely
transparent for the signal (pump) frequencies.}%
\label{fFamilies2}%
\end{center}
\end{figure}

The idea comes from the following fact. The only difference between the
nonlinear couplings (\ref{Chil}) of the different OAM pairs is the three--mode
overlapping integral%
\begin{align}
\int_{0}^{+\infty}rdrG(k_{\mathrm{p}};r)\left[  \mathcal{R}_{(f-l)/2}%
^{l}(k_{\mathrm{s}};r)\right]  ^{2}=
\frac{1}{2\pi}\int_{%
\mathbb{R}
^{2}}d^{2}\mathbf{r}_{\perp}G(k_{\mathrm{p}};\mathbf{r}_{\perp},z=0)
L_{(f-l)/2,l}(k_{\mathrm{s}};\mathbf{r}_{\perp},z=0)L_{(f-l)/2,l}^{\ast
}(k_{\mathrm{s}};\mathbf{r}_{\perp},z=0);
\end{align}
now, if instead of the Gaussian profile $G(k_{\mathrm{p}};\mathbf{r}_{\perp
},z=0)$ the pump mode had a simple plane profile, that is, $G(k_{\mathrm{p}%
};\mathbf{r}_{\perp},z=0)\rightarrow C\neq C(\mathbf{r}_{\perp})$, the
integral in the RHS of this expression would be transformed into the
normalization of the Laguerre--Gauss modes, and hence all the modes would have
the same nonlinear coupling to the pump field.

It is true that having a completely homogeneous pump profile is not possible
because it is not physical. However, what we can actually do is design an OPO
configuration in which the spot size of the pump mode exceeds the value
$w_{\mathrm{p}}=w_{\mathrm{s}}/\sqrt{2}$ obtained when the pump and signal
fields are resonating within the same cavity, which was the situation
considered in the previous sections.

An obvious way of doing this is by using a singly--resonant DOPO, in which the
cavity mirrors are completely transparent at the pump frequencies; was this
the case, one could pump the crystal with a Gaussian mode of arbitrary
thickness (or whatever shape one wanted!), as the pump does not have to match
the profile of any mode resonating in the cavity at the pump frequency. The
drawback of this configuration is that pump photons pass through the nonlinear
crystal only once (\textit{single--pass }down--conversion), and hence the
threshold of the system is increased (see e.g., \cite{Patera10}).

In Figure \ref{fFamilies2} we sketch another configuration based on
semi--monolithic OPOs which is interesting for the same purposes but in which
both pump and signal are resonating. An OPO is said to be semi--monolithic if
one of the faces of the nonlinear crystal acts itself as one of the mirrors of
the cavity, what is accomplished by treating that face with an appropriate
reflecting coating. Now, if we use one face of the crystal as a mirror for the
pump, and the opposite face as a mirror for the signal as shown in Figure
\ref{fFamilies2}, the cavities for the pump and signal fields become
independent, and we can tune the spot size of the pump and signal Gaussian
modes independently.%

\begin{figure}
[t]
\begin{center}
\includegraphics[
width=\textwidth
]%
{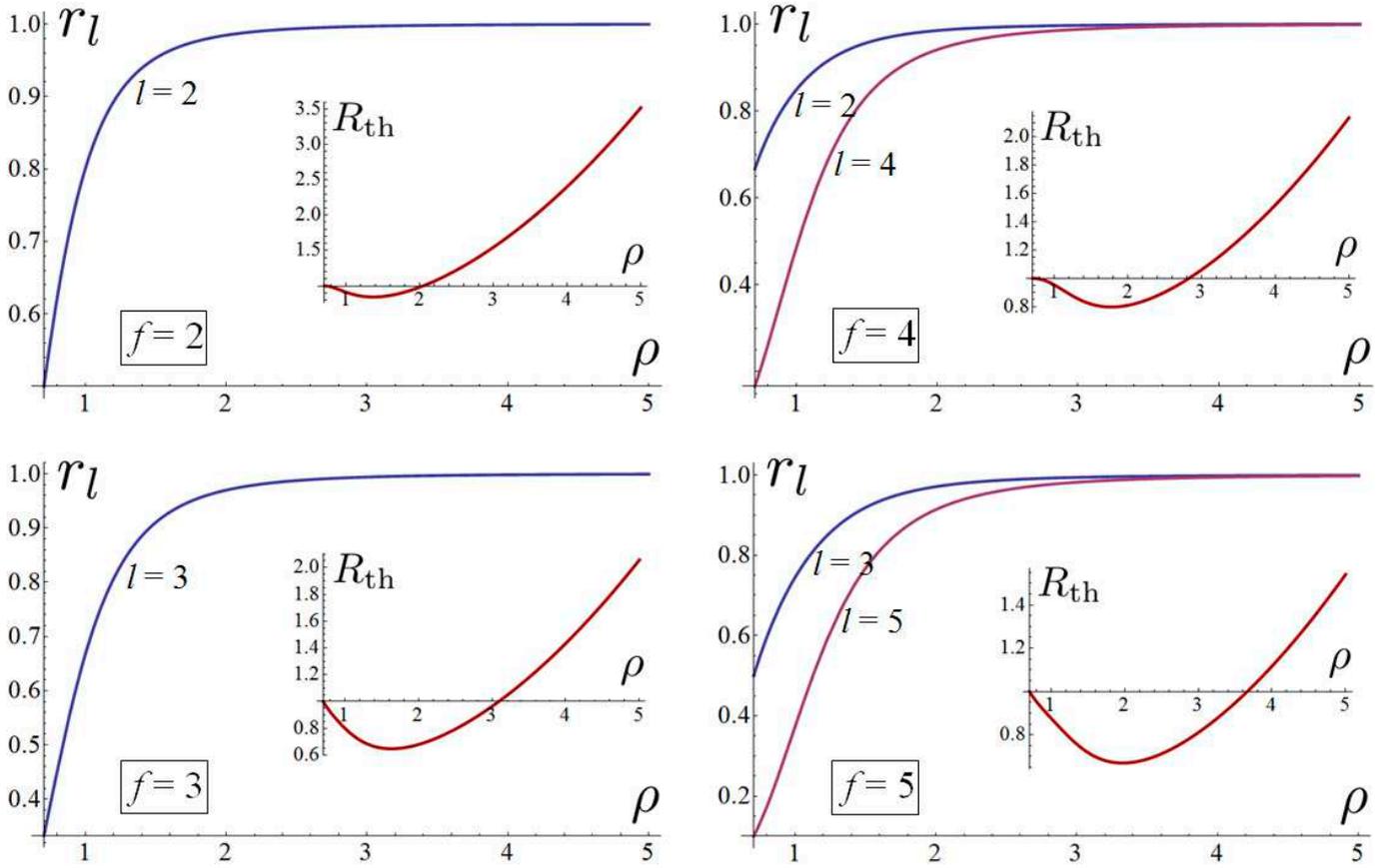}%
\caption{Dependence of the rations $r_{l}$ on the ratio between the spot sizes
of the pump and signal modes, $\rho$. The insets show how the threshold of the
system changes with $\rho$.}%
\label{fFamilies3}%
\end{center}
\end{figure}

Let us then assume that we can tune the ratio between the thickness of the
pump and signal modes at will, and define the parameter $\rho=w_{\mathrm{p}%
}/w_{\mathrm{s}}$. Under these conditions, and following (\ref{ChilIl}), the
nonlinear couplings can be written as%
\begin{equation}
\chi_{l}(\rho)=\frac{3}{2}\frac{l_{\mathrm{c}}}{w_{\mathrm{s}}}\chi
_{\mathrm{oee}}^{(2)}(2\omega_{0};\omega_{0},\omega_{0})\sqrt{\frac
{4\hbar\omega_{0}^{3}}{\pi^{3}\varepsilon_{0}n_{\mathrm{c}}^{3}L_{\mathrm{opt}%
}^{3}}}I_{l}(\rho), \label{IntToChil}%
\end{equation}
where now%
\begin{equation}
I_{l}(\rho)=\frac{1}{\rho}\frac{[(f-l)/2]!}{[(f+l)/2]!}\int_{0}^{+\infty
}due^{-u^{2}(1+1/2\rho^{2})}u^{2l+1}\left[  L_{(f-l)/2}^{l}(u^{2})\right]
^{2},
\end{equation}
and we have assumed that the lengths of the pump and signal cavities are
similar. In Figure \ref{fFamilies3} we show how the ratios $r_{l}(\rho
)=I_{l}(\rho)/I_{l_{0}}(\rho)$ change with $\rho$; note that $\rho$ does not
need to be really big in order for $r_{l}$ to approach unity (in fact,
remember that with $r_{l}>0.5$ we already find squeezing levels above 90\%).

Of course, we should worry about what happens with the threshold of the system
when changing the relation $w_{\mathrm{p}}=w_{\mathrm{s}}/\sqrt{2}$. In
particular, the ratio between the power needed to make the signal field
oscillate for an arbitrary $\rho$, and that for $\rho=1/\sqrt{2}$ is given by%
\begin{equation}
R_{\mathrm{th}}(\rho)=\frac{P_{\mathrm{th}}(\rho)}{P_{\mathrm{th}}%
(\rho=1/\sqrt{2})}=\frac{\chi_{l_{0}}^{2}(\rho=1/\sqrt{2})}{\chi_{l_{0}}%
^{2}(\rho)}=\frac{I_{l_{0}}^{2}(\rho=1/\sqrt{2})}{I_{l_0}^{2}(\rho)},
\end{equation}
as follows from (\ref{PtoE}), (\ref{FamilyParameters}), and (\ref{IntToChil}).
This expression is plotted in the insets of Figure \ref{fFamilies3} for the
different families. These figures show that the changes in the threshold are
quite reasonable, and moreover, the threshold is even lowered respect to the
$\rho=1/\sqrt{2}$ case for small enough ratios $\rho$. 

%% file: ConclusionsFO.tex
\section{Summary of the original research on OPOs}

This thesis has been devoted to the study of some new properties of optical
parametric oscillators (OPOs). Such devices consist of an optical cavity
embedding a second order nonlinear crystal, which, when pumped with an
external laser beam at frequency $\omega_{\mathrm{p}}=2\omega_{0}$, is able to
generate light at frequencies $\omega_{\mathrm{s}}$ and $\omega_{\mathrm{i}}$
such that $\omega_{\mathrm{p}}=\omega_{\mathrm{s}}+\omega_{\mathrm{i}}$,
through the process of parametric down--conversion (the $\omega_{\mathrm{p}}$,
$\omega_{\mathrm{s}}$, and $\omega_{\mathrm{i}}$ fields are called,
respectively, pump, signal, and idler). The down--conversion process is known
as degenerate when the signal and idler modes have the same frequency
$\omega_{0}$, and type I when they have the same polarization (if they are
generated in orthogonal, linearly polarized modes, we talk then about type II
down--conversion).\newline As stressed in the Introduction, the first
two--thirds of the dissertation have consisted on a self--contained discussion
about the basic physics behind these devices. Here we would like to remark the
following well known properties of OPOs (see Section \ref{QOManualOverview}
for an extensive summary of this introductory part):

\begin{itemize}
\item From a classical point of view, the interplay between parametric
down--conversion and cavity losses forces the pump power (denoted along the
thesis by the dimensionless parameter $\sigma$) to go above some threshold
level ($\sigma=1$) in order to make the signal--idler fields oscillate. Hence,
at $\sigma=1$ the system shows a bifurcation in which the OPO passes from a
steady state with the down--converted fields switched off, to a new steady
state in which these are switched on (much like in the laser).

\item Working with type I, degenerate down--conversion, we simply talk about a
degenerate OPO (DOPO). If in addition the signal and idler modes are generated
with in the same transverse mode (typically a TEM$_{00}$ mode), we talk about
a single--mode DOPO. It is possible to show that in such system the
down--converted field is in a squeezed state, that is, one of its quadratures
has fluctuations below the vacuum (shot noise) level. However, this squeezing
is only large when working close to the bifurcation: We say that it is a
critical phenomenon.

\item When the down--converted photons are distinguishable in frequency or/and
polarization, we have shown that the joint state of the signal and idler modes
corresponds to an entangled state in which these share correlations which go
beyond what is classically allowed. Again, the entanglement level is large
only when working close to threshold.\newline However, in this case there is
one highly squeezed observable at any pump level above threshold: The
signal--idler intensity difference. We say then that signal and idler are twin
beams, as they have perfectly correlated photon numbers.
\end{itemize}

These well known results were the starting point of the research developed in
this thesis; in the remaining of this section we make a summary of the main
original results that we have found (numbering of this list follows the
chapters and sections of the dissertation):

\begin{itemize}
\item[\ref{MultiOPOs}.] In the first completely original chapter we have
introduced the concept of multi--mode OPO as that having several
down--conversion channels available for a single pump mode. We have argued
that this is actually the natural way in which OPOs operate, either because it
is impossible to forbid the existence of other transverse modes close to the
one in which we would like to operate, or because the phase matching curve is
wide enough as to include more than a single signal--idler frequency
pair.\newline We have then introduced two phenomena characteristic of
multi--mode OPOs, phenomena which are considered by the PhD candidate as the
most important contribution of the thesis: Pump clamping and spontaneous
symmetry breaking. As we explain now, their main interest is that they offer
new means for generating squeezed (or entangled) light, with the incentive
that the squeezing is noncritical, that is, independent of the system parameters.

\begin{itemize}
\item[\ref{OPOfreezing}.] Let us first explain the key points describing the
phenomenon of pump clamping:

\begin{itemize}
\item In general, the different down--conversion channels of the multi--mode
OPO will have different thresholds, depending these on the coupling to the
pump, the detuning, and the loss rates of the modes involved in the particular process.

\item Classically, only the channel with the lowest threshold can be
amplified, that is, the processes with higher thresholds remain switched off
no matter how much we increase $\sigma$.

\item What is interesting is that once the lowest threshold mode is generated,
the non-amplified modes feel as if the OPO was frozen at the bifurcation, even
if we change the actual pump injection. This applies also to their quantum
properties, that is, the modes of the non-amplified channels will have the
squeezing or entanglement that they are supposed to have below their
oscillation threshold, irrespective of the system parameters.
\end{itemize}

Hence, our first prediction has been that, above threshold, apart from the
bright signal--idler modes which behave as the ones in the single--channel OPO
(bifurcation squeezing or entanglement), the output of a general OPO has a lot
more signal--idler pairs with some squeezing or entanglement, their actual
levels depending on the distance between their thresholds and the lowest
threshold of the system (the true threshold).\newline

\item[\ref{BasicSSB}.] We have introduced the basic idea behind spontaneous
symmetry breaking in an abstract way as follows:

\begin{itemize}
\item Suppose that we work with an OPO which is invariant under changes of
some continuous degree of freedom of the down--converted field, say $\epsilon
$, which we have denoted by \textquotedblleft free parameter\textquotedblright%
\ all along the dissertation. Above threshold, the down--converted field, say
$\mathbf{\bar{E}}_{\epsilon}(\mathbf{r},t)$ ---we explicitly introduce the
free parameter in the field---, is not zero, and when the OPO starts emitting
it, a particular value of the free parameter is chosen according to the
particular fluctuations from which the field is built up. We say that
$\epsilon$ is chosen through spontaneous symmetry breaking.\newline

\item Let's think now about what quantum theory might introduce to this
classical scenario. First, as variations of the FP do not affect the system
(what mathematically is reflected in the appearance of a Goldstone mode in the
matrix governing the linear evolution of the system, that is, a mode with null
eigenvalue), quantum noise is expected to make it fluctuate without
opposition, eventually making it become completely undetermined.

\item Then, invoking now the uncertainty principle, the complete
indetermination of a system's variable allows for the perfect determination of
its corresponding momentum, meaning this that we could expect perfect
squeezing in the quadrature selected by the local oscillator $-$%
\textrm{i}$\partial_{\epsilon}\mathbf{\bar{E}}_{\epsilon}(\mathbf{r},t)$ in an
homodyne detection scheme.

\item Finally, as this process relies only on the generation of the
down--converted field, this perfect squeezing is expected to be independent of
the distance to threshold, as long as we are above the bifurcation point.
\end{itemize}

The first proof of such an intuitive reasoning was offered through the study
of a DOPO with plane mirrors \cite{PerezArjona06,PerezArjona07}, where
transverse patterns as well as cavity solitons have been predicted to appear
above threshold \cite{Staliunas97,Longhi97}, hence breaking the translational
symmetry of the system. Most of the present thesis has been devoted to study
several aspects of the phenomenon of spontaneous symmetry breaking, but we
have designed a much more simple (and realistic) DOPO configuration for that matter.
\end{itemize}

\item[\ref{SRSB},\ref{2tmDOPO}.] The system we have used to study in depth the
phenomenon of spontaneous symmetry breaking is a DOPO (hence signal and idler
have the same frequency and polarization) pumped by a TEM$_{00}$ mode, and
whose cavity is tuned so that the first family of transverse modes resonates
at the down--converted frequency. Under this circumstances the down--converted
photons are created in opposite orbital angular momentum (OAM) Laguerre--Gauss
modes, and, consequently, we have called two--transverse--mode DOPO (2tmDOPO)
to this system.\newline Let us summarize now the main results that we have
found in this system.

\begin{itemize}
\item[\ref{SRSB}.] In connection with the general picture of \ref{BasicSSB},
these are the properties of the system:

\begin{itemize}
\item A weighted superposition of two $\pm1$ Laguerre--Gauss modes can be seen
as a TEM$_{10}$ mode, whose orientation in the transverse plane is given by
half the phase--difference between the underlying Laguerre--Gauss modes.
Hence, owed to the rotational invariance of the cavity, classical emission
takes place in a TEM$_{10}$ mode with an arbitrary orientation in the
transverse plane (mathematically this is clear because the 2tmDOPO classical
equations are invariant against changes in the relative phase between the
Laguerre--Gauss modes).\newline Hence, once the threshold is crossed and the
orientation is chosen according to the particular initial fluctuations, the
rotational symmetry is spontaneously broken, and we can talk about a bright
mode (the generated one which breaks the symmetry) and a dark mode (the mode
orthogonal to the generated one).

\item Quantum noise is able to make the bright TEM$_{10}$ mode rotate randomly
in the transverse plane. Mathematically, this is reflected in the appearance
of a Goldstone mode in the matrix governing the linear evolution of the
fluctuations above threshold. The null eigenvalue of this mode allows quantum
noise to change the orientation of the bright mode without opposition. Though
continuously increasing with time, this rotation of the classically excited
pattern is quite slow when working above threshold.

\item As for the squeezing properties, it has been proved that the bright mode
has the same behavior as the single--mode DOPO, i.e., large levels of
squeezing appear only close to threshold. On the other hand, accompanying the
Goldstone mode it appears another mode whose associated eigenvalue takes the
minimum possible value $\left(  -2\right)  $.\newline Together, the Goldstone
mode and the maximally damped mode are responsible of the remarkable
properties of the dark mode: Its \textsf{Y} quadrature is perfectly squeezed
at any pump level above threshold, while its \textsf{X} quadrature carries
only with vacuum fluctuations (in apparent violation of the uncertainty
principle).\newline This is in full agreement with the abstract picture of the
phenomenon offered in \ref{BasicSSB}, as the dark mode coincides (up to a
$\pi/2$ phase) with the OAM of the bright mode.
\end{itemize}

\item[\ref{OnCan}.] We have proved that the apparent violation of the
uncertainty principle is just that, apparent, as the conjugate pair of the
squeezed quadrature is not another quadrature but the orientation of the
bright mode, which in fact is completely undetermined in the long time term.

\item[\ref{FixOsci}.] Next we have pointed out that in order to measure the
quantum properties of the dark mode via homodyne detection, one has to use a
TEM$_{10}$ local oscillator that is perfectly matched to the orientation of
this mode at any time. However, the mode is rotating randomly, which seems to
make impossible the perfect matching. For this reason we have studied the
situation in which the local oscillator is matched to the dark mode's
orientation only at the initial time, remaining fixed during the detection
time.\newline We have shown that arbitrarily large levels of noise reduction
can be obtained even in this case if the phase of the local oscillator is
exactly $\pi/2$. We then considered phase deviations up to 2$%
\operatorname{{{}^\circ}}%
$ (1.5$%
\operatorname{{{}^\circ}}%
$ seems to be the current experimental limit \cite{Vahlbruch08,Takeno07}),
comparing the results with those predicted for the single--mode DOPO; similar
levels are obtained for both, with the advantage that in the 2tmDOPO this
level is independent of the distance from threshold, and hence, noncritical.

\item[\ref{BeyondApp}.] We have also shown that the assumptions made in order
to analytically solve the problem are not the responsible for the quantum
properties of the dark mode. In particular, we have used numerical simulations
to show that these analytic predictions coincide with those of the complete
nonlinear stochastic equations describing the system.
\end{itemize}

\item[\ref{2tmDOPOwithIS}.] In this chapter we have studied the classical and
quantum dynamics of the 2tmDOPO when a TEM$_{10}$ beam is injected at the
signal frequency. This study was motivated by experimental reasons: On one
hand, this injection is customarily used in real DOPOs to stabilize the
cavity, so that, in particular, the signal resonance is locked to the desired
value (active locking); on the other, in the 2tmDOPO, and given that
down--conversion is a stimulated process, we think that the injection of a
TEM$_{10}$ mode could also lock the orientation of the bright mode, what seems
highly recommendable from the detection point of view.\newline These are the
main results that we have found:

\begin{itemize}
\item[\ref{ClassiIS}.] Let's start with the classical dynamics. For signal
injections below some threshold, the system posses a unique, stable solution
in which the TEM$_{01}$ (the non-injected mode) remains switched off. However,
above this injection threshold both the TEM$_{10}$ and TEM$_{01}$ modes become
populated. Hence, the system posses a new bifurcation linked to the switching
on of the non-injected mode.\newline This bifurcation is present for any
$\sigma$, and in particular it exists also for $\sigma=0$, when the DOPO runs
as a second harmonic generator (as in this case the $2\omega_{0}$ field is
generated from the $\omega_{0}$ injection); of course in this case the
threshold is independent of the injection's phase, as there is no other phase
reference. For $\sigma>0$ this is no longer the case: The threshold is maximum
(minimum) when the signal injection is in phase (anti--phase) with the pump
injection.\newline Surprisingly, working in anti--phase and when $\sigma\geq
1$, there is no threshold for the generation of the TEM$_{01}$ mode. On the
other hand, in the in--phase case there exists a clear threshold below which
the bright mode is locked to the injected TEM$_{10}$ mode, just as we wanted
to prove.

\item[\ref{QuantumIS}.] As for the quantum properties, we have proved that the
TEM$_{01}$ mode has large levels of squeezing in all the region below their
generation threshold, hence proving that the TEM$_{10}$ injection does not
degrade too much the squeezing capabilities of the 2tmDOPO.\newline This
insensitivity to the injection (in particular we have proved that it is far
much insensitive to the injection than the single--mode DOPO, the
configuration customarily used in experiments) comes from the fact that the
TEM$_{01}$ mode shows perfect squeezing at the instability point leading to
its activation.
\end{itemize}

\item[\ref{TypeIIOPO}.] In this chapter we have studied type II OPOs tuned to
the TEM$_{00}$ mode for all the frequencies involved in the down--conversion
process. We had a twofold intention:

\begin{itemize}
\item[\ref{SPSB}.] On one hand, we wanted to generalize the phenomenon of
spontaneous symmetry breaking to the polarization degrees of freedom of light.
In particular, we have shown that when the down--conversion process is also
frequency degenerate, the system is isomorphic to the 2tmDOPO, where the role
of the opposite orbital angular momenta are now played by the orthogonal
linear polarizations of the signal--idler modes.\newline We have then proved
that the behavior expected after the general reasoning of Section
\ref{BasicSSB} applies: Classical emission takes place in an elliptically
polarized mode, and the free parameter of this system (the relative phase
between signal and idler) is related now to the eccentricity of the
corresponding polarization ellipse; quantum noise is responsible for a
diffusion process of this polarization parameter, which eventually becomes
completely undetermined; the \textsf{Y} quadrature of the dark mode, which in
this case consists on the mode orthogonally polarized to the bright one, has
perfect squeezing at any pump level above threshold.

\item[\ref{degNONdeg}.] However, we have pointed out that in real experiments
it seems not possible to obtain frequency degeneracy in OPOs. In the case of
type II operation, frequency degeneracy can be obtained by introducing a
birefringent element inside the cavity (self--phase--locked type II OPOs), but
this breaks the polarization symmetry of the system, hence destroying the
phenomenon of spontaneous symmetry breaking. Nevertheless, we have argued that
a residual squeezing should be present after this explicit symmetry breaking,
and that it was actually observed (but not interpreted a residue of the
symmetry breaking) some years ago \cite{Laurat05}.\newline We then show that
frequency degeneracy might as well be obtained by injecting an external laser
beam at the degenerate frequency (actively--phase-locked type II OPOs).
Moreover, we have studied the complete bifurcation diagram of this system in
the simple case of having opposite detunings for the signal and idler modes,
showing that there are regimes in which stationary solutions coexist with
periodic orbits.
\end{itemize}

\item[\ref{FamiliesDOPO}.] Motivated by the fact that we already studied DOPOs
tuned to the fundamental TEM$_{00}$ mode at the subharmonic (single--mode
DOPOs), as well as DOPOs tuned to the first family of transverse modes
(2tmDOPOs), in the last original chapter of the thesis we have studied what
happens when the DOPO is tuned to an arbitrary family of transverse modes at
the signal frequency. Note that family $f$ contains Laguerre--Gauss modes with
orbital angular momenta $\{\pm f,\pm(f-2),...,\pm l_{0}\}$, with $l_{0}$ equal
to $0$ for even families and $1$ for the odd ones.\newline The results have
been actually very intuitive based on the phenomena that we studied in the
previous chapters. In particular:

\begin{itemize}
\item[\ref{ClassiFamilies}.] Given that the lowest orbital angular momenta of
the family are the ones which couple the strongest to the pump (so that they
have the lowest threshold), they will be the ones classically amplified above
threshold, while the rest of modes will remain switched off.\newline For even
families this means that the bright mode will consist in a $0$ OAM mode, and
hence the rotational symmetry of the system will be preserved; on the other
hand, the bright mode for the case of odd families will consist in a
superposition of two $\pm1$ OAM modes (a Hybrid Laguerre--Gauss mode), and
hence the rotational symmetry will be spontaneously broken.

\item[\ref{QuantumFamilies}.] As for the quantum properties, they are very
simple to guess as well. First, and given that the dynamics of the
non-amplified modes decouple from the ones of the amplified ones, the $0$ OAM
mode of the even--family case will behave as the signal field of the
single--mode DOPO (bifurcation squeezing); similarly, the $\pm1$ OAM modes of
the odd--family case will behave as the ones in the 2tmDOPO (the Hybrid
Laguerre--Gauss mode generated above threshold and its orthogonal will behave,
respectively, as the bright and dark modes of the 2tmDOPO).\newline On the
other hand, and this is the most relevant result, the properties of the higher
OAM modes are dictated by the phenomenon of pump clamping, that is, the Hybrid
Laguerre--Gauss modes associated to the opposite OAM modes will show
noncritical squeezing, the particular squeezing levels depending on the
distance of their corresponding thresholds to that of the lowest OAM modes.

\item[\ref{TuningChi}.] In the last section of the chapter we have shown that
the thresholds of the higher OAM modes can be brought closer to that of the
lowest OAM modes by pumping with a TEM$_{00}$ mode wider than the one expected
for the doubly resonant cavity (in particular, in the limit of infinite
transverse thickness of the pump, the couplings of these to the different OAM
pairs become all equal). We have suggested that this might be implemented
either by using singly resonant DOPOs (in which the cavity is transparent at
the pump frequency, and therefore the shape of the pump can be arbitrary), or
by using clever monolithic cavity designs in which the pump and signal
cavities become independent.\newline This offers the possibility to generate
multi--mode beams with a large number of squeezed modes within it, what could
be interesting for quantum information protocols requiring multipartite
entanglement (quantum correlations shared between more than two parties).
\end{itemize}
\end{itemize}

\section{Outlook}

There are several ideas related to the research summarized in the previous section
which we have started to analyze, but which I have decided not to in include in
the main text either because they are not completely developed, or just to
keep the dissertation at a reasonable size. This section is devoted to
summarize some of these ideas.

\subsection{Temporal symmetry breaking}

In the thesis we have studied the phenomenon of noncritical squeezing
generated via spontaneous symmetry breaking by means of examples involving
spatial and polarization degrees of freedom. It is then natural to try
generalizing the phenomenon to the last kind of degrees of freedom that light
can hold: the temporal ones. Time is a very subtle variable in quantum
mechanics, and hence it seems interesting to find out whether our general
reasoning introduced in Section \ref{BasicSSB} applies to symmetries involving
it or not.

The idea can be explained as follows. Suppose that we work with an OPO which
posses in some parameter region a dynamic mean field or classical solution (in
the sense that the intensity of the field is not stationary, contrary to most
of the examples that we have analyzed in the thesis); assume for simplicity
that the solution is periodic. Now, as the classical equations of the OPO are
invariant under time translations, the periodic solution can start at any
point of its allowed values. We then expect quantum noise to make this
parameter fluctuate randomly, eventually making it become completely
undetermined; physically, this means that in the long time term, it won't be
possible to determined at which point of its oscillation the mean field is.
Based on the results found for the spatial and polarization symmetries, it
seems interesting to analyze whether a local oscillator with the temporal
profile of the $\pi/2$ phase--shifted time derivative of the bright field will
measure perfect, noncritical squeezing in some quadrature of the field coming
out of the cavity.

We have already analyzed these ideas in one of the most simple temporal
instabilities that can be found in OPOs: The temporal instability associated
to second harmonic generation. In particular, it is simple to show that when
only the degenerate signal field is pumped externally, there exists a
threshold value of this injection at which the stationary solution becomes
unstable through a Hopf bifurcation. Therefore, for injected intensities above
this threshold, the solution becomes time dependent, and in particular
periodic when working close enough to the bifurcation. We advance that,
unfortunately, the general reasoning explained above does not seem to apply to
this case, that is, the squeezing of the $\pi/2$ phase--shifted time
derivative of the bright field is not perfect, as happens in the case of
spontaneous spatial or polarization symmetry breaking.

We believe that the problem lies in the fact that the instability of the
stationary solution leading to the periodic solutions involves not only the
signal field, but also the pump field (that is, the eigenvectors of the
stability matrix have projections both onto the pump and signal subspaces),
and it is well known that the pump mode never has large squeezing in OPOs.

In this sense, the actively--phase--locked OPO introduced in Section
\ref{degNONdeg} seems a perfect candidate, in particular in the parameter
region where non-stationary, periodic solutions exist (see Figure
\ref{fTypeII2}), as in this case the pump fluctuations
are completely decoupled from the signal--idler fluctuations as long as one
works in the adiabatic limit $\gamma_{\mathrm{p}}\gg\gamma_{\mathrm{s}}$ (a
limit that cannot be taken in second harmonic generation because then the Hopf
bifurcation goes to infinity, and the stationary solution becomes stable for
finite external injection).

What makes the actively--phase--locked OPO even a more suited candidate for
the study of this phenomenon, is that it connects directly with the
free--running OPO when no injection at the degenerate frequency is present
($\mathcal{I}=0$ in the notation of Chapter \ref{TypeIIOPO}), and hence, one
knows what to expect at that exact point.

We have just started the analysis, and then hope to offer new results soon enough.

\subsection{Self--phase--locked two--transverse--mode type II OPO}

In order to study in depth the phenomenon of noncritical squeezing induced by
spontaneous symmetry breaking, we have made extensive use of the
two--transverse--mode DOPO, in which signal and idler are degenerate in
frequency and polarization, but have opposite orbital angular momenta. Even
though there is nothing fundamental preventing experimentalists to operate
OPOs in such configuration, it is quite hard to ensure that this particular
down--conversion channel wins the nonlinear competition, and then in real
experiments frequency--degeneracy might not be possible to achieve.

However, we believe to have found a way to bring the phenomenon of spontaneous
symmetry breaking to observable grounds. The idea is to work with a type II
non-degenerate OPO tuned to the first family of transverse modes at the signal
and idler frequencies\footnote{This system, but under the assumption of
frequency degeneracy, was recently studied in the context of
hyperentanglement, that is, entanglement between two independent degrees of
freedom (polarization and orbital angular momentum in this case) of signal and
idler \cite{dosSantos09}.}, and then use the self--phase--locking mechanism
introduced at the beginning of Section \ref{degNONdeg} to lock the signal and
idler frequencies to the degenerate one
\cite{Longchambon04a,Longchambon04b,Laurat05}; we believe that a residual
continuous symmetry in the orbital angular momentum degree of freedom of
signal and idler will be present, and hence in the frequency locking region
one should be able to study the phenomenon of spontaneous rotational symmetry breaking.

In order to show explicitly how this is so, let us reason from the Hamiltonian
of the system. Let us denote by $\hat{a}_{\mathrm{s},\pm1}$ and $\hat
{a}_{\mathrm{i},\pm1}$ the annihilation operators for signal and idler photons
with $\pm1$ orbital angular momentum, respectively. The parts of the
Hamiltonian corresponding to the down--conversion process and the mixing
induced by the $\lambda/4$ plate misaligned respect to the extraordinary axis
of the nonlinear crystal can be written for this configuration as%
\begin{subequations}
\begin{align}
\hat{H}_{\mathrm{DC}} &  =\mathrm{i}\hbar\chi\hat{a}_{\mathrm{p}}(\hat
{a}_{\mathrm{s},+1}^{\dagger}\hat{a}_{\mathrm{i},-1}^{\dagger}+\hat
{a}_{\mathrm{s},-1}^{\dagger}\hat{a}_{\mathrm{i},+1}^{\dagger})+\mathrm{H.c.}%
\\
\hat{H}_{\mathrm{P}} &  =\mathrm{i}\hbar\varepsilon(\hat{a}_{\mathrm{s}%
,+1}\hat{a}_{\mathrm{i},+1}^{\dagger}+\hat{a}_{\mathrm{s},-1}\hat
{a}_{\mathrm{i},-1}^{\dagger})+\mathrm{H.c..}%
\end{align}
\end{subequations}
As in the case in which the OPO is tuned to the fundamental Gaussian modes at
the signal--idler frequencies \cite{Longchambon04a,Longchambon04b,Laurat05},
we expect $\hat{H}_{\mathrm{P}}$ to be able to lock the frequencies of signal
and idler to the degenerate in a certain parameter region. On the other hand,
there is a continuous symmetry still present in these Hamiltonian terms,
namely%
\begin{align}
\{\hat{a}_{\mathrm{s},+1},\hat{a}_{\mathrm{s},-1},  & \hat{a}_{\mathrm{i}%
,+1},\hat{a}_{\mathrm{i},-1}\}\\
& \downarrow\nonumber\\
\{\exp(-\mathrm{i}\theta)\hat{a}_{\mathrm{s},+1},\exp(\mathrm{i}\theta)\hat
{a}_{\mathrm{s},-1},  & \exp(-\mathrm{i}\theta)\hat{a}_{\mathrm{i},+1}%
,\exp(\mathrm{i}\theta)\hat{a}_{\mathrm{i},-1}\},\nonumber
\end{align}
and hence, the phenomenon of spontaneous symmetry breaking should still be
present in the system.

In simple terms, we believe that the polarization degree of freedom will allow
us to make OPO become frequency degenerate through the self--phase--locking
mechanism, while the extra orbital angular momentum degree of freedom will
allow for the existence of a residual continuous symmetry in the system, which
will be spontaneously broken above threshold once the mean field is switched on.

We have already started the analysis of this idea, although it might take a
while to understand the parameter region where frequency locking is present,
as the classical equations related to the system involve six modes, and are
therefore hard to extract information from.

\subsection{Effect of anisotropy}

Another thing that we have been studying is how robust is the phenomenon of
squeezing induced by spontaneous symmetry breaking against imperfections of
the system's symmetry, what feels important as no perfect symmetry is present
in real physical systems. Note that we have already answered this question
partly, as in Chapter \ref{2tmDOPOwithIS} we studied the 2tmDOPO with an
injected signal beam which explicitly breaks the rotational symmetry of the
system, showing that the squeezing levels of the dark mode can still be very
large for reasonable injection intensities. This is the main reason why, in an
effort to save some space in an already lengthy dissertation as this is, we
have decided not to introduce the detailed study of symmetry imperfections in
the main text; nevertheless, let us spend now a few words on how we have made
this analysis, and what new interesting features of the phenomenon of
spontaneous symmetry breaking can we learn from it.

As usual, we have used the 2tmDOPO as the platform where studying symmetry
imperfections. In particular, we have considered two sources of anisotropy in
the transverse plane which might actually appear in experiments\footnote{In
the original paper where we considered the 2tmDOPO we already showed the
robustness of the phenomenon by allowing the TEM$_{10}$ and TEM$_{01}$ modes
to have different damping rates through the partially transmitting mirror.
However, while mathematically this anisotropy model is very easy to handle,
physically it is no justified why these modes could have different damping
rates.}. First, we have allowed the mirrors to not be perfectly spherical, in
particular taking them as ellipsoids with some finite ellipticity. Then, we
have allowed the nonlinear crystal to be misaligned respect to the cavity
axis, what may not only correspond experimental imperfections, but may even be
interesting in order to achieve angular phase--matching (see Section
\ref{SecOrderNonLinearity}).

We have already developed the 2tmDOPO model including this sources of
anisotropy, and we are currently trying to find the best way to characterize
the robustness of squeezing against the degree of anisotropy. A preliminary,
not exhaustive inspection of the resulting squeezing spectrum of the dark mode
shows that, as expected, still large levels of noise reduction can be found
for a reasonably large degree of anisotropy.

Concerning new physics, the most interesting feature that can be studied in
this context is how the phase difference $\theta$ between the Laguerre--Gauss
modes (that is, the orientation of the bright TEM$_{10}$ mode) gets locked as
the anisotropy increases, and the relation of this locking to the loss of
squeezing in the dark mode. Our preliminary analysis shows that the variance
of $\theta$, which is infinite in the long time term when the rotational
symmetry in the transverse plane is perfect, see (\ref{Theta Var}), becomes
finite when the anisotropy kicks in; of course, this variance tends to zero as
the anisotropy increases, what means that the bright and dark modes get locked
to a certain orientation in the transverse plane. Similarly, the noncritical
squeezing of the dark mode, which can be perfect when no anisotropy is
present, see (\ref{YsSpectrum}), is degraded as the anisotropy increases. What
we are trying to understand now is if the scaling of this squeezing
degradation with the degree of anisotropy is the same as that of the $\theta
$'s variance, what will be a further prove of the canonical relation existing
between $\theta$ and the quadratures of the dark mode (see Section \ref{OnCan}). 

%% file: QuantumMechanicsFO.tex
The purpose of this appendix is the introduction of the fundamental laws
(axioms) of quantum mechanics as are used throughout this thesis. In an effort
to make this thesis as self--contained as possible, the Hamiltonian formalism
of classical mechanics (which is needed to fully understand and motivate the
transition to quantum mechanics), as well as the theory of Hilbert spaces
(which is the fundamental mathematical language in terms of which quantum
theory is formulated), are briefly exposed.

\section{Classical mechanics\label{ClassicalMechanics}}

The framework offered by quantum mechanics is far from intuitive, but somehow
feels reasonable once one understands the context in which it was created.
This aims to understand (\textit{i}) the theories which were used to describe
physical systems prior to its development, and (\textit{ii}) the experiments
which did not fit in this context.

In this section we make a brief introduction to classical
mechanics\footnote{For a deeper lecture concerning this topic we recommend
Goldstein's book \cite{Goldstein01book}, as well as Greiner's books
\cite{Greiner89cbook,Greiner89dbook}}, with emphasis in analyzing the
Hamiltonian formalism and how it treats observable magnitudes. We will see
that a proper understanding of this formalism is needed to make the transition
to quantum mechanics.

\subsection{The Lagrangian formalism}

In classical mechanics the state of a system is specified by the position of
its constituent particles at all times, $\mathbf{r}_{j}\left(  t\right)
=\left[  x_{j}\left(  t\right)  ,y_{j}\left(  t\right)  ,z_{j}\left(
t\right)  \right]  $ with $j=1,2,...,N$, being $N$ the number of particles.
Defining the linear momentum of the particles as $\mathbf{P}_{j}%
=m_{j}\mathbf{\dot{r}}_{j}$ ($m_{j}$ is the mass of particle $j$), the
evolution of the system is found from a set of initial positions and
velocities by solving the Newtonian equations of motion $\mathbf{\dot{P}}%
_{j}=\mathbf{F}_{j}$, being $\mathbf{F}_{j}$ the forces acting onto particle
$j$.

Most physical systems have further constraints that have to fulfil (for
example, the distance between the particles of a rigid body cannot change with
time, that is, $\left\vert \mathbf{r}_{j}-\mathbf{r}_{l}\right\vert =const$),
and therefore the positions $\left\{  \mathbf{r}_{j}\right\}  _{j=1,...,N}$
are no longer independent, what makes Newton's equations hard to solve. This
calls for a new, simpler theoretical framework: the so-called
\textit{analytical mechanics}.

In analytical mechanics the state of the system at any time is specified by a
vector $\mathbf{q}\left(  t\right)  =\left[  q_{1}\left(  t\right)
,q_{2}\left(  t\right)  ,...,q_{n}\left(  t\right)  \right]  $; $n$ is the
number of degrees of freedom of the system (the total number of coordinates,
$3N$, minus the number of constraints), and the $q_{j}$'s are called the
\textit{generalized coordinates} of the system, which are compatible with the
constraints and related with the usual coordinates of the particles by some
smooth functions $\mathbf{q}\left(  \mathbf{r}_{j}\right)  \Leftrightarrow
\left\{  \mathbf{r}_{j}\left(  \mathbf{q}\right)  \right\}  _{j=1,...,N}$. In
the following, we will call \textit{coordinate space} to the space formed by
the generalized coordinates, and $\mathbf{q}\left(  t\right)  $ a
\textit{trajectory} on it.

The basic object in analytical mechanics is the \textit{Lagrangian}, $L\left[
\mathbf{q}\left(  t\right)  ,\mathbf{\dot{q}}\left(  t\right)  ,t\right]  $,
which is a function of the generalized coordinates and velocities, and can
even have some explicit time dependence. In general, the Lagrangian must be
built based on general principles like symmetries; however, if the forces
acting on the particles of the system are conservative, that is,
$\mathbf{F}_{j}=\boldsymbol{\nabla}_{j}V[\{\mathbf{r}_{l}\}_{l=1,...,N}%
]=\left(  \partial_{x_{j}}V,\partial_{y_{j}}V,\partial_{z_{j}}V\right)  $ for
some \textit{potential }$V[\{\mathbf{r}_{l}\}_{l=1,...,N}]$, it takes the
simple form $L=T\left(  \mathbf{\dot{q}},\mathbf{q}\right)  -V\left(
\mathbf{q}\right)  $, being $T\left(  \mathbf{\dot{q}},\mathbf{q}\right)
=\sum_{j=1}^{N}m_{j}\mathbf{\dot{r}}_{j}^{2}\left(  \mathbf{q}\right)  /2$ the
kinetic energy of the system and $V\left(  \mathbf{q}\right)  =V[\{\mathbf{r}%
_{j}\left(  \mathbf{q}\right)  \}_{j=1,...,n}]$. The dynamic equations of the
system are then formulated as a \textit{variational principle} on the
\textit{action}
\begin{equation}
S=\int_{t_{1}}^{t_{2}}dtL\left[  \mathbf{q}\left(  t\right)  ,\mathbf{\dot{q}%
}\left(  t\right)  ,t\right]  ,
\end{equation}
by asking the trajectory of the system $\mathbf{q}\left(  t\right)  $ between
two fixed points $\mathbf{q}\left(  t_{1}\right)  $ and $\mathbf{q}\left(
t_{2}\right)  $ to be such that the action is an extremal, $\delta S=0$. From
this principle, it is straightforward to arrive to the well known
Euler--Lagrange equations%
\begin{equation}
\frac{\partial L}{\partial q_{j}}-\frac{d}{dt}\frac{\partial L}{\partial
\dot{q}_{j}}=0, \label{Euler-Lagrange}%
\end{equation}
which are a set of second order differential equations for the generalized
coordinates $\mathbf{q}$, and together with the conditions $\mathbf{q}\left(
t_{1}\right)  $ and $\mathbf{q}\left(  t_{2}\right)  $ provide the trajectory
$\mathbf{q}\left(  t\right)  $.

\subsection{The Hamiltonian formalism}

As we have seen, the Euler--Lagrange equations are a set of second order
differential equations which allow us to find the trajectory $\mathbf{q}%
\left(  t\right)  $ on coordinate space. We could reduce the order of the
differential equations by taking the velocities $\mathbf{\dot{q}}$ as
dynamical variables, arriving then to a set of $2n$ first order differential
equations. This is, however, a very na\"{\i}ve way of reducing the order,
which leads to a non-symmetric system of equations for $\mathbf{q}$ and
$\mathbf{\dot{q}}$. In this section we review Hamilton's approach to
analytical mechanics, which leads to a symmetric-like first order system of
equations, and is of major importance to understand the transition from
classical to quantum mechanics.

Instead of using the velocities, the Hamiltonian formalism considers the
\textit{generalized momenta}%
\begin{equation}
p_{j}=\frac{\partial L}{\partial\dot{q}_{j}},
\end{equation}
as the dynamical variables. Note that this definition establishes a relation
between these generalized momenta and the velocities $\mathbf{\dot{q}}\left(
\mathbf{q},\mathbf{p}\right)  \Leftrightarrow\mathbf{p}\left(  \mathbf{q}%
,\mathbf{\dot{q}}\right)  $. Note also that when the usual Cartesian
coordinates of the system's particles are taken as the generalized coordinates
these momenta coincide with those of Newton's approach.

The theory is then built in terms of a new object called the
\textit{Hamiltonian}, which is defined as a Legendre transform of the
Lagrangian,%
\begin{equation}
H\left(  \mathbf{q},\mathbf{p}\right)  =\mathbf{p\dot{q}}\left(
\mathbf{q},\mathbf{p}\right)  -L\left[  \mathbf{q},\mathbf{\dot{q}}\left(
\mathbf{q},\mathbf{p}\right)  ,t\right]  ,
\end{equation}
and coincides with the total energy for conservative systems, that is,
$H\left(  \mathbf{q},\mathbf{p}\right)  =T\left(  \mathbf{q},\mathbf{p}%
\right)  +V\left(  \mathbf{q}\right)  $, with $T\left(  \mathbf{q}%
,\mathbf{p}\right)  =T\left[  \mathbf{q},\mathbf{\dot{q}}\left(
\mathbf{q},\mathbf{p}\right)  \right]  $. Differentiating this expression and
using the Euler--Lagrange equations (or using again the variational principle
on the action), it is then straightforward to obtain the equations of motion
for the generalized coordinates and momenta (the \textit{canonical
equations}),%
\begin{equation}
\dot{q}_{j}=\frac{\partial H}{\partial p_{j}}\text{ \ \ \ \ and \ \ \ \ }%
\dot{p}_{j}=-\frac{\partial H}{\partial q_{j}}, \label{CanEqs}%
\end{equation}
which together with some initial conditions $\left\{  \mathbf{q}\left(
t_{0}\right)  ,\mathbf{p}\left(  t_{0}\right)  \right\}  $ allow us to find
the trajectory $\left\{  \mathbf{q}\left(  t\right)  ,\mathbf{p}\left(
t\right)  \right\}  $ in the space formed by the generalized coordinates and
momenta, which we will call \textit{phase space}.

Another important object in the Hamiltonian formalism is the \textit{Poisson
bracket}; given two functions of the coordinates and momenta $F\left(
\mathbf{q},\mathbf{p}\right)  $ and $G\left(  \mathbf{q},\mathbf{p}\right)  $,
their Poisson bracket is defined as%
\begin{equation}
\left\{  F,G\right\}  =\sum_{j=1}^{N}\frac{\partial F}{\partial q_{j}}%
\frac{\partial G}{\partial p_{j}}-\frac{\partial F}{\partial p_{j}}%
\frac{\partial G}{\partial q_{j}}. \label{PB}%
\end{equation}
The importance of this object is reflected in the fact that the evolution
equation of any quantity $g\left(  \mathbf{q},\mathbf{p},t\right)  $ can be
written as%
\begin{equation}
\frac{dg}{dt}=\left\{  g,H\right\}  +\frac{\partial g}{\partial t},
\label{HamObsEvo}%
\end{equation}
and hence, if the quantity doesn't depend on time and \textit{commutes} with
the Hamiltonian, that is, its Poisson bracket with the Hamiltonian is zero, it
is a \textit{constant of motion}.

Of particular importance for the transition to quantum mechanics are the
\textit{fundamental Poisson brackets}, that is, the Poisson brackets of the
coordinates and momenta,%
\begin{equation}
\left\{  q_{j},p_{l}\right\}  =\delta_{jl}\text{, }\left\{  q_{j}%
,q_{l}\right\}  =\left\{  p_{j},p_{l}\right\}  =0. \label{funPB}%
\end{equation}

\subsection{Observables and their mathematical structure}

In this last section concerning classical mechanics, we would like to explain
the mathematical structure in which observables are embedded within the
Hamiltonian formalism. We will show that the mathematical objects
corresponding to physical observables form a well defined mathematical
structure, a real\ Lie algebra. Moreover, the position and momentum will be
shown to be the generators of a particular Lie group, the Heisenberg group.
Understanding this internal structure of \textit{classical observables} will
give us the chance to introduce the quantum description of observables in a
reasonable way. Let us start by defining the concept of Lie algebra.

A \textit{real} \textit{Lie algebra} is a\ real vector space\footnote{The
concept of complex vector space is defined in the next section; the definition
of a real vector space is the same, but changing complex numbers by real
numbers.} $\mathcal{L}$ equipped with an additional operation, the \textit{Lie
product}, which takes two vectors $f$ and $g$ from $\mathcal{L}$, to generate
another vector also in $\mathcal{L}$ denoted by $\left\{  f,g\right\}  $; this
operation must satisfy the following properties:

\begin{enumerate}
\item $\left\{  f,g+h\right\}  =\left\{  f,g\right\}  +\left\{  f,h\right\}  $ (linearity)

\item $\left\{  f,f\right\}  =0\overset{\mathrm{together}\text{ }%
\mathrm{with}\text{ }1}{\Longrightarrow}\left\{  f,g\right\}  =-\left\{
g,f\right\}  $ (anticommutativity)

\item $\left\{  f,\left\{  g,h\right\}  \right\}  +\left\{  g,\left\{
h,f\right\}  \right\}  +\left\{  h,\left\{  f,g\right\}  \right\}  =0$ (Jacobi identity)
\end{enumerate}

Hence, in essence a real Lie algebra is basically a vector space equipped with a linear,
non-commutative, non-associative product. They have been subject of study for
many years, and now we know a lot about the properties of these mathematical
structures. They appear in many branches of physics and geometry, specially
connected to continuous symmetry transformations, whose associated
mathematical structures are actually called \textit{Lie groups}. In
particular, it is possible to show that given any Lie group with $p$
parameters (like, e.g., the three--parameter groups of translations or
rotations in real space), any transformation onto the system in which it is
acting can be generated from a set of $p$ elements of a Lie algebra $\left\{
g_{1},g_{2},...,g_{p}\right\}  $, called the \textit{generators }of the Lie
group, which satisfy some particular relations%
\begin{equation}
\left\{  g_{j},g_{k}\right\}  =\sum_{l=1}^{p}c_{jkl}g_{l}\text{;}%
\end{equation}
these relations are called the \textit{algebra--group relations}, and the
\textit{structure constants} $c_{jkl}$ are characteristic of the particular
Lie group (for example, the generators of translations and rotations in real
space are the momenta and angular momenta, respectively, and the corresponding
structure constants are $c_{jkl}=0$ for the translation group and
$c_{jkl}=\allowbreak\mathrm{i}\epsilon_{jkl}$ for the rotation
group\footnote{$\epsilon_{jkl}$ is the Levi-Civita symbol, which has
$\epsilon_{123}=\epsilon_{312}=\epsilon_{231}=1$ and is completely
antisymmetric, that is, changes its sign after permutation of any pair of
indices.}).

Coming back to the Hamiltonian formalism, we start by noting that
\textit{observables}, being \textit{measurable} quantities, must be given by
continuous, real functions in phase space; hence they form a real vector space
with respect to the usual addition of functions and multiplication of a
function by a real number. It also appeared naturally in the formalism a
linear, non-commutative, non-associative operation between phase space
functions, the Poisson bracket, which applied to real functions gives another
real function. It is easy to see that the Poisson bracket satisfies all the
requirements of a Lie product, and hence, observables form a Lie algebra
within the Hamiltonian formalism.

Moreover, the fundamental Poisson brackets (\ref{funPB}) show that the
generalized coordinates $\mathbf{q}$ and momenta $\mathbf{p}$, together with
the identity in phase space, satisfy particular algebra--group relations,
namely\footnote{Ordering the generators as $\{\mathbf{q},\mathbf{p},1\}$, the
structure constants associated to this algebra--group relations are explicitly%
\begin{equation}
c_{jkl}=\left\{
\begin{array}
[c]{cc}%
\Omega_{jk}\delta_{l,2n+1} & j,k=1,2,...,2n\\
0 & j=2n+1\text{ or }k=2n+1
\end{array}
\right.  \text{, }%
\end{equation}
being $\Omega=\left(
\begin{array}
[c]{cc}%
0_{n\times n} & I_{n\times n}\\
-I_{n\times n} & 0_{n\times n}%
\end{array}
\right)  $, with $I_{n\times n}$ and $0_{n\times n}$ the $n\times n$ identity
and null matrices, respectively.} $\left\{  q_{j},p_{k}\right\}  =\delta
_{jk}1$ and $\left\{  q_{j},1\right\}  =\left\{  p_{j},1\right\}  =0$, and
hence can be seen as the generators of a Lie group; this group is known as the
\textit{Heisenberg group}\footnote{This group was introduced by Weyl when
trying to prove the equivalence between the Schr\"{o}dinger and Heisenberg
pictures of quantum mechanics. It was later shown to have connections with the
symplectic group, which is the basis of many physical theories.\newline We
could have taken the Poisson brackets between the angular momenta associated
to the possible rotations in the system of particles (which are certainly far
more intuitive transformations than the one related to the Heisenberg group)
as the fundamental ones; however, we have chosen the Lie algebra associated to
the Heisenberg group just because it deals directly with position and momenta,
allowing for a simpler connection to quantum mechanics.}.

Therefore, we arrive to the main result of this section:

$\bigskip$

$\lceil$The mathematical framework of Hamiltonian mechanics associates
physical observables with elements of a Lie algebra, being the phase space
coordinates themselves the generators of the Heisenberg group.$\rfloor$

\bigskip

Maintaining this structure for observables will help to develop the laws of
quantum mechanics.

\section[The mathematical language of quantum mechanics: Hilbert spaces]{The mathematical language of quantum mechanics: Hilbert spaces}

Just as classical mechanics is formulated in terms of the mathematical
language of differential calculus and its extensions, quantum mechanics takes
linear algebra (and Hilbert spaces in particular) as its fundamental grammar.
In this section we introduce the concept of Hilbert space, and discuss the
properties of some operators which will play important roles in the formalism
of quantum mechanics.

\subsection{Finite--dimensional Hilbert spaces}

In essence, a Hilbert space is a \textit{complex vector space} in which an
\textit{inner product} is defined. Let us define first these terms as will be
used in this thesis.

A \textit{complex vector space} is a set $\mathcal{V}$, whose elements will be
called \textit{vectors} or \textit{kets} and will be denoted by $\left\{
\left\vert a\right\rangle ,\left\vert b\right\rangle ,\left\vert
c\right\rangle ,...\right\}  $ ($a$, $b$, and $c$ may correspond to any
suitable label), in which the following two operations are defined: the
\textit{vector addition}, which takes two vectors $\left\vert a\right\rangle $
and $\left\vert b\right\rangle $ and creates a new vector inside $\mathcal{V}$
denoted by $\left\vert a\right\rangle +\left\vert b\right\rangle $; and the
\textit{multiplication by a scalar}, which takes a complex number $\alpha\in%
\mathbb{C}
$ (in this section Greek letters will represent complex numbers) and a vector
$\left\vert a\right\rangle $ to generate a new vector in $\mathcal{V}$ denoted
by $\alpha\left\vert a\right\rangle $.

The following additional properties must be satisfied:

\begin{enumerate}
\item The vector addition is commutative and associative, that is, $\left\vert
a\right\rangle +\left\vert b\right\rangle =\left\vert b\right\rangle
+\left\vert a\right\rangle $ and $\left(  \left\vert a\right\rangle
+\left\vert b\right\rangle \right)  +\left\vert c\right\rangle =\left\vert
a\right\rangle +\left(  \left\vert b\right\rangle +\left\vert c\right\rangle
\right)  $.

\item There exists a null vector $\left\vert null\right\rangle $ such that
$\left\vert a\right\rangle +\left\vert null\right\rangle =\left\vert
a\right\rangle $.

\item $\alpha\left(  \left\vert a\right\rangle +\left\vert b\right\rangle
\right)  =\alpha\left\vert a\right\rangle +\alpha\left\vert b\right\rangle .$

\item $\left(  \alpha+\beta\right)  \left\vert a\right\rangle =\alpha
\left\vert a\right\rangle +\beta\left\vert a\right\rangle .$

\item $\left(  \alpha\beta\right)  \left\vert a\right\rangle =\alpha\left(
\beta\left\vert a\right\rangle \right)  .$

\item $1\left\vert a\right\rangle =\left\vert a\right\rangle .$
\end{enumerate}

From these properties it can be proved that the null vector is unique, and can
be built from any vector $\left\vert a\right\rangle $ as $0\left\vert
a\right\rangle $; hence, in the following we denote it simply by $\left\vert
null\right\rangle \equiv0$. It can also be proved that any vector $\left\vert
a\right\rangle $ has a unique \textit{antivector} $\left\vert -a\right\rangle
$ such that $\left\vert a\right\rangle +\left\vert -a\right\rangle =0$, which
is given by $\left(  -1\right)  \left\vert a\right\rangle $ or simply
$-\left\vert a\right\rangle $.

An \textit{inner product} is an additional operation defined in the complex
vector space $\mathcal{V}$, which takes two vectors $\left\vert a\right\rangle
$ and $\left\vert b\right\rangle $ and associates them a complex number. It
will be denoted by $\langle a|b\rangle$ or sometimes also by $\left(
\left\vert a\right\rangle ,\left\vert b\right\rangle \right)  $, and must
satisfy the following properties:

\begin{enumerate}
\item $\langle a|a\rangle>0$ if $\left\vert a\right\rangle \neq0.$

\item $\langle a|b\rangle=\langle b|a\rangle^{\ast}.$

\item $\left(  \left\vert a\right\rangle ,\alpha\left\vert b\right\rangle
\right)  =\alpha\langle a|b\rangle.$

\item $\left(  \left\vert a\right\rangle ,\left\vert b\right\rangle
+\left\vert c\right\rangle \right)  =\langle a|b\rangle+\langle a|c\rangle.$
\end{enumerate}

The following additional properties can be proved from these ones:

\begin{itemize}
\item $\left\langle null\right.  \left\vert null\right\rangle =0.$

\item $\left(  \alpha\left\vert a\right\rangle ,\left\vert b\right\rangle
\right)  =\alpha^{\ast}\langle a|b\rangle.$

\item $\left(  \left\vert a\right\rangle +\left\vert b\right\rangle
,\left\vert c\right\rangle \right)  =\langle a|c\rangle+\langle b|c\rangle.$

\item $\left\vert \langle a|b\rangle\right\vert ^{2}\leq\langle a|a\rangle
\langle b|b\rangle$
\end{itemize}

Note that for any vector $\left\vert a\right\rangle $, one can define the
object $\left\langle a\right\vert \equiv\left(  \left\vert a\right\rangle
,\cdot\right)  $, which will be called a \textit{dual vector} or a
\textit{bra}, and which takes a vector $\left\vert b\right\rangle $ to
generate the complex number $\left(  \left\vert a\right\rangle ,\left\vert
b\right\rangle \right)  \in%
\mathbb{C}
$. It can be proved that the set formed by all the dual vectors corresponding
to the elements in $\mathcal{V}$ is also a vector space, which will be called
the \textit{dual space} and will be denoted by $\mathcal{V}^{+}$. Within this
picture, the inner product can be seen as an operation which takes a bra
$\left\langle a\right\vert $ and a ket $\left\vert b\right\rangle $ to
generate the complex number $\langle a|b\rangle$, a \textit{bracket}. This
whole \textit{bra-c-ket} notation is due to Dirac.

In the following we assume that any time a bra $\left\langle a\right\vert $ is
applied to a ket $\left\vert b\right\rangle $, the complex number $\langle
a|b\rangle$ is formed, so that objects like $|b\rangle\langle a|$ generate
kets when applied to kets from the left, $\left(  |b\rangle\langle a|\right)
|c\rangle=\left(  \langle a|c\rangle\right)  |b\rangle$, and bras when applied
to bras from the right, $\langle c|\left(  |b\rangle\langle a|\right)
=\left(  \langle c|b\rangle\right)  \langle a|$. Technically, $|b\rangle
\langle a|$ is called an \textit{outer product}.

A vector space equipped with an inner product is called an \textit{Euclidean
space}. In the following we give some important definitions and properties
which are needed in order to understand the concept of Hilbert space:

\begin{itemize}
\item The vectors $\left\{  \left\vert a_{1}\right\rangle ,\left\vert
a_{2}\right\rangle ,...,\left\vert a_{m}\right\rangle \right\}  $ are said to
be \textit{linearly independent} if the relation $\alpha_{1}\left\vert
a_{1}\right\rangle +\alpha_{2}\left\vert a_{2}\right\rangle +...+\alpha
_{m}\left\vert a_{m}\right\rangle =0$ is satisfied only for $\alpha_{1}%
=\alpha_{2}=...=\alpha_{m}=0$, as otherwise one of them can be written as a
linear combination of the rest.

\item The \textit{dimension} of the vector space is defined as the maximum
number of linearly independent vectors, and can be finite or infinite.

\item If the dimension of an Euclidean space is $d<\infty$, it is always
possible to build a set of $d$ orthonormal vectors $E=\left\{  \left\vert
e_{j}\right\rangle \right\}  _{j=1,2,..,d}$ satisfying $\langle e_{j}%
|e_{l}\rangle=\delta_{jl}$, such that any other vector $\left\vert
a\right\rangle $ can be written as a linear superposition of them, that is,
$\left\vert a\right\rangle =\sum_{j=1}^{d}a_{j}\left\vert e_{j}\right\rangle
$, being the $a_{j}$'s some complex numbers. This set is called an
\textit{orthonormal basis} of the Euclidean space $\mathcal{V}$, and the
coefficients $a_{j}$ of the expansion can be found as $a_{j}=\langle
e_{j}|a\rangle$. The column formed with the expansion coefficients, which is
denoted by $\operatorname{col}\left(  a_{1},a_{2},...,a_{d}\right)  $, is
called a \textit{representation} of the vector $\left\vert a\right\rangle $ in
the basis $E$.

Note that the set $E^{+}=\left\{  \left\langle e_{j}\right\vert \right\}
_{j=1,2,..,d}$ is an orthonormal basis in the dual space $\mathcal{V}^{+}$, so
that any bra $\left\langle a\right\vert $ can be expanded then as
$\left\langle a\right\vert =\sum_{j=1}^{d}a_{j}^{\ast}\left\langle
e_{j}\right\vert $. The representation of the bra $\left\langle a\right\vert $
in the basis $E$ corresponds to the row formed by its expansion coefficients,
and is denoted by $\left(  a_{1}^{\ast},a_{2}^{\ast},...,a_{n}^{\ast}\right)
$. Note that if the representation of $\left\vert a\right\rangle $ is seen as
a $d\times1$ matrix, the representation of $\left\langle a\right\vert $ can be
obtained as its $1\times d$ conjugate--transpose matrix.

Note finally that the inner product of two vectors $\left\vert a\right\rangle
$ and $\left\vert b\right\rangle $ reads $\langle a|b\rangle=\sum_{j=1}%
^{d}a_{j}^{\ast}b_{j}$ when represented in the same basis, which is the matrix
product of the representations of $\left\langle a\right\vert $ and $\left\vert
b\right\rangle $.
\end{itemize}

For finite dimension, an Euclidean space is a \textit{Hilbert space}. However,
in most applications of quantum mechanics (and certainly in quantum optics),
one has to deal with infinite--dimensional vector spaces. We will treat them
after the following section.

\subsection{Linear operators in finite--dimensional Hilbert
spaces\label{LinearOperators}}

We now discuss the concept of linear operator, as well as analyze the
properties of some important classes of operators. Only finite--dimensional
Hilbert spaces are considered in this section, we will generalize the
discussion to infinite--dimensional Hilbert spaces in the next section.

We are interested in maps $\hat{L}$ (operators will be denoted with
`\symbol{94}' throughout the thesis) which associate to any vector $\left\vert
a\right\rangle $ of a Hilbert space $\mathcal{H}$ another vector denoted by
$\hat{L}\left\vert a\right\rangle $ in the same Hilbert space. If the map
satisfies%
\begin{equation}
\hat{L}\left(  \alpha\left\vert a\right\rangle +\beta\left\vert b\right\rangle
\right)  =\alpha\hat{L}\left\vert a\right\rangle +\beta\hat{L}\left\vert
b\right\rangle ,
\end{equation}
then it is called a \textit{linear operator}. For our purposes this is the
only class of interesting operators, and hence we will simply call them
\textit{operators} in the following.

Before discussing the properties of some important classes of operators, we
need some definitions:

\begin{itemize}
\item Given an orthonormal basis $E=\left\{  \left\vert e_{j}\right\rangle
\right\}  _{j=1,2,..,d}$ in a Hilbert space $\mathcal{H}$ with dimension
$d<\infty$,\ any operator $\hat{L}$ has a representation; while bras and kets
are represented by $d\times1$ and $1\times d$ matrices (rows and columns),
respectively, an operator $\hat{L}$ is represented by a $d\times d$ matrix
with \textit{elements} $L_{jl}=(\left\vert e_{j}\right\rangle ,\hat
{L}\left\vert e_{l}\right\rangle )\equiv\left\langle e_{j}\right\vert \hat
{L}\left\vert e_{l}\right\rangle $. An operator $\hat{L}$ can then be expanded
in terms of the basis $E$ as $\hat{L}=\sum_{j,l=1}^{d}L_{jl}\left\vert
e_{j}\right\rangle \left\langle e_{l}\right\vert $. It follows that the
representation of the vector $\left\vert b\right\rangle =\hat{L}\left\vert
a\right\rangle $ is just the matrix multiplication of the representation of
$\hat{L}$ by the representation of $\left\vert a\right\rangle $, that is,
$b_{j}=\sum_{l=1}^{d}L_{jl}a_{l}$.

\item The\textit{ addition} and \textit{product }of two operators $\hat{L}$
and $\hat{K}$, denoted by $\hat{L}+\hat{K}$ and $\hat{L}\hat{K}$,
respectively, are defined by their action onto any vector $\left\vert
a\right\rangle $: $(\hat{L}+\hat{K})\left\vert a\right\rangle =\hat
{L}\left\vert a\right\rangle +\hat{K}\left\vert a\right\rangle $ and $\hat
{L}\hat{K}\left\vert a\right\rangle =\hat{L}(\hat{K}\left\vert a\right\rangle
)$. It follows that the representation of the addition and the product are,
respectively, the sum and the multiplication of the corresponding matrices,
that is, $(\hat{L}+\hat{K})_{jl}=L_{jl}+K_{jl}$ and $(\hat{L}\hat{K}%
)_{jl}=\sum_{k=1}^{d}L_{jk}K_{kl}$.

\item Note that while the addition is commutative, the product is not in
general. This leads us to the notion of \textit{commutator}, defined for two
operators $\hat{L}$ and $\hat{K}$ as $[\hat{L},\hat{K}]=\hat{L}\hat{K}-\hat
{K}\hat{L}$. When $[\hat{L},\hat{K}]=0$, we say that the operators
\textit{commute}.

\item Given an operator $\hat{L}$, its \textit{trace} is defined as the sum of
the diagonal elements of its matrix representation, that is, $\mathrm{tr}%
\{\hat{L}\}=\sum_{j=1}^{d}L_{jj}$. It may seem that this definition is
basis--dependent, as in general the elements $L_{jj}$ are different in
different bases. However, we will see that the trace is invariant under any
change of basis.

The trace has two important properties. It is \textit{linear} and
\textit{cyclic}, that is, given two operators $\hat{L}$ and $\hat{K}$,
$\mathrm{tr}\{\hat{L}+\hat{K}\}=\mathrm{tr}\{\hat{L}\}+\mathrm{tr}\{\hat{K}\}$
and $\mathrm{tr}\{\hat{L}\hat{K}\}=\mathrm{tr}\{\hat{K}\hat{L}\}$, as is
trivially proved.

\item We say that a vector $\left\vert l\right\rangle $ is an
\textit{eigenvector} of an operator $\hat{L}$ if $\hat{L}\left\vert
l\right\rangle =\lambda\left\vert l\right\rangle $; $\lambda\in%
\mathbb{C}
$ is called its associated \textit{eigenvalue}. The set of all the eigenvalues
of an operator is called its \textit{spectrum}.
\end{itemize}

\bigskip

We can pass now to describe some classes of operators which play important
roles in quantum mechanics.

\bigskip

\textbf{The identity operator.} The \textit{identity operator}, denoted by
$\hat{I}$, is defined as the operator which maps any vector onto itself. Its
representation in any basis is then $I_{jl}=\delta_{jl}$, so that it can
expanded as%
\begin{equation}
\hat{I}=\sum_{j=1}^{d}\left\vert e_{j}\right\rangle \left\langle
e_{j}\right\vert \text{.}%
\end{equation}
This expression is known as the \textit{completeness relation} of the basis
$E$; alternatively, it is said that the set $E$ forms a \textit{resolution of
the identity}.

Note that the expansion of a vector $\left\vert a\right\rangle $ and its dual
$\left\langle a\right\vert $ in the basis $E$ is obtained just by application
of the completeness relation from the left and the right, respectively.
Similarly, the expansion of an operator $\hat{L}$ is obtained by application
of the completeness relation both from the right and the left at the same time.

\bigskip

\textbf{The inverse of an operator.} The \textit{inverse }of an operator
$\hat{L}$, denoted by $\hat{L}^{-1}$, is defined as that satisfying $\hat
{L}^{-1}\hat{L}=\hat{L}\hat{L}^{-1}=\hat{I}$.

\bigskip

\textbf{An operator function}. Consider a real function $f\left(  x\right)  $
which can be expanded in powers of $x$ as $f\left(  x\right)  =\sum
_{m=0}^{\infty}f_{m}x^{m}$; given an operator $\hat{L}$, we define the
\textit{operator} \textit{function} $\hat{f}(\hat{L})=\sum_{m=0}^{\infty}%
f_{m}\hat{L}^{m}$, where $\hat{L}^{m}$ means the product of $\hat{L}$ with
itself $m$ times.

\bigskip

\textbf{The adjoint of an operator.} Given an operator $\hat{L}$, we define
its \textit{adjoint}, and denote it by $\hat{L}^{\dagger}$, as that satisfying
$(\left\vert a\right\rangle ,\hat{L}\left\vert b\right\rangle )=(\hat
{L}^{\dagger}\left\vert a\right\rangle ,\left\vert b\right\rangle )$ for any
two vectors $\left\vert a\right\rangle $ and $\left\vert b\right\rangle $.
Note that the representation of $\hat{L}^{\dagger}$ corresponds to the
conjugate transpose of the matrix representing $\hat{L}$, that is $(\hat
{L}^{\dagger})_{jl}=L_{lj}^{\ast}$. Note also that the adjoint of a product of
two operators $\hat{K}$ and $\hat{L}$ is given by $(\hat{K}\hat{L})^{\dagger
}=\hat{L}^{\dagger}\hat{K}^{\dagger}$.

\bigskip

\textbf{Self--adjoint operators.} We say that $\hat{H}$ is a
\textit{self--adjoint} if it coincides with its adjoint, that is, $\hat
{H}=\hat{H}^{\dagger}$. A property which will be shown to be of major
importance for the construction of the laws of quantum mechanics is that the
spectrum $\left\{  h_{j}\right\}  _{j=1,2,...,d}$ of a self--adjoint operator
is real. Moreover, its associated eigenvectors\footnote{We will assume that
the spectrum of any operator is non-degenerate, that is, only one eigenvector
corresponds to a given eigenvalue, as all the operators that appear in this
thesis have this property.} $\left\{  \left\vert h_{j}\right\rangle \right\}
_{j=1,2,...,d}$ form an orthonormal basis of the Hilbert space.

The representation of any operator function $\hat{f}(\hat{H})$ in the
\textit{eigenbasis} of $\hat{H}$ is then $\left[  \hat{f}(\hat{H})\right]
_{jl}=f\left(  h_{j}\right)  \delta_{jl}$, from which follows%
\begin{equation}
\hat{f}(\hat{H})=\sum_{j=1}^{d}f\left(  h_{j}\right)  \left\vert
h_{j}\right\rangle \left\langle h_{j}\right\vert .
\end{equation}
This result is known as the \textit{spectral theorem}.

\bigskip

\textbf{Unitary operators.} We say that $\hat{U}$ is a \textit{unitary
operator} if $\hat{U}^{\dagger}=\hat{U}^{-1}$. The interest of this class of
operators is that they preserve inner products, that is, for any two vectors
$\left\vert a\right\rangle $ and $\left\vert b\right\rangle $ the inner
product $(\hat{U}\left\vert a\right\rangle ,\hat{U}\left\vert b\right\rangle
)$ coincides with $\langle a|b\rangle$. Moreover, it is possible to show that
given two orthonormal bases $E=\left\{  \left\vert e_{j}\right\rangle
\right\}  _{j=1,2,..,d}$ and $E^{\prime}=\{|e_{j}^{\prime}\rangle
\}_{j=1,2,..,d}$, there exists a unique unitary matrix $\hat{U}$ which
connects them as $\{|e_{j}^{\prime}\rangle=\hat{U}\left\vert e_{j}%
\right\rangle \}_{j=1,2,..,d}$, and then any basis of the Hilbert space is
unique up to a unitary transformation.

We can now prove that the trace of an operator is basis--independent. Let us
denote by $\mathrm{tr}\{\hat{L}\}_{E}$ the trace of an operator $\hat{L}$ in
the basis $E$; the trace of this operator in the transformed basis can be
written then as $\mathrm{tr}\{\hat{L}\}_{E^{\prime}}=\mathrm{tr}\{\hat
{U}^{\dagger}\hat{L}\hat{U}\}_{E}$, or using the cyclic property of the trace
and the unitarity of $\hat{U}$, $\mathrm{tr}\{\hat{L}\}_{E^{\prime}%
}=\mathrm{tr}\{\hat{U}\hat{U}^{\dagger}\hat{L}\}=\mathrm{tr}\{\hat{L}\}_{E}$,
which proves that the trace is equal in both bases.

Note finally that a unitary operator $\hat{U}$ can always be written as the
exponential of $\mathrm{i}$--times a self--adjoint operator $\hat{H}$, that is,
$\hat{U}=\exp(\mathrm{i}\hat{H})$.

\bigskip

\textbf{Projection operators.} In general, any self--adjoint operator $\hat
{P}$ satisfying $\hat{P}^{2}=\hat{P}$ is called a \textit{projector}. We are
interested only in those projectors which can be written as the outer product
of a vector $\left\vert a\right\rangle $ with itself, that is, $\hat{P}%
_{a}=\left\vert a\right\rangle \left\langle a\right\vert $; when applied to a
vector $\left\vert b\right\rangle $, this gets \textit{projected} along the
`direction' of $\left\vert a\right\rangle $ as $\hat{P}_{a}\left\vert
b\right\rangle =(\langle a|b\rangle)\left\vert a\right\rangle $.

Note that given an orthonormal basis $E$, we can use the projectors $\hat
{P}_{j}=\left\vert e_{j}\right\rangle \left\langle e_{j}\right\vert $ to
extract the components of a vector $\left\vert c\right\rangle $ as $\hat
{P}_{j}\left\vert c\right\rangle =c_{j}\left\vert e_{j}\right\rangle $. Note
also that the completeness and orthonormality of the basis $E$ implies that
$\sum_{j=1}^{d}\hat{P}_{j}=\hat{I}$ and $\hat{P}_{j}\hat{P}_{l}=\delta
_{jl}\hat{P}_j$, respectively.

\bigskip

\textbf{Density operators.} A self--adjoint operator $\hat{\rho}$ is called a
\textit{density operator} if it is \textit{positive semidefinite}, that is
$\left\langle a\right\vert \hat{\rho}\left\vert a\right\rangle \geq0$ for any
vector $\left\vert a\right\rangle $, and has unit trace.

The interesting property of density operators is that they `hide' probability
distributions in the diagonal of its representation. To see this just note
that given an orthonormal basis $E$, the self--adjointness and positivity of
$\hat{\rho}$ ensure that all its diagonal elements $\left\{  \rho
_{jj}\right\}  _{j=1,2,...,d}\ $are either positive or zero, that is,
$\rho_{jj}\geq0$ $\forall j$, while the unit trace makes them satisfy
$\sum_{j=1}^{d}\rho_{jj}=1$. Hence, the diagonal elements of a density
operator have all the properties required by a \textit{probability
distribution}.

It is possible to show that a density operator can always be expressed as a
\textit{statistical} or \textit{convex mixture} of projection operators, that
is, $\hat{\rho}=\sum_{k=1}^{M}w_{k}\left\vert a_{k}\right\rangle \left\langle
a_{k}\right\vert $, where $\sum_{k=1}^{M}w_{k}=1$ and the vectors $\left\{
\left\vert a_{k}\right\rangle \right\}  _{k=1}^{M}$ are normalized to one, but
don't need not to be orthogonal (note that in fact $M$ doesn't need to be
equal to $d$). Hence, another way of specifying a density matrix is by a set
of normalized vectors together with some statistical rule for mixing them.
When only one vector $\left\vert a\right\rangle $ contributes to the mixture,
$\hat{\rho}=|a\rangle\langle a|$ is completely specified by just this single
vector, and we say that the density operator is \textit{pure}; otherwise, we
say that it is \textit{mixed}.

\subsection{Generalization to infinite dimensions\label{InfiniteHilbert}}

As shown in the thesis (see Chapters \ref{HarmonicOscillator} and
\ref{Quantization}), the natural Euclidean space for quantum optics is
infinite--dimensional. Unfortunately, not all the previous concepts and
objects that we have introduced for the finite--dimensional case are trivially
generalized to infinite dimensions; in this section we discuss this generalization.

The first problem that we meet when dealing with infinite--dimensional
Euclidean spaces is that the existence of a basis $\left\{  \left\vert
e_{j}\right\rangle \right\}  _{j=1,2,...}$ in which any other vector can be
represented as $\left\vert a\right\rangle =\sum_{j=1}^{\infty}a_{j}\left\vert
e_{j}\right\rangle $ is not granted. The class of infinite--dimensional
Euclidean spaces in which these infinite but countable bases exist are called
\textit{Hilbert spaces}, and are the ones that will be appearing in quantum mechanics.

The conditions which ensure that an infinite--dimensional Euclidean space is
indeed a Hilbert space can be found in, for example, reference
\cite{Prugovecky71book}. Here we just want to stress that quite intuitively,
any infinite--dimensional Hilbert space\footnote{An example of
infinite--dimensional complex Hilbert space consists in the vector space
formed by the complex functions of real variable, say $|f\rangle=f\left(
x\right)  $ with $x\in%
\mathbb{R}
$, with integrable square, that is%
\begin{equation}
\int_{%
\mathbb{R}
}dx|f\left(  x\right)  |^{2}<\infty,
\end{equation}
with the inner product%
\begin{equation}
\langle g|f\rangle=\int_{%
\mathbb{R}
}dxg^{\ast}\left(  x\right)  f\left(  x\right)  \text{.}%
\end{equation}
This Hilbert space is known as the $\mathrm{L}^{2}\left(  x\right)  $ space.}
is \textit{isomorphic} to the space called $l^{2}\left(  \infty\right)  $,
which is formed by the column vectors $\left\vert a\right\rangle
\equiv\operatorname{col}(a_{1},a_{2},...)$ where the set $\{a_{j}\in%
\mathbb{C}
\}_{j=1,2,...}$ satisfies the restriction $\sum_{j=1}^{\infty}|a_{j}%
|^{2}<\infty$, and has the operations $\left\vert a\right\rangle +\left\vert
b\right\rangle =\operatorname{col}(a_{1}+b_{1},a_{2}+b_{2},...)$,
$\alpha\left\vert a\right\rangle =\operatorname{col}(\alpha a_{1},\alpha
a_{2},...)$, and $\langle a|b\rangle=\sum_{j=1}^{\infty}a_{j}^{\ast}b_{j}$.

Most of the previous definitions are directly generalized to Hilbert spaces by
taking $d\rightarrow\infty$ (dual space, representations, operators,...).
However, there is one crucial property of self--adjoint operators which
doesn't hold in this case: its eigenvectors may not form an orthonormal basis
of the Hilbert space. The remainder of this section is devoted to deal with
this problem.

Just as in finite dimension, given an infinite--dimensional Hilbert space
$\mathcal{H}$, we say that one of its vectors $|d\rangle$ is an eigenvector of
the self--adjoint operator $\hat{H}$ if $\hat{H}|d\rangle=\delta|d\rangle$,
where $\delta\in%
\mathbb{C}
$ is called its associated eigenvalue. Nevertheless, it can happen in
infinite--dimensional spaces that some vector $|c\rangle$ not contained in
$\mathcal{H}$ also satisfies the condition $\hat{H}|c\rangle=\chi|c\rangle$,
in which case we call it a \textit{generalized eigenvector}, being $\chi$ its
\textit{generalized eigenvalue}\footnote{In $\mathrm{L}^{2}\left(  x\right)  $
we have two simple examples of self--adjoint operators with eigenvectors not
contained in $\mathrm{L}^{2}\left(  x\right)  $: the so-called $\hat{X}$ and
$\hat{P}$, which, given an arbitrary vector $|f\rangle=f\left(  x\right)  $,
act as $\hat{X}|f\rangle=xf\left(  x\right)  $ and $\hat{P}|f\rangle
=-\mathrm{i}df/dx$, respectively. This is simple to see, as the equations%
\begin{equation}
xf_{X}\left(  x\right)  =Xf_{X}\left(  x\right)  \text{ and }-\mathrm{i}%
\frac{d}{dx}f_{P}\left(  x\right)  =Pf_{P}\left(  x\right)  ,
\end{equation}
have%
\begin{equation}
f_{X}\left(  x\right)  =\delta\left(  x-X\right)  \text{ and }f_{P}\left(
x\right)  =\exp\left(  \mathrm{i}Px\right)  \text{,}%
\end{equation}
as solutions, which are not square--integrable, and hence do not belong to
$\mathrm{L}^{2}\left(  x\right)  $.}. The set of all the eigenvalues of the
self--adjoint operator is called its \textit{discrete }(or \textit{point}%
)\textit{ spectrum} and is a countable set, while the set of all its
generalized eigenvalues is called its \textit{continuous spectrum} and is
uncountable, that is, forms a continuous set \cite{Prugovecky71book} (see also
\cite{Galindo90book}).

In quantum optics we find two extreme cases: either the observable, say
$\hat{H}$, has a pure discrete spectrum $\{h_{j}\}_{j=1,2,...}$; or the
observable, say $\hat{X}$, has a pure continuous spectrum $\{x\}_{x\in%
\mathbb{R}
}$. It can be shown that in the first case the eigenvectors of the observable
form an orthonormal basis of the Hilbert space, so that we can build a
resolution of the identity as $\hat{I}=\sum_{j=1}^{\infty}\left\vert
h_{j}\right\rangle \left\langle h_{j}\right\vert $, and proceed along the
lines of the previous sections.

In the second case, the set of generalized eigenvectors cannot form a basis of
the Hilbert space in the strict sense, as they do not form a countable set and
do not even belong to the Hilbert space. Fortunately, there are still ways to
treat the generalized eigenvectors of $\hat{X}$ `as if' they were a basis of
the Hilbert space. The idea was introduced by Dirac \cite{Dirac30book}, who
realized that normalizing the generalized eigenvectors as\footnote{This
$\delta\left(  x\right)  $ function the so-called \textit{Dirac--delta
distribution} which is defined by the conditions%
\begin{equation}
\int_{x_{1}}^{x_{2}}dx\delta\left(  x-y\right)  =\left\{
\begin{array}
[c]{cc}%
1 & \text{if }y\in\left[  x_{1},x_{2}\right] \\
0 & \text{if }y\notin\left[  x_{1},x_{2}\right]
\end{array}
\right.  .
\end{equation}
} $\langle x|y\rangle=\delta\left(  x-y\right)  $, one can define the
following integral operator%
\begin{equation}
\int_{%
\mathbb{R}
}dx|x\rangle\left\langle x\right\vert =\hat{I}_{\mathrm{c}},
\end{equation}
which acts as the identity onto the generalized eigenvectors, that is,
$\hat{I}_{\mathrm{c}}|x\rangle=|x\rangle$; it is then assumed that $\hat
{I}_{\mathrm{c}}$ coincides with the identity in $\mathcal{H}$, so that any
other vector $|a\rangle$ or operator $\hat{L}$ in the Hilbert space can be
expanded as%
\begin{equation}
|a\rangle=\int_{%
\mathbb{R}
}dxa\left(  x\right)  |x\rangle\text{ \ \ \ \ and \ \ \ }\hat{L}=\int_{%
\mathbb{R}
^{2}}dxdyL\left(  x,y\right)  |x\rangle\left\langle y\right\vert
\end{equation}
where the elements $a\left(  x\right)  =\langle x|a\rangle$ and $L\left(
x,y\right)  =\langle x|\hat{L}|y\rangle$ of this \textit{continuous
representations} form complex functions defined in $%
\mathbb{R}
$ and $%
\mathbb{R}
^{2}$, respectively. From now on, we will call \textit{continuous basis} to
the set $\left\{  |x\rangle\right\}  _{x\in%
\mathbb{R}
}$.

Dirac introduced this continuous representations as a `limit to the continuum'
of the countable case; even though this approach was very intuitive, it lacked
of mathematical rigor. Some decades after Dirac's proposal, Gel'fand showed
how to generalize the concept of Hilbert space to include these generalized
representations in full mathematical rigor \cite{Gelfand64book}. The
generalized spaces are called \textit{rigged Hilbert spaces} (in which the
algebra of Hilbert spaces joins forces with the theory of continuous
probability distributions), and working on them it is possible to show that
given any self--adjoint operator, one can use its eigenvectors and generalized
eigenvectors to expand any vector of the Hilbert space.

Note finally that given two vectors $|a\rangle$ and $|b\rangle$ of the Hilbert
space, and a continuous basis $\left\{  |x\rangle\right\}  _{x\in%
\mathbb{R}
}$, we can use their generalized representations to write their inner product
as%
\begin{equation}
\langle a|b\rangle=\int_{%
\mathbb{R}
}dxa^{\ast}\left(  x\right)  b\left(  x\right)  .
\end{equation}
It is also easily proved that the trace of any operator $\hat{L}$ can be
evaluated from its continuous representation on $\left\{  |x\rangle\right\}
_{x\in%
\mathbb{R}
}$ as%
\begin{equation}
\mathrm{tr}\{\hat{L}\}=\int_{%
\mathbb{R}
}dxL\left(  x,x\right)  .
\end{equation}
This has important consequences for the properties of density operators, say
$\hat{\rho}$ for the discussion which follows. We explained at the end of the
last section that when represented on an orthonormal basis of the Hilbert
space, its diagonal elements (which are real owed to its self--adjointness)
can be seen as a probability distribution, because they satisfy $\sum
_{j=1}^{\infty}\rho_{jj}=1$ and $\rho_{jj}\geq0$ $\forall$ $j$. Similarly,
because of its unit trace and positivity, the diagonal elements of its
continuous representation satisfy $\int_{%
\mathbb{R}
}dx\rho\left(  x,x\right)  =1$ and $\rho\left(  x,x\right)  \geq0$ $\forall$
$x$, and hence, the real function $\rho\left(  x,x\right)  $ can be seen as a
\textit{probability density function}.

\subsection{Composite Hilbert spaces}

In many moments of this thesis, we find the need associate a Hilbert space to
a composite system, the Hilbert spaces of whose parts we now. In this section
we show how to build a Hilbert space $\mathcal{H}$ starting from a set of
Hilbert spaces $\left\{  \mathcal{H}_{A},\mathcal{H}_{B},\mathcal{H}%
_{C}...\right\}  $.

Let us start with only two Hilbert spaces $\mathcal{H}_{A}$ and $\mathcal{H}%
_{B}$ with dimensions $d_{A}$ and $d_{B}$, respectively (which might be
infinite); the generalization to an arbitrary number of Hilbert spaces is
straightforward. Consider a vector space $\mathcal{V}$ with dimension
$\mathrm{dim}(\mathcal{V})=d_{A}\times d_{B}$. We define a map called the
\textit{tensor product} which associates to any pair of vectors $|a\rangle
\in\mathcal{H}_{A}$ and $|b\rangle\in\mathcal{H}_{B}$ a vector in
$\mathcal{V}$ which we denote by $|a\rangle\otimes|b\rangle\in\mathcal{V}$.
This tensor product must satisfy the following properties:

\begin{enumerate}
\item $\left(  |a\rangle+|b\rangle\right)  \otimes|c\rangle=|a\rangle
\otimes|c\rangle+|b\rangle\otimes|c\rangle.$

\item $|a\rangle\otimes\left(  |b\rangle+|c\rangle\right)  =|a\rangle
\otimes|b\rangle+|a\rangle\otimes|c\rangle.$

\item $\left(  \alpha|a\rangle\right)  \otimes|b\rangle=|a\rangle
\otimes\left(  \alpha|b\rangle\right)  .$
\end{enumerate}

If we endorse the vector space $\mathcal{V}$ with the inner product $\left(
|a\rangle\otimes|b\rangle,|c\rangle\otimes|d\rangle\right)  =\langle
a|c\rangle\langle b|d\rangle$, it is easy to show it becomes a Hilbert space,
which in the following will be denoted by $\mathcal{H}=\mathcal{H}_{A}%
\otimes\mathcal{H}_{B}$. Given the bases $E_{A}=\{|e_{j}^{A}\rangle
\}_{j=1,2,...,d_{A}}$ and $E_{B}=\{|e_{j}^{B}\rangle\}_{j=1,2,...,d_{B}}$ of
the Hilbert spaces $\mathcal{H}_{A}$ and $\mathcal{H}_{B}$, respectively, a
basis of the \textit{tensor product Hilbert space }$\mathcal{H}_{A}%
\otimes\mathcal{H}_{B}$ can be built as $E=E_{A}\otimes E_{B}=\{|e_{j}%
^{A}\rangle\otimes|e_{l}^{B}\rangle\}_{l=1,2,...,d_{B}}^{j=1,2,...,d_{A}}$
(note that the notation in the first equality is symbolical).

We will use a more economic notation for the tensor product, namely
$|a\rangle\otimes|b\rangle=|a,b\rangle$, except when the explicit tensor
product symbol is needed for any reason. With this notation the basis of the
tensor product Hilbert space is written as $E=\{|e_{j}^{A},e_{l}^{B}%
\rangle\}_{l=1,2,...,d_{B}}^{j=1,2,...,d_{A}}$.

The tensor product also maps operators acting on $\mathcal{H}_{A}$ and
$\mathcal{H}_{B}$ to operators acting on $\mathcal{H}$. Given two operators
$\hat{L}_{A}$ and $\hat{L}_{B}$ acting on $\mathcal{H}_{A}$ and $\mathcal{H}%
_{B}$, the \textit{tensor product operator} $\hat{L}=\hat{L}_{A}\otimes\hat
{L}_{B}$ is defined in $\mathcal{H}$ as that satisfying $\hat{L}%
|a,b\rangle=(\hat{L}_{A}|a\rangle)\otimes(\hat{L}_{B}|b\rangle)$ for any pair
of vectors $|a\rangle\in\mathcal{H}_{A}$ and $|b\rangle\in\mathcal{H}_{B}$.
When explicit subindices making reference to the Hilbert space on which
operators act on are used, so that there is no room for confusion, we will use
the shorter notations $\hat{L}_{A}\otimes\hat{L}_{B}=\hat{L}_{A}\hat{L}_{B}$,
$\hat{L}_{A}\otimes\hat{I}=\hat{L}_{A}$, and $\hat{I}\otimes\hat{L}_{B}%
=\hat{L}_{B}$.

Note that the tensor product preserves the properties of the operators; for
example, given two self--adjoint operators $\hat{H}_{A}$ and $\hat{H}_{B}$,
unitary operators $\hat{U}_{A}$ and $\hat{U}_{B}$, or density operators
$\hat{\rho}_{A}$ and $\hat{\rho}_{B}$, the operators $\hat{H}_{A}\otimes
\hat{H}_{B}$, $\hat{U}_{A}\otimes\hat{U}_{B}$, and $\hat{\rho}_{A}\otimes
\hat{\rho}_{B}$ are self--adjoint, unitary, and a density operator in
$\mathcal{H}$, respectively. Note that this doesn't mean that any
self--adjoint, unitary, or density operator acting on $\mathcal{H}$ can be
written in a simple tensor product form $\hat{L}_{A}\otimes\hat{L}_{B}$.

\section{The laws of quantum mechanics}

\subsection{A brief historical introduction}

By the end of the XIX century there was a great feeling of security among the
physics community: analytical mechanics (together with statistical mechanics)
and Maxwell's electromagnetism,\ (in the following \textit{classical physics}
altogether) seem to explain the whole range of physical phenomena that one
could observe, and hence, in a sense, the foundations of physics were
completed. There were, however, a couple of experimental observations which
lacked explanation within this `definitive' framework, which actually lead to
the construction of a whole new way of understanding physical phenomena:
quantum mechanics.

Among these experimental evidences, the shape of the spectrum of the radiation
emitted by a black body, the photoelectric effect which showed that only light
exceeding some frequency can release electrons from a metal irrespective of
its intensity, and the discrete set of spectral lines of hydrogen, were the
principal triggers of the revolution to come in the first quarter of the XX
century.\ The first two lead Planck and Einstein to suggest that
electromagnetic energy is not continuous but divided in small packages of
energy $\hbar\omega$ ($\omega$ being the frequency of the radiation), while
Bohr succeed in explaining the latter by assuming that the electron orbiting
the nucleus can occupy only a discrete set of orbits with angular momenta
proportional to $\hbar$. The constant $\hbar=h/2\pi\sim10^{-34}\mathrm{J\cdot
s}$, where $h$ is now known as the Planck constant, appeared in both cases as
somehow the `quantization unit', the value separating the quantized values
that energy or angular momentum are able to take.

Even though the physicists of the time tried to understand this quantization
of the physical magnitudes within the framework of classical physics, it was
soon realized that it would need a complete new theory to do it. The first
attempts to build such a theory (which actually worked for some particular
scenarios) were based on applying ad--hoc quantization rules to various
mechanical variables of systems, but with a complete lack of physical
interpretation for such rules \cite{Waerden68book}. However, between 1925 and
1927 the first real formulations of the needed theory were developed: the
\textit{wave mechanics} of Schr\"{o}dinger \cite{Schrodinger26} and the
\textit{matrix mechanics} of Heisenberg, Born and Jordan
\cite{Heisenberg25,Born25,Born26} (see \cite{Waerden68book} for English
translations), which received also independent contributions by Dirac
\cite{Dirac26}. Even though in both theories the quantization of various
observable quantities appeared naturally and in correspondence with
experiments, they seem completely different, at least until Schr\"{o}dinger
showed the equivalence between them both.

The new theory was later formalized mathematically using vector spaces by
Dirac \cite{Dirac30book} (though not very rigorously), and a little after by
von Neumann with full mathematical rigor using Hilbert spaces
\cite{vonNeumann32book} (\cite{vonNeumann55book} for an English version); they
developed the laws of \textit{quantum mechanics} basically as we know it now
\cite{Cohen77abook,Cohen77bbook,Greiner89abook,Greiner89bbook,Basdevant02book}.

\subsection{The axioms of quantum mechanics\label{Axioms}}

In this section we will introduce the basic postulates which describe how
quantum mechanics treats physical systems. Being the basic blocks that build
the theory, these axioms cannot be \textit{proved}; they can only be
formulated following \textit{plausibility arguments} based on the
\textit{observation} of physical phenomena and the \textit{connection} of the
theory with previous theories which are known to work in some limit. We will
try to motivate (and justify to a point) these axioms as much as possible.

Along these lines, the experimental evidence for the tendency of observable
physical quantities to be quantized at the microscopic level motivates the
first axiom:

\bigskip

\textbf{Axiom I. }$\lceil$Any physical observable quantity \textsf{A}
corresponds to a self--adjoint operator $\hat{A}$ acting on an abstract
Hilbert space; after a measurement of \textsf{A}, the only possible outcomes
are the eigenvalues of $\hat{A}$.$\rfloor$

\bigskip

The quantization of physical observables is therefore directly introduced in
the theory by this postulate. Note that it doesn't say anything about the
dimension $d$ of the Hilbert space corresponding to a given observable, and it
even leaves open the possibility of observables having a continuous spectrum,
not a discrete one. The problem of how to make the proper correspondence
between observables and self--adjoint operators will be addressed in an axiom
to come.

In the following we will use the name `observable' for both the physical
quantity \textsf{A} and its associated self--adjoint operator $\hat{A}$
indistinctly. Observables having pure discrete spectrum or pure continuous
spectrum will be referred to as \textit{countable} and \textit{continuous
observables}, respectively.

Now we introduce the second axiom. It follows from the following question: the
eigenvalues of an observable are the only values that can appear when
measuring it, but what about the statistics of such a measurement? We know a
class of operators in Hilbert spaces which act as probability distributions
for the eigenvalues of any self--adjoint operator, density operators. This
motivates the second axiom:

\bigskip

\textbf{Axiom II. }$\lceil$The state of the system is completely specified by
a density operator $\hat{\rho}$. When measuring an observable \textsf{A}
having a complete orthonormal set of eigenvectors $\left\{  |a_{j}%
\rangle\right\}  _{j=1,2,...,d}$ ($d$ might be infinite), it is associated to
the possible outcomes $\left\{  a_{j}\right\}  _{j=1,2,...,d}$ a probability
distribution $\left\{  p_{j}=\rho_{jj}\right\}  _{j=1,2,...,d}$ which
determines the statistics of the experiment.

Similarly, when measuring an observable \textsf{X} having a complete set of
generalized continuous eigenvectors $\left\{  |x\rangle\right\}  _{x\in%
\mathbb{R}
}$, the probability density function $P\left(  x\right)  =\rho\left(
x,x\right)  $ is associated to the possible outcomes $\{x\}_{x\in%
\mathbb{R}
}$ in the experiment.$\rfloor$

\bigskip

This postulate has deep consequences that we analyze now. Contrary to
classical mechanics (and intuition), even if the system is in a fixed state,
the value of an observable is in general not well defined; we can only specify
with what probability a given value of the observable will come out in a
measurement. Hence, this axiom proposes a change of paradigm, determinism must
be abandoned: theory is no longer able to predict with certainty the outcome
of a single experiment in which an observable is measured, but rather gives
the statistics that will be extracted after a large number of such experiments.

To be fair, there is a case in which the theory allows us to predict the
outcome of the measurement of an observable with certainty: When the system is
prepared such that its state is an eigenvector of the observable. This seems
much like when in classical mechanics the system is prepared with a given
value of its observables. One can show however that it is impossible to find a
common eigenvector to \textit{all} the available observables of a system, and
hence the difference between classical and quantum mechanics is that in the
later it is impossible to prepare the system in a state which would allow us
to predict with certainty the outcome of a measurement of each of its
observables. Let us try to elaborate on this in a more rigorous fashion.

Let us define the \textit{expectation value} of a given operator $\hat{B}$ as%
\begin{equation}
\langle\hat{B}\rangle\equiv\mathrm{tr}\{\hat{\rho}\hat{B}\}\text{.}%
\end{equation}
In the case of a countable observable $\hat{A}$ or a continuous observable
$\hat{X}$, this expectation value can be written in their own eigenbases as%
\begin{equation}
\langle\hat{A}\rangle=\sum_{j=1}^{d}\rho_{jj}a_{j}\text{ and }\langle\hat
{X}\rangle=\int_{-\infty}^{+\infty}dx\rho\left(  x,x\right)  x\text{,}%
\end{equation}
which correspond to the mean value of the outcomes registered in large number
of measurements of the observables. We define also the \textit{variance} of
the observable as the expectation value of the square of its
\textit{fluctuation operator }$\delta\hat{A}=\hat{A}-\langle\hat{A}\rangle$,
that is,%
\begin{equation}
V(A)\equiv\mathrm{tr}\left\{  \hat{\rho}(\delta\hat{A})^{2}\right\}
=\langle\hat{A}^{2}\rangle-\langle\hat{A}\rangle^{2}\text{,}%
\end{equation}
from which we obtain the \textit{standard deviation} or \textit{uncertainty
}as\textit{ }$\Delta A=V(A)^{1/2}$, which measures how much the outcomes of
the experiment deviate from the mean, and hence, somehow specifies how `well
defined' is the value of the observable \textsf{A}.

Note that the probability of obtaining the outcome $a_{j}$ when measuring
\textsf{A} can be written as the expectation value of the projection operator
$\hat{P}_{j}^{\mathsf{A}}=|a_{j}\rangle\langle a_{j}|$, that is $p_{j}%
=\langle\hat{P}_{j}^{\mathsf{A}}\rangle$. Similarly, the probability density
function associated to the possible outcomes $\{x\}_{x\in%
\mathbb{R}
}$ when measuring \textsf{X} can be written as $P\left(  x\right)
=\langle\hat{P}_{x}^{\mathsf{X}}\rangle$, where $\hat{P}_{x}^{\mathsf{X}%
}=|x\rangle\langle x|$.

Having written all these objects (probabilities, expectation values, and
variances) in terms of traces is really important, as traces can be evaluated
in any basis we want to work with, see Section \ref{LinearOperators}.

These axioms have one further unintuitive consequence. It is possible to prove
that irrespective of the state of the system, the following relation between
the variances of two non-commuting observables \textsf{A} and \textsf{B} is
satisfied:%
\begin{equation}
\Delta A\Delta B\geq\frac{1}{2}|\langle\lbrack\hat{A},\hat{B}]\rangle|\text{.}%
\end{equation}
According to this inequality, known as the \textit{uncertainty principle}
(which was first derived by Heisenberg), the only way in which the observable
\textsf{A} can be perfectly defined ($\Delta A\rightarrow0$) is by making
completely undefined observable \textsf{B} ($\Delta B\rightarrow\infty$), or
vice-versa. Hence, in the quantum formalism one cannot, in general, build a
state of the system such that all its observables are well defined, what is
completely opposite to our everyday experience. This property of quantum
mechanics will play a major role in this thesis.

Before going to the third axiom, let us comment one last thing. When the state
of the system is given by a pure density operator $\hat{\rho}=|\psi
\rangle\langle\psi|$, we say that the system is in a \textit{pure state}
$|\psi\rangle$ (if the density operator is mixed we then say that the system
is in a \textit{mixed state}). In this case, the expectation value of an
operator $\hat{B}$ takes the simple form $\langle\psi|\hat{B}|\psi\rangle$.
Moreover, the pure state can be expanded in the countable and continuous bases
of the observables $\hat{A}$ and $\hat{X}$ as%
\begin{equation}
\left\vert \psi\right\rangle =\sum_{j=1}^{d}\psi_{j}\left\vert a_{j}%
\right\rangle \text{ and }\left\vert \psi\right\rangle =\int_{-\infty
}^{+\infty}dx\psi\left(  x\right)  |x\rangle,
\end{equation}
respectively, being $\psi_{j}=\langle a_{j}|\psi\rangle$ and $\psi\left(
x\right)  =\langle x|\psi\rangle$. In this case, the probability distribution
for the discrete outcomes $\left\{  a_{j}\right\}  _{j=1,2,...,d}$ and the
probability density function for the continuous outcomes $\{x\}_{x\in
\mathcal{D}}$ are given by $\left\{  p_{j}=|\psi_{j}|^{2}\right\}
_{j=1,2,...,d}$ and $P\left(  x\right)  =|\psi\left(  x\right)  |^{2}$, respectively.

The introduction of the third axiom is motivated by the following fact. The
class of self--adjoint operators forms a real vector space with respect to the
addition of operators and the multiplication of an operator by a real number.
Using the commutator we can also build an operation which takes two
self--adjoint operators $\hat{A}$ and $\hat{B}$ to generate another
self--adjoint operator $\hat{C}=\mathrm{i}[\hat{A},\hat{B}]$, which in
addition satisfies all the properties required by a Lie product. Hence, even
if classical and quantum theories seem fundamentally different, it seems that
observables are treated similarly within their corresponding mathematical
frameworks: They are elements of a Lie algebra.

On the other hand, we saw that the generalized coordinates and momenta have a
particular mathematical structure in the Hamiltonian formalism, they are the
generators of the Heisenberg group. It seems then quite reasonable to ask for
the same in the quantum theory, so that at least in what concerns to
observables both theories are equivalent. This motivates the third axiom:

\bigskip

\textbf{Axiom III. }$\lceil$Consider a physical system described classically
within a Hamiltonian formalism by a set of generalized coordinates
$\mathbf{q}=\left\{  q_{j}\right\}  _{j=1}^{n}$ and momenta $\mathbf{p}%
=\left\{  p_{j}\right\}  _{j=1}^{n}$; within the quantum formalism, the
corresponding observables $\mathbf{\hat{q}}=\left\{  \hat{q}_{j}\right\}
_{j=1}^{n}$ and $\mathbf{\hat{p}}=\left\{  \hat{p}_{j}\right\}  _{j=1}^{n}$
satisfy the commutation relations%
\begin{equation}
\left[  \hat{q}_{j},\hat{p}_{l}\right]  =\mathrm{i}\hbar\delta_{jl}\text{
\ \ \ \ and \ \ \ }\left[  \hat{q}_{j},\hat{q}_{l}\right]  =\left[  \hat
{p}_{j},\hat{p}_{l}\right]  =0\text{.}\rfloor\label{CanComPosMom}%
\end{equation}

\bigskip

The constant $\hbar$ is included because, while the Poisson bracket $\left\{
q_{j},p_{l}\right\}  $ has no units, the commutator $\left[  \hat{q}_{j}%
,\hat{p}_{l}\right]  $ has units of action; that the proper constant is
$\hbar$ is seen only once the theory is compared with experiments.

We can now discuss how to build the self--adjoint operator corresponding to a
given observable. Suppose that in the Hamiltonian formalism the observable
\textsf{A} is represented by the phase space function $A\left(  \mathbf{q}%
,\mathbf{p}\right)  $. It seems quite natural to use then $A\left(
\mathbf{\hat{q}},\mathbf{\hat{p}}\right)  $ as the corresponding quantum
operator; however, this correspondence faces a lot of troubles derived from
the fact that while coordinates and momenta commute in classical mechanics,
they do not in quantum mechanics. For example, given the classical observable
$A=qp=pq$, we could be tempted to assign it any of the quantum operators
$\hat{A}_{1}=\hat{q}\hat{p}$ or $\hat{A}_{2}=\hat{p}\hat{q}$; these two
operators are different and they are not even self--adjoint, and hence, cannot
represent observables. One possible solution to this problem, at least for
observables with a series expansion, is to always symmetrize the classical
expressions with respect to coordinates and momenta, so that the resulting
operator is self--adjoint. Applied to our previous example, we should take
$\hat{A}=(\hat{p}\hat{q}+\hat{q}\hat{p})/2$ according to this rule. This
simple procedure leads to the correct results most of the times, and when it
fails (for example, if the classical observable doesn't have a series
expansion) it was proved by Groenewold \cite{Groenewold46} that it is possible
to make a faithful systematic correspondence between classical observables and
self--adjoint operators by using more sophisticated correspondence rules.

Of course, when the observable corresponds to a degree of freedom which is not
defined in a classical context (like spin), it must be built from scratch
based on observations and first principles. Nevertheless, we won't be working
with observables having no classical analog in this thesis.

Note that the commutation relations between coordinates and momenta makes them
satisfy the uncertainty relation $\Delta q\Delta p\geq\hbar/2$, and hence if
one of them is well defined in the system, the other must have statistics very
spread around the mean. We will examine this relation in depth when studying
the harmonic oscillator in the next section.

The three previous axioms have served to define the mathematical structure of
the theory and its relation to physical systems. We haven't said anything yet
about how quantum mechanics treats the evolution of the system. Just as with
the last axiom, it feels pretty reasonable to keep the analogy with the
Hamiltonian formalism, a motivation which comes also from the fact that, as
stated, quantum mechanics must converge to classical mechanics in some limit.
In the Hamiltonian formalism, observables evolve according to (\ref{HamObsEvo}%
), so that making the correspondence between the classical and quantum Lie
products as in the third axiom, we enunciate the fourth axiom (for simplicity,
we assume no explicit time dependence of observables):

\bigskip

\textbf{Axiom IV. }$\lceil$The evolution of an observable $\hat{A}$ is given
by%
\begin{equation}
\mathrm{i}\hbar\frac{d\hat{A}}{dt}=[\hat{A},\hat{H}],
\end{equation}
which is known as the Heisenberg equation, and where $\hat{H}$ is the
self--adjoint operator corresponding to the Hamiltonian of the system.$\rfloor
$

\bigskip

For the case of a time--independent Hamiltonian, this evolution equation
admits the explicit solution%
\begin{equation}
\hat{A}\left(  t\right)  =\hat{U}^{\dagger}\left(  t\right)  \hat{A}\left(
0\right)  \hat{U}\left(  t\right)  \text{, being }\hat{U}\left(  t\right)
=\exp[\hat{H}t/\mathrm{i}\hbar],
\end{equation}
a unitary operator called the \textit{evolution operator}. For time--dependent
Hamiltonians it is still possible to solve explicitly the Heisenberg equation,
but we won't worry about this case, as it won't appear throughout the thesis.

Note that within this formalism the state $\hat{\rho}$ of the system is fixed
in time, the observables are the ones which evolve. On the other hand, we have
seen that on what concerns to observations (experiments), only expectation
values of operators are relevant, and for an observable $\hat{A}$ at time $t$,
this can be written as%
\begin{equation}
\langle\hat{A}\left(  t\right)  \rangle=\mathrm{tr}\{\hat{\rho}\hat{A}\left(
t\right)  \}=\mathrm{tr}\{\hat{U}\left(  t\right)  \hat{\rho}\hat{U}^{\dagger
}\left(  t\right)  \hat{A}\left(  0\right)  \}\text{,}%
\end{equation}
where in the last equality we have used the cyclic property of the trace. This
expression shows that, instead of treating the observable as the evolving
operator, we can define a new state at time $t$ given by%
\begin{equation}
\rho\left(  t\right)  =\hat{U}\left(  t\right)  \hat{\rho}\left(  0\right)
\hat{U}^{\dagger}\left(  t\right)  ,
\end{equation}
while keeping fixed the operator. In differential form, this expression reads%
\begin{equation}
\mathrm{i}\hbar\frac{d\hat{\rho}}{dt}=[\hat{H},\hat{\rho}],
\end{equation}
which is known as the \textit{von Neumann equation}. When the system is in a
pure state $|\psi\rangle$, the following evolution equation is derived for the
state vector itself%
\begin{equation}
\mathrm{i}\hbar\frac{d}{dt}|\psi\rangle=\hat{H}|\psi\rangle\text{,}%
\end{equation}
which is known as the \textit{Schr\"{o}dinger equation}, from which the state
at time $t$ is found as $|\psi\left(  t\right)  \rangle=\hat{U}\left(
t\right)  |\psi\left(  0\right)  \rangle$.

Therefore, we have two different but equivalent evolution formalisms. In one,
which we shall call \textit{Heisenberg picture}, the state of the system is
fixed, while observables evolve according to the Heisenberg equation. In the
other, which we will denote by \textit{Schr\"{o}dinger picture}, observables
are fixed, while states evolve according to the von Neumann equation.

We can even define intermediate pictures in which both the state and the
observables evolve, the so-called \textit{interaction pictures}. To show how
this is done, let us denote by $\hat{A}_{\mathrm{S}}$ and $\hat{\rho
}_{\mathrm{S}}\left(  t\right)  $, an observable and the state of the system
in the Schr\"{o}dinger picture.

Suppose that we are in the Schr\"{o}dinger picture, so that the expectation
value of an observable \textsf{A}\ is written as $\mathrm{tr}\{\hat{\rho
}_{\mathrm{S}}(t)\hat{A}_{\mathrm{S}}\}$, and want to go to a new picture in
which both the state and the operator evolve; all we need to do is define a
unitary operator $\hat{U}_{\mathrm{c}}=\exp[\hat{H}_{\mathrm{c}}%
t/\mathrm{i}\hbar]$, with $\hat{H}_{\mathrm{c}}$ some self--adjoint operator,
and then a transformed state $\hat{\rho}_{\mathrm{I}}=\hat{U}_{\mathrm{c}%
}^{\dagger}\hat{\rho}_{\mathrm{S}}\hat{U}_{\mathrm{c}}$ and a transformed
observable $\hat{A}_{\mathrm{I}}=\hat{U}_{\mathrm{c}}^{\dagger}\hat
{A}_{\mathrm{S}}\hat{U}_{\mathrm{c}}$. This transformation leaves invariant
the expectation value, which can be evaluated as $\mathrm{tr}\{\hat{\rho
}_{\mathrm{I}}\left(  t\right)  \hat{A}_{\mathrm{I}}\left(  t\right)  \}$, but
now the evolution equations of the state and the observable read%
\begin{equation}
\mathrm{i}\hbar\frac{d\hat{\rho}_{\mathrm{I}}}{dt}=[\hat{H}_{\mathrm{I}}%
,\hat{\rho}_{\mathrm{I}}]\text{ \ \ \ \ and \ \ \ \ }\mathrm{i}\hbar
\frac{d\hat{A}_{\mathrm{I}}}{dt}=[\hat{A}_{\mathrm{I}},\hat{H}_{\mathrm{c}}],
\end{equation}
so that within this new picture states evolve according to the
\textit{interaction Hamiltonian} $\hat{H}_{\mathrm{I}}=\hat{U}_{\mathrm{c}%
}^{\dagger}\hat{H}\hat{U}_{\mathrm{c}}-\hat{H}_{\mathrm{c}}$, while
observables evolve according to the \textit{transformation Hamiltonian}
$\hat{H}_{\mathrm{c}}$.

The last axiom specifies how the theory accommodates dealing with composite
systems within its mathematical framework. Of course, a composition of two
systems is itself another system subject to the laws of quantum mechanics; the
question is how can we build it.

\bigskip

\textbf{Axiom V. }$\lceil$Consider two systems A and B with associated Hilbert
spaces $\mathcal{H}_{A}$ and $\mathcal{H}_{B}$; then, the state of the
composite system $\hat{\rho}_{AB}$ as well as its observables act onto the tensor
product Hilbert space $\mathcal{H}_{AB}=\mathcal{H}_{A}\otimes\mathcal{H}_{B}%
$.$\rfloor$

\bigskip

This axiom has the following consequence. Imagine that the systems \textit{A}
and \textit{B} interact during some time in such a way that they cannot be
described anymore by independent states $\hat{\rho}_{A}$ and $\hat{\rho}_{B}$
acting on $\mathcal{H}_{A}$ and $\mathcal{H}_{B}$, respectively, but by a
state $\hat{\rho}_{AB}$ acting on the joint space $\mathcal{H}_{AB}$. After
the interaction, system \textit{B} is kept isolated of any other system, but
system \textit{A} is given to an observer, who is therefore able to measure
observables defined in $\mathcal{H}_{A}$ only, and might not even know that
system \textit{A} is part of a larger system. The question is, is it possible
to reproduce the statistics of the measurements performed on system \textit{A}
with some state $\hat{\rho}_{A}$ defined in $\mathcal{H}_{A}$ only? This question
has a positive and \textit{unique} answer: this state is given by the
\textit{reduced density operator} $\hat{\rho}_{A}=\mathrm{tr}_{B}\{\hat{\rho
}_{AB}\}$, that is, by performing the partial trace respect system's
\textit{B} subspace onto the joint state.

\bigskip

These five axioms (together with the definitions of the previous sections)
define quantum mechanics as we use it throughout the thesis.

We would like to note finally, that in most of the textbooks about quantum
mechanics one can find two more axioms. The first one refers to how
multi--particle states have to be built: For indistinguishable integer
(half--integer) spin particles the state must be symmetric (antisymmetric)
with respect to the permutation of any two particles; we haven't incorporated
this axiom mainly because we don't use it. In any case it finds full
justification in the context of quantum field theory via the
\textit{spin--statistics theorem}, so we do not find it a true foundational axiom

The second one is more controversial \cite{Basdevant02book} (see also the
Appendix E of \cite{Galindo90book}): It states that if the value $a_{j}$ is
observed for observable \textsf{A} in an experiment, then immediately after
the measurement the state of the system \textit{collapses} to $|a_{j}\rangle$.
Even though this axiom leads to predictions in full agreement with
observations, it somehow creates an inconsistency in the theory because of the
following argument. According to Axiom IV the evolution of a closed system is
\textit{reversible} (unitary); on the other hand, the \textit{collapse} axiom
states that when the system is put in contact with a measurement device and an
observable is measured, the state of the system collapses to some other state
in a \textit{non-reversible} way. However, coming back to Axiom IV, the whole
measurement process could be described reversibly by considering, in addition
to the system's particles, the evolution of all the particles forming the
measurement device (or even the human who is observing the measurement
outcome!), and a Hamiltonian for the whole `observed system + measurement
device' system. Hence, it seems that, when including the collapse axiom,
quantum mechanics allows for two completely different descriptions of the
measurement process, one reversible and one irreversible, without giving a
clear rule for when to apply each. This is the sense in which there is an
inconsistency in the theory.

There are three main lines of thought regarding to how this inconsistency
might be solved. First, there are the ones who believe that the collapse is
real, and that, even though we still don't know it, there exists a rule
explaining under which circumstances one has to apply unitary evolution or the
collapse of the state. The second line of thought suggests that the collapse
should be derivable from unitary evolution according with some procedure yet
to be devised; for example, when the density of particles in the system
exceeds some value, one cannot expect to track the reversible evolution of
each single particle, and this \textquotedblleft missing
information\textquotedblright\ could give rise to the irreversible collapse. A
third possible approach states that the collapse is just an operationally
convenient way of describing measurements, but it is far from real; instead,
the measurement is described as a joint unitary transformation onto the system
and the measurement apparatus, leading to a final entangled state of these in
which the eigenstates of the system's observable are in one--to--one
correspondence with a set of macroscopic states of the measurement device
\cite{Basdevant02book}.

Real or not real,\ the collapse postulate offers the easiest successful way of
analyzing schemes involving measurements, and hence we apply it when needed.
Nevertheless, during practically all the thesis we consider only measurements
in which the observed system is destroyed after its observation, the so-called
\textit{destructive measurements}, so we won't need to worry about the state
of the system after the measurement. 

%% file: LinearStochasticFO.tex
In this Appendix we explain how to deal with one of the most simple examples
of an stochastic equation: A one--variable linear equation with additive
noise. This type of equation appears all along the thesis, see equations
(\ref{ProjLinLanBelowDOPO}) and (\ref{ProjLinLanBelowOPO}) for example, which
indeed match the general form
\begin{equation}
\dot{c}=-\lambda c+\Gamma\eta(\tau),
\end{equation}
where $c(\tau)$ is the stochastic variable, $\lambda$ is a positive real
parameter, $\Gamma$ might be complex, and $\eta(\tau)$ is a real noise with
zero mean and two--time correlation $\langle\eta(\tau)\eta(\tau^{\prime
})\rangle=\delta(\tau-\tau^{\prime})$.

The solution of the equation is readily found by making the variable change
$z\left(  \tau\right)  =c\left(  \tau\right)  \exp\left(  \lambda\tau\right)
$, what leads to $\dot{z}\left(  \tau\right)  =\Gamma\exp\left(  \lambda
\tau\right)  \eta(\tau)$, and hence to%
\begin{equation}
c\left(  \tau\right)  =c\left(  0\right)  e^{-\lambda\tau}+\Gamma\int
_{0}^{\tau}d\tau_{1}\eta\left(  \tau_{1}\right)  e^{\lambda\left(  \tau
_{1}-\tau\right)  },
\end{equation}
or in the $\tau\gg\lambda^{-1}$ limit%
\begin{equation}
c\left(  \tau\right)  =\Gamma\int_{0}^{\tau}d\tau_{1}\eta\left(  \tau
_{1}\right)  e^{\lambda\left(  \tau_{1}-\tau\right)  }.
\end{equation}
Note that in this limit the solution does not depend on its initial value, and
hence we can expect the process to be stationary in this limit
\cite{Mandel95book}.

The stochastic variable $c(\tau)$ has then zero mean in the stationary limit,
that is, $\left\langle c\left(  \tau\gg\lambda^{-1}\right)  \right\rangle =0$.
On the other hand, the two--time correlation function of $c(\tau)$ can be
written for $\tau\gg\lambda^{-1}$ as%
\begin{equation}
\left\langle c\left(  \tau\right)  c\left(  \tau^{\prime}\right)
\right\rangle =\Gamma^{2}e^{-\lambda\left(  \tau+\tau^{\prime}\right)  }%
\int_{0}^{\tau}d\tau_{1}\int_{0}^{\tau^{\prime}}d\tau_{2}\delta\left(
\tau_{1}-\tau_{2}\right)  e^{\lambda\left(  \tau_{1}+\tau_{2}\right)  }.
\end{equation}
Considering separately the cases $\tau^{\prime}>\tau$ and $\tau^{\prime}<\tau
$, this integral is easily carried out, yielding (again in the limit $\tau
\gg\lambda^{-1}$)%
\begin{equation}
\left\langle c\left(  \tau\right)  c\left(  \tau^{\prime}\right)
\right\rangle =\frac{\Gamma^{2}}{2\lambda}e^{-\lambda|\tau^{\prime}-\tau|},
\label{Ccorr}%
\end{equation}
where we see that this function depends only on the time difference
$|\tau^{\prime}-\tau|$, and hence $c(\tau)$ arrives indeed to a stationary
state for large enough times \cite{Mandel95book} (it is invariant under
changes of the time origin).

The quantity we are usually interested in is the correlation spectrum of
$c(\tau)$, which is finally evaluated in the stationary limit as the integral%
\begin{align}
\tilde{C}(\tilde{\Omega})=\int_{-\infty}^{+\infty}d\tau^{\prime
}e^{-i\tilde{\Omega}\tau^{\prime}}\left\langle c\left(  \tau\right)  c\left(
\tau+\tau^{\prime}\right)  \right\rangle \label{CorrSpectrum}=\frac{\Gamma^{2}}{2\lambda}\left[  \int_{0}^{+\infty}d\tau^{\prime
}e^{-(\lambda+i\tilde{\Omega})\tau^{\prime}}+\int_{-\infty}^{0}d\tau^{\prime
}e^{(\lambda-i\tilde{\Omega})\tau^{\prime}}\right]=\frac{\Gamma^{2}}{\lambda^{2}+\tilde{\Omega}^{2}}.
\end{align}
We make extensive use of this last result all along the thesis.

%% file: Lin2tmDOPOFO.tex
Following the linearization procedure we explained in \ref{Quantum2tmDOPO}
leads not to Eq. (\ref{LinLan}) directly, but to the following one%
\begin{equation}
\mathrm{i}\left(  \mathcal{G}\mathbf{b}-2\rho\mathbf{w}_{0}\right)
\dot{\theta}+\mathbf{\dot{b}=}\mathcal{L}\mathbf{b}+g\mathcal{K}\left(
\theta\right)  \boldsymbol{\zeta}\left(  \tau\right)  \text{,}%
\label{FullLinLan}%
\end{equation}
where%
\begin{align*}
\mathcal{K}\left(  \theta\right)   &  =\operatorname{diag}\left(
e^{\mathrm{i}\theta},e^{-\mathrm{i}\theta},e^{-\mathrm{i}\theta}%
,e^{\mathrm{i}\theta}\right)  \text{,}\\
\mathcal{G} &  =\operatorname{diag}\left(  -1,1,1,-1\right)  ,
\end{align*}
and the rest of vectors and symbols were defined in the corresponding section.
Note that the differences between this system of equations and the one used
throughout the thesis (\ref{LinLan}) are the matrix $\mathcal{K}\left(
\theta\right)  $ and the $\mathrm{i}\mathcal{G}\mathbf{b}\dot{\theta}$ term.
The latter is of order $g^{2}$ (as the $b$'s and $\dot{\theta}$ are of order
$g$), and hence it can be simply removed within the linearized theory.

Understanding why $\mathcal{K}\left(  \theta\right)  $ can be removed from the
linearized equations is a little more involved. Projecting these equations
onto the eigensystem of $\mathcal{L}$ (\ref{Eigensystem}) and defining the
vector $\mathbf{c}=\operatorname{col}\left(  \theta,c_{1},c_{2},c_{3}\right)
$ leads to the following system of equations (remember that we set $c_{0}=0$)
\begin{equation}
\mathbf{\dot{c}=-}\Lambda\mathbf{c}+g\mathcal{BR}\left(  \theta\right)
\boldsymbol{\eta}\left(  \tau\right)  \label{cLinLan}%
\end{equation}
with%
\begin{align*}
\Lambda &  =2\operatorname{diag}\left(  0,1,\sigma-1,\sigma\right)  ,\\
\mathcal{B} &  =\operatorname{diag}\left(  1/2\rho,\mathrm{i},1,1\right)  ,\\
\mathcal{R}\left(  \theta\right)   &  =\mathcal{R}_{1,3}\left(  \theta\right)
\mathcal{R}_{2,4}\left(  -\theta\right)  ,
\end{align*}
being $\mathcal{R}_{i,j}\left(  \theta\right)  $ the two--dimensional rotation
matrix of angle $\theta$ acting on the $i-j$ subspace, and where the
components of vector $\boldsymbol{\eta}\left(  \tau\right)  $ are real,
independent noises satisfying the usual statistical properties
(\ref{RealGaussStat}). Now, we will prove that this system, and the same with
$\mathcal{R}\left(  \theta=0\right)  $ are equivalent within the linearized
theory, and hence (\ref{FullLinLan}) and (\ref{LinLan}) are equivalent too.

To show this, we just write the Fokker--Planck equation (\ref{FP})
corresponding to this stochastic system (which we remind is in Stratonovich
form), whose drift vector and diffusion matrix are found to be%
\begin{equation}
\mathcal{\vec{A}}=\Lambda\mathbf{c}+\frac{g^{2}}{4\rho}\operatorname{col}%
\left(  0,0,1,0\right)  ,
\end{equation}
and%
\begin{equation}
\mathcal{D}=g^{2}\mathcal{BB}^{T},
\end{equation}
respectively. Note that in the last equation we have used that $\mathcal{R}%
\left(  \theta\right)  $ is an orthogonal matrix.

The proof is completed by writing the stochastic system corresponding to this
Fokker-Planck equation up to the linear order in $g$, which reads%
\begin{equation}
\mathbf{\dot{c}=-}\Lambda\mathbf{c}+g\mathcal{B}\boldsymbol{\eta}\left(
\tau\right)  ,
\end{equation}
corresponding to (\ref{cLinLan}) with $\mathcal{R}\left(  \theta=0\right)  $
as we wanted to prove.

Hence, removing $\mathcal{K}\left(  \theta\right)  $ and neglecting the
$i\mathcal{G}\mathbf{b}\dot{\theta}$ term\ from (\ref{FullLinLan}) doesn't
change its equivalent Fokker-Planck equation within the linearized
description, and thus equation (\ref{LinLan}) must lead to the same
predictions as (\ref{FullLinLan}).

%% file: SinCosCorrFO.tex
In this appendix we evaluate the correlation functions $S\left(  \tau_{1}%
,\tau_{2}\right)  =\left\langle \sin\theta\left(  \tau_{1}\right)  \sin
\theta\left(  \tau_{2}\right)  \right\rangle $ and $C\left(  \tau_{1},\tau
_{2}\right)  =\left\langle \cos\theta\left(  \tau_{1}\right)  \cos
\theta\left(  \tau_{2}\right)  \right\rangle $ starting from%
\begin{equation}
\theta(\tau)=\sqrt{D}\int_{0}^{\tau}d\tau^{\prime}\eta_{0}\left(  \tau
^{\prime}\right)  ,
\end{equation}
which is found by integrating equation (\ref{ThetaEvo}), and assuming that
$\theta(0)=0$ at any stochastic realization, as explained in the main text.

The correlation functions are easy to find by noticing that, as $\eta_{0}%
(\tau)$ is a Gaussian noise, $\theta(\tau)$ is a Gaussian variable, that is,
all its moments are determined from the first and second ones. Under these
conditions, it is straightforward to show (for example by performing a series
expansion) that the following property holds%
\begin{equation}
\left\langle e^{\pm\mathrm{i}\theta(\tau)}\right\rangle _{P}=e^{-\left\langle
\theta^{2}(\tau)\right\rangle _{P}/2},
\end{equation}
or, as the combinations $\theta(\tau_{1})\pm\theta(\tau_{2})$ are also
Gaussian variables,%
\begin{equation}
\left\langle e^{\pm\mathrm{i}\left[  \theta(\tau_{1})\pm\theta(\tau
_{2})\right]  }\right\rangle _{P}=e^{-\left\langle \left[  \theta(\tau_{1}%
)\pm\theta(\tau_{2})\right]  ^{2}\right\rangle _{P}/2}.
\end{equation}

Now, given that%
\begin{equation}
\left\langle \theta(\tau_{1})\theta(\tau_{2})\right\rangle _{P}=D\int
_{0}^{\tau_{1}}d\tau\int_{0}^{\tau_{2}}d\tau^{\prime}\delta(\tau-\tau^{\prime
})=D\min(\tau_{1},\tau_{2}),
\end{equation}
we get%
\begin{equation}
\left\langle \left[  \theta(\tau_{1})\pm\theta(\tau_{2})\right]
^{2}\right\rangle =D\left[  \tau_{1}+\tau_{2}\pm2\min(\tau_{1},\tau
_{2})\right]  ,
\end{equation}
and finally%
\begin{align}
S\left(  \tau_{1},\tau_{2}\right)=\left\langle \sin\theta\left(
\tau_{1}\right)  \sin\theta\left(  \tau_{2}\right)  \right\rangle =-\frac
{1}{4}\left\langle e^{\mathrm{i}\left[  \theta(\tau_{1})+\theta(\tau
_{2})\right]  }\right\rangle _{P}+\frac{1}{4}\left\langle e^{\mathrm{i}\left[
\theta(\tau_{1})-\theta(\tau_{2})\right]  }\right\rangle _{P}+\mathrm{c.c.}=e^{-\frac{D}{2}(\tau_{1}+\tau_{2})}\sinh[D\min(\tau_{1},\tau_{2})],
\label{SinCorr}%
\end{align}
and%
\begin{align}
C\left(  \tau_{1},\tau_{2}\right)=\left\langle \sin\theta\left(
\tau_{1}\right)  \sin\theta\left(  \tau_{2}\right)  \right\rangle =\frac{1}%
{4}\left\langle e^{\mathrm{i}\left[  \theta(\tau_{1})+\theta(\tau_{2})\right]
}\right\rangle _{P}+\frac{1}{4}\left\langle e^{\mathrm{i}\left[  \theta
(\tau_{1})-\theta(\tau_{2})\right]  }\right\rangle _{P}+\mathrm{c.c.}=e^{-\frac{D}{2}(\tau_{1}+\tau_{2})}\cosh[D\min(\tau_{1},\tau_{2})].
\label{CosCorr}%
\end{align}

%% file: Numerical2tmDOPOFO.tex
In this appendix we want to briefly summarize the details concerning the
numerical simulation of the Langevin equations (\ref{ScaledLangevin2tmDOPO})
which model the 2tmDOPO.

The first important property of these equations is that, irrespective of the
initial conditions, the amplitudes corresponding to opposite orbital angular
momentum modes become complex--conjugate after a short transitory time, i.e.,
$\left(  \beta_{-1},\beta_{-1}^{+}\right)  \rightarrow\left(  \beta_{+1}%
^{\ast},\left[  \beta_{+1}^{+}\right]  ^{\ast}\right)  $. Hence, if the
initial conditions are chosen so this property is already satisfied, we can be
sure that these amplitudes will remain complex--conjugate during the
evolution. In particular, we have chosen the above threshold stationary
solution (\ref{2tmDOPOabove}) with $\theta=0$ as the initial condition. Under
these conditions, the 6 Langevin equations (\ref{ScaledLangevin2tmDOPO}) get
reduced to the following 4 (which we write in matrix form):%
\begin{equation}
\boldsymbol{\dot{\beta}}=\mathbf{A}\left(  \boldsymbol{\beta}\right)
+\mathcal{B}\left(  \boldsymbol{\beta}\right)  \cdot\boldsymbol{\zeta}\left(
\tau\right)  \text{,}\label{ReducedEqs}%
\end{equation}
with%
\begin{align}
\boldsymbol{\beta} &  =%
\begin{pmatrix}
\beta_{0}\\
\beta_{0}^{+}\\
\beta_{+1}\\
\beta_{+1}^{+}%
\end{pmatrix}
,\text{ }\boldsymbol{\zeta}\left(  \tau\right)  =%
\begin{pmatrix}
0\\
0\\
\zeta\left(  \tau\right)  \\
\zeta^{+}\left(  \tau\right)
\end{pmatrix}
,\\
\mathbf{A}\left(  \boldsymbol{\beta}\right)   &  =%
\begin{pmatrix}
\sigma-\beta_{0}-\left\vert \beta_{+1}\right\vert ^{2}\\
\sigma-\beta_{0}^{+}-\left\vert \beta_{+1}^{+}\right\vert ^{2}\\
-\beta_{+1}+\beta_{0}\left[  \beta_{+1}^{+}\right]  ^{\ast}\\
-\beta_{+1}^{+}+\beta_{0}^{+}\beta_{+1}^{\ast}%
\end{pmatrix}
,\nonumber\\
\mathcal{B}\left(  \boldsymbol{\beta}\right)   &  =g\operatorname{diag}\left(
0,0,\sqrt{\beta_{0}},\sqrt{\beta_{0}^{+}}\right)  .\nonumber
\end{align}

In order to solve numerically these equations we use the semi--implicit
algorithm developed in \cite{Drummond91}. This algorithm is a
finite--differences based method in which the total integration time
$\tau_{\mathrm{end}}$ (the integration is supposed to begin always at $\tau
=0$) is divided in $N$ segments, creating hence a lattice of times $\left\{
\tau_{n}\right\}  _{n=0,1,...,N}$ separated by time steps $\Delta\tau
=\tau_{\mathrm{end}}/N$. Then, a recursive algorithm starts in which the
amplitudes at time $\tau_{n}$, say $\boldsymbol{\beta}_{n}$, are found from
the amplitudes $\boldsymbol{\beta}$$^{n-1}$ at an earlier time $\tau_{n-1}$
from%
\begin{equation}
\boldsymbol{\beta}^{n}=\boldsymbol{\beta}^{n-1}+\Delta\tau\mathbf{A}\left(
\boldsymbol{\tilde{\beta}}^{n}\right)  +\mathcal{B}\left(  \boldsymbol{\tilde
{\beta}}^{n}\right)  \cdot\mathbf{W}^{n},
\end{equation}
where $\boldsymbol{\tilde{\beta}}$$^{n}$ is an approximation to the amplitudes
at the mid--point between $\tau_{n-1}$ and $\tau_{n}$ (hence the name
\textquotedblleft semi--implicit\textquotedblright\ for the algorithm) and the
components of $\mathbf{W}^{n}$ are independent discrete noises $W_{j}^{n}$
with null mean and satisfying the correlations%
\begin{equation}
\left\langle W_{j}^{m},\left[  W_{k}^{n}\right]  ^{\ast}\right\rangle
=\Delta\tau\delta_{mn}\delta_{jk}.
\end{equation}

The mid--point approximation is found from the following iterative algorithm%
\begin{equation}
\boldsymbol{\tilde{\beta}}^{n,p}=\boldsymbol{\beta}^{n-1}+\frac{1}{2}\left[
\Delta\tau\mathbf{A}\left(  \boldsymbol{\tilde{\beta}}^{n,p-1}\right)
+\mathcal{B}\left(  \boldsymbol{\tilde{\beta}}^{n,p-1}\right)  \cdot
\mathbf{W}^{n}\right]  ,
\end{equation}
where $\boldsymbol{\tilde{\beta}}$$^{n,0}=$$\boldsymbol{\beta}$$^{n-1}$, being
$p$ the iteration index (two iterations are carried in all our simulations),
while the discrete noises can be simulated at any step as \cite{Fox88}%
\begin{equation}
W_{j}^{n}=\sqrt{\Delta\tau}\left[  r\left(  z_{j},z_{j}^{\prime}\right)
+\mathrm{i}r\left(  y_{j},y_{j}^{\prime}\right)  \right]  ,
\end{equation}
with%
\begin{equation}
r\left(  z,z^{\prime}\right)  =\sqrt{-\log z}\cos\left(  2\pi z^{\prime
}\right)  ,
\end{equation}
being $z_{j}$, $z_{j}^{\prime}$, $y_{j}$ and $y_{j}^{\prime}$ independent
random numbers uniformly distributed along the interval $\left[  0,1\right]  $.

This algorithm allows us to simulate one stochastic trajectory. Then, by
repeating it $\Sigma$ times, the stochastic average of any function can be
approximated by the arithmetic mean of its values evaluated at the different
stochastic trajectories.